\documentclass[aps,prx,showpacs,notitlepage,twocolumn,superscriptaddress,nofootinbib,preprintnumbers,floatfix,longbibliography,10pt]{revtex4-2}

\usepackage{array}

\usepackage[T1]{fontenc}
\usepackage{bbm}
\usepackage{bm}
\usepackage{mathrsfs}
\usepackage{siunitx}

\usepackage{epsfig}
\usepackage{multirow}
\usepackage{widetable}
\usepackage{afterpage}

\usepackage{soul,color}
\usepackage{graphicx}

\graphicspath{{./FIGURES/}}
\usepackage{amsmath}

\usepackage{braket}
\usepackage{enumerate}
\usepackage{appendix}
\usepackage{makecell}
\usepackage{diagbox}
\usepackage{natbib}
\usepackage{textcase}

\usepackage{xcolor}
\definecolor{verylightgray}{gray}{0.97}

\usepackage[commandnameprefix=always]{changes}
\usepackage{complexity}

\usepackage{amsfonts}
\usepackage{dsfont}
\usepackage{amsthm}
\usepackage[figuresright]{rotating}
\usepackage{xcolor}
\usepackage{amssymb}
\usepackage{amsmath}
\usepackage{dcolumn}
\usepackage{physics}
\usepackage{float}
\usepackage{bm}
\usepackage{verbatim}
\usepackage[normalem]{ulem}
\usepackage[ruled,vlined,linesnumbered]{algorithm2e}
\usepackage{setspace}
\usepackage{stackengine}
\usepackage{qcircuit}

\newcommand{\subsubsubsection}[1]{{\textbf{#1}}}

\usepackage{fancyhdr}
\usepackage{longtable,tabularx}
\usepackage{url}

\fancypagestyle{refstyle}{
    \fancyhf{}

    \fancyhead[LO,RE]{References} 
    \fancyhead[LE,RO]{\thepage}

}

\usepackage{hyperref}
\hypersetup{colorlinks,linkcolor=blue,anchorcolor=blue,citecolor=blue,urlcolor=blue}

\usepackage{chngcntr}
\counterwithin{table}{section}

\usepackage{mdframed}
\mdfsetup{skipabove=\topskip,skipbelow=\topskip}
\global\mdfdefinestyle{exampledefault}{
	linecolor=black,linewidth=1.5pt,
	leftmargin=1cm,rightmargin=1cm
}

\newtheorem{cbox}{{\bf Box}}[section]
\newcommand{\abox}[2]{\medskip\medskip\noindent\fcolorbox{black}{verylightgray}{\begin{minipage}[h]{0.469\textwidth}
	\begin{cbox}{\bf #1} #2\end{cbox}\end{minipage}}\medskip\medskip}

\newcommand{\nn}{\nonumber}

\newcommand{\kcs}[1]{{\color[RGB]{148, 3, 252}{Kevin: #1}}}

\newcommand{\YaleQI}{Yale Quantum Institute, Yale University, New Haven, CT 06511, USA}
\newcommand{\YaleAP}{Departments of Applied Physics and Physics, Yale University, New Haven, CT 06520, USA}
\newcommand{\MITPh}{Department of Physics, Massachusetts Institute of
Technology, Cambridge, MA 02139, USA}
\newcommand{\MITEECS}{Department of Electrical Engineering and Computer Science, Massachusetts Institute of Technology, Cambridge, MA 02139, USA}
\newcommand{\NCSUECE}{Department of Electrical and Computer Engineering,\\ North Carolina State University, Raleigh, NC 27606, USA}
\newcommand{\NCSUCS}{Department of Computer Science, North Carolina State University, Raleigh, NC 27606, USA}
\newcommand{\UWM}{Department of Chemistry, University of Wisconsin-Madison, Madison, WI 53706, USA}
\newcommand{\UT}{Department of Computer Science, University of Toronto, Toronto, ON M5S1J7, Canada}
\newcommand{\JQI}{Joint Quantum Institute and QuICS, NIST/University of Maryland, College Park, MD 20742 USA}
\newcommand{\BNL}{Brookhaven National Laboratory, Upton, NY 11973, USA}
\newcommand{\PNNL}{Pacific Northwest National Laboratory, Richland WA, USA}
\newcommand{\CIAR}{Canadian Institute for Advanced Research, Toronto ON, Canada}

\usepackage{orcidlink}

\begin{document}

\title{Hybrid Oscillator-Qubit Quantum Processors:\\
Instruction Set Architectures, Abstract Machine Models, and Applications}

\author{Yuan Liu\orcidlink{0000-0003-1468-942X}}\thanks{These authors contributed equally (ordered alphabetically).}
\affiliation{\MITPh}
\affiliation{\NCSUECE}
\affiliation{\NCSUCS}

\author{Shraddha Singh\orcidlink{0000-0002-4921-1410}}\thanks{These authors contributed equally (ordered alphabetically).}
\affiliation{\YaleQI}
\affiliation{\YaleAP}

\author{Kevin C. Smith\orcidlink{0000-0002-2397-1518}}\thanks{These authors contributed equally (ordered alphabetically).}
\affiliation{\BNL}
\affiliation{\YaleQI}
\affiliation{\YaleAP}

\author{Eleanor Crane\orcidlink{0000-0002-2752-6462}}
\affiliation{\MITPh}

\author{John M. Martyn\orcidlink{0000-0002-4065-6974}}
\affiliation{\MITPh}

\author{Alec Eickbusch\orcidlink{0000-0001-5179-0215}
}\thanks{Present address: Google Quantum AI, Santa Barbara, CA}
\affiliation{\YaleQI}
\affiliation{\YaleAP}

\author{Alexander Schuckert\orcidlink{0000-0002-9969-7391}}
\affiliation{\JQI}

\author{Richard D. Li\orcidlink{0009-0001-3361-3431}}
\affiliation{\YaleQI}
\affiliation{\YaleAP}

\author{Jasmine Sinanan-Singh\orcidlink{0000-0002-8637-1413}}
\affiliation{\MITPh}

\author{Micheline B. Soley\orcidlink{0000-0001-7973-2842}}
\affiliation{\UWM}

\author{Takahiro Tsunoda\orcidlink{0000-0002-0869-2991}}
\affiliation{\YaleQI}
\affiliation{\YaleAP}

\author{Isaac L. Chuang\orcidlink{0000-0001-7296-523X}}
\affiliation{\MITPh}
\affiliation{\MITEECS}

\author{Nathan Wiebe \orcidlink{0000-0001-6047-0547}}
\affiliation{\UT}
\affiliation{\CIAR}
\affiliation{\PNNL}

\author{Steven M. Girvin\orcidlink{0000-0002-6470-5494}}\thanks{Corresponding author.}
\affiliation{\YaleQI}
\affiliation{\YaleAP}

\begin{abstract}

This tutorial offers a pedagogical guide to hybrid quantum processors that integrate discrete-variable (DV) qubits and continuous-variable (CV) oscillators. Aimed at computer scientists, engineers, and physicists, it provides an overview of the experimental, algorithmic, and architectural aspects of this novel and rapidly developing hardware model.   Experimental realizations of this model include superconducting, trapped ion, and neutral atom platforms. By combining DV and CV components, hybrid oscillator-qubit processors enable a powerful new paradigm that offers complementary strengths for quantum control, error correction, computation, and simulation.

Working towards the goal of a full-stack system connecting applications to CV-DV hardware, we define and formulate Abstract Machine Models and Instruction Set Architectures.  These essential abstractions enable co-design of hardware and software, and resource estimation for exploring the potential of current and future hardware for computational and simulation tasks.

 Using these abstractions, we present both new and existing examples that illustrate the benefits of hybrid CV-DV processors relative to traditional DV-only hardware in computation as well as quantum simulation of physical models. Examples include
 algorithms for transferring states between DV and CV systems, performing the quantum Fourier transform, 
 and simulation of lattice gauge theories. Relative to qubit-only hardware, the bosonic degrees of freedom natively available in hybrid architectures can substantially reduce the circuit complexity of simulations for physical models containing bosons. 
 A key technique is the extension of quantum signal processing ideas to CV-DV systems.

This work is intended to serve as a timely and comprehensive guide to this relatively unexplored yet promising approach to quantum computation and to provide  

a road map to guide future development.

\end{abstract}

\maketitle

\makeatletter

\def\l@subsubsection#1#2{}
\makeatother

\tableofcontents

\newpage

\section{Introduction}
\label{sec:introduction}
The ability to harness quantum resources in physical systems has revolutionary potential for computation, information processing, and communication \cite{Nielsen_Chuang,PRXQuantum.2.017001}. Most work to date has largely focused on leveraging two-state, discrete-variable (DV) systems acting as qubits for quantum computation. These include nuclear \cite{vandersypen2005nmr,jones2024controllingNMR} and electron spins \cite{dijkema2023twoqubit}, pairs of levels in neutral atoms \cite{Jaksch2000,Schlosser2001,Bluvstein2022,Graham2022}, ions \cite{Cirac1995,Monroe1995,RevModPhys.75.281,HAFFNER2008155,HomeCompleteMethodsSetIonTrapQC,MonroeKimScalingIonTrapQC,bruzewicz2019trapped,Quantinuum_moses2023race,hemery2023,Katz_2023_prxq,Iqbal2024} and color centers \cite{ColorCenters10.1063/5.0007444}, as well as synthetic atoms such as quantum dots \cite{RevModPhys.95.025003} and superconducting qubits \cite{Reagor2013,Reagor_memory_2016,PhysRevApplied.13.034032,
Rosenblum_TensofMilliseconds_PRXQuantum.4.030336,YYGao_QIP_bosonic_cQED,BlaiscQEDReviewRMP2020,Blais2020}, and various possible hybrid systems 
\cite{RevModPhys.85.623,andersen2015hybrid}. 
Separately, quantum resources that fundamentally differ from two-level spin systems have been proposed for use, including bosons~\cite{LloydQuantumComputation1999,dutta2025solvingconstrainedoptimizationproblems} and fermions \cite{Bravyi2002,obrien2018majorana,GonzlezCuadra2023,rad2024,schuckert2024fermionqubitfaulttolerantquantumcomputing}. Of particular importance is the fact that a single bosonic mode, also known as a harmonic oscillator, has a countable infinity of states and can therefore be characterized by a continuous variable (CV), in stark contrast to both qubits and fermions. 

In this tutorial, we focus our discussion on \emph{hybrid quantum CV-DV systems} composed of both oscillators and qubits, a paradigm that offers great potential for quantum computation as well as quantum simulation of physical models of interacting bosons (e.g., the boson Hubbard model \cite{Girvin_Quantum_Fluids_of_Light}) or bosons coupled to spins or fermions (e.g., lattice gauge theories \cite{C2QA-LGTpaper} and  polaritonic chemistry \cite{chiari2025abinitiopolaritonicchemistry}).
Thus far, a systematic discussion of such systems has not been established in the literature. This is largely due to several theoretical challenges. Firstly, the physics of bosonic modes is very different from qubits, and the creation and annihilation operators (or equivalently the position and momentum operators) used to describe the Hamiltonian of oscillators satisfy completely different commutation relations than the familiar Pauli matrices in the qubit case. This can make it challenging to treat oscillators and qubits in the same mathematical framework for quantum computation, while hinting that bosonic modes can be a resource of considerable computational power. Secondly, going beyond the physical systems to a more abstract computer science point of view, answers to crucial questions have been lacking.  How should such hybrid quantum CV-DV processors be programmed?  How can their computational power be reasoned about?  How can the required resources for such processors be estimated?  

While the first set of challenges is more fundamental, the key to overcoming the computer science challenges highlighted above is to construct \emph{abstract machine models} (AMMs) and corresponding \emph{universal instruction set architectures} (ISAs) for such hybrid processors. An ISA \cite{hennessy2012computer} contains an inventory of the discrete set of fundamental operations and measurements that are possible in the hardware. As is well-known on the classical side, AMMs such as Turing machines, random access machines, finite automata, and even lambda calculus \cite{davis1994computability} facilitate formal reasoning about the computational power of classical computers (e.g., the computability of specified functions such as the halting function). For qubit-only quantum processors, AMMs and ISAs have been useful for quantifying the quantum advantage of qubit-based quantum computers \cite{beverland2022assessing,bravyi2022future}.
Similarly, the establishment of AMMs and ISAs for hybrid CV-DV quantum processors will be crucial for quantifying their computational power and performing resource estimation 
as well as for co-design of algorithms, software, and hardware~\cite{hall1992microprocessors, stallings2003computer} to demonstrate the potential quantum advantage of not only quantum versus classical computers but also hybrid oscillator-qubit versus qubit-only quantum computers.

\if 0
The goals of the present work are to:
\begin{itemize}
\item Provide a pedagogical introduction to hybrid CV-DV systems and their applications.
\item Define several distinct AMMs that present different abstractions of hybrid hardware to the compiler layer of the full software stack. 
\item Define several distinct and experimentally realistic ISAs based on the current state-of-the-art and show how they provide universal control of hybrid CV-DV systems.
\item Extend the concept of Quantum Signal Processing \cite{GrandUnificationAlgos} to hybrid CV-DV systems \cite{sinanansingh2023singleshot} and show how to use this framework to develop compilation techniques useful for mapping subroutines into different ISAs, a necessary ingredient for high-level computation and simulation applications.
\item Present compilation techniques for bosonic quantum error correction.
\item Describe efficient techniques for state and process tomography in hybrid CV-DV systems.
\item Discuss a variety of physical models that are well-suited for quantum simulation using hybrid oscillator-qubit processors.
\item Summarize the current state of theory and experiment and open challenges in the field.
\end{itemize}
\fi

The goal of the present work is to provide a compelling tutorial overview of hybrid CV-DV quantum computation through a complete, bottom-up picture spanning seven main layers, from physical mechanisms through models, architectures, programming languanges \& compilation, and algorithms \& applications:

\begin{itemize}
\item {\bf Physical Mechanisms}: foundational physical primitives and formalism for hybrid quantum CV-DV systems and their physical hardware realizations.
\item {\bf Abstract Machine Models}:  theoretical abstractions of hybrid quantum CV-DV hardware allowing exploration of design decisions, bridging between computational complexity and architecture design.
\item {\bf Instruction Set Architectures}: hardware-software interface that allows translation of algorithms into well-defined primitive and abstract operations.

\item {\bf Programming Languages and Compilation}: languages for algorithmic design and techniques to map these instructions onto circuits executable within CV-DV ISAs, including
\begin{itemize}
\item Analytic techniques for compiling multi-qubit and multi-oscillator entangling gates,
\item Extension of quantum signal processing to hybrid CV-DV systems, and
\item Methods for realizing bosonic quantum error correction through compilation. \end{itemize}
\item {\bf Algorithms and Applications}: computational problems (described in a high-level programming language) naturally compilable into this model, their resource requirements, and debugging procedures, including such examples as:

\begin{itemize}
\item Quantum state transfer between CV and DV systems, with application to performing the quantum Fourier transform,
\item Efficient techniques for state and process tomography in hybrid CV-DV systems.
\item Quantum simulation of physical models involving both oscillators and qubits, and
\end{itemize}
\end{itemize}

\subsection{Outline of this Manuscript}
\label{ssec:intro_outline}

The remainder of this paper is structured as follows. Sec.~\ref{sec:architecture} presents the hybrid CV-DV architecture discussing the seven layers of abstraction in detail.  We then dive deeper into the theoretical aspects and experimental prospects for each architecture layer. Beginning with the physical layer, in Sec.~\ref{sec:basics} we provide a pedagogical introduction to the representation of bosonic states intended for non-physicists who are familiar with qubits but not oscillators.
In Sec.~\ref{sec:gates}, we present the hierarchy of bosonic and hybrid gates beginning with simple Gaussian operations on bosonic modes (analogous to Clifford operations on qubits), followed by non-Gaussian hybrid oscillator-qubit gates, and ending with descriptions of various universal instruction sets. In Sec.~\ref{sec:bosonic-QEC}, we give an overview of quantum error correction in hybrid CV-DV processors. In Sec.~\ref{sec:compilation}, we present exact and approximate analytic and numerical compilation and circuit synthesis techniques, including the extension of the quantum signal processing framework to hybrid CV-DV systems. In Sec.~\ref{sec:apps}, we leverage the compilation techniques developed in the preceding section for small-scale applications including useful computational subroutines and quantum simulation of important physical models. In Sec.~\ref{sec:advantages}, we discuss the advantages of oscillator-qubit hardware over qubit-only hardware and, en-route, introduce novel compilation methods to simulate native oscillator gates in the latter setting.
Finally, Sec.~\ref{sec:ConclusionsOutlook} presents conclusions, outlooks, and open research problems.
This is followed by a series of technical appendices on physical implementations and microscopic Hamiltonians, realistic noise models, measurement and tomography, circuit synthesis techniques, universality and cross-compilation of different ISAs, and other details relevant to the main text.

{\em About vocabulary and notation}: Throughout the text, we use the terms bosonic mode, qumode, oscillator, resonator, and cavity mode interchangeably.  To indicate the identity and the Pauli matrices, we use both the physics notation $\sigma_0,\sigma_x,\sigma_y,\sigma_z$ and the quantum computer science notation $I,X,Y,Z$. It is also important to note that we use the Nielsen and Chuang \cite{Nielsen_Chuang} convention that $|0\rangle$ represents the excited state of the qubit and $|1\rangle$ represents the ground state.  This convention is carefully defined in Sec.~\ref{sec:review-qubits}. For clarity, we often use $|\cdot\rangle_\mathrm{osc}$ to distinguish the oscillator subspace from the qubit subspace.

\subsection{Reader's Guide}
\label{ssec:readersguide}

\begin{figure}[htb]
   \centering
  \includegraphics[width=\linewidth]{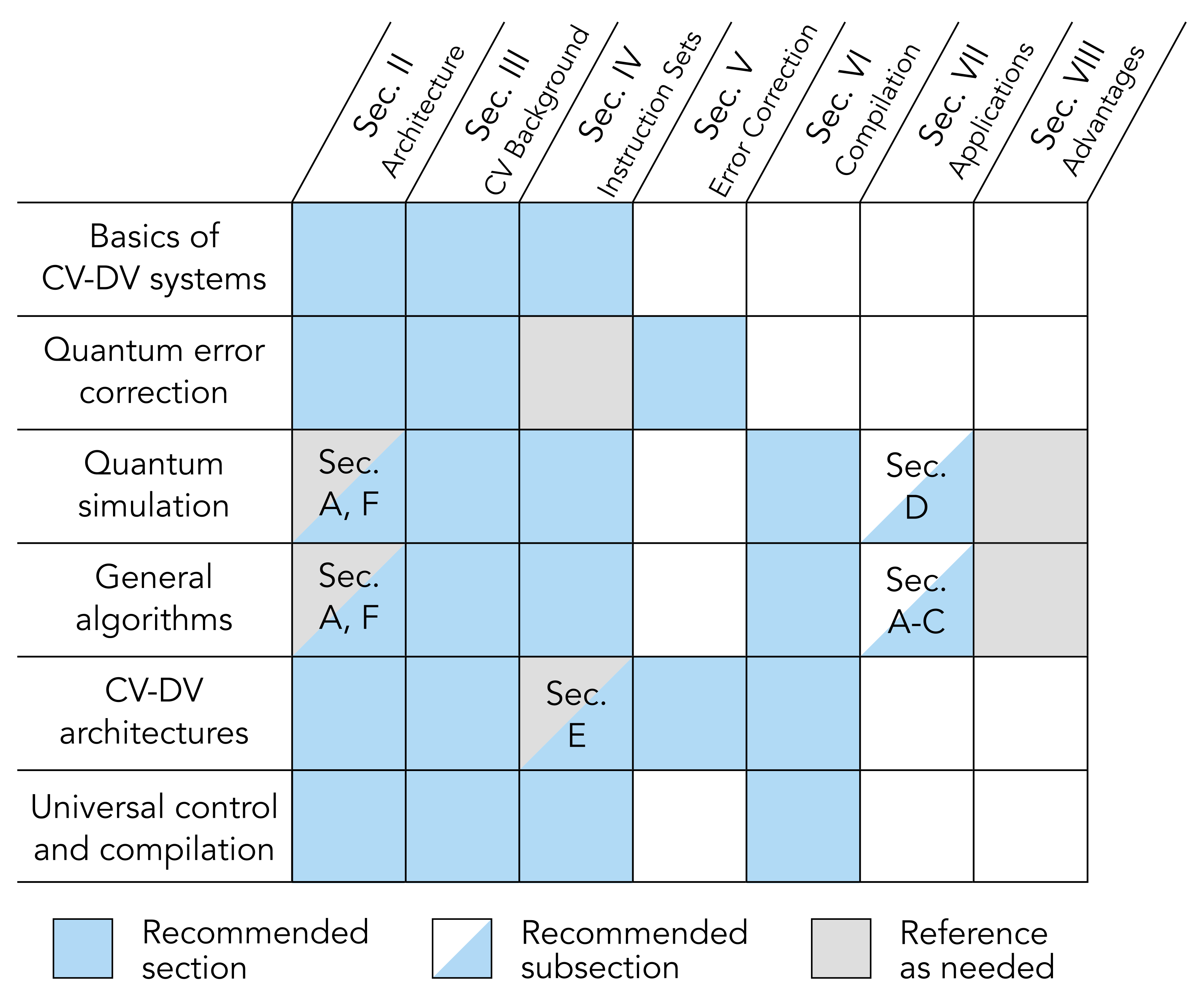}
  \caption{A reader's guide for this work based on particular subtopics. We note that the indicated recommended sections and subsections are non-exhaustive, but are intended to give a rough outline of possible trajectories in reading this work. The introductory (Sec.~\ref{sec:introduction}) and concluding sections (Sec.~\ref{sec:ConclusionsOutlook}) are recommended for all readers.}
   \label{fig:readersguidetotopics}
\end{figure}

Given the length of this manuscript and the interdisciplinary nature of the material covered,  it is useful to augment the above outline with a `reader's guide.' Fig.~\ref{fig:readersguidetotopics} provides a simple map for readers interested in navigating to particular topics.  Below, we give recommendations for how readers with different backgrounds can engage with the material.

\paragraph{Beginners:} For non-expert readers, following the architectural overview in Sec.~\ref{sec:architecture},  we begin in Sec.~\ref{sec:basics} with an introduction to quantum harmonic oscillators and the language from the field of quantum optics used to describe them. It is essential to understand this foundational material. For a first reading, the subsequent sections are best covered in the order presented. 

\paragraph{CV/Quantum Optics Background:} For readers with a strong background in quantum optics but less background in quantum computing, we suggest using this document as follows. We  recommend that these readers familiarize themselves with the discussion on instruction set architectures and abstract machine models in Sec.~\ref{sec:architecture}, as these ideas are likely to be new.   We recommend skimming through the majority of Sec.~\ref{sec:basics}, but paying attention to the truncation discussion in Sec.~\ref{sec:truncation-hilbert-space} and the mapping between oscillators and qubits in Sec.~\ref{sec:mode-to-qubit}.  Unlike traditional quantum optics, circuit QED (the microwave quantum optics of superconducting circuits) and trapped-ion systems offer powerful gate-based control of qubits and oscillators and novel tomographic techniques. We recommend closely reading the discussion in Sec.~\ref{sec:gates} to understand how the specific features of the programming model we consider work. We further recommend paying close attention to Sec.~\ref{sec:qec-compilation}, which discusses compiling for bosonic quantum error correction schemes that may be of interest. The following Sec.~\ref{sec:compilation} will also be of great use, but we recommend paying particular attention to the methods of Sec.~\ref{ssec:approximate-1-qubit-oscillator-unitary} which give systematic compilation techniques for gates including BCH and Trotter-based approaches and quantum signal processing-based strategies.   We then recommend reading the discussion in Sec.~\ref{sec:application-ham-sim}, which covers the application to Hamiltonian simulation and is likely to be of great interest to the reader. 

\paragraph{Computer Science Background:} For computer scientists familiar with qubits, but unfamiliar with quantum oscillators, we recommend paying close attention to Sec.~\ref{sec:basics} to understand the basic formalism of bosonic quantum computation as well as methods to encode qubits in such systems. As a basic motivation of the gate complexity advantages of oscillators over qubits, we recommend reading Sec.~\ref{sec:advantages}, which shows the overheads encountered when simulating oscillators with qubits. The discussion in Sec.~\ref{sec:gates} is also vital to understanding the gates used in our instruction sets, and special attention should be paid to the boxes given within that describe the fundamental operations included within our various ISAs. The abstract machine models and ISAs discussed in Sec.~\ref{sec:instruction_set} will be of great interest to computer scientists approaching this from an architecture perspective, although the discussion in Sec.~\ref{ssec:controlflow} can be skimmed over. The discussion in Sec.~\ref{sec:qec-compilation} and Sec.~\ref{sec:compilation} is very important for the computer science reader, as it shows strategies of interest for those with algorithms and error-correction backgrounds. The discussion of the algorithms in Sec.~\ref{sec:apps} provides several novel state transfer protocols, and a novel approach for performing the quantum Fourier transform using properties of quantum harmonic oscillators.

\paragraph{The Expert Reader:} For readers who are familiar with the concepts of bosonic quantum computation as well as error correction on these platforms, much of the early material can be used for reference. For the benefit of these readers, we summarize below our key results which are novel either because they are original, or because they have recently appeared in narrow venues and are synthesized in this article into a broader context for a wider audience.

The discussion on control flow and benchmarking in Sec.~\ref{ssec:controlflow} contains several important results including an elaboration and formal error-bound analysis on Wigner process tomography on hybrid oscillator-qubit devices. The highly developed experimental state-of-the-art for hybrid CV-DV Wigner function and characteristic function tomography in ion-trap and circuit QED systems is not widely known outside these communities. App.~\ref{app:measurement-tomography} is devoted to general tomographic methods.

Sec.~\ref{sec:compilation} contains a host of new and recently reported results. Specifically, the discussions of synthesizing unitaries in Secs.~\ref{ssec:exact-analytical-qubit-gates} and \ref{ssec:approximate-1-qubit-oscillator-unitary} provide several useful ways to perform such synthesis that may be of great interest to an expert reader. In particular, the extension of quantum signal processing (QSP) approaches to hybrid CV-DV systems presented therein is new and demonstrates a connection between block encodings and qubit-controlled oscillator gates.
In traditional qubit-only QSP \cite{GrandUnificationAlgos}, one deals with sequences of qubit rotations through specified angles to achieve the desired transformation of a block-encoded operator. We have extended this to the case where the rotation angles are non-commuting quantum operators associated with the position and momentum of an oscillator. This section should be reviewed carefully even if methods such as group-commutator-based approaches to unitary synthesis are already well known to the reader. Specific pedagogical and graphically illustrated examples of the application of hybrid CV-DV QSP are also presented, which we hope will be beneficial to theorists and experimentalists alike.

Sec.~\ref{ssec:exact-analytical-qubit-gates} presents new results for exact synthesis (i.e., without Trotter error) for a variety of long-distance qubit-qubit and oscillator-oscillator entangling gates for our hybrid hardware layout. The application of these ideas to quantum simulation, state transfer, and the quantum Fourier transform given in Sec.~\ref{sec:apps} are also of great interest. Sec.~\ref{sec:application-ham-sim} for example, outlines new compilation techniques that take advantage of bosonic modes natively available in the hardware for efficient simulation of strongly correlated many-body systems and lattice gauge theories containing bosons. For describing oscillators with qubit-only hardware, Sec.~\ref{sec:advantages} contains novel, scalable  compilation techniques and a costing of the gate count. Most of the remaining sections not mentioned above can be skimmed or used as a reference for an expert reader, but the aforementioned results are novel and may be of great interest.

In addition to formal circuit synthesis techniques, we place into context tools that have somewhat narrowly arisen in computer science and quantum computer science and should be of broader interest to quantum information physicists. Examples include Strawberry Fields \cite{Killoran_2019}, a compiler for photonic circuits, and Bosehedral \cite{zhou2024bosehedral}, a compiler for beam-splitter networks relevant to boson sampling circuits. Another is Bosonic Qiskit \cite{Biskit,BiskitGitHub,BiskitBlogPost,BiskitBlogPost2}, an extension of Qiskit that provides a quantum intermediate representation for hybrid CV-DV systems with standardized naming and classification schemes for the bosonic gate sets used in the ISAs we present in this work.  Finally, we mention Genesis \cite{chen2025genesiscompilerframeworkhamiltonian} and the Symbolic Hamiltonian Compiler~\cite{decker2025symbolichamiltoniancompilerhybrid}, which are compilers designed to support Hamiltonian Simulation on hybrid CV-DV quantum computing systems.

\section{Hybrid CV-DV Quantum Computing Architecture}
\label{sec:architecture}

This section presents an overview of hybrid CV-DV architectures, beginning  with a perspective on the physical model of hybrid CV-DV quantum computation in Sec.~\ref{ssec:intro_phys}, followed by principled definitions of three main abstract machine models in Sec.~\ref{ssec:amms-detail}, an introduction to instruction set architectures in Sec.~\ref{ssec:intro_isa}, the concept of compilation and error correction in Secs.~\ref{sec:intro_qec} and~\ref{ssec:intro_compilation}, and potential applications and opportunities to be expected for hybrid CV-DV quantum computation in Sec.~\ref{ssec:intro_apps}.

\subsection{Physical Model of Hybrid CV-DV Quantum Computation}

\label{ssec:intro_phys}

\begin{figure}
\centering
\includegraphics[width=\linewidth]{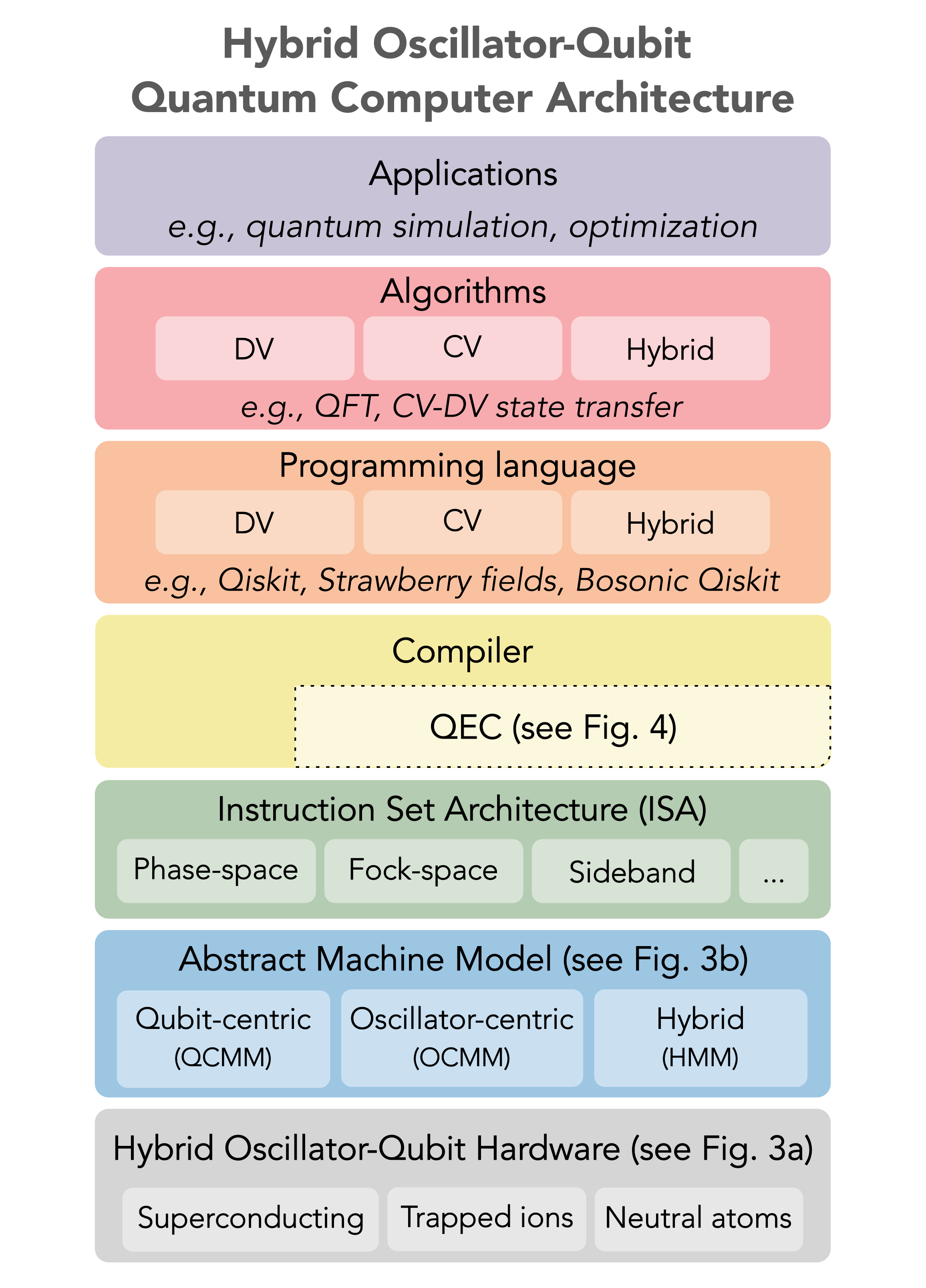}
\caption{The seven layers of hybrid oscillator-qubit quantum computer architecture. Each horizontal layer represents an abstraction of the one below it, while inset boxes provide options and examples for each layer. From top-to-bottom, a user first decides on an application to which they can employ DV, CV, or hybrid CV-DV algorithms. A program is written using an appropriate programming model for the user-level AMM (see Sec.~\ref{sssec:AlternativeTaxonomy}). The compiler takes high-level program instructions and returns a sequence of logical gates expressed in a low-level language, taking into account the chosen quantum error correction options (see Fig. \ref{fig:QEC-stack}). The logical gates are then transpiled into a physical gate sequence that is interpreted by the instruction set architecture. Depending on the nature of the instructions, they are implemented using a qubit-centric, oscillator-centric, or hybrid machine model. Finally, the instruction sequence is translated into a pulse sequence executed on the physical hardware (see Fig. \ref{fig:hardware_layout}).}
\label{fig:hybrid-processor-arch}
\end{figure}

Modern computing systems are assembled from multiple abstractions, and hybrid CV-DV quantum processors naturally share this approach, as illustrated in Fig.~\ref{fig:hybrid-processor-arch}. This abstraction stack draws from pure DV quantum processors \cite{beverland2022assessing}, but also possesses important differences. While other concepts for hybrid CV-DV system architectures are known~\cite{Pan_2023}, and while current hybrid CV-DV quantum systems do not yet have mature realizations of all the layers depicted, a natural starting point is how such machines are physically realized.

\begin{figure*}[htb]
  \centering
  \includegraphics[width=\linewidth]{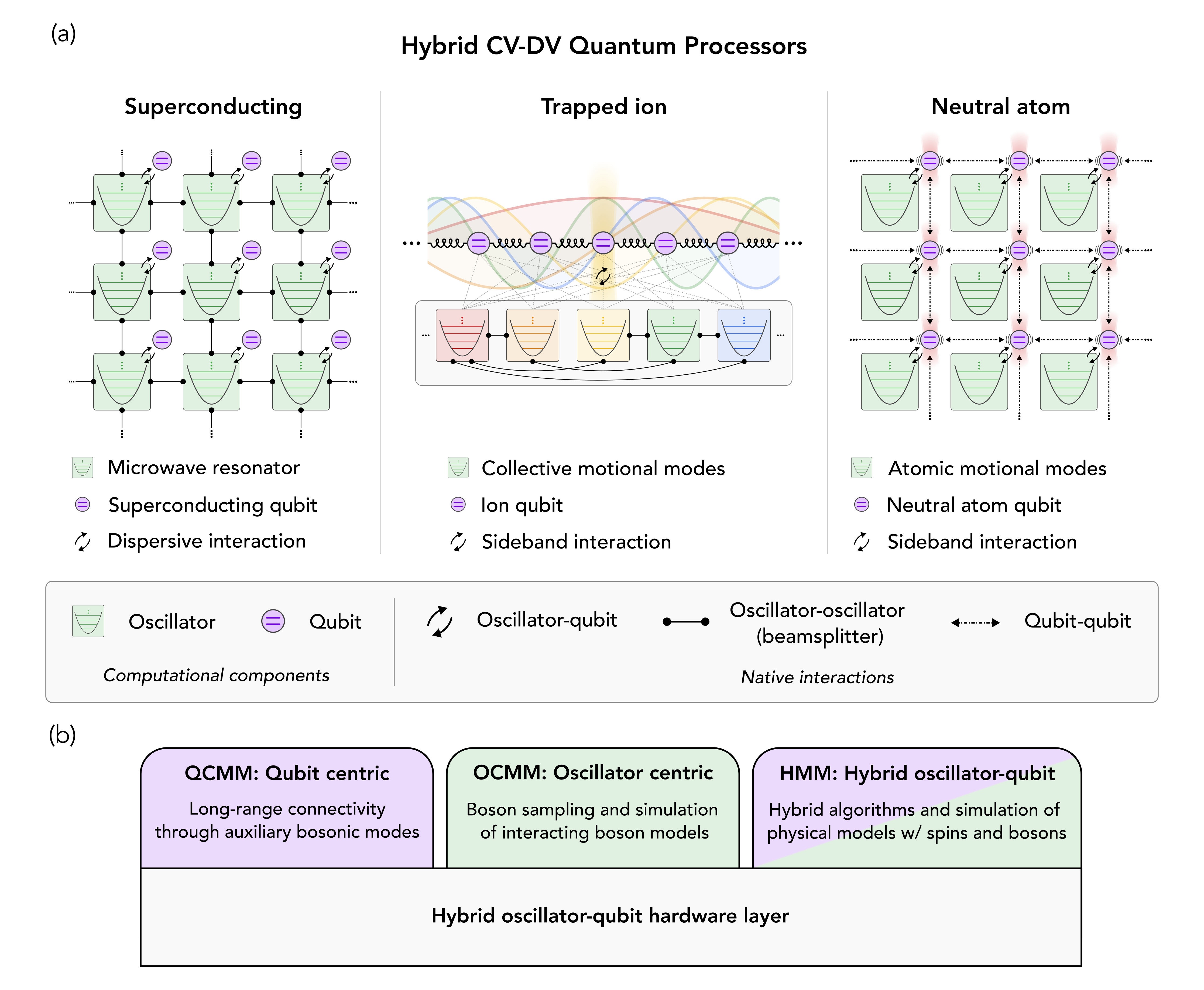}
  \caption{Schematic illustration of hybrid CV-DV hardware and its abstraction. (a) \textit{Left panel:} A hybrid CV-DV quantum processor composed of superconducting microwave resonators dispersively coupled to individual superconducting qubits. Adjacent microwave resonators are coupled via microwave-controlled beam-splitters that permit fast, high fidelity SWAP gates ($\sim100$ ns, $99.92\%$ fidelity in the single photon subspace~\cite{lu2023highfidelity}) that enable remote communication and entanglement of both oscillators and qubits. As each qubit is coupled to only a single bosonic mode, control and cross-talk issues are minimized. (Ref.~\cite{leppäkangas2025quantumalgorithmssimulatingsystems} considers a similar architecture but allows for native two-qubit gates between neighboring qubits.) This superconducting architecture will be our primary focus throughout this work, though much of our discussion is extendable to hybrid CV-DV trapped ion and neutral atom platforms. 
  \textit{Center panel:} In contrast, an ion-trap implementation of a hybrid CV-DV processor has ionic-spin qubits each coupled to all $3N$ collective motional modes of the ion array (as indicated by the gray dashed lines). Native beam-splitters between selected motional modes can be realized through, e.g., laser-based methods~\cite{sutherland2021universal,gan2020hybrid}, laser-free methods such as modulation of trap electrode potentials~\cite{gorman2014twomode,sutherland2021universal,Slichter_coherent_2024}, or via mutual coupling to an ionic qubit~\cite{katz2023programmable}.
  \textit{Right panel:} A hybrid CV-DV quantum processor using neutral atoms in an optical tweezer array. Here, DV components correspond to atomic qubits, while CV degrees of freedom correspond to the motional modes of individual atoms. Similar to the superconducting architecture, each qubit is coupled to a single motional mode/oscillator, minimizing cross-talk. In place of beam-splitters, however, connectivity between oscillator-qubit `units' is realized using qubit-qubit Rydberg interactions.
   (b) As discussed in the main text, the flexibility of the hybrid oscillator-qubit hardware layer enables three distinct abstract machine models visible to the next layer of the stack: the qubit centric machine model (QCMM) presents a standard qubit architecture to the compiler layer, abstracting away the bosonic modes which play a secondary but powerful role as the communication fabric for remote entanglement and gates; the oscillator centric machine model (OCMM) presents the bosonic modes to the compiler layer with the quantum information encoded in the oscillators (possibly via bosonic error correction codes), abstracting away the qubits which play a secondary role as non-linear elements needed for universal control of the bosonic modes; finally, the hybrid machine model (HMM) exposes both CV and DV degrees of freedom on equal footing to the transpiler layer, enabling applications such as quantum simulation of physical models containing both spins (and/or fermions) and oscillators. This taxonomy and an alternative taxonomy are discussed in Sec.~\ref{ssec:amms-detail}.
  \label{fig:hardware_layout}
  }
\end{figure*}

 Tremendous theoretical and experimental progress has been made in the past two decades on quantum control \cite{Krastanov2015,EickbuschECD,qudit2013mischuck,QuditsfromOscillatorsPhysRevA.104.032605} and quantum error-correction \cite{Ofek2016,Campagne-Ibarcq2020,Sivak_GKP_2022,GKP_qudit_2024,LuyanSun2020,ni2022beating,ChenWangAutonomous} based on small versions of a hybrid device architecture in which quantum information can be stored and manipulated in both superconducting qubits and photonic states of microwave oscillators (bosonic modes) within the `circuit QED' paradigm, notionally illustrated in Fig.~\ref{fig:hardware_layout}a and outlined in App.~\ref{app:cQED}.   Circuit QED is the microwave analog of optical cavity quantum electrodynamics with natural atoms replaced by synthetic atoms (superconducting qubits) whose macroscopic size permits extremely strong coupling between the atom and the microwave optical field in a 2D or 3D superconducting resonator, yielding cooperativities many orders of magnitude larger than is possible in traditional cavity QED \cite{Blais2004,Wallraff_cQED_2004,Blais2020,BlaiscQEDReviewRMP2020,Pan_2023}.  This strong coupling permits high-fidelity quantum non-demolition (QND) measurement of the qubit state using the dispersive coupling to a readout resonator.  The same coupling permits QND measurement of the photon number (and binary functions of the photon number such as the parity) in a storage resonator.  As we will discuss in detail, the strong coupling also opens up remarkable opportunities for quantum control and the creation of exotic hybrid quantum states of oscillators and qubits. For example, it is relatively easy to displace the oscillator conditioned on the state of the qubit (as is done in trapped ion systems) and to rotate the qubit on its Bloch sphere conditioned on the photon number in the resonator (which is not readily achievable in trapped ions). Such gates will play important roles in our discussion.  Another feature of circuit QED, shared to an extent with cold atoms in optical lattices, is the ability to engineer parametric processes which give experimenters control of the phase and amplitude of the hopping of quanta between lattice sites; physically, such quanta may be photons in circuit QED and atoms in optical lattices.  This turns out to be important for simulating physical models in which neutral bosons are coupled to either static gauge fields as if they were charged particles in a magnetic field, or to dynamical gauge fields in simulations of lattice gauge theories.

 There have likewise been rapid advances in trapped ion systems \cite{molmer1999quantum,molmer1999multiparticle,milburn2000iontrap,sorensen2000entanglement,katz2022nbody,buazuavan2024squeezing,bruzewicz2019trapped,katz2023programmable,Katz_2023,Whitlow2023,Iqbal2024,Serafini_2009,burd_quantum_2021,gan2020hybrid,so2024trappedion,Slichter_coherent_2024,sutherland2021universal,mielenz2016arrays,jain2020scalable,jain2024penning,hakelberg2019interference,sterling2014fabrication,palani2023high,olsacher2020scalable,schwerdt2024scalable} that combine DV ion qubits with CV motional degrees of freedom (notionally illustrated in Fig.~\ref{fig:hardware_layout}b and outlined in App.~\ref{app:trapped-ions}). These advances include the realization of scalable 2D ion arrays \cite{mielenz2016arrays,hakelberg2019interference,sterling2014fabrication, palani2023high}, Penning micro-traps \cite{jain2020scalable, jain2024penning}, the integration of optical tweezers and dynamic potentials \cite{olsacher2020scalable, schwerdt2024scalable}, and the implementation of strong higher-order nonlinear interactions, including trisqueezing and quadsqueezing \cite{buazuavan2024squeezing,sutherland2021universal}, opening new opportunities for non-Gaussian resource generation. These technological advancements have enabled cutting-edge experimental demonstrations including the analog simulation of chemistry \cite{Whitlow2023,so2024trappedion} and other fermionic models \cite{hemery2023}, the observation of finite-energy transitions \cite{schuckert2025} and emergent hydrodynamics ~\cite{joshi_observing_2022}, and the realization of non-abelian topological order 
\cite{Iqbal2024}. Trapped ion systems tend to be used in a qubit-centric paradigm with the bosonic motional modes used to realize all-to-all connectivity (QCMM in Fig.~\ref{fig:hardware_layout}b). However, there are notable exceptions focusing on the creation and measurement of novel bosonic (phonon) operations and states \cite{Chen_2021,chen_scalable_2023,buazuavan2024squeezing,saner2024,toyoda_hongoumandel_2015,nguyen2021experimental,RingerTrapped-ion}, the preparation of GKP code words for error correction \cite{HOME-GKP2019,HomeGKPQEC2022,HomeCharacteristicFunction,matsos2023robust}, bosonic simulations of quantum chemistry \cite{Whitlow2023,kang_seeking_2024,valahu_direct_2023}, and the use of bosonic modes for machine-learning tasks \cite{nguyen2021quantum}. 

Neutral atoms in optical tweezers (illustrated in Fig.~\ref{fig:hardware_layout}a and outlined in App.~\ref{sct:neutralatoms}) have traditionally been qubit-centric in their approach, viewing the  mechanical oscillations of the atoms within the tweezer trap as a source of decoherence rather than as a resource.  This is beginning to change as experiments have  started to explore hybrid qubit-boson operations \cite{Schlosser2001, sortais2007,Kaufman2012}. In particular, Ref.~\cite{scholl2023} recently realized mode lifetimes exceeding those of the qubits, and further demonstrated hyper-entanglement of two tweezers in both the qubit and motional degrees of freedom. In addition, bosonic neutral atoms in conjunction with precision placement of atoms using optical tweezers have been used to realize the bosonic Hong-Ou-Mandel effect \cite{Kaufman2014} and, more recently, boson sampling \cite{young2023atomic}. Theory proposals have explored realizing bosonic quantum error-correcting codes \cite{bohnmann2024bosonicquantumerrorcorrection} and exploiting the intrinsic non-linearity of the potential \cite{Grochowski2023}.

At a high level, this tutorial envisions a system composed of oscillators (see Sec.~\ref{sec:basics}) and qubits coupled via the hybrid CV-DV gates introduced in Sec.~\ref{sec:gates}, while abstracting away physical implementation details wherever possible. We assume that only the qubit subsystems are directly measurable, and that beam-splitters enable interactions between oscillators. Throughout, we primarily focus on architectures where beam-splitter operations serve as the main mechanism for building connectivity. However, we note that in some platforms such as neutral atom arrays, direct two-qubit gates (e.g., Rydberg interactions) may be more natural for connecting nearby qubits (see Fig.~\ref{fig:hardware_layout}a). All coupling operations are constrained by a specified interaction graph, and algorithms must be designed to respect this hardware connectivity.

To ground the discussion, we will often specialize to the simple yet experimentally realistic hardware layout shown in Fig.~\ref{fig:hardware_layout}a and featuring near-neighbor connectivity. In this setup, each qubit is tightly integrated with a nearby oscillator, enabling fast hybrid operations. The qubits are conceptually arranged in a two-dimensional array, with one oscillator paired to each qubit. While this layout is fixed in circuit-QED architectures, ion-trap and optical tweezer platforms allow for dynamic reconfiguration, as qubits can be physically moved to reconfigure the geometry on the fly.

\subsection{Abstract Machine Models}
\label{ssec:amms-detail}

An abstract machine model (AMM) plays an important role between the hardware architecture and an instruction set architecture. For classical computers, AMMs provide valuable theoretical models that bridge between computational complexity and practical hardware architecture design, allowing exploration of design decisions and analyses of their impact before physical implementation. For example, several AMMs have played important roles in classical computer architectures, providing a mechanism to understand and optimize the balance between computation, communication, and storage resources allocated in hardware to best suit algorithmic and application needs.  These seminal AMMs include \cite{hennessy2012computer} the random access machine (RAM), which models the cost of accessing memories with different latencies and throughputs; the message passing interface (MPI), which models the tradeoff between communication and local computation when no global shared memory is provided; and the Java virtual machine (JVM), which abstracts away the underlying processor and operating system, providing a model for ``write once, run anywhere'' software development.   Other important early AMMs \cite{aho2007compilers} include finite state machines, Turing machines, and even cellular automata.  All these models have provided invaluable mechanisms for analytical reasoning about computational complexity given constraints on memory, concurrency, connectivity, and other fundamental resources such as energy, space, and time.

Compared with the sophistication of AMMs for classical computation, AMMs for quantum computation are still in their infancy but they certainly aspire to satisfy many of the same goals.  Quantum AMMs seek to allow analytical study of computational complexity given constrained resources, just as classical AMMs do.  But these must extend to include quantum-specific resources, such as entanglement and coherence, as is apparent from considering uniquely quantum aspects of communication, storage, and computation:
\begin{itemize}
    \item {\bf Communication} can be performed using qubits instead of just with bits.  In certain hardware schemes such as trapped ion systems, physically moving qubits is very slow \cite{khatri2020principles,10.1109/ISCA45697.2020.00051} compared with gate speeds.  But if entangled quantum states are available, then they can be employed to teleport quantum states over long distances at the speed of classical communication; quantum teleportation can thus serve as a kind of ``quantum wire''  \cite{oskin2003building}, with its own latency and throughput trade-offs. 
    \item {\bf Storage} of quantum states can also be much more challenging than storing bits. Good quantum memories are not yet available and may require costly encoding in error correction codes and constant refreshing \cite{xu2023QRAMsystems}.  Quantum memories also come with many different access models, including quantum random-access memories (QRAM) with parallel or serial access \cite{PhysRevLett.100.160501,PhysRevA.78.052310,jaques2023qram,ConnorQRAM_PRXQuantum.2.020311,xu2023QRAMsystems,PhysRevA.86.010306,PhysRevA.108.032610,8962352,weiss2024qram,phalak2023quantum}. 

    \item {\bf Computation} is often expressed in the language of gates and circuits, and while quantum computation may use this language much like classical computation, alternative and distinctly non-classical models are also available.  This notably includes {\em measurement-based} quantum computation (MBQC), in which computation proceeds through a sequence of measurements alone, with no gates being applied \cite{briegel2009measurement}.  Remarkably, quantum states can embody the ability to perform gates of complexity beyond what is possible with native hardware \cite{preskill1999plugin}; this is the idea behind the magic states used with gate teleportation to enable universal fault-tolerant quantum computation with hardware that performs only Clifford gates \cite{bravyi2005magic}.  
\end{itemize}

Quantum AMMs have arisen to capture many of these uniquely quantum aspects, but these models largely build on qubits (DV quantum systems) as the primitives and eschew a distinct role for bosonic (CV) quantum hardware.  Naively, a quantum computation consists of only three steps: state preparation of the input quantum registers, application of a unitary transformation representing the algorithm, and readout (measurement) of the output registers.  This native model is the dominant one for nearly all modern quantum programming models such as Qiskit \cite{QiskitWebSite}.  However, with the rise of bosonic quantum hardware, programming models are beginning to emerge for such systems, including Bosonic Qiskit \cite{Biskit} and Strawberry Fields \cite{Killoran_2019}.

Most interesting is not just an AMM that captures classical computation, or CV or DV quantum computation, but instead, all of these combined.  A compelling example of the potential for combining qubit-based quantum computation with classical computation is the ``Local Alternating Quantum-Classical Computation'' (LAQCC) model \cite{buhrman2023LAQCC,Niels_Neumann_Thesis2025,pham20132d} of tightly coupled quantum and classical computation and communication with frequent measurements and feed-forward.  In other words, unitary quantum computation is followed by further unitary operations conditioned on classically processed measurement results from the first stage, and so on.  This is a substantially more powerful computational model than quantum alone, or classical alone, and strongly indicates the desirability of including classical computation aspects in quantum AMMs.  We note that the communication via quantum `wires,' the gate teleportation, and MBQC mentioned above all require the mixed quantum-classical operations and feed-forward of the LAQCC model.

In discussing quantum AMMs, a key high-level question arises: what is the appropriate level of abstraction for a quantum AMM, given the infancy of quantum hardware?  Specifically, two contrasting viewpoints arise naturally, as suggested in Fig.~\ref{fig:hybrid-processor-arch}.  One is the viewpoint from the perspective of a user who wishes to compile efficient applications utilizing a given computational model.  A {\em user-visible quantum AMM} (and its associated high-level instruction sets) for CV-DV hybrid computation should abstract away all details of the hardware and its connectivity and simply present a set of abstract qubits and abstract oscillator objects.  Such an AMM should hide details of the fact that the abstract qubit might be a physical qubit or a logical qubit comprising many physical qubits or a logical qubit encoded in a physical oscillator.

In contrast, consider the viewpoint from the perspective of the transpiler, which must deal with the task of mapping applications into primitive operations on physical systems.  A {\em transpiler-visible quantum AMM} (and its associated low-level ISAs) is an abstraction of the physical layer (at the level shown schematically in Fig.~\ref{fig:hardware_layout}).  Transpiler-visible AMMs can be very helpful for resource estimation.  The ISAs that we primarily focus on in this paper sit between the transpiler and the physical hardware layer and allow the transpiler to convert high-level instructions into low-level instructions that are consistent with the connectivity and error properties of the physical hardware.  Given this, and also since resource estimations are a goal of the present work,  we largely focus here on transpiler-visible quantum AMMs, but we do provide some perspective on user-visible quantum AMMs as well.

Specifically, we  elaborate  below on three types of transpiler-visible quantum AMMs, conceptualizing qubit-centric (QCMM), oscillator-centric (OCMM), and hybrid quantum CV-DV (HMM) models.  

These three make distinct use of the relative strengths of qubits and oscillators to serve different architectural purposes, in particular for quantum computation, internal communication, and storage. We  follow this with brief remarks on the higher level of abstraction in user-visible AMMs.

\begin{table*}[htb!]
\footnotesize
\centering

\begin{tabular}{|c|c|c|c|c|c|}
\hline
\parbox[t]{2.25cm}{~\\ \textbf{AMM Type}} \rule{0pt}{4.4ex} &  \parbox[t]{2.25cm}{~\\ \textbf{Subtype}} & \parbox[t]{2.25cm}{\textbf{Physical\\ Qubit \\ Role}} & ~~ \parbox[t]{2.25cm}{\textbf{Physical\\ Oscillator \\ Role}} & \parbox[t]{2.25cm}{\textbf{Physical \\ Qubit\\ Instructions \rule[-2ex]{0pt}{2.7ex} }}  & \parbox[t]{2.5cm}{\textbf{Physical\\ Bosonic\\ Instructions \rule[-2ex]{0pt}{2.7ex} }} \\ \hline

\multirow{3}{*}{~\parbox[t]{2.5cm}{{\bf QCMM} \rule[2ex]{0pt}{2.17ex}  \\ (qubit-centric)}} & ~ \parbox[t]{2.25cm}{Qubits \rule[2ex]{0pt}{2.17ex}  \\ + Bosonic \\ Memory \rule[-2ex]{0pt}{2.7ex}} & Computation & ~ \parbox[t]{2.25cm}{Memory \\ (error corrected)} & Universal & ~ \parbox[t]{3.15cm}{Single-oscillator \\ error correction} \\ \cline{2-6} 
 & \parbox[t]{2.15cm}{Bosonic \rule[2ex]{0pt}{2.17ex}  \\ Bus} & Computation & Communication & Universal & ~\parbox[t]{3.15cm}{Entanglement\\ swapping \rule[-2ex]{0pt}{2.7ex} } \\ \cline{2-6} 
 & \parbox[t]{2.15cm}{Bosonic \rule[2ex]{0pt}{2.17ex} \\ Sensing} & Computation and I/O & Sensing & Universal & ~ \parbox[t]{3.15cm}{Displacement, \\ Jaynes-Cummings \rule[-2ex]{0pt}{2.7ex} } \\ \hline

\multirow{2}{*}{~\parbox[t]{2.5cm}{{\bf OCMM} \rule[2ex]{0pt}{2.17ex}  \\ (oscillator-centric)}} & \parbox[t]{2.5cm}{\rule[2ex]{0pt}{2.17ex} Boson\\ Sampling}  & I/O & Computation & \parbox[t]{2.25cm}{Boson State\\ preparation\\ and\\ measurement} & \parbox[t]{3.15cm}{Displacement, \\ Multi-mode  Squeezing, \\ Multi-mode Beam-splitter\rule[-2ex]{0pt}{2.7ex} } \\ \cline{2-6} 
 & \parbox[t]{2.15cm}{Bosonic\rule[2ex]{0pt}{2.17ex}  \\ System\\ Simulation \rule[-2ex]{0pt}{2.7ex} } & I/O & Simulation & \parbox[t]{2.25cm}{Boson State\\ preparation,\\evolution, and\\ measurement \rule[-2ex]{0pt}{2.7ex} } & Universal \\ \hline

\parbox[t]{2.5cm}{{\bf HMM} \rule[2ex]{0pt}{2.17ex} \\ (hybrid CV-DV) \rule[-2ex]{0pt}{2.7ex} } & Hybrid &\parbox[t]{2.25cm}{ Computation \\Simulation of\\ Spins/Fermions} &\parbox[t]{2.25cm}{Computation\\ Simulation} & Universal & Universal \\ \hline
\end{tabular}

\caption{Three types of transpiler-visible AMMs, with examples for each, specifying the roles of the physical qubits and oscillators, and the nature of instruction sets employed. The QCMM gives an abstraction for settings where the bosonic modes and qubits are both used to harness physical qubits.  The OCMM provides a model for bosonic experiments where the qubit is used to facilitate state preparation, Hamiltonian evolution, and measurement in the bosonic modes.  The HMM describes models which seek to utilize both continuous variable and discrete operations for computation.  Applications utilizing the HMM have been under explored to date, but we argue in Sec.~\ref{sec:application-ham-sim} that there may be circumstances where the devices modelled by the HMM permit simulations that would otherwise be out of reach with present-day technology.   Complementing these three transpiler-visible AMMs, a different but related set of user-visible AMMs could also be formulated; such AMMs may be appropriate for describing logical qubits that are built on top of the qubits and qumodes exposed by the above models, as described in the text.
}
\label{tab:three_AMMs}
\end{table*}

\subsubsection{Qubit-centric Abstract Machine Model (QCMM)}
\label{sssec:QCMM}

The qubit-centric model, QCMM in Table \ref{tab:three_AMMs}, utilizes qubits for computation, but takes advantage of bosonic subsystems for key tasks they may perform better than qubits.  

In particular, bosonic superconducting resonators are known to offer long coherence times which exceed transmon qubit lifetimes by an order of magnitude or more.  As will be discussed in Sec.~\ref{sec:bosonic-QEC}, bosonic quantum error correction can be employed to lengthen coherence times of states encoded into a resonator, with a hardware cost overhead that is significantly lower than that required for qubits.  This is because the codes can focus on correcting photon loss, the single dominant error mechanism faced by resonators; in contrast, qubits generically suffer from more error mechanisms, including both bit flip and phase flip errors, and in addition, multiple qubits are required to achieve the larger Hilbert space dimensions needed for redundant encoding.  Hardware-efficient bosonic memories may thus be employed in aid of qubit-based quantum computation.  In this ``qubits + bosonic memory'' model, a universal set of gates is chosen for qubits but a limited set of gates are chosen for the oscillators, enabling error correction for their use as quantum memories. Qubit states can be written into and readout from the bosonic memories, but direct logic gates on the quantum memories are not allowed.  Different QEC codes could be employed for different resonators, enabling realization of a hierarchy of quantum memories \cite{10.1109/ISCA.2006.32}; for example, a short code that is easy to decode could provide a short-term cache memory, while a large code (with longer read-write times for the memory) could serve as a longer-term static memory.

Bosonic systems also offer the potential for long-range interconnection between qubit subsystems.  Electrical transmission lines, co-planar waveguides, optical cavities, and optical fibers are all bosonic systems.  When patches of qubits are coupled into such bosonic systems, these resonators and transmission lines can provide a kind of ``bus'' interconnection between the patches \cite{majer_quantum_bus_2007}.  Such bosonic interconnects have been envisioned as being the glue enabling modular quantum computers \cite{schoelkopf-modular-qc-scientific-american}, which may be many quantum processing units tied together within a single chip, or superconducting processors connected between different dilution refrigerators, or trapped ion processors linked across a network spanning a laboratory or even a metropolitan area.  As we discuss in Sec.~\ref{ssec:exact-analytical-qubit-gates}, in this ``bosonic bus'' model, the gates needed for qubit entanglement swapping used to realize quantum communication include qubit-oscillator entanglement (as in trapped-ion systems) and oscillator mode SWAPs (for remote communication).  Other qubit operations can also be implemented using the bosonic modes to implement gates on the qubits.  This reinforces the fact that in such models, the qumodes need not be solely used for memory or encoding qubits within such AMMs.

Another unique feature of bosonic systems is their utility for sensing physical phenomena.  Magnetic and electric fields are all bosonic, and if quantum processing is to be applied to information arising from such fields, then the qubits must couple to them.  Such a coupling between qubits and bosonic systems is precisely the native physical Jaynes-Cummings interaction presented in App.~\ref{app:PhysImp} and underlies the oscillator and hybrid qubit-oscillator gates defined in  Sec.~\ref{sec:gates}.  The oscillator displacement gate embodies a prototypical signal to be sensed.  In the ``bosonic sensing'' model, a bosonic subsystem is explicitly introduced for transducing physical fields into qubit systems used for further quantum processing.  This model is exemplified by the quantum logic spectroscopy performed using trapped ions \cite{schmidt2005spectroscopy} and digital homodyne/heterodyne \cite{strandberg2023digital} and digital photon number detection \cite{johnson_quantum_2010,Wang2020FCFs,PhysRevA.98.022305} of microwave (and axion dark matter \cite{Chou_Axion_Qubit}) signals in resonators, for example. As another example, by using the ISAs we present, it is straightforward to construct large two- and four-legged Schr{\"o}dinger cat states that are sensor states capable of detecting displacements of an oscillator by distances much less than the zero-point uncertainty in the position  through the rapid oscillations of the photon number parity as a function of the displacement \cite{Vlastakis100photoncat,Rosenblum_TensofMilliseconds_PRXQuantum.4.030336}.  Such small displacements can also be detected by using entanglement-enhanced interferometery \cite{YURKE_SU(11)_PhysRevA.33.4033,Backes2021,oh2024entanglement} based on two-mode squeezing operations.

\subsubsection{Oscillator-centric Abstract Machine Model (OCMM)}
\label{sssec:OCMM}

The oscillator-centric model, OCMM in Table \ref{tab:three_AMMs}, utilizes bosonic resonators for computation, and employs qubits only sparsely in support of the computation.  This is natural to do when the computation closely matches the dynamics of coupled resonators.  For example, this is the case for the well-known boson sampling problem \cite{aaronson2011computational}, which samples the output of a large system of coupled beam-splitters and phase-shifters, when fed (ideally) with single photon inputs.  In principle, sampling of the output distribution is very difficult to compute classically \cite{aaronson2011computational}, but in realistic (optical) implementations, the input is a squeezed state of light instead of a single photon, and imperfect squeezing and photon losses \cite{PhysRevLett.131.150601,tillmann2013experimental,spring2013boson,thekkadath2022experimental,zhong2019experimental,huh_boson_2015,Wang2020FCFs} can simplify classical simulation of the system \cite{oh2023tensor,Qi2020Regimes,GarciaPatron2019simulatingboson}.  Nevertheless, it is known that coupled oscillator systems can be very hard to classically simulate  \cite{wiebe-oscillators}, so their computational potential is of great interest.  In the microwave domain, auxiliary qubits allow high-fidelity preparation of the requisite input single-photon states, and with processing of the output states for sampling.  In addition, auxiliary qubits can be used to induce interactions among the oscillators for quantum simulations of, for example, the Bose Hubbard model (see Sec.~\ref{BoseHubbard}). We note that the Bosehedral compiler \cite{zhou2024bosehedral} has recently been developed to simply the task of efficiently compiling beam-splitter networks for the Gaussian boson sampling problem that has applications in solving graph-theoretic problems.

\subsubsection{Hybrid Abstract Machine Model (HMM)}
\label{sssec:HMM}

HMM in Table \ref{tab:three_AMMs}, the hybrid quantum CV-DV model, employs qubits as well as bosonic resonators for computation. A full complement of qubit gates and bosonic gates is also utilized, to enable arbitrary transforms to be performed on both systems.  This hybrid model is the main focus of this paper.  After a brief discussion of user-visible AMMs in the next subsection, we continue in the rest of this section by describing several instruction set architectures natural to the HMM.

\subsubsection{User-visible AMMs as an Alternative Taxonomy}
\label{sssec:AlternativeTaxonomy}

The above discussion presents one possible taxonomy of transpiler-visible abstract machine models which explicitly makes reference to the relative roles of the physical components (qubits and bosonic modes).  An alternative AMM taxonomy at a slightly higher user-visible level of abstraction could be based solely on the (effective) quantum degrees of freedom (defined by their Hilbert spaces) exposed to the software stack.  Thus for example, the qubit-centric model QCMM  would only provide access to two-level systems without distinguishing between physical qubits (i.e., transmons), or logical qubits comprising either multiple physical qubits, or logical bosonic qubit encodings in oscillators.  In this model, the only operations visible to the upper levels of the stack might be logical qubit rotations and two-qubit gates (independent of the physical hardware involved). An example of how bosonic systems can model DV quantum computation is given by Perceval, a quantum programming framework for DV quantum computation with photonic systems \cite{Heurtel_2023}.  The QCMM might additionally expose error models and connectivity for these objects which would of course depend on their underlying physical hardware. The oscillator-centric machine model, or OCMM, exposes all of the natively available Gaussian operations as well as the non-Gaussian oscillator gates discussed in Sec.~\ref{sec:gates}.  The non-Gaussian oscillator gates require the use of auxiliary physical qubits, but the OCMM would abstract away the details of the qubit portion of the associated hybrid gates.  Finally, the CV-DV HMM would expose to the user both the CV and DV degrees of freedom along with one of the hybrid ISAs presented in Sec.~\ref{sec:gates}, but might, for example, allow the user to assume all-to-all connectivity or abstract away the physical qubits comprising logical qubits if error correction is invoked.

\subsection{Instruction Set Architectures}
\label{ssec:intro_isa}

Next above the physical level of the architecture stack of Fig.~\ref{fig:hybrid-processor-arch} is the {\em instruction set architecture} (ISA). 
Across the entire hybrid quantum computing stack, the ISA plays a pivotal role as the hardware-software interface that translates quantum algorithms/circuits into a well-defined discrete (but continuously parameterized) instruction set. An ISA builds upon a choice of AMM and specifies a universal instruction set.  There may be a high-level ISA suitable for programming, as well as a low-level ISA suitable as the target for compilation and hardware implementation.
While the mathematically detailed construction of a hybrid CV-DV ISA is presented in Sec.~\ref{sec:gates}, here we begin with a broader perspective about the intent of such ISAs, design choices involved, and differences compared with classical ISAs.  

{\bf Goals for a good ISA}: Traditionally the ISA plays a vital role in classical computer design by defining how software controls the hardware.  The ISA can be considered to provide two viewpoints: one to a programmer interacting with the hardware at a low level, and in this sense the ISA serves as a programmer's manual to the hardware.  The ISA also provides a crucial viewpoint to compilers, which serve to translate high-level languages to hardware operations via the ISA.  An ISA provides such viewpoints by specifying what instructions can be performed by an information processing unit, what data types are provided, the memory and memory hierarchy provided for information storage, and how the processor accesses memory and with what costs.

A well-designed ISA offers many important features.  For example, a portable ISA may allow software to be written once and compiled to work on many different hardware platforms implementing the same ISA.  Good ISA designs seek to also be complete, in terms of providing a sufficient set of instructions that capture capabilities provided by hardware implementations.  Instructions should ideally also be efficient, enabling fast execution of desirable operations while also capturing some sense of regularity and consistency such that high-level languages can be optimally compiled into ISA operations.  At the same time, a well-designed ISA should remain sufficiently simple that it is reasonably easy to implement with hardware, and understandable to programmers who must implement compilers.

Designing an ISA thus entails a multi-faceted balancing act, trading off between simplicity versus complexity, and optimality versus universality.  We seek to design a thoughtful ISA for hybrid CV-DV quantum computers, and ideally this should inherit lessons learned from ISAs designed for classical computers.  Two well known examples are the reduced instruction set computer (RISC) and the complex instruction set (CISC) ISAs, exemplified by Arm and Intel processors, respectively.  These two very different examples provide a crude map for locating quantum ISAs in perspective.  Because quantum hardware is far less mature, it is evident that the ISAs we present here share similarities with the RISC model.  At the same time, the quantum ISAs lack features that are common in almost all rudimentary classical computer ISAs.

{\bf ISA design assumptions}: The quantum ISAs presented here assume an architecture that tightly integrates qubits and oscillators, where each qubit is located near another oscillator (as notionally illustrated in Fig.~\ref{fig:hardware_layout}a) with fast hybrid instructions available between the qubit and oscillator.  Qubits may be directly linked to each other, but we assume here they are not. In the pedagogically oriented model employed in this tutorial, each oscillator is linked to their paired qubit and connected to neighboring oscillators via microwave activated couplers. Moreover, depending on whether quantum error correction is employed or not, the architecture model may capture either NISQ (no QEC) or fault-tolerant (with QEC) applications.

Building on the physical model for hybrid CV-DV quantum processors,
our execution model assumes that the qubits and the qumodes are tightly coupled, meaning that the operations are synchronized.  This means that we do not need to consider message passing to execute quantum code in this setting.  We assume that (most) gates may be executed in parallel on different qumodes and qubits.  We further assume that measurement results can be used to conditionally control subsequent quantum operations.  This is needed for quantum error correction, linear optical quantum computing, as well as measurement-based quantum computing, and more broadly for the general LAQCC model mentioned earlier.

Finally, the programming model that we take assumes that the sequence of raw gates is represented using a gate description language that describes the precise sequences of gates applied to the qumodes and qubits in the system.  This language is interpreted by a compiler and translated into physical gates that the quantum computer then executes.  There is no assumption that the code provided is stored within the quantum computer; rather we assume that it is held by the external classical computer.

Note that, at present, memory architecture for quantum computers is often not discussed (though see Refs. \cite{10.1109/ISCA.2006.32,liu2023quantum__memory,stein2023microarchitectures} on this topic), nor are specialized processing elements such as Arithmetic Logic Units (ALUs). 

As a result, the role of data structures is thus far limited in quantum computation, particularly when compared to classical computation where the power of these ideas often stems from the assumption that cheap memory access is available.  The closest quantum analog is perhaps the QRAM structure, which provides a low-depth circuit implementation for a lookup table in a quantum computer~\cite{PhysRevLett.100.160501,PhysRevA.78.052310,arunachalam2015robustness,ConnorQRAM_PRXQuantum.2.020311,8962352,PhysRevA.86.010306,PhysRevA.108.032610,xu2023QRAMsystems}.  However, even here the assumption of constant cost unit memory access may be questionable.  For these reasons, we do not discuss the implementation of such computational elements in hybrid oscillator-qubit hardware; however, bosonic operations may aid in the replacement of some costly arithmetic functions and we thus note the possibility that bosonic hardware may play a role in future quantum architectures with ALUs and potentially even QRAM structures \cite{weiss2024qram}.

\if 0

\chdeleted{Below we discuss the properties of the various hybrid instruction sets. In particular, we present discussions of the universality of, the primary focus of this work, the phase-space hybrid system instruction set in Sec.~\ref{sssec:phasespaceISA}.  We remind the readers that the phase-space instruction set deals with the CV space of the oscillator, a non-trivial concept to theorists familiar with DV qubit/qudit architecture. This is the reasoning behind our focus on this particular instruction set. In App.~\ref{app:cross-compilation}, we present a discussion of cross-compilation amongst the various universal instruction sets.}
\fi

The choice of instruction set can be made using two different approaches, as is elaborated further upon in Sec.~\ref{sec:gates}. In the top-down approach, the algorithms/user-visible abstract machine models choose a convenient option from the various instruction sets. For example, 
{\if 0  as shown in the left part of Fig.~\ref{fig:hybrid-q-processor-resource-est}, \fi} a general application needs to be compiled into the native ISA on the hybrid processor. In the bottom-up approach, we choose a hardware-convenient instruction set and then decide the algorithms which can be compiled efficiently with this instruction set. Information about the physical layers of the oscillator-qubit device enables the estimation of resource requirements for execution within the ISA (possibly with QEC) for a particular application/algorithm of interest. Sec.~\ref{sec:gates} thus prepares the way for compilation methods for logical operations of an error-corrected encoding in an oscillator (see Sec.~\ref{sec:bosonic-QEC}), or quantum circuits used in various CV-DV applications (see Secs.~\ref{sec:compilation} and~\ref{sec:apps}).

\label{sssec:hybrid_isa}

{\bf Hybrid CV-DV ISA versus qubit and classical ISAs}: Hybrid oscillator-qubit mode hardware systems are not of wide familiarity to the quantum computer science community, but such systems are rapidly becoming experimentally accessible and deserve increased attention given their potential advantages.

As these advantages are now starting to be realized experimentally, the time is ripe to develop and analyze formal AMMs and ISAs for existing and future hardware, but there are challenges to be overcome.
We are in the very earliest stages of learning how to program such hardware, and ISAs are needed to give developers (from both computer science and quantum physics backgrounds) an understanding of what is possible with current and future hardware for creating fault-tolerant circuits, modules, and processors, and for compiling efficient algorithms. Also, while the present-day development of digital-analog hybrid computing in classical computer architectures \cite{guo2016energy,huang2017hybrid} shares some common themes with hybrid CV-DV quantum systems, such classical hybrids are yet fundamentally different from the quantum case.

One feature of quantum ISAs that is important to understand is that even for hardware based solely on DV objects (i.e., qubits), quantum machines share some features with analog computers (but unlike analog computers, do permit error correction) \cite{Girvin_LesHouches_QEC}. For example, the family of superposition states of a single qubit is continuous, much like the continuous states of an analog computer. Any single-qubit superposition state can be obtained from a fixed starting state ($|0\rangle$, say) using only a discrete set of two gates describing rotations about the $x$ and $y$ axes of the single-qubit Bloch sphere
\begin{align}
  R_x(\theta)&=e^{-i\frac{\theta}{2}X},\\
  R_y(\theta)&=e^{-i\frac{\theta}{2}Y}\,,
\end{align} 
using notation standard in the field \cite{Nielsen_Chuang}, with $X$ and $Y$ being the usual Pauli matrices.
We see however that, unlike the discrete gate sets used in classical computation, quantum gates are generally parametrized by one or more continuous variables (rotation angles).
To develop and formally reason about circuit synthesis and fault tolerance, these continuous parameters may themselves be discretized (e.g., into Clifford operations plus one fixed non-Clifford gate \cite{Nielsen_Chuang}). 
Hybrid architectures are `even more analog' in the sense that bosonic modes are CV degrees of freedom, as described above. Techniques for general circuit synthesis and fault-tolerance analysis in CV systems are only beginning to be developed, and our ability to formally reason about the properties of such circuits \cite{Arrazola_2019,PhysRevX.11.031044,Krenn_Zeilinger,CerveraLierta2022designofquantum,Nichols_2019,fosel2020efficient,EickbuschECD} is still quite limited, and the complexity classification of bosonic states and operations is only beginning to be explored \cite{chabaud2023resources,chabaud2024bosonicquantumcomputationalcomplexity}.

Two principal hybrid CV-DV quantum computation instruction set architectures which go beyond qubit-only and classical ISAs are the Fock-space ISA and the phase-space ISA, as introduced in Sec.~\ref{sec:gates}.   The Fock-space instruction set is based on primitive gates conditioned on the number of excitations held in the oscillator. This is in contrast with the phase-space instruction set, in which qubit operations are conditioned on the position or momentum of the oscillator. A discussion of cross-compilation among the various universal instruction sets is given in App.~\ref{app:cross-compilation}.

\subsection{Quantum Error Correction}
\label{sec:intro_qec}

Above the ISA layer of Fig.~\ref{fig:hybrid-processor-arch}, as an optional part of the compilation process we have an optional step of quantum error correction (QEC).
Unlike modern classical computing systems, quantum computers remain highly prone to errors at their physical layers.  For example, a typical CMOS logic gate may fail with probability perhaps much less than one in $10^{20}$ operations, whereas current two-qubit quantum gates currently fail one in $10^5$ operations at best.  Unfortunately, many quantum applications call for $10^{15}$ to $10^{20}$ nontrivial quantum operations.  It is widely believed that such levels of high-fidelity operation may be achieved by encoding a ``logical'' qubit using multiple physical qubits using quantum error correction and by employing methods of fault-tolerant quantum computation \cite{Nielsen_Chuang}. Some overhead of physical gates and qubits is needed to achieve such fault-tolerant quantum computation, and in principle the required overhead is only polynomial in the scale of the ideal circuit being simulated.  However, the constant prefactors and other scaling terms can be prohibitive in view of physically realistic numbers of qubits and control circuits accessible today.  Thus, the employment of such QEC is not always desired, so we envision the architecture providing a flag to determine whether or not QEC is used before the next layer. 
Sec.~\ref{sec:bosonic-QEC} is dedicated to the discussion of QEC, and provides definition of a ``logical'' level of instructions which arises when QEC is included.

For hybrid processors, bosonic quantum error correcting codes can be used to encode (DV) logical qubits into physical oscillators to realize DV-only logical machines using hybrid oscillator-qubit physical resources. These logical DV machines will have a particular layout and connectivity between logical qubits depending on specific physical layer hardware connectivity as well as the QEC codes used. They will also have particular (residual, uncorrected) errors that may vary with the particular gate or measurement being performed--abstract machine model information that can be passed up the stack to help with noise-aware compilation. For bosonic Hamiltonian simulation applications, oscillator-to-oscillator QEC encodings may be possible (Sec.~\ref{sssec:oscillator-to-oscillator}) but would be challenging.  Alternatively, we can omit the QEC stage to retain the full CV characteristics of the modes that are essential to bosonic simulations.  In this case, physical layer oscillator-qubit instruction sets are directly used as `logical' level instruction sets with connectivity and layout determined by the hardware implementation.

For hybrid qubit-boson Hamiltonian simulation, QEC can be employed in both the Fock-space ISA and the phase-space ISA in a semi-fault-tolerant way by using the higher levels of an auxiliary transmon qubit mode while keeping the main cavity mode intact. For example, in \cite{ReinholdErrorCorrectedGates}, it was shown that the $\ket{g}$-$\ket{f}$ (ground and second excited state) manifold of a transmon can be used to perform SNAP gates (see Box~\ref{Box:SNAP}) fault-tolerantly on a rotationally symmetric code. Similarly, controlled displacement gates could be performed using the $\ket{g}$-$\ket{f}$ manifold, using the $\ket{e}$ state to flag decay errors. We discuss the details of the error correction layer in 
Sec.~\ref{sec:bosonic-QEC}. 

\begin{figure}[!htb]
    \centering
    \includegraphics[width=0.95\linewidth]{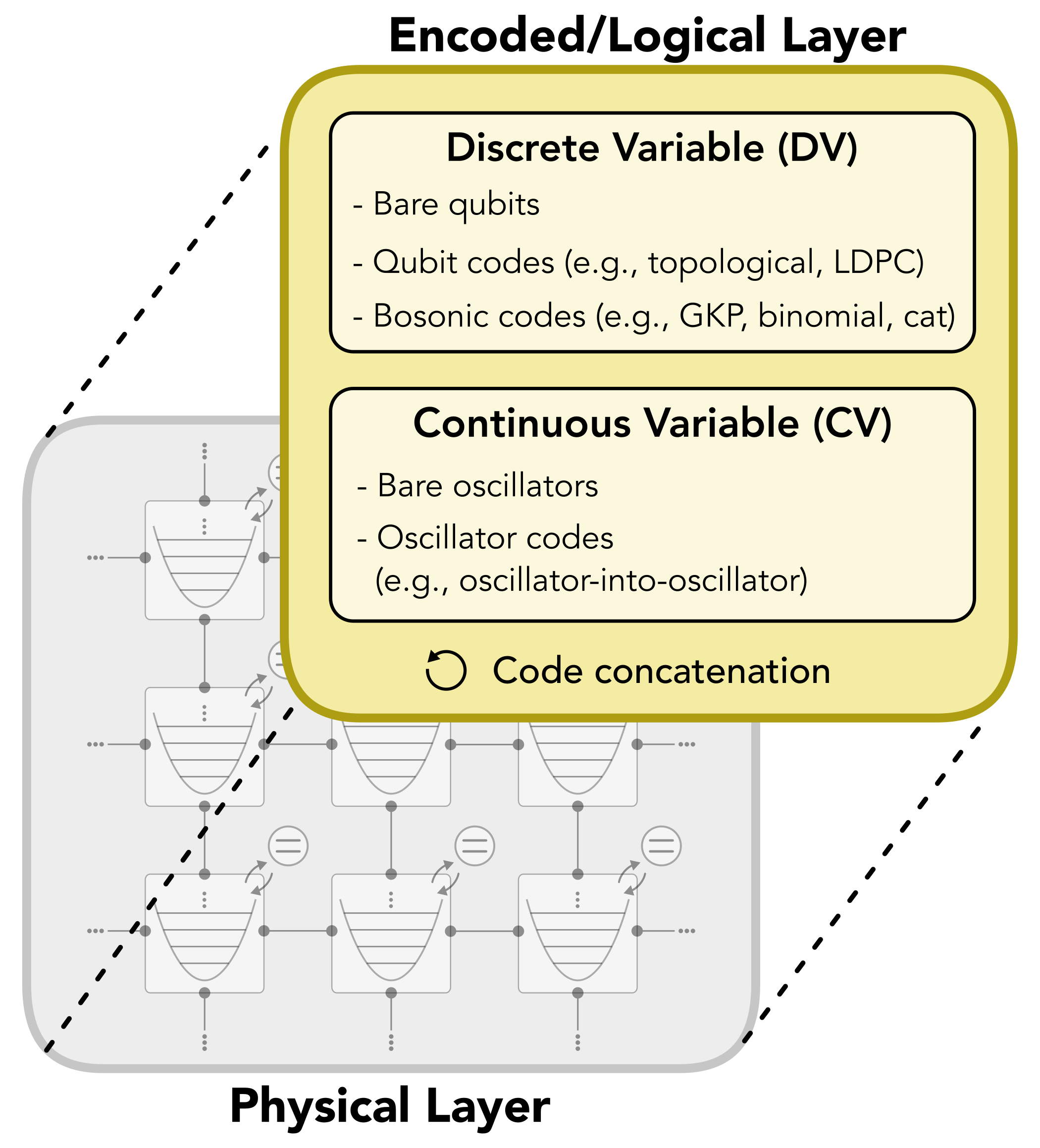}
    \caption{The QEC Stack. In place of computation using physical CV and DV elements, it is beneficial to instead compute using encoded CV and DV computational objects that enable quantum error detection and/or correction  (also see Fig.~\ref{fig:hybrid-processor-arch}). Possibilities for DV codes include both qubit codes (DV to DV) and bosonic codes (CV to DV), while CV codes include oscillator-to-oscillator codes. These codes can be recursively ``stacked'' using what is known as code concatenation.}
    \label{fig:QEC-stack}
\end{figure}

In the noisy intermediate-scale quantum (NISQ) era, we skip QEC and directly transition from the compiler to the hardware stack. In this scenario, the compiler outputs gate sequences composed of physical gates that can be processed directly by the control pulse generator.

\subsection{Programming Languages and Compilation}
\label{ssec:intro_compilation}
The next two integral layers of the architecture of Fig.~\ref{fig:hybrid-processor-arch}  are the compilation and programming language layers.  Compilation involves the mapping of programming languages into the instruction set. The programming language takes as input the algorithmic description from the layer above and outputs to the compiler layer a high-level description of the underlying quantum algorithms \cite{chong2017programming,heim2020quantum,javadiabhari2014scaffcc,Cirq,QDK} in the associated software development kit (SDK) such as Qiskit \cite{QiskitWebSite} or Bosonic Qiskit \cite{Biskit,BiskitGitHub} with proper circuit language descriptions and syntax.

Any programming language specialized to express algorithms that we wish to deploy on bosonic hardware needs to have appropriate abstractions to reflect the architecture of the system that we wish to program at a level of abstraction appropriate for the problem.  
For any programming language that is well suited for our architecture, we need to have appropriate abstractions for registers of oscillators (or qumodes) while also retaining appropriate abstractions for registers and operations on ordinary qubits, as well as for hybrid CV-DV operations.  

As an example, Bosonic Qiskit \cite{Biskit,BiskitGitHub} achieves this by building on top of Qiskit \cite{QiskitWebSite}, which is a leading low-level programming language embedded in Python that can be used to program qubit-based quantum computers.  Specifically, Bosonic Qiskit inherits the great strength of Qiskit which is the ease with which it can be used to provide a low-level description of a quantum circuit and merges this with new functionality that allows it to describe bosonic modes and operations.  Bosonic Qiskit provides a new class for describing oscillators called a qumode register.  This class provides information about the state of a qumode within a specified mode occupation number cutoff $2^n-1$ for integer $n$.  This choice is made to allow easy simulation within existing Qiskit simulators which work with qubit register Hilbert space dimensions that are a power of 2.  A collection of associated qumodes is brought together into a concept called a qumode register.  Qumodes, qubits, and classical memory can all be wrapped together in a single class known as a $\mathtt{CVCircuit}$.  This class can be initialized with the following instructions for the case of a single qumode (qmr) with a cutoff of $2^6-1$, one qubit (qbr), and a classical memory register (cr) \cite{Biskit,BiskitGitHub}
\begin{flalign*}
&\text{\color{blue} \% Initialize qumode register} \\
&\mathtt{qmr = c2qa.QumodeRegister(
num\_qumodes=1,} \\
& ~~~~~~~~~~~~~~~~~\mathtt{num\_qubits\_per\_qumode=6)} \\
&\text{\color{blue} \% Initialize qubit register} \\
&\mathtt{qbr = qiskit.QuantumRegister(1)} \\
& \text{\color{blue} \%  Initialize classical register}  \\
&\mathtt{cr = qiskit.ClassicalRegister(1)} \\
&\text{\color{blue} \%  Initialize hybrid CV-DV quantum circuit} \\
&\mathtt{circuit = c2qa.CVCircuit(qmr, qbr, cr)}
\end{flalign*}
We can then apply a Hadamard gate to the qubit and a controlled displacement for $\alpha=1$ to the qumode followed by a measurement of the qubit register that stores the result in the classical register via the following code

\begin{flalign*}
&\text{\color{blue} \% Assign gate parameters for CV gates} \\
&\mathtt{alpha = 1}\hphantom{circuit = c2qa.CVCircuit(qmr, qbr, cr}  \\
&\text{\color{blue} \% Append a DV gate} \\
&\mathtt{circuit.h(qbr[0])} \\
&\text{\color{blue} \% Append a hybrid CV-DV gate}\\
&\mathtt{circuit.cv\_c\_d(alpha, qmr[0], qbr[0])}  \\
&\text{\color{blue} \% Measure CV and DV components} \\
&\mathtt{circuit.measure(qbr[0],cr[0])}
\end{flalign*}
This code generates the sequence of bosonic and qubit gates illustrated in the quantum circuit diagram below:
\[
\Qcircuit @C=1em @R=.7em {
  \lstick{\ket{0}_{\rm qubit}} & \gate{H} & \ctrl{1} & \meter{}& \cw&\rstick{M} \\
  \lstick{\ket{0}_{\rm osc}} \quad & \qw & \gate{D(\alpha)} & \qw& \\
}
\]
This provides an example of the syntax and circuit description for a simple hybrid algorithm expressed through Bosonic Qiskit \cite{Biskit,BiskitGitHub}.

Sec.~\ref{sec:compilation} describes how the compiler takes as input the high-level language description and connectivity graph of the hardware (an potentially also the heterogeneous hardware noise model) and outputs a low-level language description of the program as logical gate sequences. This can be a challenging task.  Translation from high- to low-level languages often proceeds hierarchically. More specifically, the quantum program description in a high-level language can first be translated into the so-called quantum intermediate representation (QIR), which is a representation of the original program independent of different source programming languages and is also agnostic to hardware instruction details. The QIR serves as a middle-ground to connect high-level programming languages to instruction sets to enable the modular design of quantum compilers. 
The QIR can be further compiled to the elementary logical gate set, given the connectivity graph of the device. 

For DV-only applications (including bosonically encoded logical DVs), a standard DV QIR may be used \cite{haner2018software,cross2022openqasm,QIR,QIRAlliance,smith2016practical},
while for Hamiltonian simulation of bosonic systems, an additional ``bosonic QIR'' can be helpful. On a CV mode, the most general unitary operation is a phase-space volume-preserving diffeomorphism generated by a control Hamiltonian that is a polynomial in the oscillator positions and momenta multiplied by Pauli operators (or the identity) (See Sec.~\ref{sec:algorithm-polynomial-description}). Such polynomials naturally serve as a QIR for bosonic system gates. As an example, such a bosonic QIR might contain terms such as the conditional beam-splitter operation (see Table \ref{tab:gates-qubit-osc})
\begin{align}
    e^{i \theta Z (a^\dagger b + b^\dagger a)} = e^{i 2\theta Z (\hat{x}_a \hat{x}_b +  \hat{p}_a \hat{p}_b)} 
\end{align}
 which has linear couplings between the position $x$ and momentum $p$ of two different oscillators $a$, $b$ multiplied by a single Pauli operator $Z$.  As noted in Sec.~\ref{Z2section}, this gate is particularly useful for $\mathbb{Z}_2$ lattice gauge theory simulation (see Ref.~\cite{C2QA-LGTpaper} for more details). 
 More generally, higher-order polynomials can be used as well. 

 The programming languages and compilation stack layer process the instruction set gates to enact circuits for the layer above it that specifies the algorithms or applications in use.

\subsection{Algorithms and Applications}

\label{ssec:intro_apps}

At the top of the stack of Fig.~\ref{fig:hybrid-processor-arch} sit \emph{applications} and \emph{algorithms}.

Of course, hybrid CV-DV quantum computers should be able to realize general quantum algorithms specific to DV-only hardware, which is a large topic of its own right.  We instead focus here on just the special quantum algorithms and applications particular to the hybrid architecture.  Specifically, what advantages might be found for applications utilizing hybrid CV-DV hardware, instruction sets, and compilers?
Sec.~\ref{sec:apps} details how one likely advantage of hardware natively containing bosonic modes is that each such mode (formally) has a countably infinite Hilbert space dimension while having a relatively simple error model\footnote{Superconducting microwave cavity modes have weak amplitude damping (photon loss) but very little intrinsic dephasing. The mechanical modes in ion traps tend to suffer from dephasing but have little amplitude damping.}, because there are very few `moving parts' relative to a collection of qubits with a large composite Hilbert space dimension.  This hardware efficiency is one of the key reasons that the superconducting cavity architecture was the first one to achieve memory quantum error correction operating above the break-even point\footnote{We conservatively define memory break-even as the logical qubit lifetime exceeding the lifetime of the best quantum component among the physical qubits comprising the logical qubit.} using bosonic codes \cite{Ofek2016,hu_quantum_2019,LuyanSun2020,ni2022beating,Sivak_GKP_2022,Campagne-Ibarcq2020,HomeGKPQEC2022,ChenWangAutonomous}.

A second potential advantage of hardware natively containing bosonic modes is in the construction of programmable quantum simulators for physical models containing bosons (e.g., the bosonic Hubbard model from condensed matter physics and lattice gauge theories from particle physics). Qubit-only systems are naturally efficient for the simulation of quantum magnets (especially spin-1/2 magnets).  Qubits can also simulate fermions using the Jordan-Wigner transformation to insert the appropriate fermion minus signs \cite{JordanWigner1928,Bravyi2002,Ortiz2001,Ortiz2001Erratum,Seeley2012,Troyer2017} but this is expensive since it involves very high-weight operators, especially in greater than one spatial dimension.  It is less well known that there are also significant difficulties in efficiently representing bosons using qubits~\cite{sawaya2019quantum,shaw2020quantum}.  As we discuss, this is partially because each bosonic mode can contain many bosons.
 
However, if we invoke a cutoff $N_\mathrm{max}$ on the number of bosons per mode, we can represent a Hilbert space of the same dimension using only $\sim\log_2[N_\mathrm{max}+1]$ qubits (per mode), so this is not a severe problem if the maximum boson number is below this cutoff, both in the initial state and throughout the simulation. A more serious difficulty lies in the matrix elements of the field operators $a,a^\dagger$ which contain factors related to the square root of the boson number. These field operators are natively available in hardware containing bosonic modes but are quite difficult to synthesize in qubit representations of bosonic modes because a very large number of quantum gates needed to accurately synthesize these square root factors~\cite{soeken2017hierarchical,bhaskar2016quantum} (see Secs.~\ref{sec:application-ham-sim} and \ref{app:compilation-cv-to-qubits}). The practical benefits of the ability to efficiently represent bosons have recently been experimentally demonstrated by the construction of a highly hardware-efficient quantum simulation of the Franck-Condon vibrational spectra of small molecules using Gaussian boson sampling with optical modes \cite{huh_boson_2015,sparrow_simulating_2018,Kihwan_Kim_C7SC04602B} and with microwave modes \cite{HU2018293,Wang2020FCFs} representing the mechanical vibrational modes.  These simple devices with natively available bosonic modes accurately carried out simulations that would be impossible on any currently existing superconducting qubit-only quantum computer \cite{Wang2020FCFs}.  A related experiment has recently simulated dissipative molecular quantum dynamics near a conical intersection \cite{WangConicalIntersection}.

A third potential advantage of hybrid hardware we discuss 
is that state and process tomography for CV systems (see App.~\ref{app:measurement-tomography}) can be straightforwardly performed through simple protocols for measurement of either the Wigner function or the characteristic function \cite{Lutterbach_DavidovichPhysRevLett.78.2547,Bertet2002,hofheinz2009synthesizing,HomeCharacteristicFunction,Haroche_Raimond,Walls_Milburn,Vlastakis2015,Campagne-Ibarcq2020,Sivak_GKP_2022,EickbuschECD}.  State and process tomography makes it possible to understand and calibrate the error model for the particular hardware.  This in turn allows one to develop a corresponding AMM containing a description of the appropriate error models associated with the gates, measurements, and time evolution of the hardware within the framework of the ISA. This information is useful as a basis for integration into either high-level or intermediate representation languages such as Qiskit \cite{QiskitWebSite}, Q$^\#$ \cite{QSharpAzure}, QIR \cite{QIR,QIRAlliance}, and Multi-Level Intermediate Representation Compiler \cite{nguyen2021quantum}.  
And as mentioned earlier, Bosonic Qiskit~\cite{Biskit} is an extension of the Qiskit language that can represent gates, measurements, and error models for hybrid oscillator-qubit processors, and is available to the community \cite{BiskitGitHub, BiskitBlogPost, BiskitBlogPost2} as a co-design tool.

The above advantages are largely directly applicable to trapped ion systems~\cite{katz2023programmable,Whitlow2023}, neutral atoms~\cite{scholl2023,10.1063/5.0197119,bohnmann2024bosonicquantumerrorcorrection} and superconducting qubit/microwave resonator systems~\cite{krasnok2023CavityAdvancements,Krantz2019,Blais2020,BlaiscQEDReviewRMP2020}. In this work, we primarily specialize in the superconducting platform but also comment on key differences in trapped ion and neutral atom platforms where applicable.  Recent introductory reviews of superconducting circuit QED concepts can be found in Refs.~\cite{GirvinLesHouches2011,Krantz2019,Blais2020,BlaiscQEDReviewRMP2020}.  Useful articles on continuous-variable quantum information processing include Refs.~\cite{Braunstein2005,bartlett2002efficient,Gilchrist2004,Aoki2009,Weedbrook2012,Liu2016e,Albert2016}. The power of oscillators for information processing tasks such as phase estimation is surveyed in Ref.~\cite{Liu2016e}. Readers interested in background information on the microscopic Hamiltonians for hardware implementations of superconducting circuits, trapped ion systems, and neutral atoms are referred to App.~\ref{app:PhysImp}. See Ref.~\cite{stancil2022principles} for a textbook on principles of superconducting quantum computers.

The hybrid CV-DV systems notionally illustrated in Fig.~\ref{fig:hardware_layout} enjoy many of the useful features of linear optics quantum computation (LOQC) \cite{LinearOpticsQC_RevModPhys.79.135,KLMdualrail,Bourassa2021blueprintscalable,Fusion-basedQC,sahay2022tailoring,takeda2019toward,o2007optical}, but have key advantages over LOQC because of the ability to use the qubits as auxiliary controllers to create and manipulate complex and highly non-Gaussian photon states in the resonators (e.g., bosonic error-correction code words) and to deterministically perform non-trivial gate operations on the bosonic modes \cite{teoh2022dualrail,chou2023demonstrating,tsunoda2023error,deGraaf2024midcircuit} without relying on the measurement- and fusion-based protocols to supply the required non-linearity as is done in LOQC.  An additional advantage in the superconducting case (Fig.~\ref{fig:hardware_layout}a) is that each qubit is coupled to a single bosonic mode (to minimize cross talk) and the beam-splitters connecting the resonator modes are real-time controllable (microwave pulse activated) \cite{chapman2022high,lu2023highfidelity}, allowing for rapid and high-fidelity routing of the bosonic quantum information throughout the (2D or potentially 3D) hardware fabric.  With proper scheduling to avoid collisions, this routing can be efficiently parallelized.

Finally we note a critically important advantage of the superconducting architecture is that, unlike in LOQC, photon number measurements \cite{Schuster2007a,PhysRevA.98.022305,PhysRevX.10.011001,Wang2020FCFs} (and photon number parity measurements \cite{Sun2014}) are very nearly quantum non-demolition (no photons are absorbed by the detector).  This means, for example, that in the dual-rail architecture (originally developed for LOQC \cite{DualRailOriginalPhysRevA.52.3489,KLMdualrail}) where the logical qubit states involve one photon shared between two resonators, the dominant error (photon loss) can be converted with very high efficiency to an erasure error via QND measurement of the joint photon number in the dual-rail qubit \cite{teoh2022dualrail,chou2023demonstrating,koottandavida2024erasure,deGraaf2024midcircuit} which jumps from one to zero when a photon is lost, or from one to two in the less likely event that a photon is gained (see also the qubit version of the dual-rail in \cite{kubica2022erasure,levine2024demonstrating}).  Because the location of erasure errors is flagged, such errors have lower entropy and are vastly easier to correct \cite{PuriThompsonErasure,kang2023quantumerasure,PhysRevX.13.041013,ma_high-fidelity_2023}. 

Having established the details of the hybrid CV-DV architecture, we are now ready to deep dive into each layer of the stack in the next sections.

\section{States and Operators}
\label{sec:basics}
The physical hardware is the lowest layer in the CV-DV architecture stack (Fig.~\ref{fig:hybrid-processor-arch}), and in that context this section describes the basics of a CV-only architecture, i.e., oscillator states and operations. This is intended to provide a pedagogical introduction to the non-trivial aspects of a continuous-variable space from a mathematical perspective, assuming prior familiarity with standard qubit-based quantum computation.  The notation introduced in this section is used throughout this tutorial.

After summarizing standard qubit states and gates in Sec.~\ref{sec:review-qubits}, we define in Sec.~\ref{sec:hilbert-space} the Hilbert space of a multi-qubit and multi-oscillator quantum system that our bosonic quantum processor operates on. 

Sec.~\ref{ssec:rep-basis} then introduces the representation of bosonic states in two bases, the Fock basis and phase-space basis. While the former can be directly translated into the language of qudits, the latter does not have any one-to-one mapping with a discrete variable architecture. Hence, in that section, while introducing the two bases, we attempt to give a formal understanding of the CV space to computer scientists and DV algorithm experts, while using the opportunity to standardize our notation. Sec.~\ref{ssec:rep-basis} also introduces the concept of phase space and stellar-holomorphic representation. 
Next, in Sec.~\ref{sec:truncation-hilbert-space}, the issue of truncating the infinite-dimensional Hilbert space is discussed briefly. 
The various representations discussed here lead to different ISAs for hybrid CV-DV processors as discussed later in Sec.~\ref{sec:gates}.

\subsection{Qubit States and Operations}
\label{sec:review-qubits}
It is useful to review the standard representation that we employ for single- and multi-qubit states before discussing the harmonic oscillator.  To be unambiguous, we employ the following conventions here\footnote{The standard physics notation agrees that $|\uparrow\rangle=|e\rangle$ is the excited state  and $|\downarrow\rangle=|g\rangle$ is the ground state. However, in traditional quantum physics the excited states are often labeled as $|e\rangle = |1\rangle$ and $|g\rangle=|0\rangle$, in contrast to the conventions of quantum information.}
:

\begin{eqnarray}
|0\rangle=|\uparrow\rangle=|e\rangle&=&\left(\begin{array}{c}1\\0\end{array}\right)\\
|1\rangle=|\downarrow\rangle=|g\rangle&=&\left(\begin{array}{c}0\\1\end{array}\right)\\
\sigma_0=\hat I&=&\left(\begin{array}{cc}1&0\\0&1\end{array}\right)\\
\sigma_x=\sigma_1=X&=&\left(\begin{array}{cc}0&1\\1&0\end{array}\right)\\
\sigma_y=\sigma_2=Y&=&\left(\begin{array}{rr}0&-i\\+i&0\end{array}\right)\\
\sigma_z=\sigma_3=Z=|e\rangle\langle e|-|g\rangle\langle g|&=&\left(\begin{array}{rr}+1&0\\0&-1\end{array}\right)\\
\sigma^+=\frac{X+ iY}{2}=|e\rangle\langle g|&=&\left(\begin{array}{cc}0&1\\0&0\end{array}\right)\\
\sigma^-=\frac{X- iY}{2}=|g\rangle\langle e|&=&\left(\begin{array}{cc}0&0\\1&0\end{array}\right)
\end{eqnarray}
Consistent with physics conventions, $\sigma_z$ is still the energy operator, $\sigma^+$ raises the energy, and $\sigma^-$ lowers the energy.  Single-qubit gates are simply rotations 
\begin{align}
    U=e^{-i\frac{\theta}{2}\vec h\cdot\vec \sigma},
    \label{eq:singlequbitrotation}
\end{align}
parameterized by a rotation angle $\theta$ and an axis on the Bloch sphere represented by the unit vector $\hat h=(h_x,h_y,h_z)$.
Here the `spin' vector is
\begin{align}
    \vec \sigma = (\sigma_x,\sigma_y,\sigma_z).
\end{align}
We will have occasion to make frequent use of rotations about an axis in the equatorial plane lying at an angle $\varphi$ away from the $x$ axis, and so we define the following notation for this purpose
\begin{align}
    \sigma_\varphi= \sigma_x \cos\varphi+  \sigma_y \sin\varphi \,.
\end{align}

The basis for multi-qubit states that we use is also the standard one.  For example, for two qubits we have four states that are direct products of two single-qubit states:
\begingroup
    \allowdisplaybreaks
\begin{align}
|\psi_0\rangle=|00\rangle &= \left(\begin{array}{c}1\\0\end{array}\right)\otimes \left(\begin{array}{c}1\\0\end{array}\right) = \left(\begin{array}{c}1\\0\\0\\0\end{array}\right) \label{eq:psi0}    \\
|\psi_1\rangle=|01\rangle &= \left(\begin{array}{c}1\\0\end{array}\right)\otimes \left(\begin{array}{c}0\\1\end{array}\right) = \left(\begin{array}{c}0\\1\\0\\0\end{array}\right) \label{eq:psi1} \\
|\psi_2\rangle=|10\rangle &= \left(\begin{array}{c}0\\1\end{array}\right)\otimes \left(\begin{array}{c}1\\0\end{array}\right) = \left(\begin{array}{c}0\\0\\1\\0\end{array}\right) \label{eq:psi2}\\  
|\psi_3\rangle=|11\rangle &= \left(\begin{array}{c}0\\1\end{array}\right)\otimes \left(\begin{array}{c}0\\1\end{array}\right) = \left(\begin{array}{c}0\\0\\0\\1\end{array}\right)  \label{eq:psi3}    
\end{align}
\endgroup

In this standard basis, multi-qubit operators are given by (sums of) direct products of individual qubit operators.  For example, the joint parity operator for two qubits is
\begin{eqnarray}
Z\otimes Z&=&\left(\begin{array}{rr}+1&0\\0&-1\end{array}\right)\otimes \left(\begin{array}{rr}+1&0\\0&-1\end{array}\right)\\
&=&\left(\begin{array}{rrrr}+1&0&0&0\\0&-1&0&0\\0&0&-1&0\\0&0&0&+1\end{array}\right)
\end{eqnarray}
Further discussion on qubit-only gates can be found in Table~\ref{tab:gates-qubit} of App.~\ref{app:PhysImp}. 

\subsection{The Hilbert Space of Oscillators}
\label{sec:hilbert-space}

A bosonic mode is typically hosted in a superconducting microwave resonator, the mechanical oscillation of trapped ions in radio-frequency potentials or neutral atoms in optical potentials, as illustrated in Fig. \ref{fig:hardware_layout}.  Here, we represent a bosonic mode of frequency $\omega_\mathrm{R}=2\pi f$ as a simple harmonic oscillator with Hamiltonian (in units with Planck's constant $\hbar=1$)
\begin{equation}
    H_0 = \omega_\mathrm{R} \left[\hat n+\frac{1}{2}\right],
    \label{eq:SHOham}
\end{equation}
where $\hat n$ is the number operator denoting the number of photons (or phonons, i.e., bosons) in the resonator mode.  
The corresponding energy eigenstate $|m\rangle$ is called a number state or Fock state as
\begin{equation}
   \hat n|m\rangle = m|m\rangle; \hskip5mm m=0,1,2,3,\ldots 
   \label{eq:number-state}
\end{equation}
The number operator is thus
\begin{equation}
    \hat n = \sum_{m=0}^\infty m {\hat P}_m
\end{equation}
where
\begin{equation}
   \hat P_m\equiv |m\rangle\langle m|\label{eq:Fockproj}
\end{equation}
is the projector onto Fock state $|m\rangle$.  The Fock state with zero photons is referred to as the vacuum state.

A key feature of the harmonic oscillator is that the energy level spacing is constant, which means that, like its classical counterpart, the quantum harmonic oscillator is isochronous -- its period of oscillation is independent of the amplitude of the oscillation.   
 This makes the transformation to the interaction representation very simple and allows us to eliminate the rapid time-dependence generated by $H_0$ and study only the slower evolution under additional control terms (see App.~\ref{app:PhysImp}). That is, in essence, we can move to a rotating frame in which $H_0$ is effectively zero.  We therefore use the interaction representation almost exclusively throughout this work.

\subsection{Representation and Bases}
\label{ssec:rep-basis}

The fundamental difference between the state space that is often used in quantum computing and the one that we discuss here is that the oscillator state space is no longer conveniently represented in terms of two-level systems, i.e. qubits.  Instead, quantum information is stored and manipulated in the joint Hilbert space of both harmonic oscillators and qubits. Similar in spirit to the nomenclature of a ``qubit,'' the vector space of the oscillator is referred to as a ``qumode''~\cite{macridin2023qumode}. We discuss a few representations of the qumode state and operators in this section.

\subsubsection{Fock Basis}
\label{sssec:fock-basis}

The simplest basis for representing the qumode state is just the eigenstates of the number operator $\hat{n}$, the Fock basis $\ket{m}$, defined in Eq. \eqref{eq:number-state}. Two other physically important operators are the boson creation operator $a^\dagger$ and its adjoint, the destruction operator $a$ which obey
\begin{eqnarray}
    a^\dagger |m\rangle &=& \sqrt{m+1}|m+1\rangle\label{eq_sqrt1}\\
    a|m+1\rangle &=& \sqrt{m+1}|m\rangle\label{eq_sqrt2}\\
    \left[ a , a^\dagger\right] &=& 1 \\
    \hat n &=& a^\dagger a \,.
\end{eqnarray}
$a^\dagger$ and $a$ are also known as raising and lowering operators or ladder operators since they move the excitation number up and down the ladder of Fock states. Fock state $|m\rangle$ can be created from the vacuum state $|0\rangle$ by application of $m$ raising operators
\begin{align}
    |m\rangle=\frac{1}{\sqrt{m!}}(a^\dagger)^m|0\rangle.
\end{align}

The eigenstates $|\alpha\rangle$ of the lowering (boson destruction) operator are known as coherent states and are a continuous family of states labeled by a complex parameter $\alpha$
\begin{eqnarray}
    |\alpha\rangle &=& e^{\alpha a^\dagger - \alpha^* a}|0\rangle=e^{-\frac{|\alpha|^2}{2}}e^{\alpha a^\dagger}|0\rangle\label{eq:coherentstateviatranslation}\\
    &=&e^{-\frac{|\alpha|^2}{2}}\sum_{n=0}^\infty \frac{\alpha^n}{\sqrt{n!}}|n\rangle.\label{eq:coherentasFock}
\end{eqnarray}
This family of eigenstates of the lowering operator obeys
\begin{equation}
    a|\alpha\rangle=\alpha|\alpha\rangle \label{eq:eigenstateofa}
\end{equation}
with (complex) eigenvalue $\alpha$.  The operators $a,a^\dagger$ are non-Hermitian and are defective in the Jordan sense.   That is, $a^\dagger$ has no right eigenstates and $a$ has no left eigenstates.    It seems strange that there exist quantum states from which one can remove a photon and still be in the same state.  This is a `Hilbert hotel' phenomenon associated with the countable infinity of Fock states.

The fact that coherent states form a continuum means that they are uncountable and must necessarily be over-complete since the Hilbert space dimension is countably infinite.  Indeed, it follows from Eq.~(\ref{eq:coherentasFock}) that the resolution of the identity is
\begin{equation}
    \hat I = \frac{1}{\pi}\int d^2\alpha\,\, |\alpha\rangle\langle\alpha|=\sum_{n=0}^\infty |n\rangle\langle n|.
\end{equation}
Coherent states are thus our first hint that, unlike `discrete-variable' systems (qubits), oscillators have `continuous-variable' features.  

It can be argued \cite{Goldberg2024covariantoperator} that coherent states are the least `quantum' (since, as we show in Box~\ref{Box:UncondDispGate}, they are simple classical displacements of the Gaussian vacuum) and Fock states are the most `quantum' (since they are highly non-Gaussian).  Gaussianity is explained in the next section.

\subsubsection{Position-Momentum Basis}
\label{sssec:xp-basis}

The fact that oscillators admit a continuous-variable description becomes more clear when we consider the position and momentum operators for oscillators, as we now discuss.
In quantum optics, the position coordinate of an electromagnetic oscillator mode is generally taken to be the electric field of the mode at some particular spatial location, and the conjugate momentum is then related to the magnetic field at that point.  As we see below, the ground state wave function in the position basis is a Gaussian so the electric field has a Gaussian distribution even in the absence of excitations.  These ground state fluctuations in the electric field are referred to as \emph{vacuum noise} (even though the underlying state is pure) \cite{Clerk2010}.
The wave function is also a Gaussian in the momentum representation as the Fourier transform serves to convert between the position and momentum representations, and Gaussians are mapped to Gaussians under Fourier transforms.

We can form dimensionless Hermitian position ($\hat x$) and momentum ($\hat p$) operators from the ladder operators via
\begin{eqnarray}
\hat x&=&{\phantom{-i}}\lambda_x (a+a^\dagger)\label{eq:x}\\
\hat p&=&-i\lambda_p(a-a^\dagger),\label{eq:p}
\end{eqnarray}
where $\lambda_x,\lambda_p$ are real constants.  One conventional choice is $\lambda_x=\lambda_p=\frac{1}{\sqrt{2}}$ which yields the usual commutator (again with $\hbar=1)$
\begin{equation}
    [\hat x,\hat p]=+i.
    \label{eq:usual_xp_Commutator}
\end{equation}
We refer to this case as employing `standard units.'
We, however, also use a different convention $\lambda_x=\lambda_p=\frac{1}{2}$ which we refer to as employing `Wigner units' and which yields
\begin{equation}
    [\hat x,\hat p]=+\frac{i}{2}.
    \label{eq:Wignerunits_xp_Commutator}
\end{equation}
We alert the reader as to which unit choice is being made as needed.

The choice of Wigner units has the advantage that we may write the ladder operators in simple form
\begin{eqnarray}
a^{\phantom{\dagger}}&=&\hat x+i\hat p\label{eq:aisx+ip}\\
a^\dagger&=&\hat x-i\hat p,
\end{eqnarray}
and more importantly, we have the convenient feature that for coherent states the mean position and momentum obey
\begin{eqnarray}
\langle\alpha|\hat x|\alpha\rangle &=& \mathrm{Re}\,\alpha\\
\langle\alpha|\hat p|\alpha\rangle &=& \mathrm{Im}\,\alpha,
\end{eqnarray}
and the corresponding variances are given by
\begin{eqnarray}
\sigma_x^2 = \langle\alpha|{\hat x}^2|\alpha\rangle - \langle\alpha|\hat x|\alpha\rangle^2  &=& \frac{1}{4},\label{eq:cohvarx}\\
\sigma_p^2 = \langle\alpha|{\hat p}^2|\alpha\rangle - \langle\alpha|\hat p|\alpha\rangle^2  &=& \frac{1}{4}.\label{eq:cohvarp}
\end{eqnarray}
With this choice of dimensionless `Wigner' units, we can now write the Hamiltonian in Eq.~(\ref{eq:SHOham}) in the form familiar from a mass-and-spring harmonic oscillator
\begin{equation}
    H_0=\omega_\mathrm{R} [\hat x^2+\hat p^2].
\end{equation}

The operator $a$ is defective, having only a single right eigenvector, namely the vacuum state which has an eigenvalue of 0
\begin{align}
    a|0\rangle=0|0\rangle.
\end{align}
The Hermitian position operator is the sum of the creation and annihilation operators and has a continuous spectrum of non-normalizable eigenvectors labeled by the position at which they are infinitely sharply peaked
\begin{align}
    \hat x|x\rangle = x|x\rangle.
\end{align}
In terms of these position eigenstates, the resolution of the identity is
\begin{align}
    I=\int_{-\infty}^{+\infty}\mathrm{d}x\, |x\rangle\langle x|.
\end{align}
From the relation $I^2=I$ it follows that the orthonormality condition on the position eigenvectors is
\begin{align}
    \langle y|x\rangle = \delta(x-y),
\end{align}
where $\delta$ is the Dirac delta function, illustrating that the position eigenstates are non-normalizable and strictly speaking, can only be defined in terms of a distribution, i.e.,  sequences of narrower and narrower normalizable wavepackets, e.g., Gaussians.

In the so-called position representation, the state vector  of the oscillator 
\begin{align}
    |\psi\rangle = \int_{-\infty}^{+\infty}\mathrm{d}x\, \psi(x)|x\rangle
\end{align}
is represented by a (possibly complex) wave function 
\begin{align}
    \psi(x)=\langle x|\psi\rangle
\end{align}
whose argument is the position of the oscillator.  In this position representation,
 the position operator $\hat{x}$ is represented by multiplication by the number $x$
\begin{equation}
    \hat x|\psi\rangle \rightarrow x\psi(x).
\end{equation}
From the commutation relation in Eq.~(\ref{eq:Wignerunits_xp_Commutator}), we see that 
the momentum operator acting on the state can be written as a derivative acting on the wave function
\begin{equation}
    \hat p|\psi\rangle \rightarrow -\frac{i}{2}\frac{d}{{d}x}\psi(x).\label{eq:momasderiv}
\end{equation}
In this representation, Eq.~(\ref{eq:eigenstateofa}) becomes the differential equation
\begin{equation}
    \left(x+\frac{1}{2}\frac{d}{{d}x}\right)\psi_\alpha(x)=\alpha\psi_\alpha(x),
\end{equation}
which has normalized solution for the coherent state wave function
\begin{equation}
   \psi_\alpha(x)  = \left({\frac{2}{\pi}}\right)^{\frac{1}{4}} e^{-(x-\alpha)^2}.
   \label{eq:coherentstatewavefunction}
\end{equation}
We see that for real $\alpha=x_0$, the center of the Gaussian is shifted from the origin to position $x_0$ while for imaginary $\alpha=iK$ we have
\begin{eqnarray}
     \psi_{iK}(x)  &=& \left({\frac{2}{\pi}}\right)^{\frac{1}{4}} e^{-(x-iK)^2}\\
     &=&\left({\frac{2}{\pi}}\right)^{\frac{1}{4}} e^{K^2}e^{2iKx} e^{-x^2},
\end{eqnarray}
which, because of the factor of $1/2$ in Eq.~(\ref{eq:momasderiv}), indeed corresponds to a momentum boost of $K$.

Returning to Eq.~(\ref{eq:coherentstateviatranslation}) we see that the operator that takes the vacuum state to coherent state $|\alpha\rangle$ is a unitary translation operator in phase space (a concept defined in the next section) 
\begin{eqnarray}
D(\alpha) &\equiv& e^{\alpha a^\dagger - \alpha^*a}=e^{-\frac{|\alpha|^2}{2}}e^{+\alpha a^\dagger}e^{ - \alpha^*a}\\
&=&e^{2i\left\{[-\mathrm{Re}\,\alpha] \hat p + [\mathrm{Im}\,\alpha]\hat x\right\}}.\label{eq:Wignerunitsdisp}
\end{eqnarray}
We see from this that when using Wigner units, it is $2\hat p$ that is the generator of displacements and $2\hat x$ that is the generator of momentum boosts.

\subsubsection{Introduction to Phase Space}
\label{subsec:flowsinphasespace} 
Phase space is a concept borrowed from classical mechanics: a single point in this space corresponds to a two-component position/momentum vector $\vec R=(x,p)$. While the state of a classical particle moving in one spatial dimension is fully specified by its position and momentum (position times velocity), in quantum mechanics, this point becomes a `fuzzy blob' because the operators $\hat x$ and $\hat p$ do not commute. The Heisenberg uncertainty principle guarantees a lower bound on the product of the variances of $\hat x$ and $\hat p$.  Since coherent states are Gaussians, it is straightforward to show that the variance of each is 1/4 in our dimensionless `Wigner' units.

As noted above,
CV systems have the interesting property that their states can be represented in the (countably) infinite discrete Fock basis, or in terms of smooth continuous wave functions $\psi(x)$, or continuous density `matrices' $\rho(x,x')$ in the position (or the momentum) basis. To develop intuition about these continuous representations, it is helpful to begin with a study of the classical oscillator whose state is not defined in Hilbert space, but rather by its position $\vec R=(x,p)$ in the two-dimensional `phase space' of position and momentum.  Statistical mixtures (ensembles) of classical states are then described by a probability distribution $P(x,p)$.  Quantum mechanically, the numbers ($x, p$) are replaced by non-commutative operators $(\hat{x},\hat{p})$ that satisfy the canonical commutations in Eq.~\eqref{eq:usual_xp_Commutator} and \eqref{eq:Wignerunits_xp_Commutator}. Mixtures of quantum oscillator states are described by the Wigner function  (See App.~\ref{sec:characteristic-function}), $W(x, p)$, a \emph{quasi-probability} distribution  which is real but can take negative values \cite{wigner1932on,Haroche_Raimond}. Two other commonly used representations of CV states, the characteristic function and the Husimi-Q function in Apps.~\ref{sec:characteristic-function} and \ref{sec:husimi-q}, respectively.

Universal quantum control requires the ability to execute an arbitrary unitary transformation on the Hilbert space of the combined qubit/oscillator system.  It is useful to think of the problem of unitary synthesis as one of Hamiltonian synthesis:  how we synthesize a Hermitian operator (Hamiltonian) $H$ such that it generates the desired unitary $U=e^{-iHt}$ under time evolution, or more generally for a time-dependent Hamiltonian
\begin{align}
    \frac{dU}{dt}=-iH(t)U(t).
    \label{eq:Usynthesis_Hsynthesis}
\end{align}
It is useful to begin with the classical dynamics of a single oscillator. The intuition gained from this will prove very useful in designing quantum Hamiltonians to achieve desired unitary transformations in Hilbert space. in the classical limit we evaluate the quantum equations of motion only to zeroth order in Planck's constant \cite{kim1991phase}. Rather than Hilbert space, we describe the state of the oscillator in terms of a point $\vec R=(x,p)$ in the two-dimensional phase space, where the numbers (rather than operators) $x$ and $p$ are respectively the position and the corresponding canonical momentum.  The Hamiltonian $H(x,p)$ determines how the system moves  through phase space according to Hamilton's equations of motion\footnote{Here we are using the standard definitions of position and momentum that have Poisson bracket equal to unity rather than choosing the value of one-half that would correspond in the quantum case to the choice of Wigner units.}
\begin{align}
\vec \Lambda(\vec R)=\frac{d\vec R}{dt}&=\left(\frac{\partial H(x,p)}{\partial p},- \frac{\partial H(x,p)}{\partial x}\right).
\end{align}
 Time evolution maps the phase space onto itself, and it turns out to be useful to think of this mapping as described by the velocity field $\vec\Lambda(\vec R )$ of a fluid filling all of the phase space--the position $\vec R$ of each fluid element evolves according to the Hamilton equation of motion. An interesting feature of this fluid is that it is guaranteed to be incompressible because the flow field is automatically divergenceless (Liouville's theorem)
 \begin{align}
     \vec\nabla\cdot\vec\Lambda(\vec R)&=\frac{\partial}{\partial x}\frac{\partial H(x,p)}{\partial p}-\frac{\partial}{\partial p} \frac{\partial H(x,p)}{\partial x}=0.
 \end{align}
This means that we can think of the mapping of phase space onto itself under Hamiltonian evolution as a volume-preserving diffeomorphism.  

\begin{figure}[t]
    \centering
    \includegraphics[width=0.4\textwidth]{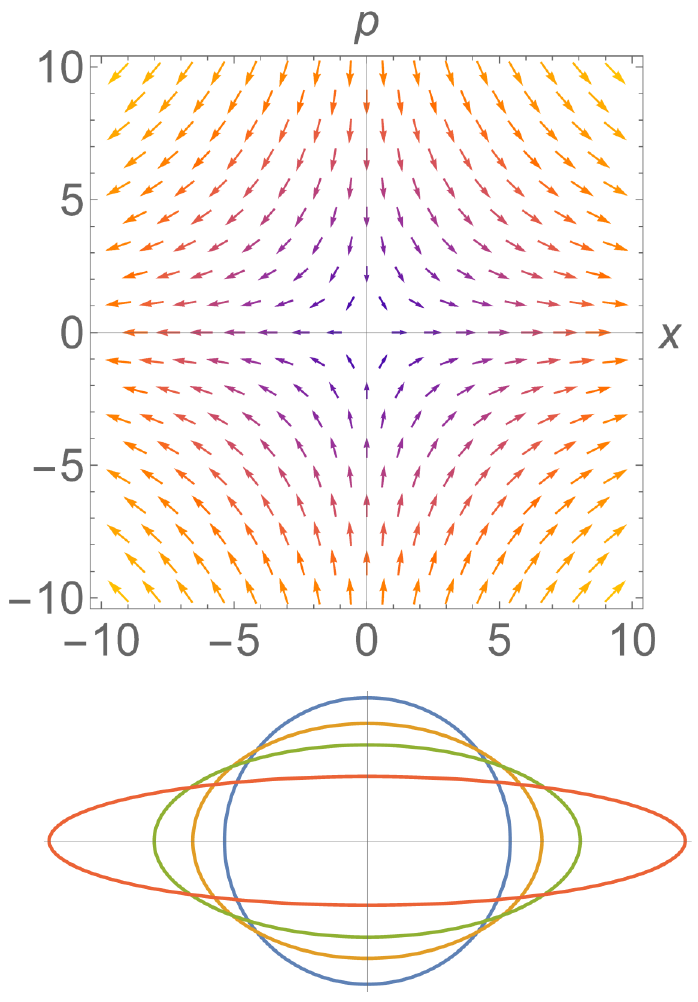}
    \caption{Upper panel:  Phase space flow field $\vec \Lambda$ under the squeezing Hamiltonian in Eq.~(\ref{eq:squeezingHam}). Lower panel: Time evolution of an initial circular region into a squeezed ellipse (of the same total area) under the phase space flow field for times $g t = 0.0,0.2,0.4$ and $0.8$ (blue, orange, green, and red).}
    \label{fig:SqueezingPlot}
\end{figure}

 As a first illustrative example, consider the simple classical Hamiltonian
\begin{align}
    H=g \hat{x} \hat{p},
    \label{eq:squeezingHam}
\end{align}
where $g$ is a real constant.  Because the Hamiltonian is quadratic, the equation of motion for the divergenceless flow field $\vec{\Lambda}$
\begin{align}
    \vec\Lambda(x,p) = \left(\frac{{d}x}{dt},\frac{dp}{dt}\right)=g\left(x,-p\right),
\end{align}
is linear and thus readily solved
\begin{align}
    x(t) &=x(0)e^{g t}\\
    p(t) &= p(0)e^{-g t},
\end{align}
where $x(0), p(0)$ are the initial position and momentum at time $t=0$.
From this we see that $H$ is the `squeezing' Hamiltonian because under its action, a circular region of the phase space `fluid' evolves into a squeezed ellipse (of precisely the same area) as illustrated in Fig. \ref{fig:SqueezingPlot}. This is a simple example of a Gaussian operation, a concept that is discussed in more detail for the quantum case in Sec.\ \ref{subsubsec:gaussian_ops}.  

\begin{figure}[t]
    \centering
    \includegraphics[width=0.4\textwidth]{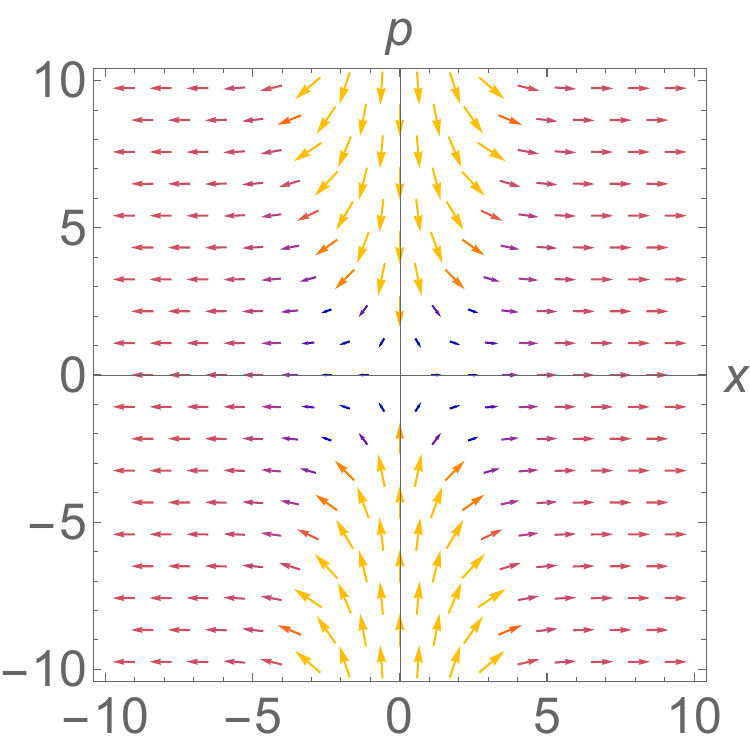}
    \caption{Flow field for the Hamiltonian given in Eq.~(\ref{eq:Tanh2-LegHam}). Notice that the flow field for large $|x|$ and small $|p|$ is much more uniform (less squeezing) than that of the squeezing Hamiltonian shown in Fig.~\ref{fig:SqueezingPlot}.}
    \label{fig:TanhFlowField-2Leg}
\end{figure}

As a second example, we consider the non-quadratic Hamiltonian
\begin{align}
    H=g\tanh\left(\frac{\hat{x}}{2}\right) \hat{p}.
    \label{eq:Tanh2-LegHam}
\end{align}
The phase-space flow field for this Hamiltonian is illustrated in Fig.~\ref{fig:TanhFlowField-2Leg}.  Notice that, because the hyperbolic tangent is a constant for a large argument, the flow field is very uniform for large $|x|$, being well approximated by 
\begin{align}
    \vec\Lambda(x,p)\sim \mathrm{sgn}(x)\,(g,0).
\end{align}

\begin{figure}[tbh]
    \centering
    \includegraphics[width=0.4\textwidth]{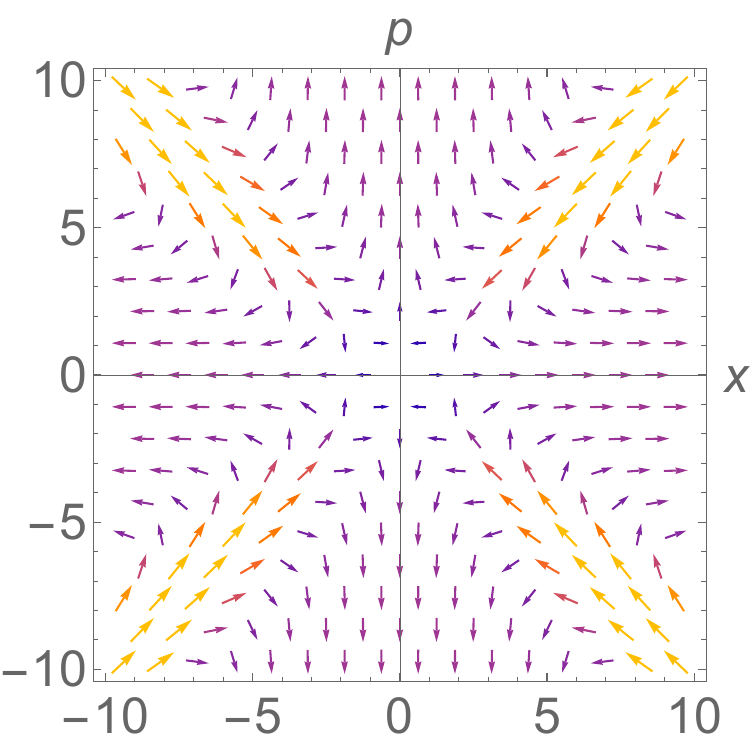}
    \caption{Flow field exhibiting the four-fold rotational symmetry of the Hamiltonian given in Eq.~(\ref{eq:4legflowfield}).}
    \label{fig:TanhFlowField-4Leg}
\end{figure}

Thus application of this Hamiltonian for time $t$ leads to a simple displacement (for large $|x|$) rather than squeezing, with the sign of the displacement matching the sign of $x$
\begin{align}
    x\rightarrow x+gt\,\mathrm{sgn}(x).
\end{align}
This particular flow will prove useful for the `two-legged cat amplifier' operation described in Sec.~\ref{ssec:compilation-bosonic-qsp-qsvt}. An elaboration of this Hamiltonian 

 \begin{align}
     H&=g\bigg[-\tanh\left(\frac{\hat{x} + \hat{p}}{2}\right) (\hat{x} - \hat{p})\nonumber\\
     &~~~~~~~~ + \tanh\left(\frac{\hat{x} - \hat{p}}{2}\right) (\hat{x} + \hat{p})\bigg]
     \label{eq:4legflowfield}
 \end{align}
produces a flow field with four-fold rotational symmetry illustrated in Fig.~\ref{fig:TanhFlowField-4Leg}.  As described in Sec.~\ref{sec:qec-compilation}, this proves useful for quantum error correction using the bosonic 4-leg cat code.

As we discuss in more detail when we present compilation techniques in Sec.~\ref{sec:compilation}, these examples illustrate the fact that the classical intuition associated with visualizing flow fields can be extremely helpful to keep in mind when designing quantum unitary transformations that perform various useful tasks. We see that the nonlinearities needed for oscillator Hamiltonians can be implemented by either non-Gaussian oscillator controls or hybrid oscillator-qubit gates as defined in Tables \ref{tab:gates-osc} and \ref{tab:gates-qubit-osc}. We also define a minimum universal gate set in Table \ref{tab:ISA_overview}.

\subsubsection{Stellar-Holomorphic Representation}
\label{sssec:stellar-rep}

A less well-known representation of CV states is via analytic functions in the complex plane (i.e., holomorphic functions), the so-called stellar representation \cite{chabaud2021PhDThesis,ChabaudStellarRep,chabaud2022holomorphic,chabaud2023resources}, which has also been also used in condensed matter physics \cite{GirvinJach} to describe quantum Hall states. In the stellar representation, the bosonic creation and annihilation operators are mapped to a complex variable and its derivative operator. This representation provides a visualization of CV state dynamics in terms of the movement of the zeros of analytic functions. Additionally, it quantifies the non-Gaussian character of CV states (a useful resource for universal CV computation; see Sec.~\ref{sec:gates} for more details) in terms of the number of zeros of the analytic functions. We provide an introduction to the stellar representation in this section and toward the end also highlight open questions.

The basic idea is to use the following mapping from bosonic operators to a complex variable $z$ and the corresponding derivative
\begin{align}
    a^\dagger \rightarrow z, ~~~ a \rightarrow \partial_z,\label{eq:holomorphictranslation}
\end{align}
such that the canonical commutation relationship $[a, a^\dagger] = [\partial_z, z] = 1$ is preserved. Then, any CV state in the Fock representation $\sum_{n=0}^\infty \frac{\alpha_n}{\sqrt{n!}} (a^\dagger)^n \ket{0}$ (with $\alpha_n$ being a complex number) can be mapped to an analytic function of $z$
\begin{align}
    \sum_{n=0}^\infty \frac{\alpha_n}{\sqrt{n!}} (a^\dagger)^n \ket{0} \rightarrow f(z) \ket{0},
\end{align}
where 
\begin{align}
    f(z) = \sum_{n=0}^\infty \frac{\alpha_n}{\sqrt{n!}} z^n,
\end{align}
and the vacuum state is replaced by the integration measure
\begin{align}
    |0\rangle \rightarrow \frac{1}{\sqrt{\pi}}e^{-\frac{1}{2}|z|^2}.
\end{align}
By convention, the derivative $a \rightarrow \partial_z$ does not act on the integration measure.  Thus for example we have the translation
\begin{align}
    \langle 0|a^n(a^\dagger)^n|0\rangle &\rightarrow \frac{1}{\pi}\int d^2z e^{-|z|^2}\,\left[\frac{\partial^n}{\partial z^n}z^n\right] =n!\\
    &\rightarrow \frac{1}{\pi}\int d^2z e^{-|z|^2}\,\left[(z^*)^nz^n\right].
\end{align}
Here the first line follows from the picture that we are calculating the vacuum expectation value of the operator $a^n(a^\dagger)^n$ using the translation in Eq.~(\ref{eq:holomorphictranslation}), while the second line follows from integration by parts or from the picture that we are computing the norm of the state $|\psi\rangle = (a^\dagger)^n|0\rangle$ using the translation $a^\dagger\rightarrow z$.

Moreover, the analytic function $f(z)$ admits the following factorization \cite{conway2012functions} 
\begin{align}
    f(z) &= e^{-\frac{1}{2}Az^2+Bz+C}\,\, g(z), \,\,\,\,\mathrm{where}, \nonumber \\
    g(z) &= \prod_{j=1}^S (z-\lambda_j),
    \label{stellar-factorization}
\end{align}
for $A,B,C \in \mathbb{C}, \lambda_j \in \mathbb{C}$. The factorization in Eq.~\eqref{stellar-factorization} naturally decomposes $f(z)$ into a Gaussian part ($e^{-\frac{1}{2}Az^2+Bz+C}$) and a non-Gaussian (polynomial) part $g(z)$ which is an entire function and hence uniquely determined\footnote{Uniquely since the overall normalization of the state is fixed and the global phase is irrelevant.} by its zeros $\{\lambda_j;j=1,\ldots S\}$.  The number of zeros, $S$, is called the stellar rank and is a measure of the non-Gaussianity of the state \cite{walschaers2021non}  (and if $S=0$ we take $g(z)=1$).  Here the parameters $A,B,C$  respectively play the role of squeezing, displacing, and adding a constant phase (and normalization) to the CV states. In particular, for $A = |A| e^{i\phi}$, the magnitude $0\le |A| <1$ controls the squeezing of the CV state, while the phase $\phi$ controls the direction of squeezing. Similarly, $|B|$ controls the displacement, and $\arg(B)$ determines the direction of the displacement.  For $\phi=0$ (real $A$), $B=0$ we have for the purely Gaussian case ($S=0, g(z)=1$)
\begin{align}
    \langle (\hat x)^2\rangle &= \frac{1}{4}\left(\frac{1+A}{1-A}\right)\\
    \langle (\hat p)^2\rangle &= \frac{1}{4}\left(\frac{1-A}{1+A}\right),
\end{align}
which demonstrates the squeezing relative to the vacuum variances of $\frac{1}{4}$ for each that were given in Eqs.~(\ref{eq:cohvarx}-\ref{eq:cohvarp}).

Interestingly, it has been established that the difficulty of preparing a CV state can be quantified by the stellar rank in the above representation, while the dynamics of a qumode can be visualized by the movement (shifting, merging, and splitting) of the roots in the complex plane \cite{chabaud2022holomorphic}. Therefore, a time-dependent CV state is associated with a corresponding time-dependent analytic function $f(z, t)$.

An arbitrary unitary operation on a qumode can always be written as a function of $a^\dagger$ and $a$. Therefore, in the stellar representation, a CV unitary operation may be written as
\begin{align}
    U(z, \partial_z) = e^{-i H(z, \partial_z)}, 
\end{align}
such that the dynamics of a CV state is governed by the following differential equation on the corresponding analytic function $f(z, t)$:
\begin{align}
    i\partial_t f(z, t) = H(z, \partial_z) f(z, t).
\end{align}
This agrees with our discussion about visualizing CV state dynamics in Sec.~\ref{subsec:flowsinphasespace} as flows in phase space.  

As a simple example to better understand the CV dynamics in stellar representation, let us examine how the SNAP gate (described in Sec.~\ref{sec:gates})  moves the roots of the stellar function. By definition, the SNAP gate unitary $e^{i\sum_n\theta_n \ket{n}\bra{n}}$ imparts a different phase $\theta_n$ to each Fock state $|n\rangle$. In the stellar picture, this means the initial stellar function $\sum_{n = 0}^\infty a_n z^n \propto \prod_j (z - \lambda_j)$ is changed to $\sum_{n = 0}^\infty a_n e^{i\theta_n} z^n \propto \prod_{k} (z - \lambda'_k)$. In the special case of $\theta_n = n \theta$ for a constant $\theta$, the resulting stellar function is
\begin{align}
    \sum_n a_n e^{in\theta} z^n = \sum_n a_n (e^{i\theta} z)^n \propto \prod_j (e^{i\theta} z - \lambda_j).
\end{align}
This means the new roots $\lambda'_k$ are related to the old roots $\lambda_j$ via
\begin{align}
    \lambda'_k = e^{-i\theta} \lambda_k, \forall ~k
\end{align}
and the number of roots does not change. We note that this special SNAP gate is nothing but the free evolution of the oscillator, which induces a global phase transformation on all the roots, i.e., a global rotation in the phase space. It remains an open question as to how the roots of the stellar function move subject to the action of an arbitrary SNAP gate or other arbitrary unitary.

More concretely, it would be interesting and useful to develop theories and protocols to: i) shift an individual root of the stellar function while keeping all other roots unchanged; ii) delete an individual root; iii) create an additional root, by leveraging the instruction set architecture in Sec.~\ref{sec:gates}. This is an open problem at present, but developments along this line will provide insights that would allow us to better compile CV algorithms as we discuss more in Sec.~\ref{sec:compilation}. For example, a universal instruction set based on the stellar representation, i.e. ``stellar ISA", may be constructed composed of operations that can manipulate individual roots of the stellar function, in a similar spirit to the ISAs that we introduce in Sec.~\ref{sec:architecture}. Furthermore, when coupling to a qubit, the joint oscillator-qubit system requires two separate stellar functions to describe the oscillator state entangled with each of the two-qubit bases.
Finally, we note that  the stellar representation has been generalized to the multi-mode setting in \cite{ChabaudStellarRep}. For a more detailed discussion of non-Gaussian quantum states and stellar representations, we refer the reader to the tutorial \cite{walschaers2021non} and to a recent paper on `photon catalysis' for multi-mode non-Gaussian state preparation \cite{aralov2025photoncatalysisgeneralmultimode}.

\subsection{Truncation of the Hilbert Space}
\label{sec:truncation-hilbert-space}

The representation discussed in the previous section concerns an infinite Hilbert space which is not feasible in practice. Here, we discuss possible truncations that can be applied to the oscillator space in realistic settings while keeping the advantages of a bosonic system. The Hilbert space of an ideal harmonic oscillator has infinite dimension, which may lead to technical difficulties in theoretical treatments. Of course, for physically realizable states of an oscillator in real hardware, we can never approach infinite energy, and the inevitable effects of even small anharmonicities increase as the number of bosons increases. For theoretical discussion of an abstract machine model in terms of an ISA, truncation is necessary to avoid any notion of infinity as well as the possible inception of anharmonicity in physical hardware.  This former issue is particularly relevant as it prevents us from proposing algorithms that would use systems with infinite energy to instantaneously solve problems and also to ensure that standard algorithmic analysis techniques such as remainder estimates from Taylor's theorem can be rigorously used to bound approximation errors.

Truncations may be introduced in various forms. In the phase space picture, truncation can be applied by a Gaussian damping envelope centered at the origin of the phase space. It may also be implemented by limiting the number of roots of the stellar function (or equivalently the degree of the polynomial $g(z)$) throughout the computation. Within the Fock basis, truncation can be defined by omitting all Fock levels with energy higher than a certain threshold. Such a sharp energy cutoff in the Fock basis can also be replaced by a smoother damping function. A rigorous discussion of truncation errors is presented in Ref.~\cite{arzani2025can}.

In the following discussion, for simplicity, we use a sharp cutoff in Fock levels to truncate the Hilbert space.
This is achieved by defining a maximum allowed boson number, $N_\mathrm{max}$, to restrict the oscillator Hilbert space dimensionality to $d=N_\mathrm{max}+1$. Physically, this places an upper bound on the energy.  For example, for $N_\mathrm{max}=3$ we have for the matrix representation of the destruction operator 
\begin{equation}
    a=\left(\begin{array}{cccc}
    0&1&0&0\\
    0&0&\sqrt{2}&0\\
    0&0&0&\sqrt{3}\\
    0&0&0&0
    \end{array}
    \right)\label{eq:amatrix}
\end{equation}
and for the creation operator
\begin{equation}
    a^\dagger=\left(\begin{array}{cccc}
    0&0&0&0\\
    1&0&0&0\\
    0&\sqrt{2}&0&0\\
    0&0&\sqrt{3}&0
    \end{array}
    \right)\label{eq:adaggermatrix}
\end{equation}
Notice that these truncated representations correctly reproduce
\begin{equation}
a^\dagger a = \hat n = 
\left(
\begin{array}{cccc}
 0 & 0 & 0 & 0 \\
 0 & 1 & 0 & 0 \\
 0 & 0 & 2 & 0 \\
 0 & 0 & 0 & 3 \\
\end{array}
\right),
\end{equation}
but incorrectly yield
\begin{equation}
a a^\dagger  = 
\left(
\begin{array}{cccc}
 1 & 0 & 0 & 0 \\
 0 & 2 & 0 & 0 \\
 0 & 0 & 3 & 0 \\
 0 & 0 & 0 & 0 \\
\end{array}
\right)\ne \hat n + \hat I.
\end{equation}
Thus the commutator is also not the identity
\begin{equation}
[a, a^\dagger]
= 
\left(
\begin{array}{cccc}
 1 & 0 & 0 & 0 \\
 0 & 1 & 0 & 0 \\
 0 & 0 & 1 & 0 \\
 0 & 0 & 0 & -3 \\
\end{array}
\right)\ne  \hat I.
\end{equation}
It is therefore best if, before the numerical evaluation of any expression, one moves all the creation operators to the left of all the destruction operators (putting them in ``normal order'') taking into account the correct commutation relations.  In any case, $N_\mathrm{max}$ must be set large enough that the quantum amplitudes in the largest boson number states are extremely small throughout the entire calculation.

For low-energy states, it has been shown~\cite{tong2022provably} that a value of $N_{\max}$ that is logarithmic in the truncation error desired will often suffice for the simulation.  However, for dynamical simulation with a large number of bosons,  cruder but more widely applicable bounds can be used.  Specifically, if we assume that we have a state $\ket{\psi(0)}$ such that 
\begin{equation}
    \bra{\psi(0)} H\ket{\psi(0)} = E,
\end{equation}
it is easy to see that for any time-independent Hamiltonian $H$ this mean energy is a constant of motion, meaning that the time-evolved state satisfies $\bra{\psi(t)} H\ket{\psi(t)} = E$ for all $t\ge 0$.  We then have if $H= \omega a^\dagger a + V$ for positive semi-definite $V$ that
\begin{align}
    E &= \omega\bra{\psi(t)} a^{\dagger} a \ket{\psi(t)} + \bra{\psi(t)} V \ket{\psi(t)}\nonumber\\
    &\ge \omega \bra{\psi(t)} a^{\dagger} a \ket{\psi(t)}:=\omega \mathbb{E}({N}(t)) \,,
\end{align}
where $\mathbb{E}$ denotes an expectation value over the state, and $N(t)$ is the time-dependent number operator.
Markov's inequality states that the probability that a non-negative random variable achieves a value of $\aleph$ times its mean value is at most $1/\aleph$.  From this, it follows that
for any constant $\aleph\ge 1$
\begin{equation}
\sum_{j \ge \aleph E/\omega} |\braket{\psi(t)}{j}|^2 \le \frac{1}{\aleph}, \,\,\forall~t\ge 0.
\end{equation}
Thus if we wish for  the total probability of observing a boson number greater than the cutoff $N_{\max} = \aleph E/\omega$ to be at most $\delta_{\rm cut}$ for all times, then it suffices to choose
\begin{equation}
    N_{\max}\ge \frac{E}{\delta_{\rm cut}\omega}.
\end{equation}
Tighter bounds are possible given more information, but the simplicity and general applicability of this bound make it a useful if crude estimate of the cutoff for most cases with quadratic Hamiltonians in the field operators.

Much later in this tutorial, in Sec.~\ref{sec:mode-to-qubit}, various mappings between the oscillator representation and the qubit representation are discussed, to bridge the gap between these continuous-variable and the familiar discrete-variable systems. The non-triviality of the CV space is an exemplary resource that is hard to beat via a discrete architecture and hence becomes the central focus of this work in later sections. To highlight this non-triviality on the unique features of CV architecture, Sec.~\ref{app:compilation-cv-to-qubits} represents a quantitative compilation of simple bosonic operators into DV gate sets.

\section{Instruction Sets for Oscillator-Qubit Systems}
\label{sec:gates}

With the basics of oscillators and qubits laid out in the previous section, we are ready to present the important elementary gates that are  natively available in current experiments on superconducting and/or ion-trap hybrid qubit-bosonic systems. Given an understanding of the natively available gates, we can select subsets of the gates to define instruction set architectures and learn how to compile these to synthesize arbitrary unitaries 

required to execute quantum error correction, computation, and simulation circuits.  We relegate the discussion of the microscopic Hamiltonians and physical implementation details behind the natively available gates to App.~\ref{app:PhysImp}.

We organize our discussion of the natively available and more complex synthesized gates according to the complexity of their corresponding Hamiltonians. The coarsest classification of gates and their corresponding Hamiltonians is whether they are qubit-only (Table \ref{tab:gates-qubit}), oscillator-only (Table \ref{tab:gates-osc}),  or hybrid entangling gates (Table \ref{tab:gates-qubit-osc}). Tables~\ref{tab:gates-osc} and ~\ref{tab:gates-qubit-osc} list commonly available oscillator and qubit-oscillator hybrid gates arranged, respectively, and will be frequently referred to throughout this section. For completeness we also list the standard qubit gates in Table~\ref{tab:gates-qubit} of App.~\ref{app:PhysImp}.

To guide the reader familiar with qubits and Bloch spheres, we have introduced the notion of phase space for continuous-variable systems and provided intuition for CV unitaries in terms of visualization of Hamiltonian flows in phase space in Sec.~\ref{subsec:flowsinphasespace}. 
In Sec.~\ref{subsubsec:gaussian_ops}, we leverage this intuition to explain the parallels between rotations on the discrete-variable Bloch sphere and simple displacements in the continuous-variable phase space generated by Hamiltonians that are linear in $\hat x,\hat p$. As mentioned previously, we approach general unitary synthesis as a Hamiltonian synthesis problem by writing the unitary in the form of Eq.~(\ref{eq:Usynthesis_Hsynthesis}) or more simply, for the case of a time-independent Hamiltonian, in the form
\begin{align}
    U(t)=e^{-iHt}
\end{align}
where the Hermitian operator $H$ is the Hamiltonian associated with the gate.  In general, this is not simply the physical Hamiltonian, but rather an effective Hamiltonian synthesized in the hardware layer through the application of various microwave pulses.  We refer the interested reader to the following experimental papers  which discuss these issues \cite{Heeres2015,ReinholdErrorCorrectedGates,Campagne-Ibarcq2020,Sivak_GKP_2022,EickbuschECD,Kundra_2022_Robust,HomeCharacteristicFunction,lo2015spin,kienzler2015quantum,HOME-GKP2019,HomeGKPQEC2022} and to App.~\ref{app:PhysImp}.  Within the oscillator-only gate family, we can express the Hamiltonian $H(\hat x,\hat p)$ as a polynomial in the position ($\hat x$) and momentum ($\hat p$) operators of the oscillator mode(s) and classify gates according to the degree of the polynomial \cite{Braunstein2005}.

Sec.~\ref{sec:gaussian gates} presents standard Gaussian gate operations, whose corresponding Hamiltonians are at most quadratic in $\hat x,\hat p$. In Sec.~\ref{subsec:GaussiansAsCliffords}, we also discuss parallels (summarized in Table~\ref{table:qubitsvsCV}) between the group of Clifford gates for qubits and Gaussian gates for oscillators.  Like their discrete-variable cousins, Gaussian gates are non-universal, and Gaussian circuits can be efficiently simulated classically (assuming the initial state is Gaussian). The efficiency for qubit Clifford circuit simulation comes from being able to easily track the transformation of the stabilizers \cite{Gottesman-Knill-Theorem,AaronsonGottesmanCliffordSims}.  For Gaussian circuits, one simply follows the evolution of the means and the covariance matrix for the oscillator position and momentum \cite{PhysRevLett.109.230503,Veitch_2013}. 

Universal oscillator control requires being able to generate polynomials of arbitrary order using combinations of gates \cite{LloydQuantumComputation1999, Braunstein2005}, as discussed in App.~\ref{app:cubic_IS}. In Sec.~\ref{sec:non-Gaussian_and_Hybrid}, we explain that this can be achieved given access to a single non-Gaussian gate having a Hamiltonian of degree three or higher.  Note that this definition assumes no decoherence; in a realistic setting, decoherence needs to be accounted for, leading to a trade-off between using more gates for higher unitary precision and the corresponding increase in infidelity due to decoherence increasing with circuit depth. Sec.~\ref{sec:non-Gaussian_and_Hybrid} also introduces useful hybrid oscillator-qubit gates.

Importantly, we show in Sec.~\ref{sec:instruction_set} that hybrid oscillator-qubit gates with Hamiltonians of the form  
\begin{align}
    H=\vec h(\hat x,\hat p)\cdot\vec \sigma
\end{align}
can be compiled to generate universal unitaries for the full hybrid system even when the oscillator part $\vec h(\hat x,\hat p)$ is only of maximum degree one.  When $\vec h$ is homogeneous of degree zero we have the ordinary single-qubit rotation gate in Eq.~(\ref{eq:singlequbitrotation}).  When $\vec h$ is homogeneous of degree one, we see that can interpret the unitary as a rotation of the qubit that depends linearly on the position and momentum of the oscillator. We also see that we can alternatively view the same unitary as a displacement of the oscillator in phase space, conditioned on the state of the qubit.  By combining these viewpoints we hope to give the reader intuition about how such hybrid gates open up new possibilities for quantum signal processing \cite{low2017optimal,rossi2021mqsp,GrandUnificationAlgos,martyn2023efficient,rossi2023quantum,qspi2023} beyond those available in qubit-only and oscillator-only systems. 

\begin{figure}[t]
    \centering
\includegraphics[width=0.3\textwidth]{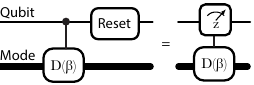}
    \caption{Conditional displacement (see Table~\ref{tab:gates-qubit-osc}) followed by a reset of the qubit is equivalent to the measurement of the qubit followed by feedback displacement conditioned on the measurement result.  In the first circuit, the information about which state was produced ends up in the environment (due to the reset).  In the second circuit, the experimentalist may have access to the information from the measurement result.  The equivalence is strictly true only if the measurement result is not available for subsequent use in the circuit.}
    \label{fig:meas+feedback}
\end{figure}

We note that by including measurements in a circuit, the concept of universal control can be extended to quantum channels that describe both unitary as well as non-unitary transformations of oscillators. Non-unitary or irreversible maps can be achieved by entangling the oscillator with an auxiliary qubit followed by measurement or reset of the qubit. 
 In Ref.~\cite{shen2017quantum} the authors show that efficient universal channel construction is possible using a single auxiliary qubit  with quantum non-demolition readout and adaptive control. It is admissible to replace readout and adaptive control with auxiliary qubit-assisted gates followed by a reset of the qubits (see for example,  Fig.~\ref{fig:meas+feedback}).
The same technique can be used to efficiently construct channels for oscillator systems.

\subsection{Gaussian Oscillator Control}
\label{subsubsec:gaussian_ops}

Building upon the visualization of classical flows in phase space in  Sec.~\ref{subsec:flowsinphasespace}, we begin our discussion of Gaussian control with its extension to the quantum case which involves unitary transformations in Hilbert space. Sec.~\ref{subsubsec:gaussian_ops_basics} gives a general discussion about Gaussian oscillator control using quadratic Hamiltonians, which induce linear (symplectic) transformations on the phase space; Sec.~\ref{compose_gauss} discusses composition of such CV transformations.

\subsubsection{Basics of Gaussian Control}
\label{subsubsec:gaussian_ops_basics}

Gaussian states are defined to be those whose wave functions $\Psi(x_1,\ldots,x_N)$ are Gaussians. The ground state of a harmonic oscillator is a prototypical Gaussian state.
Gaussian unitaries are defined to be those generated by Hamiltonians that are at most quadratic in the position and momentum operators $\hat x_i,\hat p_i$ for the various modes (or equivalently at most quadratic in the creation and annihilation operators $a_i,a^\dagger_i$) such as the quantum version of the classical squeezing Hamiltonian given in Eq.~(\ref{eq:squeezingHam}).  We return to this specific example shortly and see that all Gaussian unitaries applied to Gaussian states preserve the Gaussian nature of such states.  But first, we begin with some general remarks on the quantum version of phase space where the position and momentum operators do not necessarily commute. 

Given $N$ oscillators, let us define the quantum operator phase space coordinate vector of length $2N$
\begin{align}
    \hat{\mathbf{R}} = \left(\hat x_1,\hat p_1, \hat x_2, \hat p_2,..., \hat x_N,\hat p_N\right).
    \label{eq:phasespacecoordinatevector}
\end{align}
The most general quadratic Hamiltonian $H_G$ can be written as
\begin{align}
    H_\mathrm{G}=\sum_{i=1}^{2N}\lambda_i \hat R_i+\frac{1}{2}\sum_{i,j}^{2N} \hat R_i\Lambda_{ij}\hat R_j, \label{eq:gaussianH}
\end{align}
where the vector  ${\bm{\lambda}}$ is real and the matrix $\mathbf\Lambda$ is real and symmetric.

It is useful to define the antisymmetric matrix
\begin{align}
    \mathbf{\Omega}=\bigoplus_N\left(  \begin{array}{cc} ~~0&+1\\-1& ~~0  \end{array}\right),
\end{align}
which allows us to represent the commutator of different elements of $\hat{\mathbf{R}}$ as
\begin{align}
    [\hat R_j,\hat R_k]=+i\Omega_{jk}.
    \label{eq:symplecticcommutator}
\end{align}
All Gaussian unitaries take the form 
\begin{align}
    U_G(t) = \exp\left(-i H_\mathrm{G}t\right)
    \label{eq:GaussianUnitary}
\end{align}
 and (since the Heisenberg equation of motion is linear) produce a linear transformation of the form
\begin{align}
    \hat{\mathbf{R}}(t) &=U_G^\dag(t)\hat{\mathbf{R}}(0) U_G(t)\label{eq:GaussianXform0}\\
    &= \mathbf{\Delta}(t)+\mathbf{M}(t)\hat{\mathbf{R}}(0),\label{eq:GaussianXform}
\end{align}
where $\mathbf\Delta$ is a real valued phase-space displacement vector and $\mathbf{M}(t)$ is a real matrix. It is important to note that Eq.~(\ref{eq:GaussianXform0}) describes element-wise conjugation of $\hat{\mathbf{R}}(0)$ by the unitary.  It does not represent matrix multiplication of matrices with compatible dimensions. By contrast, the second term in Eq.~(\ref{eq:GaussianXform}) does represent ordinary multiplication of the column vector $\hat{\mathbf{R}}(0)$ by the matrix $\mathbf{M}(t)$.

In order to preserve the commutation relation in Eq.~(\ref{eq:symplecticcommutator}),  $\mathbf{M}(t)$ must obey
\begin{align}
     \mathbf{M}^\mathrm{T}(t)\mathbf{\Omega}\mathbf{M} (t)=\mathbf{\Omega},
     \label{eq:symplectic}
\end{align}
meaning that $\mathbf{M}(t)\in \mathrm{Sp}(2N,\mathbb{R})$ must be a real symplectic matrix. 

Eq.~(\ref{eq:GaussianXform0}) is solved using the Heisenberg equation of motion  
    \begin{align}
        \frac{d\hat{\bm R}(t)}{dt}&=+i[H_\mathrm{G},\hat{\bm R}(t)].
    \end{align}
Two limiting cases are particularly simple
\begin{enumerate}
    \item[]Case I:  $H_\mathrm{G}={\bm \lambda}\hat{\bm R}$
    \begin{align}
        \frac{d\hat{\bm R}(t)}{dt}&={\bm \Omega}{\bm\lambda},\\
        \hat{\bm R}(t)&=\hat{\bm R}(0)+{\bm\Delta(t)},\\
        {\bm\Delta(t)}&={\bm \Omega}{\bm\lambda}t.\label{eq:lineardisplacement_t}
    \end{align}
    \item[]Case II: $H_\mathrm{G}=\frac{1}{2}\hat{\bm R}{\bm\Lambda}\hat{\bm R}$
    \begin{align}
         \frac{d\hat{\bm R}(t)}{dt}&={\bm \Omega\bm\Lambda}\hat{\bm R}(t),\\
         \hat{\bm R}(t)&={\bm M}(t)\hat{\bm R}(0),\\
         {\bm M}(t)&=\exp\left({\bm\Omega\bm\Lambda}t\right).
    \end{align}
\end{enumerate}
It is straightforward to show that this exponential form of ${\bm M}(t)$ is indeed symplectic and obeys Eq.~(\ref{eq:symplectic}).  The general case of the Hamiltonian in Eq.~(\ref{eq:gaussianH}) can be reduced to Case II by simply completing the square, yielding
\begin{align}
    \hat{\bm R}(t) &=-{\bm\Lambda}^{-1}{\bm\lambda}+\exp\left({\bm\Omega\bm\Lambda} t\right)\left(\hat{\bm R}(0)+{\bm\Lambda}^{-1}{\bm\lambda}\right).
\end{align}

It is sometimes convenient to represent the symplectic transformations not in terms of the Hermitian position and momentum operators, $\hat x,\hat p$, but rather in terms of the complex mode operators, $a$ and $a^\dagger$. In this basis, the symplectic transformation matrices can be complex. We note that the displacement vector $\mathbf{\Delta}$ does not affect the commutation relations. Further details can be found in Refs.~\cite{Weedbrook2012} and \cite{royer2022encoding}.  

\begin{figure}
\includegraphics[width=0.98\linewidth]{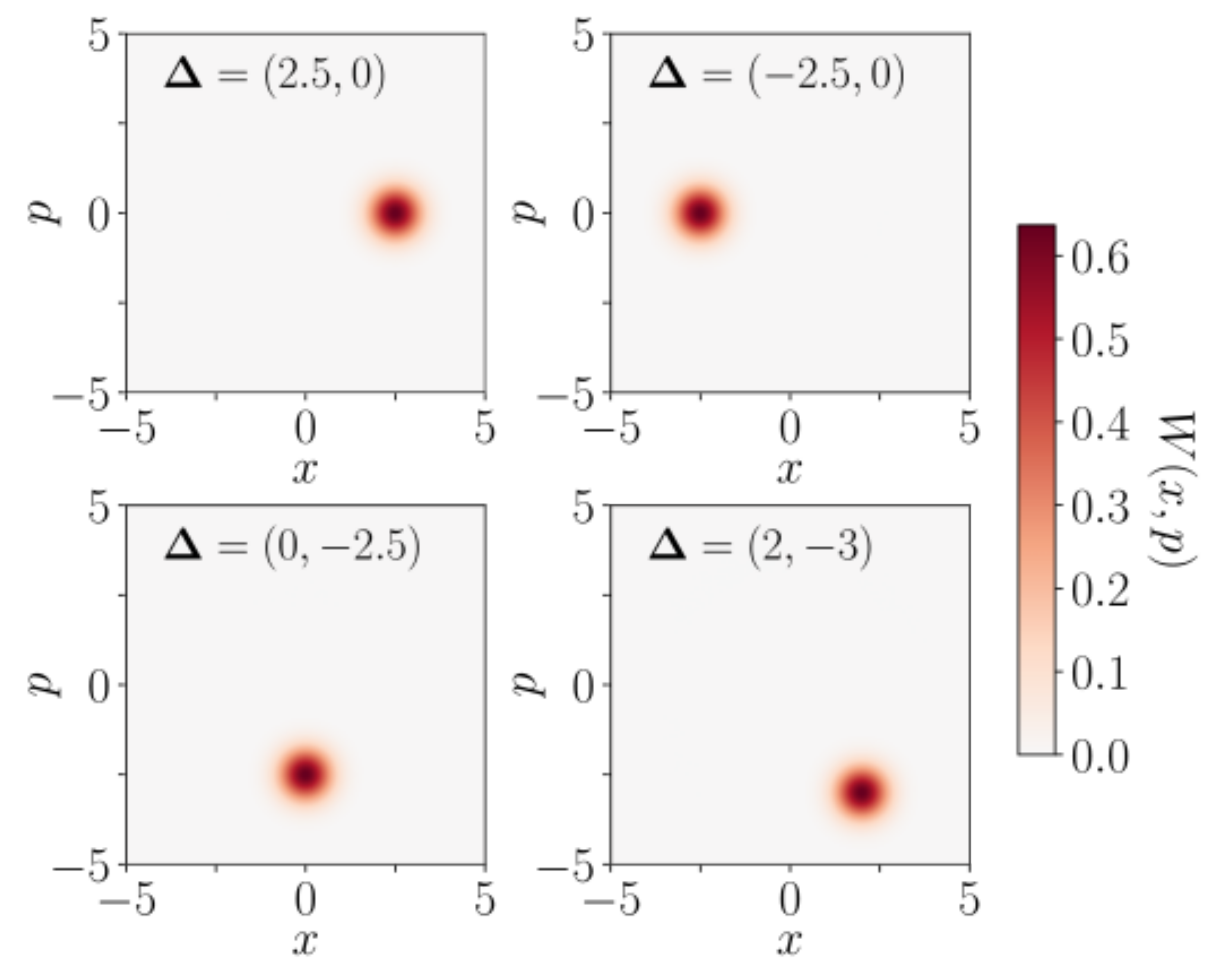}
\caption{Wigner quasiprobability distribution after evolution under the linear displacement Hamiltonian in Eq.~(\ref{eq:quantum_displacement_ham}) for various values of ${\bf \Delta}=(+\lambda_p t, -\lambda_x t)$.}
\label{fig:displacement_examples}
\end{figure}

As an example of Case I for a single oscillator, the simplest nontrivial control Hamiltonian is the linear Hamiltonian
\begin{align}
    H_G=\lambda_p\hat p + \lambda_x\hat x,
    \label{eq:quantum_displacement_ham}
\end{align}
for which the Heisenberg equations of motion are (in standard units)
\begin{align}
    \frac{d}{dt}\hat x &= +\lambda_p \\
    \frac{d}{dt}\hat p &= -\lambda_x ,
\end{align}
which yields the displacement canonical transformation with
\begin{align}
    \mathbf{\Delta}(t)&={\mathbf\Omega}{\bm{\lambda}} t=(+\lambda_p,-\lambda_x)t,\label{eq:DeltaisOmegalambda}\\
    \mathbf{M}(t)&=\hat{I},
\end{align}
in Eq.~(\ref{eq:GaussianXform}), as illustrated in Fig.~\ref{fig:displacement_examples}.

As a second simple example, this time for Case II, let us examine (in standard units) the quantum version of the squeezing Hamiltonian in Eq.~(\ref{eq:squeezingHam})
\begin{align}
    H_G = \frac{g}{2}[\hat x\hat p + \hat p\hat x],\label{eq:QsqueezeH}
\end{align}
which must be symmetrized for it to be Hermitian and corresponds to the case
\begin{align}
    \bm{\lambda}&=\bm{0},\\
    \bm{\Lambda}&=g\left(\begin{array}{cc}0&1\\1&0\end{array}\right).
    \label{eq:quantum_squeezing_ham}
\end{align}
The solution of the Heisenberg equation of motion is identical to the classical case discussed above
\begin{align}
    \mathbf{\Delta} &=\mathbf{0},\\
    \mathbf{M} &=\exp\left({\bm\Omega\bm\Lambda}t\right)=\left(\begin{array}{cc}e^{gt}&0\\0&e^{-gt}    \end{array}\right),\\
    \hat x(t)&=e^{+gt}\hat x(0),\\
    \hat p(t)&=e^{-gt}\hat p(0),
\end{align}
and preserves the (standard units) commutation relation
\begin{align}
    [\hat x(t),\hat p(t)]=[\hat x(0),\hat p(0)]=+i.
\end{align}

\begin{figure}
    \includegraphics[width=0.98\linewidth]{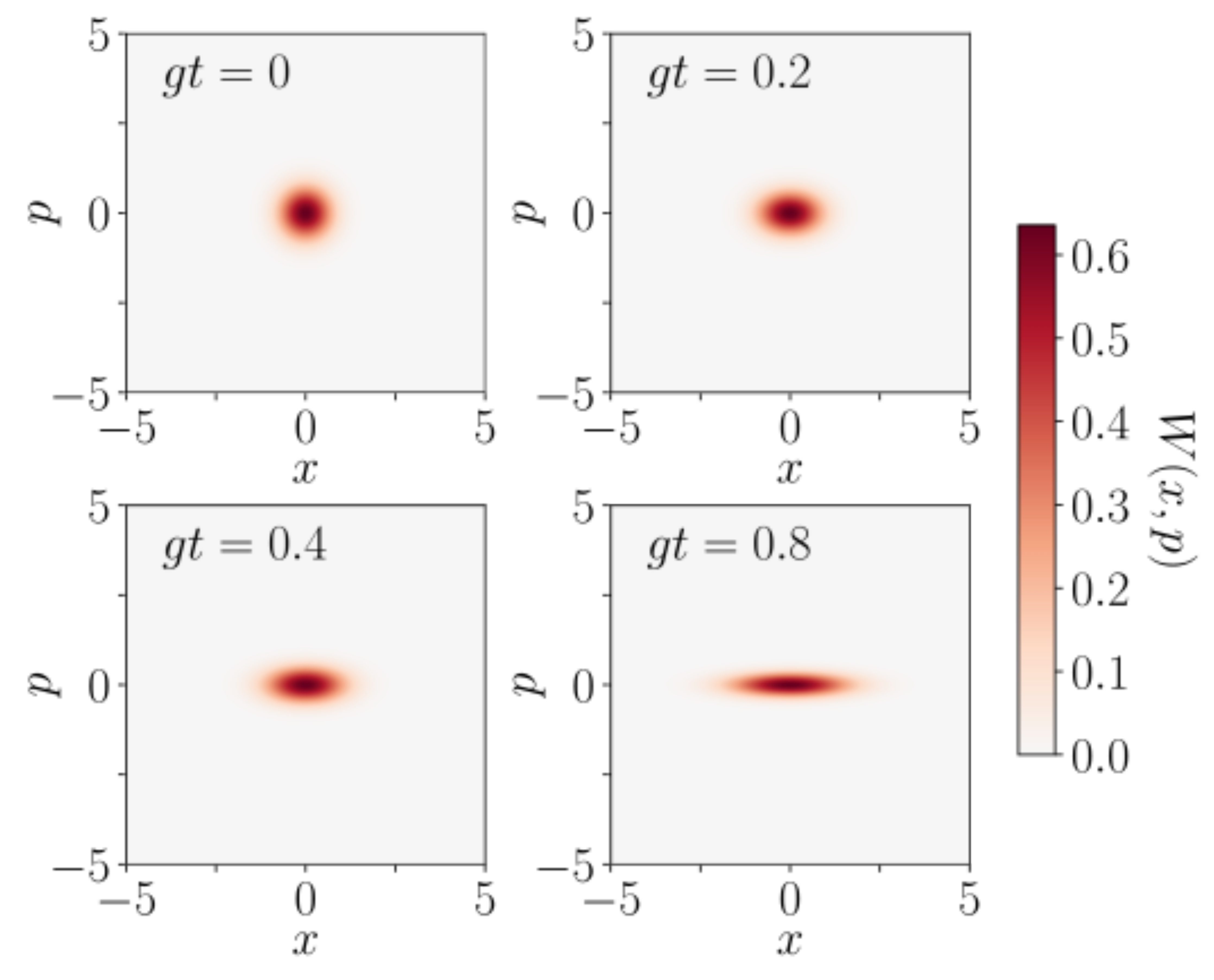}
    \caption{Wigner quasiprobability distribution of an initial vacuum state evolved under the squeezing Hamiltonian in Eq.~(\ref{eq:quantum_squeezing_ham}) for the same values of $gt$ used in the lower panel of Fig.~\ref{fig:SqueezingPlot}.}
    \label{fig:squeezing_examples_plot}
\end{figure}

Reverting to the Schr{\"o}dinger picture and time evolving the vacuum state under the squeezing Hamiltonian yields the so-called `squeezed vacuum' 
\begin{align}
    |\psi(t)\rangle = e^{-iHt}|0\rangle.
\end{align}
The Wigner (quasi-probability) function (defined in App.~\ref{sec:characteristic-function}) of this state is shown in Fig.~\ref{fig:squeezing_examples_plot} for the same parameter values as in Fig.~\ref{fig:SqueezingPlot}.  Notice the close similarity of the shape of the Wigner function to the shape of the circular region of classical phase space that has been squeezed into an ellipse in the lower panel of Fig.~\ref{fig:SqueezingPlot}.  In the classical case, the initial circular region was chosen arbitrarily.  For the Wigner function, the area of the initial circular region and the subsequent squeezed regions is fixed by the size of the (squeezed) vacuum fluctuations given in Eqs.~(\ref{eq:Sq1}, \ref{eq:Sq2}).

\subsubsection{Composition of Two Gaussian Transformations}\label{compose_gauss}
A subtlety arises when evaluating Eq.~(\ref{eq:GaussianXform}) for the case of applying two successive Gaussian transformations $U_1$ and $U_2$ that individually obey
\begin{align}
    \hat{\mathbf{R}}_1 &=U_1^\dag\hat{\mathbf{R}}_0 U_1=\mathbf{\Delta}_1+\mathbf{M}_1\hat{\mathbf{R}}_0\\
    \hat{\mathbf{R}}_2 &=U_2^\dag\hat{\mathbf{R}}_0 U_2=\mathbf{\Delta}_2+\mathbf{M}_2\hat{\mathbf{R}}_0.\label{eq:GaussianComposition}
\end{align}
To illustrate this point, for simplicity, we specialize to the case $\mathbf{\Delta}_1=\mathbf{\Delta}_2=\mathbf{0}$ and consider the sequential transformation
\begin{align}
    \hat{\mathbf{R}}_3 &=U_1^\dag U_2^\dag\hat{\mathbf{R}}_0U_2 U_1.\label{eq:GaussianComposition2}
\end{align}
In the Schr{\"o}dinger picture, we can understand this sequential transformation as an initial active transformation of the state $|\Psi\rangle$ to a new state $U_1|\Psi\rangle$, followed by a subsequent active transformation resulting in the final state $U_2(U_1|\Psi\rangle)$.  In the Heisenberg picture, however, one can understand
$U_1^\dagger U_2^\dag\hat{\mathbf{R}}_0 U_2 U_1$ as carrying out a sequence of passive transformations to change the basis.  Naively, Eq.~(\ref{eq:GaussianComposition2}) appears to tell us to carry out the passive transformations in reverse order: $U_2$ first and then $U_1$ so that the net result would be $\hat{\mathbf{R}}_3=\mathbf{M}_1\mathbf{M}_2\hat{\mathbf{R}}_0$.  However, this is incorrect as can be seen by rewriting Eq.~(\ref{eq:GaussianComposition2}) in the form recommended by Schwinger \cite{SchwingerPassiveU}
\begin{align}
    \hat{\mathbf{R}}_3 &=[U_1^\dag U_2^\dag U_1] [ U_1^\dag\hat{\mathbf{R}}_0 U_1] [U_1^\dag U_2 U_1].\label{eq:GaussianComposition3}
\end{align}
Here, the interpretation is that one first applies the passive transformation $U_1$ followed by the passive transformation $U_2$ (in the new basis generated by the first passive transformation).  Thus the correct final expression is
\begin{align}
    \hat{\mathbf{R}}_3 &=\mathbf{M}_2\mathbf{M}_1\hat{\mathbf{R}}_0,\label{eq:GaussianComposition4}
\end{align}
showing that the symplectic transformations of the operators are applied in the same order as the (active) unitary transformations $U_2U_1$ are applied to the state.

\subsection{Gaussian Operations}
\label{sec:gaussian gates}
A generating set of Gaussian gates \cite{GKP2001} is given by displacement (Sec.~\ref{subsec:displacement gate}), phase-space rotation which turns out to be related to the Fourier transform (Sec.~\ref{subsec:phase space rotation gate}), single-mode squeezing (Sec.~\ref{subsec:SMSqueezingGate}), and beam splitting (Sec.~\ref{sec:TMBSGates}) or two-mode squeezing (Sec.~\ref{subsec:tms gate}); 
see Table \ref{tab:gates-osc}. As noted in the caption of Table \ref{tab:free_interaction},  all of the Gaussian operations on oscillators that we discuss here are natively available in the superconducting platform.  Compilation of these primitive gates into arbitrary Gaussian unitaries can be efficiently performed using the Bloch-Messiah decomposition of the corresponding symplectic transformation matrices (see App.~\ref{app:bloch-messiah}) and other standard techniques \cite{Braunstein2005squeezing,Weedbrook2012,PhysRevA.94.062109}.
An over-complete list of primitive operations that can be used to generate arbitrary multi-mode Gaussian operations is the following:

\begin{itemize}
    \item Displacement
    \item Fourier transform 
    \item Single-mode squeezing
    \item Beam-splitter
    \item Two-mode squeezing
\end{itemize}

\noindent Displacements are used to generate the elements of the Heisenberg-Weyl group, $\mathrm{H}(n)$, while the next three are used to generate unitaries corresponding to the $\mathbf{M}\in \mathrm{Sp}(2n,\mathbb{R})$ in Eq.~(\ref{eq:symplectic}) generated by the algebra of the Lie brackets of the homogeneous quadratic polynomials.  In this section, we elaborate on these operations and the decomposition of a few other well-known Gaussian operations mentioned in Table~\ref{tab:gates-osc} in terms of the primitives.  For convenience, we switch here from using position and momentum operators to creation and annihilation operators.

\subsubsection{Displacement Gate}\label{subsec:displacement gate} The phase-space translation operation described above and listed in Box \ref{Box:UncondDispGate} is a very important primitive for any bosonic mode instruction set.  The displacement in phase space is parametrized by the complex number $\alpha$ and the gate has the following properties 
\begin{eqnarray}
D^\dagger(\alpha) &=& D(-\alpha)\label{eq:dispdagisneg}\\
    D^\dag(\alpha)aD(\alpha) &=&a+\alpha.
    \label{eq:translatea}
\end{eqnarray}
From this and Eq.~(\ref{eq:aisx+ip}) it follows that $D(+\alpha)$ is a translation in phase space by an amount (in Wigner units) $\Delta x=\mathrm{Re}\,\alpha$ and $\Delta p=\mathrm{Im}\,\alpha$. From Eq.~(\ref{eq:translatea}) it follows that the coherent state $|\alpha\rangle$ is simply the displaced vacuum state
\begin{align}
    |\alpha\rangle = D(\alpha)|0\rangle,
\end{align}
and more generally, that displacement maps coherent states to coherent states
\begin{align}
    D(\alpha)|\beta\rangle=|\alpha+\beta\rangle.
\end{align}

The normal-ordered form of the displacement operator in Eq.~(\ref{eq:dispopnormalordered}) is preferred for numerical calculations where the Hilbert space of the oscillator is truncated to the lowest $d=N_\mathrm{max}+1$ states.

\abox{Displacement Gate}{
\begin{eqnarray}
    D(\alpha) &\equiv& e^{\alpha a^\dagger - \alpha^*a}\label{eq:dispop}\\
    &=& e^{-\frac{\alpha^*\alpha}{2}}
    e^{\alpha a^\dagger}e^{- \alpha^* a},\label{eq:dispopnormalordered}
    \end{eqnarray}
    Composition of Displacement Gates
\begin{eqnarray}
    D(\beta) D(\alpha) &=& D(\alpha+\beta)e^{+\frac{1}{2}[\beta\alpha^*-\beta^*\alpha]},\label{eq:UCDcompos1}\\
    D(\alpha) D(\beta) &=& D(\alpha+\beta)e^{-\frac{1}{2}[\beta\alpha^*-\beta^*\alpha]},\label{eq:UCDcompos2}\\
    &=&   D(\beta) D(\alpha)e^{-[\beta\alpha^*-\beta^*\alpha]},\label{eq:UCDcompos3}\\
    &=&D(\beta) D(\alpha)e^{-2iA(\alpha,\beta)},\label{eq:UCDcompos4}
\end{eqnarray} 
where $A(\alpha,\beta)$ is the oriented area of the parallelogram formed by the displacements $\alpha,\beta$.
\\ \\
In terms of the phase-space coordinates, the multi-mode displacement gate is given by Eq.~(\ref{eq:GaussianUnitary}) with only linear terms in the Hamiltonian,
and the resulting displacement is given by Eq.~(\ref{eq:lineardisplacement_t}).
\label{Box:UncondDispGate}
} 

\begin{figure}[htb]
    \centering
    \includegraphics[width=0.35\textwidth]{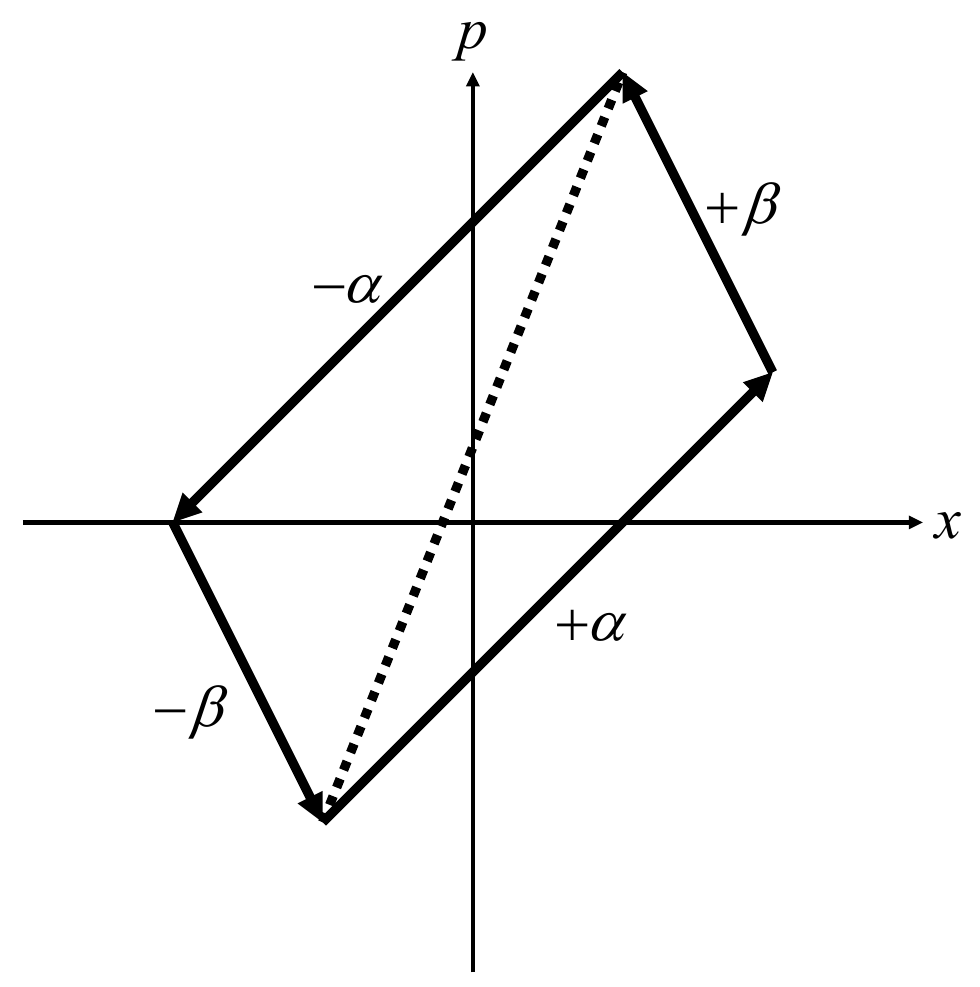}
    \caption{Traversing a right-handed closed path of area $A$ in phase space yields a geometric phase factor $e^{2iA}=e^{\alpha^*\beta-\alpha\beta^*}$ (in the dimensionless Wigner units we are using).}
    \label{fig:parallelogram}
\end{figure}

Mathematically, the composition (multiplication) of two translations gives another translation, and hence these operations form a group known as the Weyl-Heisenberg group.
Because the generators $\hat x,\hat p$ (or equivalently $a,a^\dagger$) do not commute, the group is non-Abelian. In analyzing the structure of the group of translations, it is helpful to use the following identity
\begin{equation}
   e^{B+C} =e^{-\frac{1}{2}[B,C]} e^Be^C
\end{equation}
which is valid provided that both $B$ and $C$ commute with their commutator $[B,C]$. 
From this, the displacement composition rules shown in Eqs.~(\ref{eq:UCDcompos1}-\ref{eq:UCDcompos4}) follow, with
$A(\alpha,\beta)$ being the oriented area of the parallelogram in the complex plane defined by $\alpha$ and $\beta$ as shown in Fig.~\ref{fig:parallelogram}.  In the Wigner units we have chosen, if this area is an integer multiple of $\pi$ ($2\pi$ in standard dimensionless units), the two displacements commute.  Using Eq.~(\ref{eq:dispdagisneg}) and the composition rules,
we see that if we take the system along a right-handed loop around the parallelogram the system returns to its original state up to a geometric phase
\begin{equation}
      D^\dagger(\beta) D^\dagger(\alpha) D(\beta) D(\alpha)= e^{+2iA(\alpha,\beta)}\,\hat I.
      \label{eq:phase_space_gates}
\end{equation}

This geometric phase plays a crucial general role in the control of bosonic modes and a particular role in the Gottesman-Kitaev-Preskill (GKP) quantum error correction code \cite{GKP2001,GrimsmoPuri_GKP_2021}. Separately, it can be used to compile entangling gates between physical qubits, as we explain in Sec.~\ref{ssec:compilation-entangling}.

\subsubsection{Phase-Space Rotation Gate}\label{subsec:phase space rotation gate} Another simple gate for a cavity is the phase-space rotation gate given in Box~\ref{Box:phase-space-rotation}.

The phase-space rotation gate has the following properties, 
\begin{eqnarray}
R(\theta)|\alpha\rangle &=& |e^{-i\theta}\alpha\rangle,\\
R^\dagger(\theta)&=&R(-\theta),\\
R^\dagger(\theta)a R(\theta) &=& e^{-i\theta}a,\\
R^\dagger(\theta)a^\dagger R(\theta) &=& e^{+i\theta}a^\dagger,\\
R^\dagger(\theta)\hat x R(\theta) &=& \cos\theta\, \hat x + \sin\theta\,\hat p,\\
R^\dagger(\theta)\hat p R(\theta) &=& \cos\theta\, \hat p - \sin\theta\,\hat x,
 \end{eqnarray}
corresponding to a clockwise 
rotation of the state in phase space by angle $\theta$ via the symplectic transformation
\begin{align}
\left(\begin{array}{c}
\hat x(\theta)\\ \hat p(\theta)
\end{array}\right)=\left(\begin{array}{cc}\cos\theta&+\sin\theta\\-\sin\theta&\cos\theta\end{array}\right)\left(\begin{array}{c}
\hat x\\ \hat p
\end{array}\right).
\end{align}

The rotation gate $R(\pi/2)$ executes a quantum Fourier transform
from the position basis to the momentum basis by simply rotating the state in phase space \cite{Liu2016e}.  If we view the oscillator motion in the non-rotating lab frame, we see that the fact that the potential energy of an oscillator is quadratic in the position coordinate means that if one simply waits for a quarter cycle of the oscillatory motion, the position is precisely converted into (negative) momentum.  This is true quantum mechanically as well as classically.  For example a position-squeezed state (described in Sec.\ \ref{subsec:SMSqueezingGate}) that has highly uncertain momentum is converted upon rotation in phase space by $\pi/2$ into a state with
definite momentum and highly uncertain position. See Sec.~\ref{sssec:qft} for applications of the quantum Fourier transform gate.

\abox{Phase-Space Rotation Gate}{
\begin{equation}
 R(\theta)=e^{-i\theta\hat n}.   
\end{equation}
Quantum Fourier Transform Gate
\begin{equation}
 F=R\left(\frac{\pi}{2}\right)=e^{-i\frac{\pi}{2}\hat n}.
\end{equation}
\begin{align}
    F^\dagger \hat x F &=+\hat p\\
    F^\dagger \hat p F &=-\hat x.
\end{align}
\label{Box:phase-space-rotation}
}

Notice that free evolution in the lab frame under Hamiltonian $H_0$ in Eq.~(\ref{eq:SHOham}) for a time $t$ yields a clockwise rotation by $\theta=\omega_R t$.  Consequently, it is not necessary to apply any perturbation or drive to the cavity to carry out the gate. Instead, one simply needs to shift the control system's time reference. This shift of the time reference is equivalent to a shift in the phase of the carrier wave at the resonator frequency used to effect displacement gates.  
Thus one simply has to keep track at the classical level (i.e., `in software') that any subsequent
drive amplitude applied to the resonator is phase shifted by $\theta$ and 
any instruction set can readily capture this type of \emph{in silico} gate (also known as a `virtual' gate).

\subsubsection{Single-Mode Squeezing Gate}\label{subsec:SMSqueezingGate} Squeezing is a fundamental quantum resource for CV systems \cite{Braunstein2005}. The single-mode squeezing gate in Box~\ref{Box:1-mode-squeezing} can be used to reduce the quantum fluctuations in the oscillator position by carrying out the transformation
\begin{eqnarray}
    \hat x_r &=& e^{-r}\hat x\label{eq:xsqueezex}\\
    \hat p_r &=& e^{+r}\hat p.\label{eq:xsqueezep}
\end{eqnarray}
It is clear that this is a canonical(i.e., symplectic) transformation because it anti-squeezes the momentum by just the right amount to preserve the commutation relation (here in Wigner units),
\begin{equation}
    [\hat x_r,\hat p_r]=[\hat x,\hat p]=+\frac{i}{2}.
\end{equation}
As a result, squeezed states saturate the Heisenberg inequality by minimizing the product of the variances of position and momentum
\begin{eqnarray}
\langle 0|\hat x_r^2|0\rangle -\langle 0|\hat x_r|0\rangle^2 &=&\frac{1}{4}e^{-2r},\label{eq:Sq1}\\
\langle 0|\hat p_r^2|0\rangle - \langle 0|\hat p_r|0\rangle^2 &=&\frac{1}{4}e^{+2r},\label{eq:Sq2}
\end{eqnarray}
just as coherent states do in Eqs.~(\ref{eq:cohvarx}-\ref{eq:cohvarp}).

Unlike the displacement transformation which is generated by terms linear in $a$ and $a^\dagger$, the squeezing transformation is generated by quadratic terms 
\begin{align}
	s(r) &= e^{\frac{1}{2}r(a^2 - a^{\dag 2})},\\
 &=e^{ir(\hat x\hat p+\hat p\hat x)}.
	\label{eq:xsqueezingunitary}
\end{align}
Note that the latter form matches the time evolution using the Hamiltonian in Eq.~(\ref{eq:QsqueezeH}) with $r=-2gt$. 
These quadratic terms (`two-photon driving') can be generated by the nonlinear quantum optics technique of three- or four-wave mixing, or through dissipation engineering \cite{Castellanos-Beltran_2008,Dassonneville_2021,EickbuschECD,YYGao_protecting_cats_squeezing_PhysRevX.13.021004,kienzler2015quantum,lo2015spin}. The current state-of-the-art in superconducting circuits yields squeezing up to factors $e^{2r} \sim 10$~dB~\cite{EickbuschECD}, where 1~dB means a factor of $10^{1/10}$ and $n$~dB represents a factor of $10^{n/10}$. 
In superconducting circuits, the maximum squeezing is limited in part by the self-Kerr effect (weak anharmonicity induced in the resonator by the presence of the qubit), an effect we have neglected here. 

\abox{Single-mode Squeezing Gate}{
\begin{eqnarray}
    S(r,\theta)&=&e^{+i\theta\hat n}
    s(r)e^{-i\theta\hat n}\\
    &=& e^{\frac{1}{2}(\zeta^* a^2 - \zeta a^{\dag 2})}
\end{eqnarray}
where $\zeta=re^{i2\theta}$ and $s(r)$ is defined in Eq.~(\ref{eq:xsqueezingunitary}).  This operation performs the transformation
\begin{align}
(+\cos\theta \hat x + \sin\theta \hat p)  & \rightarrow  e^{-r}(+\cos\theta \hat x + \sin\theta \hat p),\\
(-\sin\theta \hat x + \cos\theta \hat p) & \rightarrow  e^{+r}(-\sin\theta \hat x + \cos\theta \hat p).
\end{align}
The symplectic transformation matrix (in the mode basis) induced by $s(r)$ is given in Eq.~(\ref{eq:smsqueezesymplectic}).
\label{Box:1-mode-squeezing}
}

To verify that the unitary in Eq.~(\ref{eq:xsqueezingunitary}) is correct, let 
\begin{eqnarray}
a^{\phantom{\dagger}}_r&=& s^\dagger(r)as(r),\\
a^\dagger_r&=& s^\dagger(r)a^\dagger s(r),
\end{eqnarray}
which obey
\begin{align}
    \frac{d}{dr}\left(\begin{array}{c}a_r\\a^\dag_r\end{array}\right) &=-\left(\begin{array}{cc}0&1\\1&0\end{array}\right)\left(\begin{array}{c}a_r\\a^\dag_r\end{array}\right),
    \end{align}
    and has solution
    \begin{align}
        \left(\begin{array}{c}a_r\\a^\dag_r\end{array}\right) &=
        \mathbf{M}_s(r)
        \left(\begin{array}{c}a\\a^\dag\end{array}\right),
    \end{align}
    where the symplectic mode transformation matrix is given by
    \begin{align}
    \mathbf{M}_s(r)=\left(\begin{array}{cc}
        \cosh r&-\sinh r\\
        -\sinh r&\cosh r
        \end{array} \right).\label{eq:smsqueezesymplectic}
        \end{align}
These results in turn yield Eqs.~(\ref{eq:xsqueezex}-\ref{eq:xsqueezep}) showing that the symplectic transformation matrix in the position/momentum basis is diagonal.
The unitary transformation in Eq.~(\ref{eq:xsqueezingunitary}) squeezes the position fluctuations of the oscillator.  As shown in Box~\ref{Box:1-mode-squeezing}, this transformation can be generalized to squeeze any quadrature (that is, any linear combination of $\hat x$ and $\hat p$) by simply applying a rotation transformation.  Notice that rotating the squeezing axis by $\theta=\pi$ returns us to the original squeezed state with $\theta=0$ as it should.

\subsubsection{Two-Mode Beam-Splitter Gates}\label{sec:TMBSGates} Recent experimental progress using parametric modulation of the coupling between two bosonic modes \cite{chapman2022high,lu2023highfidelity} has allowed the high-fidelity realization of Hamiltonians which are bilinear in operators for the two modes.  The first of these is the beam-splitter Hamiltonian
\begin{equation}
    \hat V_\mathrm{BS}= \frac{1}{2}(\chi^* ab^\dagger+\chi a^\dagger b)\label{eq:hatVBS}.
\end{equation}
Evolution under this Hamiltonian for time $t$ with $\chi t=\theta e^{i\varphi}$ yields the beam-splitter unitary gate given in Box \ref{Box:beam-splitter}.
The corresponding $4\times 4$ symplectic mode operator transformation matrix is given in Eq.~(\ref{eq:BSsymplec}) found in App.~\ref{app:bloch-messiah}.

\abox{Beam-splitter Gate}{

\begin{align}
    \mathrm{BS}(\theta,\varphi)=e^{-i\frac{\theta}{2}\left[e^{i\varphi} a^\dag b+e^{-i\varphi} ab^\dag\right]}.\label{eq:BSgateunitary}
\end{align}
In the interaction picture, the beam-splitter gate transforms the two bosonic mode operators as follows
\begin{align}
a_{\theta,\varphi}&=\mathrm{BS}^\dagger(\theta,\varphi)\,a\,\mathrm{BS}(\theta,\varphi)=\cos\frac{\theta}{2}a - i\sin\frac{\theta}{2}e^{+i\varphi}b\label{eq:athetaphi}\\
b_{\theta,\varphi}&=\mathrm{BS}^\dagger(\theta,\varphi)\,b\,\mathrm{BS}(\theta,\varphi)=\cos\frac{\theta}{2}b - i\sin\frac{\theta}{2}e^{-i\varphi}a.\label{eq:bthetaphi}
\end{align}
\label{Box:beam-splitter}
}

\noindent
A key feature of the beam-splitter gate is that it preserves the total excitation number in the two cavities
\begin{equation}
    [\mathrm{BS}(\theta,\varphi),\hat n_a + \hat  n_b]=0.
\end{equation}
This fact plays an important role in the Schwinger boson representation of spins discussed in Sec.~\ref{sssec:many-spins}.

Notice from Eqs.~(\ref{eq:athetaphi}-\ref{eq:bthetaphi}) that for $\theta=\pi$ and $\varphi=0$
\begin{eqnarray}
a_{\pi,0}&=& - ib\label{eq:aSWAP}\\
b_{\pi,0}&=& - ia,\label{eq:bSWAP}
\end{eqnarray}
the beam-splitter swaps the two mode operators. Reverting to the Schr{\"o}dinger picture we find that this gate swaps the states of the two cavities
\begin{eqnarray}
    \mathrm{BS}(\pi,0)|\Psi_a,\Psi_b\rangle&=&e^{-i\frac{\pi}{2}[\hat n_a + \hat  n_b]}\mathrm{SWAP}|\Psi_a,\Psi_b\rangle\nonumber\\&=&e^{-i\frac{\pi}{2}[\hat n_a + \hat  n_b]}|\Psi_b,\Psi_a\rangle,
\end{eqnarray}
where the exponential of the total number operator in front accounts for the $(-i)$ factor in Eqs.~(\ref{eq:aSWAP}-\ref{eq:bSWAP}).  For convenience of notation, we have assumed here that the two cavities are in a product state, but by the linearity of quantum mechanics, the gates perform equally well on all states (including states of indefinite total photon number) since any two-mode state can be written as a linear superposition of product states.  
A key and powerful feature of the bosonic SWAP gate is that it is agnostic as to the particular state of the cavities (e.g., the particular bosonic code being utilized to hold the quantum information). This is highly useful for routing and transporting bosonic qubits across the hardware fabric, a possibility that is used for compilation in Sec.~\ref{ssec:exact-analytical-qubit-gates}.

The reader should not confuse this gate with the so-called iSWAP gate which is an entangling gate for two-level qubits. Importantly, one might think that $\mathrm{BS}(\pi/2,0)$ is related to the entangling gate $\sqrt{\mathrm{iSWAP}}$, but it is not.  $\mathrm{BS}(\pi/2,0)$ is the 50:50 beam-splitter familiar from optics.  This gate has been used experimentally to demonstrate the Hong-Ou-Mandel effect for microwave photons \cite{Lang2013,GaoYY2018} in which two identical photons incident in each input port of a 50:50 beam-splitter always exit from the same port (or in this case end up in the same resonator)
\begin{equation}
    \mathrm{BS}(\pi/2,0)|1\rangle|1\rangle=-\frac{1}{\sqrt{2}} \left [  
 \rule{0pt}{2.4ex} |0\rangle|2\rangle+|2\rangle|0\rangle \right] \, .
\end{equation}

If one encodes quantum information (i.e., a qubit) in the $n=0,1$ Fock states of each cavity
\begin{equation}
    |\Psi\rangle=\alpha|0\rangle+\beta|1\rangle,
\end{equation}
we see that the Hong-Ou-Mandel effect can take the system out of that code space by populating the $n=2$ Fock states (while still preserving the total photon number in the two modes).  Thus, this cannot be related to $\sqrt{i\mathrm{SWAP}}$ despite the fact that applying $\mathrm{BS}(\pi/2,0)$ twice does indeed yield $\mathrm{BS}(\pi,0)$, which, up to phase factors, is SWAP.

\subsubsection{Two-Mode Squeezing Gate}\label{subsec:tms gate} Another useful  bilinear interaction is the two-mode squeezing (TMS) interaction \cite{PhysRevLett.109.183901}
\begin{align}
    \hat V_\mathrm{TMS} &= ig_\mathrm{TMS}\left(e^{i\varphi}\, a^\dagger b^\dagger - e^{-i\varphi}\,ab\right).
\end{align}

Time evolution under this interaction yields the two-mode squeezing gate in Box~\ref{Box:2-mode-squeezing}.
From the Heisenberg equations of motion
\begin{eqnarray}
\frac{d}{dt}\left(\begin{array}{c}a(t)\\b^\dagger(t)\end{array}\right) &=& g_\mathrm{TMS}\left(\begin{array}{cc}0&e^{i\varphi}\\e^{-i\varphi}&0\end{array}\right)\left(\begin{array}{c}a(t)\\b^\dagger(t)\end{array}\right)
\end{eqnarray}
we see that the mode operators transform as
\begin{align}
\left(\begin{array}{c}a(t)\\b^\dagger(t)\end{array}\right)
=&\left\{\cosh r\,\, \hat I + \sinh r\,\left(
\begin{array}{cc}0&e^{i\varphi}\\e^{-i\varphi}&0\end{array}\right)\right\}
\left(\begin{array}{c}a(0)\\b^\dagger(0)\end{array}\right),
\end{align}
 where $r=g_\mathrm{TMS}t$.  From this result we can obtain the symplectic transformation matrix $\mathrm{\bf TMS}(r,\varphi)$ (here expressed in the basis of mode creation and annihilation operators rather than positions and momenta) corresponding to the TMS unitary
 \begin{align}
     \left(\begin{array}{c}a^{\phantom{\dag}}(t)\\a^\dag(t)\\b^{\phantom{\dag}}(t)\\b^\dag(t) \end{array}\right)&=
     \mathrm{\bf TMS}(r,\varphi)\left(\begin{array}{c}a^{\phantom{\dag}}\\a^\dag\\b^{\phantom{\dag}}\\b^\dag \end{array}\right),
     \label{eq:TMS_mode_ops_transformation}
 \end{align}
 where
 \begin{align}
     &\mathrm{\bf TMS}(r,\varphi)=\nonumber\\
     &\left( \begin{array}{cccc}\cosh r&0&0&e^{i\varphi}\sinh r\\
     0&\cosh r&e^{-i\varphi}\sinh r &0\\
     0&e^{i\varphi}\sinh r&\cosh r&0\\
     e^{-i\varphi}\sinh r&0&0&\cosh r
     \end{array}\right).
     \label{eq:TMSsymplectic}
 \end{align}

\abox{Two-Mode Squeezing Gate }{
\begin{align}
\mathrm{TMS}(r,\varphi)=&e^{-i\hat V_\mathrm{TMS} t}=e^{r\left(e^{i\varphi}\, a^\dagger b^\dagger - e^{-i\varphi}\,ab\right)},
\end{align}
where $r=g_\mathrm{TMS}t$.
  The corresponding symplectic mode transformation matrix $\mathrm{\bf TMS}(r,\varphi)$ is given in Eq.~(\ref{eq:TMSsymplectic}).
\\
The two-mode squeezed vacuum state \cite{gerry_knight_2004}
\begin{align}
    \mathrm{TMS}(r,\varphi)|0,0\rangle &= \frac{1}{\cosh r}\sum_{m=0}^\infty \left [e^{i\varphi}\tanh r \rule{0pt}{2.4ex} \right]^m\,|m,m\rangle,
\end{align}
is an entangled state of two modes characterized by having equal photon numbers in both modes.  For $\varphi=0$ one can think of this entangled state as the CV analog of the DV Bell state
\begin{align}
    |\mathrm{Bell}\rangle=\frac{1}{\sqrt{2}} \left [ \rule{0pt}{2.4ex} |00\rangle + |11\rangle \right],
\end{align}
which is characterized by an equal excitation number in both qubits.
\label{Box:2-mode-squeezing}
}

\begin{figure}[tb]
    \centering
    \includegraphics[width=0.48\textwidth]{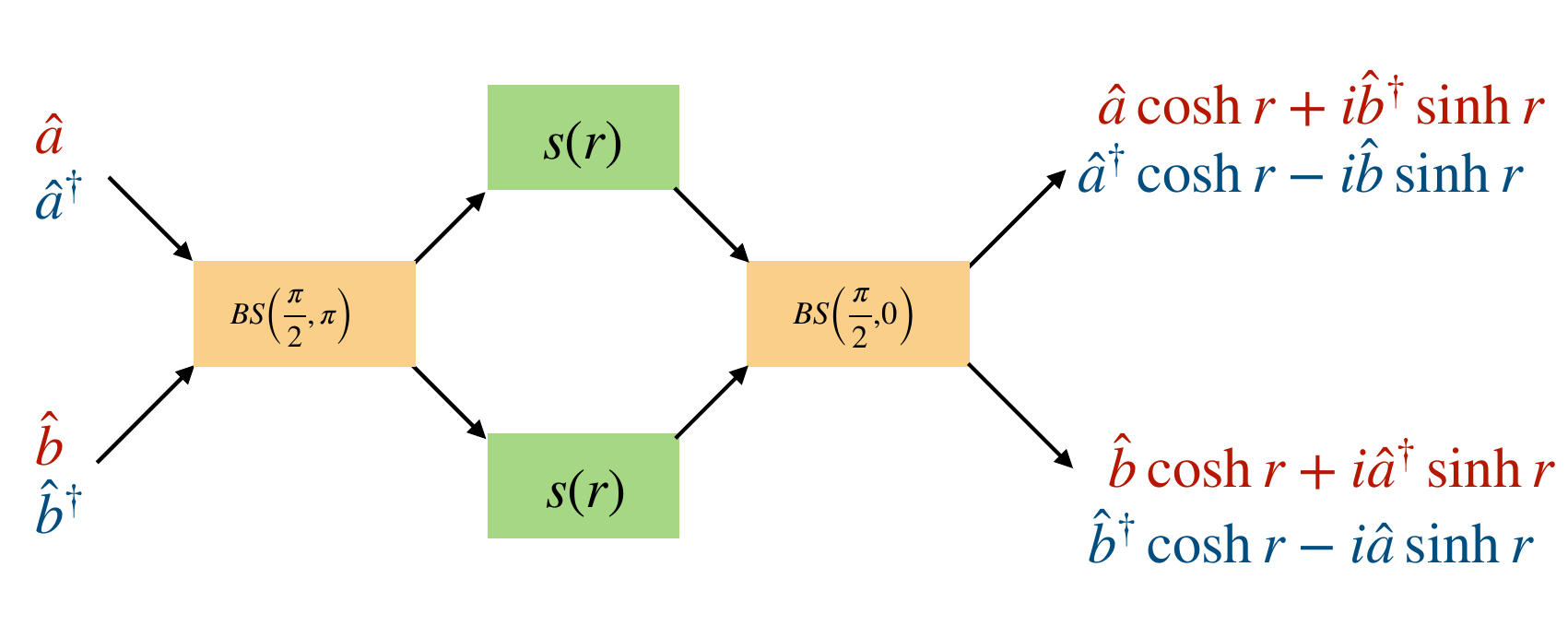}
    \caption{Bloch-Messiah decomposition of the two-mode squeezing $\mathrm{ TMS}\Big(r,\frac{\pi}{2}\Big)$ gate using photon-number preserving beam-splitter gates and single-mode squeezing gates derived in App.~\ref{app:bloch-messiah}. The transformed mode operators on the right are given by the equations in Box.~\ref{Box:2-mode-squeezing}. Here $a$ and $b$ denote different oscillators. The effect of this operation is described by how the annihilation and creation operators are transformed. The red (blue) operators on the left are transformed to red (blue) operators on the right for each oscillator. Note that the terms in Eq.~(\ref{eq:BMdecompTMS}) are applied right to left while the circuit diagram should be read left to right.
    }
    \label{fig:TMS}
\end{figure}

The interferometer circuit in Fig.~\ref{fig:TMS} realizes the Bloch-Messiah decomposition \cite{Braunstein2005squeezing,Weedbrook2012,PhysRevA.94.062109} of the two-mode squeezing gate using a pair of beam-splitters and single-mode squeezers as follows,
\begin{equation}
   \mathrm{TMS}(r,\pi/2)=\mathrm{BS}(\pi/2,0)[s(r)\otimes s(r)]\mathrm{BS}(\pi/2,\pi).
   \label{eq:BMdecompTMS}
\end{equation}
Here the gate symbols represent the full unitary operators acting on the Hilbert space, not the symplectic matrix representation of the gates. The direct product $s(r)\otimes s(r)$ represents single-mode squeezing applied to each arm of the interferometer in Fig.~\ref{fig:TMS}.   
The derivation of this circuit is given in App.~\ref{app:bloch-messiah} as is the symplectic transformation of the quadrature coordinates. 

In the microwave domain, two-mode squeezing has been performed with parametric pumping of a nonlinear Josephson mixer to realize $\hat{V}_\text{TMS}$ directly. To the best of our knowledge, the largest TMS coupling rate between two superconducting microwave oscillators achieved to date is $g_\text{TMS}/2\pi = \SI{12.6}{MHz}$ using a pumped Josephson ring modulator \cite{Markovi_2018}. Two-mode squeezing has also been demonstrated between phonons in surface-acoustic-wave (SAW) resonators \cite{Andersson_2022}. We also note that two-mode squeezing of itinerant microwave photons is commonly used for quantum-limited parametric amplification \cite{Eichler_2011,Roy_2016, Chien_2020, Esposito_2022}.  An entanglement-enhanced interferometer can be constructed from a pair of two-mode squeezers and used to detect small displacements of the electromagnetic field \cite{YURKE_SU(11)_PhysRevA.33.4033,Backes2021,oh2024entanglement} below the vacuum fluctuation limit.

\subsubsection{Two-Mode Sum Gate}\label{SUM-gate} Let $|x_a\rangle|x_b\rangle$ be position eigenstates for two oscillators, a and b.  As shown in Box~\ref{Box:2-mode-sum}, the two-mode sum gate displaces one oscillator by an amount determined by the position of the other oscillator. 
The SUM gate and any other gate generated by bilinears in the oscillator operators can be created by simultaneous use of the beam-splitter and two-mode squeezing interactions
\begin{eqnarray}
\hat V&=& \left [ \rule{0pt}{2.4ex} \chi ab^\dagger+\chi^* a^\dagger b \right]
+i \left [ \rule{0pt}{2.4ex} \zeta^* a^\dagger b^\dagger-\zeta ab \right] \,.
\end{eqnarray}
For example, we see from Eq.~(\ref{eq:SUMGate}) that required Hamiltonian has the form
\begin{align}
    \hat V &= -ig[a+a^\dagger][b-b^\dagger]\nonumber\\
    &=+ig[ab^\dagger-a^\dagger b]+ig[a^\dagger b^\dagger-ab],
\end{align}
for which the beam-splitter and two-mode squeezing rates are identical.
Examples of other gates that can be realized in this way include 
\begin{eqnarray}
S_{xx}&=&e^{i2\lambda \hat x_a \hat x_b}\\
S_{pp}&=&e^{i2\lambda \hat p_a \hat p_b}\\
S_{xp-px}&=&e^{i2\lambda[\hat x_a \hat p_b-\hat x_b \hat p_a]}.
\end{eqnarray}

\abox{Two-Mode Sum Gate}{
\begin{eqnarray}
\mathrm{SUM}(\lambda)=e^{-i2\lambda \hat x_a \hat p_b}&=&e^{\frac{\lambda}{2}(a+a^\dagger)(b^\dagger-b)}\label{eq:SUMGate}\\
\mathrm{SUM}(\lambda)|x_a\rangle|x_b\rangle &=&|x_a\rangle|x_b+\lambda x_a\rangle
\end{eqnarray}
where $\lambda$ is an arbitrary real scale factor. The factor of 2 comes from the choice of Wigner units in which the generator of displacements is $2\hat p$ as shown in Eq.~(\ref{eq:Wignerunitsdisp}).

In the momentum representation, the action of the SUM gate is
\begin{align}
   \mathrm{SUM}(\lambda)|p_a\rangle|p_b\rangle &=|p_a-\lambda p_b\rangle|p_b\rangle 
\end{align}

The SUM gate can be used to create Bell-like entangled states of the following form in the position representation
\begin{align}
    &\mathrm{SUM}(\lambda\!=\!1) \int_{-\infty}^{+\infty}\mathrm{d}x\, \Psi(x)\left[| \rule{0pt}{2.4ex} x\rangle\otimes S(r,0)|0\rangle\right]\nonumber\\
    &\approx \int_{-\infty}^{+\infty}\mathrm{d}x\, \Psi(x)\left[\rule{0pt}{2.4ex} |x\rangle\otimes|x\rangle\right],
\end{align}
where we have assumed $r\gg 1$ so that the squeezed vacuum state $S(r,0)|0\rangle$ is approximately a position eigenstate centered at position $0$.
\label{Box:2-mode-sum}
}

\begin{figure}[tb]
    \centering
    \includegraphics[width=0.48\textwidth]{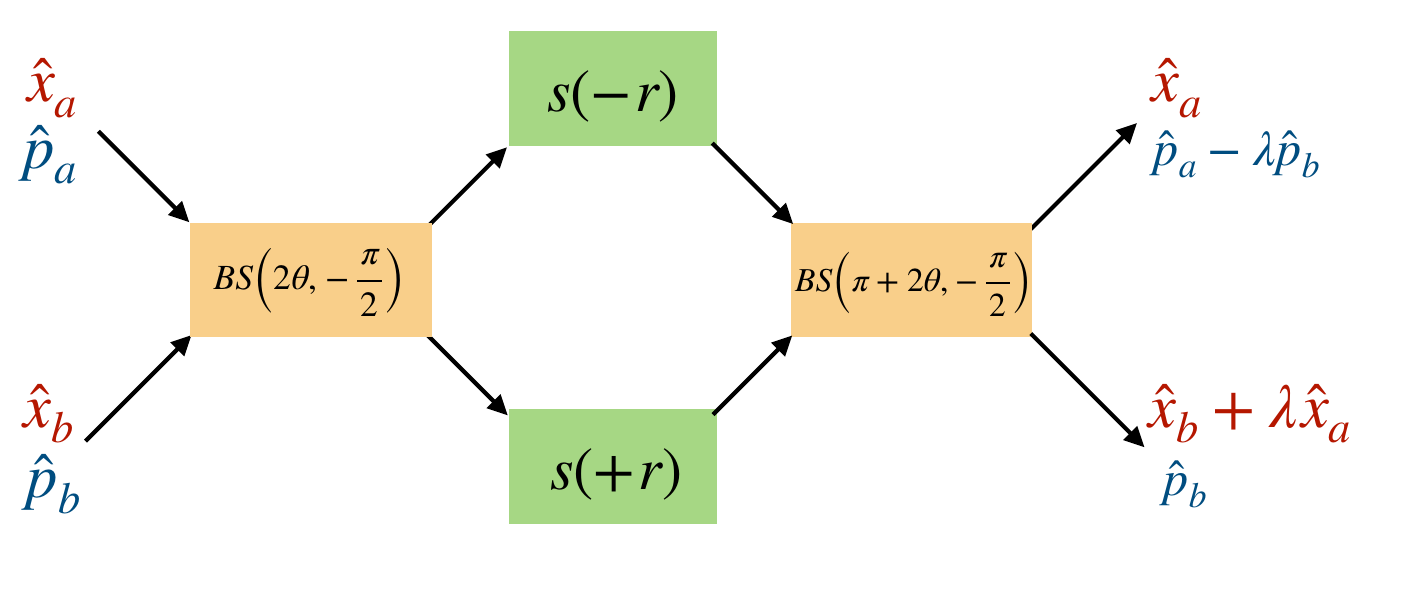}
    \caption{Bloch-Messiah decomposition of SUM$(\lambda)$ gate \cite{tzitrin2020progress} using photon-number preserving  beam-splitter gates and single-mode squeezing gates. The transformation from the initial mode quadrature operators on the left to the final quadratures on the right is given by the equations in Box.~\ref{Box:2-mode-sum}. Here $a$ and $b$ denote different oscillators. The effect of this operation is described by how the position and momentum operators are transformed. The red (blue) operators on the left are transformed into red (blue) operators on the right for each oscillator.}
    \label{fig:SUM}
\end{figure}

The Bloch-Messiah decomposition for the SUM gate \cite{tzitrin2020progress} has the advantage that two-mode squeezing can be replaced by simpler single-mode squeezing,
\begin{align}
   \mathrm{SUM}(\lambda)
   &=\mathrm{BS}(\pi+2\theta,-\pi/2)[s(r)\otimes s(-r)]\mathrm{BS}(2\theta,-\pi/2)\label{eq:SUMBlochMessiah}\\
   \sinh{r}&=\frac{\lambda}{2},\label{eq:SUMBM2}\\
   \cos(2\theta)&=\tanh(r),\label{eq:SUMBM3}\\
   \sin(2\theta)&=-\sech(r),\label{eq:SUMBM4}
\end{align}
where we use the tensor product ordering convention that $B\otimes A$ means that $A$ is applied to the upper arm of the interferometer and $B$ is applied to the lower arm.  Thus, $s(-r)$ is applied to the upper arm and $s(+r)$ is applied to the lower arm of the interferometer in Fig.~\ref{fig:SUM}.
The derivation of this decomposition can be found in App.~\ref{app:bloch-messiah}.

\subsection{Gaussian Operations are Analogous to Clifford Gates}
\label{subsec:GaussiansAsCliffords}
In this section, we discuss the analogies summarized in Table~\ref{table:qubitsvsCV} between Gaussian operations on oscillators and Clifford operations on qubits. The analogies are close but not perfect as we shall see in this pedagogical discussion which combines some previously known facts and new insights.
\begin{table*}[tb]   
\centering
\def\arraystretch{2}
\renewcommand\theadfont{}
\begin{tabular}{|c|c|c|}
 \hline 
 {\bf Aspect} & {\bf Qubits} & {\bf Bosonic Modes}   \\
  \hline
 Standard Basis & $Z$, $\{0,1\}$ & $\hat{x}$, $|x\rangle$ \\
  \hline
 Conjugate Basis & $X$, $|0\rangle\pm |1\rangle$ & $\hat{p}$, $|p\rangle = \int \mathrm{d}x\,e^{ixp}|x\rangle$ \\
 \hline
 Basis Transformation & Hadamard & Fourier Transform (i.e., $e^{i\frac{\pi}{2} \hat{n}}$) \\
   \hline
\multirow{2}{*}{Group} & Pauli & Weyl-Heisenberg (e.g., $e^{i\alpha \hat{x}},\, e^{i\beta \hat{p}} )$ \\
  \cline{2-3}
  & Clifford  & Gaussian unitaries (e.g., $e^{i\theta \hat n},e^{i\lambda{\hat x}^2},e^{i \hat x_1\hat p_2}) $ \\
  \hline
  Channel & Pauli/Clifford & Displacement/Gaussian\\
 \hline
\multirow{3}{*}{Measurement} &  
\makecell{Pauli basis\\Measurement of $X,Z$\\ requires Clifford operations} & \makecell{Homodyne\\ Measurement of $\hat{x},\hat{p}$\\ requires Gaussian operations }\\
  \cline{2-3}&\makecell{ Non-Clifford\\ ($\vec b\cdot\vec\sigma$)}  & \makecell{Non-Gaussian\\ ($a^\dagger a,|n\rangle\langle n|,aa,\hat x^2,\hat p^3,\cos k\hat x$)}\\
\hline
 Representation & State Tomography &  Wigner (or Characteristic) Function Tomography\\
 \hline
 Benchmarking & Clifford operations are 2-design & Gaussian operations are not 2-design\\
 \hline
 Example Entangled State &Bell: $|00\rangle+|11\rangle$ & EPR (TMS, SUM): $\sum_n|n\rangle_1|n\rangle_2,\,\, \int \mathrm{d}x\, |x\rangle_1|x\rangle_2$ \\
  \hline
 Universal Computation& 
\makecell{a non-Clifford gate, e.g., $T= e^{i(\pi/8) Z}$,\\augments the Clifford group\\ with magic (non-Cliffordness)}
 & \makecell{a non-Gaussian unitary, cubic gate: $e^{i\chi\hat x^3}$\\augments Gaussian operations\\ with Wigner negativity}\\
  \hline
\end{tabular}
\caption{\footnotesize Qubit Clifford Circuits vs.\ Gaussian Bosonic Circuits \cite{bartlett2002efficient,Braunstein2005}.
TMS stands for two-mode squeezing. The Weyl-Heisenberg group is the group of (non-commuting) translations in phase space effected (in standard dimensionless units) via $X(s)=e^{-is\hat p}$ which translates by $s$ in position and $Z(t)=e^{it\hat x}$ which produces translations by $t$ in momentum. Under non-Gaussian measurements, $|n\rangle\langle n|$ is the projector onto the Fock state with $n$ photons, and $a^\dagger a$ stands for photon number resolving detection. Note in general, there is no unique one-to-one mapping between the qubit gates in the second column to the bosonic gates in the third column. Note the difference between Fourier transform $F$ and Hadamard operation: $F^\dagger \ne F$ while $H^\dagger  = H$. As a result, Fourier transform maps $\hat{x}$ into $\hat{p}$ and $\hat{p}$ into $-\hat{x}$, while Hadamard maps $X$ into $Z$ and $Z$ back into $X$ (with no minus sign).
}
\label{table:qubitsvsCV}
\end{table*}

Before considering CV gates, let us briefly review the classification of qubit gates according to the Clifford hierarchy which is defined as follows.
\begin{align}
C_1 P C_1^\dagger &\in C_1\\
    C_nPC_n^\dagger &\in C_{n-1}
    \quad 
    \text{for }n>1\\
    C_n&\subset C_{n+1},\quad\forall n,
\end{align}
where $C_1$ is the set of \emph{Pauli gates} $\{P\}$. The second level of the Clifford hierarchy $C_2$ is the \emph{Clifford group}  whose elements conjugate a Pauli gate (i.e., a string of Pauli operators) into another Pauli gate; that is, the Clifford group is the \emph{normalizer} of the Pauli group. If a gate does not belong to $C_2$ it falls under the class of non-Clifford gates.

The Clifford group for an arbitrary number of qubits, $N$, can be generated by just three types of gates: $\mathrm{H}$ (the Hadamard gate), $S=Z^{\frac{1}{2}}$, and CNOT.  Even though Clifford circuits can generate large superposition states such as maximally entangled Bell and GHZ states (using the CNOT gate), the Gottesman-Knill theorem   \cite{Gottesman-Knill-Theorem,AaronsonGottesmanCliffordSims} tells us that they are easy to simulate classically.   To see this, consider an $N$-qubit starting state $|000\ldots 000\rangle$.  This state is `stabilized' (i.e., is a +1 eigenstate of) the set of $N$ single-qubit Pauli $Z$ operators, $\{Z_N,Z_{N-1},\ldots,Z_1,Z_0\}$, and is uniquely defined by this list of stabilizers.  Under the action of an arbitrary Clifford circuit, this set of stabilizers is mapped (in the Heisenberg picture) by conjugation to new Pauli strings (generally of weight higher than one under the action of CNOT gates that create entanglement) and the list of transformed stabilizers continues to uniquely define the quantum state at the output of the circuit.  Consider, for example, the Bell state generation circuit 
\begin{align}
    \mathrm{CNOT}_{10} \left [\rule{0pt}{2.4ex} \mathrm{H}\otimes I \right]|00\rangle = \frac{1}{\sqrt{2}} \left [ \rule{0pt}{2.4ex} |00\rangle+|11\rangle\right ],
\end{align} where the notation $\mathrm{CNOT}_{10}$ is used to indicate that qubit 1 is the control and qubit 0 is the target (and the qubits are numbered right-to-left starting with ordinal number zero).  This circuit transforms the initial stabilizer set $\{Z_1,Z_0\}$ to the set of weight-two stabilizers $\{X_1X_0,Z_1Z_0\}$ that uniquely defines the Bell state.    

The efficiency advantage of the stabilizer formalism can be seen by noting that if we had a set of $N=2M$ qubits and a Clifford circuit that produced $M$ randomly selected Bell pairs, the quantum state would be described by a large superposition of $2^N$ quantum amplitudes, whereas the set of stabilizers would still only be of size $N$.  Simple classical algorithms exist \cite{AaronsonGottesmanCliffordSims} to update the list of stabilizers of the state according to the Clifford transformations and thus Clifford circuits can be efficiently simulated classically and therefore do not represent the full power of quantum computation.  We define a `stabilizer state' as any state that can be produced from the all-zero state using a Clifford circuit.  Equivalently, a stabilizer state for $N$ qubits is a +1 eigenstate of $N$ independent generators (Pauli strings) of the stabilizer group. 

An analogous classification of CV gates \cite{bartlett2002efficient,Braunstein2005} can be carried out following the structure of the Clifford hierarchy~\cite{bartlett2003quantum,singh2025quantum}. Table~\ref{table:qubitsvsCV} summarizes the overall analogy we discuss here.
The analog of qubit Pauli operations are oscillator displacements (in position and momentum). The analog of the Pauli group for oscillators is the Heisenberg-Weyl group $\mathrm{H}(1)$ which is a continuous Lie group composed of displacements in phase space.  We have already seen that the product of two displacements is (up to a global phase) another displacement (see Box~\ref{Box:UncondDispGate}) and that every displacement has an inverse. The algebra $\mathfrak{h}(1)$ that generates this group is spanned by the canonical position and momentum operators (see Eq.~\eqref{eq:phasespacecoordinatevector}) and the identity operator, $\{\hat{\bm R}, \hat I\}$, satisfying the required commutation relations. 

By analogy with the DV case, the CV analog of the Clifford group $C_2$ is associated with those CV operations that preserve (up to overall phases) the group of displacements under conjugation. The only such operations are the set of Gaussian operations defined in Eq.~(\ref{eq:GaussianUnitary}).  As shown in Eq.~(\ref{eq:GaussianXform}), Gaussian operations induce a linear transformation (rotation, translation, and symplectic rescaling) on the phase-space coordinates.  It follows from this that a phase-space displacement is mapped into a different displacement under conjugation by any Gaussian operation
\begin{align}
    U_{\mathrm{G}}^\dagger e^{i{\bm\lambda} \hat{\bm R}}U_{\mathrm{G}} &= e^{i\theta}e^{i{\bm \lambda}^\prime\hat{\bm R}},\\
    {\bm \lambda}^\prime &={\bm\lambda} {\bm M},\\
    \theta &= {\bm\lambda} {\bm\Delta},
\end{align}
where ${\bm M}$ and ${\bm \Delta}$ are defined in Eq.~(\ref{eq:GaussianXform}).

From the definition above of a stabilizer state  in the DV setting, one might imagine that the the CV analog of a DV stabilizer state is a Gaussian state--any state created from the CV vacuum state (Fock state $|0\rangle$) via Gaussian operations. We will see shortly that this definition has some problems, but for the moment we pursue it further to see where it leads.  The vacuum (ground state, or a coherent state with displacement $\alpha=0$) wave function of a harmonic oscillator is of course a Gaussian function of the coordinate as shown in Eq.~(\ref{eq:coherentstatewavefunction}). 
We have seen that under the influence of any single- or multi-mode Hamiltonian that is quadratic in position and momentum (or equivalently quadratic in the creation and annihilation operators), the Heisenberg equations of motion for the operators are linear.  This immediately implies that Gaussian states evolve into other Gaussians. Time evolution under the application of quadratic control Hamiltonians can displace the position, boost the momentum, and/or squeeze the Gaussian, but the wave function always remains a Gaussian. 
This means that it is easy to classically compute and represent the time-evolution in terms of a few parameters representing the mean values and covariances of the positions and momenta, $\vec R$, or equivalently of the creation and annihilation operators, $\langle a\rangle$, $\langle a^\dagger a\rangle$, $\langle aa\rangle$, $\langle a^\dagger b^\dagger\rangle$, etc.  We do not need to keep track of higher-order correlators, because for Gaussian states, Wick's theorem \cite{Wick_PhysRev.80.268} guarantees that they are simple products of lower-order correlators.  For example (assuming for simplicity that $\langle a\rangle=\langle b\rangle=0$), Wick's theorem tells us that
\begin{equation}
\langle a^\dagger a b^\dagger b\rangle = \langle a^\dagger a\rangle\langle b^\dagger b\rangle
+\langle a^\dagger b^\dagger\rangle\langle ab\rangle
+ \langle a^\dagger b\rangle\langle ab^\dagger\rangle.
\end{equation}
In condensed matter physics and quantum field theory, terms like $\langle aa\rangle$ and $\langle ab\rangle$ are known as anomalous correlators.  In quantum optics, they are described as representing one- and two-mode squeezing respectively.

We can draw an analogy \cite{bartlett2002efficient} between the simplicity of computing time evolution under quadratic bosonic Hamiltonians (Gaussian operations) and the Gottesman-Knill theorem \cite{Gottesman-Knill-Theorem,AaronsonGottesmanCliffordSims} that a quantum computer based on qubits and using only Clifford group operations is easy to simulate classically.  
The fact that DV Clifford circuits can be efficiently simulated classically is related not just to the ability to do stabilizer updates but also to the fact that the full non-classical correlations inside Bell states cannot be revealed without making non-Clifford (e.g., T-gate 45-degree) rotations on the Bloch sphere (or rotating the measurement axis by 45 degrees) to violate the Bell inequalities. Such non-Clifford rotations are sometimes described as introducing `magic' (or `non-stabilizerness') to a state \cite{Emerson_Magic_Resource,PhysRevLett.115.070501,doi:10.1098/rspa.2019.0251,Bravyi2019simulationofquantum,PRXQuantum.4.010301}.  Similarly, CV transformations by Gaussian operations only update the mean and variance of Gaussian states (analogous to stabilizer updates), and fail to introduce any negativity in the Wigner function (defined in App.~\ref{sec:characteristic-function}). If we start with a state with a non-negative Wigner function, then that property is preserved under all Gaussian operations. Hence the expectation value of bosonic observables can be readily obtained by classical importance sampling of the wave function, without suffering from any sign problems \cite{PhysRevLett.109.230503,Veitch_2013}. This is not generically the case for non-Gaussian states and thus Wigner negativity for bosonic systems is akin to `magic' in qubit states. Ref.~\cite{Jaffe_doi:10.1073/pnas.2304589120} highlights the similarity between Gaussian CV states and stabilizer DV states.

%\afterpage{\clearpage}
%{
\def\arraystretch{2}
\renewcommand\theadfont{}
\begin{longtable*}[t!]{|c|c|c|l|l|c|} 

\hline  &                                    & \textbf{Gate Name}                            & \thead{\textbf{Parameters}} & \thead{\textbf{Definition}} & \makecell{\textbf{Reference} \\ \textbf{Theory/Exp.}} \\ \hline
\multirow{16}{*}{\rotatebox[origin=c]{90}{\textbf{Oscillator-only Gates}}}             & \multirow{8}{*}{\rotatebox[origin=c]{90}{Gaussian (Sec.~\ref{sec:gaussian gates})}}          & \makecell{Displacement\\ (Box \ref{Box:UncondDispGate})}                         & $\alpha \in \mathbb{C}$   & $D(\alpha) = \exp\left[\alpha a^\dag - \alpha^* a\right] $   & \cite{gerry_knight_2004}/ \cite{Blais2020}    \\ \cline{3-6} 
&  & \makecell{Phase-space rotation\\ (Box \ref{Box:phase-space-rotation})}   &  $\theta\in [0, 2\pi)$   & $R(\theta) = \exp\left[-i\theta a^\dag a\right] $   & \cite{Weedbrook2012}/    \\  
 &  &Fourier transform    & $\theta=\pi/2$ & $F=\exp\left[-i\frac{\pi}{2} a^\dag a\right]$&     \\ \cline{3-6}     
&             ,                      & \makecell{(Single-mode) squeezing\\ (Box \ref{Box:1-mode-squeezing})}             & $\zeta \in \mathbb{C}$    & $S(\zeta) = \exp\left[\frac{1}{2}(\zeta^* a^2 - \zeta a^{\dag 2})\right]$   &   \cite{gerry_knight_2004}/ \cite{Dassonneville_2021}   \\ \cline{3-6} 
 &   & \makecell{Beam-splitter\\ (Box \ref{Box:beam-splitter}) } & $\theta\in[0,4\pi),\varphi\in[0,\pi)$    &  $\mathrm{BS}(\theta,\varphi)=\exp\left[-i\frac{\theta}{2}\left(e^{i\varphi} a^\dag b+e^{-i\varphi} ab^\dag\right)\right]$    &\makecell{ \cite{gerry_knight_2004,Zhang_engineering_bilinear_2019}/ \\ \cite{lu2023highfidelity, chapman2022high}}     \\ 
 &  & SWAP        &  $\theta=\pi,\varphi=\pi/2$   & $\mathrm{SWAP} = \exp\left[\frac{\pi}{2}(a^\dag b - a b^\dag)\right]$  &     \\ \cline{3-6}                                              
 &   & \makecell{Two-mode squeezing\\ (Box \ref{Box:2-mode-squeezing})} & $r \in \mathbb{R}, \varphi \in [0, \pi)$    & $\text{TMS}(r,\varphi) =\exp \left[r\left(e^{i\varphi} a^\dag b^\dag - e^{-i\varphi} a b\right)\right] $   &  \cite{gerry_knight_2004}/ \cite{Chien_2020}     \\ \cline{3-6} 
 &  & \makecell{Two-mode summing\\ (Box \ref{Box:2-mode-sum})  } & $\lambda \in \mathbb{R}$    & $\text{SUM}(\lambda) = \exp \left[\frac{1}{2}\lambda \left(a+a^\dagger \right)\left( b^\dagger - b \right)\right] $   &   \makecell{\cite{bartlett2002efficient}/\\ \cite{PhysRevLett.101.250501,Markovi_2018} }   \\ \cline{2-6} 
 & \multirow{7}{*}{\rotatebox[origin=c]{90}{Non-Gaussian (Sec.~\ref{sec:non-Gaussian_and_Hybrid})}}      & \makecell{Generalized \\ single-mode squeezing\textsuperscript{1} }                      & $z\in\mathbb{C},  N\geq 3$    & $U_N(z)  = \exp\left[za^{\dag N} - z^* a^N\right]$    & \makecell{\cite{Braunstein1987generalized}/\\ \cite{chang2020observation,eriksson2023universal}  }  \\ \cline{3-6}
                                            & & \makecell{Generalized $k$-mode\\ squeezing}                        & \makecell[l]{$z\in\mathbb{C}, k\ge 3$ \\ $\vec{n} = \{n_1,\cdots, n_k \}$}    & $U_{\vec{n}}(z) = \exp\left[z \prod_{p=1}^k a_{p}^{\dag n_p} - z^* \prod_{p=1}^k a_{p}^{n_p} \right] $    &   \makecell{\cite{zhang2023genuine}/\\ \cite{chang2020observation,agusti2020tripartite}}  \\ \cline{3-6}
                                             & & Cubic-phase                        & $r\in\mathbb{R}$    & $C(r) = \exp\left[-ir\hat{x}^3\right]$    &   \makecell{\cite{marshall2015repeat,zheng2021gaussian}/ \\ \cite{chang2020observation,Kundra_2022_Robust,eriksson2023universal}}  \\ \cline{3-6}
                                             &                                    & Self-Kerr                                 &  $\theta\in\mathbb{R}$ & $K(\theta) = \exp\left[-i\frac{\theta}{2} a^{\dag 2}a^2\right]$   & \makecell{\cite{gerry_knight_2004,GirvinLesHouches2011} /\\ \cite{collapse-revivalKirchmair} }  \\ \cline{3-6} 
                                             &                                    & SNAP\textsuperscript{2} (Box \ref{Box:SNAP})                              & $\vec{\varphi} = \{\varphi_n \}, \varphi_n \in [0,2\pi)$   & $\text{SNAP} (\vec{\varphi}) = \sum_{n}e^{-i\varphi_n}\ket{n}\bra{n}$   &  \cite{Krastanov2015}/\cite{Heeres2015}   \\ \cline{3-6}
                                             &                                    & eSWAP (Box \ref{Box:e-swap})       & $\theta\in [0,4\pi)$  & $\exp\left[-i\frac{\theta}{2} \left(\text{SWAP}\right)\right]$  &  \cite{LauPlenio}/\cite{Gao2019}    \\ \hline 
\caption{ \small\textbf{Common oscillator-only gates.} This table is not comprehensive; however, it contains some of the most common gates on bosonic oscillators. While for simplicity we list these as oscillator-only gates, the experimental realization of many of them requires the use of an auxiliary non-linear system (e.g., a qubit) for their execution. \textsuperscript{1}Generalized squeezing is non-Gaussian for $N\geq 3$. $U_3$ is often called \textit{Tri-squeezing}, and $U_4$ can be called \textit{quartic squeezing}. 
 \textsuperscript{2}SNAP stands for \textbf{S}elective \textbf{N}umber-dependent \textbf{A}rbitrary \textbf{P}hase  gate. The SNAP gate is implemented with a sequence of SQR gates (defined in Box \ref{box:SQRgate}) using an auxiliary qubit, here not included for clarity as the oscillator and qubit are left unentangled if the qubit starts in the computational basis. However, the sign of the phase imparted is dependent upon the qubit state, and as such the SNAP gate is more completely expressed as $\text{SNAP} (\vec{\varphi}) = \sum_{n}e^{-i\varphi_n \sigma_z}\ket{n}\bra{n}$. 
 The simple phase space rotation gate $R(\theta)$ can be performed `virtually' in software by simply resetting the phase of subsequent drives on the oscillator in analogy with the virtual Z gate used in qubit systems \cite{EfficientZgates_PhysRevA.96.022330}.
}
\label{tab:gates-osc}
\end{longtable*}
%}
%}

Even though Gaussian operations on Gaussian states cannot produce Wigner negativity, they \emph{can} produce entanglement \cite{Braunstein2005}, just as Clifford operations on qubits can.  Box~\ref{Box:2-mode-squeezing} shows that two-mode squeezing is a Gaussian operation that produces entanglement (most readily apparent in the Fock basis) and Box~\ref{Box:2-mode-sum} shows that the SUM gate is a Gaussian operation that produces entanglement (most readily apparent in the position basis).

Despite the rough analogy described above between DV stabilizer states and Clifford gates on  the one hand and CV Gaussian states and gates, there are important differences between them. For the DV case, Clifford gates are (for the case of a single qubit) a subgroup of the compact group $\mathrm{SU}(2)$, whereas for the CV case, Gaussian operations on a single oscillator mode belong to the non-compact group $\mathrm{SU}(1,1)$.  The reason for this difference originates with the Bogoliubov transformation of the mode operators effected by conjugation with the squeezing transformation (see  Eqs.~(\ref{eq:smsqueezesymplectic}, \ref{eq:TMS_mode_ops_transformation}, \ref{eq:TMSsymplectic}) and Boxes~\ref{Box:1-mode-squeezing}, \ref{Box:2-mode-squeezing}). We have seen already that Gaussian operations can be represented by symplectic transformations of the $N$ coordinates and momenta so that $\mathrm{SU}(1,1)$ is isomorphic to $\mathrm{Sp}(2N,\mathbb{R})$. 

At this point we must note one glaring problem with taking the CV analog of stabilizer states to be Gaussian states--namely, Gaussian states are not eigenstates of translation operators (which are the analog of Pauli operators).

Working in standard dimensionless units, consider a displacement operation by a distance $\Delta$ in position applied to a position eigenstate $|x\rangle$
\begin{align}
    D_x(\Delta)|x\rangle=|x+\Delta\rangle.
\end{align}
The (unnormalizable) eigenvectors of this displacement operation is a highly non-Gaussian GKP-like (see Sec.~\ref{sec:qec-compilation}) states \cite{GKP2001,royer2022encoding}
\begin{align}
    |\psi_{k,x_0}\rangle &=\sum_{j=-\infty}^{+\infty} e^{+ikj\Delta }|x_0-j\Delta\rangle,\,\,\,\,\,\textrm{obeying}\\
    D_x(\Delta) |\psi_{k,x_0}\rangle &= e^{ik\Delta} |\psi_{k,x_0}\rangle,
\end{align}
where the fiducial position $x_0$ is arbitrary.  

From the above, we see that we need to rethink our definition of stabilizer states.  One reasonable possibility is to restrict our attention to the two GKP code states
\begin{align}
    |0\rangle_\mathrm{L} &= \sum_{m=-\infty}^{+\infty}|2m\sqrt{\pi}\rangle_x,\label{eq:GKP0L}\\
    |1\rangle_\mathrm{L} &= \sum_{m=-\infty}^{+\infty}|(2m+1)\sqrt{\pi}\rangle_x,
    \label{eq:GKP1L}
\end{align}
and to define the Pauli group to be the subgroup of the Gaussian operations which act like DV Pauli operators within this restricted code space.  These are the phase space translations
\begin{align}
    X_\mathrm{L} &= D_x(\sqrt{\pi})=e^{-i\sqrt{\pi}\hat p},\\
    Z_\mathrm{L} &= D_p(\sqrt{\pi})=e^{+i\sqrt{\pi}\hat x},\\
    Y_\mathrm{L} &= +iX_\mathrm{L}Z_\mathrm{L}.
\end{align}
It is straightforward to show from the commutation properties of these translations given in Eq.~(\ref{eq:phase_space_gates}) that they precisely obey the Pauli algebra
\begin{align}
    X_\mathrm{L} Y_\mathrm{L} = i Z_\mathrm{L},\,\, \textrm{et cyc.,}
\end{align}
and show that
\begin{align}
     X_\mathrm{L}|0\rangle_\mathrm{L}&= |1\rangle_\mathrm{L},\\
     X_\mathrm{L}|1\rangle_\mathrm{L}&= |0\rangle_\mathrm{L},\\
     Z_\mathrm{L}|0\rangle_\mathrm{L}&= +|0\rangle_\mathrm{L},\\
     Z_\mathrm{L}|1\rangle_\mathrm{L}&= -|1\rangle_\mathrm{L},
\end{align}
as expected.  Furthermore, both GKP code words are stabilizer states that are $+1$ eigenstates of the two stabilizers
\begin{align}
    S_x &= X_\mathrm{L}^2,\\
    S_p &= Z_\mathrm{L}^2.
\end{align}
As a result, the Pauli operators effectively square to the identity as they do in the usual DV case.  We see that the GKP code effectively compactifies the phase space plane into a torus of surface area $4\pi$ that contains a Hilbert space of dimension 2.  One might ask how it is that only two stabilizers are needed to reduce the infinite oscillator Hilbert space dimension down to two.  This is because, unlike ordinary DV Pauli string stabilizers that only take on two values, $\pm$, $S_x$ and $S_p$ are unitaries with a continuum of eigenvalues on the unit circle in the complex plane, and continuous displacements of the code words in phase space cause continuous changes in the phase of the stabilizer eigenvalues.  Thus, specifying the phase of the stabilizer requires an infinite number of bits of information, and it is this fact that causes the infinite reduction in Hilbert space dimension.

We do not delve into the more general Gaussian operations that are analogous to Clifford group operations on GKP bosonic qubits other than to note that the CNOT gate is the following Gaussian operation on two oscillators (denoted $a$ and $b$)
\begin{align}
\mathrm{CNOT} &= e^{-i\hat p_b\hat x_a},
\end{align}
which is nothing but the SUM gate defined in Box~\ref{Box:2-mode-sum}.  Mode $b$ is displaced by an odd number of half lattice constants (i.e., and $X_\mathrm{L}$ operation is performed) if and only if mode $a$ is in state $|1_\mathrm{L}\rangle$.
Finally, we mention that experimentally realistic, finite-energy (normalizable) approximations to the GKP code words are discussed in Sec.~\ref{sec:qec-compilation}.

Note that in Table \ref{table:qubitsvsCV} we compare qubit Pauli measurements to homodyne measurements of the oscillator phase space coordinates $\hat x,\hat p$.   The latter are the generators of displacements, but they are not the displacements themselves (which are the objects that are analogous to the Pauli operators).  While homodyne measurements are considered to be Gaussian measurements (typically performed by amplifiers involving one- or two-mode squeezing or using beam splitters and local oscillators as described in App.~\ref{sec:homodyne-detection}), measurement of displacement operators requires non-Gaussian operations (see App.~\ref{sec:characteristic-function}).

Another point related to measurement is the following: one important task for quantum computation is to benchmark the fidelity of different operations. Given the analogy between Clifford operations and Gaussian operations, it may be tempting to conclude that techniques such as randomized benchmarking \cite{knill2008randomized} for Clifford gates can be directly applied to CV systems. However, this is not the case because Gaussian operations do not form a unitary $t$-design. We will come back to this in more detail in Sec.~\ref{sssec:benchmarking}.

 The fact that Gaussian states are not eigenstates of displacements also leads to a major point of difference found in the context of quantum error correction. For the case of qubits, even though our adversary, the `noise demon,' may have the power to do non-Clifford operations, we can (under certain conditions) defeat it and carry out QEC using only Clifford circuits and measurements~\cite{Girvin_LesHouches_QEC}.  The analogous statement for continuous variable systems is not true.  There is an important no-go theorem stating that Gaussian resources (i.e., a Gaussian channel comprising Gaussian operations plus homodyne measurement) are insufficient to do continuous variable QEC \cite{Gaussian-no-go}.  This is discussed further in Sec.~\ref{sssec:oscillator-to-oscillator}.

In addition to Table \ref{table:qubitsvsCV} which summarizes the overall analogies we have discussed here, the reader is directed to Table~\ref{tab:gates-osc} which summarizes experimentally relevant Gaussian and non-Gaussian gates for oscillators.

\subsection{Hybrid Non-Gaussian Operations}
\label{sec:non-Gaussian_and_Hybrid}
 We present a discussion of the gate operations (listed in Table~\ref{tab:gates-qubit-osc}) 
that are possible (under certain physics approximations) for the hybrid oscillator-qubit systems, from a computer-science perspective, leaving details of practical implementation in different hybrid CV-DV platforms to App.~\ref{app:PhysImp}.  The reader is also directed to important prior work in Ref.~\cite{PhysRevLett.99.117203} describing control techniques to create non-Gaussian states in a hybrid system comprising a nano-mechanical resonator and a charge-based qubit.

In the context of universal quantum channel construction, combining the qubit-assisted universal oscillator control, discussed in this section, with qubit readout or reset can thus yield universal channel control for oscillators using single auxiliary qubits. There have been several experimental demonstrations of channel constructions using oscillator-qubit coupling. Dissipative cooling of an oscillator to a squeezed vacuum state was achieved in trapped ions \cite{lo2015spin,kienzler2015quantum} and in microwave resonators \cite{Dassonneville_2021}. Dissipation to a superposition of squeezed states, demonstrated in trapped ions as well as superconducting circuits~\cite{RoyerGKP1_2020,Campagne-Ibarcq2020,Sivak_GKP_2022}, was used in the context of quantum error correction. Error correction maps are irreversible quantum channels and thus such universal channel construction methods can be used to remove the excess entropy/excitations of a bosonic code. We refer the readers to Sec.~\ref{sec:qec-compilation} for details regarding these examples.

%\afterpage{clearpage}{
\begin{center}
\renewcommand\theadfont{}
\def\arraystretch{2}
\begin{longtable*}[tb]{|c|c|c|c|l|c}  
                                              \cline{1-5}&                                    & \textbf{Gate Name}                            & \thead{\textbf{Parameters}} & \thead{\textbf{Definition}} \\ \cline{1-5}
 \cline{1-5}
\multirow{11}{*}{\rotatebox[origin=c]{90}{\textbf{Hybrid Oscillator-Qubit Gates}}} & \multirow{8}{*}{\rotatebox[origin=c]{90}{Single-Oscillator (Sec.~\ref{sec:non-Gaussian_and_Hybrid})}} & Conditional rotation  (Box \ref{Box:CRotation})            & $\theta \in [0, 4\pi)$     & $\text{CR}(\theta) = \exp\left[-i\frac{\theta}{2} \sigma_z a^\dagger a \right]$  &    \\  
                                             &                                    &Conditional parity (Box \ref{Box:CRotation})   & $\theta =\pi$
                                                & $\text{CP}= \exp\left[-i\frac{\pi}{2} \sigma_z a^\dagger a \right]$     \\  \cline{3-5}
&                                    & SQR\textsuperscript{1} \cite{wang2021photon} (Box \ref{box:SQRgate})      & \makecell{$\theta_n \in [0, 4\pi)$ \\ $\varphi_n \in [0, 2\pi)$}      & $\text{SQR}(\vec{\theta},\vec{\varphi}) = \sum_n R_{\varphi_n}(\theta_n) \otimes \ket{n}\bra{n}$      \\ \cline{3-5}
                                             &                                    & Jaynes-Cummings\textsuperscript{2}       & \makecell{$\theta,\varphi \in [0, 2\pi)$}   & $\text{JC}(\theta,\varphi) = \exp\left[-i\theta\left( e^{i\varphi} \sigma_- a^\dag  + e^{-i\varphi}\sigma_+ a\right)\right]$          \\ \cline{3-5} 
                                             &                                    & Anti-Jaynes-Cummings\textsuperscript{3} &  \makecell{$\theta,\varphi \in [0, 2\pi)$}    & $\text{AJC}(\theta,\varphi) = \exp\left[-i\theta\left( e^{i\varphi} \sigma_+ a^\dag  + e^{-i\varphi}\sigma_- a\right)\right]$           \\ \cline{3-5} 
                                             &                                    & Conditional displacement \cite{EickbuschECD, Campagne-Ibarcq2020} (Box \ref{Box:c-displacement})           & $\alpha\in\mathbb{C}$     & $\mathrm{CD}(\alpha) = \exp\left[\sigma_z\left(\alpha a^\dag - \alpha^* a\right)\right]$        \\ \cline{3-5} 
                                             &                                    & Rabi interaction      
                                             & $\theta \in \mathbb{R}$    & $\text{RB}(\theta) = \exp\left[ -i \sigma_x(\theta a^\dagger + \theta^* a)   \right]$          \\ \cline{3-5}
                                             &                                    & Conditional squeezing                & $\zeta\in\mathbb{C}$     & $\text{CS}\left( \zeta \right) = \exp\left[\frac{1}{2}\sigma_z(\zeta^* a^2 - \zeta a^{\dag 2}) \right]$          \\  \cline{2-5}
                                             & \multirow{3}{*}{\rotatebox[origin=c]{90}{\makecell{Multi-\\ Oscillator \\(Sec.~\ref{sec:non-Gaussian_and_Hybrid})}}}  & Conditional beam-splitter \cite{chapman2022high}           & \makecell{$\theta \in [0, 4\pi)$ \\ $\varphi\in [0, \pi)$}    & $\text{CBS}(\theta,\varphi) = \textrm{exp}\left[-i\frac{\theta}{2}\sigma_z(e^{i\varphi}a^\dagger b + e^{-i\varphi}a b^\dagger)\right ]$         \\ \cline{3-5} 
                                             &                                    & Conditional two-mode squeezing       & $\xi\in\mathbb{C}$   & $\text{CTMS}(\xi) = \exp\left[\sigma_z(\xi a^\dagger b^\dagger - \xi^* a b)  \right]$        \\ \cline{3-5}
                                             &                                    & Conditional SUM gate       & $\lambda\in\mathbb{R}$   & $\text{CSUM}(\lambda) =\exp\left[ \frac{\lambda}{2} \sigma_z(a^\dagger + a)(b^\dagger - b) \right]$        \\ \cline{1-5}                      
                                             
\caption{\textbf{Common hybrid oscillator-qubit gates.} This table is not comprehensive, however, it contains some of the most common gates for hybrid oscillator-qubit systems.  \textsuperscript{1}SQR stands for number-\textbf{S}elective \textbf{Q}ubit \textbf{R}otation. The single-qubit rotation $R_\varphi(\theta)$ is defined in Table \ref{tab:gates-qubit} of App.~\ref{app:PhysImp}. \textsuperscript{2}Also called Red-Sideband. 
 \textsuperscript{3}Also called Blue-Sideband.
}
\label{tab:gates-qubit-osc} 
\end{longtable*}
\end{center}
%}

In the case of hybrid oscillator-qubit gates, throughout this work, we interchangeably use the prefixes ``conditional'' and ``controlled'' to indicate an oscillator gate that depends on the state of the qubit (or vice-versa). We emphasize that this use of ``controlled'' differs from the conventional usage for multi-qubit gates, where a controlled gate corresponds to an operation that is carried out if and only if the control qubit is in a particular state. In contrast, here these prefixes indicate a gate that more generally carries out a different operation for each possible state of the control system.

\subsubsection{Conditional Displacement} 
Qubit-controlled cavity displacement operations (see Box \ref{Box:c-displacement}) have been discussed in the context of superconducting platforms in~\cite{qcMAP,EickbuschECD}. In trapped ions and neutral atoms, these gates are realized natively using side-band interactions in~\cite{haljan2005spin,HOME-GKP2019,HomeGKPQEC2022,scholl2023}. 

\allowdisplaybreaks
{
\abox{Conditional Displacement}{
\begin{equation}
    D_\mathrm{c}(\alpha,\beta)= |0\rangle\langle 0| \otimes \,D(\alpha) +
 |1\rangle\langle 1| \otimes \,D(\beta),  \label{eq:condDisp}
\end{equation}
where $D(\alpha)$ is the unconditional displacement gate given in Eq.~(\ref{eq:dispopnormalordered}).  We also define the symmetric conditional displacement gate
\begin{align}
    \mathrm{CD}(\alpha)&=D_\mathrm{c}(+\alpha,-\alpha)\\
    &= \exp\left[\sigma_z \otimes \left(\alpha a^\dag - \alpha^* a\right)\right]. 
\end{align}
For $\beta$ real, we have (in Wigner units)
\begin{align}
    \mathrm{CD}(\beta)=e^{-2i\beta\sigma_z\hat p},\\
    \mathrm{CD}(i\beta)=e^{+2i\beta\sigma_z\hat x}.
\end{align}
which are respectively conditional position displacement and momentum boost by $\beta$.

\phantom{XXX}

A more general form of the conditional displacement can be obtained by conjugating $\mathrm{CD}(\alpha)$ (or its asymmetric variant $D_c(\alpha,\beta)$) with auxiliary single qubit rotations, yielding
    \begin{equation} \mathrm{CD}_{\hat{b}\cdot\vec\sigma}(\alpha)=e^{\hat b\cdot\vec\sigma \otimes[\alpha a^\dagger -\alpha^* a]}
    \label{eq:gencondisp1}
\end{equation}
in the former case, where $\hat b$ is a unit vector defining an axis on the auxiliary qubit Bloch sphere.

\phantom{XXX}

We will also utilize still further generalizations of these gates to the multi-mode/multi-qubit case where the displacement of mode $j$ is controlled by qubit $k$.  These are denoted $\mathrm{CD}^{(j,k)}(\alpha)$ (see Eq.~(\ref{eq:cond_disp_compilation})) and $D_c^{(j,k)}(\alpha,\beta)$ and are discussed in Sec.~\ref{ssec:exact-analytical-qubit-gates}.

\label{Box:c-displacement}
}
}

These gates are used to displace oscillators in phase space along a vector defined by the complex number $\alpha$ or $\beta$, conditioned on the qubit state. For example, if $\alpha,\beta\in\mathbb{R}$, then the above-defined gate displaces the oscillator by an amount $\alpha$ ($\beta$) from the origin along $\hat{x}$ if the qubit is in state $\ket{0}$ ($\ket{1}$). Commonly, we realize conditional displacement gates with $\alpha=-\beta$ which yields the gate defined in Table~\ref{tab:gates-qubit-osc}. Throughout this manuscript, we most often use the term ``conditional displacement'' to refer to this latter symmetric case, and this is made explicit when referencing the more general form. Furthermore, we note the useful algebraic identity for composing two conditional displacements given in Sec.~\ref{sssec:composingCDs}. This section also provides some intuitive pictures to aid in understanding the application of these important gates.

\begin{figure}[htpb!]
    \centering
    \includegraphics[width=0.4\textwidth]{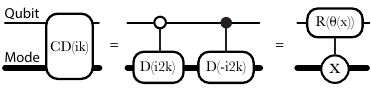}
    \caption{Circuit representations of the two equivalent interpretations of the conditional displacement gate ${\rm CD}(i2k)$ in Eq.~\eqref{eq:twoviewscD}. On the left, a displacement is applied to the oscillator conditioned on the qubit state; this is equivalently viewed on the right as a qubit $z$-rotation where the rotation angle depends on the position of the oscillator. 
    }
    \label{fig:CD-control-equiv}
\end{figure}

Note that from a control perspective, the conditional displacement gate can also be viewed as an oscillator-controlled qubit rotation. For example, The LHS of the controlled momentum boost 
\begin{equation}
\begin{split}
\mathrm{CD}(ik)&=
    e^{i2k\hat x\sigma_z}\\ &=\int \mathrm{d}x\, |x\rangle\langle x| e^{-i\frac{\theta(x)}{2}\sigma_z},\quad \theta(x)=-4kx.
\end{split}
\label{eq:twoviewscD}
\end{equation}
can be viewed as a momentum boost of the oscillator by $\pm k$ controlled by the state of the qubit and the RHS can be viewed as a rotation by angle $\theta(x)$ controlled by the oscillator position $x$.  These two complementary views are represented in circuit language in Fig.~\ref{fig:CD-control-equiv}.

\subsubsection{Conditional Rotation and Parity Gates} In addition to displacements of the bosonic mode conditioned on the state of an auxiliary qubit, we can have conditional rotations of the bosonic mode in phase space (see Box~\ref{Box:CRotation}).

\abox{Conditional Rotation and Parity Gates}{
This is the conditional version of the phase-space rotation gate defined in Box~\ref{Box:phase-space-rotation}
\begin{align}
    \mathrm{CR}(\theta)&=e^{-i\frac{\theta}{2}\sigma_z a^\dagger a}
\end{align}
This gate is natively available through the dispersive interaction between the bosonic mode and the auxiliary qubit (see Eq.~(\ref{eq:dispersiveHD}) of App.~\ref{app:PhysImp}).  The natively available dispersive interaction can be synthetically enhanced or compensated using a scheme utilizing the SNAP gate (defined in \ref{Box:SNAP}), as described in Sec.~\ref{sssec:SNAPgatesApp}.\\

The conditional photon number parity gate can be defined either as
\begin{align}
    \mathrm{CP} &=\mathrm{CR}(\pi)
\end{align}
or, using a different phase convention, as
\begin{align}
    \mathrm{CP}^\prime &=e^{-i\pi\frac{I-\sigma_z}{2} a^\dagger a}.
\end{align}
Either way, the phase kickback from the controlled parity gate (see Fig.~\ref{fig:CavityParityMeas}) can be used to make a QND measurement of the photon number parity \cite{Sun2014} and even the full photon number distribution via phase estimation or related methods \cite{Krastanov2015,Heeres2017,Wang2020FCFs}.

A generalization of these gates, $\mathrm{CR}^{k,j}(\theta)$, to the multi-qubit, multi-mode case in which oscillator $k$ is rotated conditioned on the state of qubit $j$ is defined in Sec.~\ref{ssec:exact-analytical-qubit-gates}. See Eq.~(\ref{eq:CRjk}).

\label{Box:CRotation}
}

\subsubsection{Photon Number Selective Qubit Rotation (SQR) Gate}\label{sssec:SQRgate} The conditional rotation gate in Box \ref{Box:CRotation} can be interpreted as a $Z$-rotation on the qubit by a rotation angle that is proportional to the boson  number (operator). However, it would be even more powerful if a different arbitrary qubit rotation could be applied conditioned on each oscillator Fock state.  This is accomplished with the photon number selective qubit rotation (SQR) gate (see Box \ref{box:SQRgate}). Combined with the displacement gate, the SQR gate suffices for universal hybrid control of the oscillator-qubit system (Table \ref{tab:ISA_overview}).

\abox{SQR Gate}{The cavity-conditioned qubit rotation gate, also known as the photon-number-Selective Qubit Rotation (SQR) gate, is defined as
\begin{align}
    \mathrm{SQR}(\vec{\theta},\vec{\varphi}) &= \sum_{n=0}^{N_\mathrm{max}} R_{\varphi_n}(\theta_n) \otimes \ket{n}\bra{n}\\
    &=e^{-i\sum_{n=0}^{N_{\mathrm{max}}}\frac{\theta_n}{2}\vec b_n\cdot\vec\sigma \otimes\hat P_n},\label{eq:U_mgate}
\end{align}
where $R_{\varphi_n}(\theta_n)$ (defined in Table \ref{tab:gates-qubit} in App.~\ref{app:PhysImp}) is a qubit rotation by angle $\theta_n$ about the axis $\vec b_n=(\cos\varphi_n,\sin\varphi_n,0)$ (lying on the equator of the Bloch sphere), $\hat P_n$ is the Fock state projector defined in Eq.~(\ref{eq:Fockproj}), and the total rotation angle of the qubit is given by $\theta_n$ for cavity Fock state $n$. Different qubit rotations conditioned on  different photon numbers can be executed in parallel by supplying multiple simultaneous drive tones to achieve this gate. See App.~\ref{app:PhysImp}.
\label{box:SQRgate}}

\subsubsection{Photon Number Selective Phase (SNAP) Gate}
\label{sssec:SNAP} In some cases, one may be interested in only controlling the oscillator while keeping the qubit state unchanged. This can be realized by a reduced version of the SQR gate, the SNAP (photon number selective phase gate) \cite{Heeres2015} gate, which applies a different Berry phase $\varphi_n$ to each photon Fock state $|n\rangle$ as described in Box~\ref{Box:SNAP}.
Together with (unconditional) cavity displacements, this is sufficient for universal control of the oscillator (Table \ref{tab:ISA_overview}). An efficient numerical gate synthesis protocol has been given to optimize short SNAP gate sequences to implement any given unitary on the oscillator \cite{fosel2020efficient}.  There has also been recent theoretical progress in analytically solving for efficient gate sequences \cite{job2023efficientSNAP_compilation}.  The use of SNAP gates has also been proposed to enhance non-linearities in optomechanical systems \cite{park2023efficient}.

\abox{SNAP Gate}{The SNAP gate \cite{Krastanov2015} is a powerful gate that imparts a different phase (chosen by the user) on each Fock state
\begin{align}
    \mathrm{SNAP}(\vec\varphi)&=e^{-i\sum_n\varphi_n |n\rangle\langle n|}\label{eq:SNAPU1}\\
    \mathrm{SNAP}(\vec\varphi)\sum_n\Psi_n|n\rangle&=\sum_n e^{-i\varphi_n}\Psi_n|n\rangle,
    \label{eq:SNAPU2}
\end{align}
where $\vec\varphi=(\varphi_0,\varphi_1,\varphi_2,\ldots,\varphi_{N_\mathrm{max}})$. \\

While frequently used as an oscillator-only gate, in practice this gate is achieved using an auxiliary qubit dispersively coupled to the bosonic mode and hence is more realistically written as a hybrid gate
\begin{align}
    \mathrm{SNAP}(\vec\varphi)&=e^{-i\sum_n\sigma_z\varphi_n |n\rangle\langle n|}.
        \label{eq:SNAPU3}
\end{align}
\label{Box:SNAP}}

To understand how to synthesize the SNAP gate, we make use of the cavity-conditioned qubit rotation (SQR) gate in Box~\ref{box:SQRgate} and the identity \cite{Krastanov2015}
\begin{align}
   \mathrm{SNAP}(\vec\varphi)&= \mathrm{SQR}(-\vec\pi,\vec\varphi)\mathrm{SQR}(+\vec\pi,\vec 0), 
   \label{eq:SNAPSQRSQR}
\end{align}
where $\vec \pi=(\pi)(1,1,\ldots,1)$.
This identity follows from the fact that if we start the auxiliary qubit in state $|0\rangle$, this pair of gates causes the qubit trajectory to return to the initial state but the enclosed area on the Bloch sphere imparts a Berry phase $\varphi_n$ on the cavity if it is in Fock state $n$.  For simplicity, we have written the SNAP gate as if it were a cavity-only non-Gaussian gate.  However as we have seen, the auxiliary qubit is used as a `catalyst' to effect the gate.  If the qubit starts out in state $|1\rangle$, the sign of the Berry phase is reversed.  Thus there is an implicit factor of $\sigma_z$ associated with the factor $\varphi_n$ in Eqs.~(\ref{eq:SNAPU1}, \ref{eq:SNAPU2}) and this is actually a hybrid qubit-cavity gate.

\subsubsection{Exponential-SWAP Gate} Another important hybrid gate that has been realized \cite{Gao2019} is the exponential SWAP (eSWAP) gate, which can be used as an entangling gate for universal computation that is bosonic code agnostic \cite{LauPlenio,Gao2019}. As defined below, the eSWAP gate produces a linear combination of the original and the swapped bosonic states on two modes with the relative fraction defined by $\theta$, conditioned on the qubit state. 

\abox{eSWAP Gate}{
    \begin{align}
        {\rm eSWAP}(\theta) &= e^{i \frac{\theta}{2}\, {\rm SWAP} }\nonumber\\
         &= \cos\left(\frac{\theta}{2}\right) I + i\sin\left(\frac{\theta}{2}\right) {\rm SWAP},
    \end{align}
    where the SWAP gate is defined as a special case of the beam-splitter gate in Box \ref{Box:beam-splitter} which swaps the states in two bosonic modes.

    Similar to the SNAP gate, the eSWAP gate is often used as an oscillator-only gate, but in practice is achieved using an auxiliary qubit that is initialized to $\ket{0}$. Generalizing to an arbitrary initial ancillary state, the eSWAP gate is written as
\begin{align}
    {\rm eSWAP}(\theta) &= e^{i \frac{\theta}{2} \sigma_z \otimes\, {\rm SWAP} }.
\end{align}
\label{Box:e-swap}
}

One way to look at the SWAP operation in a two-mode encoding (for example, dual rail encoding), is to view it as a logical Pauli-X operation which flips between $\ket{0}_\mathrm{L} = \ket{01}$ and $\ket{1}_\mathrm{L} = \ket{10}$. In this sense, when $\theta = \pi/2$, ${\rm eSWAP}(\pi/2) = \frac{I + i {\rm SWAP}}{\sqrt{2}}$  can be used as the $\sqrt{i{\rm SWAP}}$ gate to entangle between two modes.  Remarkably this remains true even if $\ket{01}$ refers to arbitrary logically encoded states (e.g., binomial code states) rather than Fock states in the two cavities.  This is because both identity and SWAP are agnostic to the contents of the two cavities, a powerful feature of the SWAP and eSWAP gates.

\begin{table*}[tbh]
\def\arraystretch{2}
\begin{tabular}{|c|c|c|c|}
\hline
                                                    & \textbf{Instruction Set Name}    & \textbf{Minimum gate set}    &  \textbf{Sections}            \\ \hline
\rotatebox[origin=c]{90}{\parbox[c]{2cm}{\textbf{Linear oscillator control}}}                           & Gaussian         & $ \mathcal{G} = \left\{D(\alpha), S(\zeta), \mathrm{BS}(\theta,\varphi) \text{ or } \text{TMS}(r,\varphi)\right\}$    & \multirow{4}{*} {  \makecell{Sec.~\ref{sec:gaussian gates}\\ and\\  Table \ref{tab:gates-osc} }   }          \\ 
\cline{1-3}
\multirow{3}{*}{\rotatebox[origin=c]{90}{\parbox[c]{2cm}{\textbf{Universal oscillator control}}}}  
                                                    & Cubic            & $\mathcal{G} + U_3\left(z\right) $   &                         \\ \cline{2-3} 
                                                    & Quartic          & $\mathcal{G} + U_4\left(z\right)$    &                   \\ \cline{2-3} 
                                                    & SNAP             & $\left\{D(\alpha), \text{SNAP}(\vec\varphi ),  \mathrm{BS}(\theta,\varphi) \text{ or }
                                                    \text{TMS}(r,\varphi)\right\}$               &          \\ 
\hline
\multirow{3}{*}{\rotatebox[origin=c]{90}{\parbox[c]{2cm}{\textbf{Universal hybrid control}}}}  
                                                    & Phase-Space      & $\left\{{\rm CD}(\beta), R_\varphi\left(\theta\right), \mathrm{BS}(\theta,\varphi)  \right\}$ & \multirow{3}{*}{ \makecell{Sec.~\ref{sec:non-Gaussian_and_Hybrid}\\ and\\  Table \ref{tab:gates-qubit-osc} } }     \\ \cline{2-3} 
                                                    & Fock-Space & $\left\{\text{SQR}( \vec{\theta},\vec{\varphi}), D(\alpha), \mathrm{BS}(\theta,\varphi)\right\}$  &                    \\ \cline{2-3}
                                                    & Sideband         &   $\left\{R_\varphi\left(\theta\right), {\rm JC}(\theta), \mathrm{BS}(\theta,\varphi)\right\}$   &                                         \\ \hline
\end{tabular}
\caption{\label{tab:ISA_overview} Common instruction sets for oscillator and hybrid oscillator-qubit architectures using gates defined in Tables \ref{tab:gates-osc}, \ref{tab:gates-qubit-osc}, and \ref{tab:gates-qubit}. Gaussian-only operations can achieve linear oscillator control. Adding non-Gaussian operations such as the general squeezing operation or SNAP gate can allow universal oscillator control. Moreover, to enable universal hybrid (oscillator-qubit) control, entangling gates (such as ${\rm CD}(\beta)$ or ${\mathrm{SQR}}( \vec{\theta},\vec{\varphi})$) between oscillators and qubits are needed. See App.~\ref{app:cross-compilation} for proof of the universality of the hybrid instruction set and cross-compilation between different instruction sets. $U_N(z)$ ($N \ge 3$) is the generalized squeezing gate as defined in Table \ref{tab:gates-osc}. $R_{\varphi}(\theta)$ is a single-qubit rotation gate defined in Table \ref{tab:gates-qubit}. The above constitute the minimal instruction sets, but we often have access to more gates such as dispersive interactions, phase rotations, etc. 
}
\end{table*}

\subsection{Instruction Sets for Universal Control of
Hybrid Oscillator-Qubit Systems}
\label{sec:instruction_set}

Having laid out in the previous section a catalog of hybrid CV-DV gates, we turn next to assembling different subgroups of these gates into instruction set architectures.  We will consider six distinct ISAs.  The first two are oscillator-only ISAs. The Gaussian ISA is easy to realize experimentally since it is generated by polynomial Hamiltonians linear or quadratic in the oscillator position and momentum operators.  It is, however, non-universal.  Universal control can be obtained by adding any higher-order polynomial, the simplest of which yields the Cubic ISA.

Turning to hybrid CV-DV ISAs that give universal control over both the oscillators and the qubits, we begin with the phase-space ISA based on oscillator displacements in phase-space controlled by the state of an ancilla qubit. This ISA is useful in hardware that has moderate dispersive coupling between the ancilla and the oscillator.  The Fock-Space ISA focuses on SQR gates that rotate the ancilla conditioned on the number of quanta in the oscillator.  This is most useful in hardware with stronger dispersive coupling between the oscillator and the ancilla.  The Sideband ISA focuses on gates generated by the Jaynes-Cummings (or anti-Jaynes-Cummings) Hamiltonian under which control pulses can add or subtract a photon from a resonator while simultaneously flipping the state of the ancilla qubit.  Finally, we will return to the Gaussian ISA which can be made universal if a supply of non-Gaussian resource states is available.

\subsubsection{Oscillator-only ISAs}
The Gaussian operations comprising displacements (in position and momentum), single-mode squeezing, and beam-splitters together form a parametrized instruction set, known as the Gaussian instruction set. The availability of this instruction set allows one to construct arbitrary Hamiltonians quadratic in $\hat a,\hat a^\dagger$. For convenience, we may also include two-mode squeezing (TMS), but it is redundant since, as illustrated in Fig.~\ref{fig:TMS}, TMS can be synthesized from single-mode squeezing using the Bloch-Messiah decomposition.  The Gaussian instruction set contains only linear ($a^\dagger,a$), quadratic ($a^\dagger a, a^\dagger a^\dagger, aa$), and bi-linear ($a^\dagger b, ab^\dagger$)  control Hamiltonians and hence, as discussed above, is analogous to a non-universal Clifford-only gate set. It maps Gaussian states to Gaussian states and thus cannot create Wigner negativity, but can create entanglement between modes (just as Clifford gates can create Bell states but not magic states for qubits). The universal control of oscillators requires a nonlinear resource beyond the quadratic Hamiltonians that generate the Gaussian instruction set. One way to achieve this is with a unitary realized via a cubic Hamiltonian (see App.~\ref{app:cubic_IS}), using, for example, specialized non-linear elements such as the SNAIL \cite{Frattini_SNAIL,hillmann2020universal,eriksson2023universal}.  In hybrid CV-DV processors, we have instruction sets that can achieve universality for oscillators via linear (or quadratic) coupling to DV ancillas -- another nonlinear resource. In fact, these instruction sets allow  for universal control over the full joint states of hybrid CV-DV processors.

\subsubsection{Phase-Space Instruction Set}
\label{sssec:phasespaceISA}
We turn next to hybrid system control.  Hybrid cavity-qubit gates can be used to  compile an effectively non-quadratic Hamiltonian using the BCH rules \cite{sefi2011how,borneman2012parallel,childs2013product,chen2022efficient,childs2021theory,kang2023leveraging,concentration2021chen}  and Trotter-Suzuki expansion (see Sec.~\ref{sec:compilation}). Remarkably, the coupling term between qubits and oscillators only needs to be linear in the oscillator mode operators (and is therefore easy to implement experimentally) to generate arbitrary nonlinear controls on the oscillators. 

The phase-space instruction set is simple and experimentally relevant \cite{EickbuschECD} for universal control of hybrid systems containing both oscillators and qubits.  It comprises qubit-controlled oscillator displacements, qubit rotations, and beam-splitters if there are multiple oscillators.  We begin our analysis of the universality with the case of a single qubit coupled to a single resonator.  The most general unitary for such a hybrid system is defined by the (effective) Hamiltonian
\begin{align}
    H=h_0(\hat x,\hat p)+\vec h_1(\hat x,\hat p)\cdot\vec\sigma,
     \label{eq:osc-qubit-control-H}
\end{align}
 where $\vec\sigma=(X,Y,Z)$ is the vector of Pauli matrices.
The first term $h_0$ above is a resonator-only Hamiltonian, while the second is a joint resonator-qubit Hamiltonian.  As we discuss further below (see also \cite{EickbuschECD}), $h_0$ is redundant since it can be generated from $\vec{h}_1$.  We therefore focus solely on $\vec{h}_1$.
Remarkably,  any arbitrary polynomial Hamiltonian of the form given in Eq.~\eqref{eq:osc-qubit-control-H}
can be generated from the Lie brackets of a very small control set
\begin{align}
    G_\mathrm{phase-space}=\{X,Z,\hat x Z,\hat p Z\},
    \label{eq:CD-ISAcontrolset}
\end{align}
which can then be used for universal control. The first two terms generate universal control of the qubit.  The second two terms individually generate controlled displacements in phase space defined in Box~\ref{Box:c-displacement}.  As noted previously, controlled Gaussian operations are (roughly) analogous to controlled-Clifford operations (e.g., the Toffoli gate) which are non-Clifford and yield universality. 
Using rotations of the qubit, we can generate $\hat x X$ from $\hat x Z$ and $\hat p Y$ from $\hat p Z$  
\begin{align}
   -i [\hat x X,\hat p Y]=Z(\hat x\hat p+\hat p\hat x).
    \label{eq:conditionalSqueeze1}
\end{align}
Notice that we obtain the anti-commutator for the cavity operators instead of the commutator as before, thereby increasing the degree of the polynomial to yield a controlled squeezing operation (which is discussed further in Sec.~\ref{sssec:conditionalsqueeze}). This Hamiltonian generates controlled squeezing of position and anti-squeezing of momentum, or vice-versa.   An alternative standard form of the (conditional) squeezing Hamiltonian can be obtained by rotating the axes of the phase-space displacements by $\pi/4$ to yield
\begin{align}
   -i\left[\frac{(\hat x +\hat p)}{\sqrt{2}}X,\frac{(\hat x - \hat p)}{\sqrt{2}}Y\right]=Z\left({\hat x}^2-{\hat p}^2\right).
   \label{eq:conditionalSqueeze2}
\end{align}

Iteration of the above commutator procedure (with appropriate qubit rotations) on the control set in Eq.~(\ref{eq:CD-ISAcontrolset}) allows us to use polynomials of degree $m,n$ to achieve higher-order terms of degree $m+n$
    \begin{align}
        [\hat{x}^n X,\hat{p}^m Y]&=  i[\hat x^n \hat p^m + \hat p^m\hat x^n] Z,
        \label{eq:controlledGauss}
    \end{align}
and thus we can generate arbitrary polynomials for $\vec h_1(\hat x,\hat y)$ giving universal qubit-controlled operations on the oscillator.  Examples of conditional Gaussians that can be achieved in superconducting cavity-qubit circuits are described in \cite{EickbuschECD,teoh2022dualrail}. In trapped ions and neutral atoms, conditional displacement using simultaneous red and blue side-band interactions are native gates \cite{HomeGKPQEC2022,HOME-GKP2019,bruzewicz2019trapped,haljan2005spin,scholl2023}.

How do we create the oscillator-only Hamiltonian, $h_0(\hat x,\hat p)$?  We can simply generate two polynomial terms of the form $A(\hat x,\hat p)Z$ and $B(\hat x,\hat p)Z$ such that their commutator yields the identity operator on the qubit and
\begin{align}
  h_0(\hat x,\hat p) &=  [A(\hat x,\hat p)Z,B(\hat x,\hat p)Z]=[A(\hat x,\hat p),B(\hat x,\hat p)],
\end{align}
acting only on the oscillator.

Again, the use of beam-splitter gates allows us to extend this to the multi-oscillator, multi-qubit case.   In summary, we see that the phase-space instruction set yields universal hybrid system control (see App.~\ref{app:cross-compilation}).   Example applications of this is presented in Sec.~\ref{sec:compilation}. Another extension of the phase-space ISA to multiple modes can also be achieved using displacements conditioned on a single-qubit ancilla state that is coupled to multiple oscillators~\cite{PhysRevX.14.011055}.

\subsubsection{Fock-Space Instruction Set}

The Fock-space instruction set comprises single-qubit rotations conditioned on the cavity photon number (SQR), simple oscillator displacements ($D(\alpha)$, \emph{not} conditioned on the state of the qubit), and beam-splitters.  This should be contrasted with the phase-space instruction set in which the operations on the cavity are conditioned on the state of the qubit. 
The Fock-space instruction set is thus dual to the phase-space instruction set in the sense that it interchanges the control and target for conditional gates.

Unconditional displacements are relatively simple in both circuit QED and trapped ion platforms.
The SQR and closely related SNAP gates were developed for superconducting systems to take advantage of the natively available dispersive coupling \cite{Heeres2015,Kundra_2022_Robust} but have not been reported for the trapped ion and neutral atom platforms where it would likely require  synthesis using side-band gates \cite{um2016phonon,Kihwan_Kim_an_experimental_2015}. Conversely, the spin-dependent forces used to create conditional displacements are natively available in ion traps but must be synthesized \cite{EickbuschECD} from the dispersive coupling in superconducting systems, but as discussed in Sec.~\ref{ssec:exact-analytical-qubit-gates}, offers greater control over the individual bosonic modes. Numerically \cite{fosel2020efficient} and analytically \cite{job2023efficientSNAP_compilation} optimized sequences of Fock-Space instructions have been theoretically demonstrated to be highly expressive and efficient for state preparation. 

\subsubsection{Sideband Instruction Set}\label{choiceOfISA}
The Sideband instruction set comprises qubit rotations, Jaynes-Cummings Hamiltonian sideband transitions (see Table \ref{tab:gates-qubit-osc}) which convert a Fock state excitation between the qubit mode and the cavity mode, and beam-splitters.  The Jaynes-Cummings interaction is natively available (but `always on') in superconducting qubit systems \cite{Blais2004,Wallraff_cQED_2004,beaudoin2012first,using2009leek}. 
Universal control of oscillator-qubit systems has been achieved using the sideband instruction set in trapped ions (see App.~\ref{subsec:sidebandgates}),  both approximately \cite{qudit2013mischuck} and analytically \cite{QuditsfromOscillatorsPhysRevA.104.032605} by strictly confining the control protocol within a certain energy budget. This is achieved by carefully choosing a subset of side-band rotations such that no leakage is possible beyond a given Fock level. Similar techniques of restricting the sideband rotations in the Jaynes-Cummings interaction to a subset have been employed to generate non-trivial qubit states and rotations without residual entanglement between the qubit and the oscillator \cite{goldberg2020transcoherent,goldberg2023beyondtranscoherent}. Numerical synthesis methods have been presented in \cite{NumericalGateSynthesis}. With the Sideband instruction set, the Eberly-Law protocol can be readily employed to prepare Fock states~\cite{hofheinz2009synthesizing}. Sideband operations have also been demonstrated in neutral atoms in the context of sideband cooling~\cite{Kaufman2012} and manipulation of the lowest two Fock levels~\cite{scholl2023}, which could be straightforwardly generalized to control of the higher levels (see App.~\ref{sct:neutralatoms}).

\subsubsection{Gaussian Instruction Set with Non-Gaussian Resource States}

An alternative model for universal computation is to consider having only Gaussian operations plus a supply of fixed non-Gaussian resource states (which would of course have to be created using non-Gaussian gates such as those described above).  The non-Gaussian `state factory' is a specialized system that handles the resource production (possibly in a non-deterministic way that utilizes distillation and `repeat-until-success' strategies).  The algorithm is then run using only relatively simple Gaussian operations on these resource states. This CV model is analogous to universal DV computation with Clifford circuits using a T-state factory as the non-Clifford resource. This model will not be discussed further here, but the reader is directed to \cite{menicucci2006universal,PhysRevA.79.062318,Weedbrook2012,Ghose-Sanders,marshall2015repeat,Baragiola,calcluth2024sufficient} for an entr{\'e}e into the literature on this subject.

\subsubsection{Choice of Instruction Set}
\begin{figure*}[!htb]
    \centering
    \includegraphics[width=0.96\textwidth]{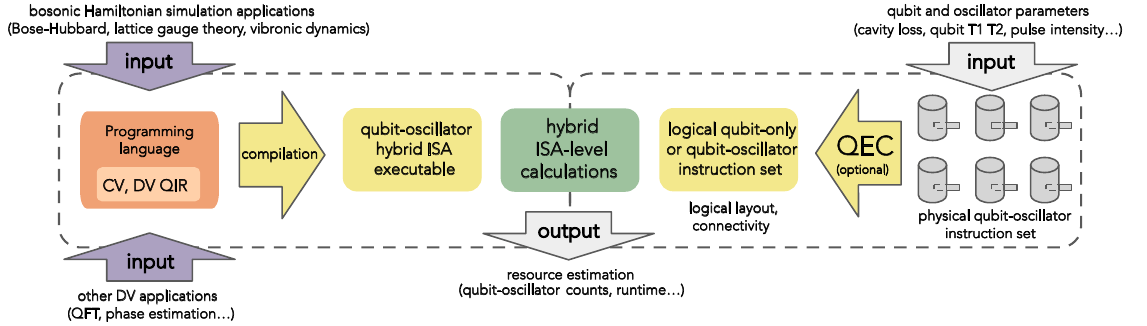}
    \caption{ 
    Resource estimation workflow for a hybrid oscillator-qubit quantum processor, inspired by Ref.~\cite{beverland2022assessing}. From left to the center: a native hybrid qubit-bosonic quantum application such as bosonic Hamiltonian simulation or a DV application such as phase estimation or quantum Fourier transform (QFT) is given as input and is described by proper programming languages including bosonic QIR (for bosonic Hamiltonian simulation) or standard DV QIR (for DV-only applications). The QIR program is then compiled into oscillator-qubit hybrid ISA executables which are agnostic to any physical device information. From right to the center: physical parameters of the hardware (such as cavity loss or qubit $T_1$ and $T_2$ coherence times) are taken as input which is used to estimate the cost of physical layer oscillator-qubit instruction sets (the grey cylinders are cartoons of 3D post cavities with transmon qubits). Optional quantum error correction is then applied to synthesize logical-level qubit-only or oscillator-qubit instruction sets from the physical ones. The two sides are combined in the middle to output the overall resource requirements for a given application.
    }
    \label{fig:hybrid-q-processor-resource-est}
\end{figure*}

The choice of the instruction set depends on the parameter details of the available hardware.  It also depends on the nature of specific applications, because 
the underlying algorithm for different applications may favor certain gates over others. In Sec.~\ref{sec:compilation}, we use single-oscillator-qubit instruction sets to discuss examples of state transfer, channel construction, and unitary applications. We use the phase-space instruction set when discussing examples related to states with a concise description in terms of the phase-space variables $\hat{x},\hat{p}$, for example, GKP logical code states \cite{GKP2001}.  We use the Fock-Space instruction set for tasks that are better represented using the photon number creation and annihilation operators $\hat{a},\hat{a}^\dagger$. Sec.~\ref{sec:application-ham-sim} presents multi-oscillator-qubit instruction set applications in terms of the advantage of using hybrid hardware over qubit-only systems for Hamiltonian simulations. The SNAP gate from the SQR gate in the Fock-space instruction set is useful for gates that conserve photon number, as shown in Table~\ref{tab:simulation}. From a computer architecture point of view, choosing different instruction sets is essentially using different hybrid operations in the bosonic QIR (Quantum Intermediate Representation, see Fig.~\ref{fig:hybrid-q-processor-resource-est}) for a hybrid CV-DV quantum computer. See Sec.~\ref{sec:architecture} for more architecture discussions.

\subsection{Control Flow and Benchmarking}
\label{ssec:controlflow}

In this section, we discuss two important aspects of architectures: control flow (Sec.~\ref{sssec:control-flow}) and benchmarking (Sec.~\ref{sssec:benchmarking}) of hybrid quantum processors.   These aspects are vitally important to understanding the capabilities of hybrid qubit-bosonic Hardware because they allow us to assess the performance of such devices and predict whether such hardware is adequately calibrated for a given computational task and how classical controls can be used to allow adaptive protocols to be run on such systems.  We discuss both of these issues in turn below.  

\subsubsection{Control Flow and Feedback}
\label{sssec:control-flow}

In quantum computers, control is realized using some (classical or quantum) information from prior circuit execution to better guide the subsequent quantum computation.  Quantum feedback control has been proposed and demonstrated experimentally \cite{lloyd2000coherent,nelson2000experimental}, and the essence of it can be most simply understood by combining the control and the target systems as part of a large interacting quantum system upon which the usual analysis can be performed without resorting to control theory.

More commonly, control signals appear as classical information, which is obtained from the measurement of part of the quantum system from prior execution or during runtime. The measurement outcomes can either be used to design better measurement schemes in later parts of the circuits or as classical control signals that parameterize (i.e., modify) later unitary quantum gates. As the most evident example, classical controls have been extensively explored for the construction of variational quantum algorithms \cite{tilly2022variational,cerezo2021variational,bharti2022noisy}. Another good use of classical control is to improve the robustness of the hybrid CV-DV system for autonomous quantum error correction as discussed in Sec.~\ref{sec:bosonic-QEC}. Error mitigation techniques \cite{temme2017error,suzuki2020quantum,endo2018practical,cai2023quantum,steckmann2025} in general and particularly on bosonic systems \cite{su2020error} are examples that demonstrate the power of classical feedback controls as well. 

At a broader scale, control flow will play an increasingly important role as hybrid CV-DV quantum processors acquire more complicated designs in the near future, especially for aspects regarding modular architectures and distributed quantum computation \cite{stein2023microarchitectures,ang2022architectures,Bourassa2021blueprintscalable,kumar2020optimal}, since a significant amount of information and signals need to be communicated across the entire computing hardware fabric.
Specifics concerning what properties of bosonic quantum systems can be measured and how measurements are performed are detailed in App.~\ref{app:measurement-tomography}. 

Finally, control flow can be used to overcome connectivity limitations in quantum hardware for which two-qubit gates can typically only be natively performed between physically adjacent qubits in the absence of a quantum bus that can transmit photons across the device.  The \emph{Locally Alternating Quantum Classical Computation} (LAQCC)  model \cite{buhrman2023LAQCC,pham20132d,friedman2023localityLAQCC,PhysRevLett.127.220503} uses runtime measurements, classical computation, communication, and feedforward conditional quantum operations to enhance the power of purely unitary circuits for tasks such as constant-time quantum state preparation \cite{PhysRevLett.127.220503,PRXQuantum.4.020339, KevinAKLT_PRXQuantum.4.020315, Iqbal2024, Smith2024} and quantum communication and entanglement distribution \cite{buhrman2023LAQCC,friedman2023localityLAQCC} for, e.g., the realization of QRAM \cite{xu2023QRAMsystems} and fault-tolerant quantum circuits with local operations \cite{Choe2024}.

\subsubsection{Benchmarking and Certification}
\label{sssec:benchmarking}

Benchmarking and certification of hybrid CV-DV quantum processors are challenging due to the infinite-dimensional nature of the oscillator Hilbert space (see \cite{PRXQuantum.6.020305} and references therein  for a discussion of these issues). 
Some protocols developed for benchmarking generic quantum systems \cite{kliesch2021theory,eisert2020quantum} may be applied to CV systems, including device-independent protocols \cite{vsupic2020self} for certifying entanglement between qubits and oscillators or between oscillators. Benchmarking protocols specifically designed for CV states have also been proposed by leveraging Gaussian measurements such as heterodyne detection \cite{chabaud_et_al:LIPIcs:2020:12062} to directly measure the Husimi-Q function (App.~\ref{app:measurement-tomography}), which has been applied to verify Gaussian Boson sampling experiments \cite{Chabaud2021efficient}. Such methods are suitable in the optical domain but do not take full advantage of the modern gate (see Sec.~\ref{sec:gates}) and measurement (see App.~\ref{app:measurement-tomography}) capabilities now available in the microwave domain of circuit QED.  In circuit QED \cite{Cat2Boxes} and in ion traps \cite{jeon2025multimodebosonicstatetomography}, multi-mode Wigner tomography (see discussion below and in App.~\ref{app:measurement-tomography}) based on joint photon number parity measurements have been demonstrated.

A major open challenge that remains in the field is determining whether there exists a protocol like randomized benchmarking for bosonic systems without Hilbert space truncation. Such a protocol would be invaluable because randomized benchmarking (and related protocols) gives a way to efficiently determine average gate fidelity for gates implemented within a gate set.   

Clifford randomized benchmarking works by drawing a random sequence of Clifford operations followed by the inverse of the sequence (which can be efficiently computed using the tableau algorithm of Gottesman~\cite{gottesman_thesis_1997stabilizer}). The state is then measured to see if it successfully matches the starting state.  The rate of decay of the success probability as a function of the length of the random Clifford circuit then yields the average gate fidelity for the Clifford operations.  Similar extensions to non-Clifford gates such as interleaved randomized benchmarking also exist and can be used to benchmark an entire gateset~\cite{magesan2012efficient,harper2017estimating}.

There is a strong mathematical analogy between the Clifford operations and Gaussian operations in optics (see Sec.~\ref{subsec:GaussiansAsCliffords}) since both have positive Wigner representations. The Wigner transform of a bosonic operator $G$ is 
\begin{equation}
g(x,p) = \int \mathrm{d}y \,  e^{ipy}  \bra{x-y/2} G \ket{x+y/2}  
\end{equation}
 and can be classically simulated (when coupled with stabilizer measurements/Gaussian measurements).  Furthermore, Gaussian operations are also closed under multiplication, which suggests that the randomized benchmarking protocol could work immediately out of the box.  Unfortunately, however, this is not the case.

The major problem facing this idea is that Gaussian operations do not form a unitary $t$-design, which is to say that averaging over the set of Gaussian gates does not necessarily match all $(t,t)$ moments (i.e., $t$ moments in the real and imaginary components) of the marginal distributions of the phase-space quadratures.  This inability to quickly scramble uniformly over the entire continuous variable space means that the randomization process will not be able to give an estimate of the average gate fidelity over the set of all Gaussian operations in this context. This means that, at present, randomized benchmarking-like protocols do not exist in such settings, and finding a suitable alternative benchmarking procedure remains an open challenge in the field. 
We note that recent progress has been made toward this goal, however, with regularized-rigged designs (i.e., in a rigged Hilbert space) offering an approximate solution for CV systems \cite{iosue2024continuous}.

A more ambitious goal is to benchmark a generic hybrid CV-DV quantum computation or simulation via \emph{process tomography}~\cite{altepeter2003ancilla} which may involve many non-Gaussian operations. One approach towards addressing this is to use the stellar hierarchy \cite{chabaud2022holomorphic}, mentioned in Sec.~\ref{sssec:stellar-rep}, to quantify the non-Gaussian character of CV states from zeros of the Husimi-Q function.  This allows us to use heterodyne measurements to perform fidelity estimation and witness Wigner negativity (which is equivalent to non-Gaussianity for pure states) \cite{chabaud2021certification} for general CV states.

Benchmarking necessarily involves measurement. One interesting question is what measurement protocols beyond simple Gaussian measurements such as homodyne and heterodyne detection can be useful to improve the benchmarking efficiency, as simple non-Gaussian gates can be compiled easily from our ISA gate sets. In addition, progress on the complexity of classical shadow tomography of CV systems can shed light on benchmarking as well \cite{becker2022classical,iosue2024continuous,gandhari2022continuous}. A rigorous analysis of measurement complexity, computation complexity, sampling complexity, and post-processing complexity is needed to fully characterize each benchmarking protocol \cite{eisert2020quantum}. 

For completeness, let us consider how to characterize a state using Wigner function tomography (also described in App.~\ref{app:measurement-tomography}, App.~\ref{sec:characteristic-function}).  $W(\alpha)$, the Wigner function at an arbitrary point in phase space represented by the complex number $\alpha$,  can be found by measuring 
\begin{equation}
W(\alpha) = {\rm Tr}[O(\alpha) \rho],\label{eq:wigner1mainbody}
\end{equation}
where $\rho$ is a quantum state associated with the system whose Wigner function we wish to identify, and we have defined  \cite{banaszek1999direct}
\begin{equation}
O(\alpha) \equiv \frac{2}{\pi} \sum_{n=0}^\infty (-1)^n D(\alpha) \ketbra{n}{n}D^\dagger(\alpha).\label{eq:wigner2mainbody}
\end{equation}
For the case of a hybrid system, we need to characterize correlations between the qubits and the resonator modes.  To achieve this it is useful to consider a discrete operator basis for the qubit portion of the system.  For example, using $n$-qubit Pauli operators $P_i$ and the orthonormality of the Pauli basis, we can always write the joint CV-DV density matrix as
\begin{equation}
\rho = \frac{1}{2^n}\left(\sum_i {\rm Tr}_\mathrm{qubit}\left [ \rule{0pt}{2.4ex}\rho\,\,(I\otimes P_i )\right]\otimes P_i\right).
\label{eq:Paulibasisforrho}
\end{equation}

From this, we can then define the conditional Wigner function to be 
\begin{equation}
W(\alpha|P_i) = {\rm Tr}\left [ \rule{0pt}{2.4ex} (O(\alpha)\otimes P_i)\rho \right],
\label{eq:conditional-wigner}
\end{equation}
the expectation values of which can be used to compute the expectation values of $\rho$.  Fig.~\ref{fig:lcu-cat-example} presents model examples of conditional Wigner functions associated with oscillator cat state production using an LCU (linear combination of unitaries protocol).     Experimental realization of such conditional Wigner functions can be found in \cite{Vlastakis2015}.

A useful aim is to understand the complexity of learning $W(\alpha|P_i)$ within error $\epsilon$ and the probability of failure at most $\delta$.  Each experiment to learn a particular value of the conditional density matrix requires a preparation of a density operator $ \rho $ and then we need to measure $O(\alpha)\otimes P_i$ for each prepared state.  For simplicity, let us assume that we are promised that $|{\rm Tr}[(D(\alpha) \ketbra{n}{n}D^\dagger(\alpha)\otimes P_i) \rho]| \le e^{\gamma_i-\lambda_i n}$ for constants $\gamma_i$ and $\lambda_i$.  Let us further define $P(n|i)=(1-e^{-\lambda_i})e^{-\lambda_i n}$ which corresponds to an exponentially decreasing probability distribution that we aim to introduce to approximately model the distribution's support on large photon numbers.  Under these circumstances, we can compute importance functions for each of the variables such that
\begin{align}
&{\rm Tr}\left(\left [ \rule{0pt}{2.4ex} O(\alpha)\otimes P_i\right ]\rho\right) = \frac{2}{\pi}\sum_n (-1)^n P(n|i)\nonumber\\
&\qquad\times\left[ {\rm Tr} \left ( \rule{0pt}{2.4ex} \left [ \rule{0pt}{2.4ex} D(\alpha) \ketbra{n}{n}D^\dagger(\alpha)\otimes P_i \right] \rho \right)/P(n|i) \right] \,.
\end{align}
We see that this takes the form of an expectation value of a random variable $(-1)^n ({\rm Tr}([D(\alpha) \ketbra{n}{n}D^\dagger(\alpha)\otimes P_i] \rho)/P(n|i))$.  We can estimate the sum then by computing the mean of this weighted quantity.  The variance in the mean is a simple function of the population variance, which is given by
\begin{align}
&\sigma_i^2 := \frac{4}{\pi^2}\sum_n \left [ \rule{0pt}{2.4ex} {\rm Tr}([D(\alpha) \ketbra{n}{n}D^\dagger(\alpha)\otimes P_i ] \rho)^2/P(n|i) \right ]\nonumber\\
&\qquad - {\rm Tr}\left ( \rule{0pt}{2.4ex} (O(\alpha)\otimes P_i)\rho \right)^2\nonumber\\
&\le \frac{4}{\pi^2}\left(\sum_n e^{2\gamma_i -\lambda_i n}/(1-e^{-\lambda_i}) \right)\nonumber\\
&= \frac{4e^{2\gamma_i}}{\pi^2(1-e^{-\lambda_i})^2}.
\end{align}
It follows that the number of samples needed to estimate the error within sample variance $\epsilon^2$ is
\begin{equation}
    N_{\rm samp} \le \frac{\sigma_i^2}{\epsilon^2}.
\end{equation}
From the Chernoff bound, it is straightforward to see that the scaling of the number of samples needed to learn just one point in the phase-space density within error $\epsilon$ and probability of failure at most $\delta$ under these assumptions scales as $O(\sigma_i^2 \log(1/\delta)/\epsilon^2)$.

Unfortunately, as the space of the oscillator is formally unbounded unless cutoffs are imposed, we would need to repeat this an infinite number of times to learn the Wigner function even inside some compact region of phase space.  Instead, we can assume that the underlying distribution is twice differentiable with the Hessian matrix of the Wigner function satisfying $\max_i\|H_W \| \le \Gamma$.  This assumption can be justified when the boson number is strictly bounded above by a constant because the Hermite polynomials used to represent a photon number eigenstate are each twice differentiable and a finite sum of twice differentiable functions is differentiable.  This also implies that the wave function vanishes super-exponentially quickly outside a finite phase space volume because the eigenfunctions decay as a Gaussian multiplied by a polynomial of bounded degree.  Under these circumstances, we can bound the error in any phase-space integral over a finite region $A_i$ of area $|\Delta_A|^2$ and further assume that each of the lengths of the regions is at most a constant multiple of all of the others for simplicity (otherwise we would need to specifically maximize over the maximum lengths rather than stating the bound more simply in terms of the area).  From the midpoint rule, if $\alpha$ is the midpoint of the region then
\begin{equation}
\bigg|\int_{x \in A_i} W(x) \mathrm{d}x - W(\alpha)|\delta|^2\bigg| \in \mathcal{O}(|\Delta_A|^4 \Gamma).
\end{equation}
Thus if we are promised that $\rho$ only has support within a bounded region of phase space with area $A$, then there are at most $A/|\Delta_A|^2$ such regions that we need to evaluate.  Thus if we wish to obtain an error of $\epsilon$ in any phase-space volume then the area of each of the small phase-space regions, $A_i$, is
\begin{equation}
    |\Delta_A|^2 \in \mathcal{O}\left(\frac{\epsilon}{\Gamma A} \right)
\end{equation}
Then in turn the number of samples is simply the number of points in the grid multiplied by the required samples per point.  This gives us

\begin{equation}
    N_{\rm samp} \in {\widetilde{\mathcal{O}}}\left(\frac{A^3\Gamma^2 \max_i \sigma_i^2\log(1/\delta)}{\epsilon^4} \right).
\end{equation}

Here we say that $f\in \tilde{\mathcal{O}}(g)$ if $f$ is asymptotically less than or equal to $g$ up to polylogarithmic multiplicative factors.
This process needs to be repeated for each of the $P_i$ in the Pauli basis.  So if we wish to compute each of the conditional Wigner functions within this fixed error budget then from the union bound we need to perform each measurement within error $\epsilon/4^n$ to ensure that the total error is at most $\epsilon$.  Thus the total number of measurements needed is in
\begin{equation}
N_{\rm samp} \in {\widetilde{\mathcal{O}}}\left(4^n\frac{A^3\Gamma^2\max_i \sigma_i^2\log(1/\delta)}{\epsilon^4} \right).
\end{equation}
This shows that the error in reconstructing a tomographically complete set of conditional Wigner functions for the state is comparable to the cost of a tomographic reconstruction of a state under similar assumptions if the variance is modest and the second derivative is also small.  In contrast, even a small qubit system may be intractable to characterize if these continuity and variance assumptions are not met.

Next, let us consider specializing this problem to the problem of ancilla-assisted process tomography.  Ancilla-assisted process tomography has long been used in qubit systems and the idea behind it
is to use the Choi-Jamiolkowski isomorphism \cite{choi1975completely,jamiolkowski1972linear}
to construct a state that is equivalent to the quantum channel that we wish to characterize.  We then can learn the channel by tomographically reconstructing the corresponding state.   The Choi state is operationally constructed for a qudit system of dimension $D_c$ in the following manner, first an entangled state of the form
\begin{equation}
\ket{\psi}_c := \frac{1}{\sqrt{D_c}} \sum_{j=0}^{D_c-1} \ket{j}\ket{j},
\end{equation}
is prepared.  Next, we apply a quantum channel, $\Lambda$, that we wish to characterize to one of the two subspaces.
\begin{equation}
    \frac{1}{D_c} \sum_{jk} \Lambda(\ketbra{j}{k}) \otimes \ketbra{j}{k}
\end{equation}
Each $j,k$ on the right-most qudit then flags the density matrix found by acting on a complete operator basis, which using the linearity property of quantum channels maps is sufficient to reconstruct the action of the channel on any density matrix.

We aim to repeat the same logic here for hybrid qubit/oscillator systems.  First, we need to construct an analog of the Choi state.  The most natural input would be to consider the following un-normalizable state acting on two copies of the oscillator space and the qubit space:
\begin{equation}
\ket{\psi_c} = \sum_i\int_{-\infty}^\infty dq\, (\ket{q}\ket{q})(\ket{i} \ket{i}),
\end{equation}
where the integral is over the oscillator position $q$.
Conceptually, this state is nearly identical to the previous proposal but the fact that the state needs to be distributed over all values of $q$ necessitates infinite squeezing, which is unphysical.  The uniform superposition over the qubit states $\ket{i}$ can be efficiently prepared using Hadamard transformations which are conceptually much less problematic. 
 We can address this issue with the oscillator states by considering the use of states with finite squeezing and aim only to reconstruct the state over a region of fixed cutoff.  Specifically, we replace the state $\ket{\psi_c}$ with the state
 \begin{equation}
 \int_{-\infty}^\infty \frac{e^{-x^2/2\zeta^2}}{{\pi}^{1/4}\sqrt{\zeta}} \ket{x}\ket{0} \mathrm{d}x \otimes  \left(\frac{1}{\sqrt{2^n}} \sum_i \ket{i}\right)  \ket{0}.
 \end{equation}
 Next by performing the two-mode sum gate (Box \ref{Box:2-mode-sum}) and $n$ controlled-NOT operations on the qubits we can transform this to
 \begin{equation}
\ket{\psi}'_c =\int_{-\infty}^\infty \frac{e^{-x^2/2\zeta^2}}{{\pi}^{1/4}\sqrt{\zeta}} \ket{x}\ket{x} \mathrm{d}x \otimes \left(\frac{1}{\sqrt{2^n}} \sum_i \ket{i}\ket{i}\right) .
 \end{equation} 
 Then we can follow the same reasoning employed before: we apply a channel to the first qubit oscillator pair to produce a state of the form
\begin{align}
\mathcal{C}:=\frac{1}{{2^n}} &\int_{-\infty}^\infty \mathrm{d}x \, \mathrm{d}y\frac{e^{(-x^2-y^2)/2\zeta^2}}{\sqrt{\pi} \zeta}\nonumber\\
\times &\sum_{ij} \Lambda(\ketbra{x}{y} \otimes \ketbra{i}{j}) \otimes \ketbra{x}{y} \otimes \ketbra{i}{j}.
\end{align}
In the limit as $\zeta\rightarrow \infty$ we see that this gives the original definition of the Choi state; however, in this context, the resulting state is not only more physically realistic (states of infinite squeezing are experimentally impossible to prepare) but also is confined to a finite region following the analysis given previously to discuss the problem of quantum state tomography of the hybrid system.

Next, we have to argue about the truncation of the function.   We have that if we wish to truncate the positions to $(x,y) \in \kappa [-\zeta, \zeta] \times [-\zeta, \zeta]$ for $\kappa \ge 1$, the maximum error can be computed by considering the volume of the integrated Gaussian that is neglected from the truncated integral.  This integral is given by the error function and from the asymptotics of the error function,  we have that the neglected portion of the integral is in $\mathcal{O}(e^{-\kappa^2 \zeta^2})$ which if we set this error to be $\epsilon_{\rm trunc}$ then it suffices to take $\kappa \in \mathcal{O}( \zeta^{-1}\sqrt{\log(1/\epsilon_{\rm trunc})})$.  Thus we can always construct a finite region in position such that the support of the state is well defined within that space and the remainder is small.

With such a Choi state we can then compute the channel fidelity with the ideal channel.  This process discussed in~\cite{johnston2011quantum}, involves the introduction of several operators that do not have direct physical interpretations that are needed to reshape the information in the Choi state (or Choi Matrix) into an appropriate form.  The first of these operators, $F$, is the Choi representation of the transpose map $T$.  The projection onto the symmetric space is then given by $P_S = (1+F)/2$.  Finally  let $\lambda_1$ be the maximum eigenvalue of $P_S(T\otimes I)(\hat{C}) P_S$.  
If we assume that we discretize our state as $\hat{C}$ then the minimum gate fidelity is given by
\begin{align}
    \mathcal{F}_{\min} &= \quad\lambda_1 - \|[\lambda_1 P_S -P_S(T\otimes I) (\hat{C})]P_S)\|_{S_1}\nonumber\\
    &=\quad\lambda_1 - \|[\lambda_1 P_S\|_{S_1} +\|P_S(T\otimes I) (\hat{C})]P_S)\|_{S_1}
\end{align}
where $\|X\|_{S_1} := \sup_{\ket{\psi},\ket{\phi}} \bra{\psi}\bra{\phi} X \ket{\psi}\ket{\phi}$.  The first term in the computation of $\mathcal{F}_{\min}$ can be computed
If we let $[X]_W$ be the discretized Wigner representation of an operator $X$ and $\hat{C}_W$ be our discretized Wigner representation of the Choi matrix then (neglecting discretization error)
\begin{align}
    &\| P_S(T\otimes I)(\hat{C})P_S\|_{S_1}\nonumber\\
    &=\sup_{\ket{\psi},\ket{\phi}}{\rm Tr}([P_S(\ketbra{\psi}{\psi} \otimes \ketbra{\phi}{\phi})P_S ]_W[(T\otimes I)(\hat{C})]_W )
\end{align}
where here the trace is taken to be over the discretized positions of the Wigner function as well as the qubit values. This shows that if we perform discretized Wigner function tomography to learn the transposed Choi state $(T\otimes I)(\hat{C})$, we can then find the channel fidelity by optimizing overall states. This optimization can be carried out using semidefinite programming in time that is polynomial in the dimension of the discretized operator.

The next step is to show how we can estimate the discrete Wigner for a hybrid qubit/oscillator state. Specifically,  we perform the multi-dimensional Wigner transformation of the channel to learn the Wigner representation of the result.  Specifically, the Wigner representation of the operator is (see equivalent definitions of the Wigner function in App.~\ref{sec:characteristic-function})
\begin{align}
W({\bf p},{\bf q}) = \int_{\mathbf{x} \in \mathbb{R}^N} d{\bf x} \bra{\psi({\bf q+ x/2})} e^{i{\bf p}\cdot {\bf x}}  \ket{\psi({\bf q - x/2})} 
\end{align}
We see from the dot product in the exponential, that the integral computed for the multidimensional Wigner transform is simply the nested integral over each of the components individually.  Thus the generalization of the formula for the value of the conditional Wigner function for the multidimensional case is simply
\begin{equation}
W({\bm{\alpha}}|P_i) = {\rm Tr}([O({\bm{\alpha}})\otimes P_i]\rho),
\end{equation}
where $O({\bm{\alpha}}) $ is the higher-dimensional generalization of the one-dimensional analog considered earlier in Eqns. \eqref{eq:wigner1mainbody}-\eqref{eq:conditional-wigner}:
\begin{equation}
O({\bm{\alpha}}) = \bigotimes_{j=1}^{{\rm dim}({\bm{ \alpha}})}\left(\frac{2}{\pi} \sum_{n=0}^\infty (-1)^n D(\alpha_j) \ketbra{n}{n}D^\dagger(\alpha_j) \right),
\end{equation}
where dim($\cdot$) represents the dimension of a vector, and $\alpha_j$ is the $j$th component of $\bm{\alpha}$.
As before, reconstructing the channel from the Choi state is then a simple matter of repeating the previous tomographic protocol.

By repeating this process, we can perform Wigner function tomography to learn the input and output pairs of Wigner functions that describe all gate operations, fiducial preparations, and measurements.  Thus up to unobservable gauges, we can perform quantum gate set tomography on such a system to characterize every possible operation in a hybrid bosonic qubit device.  This shows that, while no randomized benchmarking-like protocol has yet been developed, protocols can be easily constructed that resemble gate set tomography~\cite{chow2012universal,nielsen2021gate}.  A full characterization of the gauge orbits that are present in such a reconstruction is left for future work.

\section{Quantum Error Correction with Oscillators}
\label{sec:bosonic-QEC}
Quantum error correction (QEC) is essential for mitigating noise, faulty gates, and decoherence in a functional quantum computer. In the hybrid CV-DV architecture (Fig.~\ref{fig:hybrid-processor-arch}), QEC sits between the compiler and hardware layers, ensuring reliable logical qubits/oscillators and operations, in contrast to the noisy physical counterparts controlled via the low-level ISA. Various error correction strategies exist, leveraging multi-level quantum systems such as oscillators to enable flexible encoding choices—qubits, qudits, or encoded oscillators—based on data requirements.  

Encoding options differ based on machine model. For example, one approach that will be discussed in this Section is to encode a two-level system in an oscillator, protecting an abstract qubit from errors locally (or ``at the hardware level''). This model, though oscillator-based, corresponds to an abstract qubit model that provides the user access to effective DV degrees of freedom corresponding to an abstraction of the underlying CV-DV hardware. This is one example of how quantum error correction can change abstractions. As another example, a qudit can be encoded in a single oscillator or an oscillator in multiple oscillators. If used for bosonic system simulations, this forms an abstract oscillator model. However, qudit encodings qualify as effective oscillators only when the encoded system has enough levels for efficient bosonic simulations; otherwise, they primarily reduce resource overhead compared to an abstract qubit model (forming, instead, an abstract qudit model).  

The remainder of this section is organized as follows. Sec.~\ref{sec:dissipation} gives an overview of the noise models for oscillators. As it is more widely known, we do not explicitly discuss noise models on qubits, and instead refer to Ref.~\cite{Nielsen_Chuang}. Sec.~\ref{ssec:qec-overview} gives a comprehensive overview of encoding options for each machine model. Next, Sec.~\ref{sec:qec-compilation} explores ISA compatibility with different encodings using schemes from previous works. As the QCMM is the most developed error-correcting model using oscillators, widely classified through translation symmetric and rotational symmetric codes, we focus on compiling logical instruction sets for qubit encodings. This discussion highlights how ISAs and AMMs influence encoding choices and vice versa. Once a QEC code is selected, encoding circuits, syndrome measurements, and logical operations are compiled into the high-level ISA as primitive resources. Fig.~\ref{fig:QEC-stack} illustrates how a hybrid architecture integrates encoded/logical layers for both DV and CV computation from the physical layer.

\subsection{Dissipation and Decoherence}
\label{sec:dissipation}
Like qubits, resonators suffer amplitude damping.  For a qubit, amplitude damping takes the excited state  to the ground state  at rate $\gamma$. For a collection of $L$ qubits, the typical energy and typical energy loss rate scales linearly in $L$.  For an oscillator, amplitude damping causes the average energy to relax at rate $\kappa$
\begin{equation}
\frac{d}{dt}\langle\hat n\rangle=-\kappa \langle\hat n\rangle.
\label{eq:SHOdampingrate}
\end{equation}
This equation demonstrates why it is not advantageous to encode many qubits in a single oscillator -- the mean boson number and therefore the excitation loss rate in a typical state scales \emph{exponentially} in the number of qubits $L$.  Nevertheless, such replacements can have significant advantages for modest $L$ because, as discussed in the next section, superconducting microwave resonators storing bosonic codes have very desirable error correction capabilities.  

Damping and decoherence of quantum systems coupled to a bath are typically described using the master equation for the density matrix derived by making the Born-Markov approximation on the assumption that the coupling to the bath is weak and the bath is memoryless.  The Lindblad form of the master equation guarantees that the time-evolution of the density matrix corresponds to a completely positive trace-preserving (CPTP) map
\begin{equation}
    \frac{d}{dt}\rho = -i[H,\rho] + \sum_j \mathcal{D}(E_j)\rho,
\end{equation}
where $\mathcal D(\mathcal{O})$ is a `superoperator' whose action on the density matrix is given by
\begin{equation}
    \mathcal{D}(\mathcal{O})\rho = \mathcal{O}\rho\mathcal{O}^\dagger - \frac{1}{2}\left[\mathcal{O}^\dagger\mathcal{O}\rho + \rho\mathcal{O}^\dagger\mathcal{O} \right],
\end{equation}
and the $E_j$ are `jump' operators acting on the Hilbert space of the oscillator and describing the effects of coupling to the bath.

For oscillator Hilbert space dimension $d=N_\mathrm{max}+1$, the density matrix $\rho$ and the jump operators are $d\times d$ matrices.  The density matrix is Hermitian, positive semi-definite, and has a unit trace, but the jump operators need not be Hermitian. In numerical solutions of the Lindblad master equation, it is sometimes convenient to `vectorize' the density matrix turning it into a vector of length $d^2$.  The advantage of this is that it allows the superoperators to be written as ordinary (very large, typically sparse) $d^2\times d^2$ matrices and $\mathcal{D}(\mathcal{O})\rho$ becomes ordinary matrix multiplication into a vector.

Typical jump operators for a microwave resonator include amplitude damping
\begin{equation}
    \sqrt{\kappa(\bar n+1)} a,
\end{equation}
describing the loss of energy due to (linear) coupling to the bath,
and heating,
\begin{equation}
    \sqrt{\kappa\bar n} a^\dagger,
\end{equation}
 describing the gain of energy from the bath. Here $\kappa$ is the rate of energy (not amplitude) damping of the oscillator and $\bar n$ is a phenomenological parameter giving the steady-state mean excitation number in the oscillator caused by its interaction with the (possibly non-equilibrium) bath.  If the above two jump operators are the only ones present in the dynamics then the oscillator number distribution in the steady state becomes the Bose-Einstein distribution and we have
\begin{equation}
    \bar n = \frac{1}{e^{\beta\hbar\omega}-1},
\end{equation}
where $\omega=2\pi f$ is the oscillator (angular) frequency and $\beta=1/k_\mathrm{B}T$ is the inverse temperature.  Since $k_\mathrm{B}/h\sim 21$ GHz/Kelvin, a typical microwave resonator with $f\sim 5$ GHz should have all its excitations frozen out at a dilution refrigerator temperature of 20 mK.  This tends to be a relatively good approximation for resonators but not for qubits, as the latter suffer more strongly from non-equilibrium heating effects and can be in their first excited state with a probability as high as $\sim 1-10\%$.

Unlike superconducting qubits, superconducting resonators typically do not suffer much from intrinsic frequency fluctuations that lead to dephasing (without energy relaxation).  But if such terms are present and if the frequency fluctuations $\delta\omega(t)$ are classical Gaussian white noise having a flat spectral density\footnote{In practice this is a poor approximation since low-frequency dephasing noise often has a $1/f$-like spectrum.}  $\kappa_\varphi$ such that
\begin{align}
    \langle\langle \delta\omega(t)\delta\omega(t') \rangle\rangle = \kappa_\varphi \delta(t-t'),
\end{align}
where the double brackets refer to the ensemble average over the noise, then the corresponding jump operator would be
\begin{equation}
    \sqrt{\kappa_\varphi} \hat n,
\end{equation}
where 
$\hat n=a^\dagger a$ is the number operator.  Under the influence of pure dephasing, $\hat n$ is a constant of the motion but the amplitude autocorrelation function,
\begin{equation}
    \langle a^\dagger(t) a(0)\rangle =\langle \hat n\rangle \langle\langle e^{-i\int_0^td\tau\, \delta(\omega(\tau)}  \rangle\rangle = \langle \hat n\rangle e^{-\frac{\kappa_\varphi}{2}t},
\end{equation}
decays exponentially in time.

In practice, the primary source of dephasing errors in cavities is an extrinsic effect, namely the dispersive coupling of the cavities to the transmon qubits used to control them.  An unexpected change in the transmon state due to coupling to its environment produces a large sudden change in the resonator frequency.  This dispersive coupling is discussed further below in Sec.~\ref{sec:non-Gaussian_and_Hybrid}. For a more elaborate discussion on noise mechanisms (including amplitude damping, dephasing, and heating) on different experimental platforms, we refer the readers to App.~\ref{app:Noise}.

\subsection{Overview of Bosonic QEC}\label{ssec:qec-overview}
A correctable logical qubit requires a large state space to provide the redundancy and entanglement needed to hide and protect quantum information from decoherence processes.  For physical-qubit-only models, the large state space of dimension $D=2^n$ comes from using $n$ physical qubits to form the logical qubit.  If a single logical qubit is encoded, the logical code words span a two-dimensional (entangled) subspace of the large state space. The formally infinite-dimensional Hilbert space of an oscillator offers a distinctly different model  \cite{Joshi_2021,cai_bosonic_2021,GrimsmoPuri_GKP_2021,DissipativeCatproposal,albert_performance_2018,grimsmo_quantum_2020,BinomialCodes,QuditsfromOscillatorsPhysRevA.104.032605}
in which logical code words are superpositions of various (low-lying) photon number states  within just a single mode or a small number of modes \cite{royer2022encoding}.\footnote{For a discussion of why one cannot achieve hardware scalability by replacing a \emph{large} number of physical qubits by a single oscillator mode, see \cite{ClimbingMountScalable}.}   
 Physical errors may be corrected without losing information if they deform the code subspace without distorting the encoded information. The control overhead, i.e., the additional cost in time and circuit depth for creating and correcting logical qubits, together with errors in the imperfect control processes, are the primary limitations in achieving useful QEC (bosonic or qubit-based) in many experiments today. In addition to control hardware,  auxiliary registers are required for logical state preparation, error syndrome measurement, and logical state readout. Error correction of logical qubits encoded into bosonic modes has recently been achieved at and beyond the  break-even point (for memory) where the quantum coherence of the error-corrected logical qubit exceeds the coherence of the best of the available uncorrected qubits in the system \cite{Ofek2016, ni2022beating, Sivak_GKP_2022,lachance2024autonomous} so that the error correction gain $G>1$ exceeds unity.

 Protection of quantum memories is an important goal but large-scale fault-tolerant (FT) computation will also require noise-resilient logical operations as well as minimal overhead. An FT architecture is one in which errors in all inherently noisy procedures such as state preparation, gates, measurement, and the error correction process itself, propagate only to a limited extent, such that they can be corrected in subsequent rounds of QEC. The road to fully FT computation with superconducting circuits is long, but there is steady progress in creating better bosonic circuit components and protocols such as fault-tolerant detection of quantum errors \cite{rosenblum_fault-tolerant_2018}, `error-transparent' logical gates \cite{ErrorTransparentVy_2013,KapitErrorTransparentPhysRevLett.120.050503,LuyanSun2020}, and `path-independent' gates \cite{PathIndependentGatesPhysRevLett.125.110503,PathIndependentAlgebraicStructure,ReinholdErrorCorrectedGates}.

The dominant noise source in microwave oscillators is photon loss, which provides a simple error model, particularly since, unlike the multi-qubit case, all the information and the errors occur in a single mode. Other noise such as heating, dephasing, and displacements may also be corrected with QEC. Noise is inherently platform-dependent, and the reader is directed to App.~\ref{app:PhysImp} and \ref{app:Noise} for more details about platform-specific noise and noise models.

Discrete variable QEC requires at a minimum $n=5$ physical qubits to construct one logical qubit \cite{FiveQubitCode} that can correct a general single-qubit error.  For the case of only amplitude damping noise, $n=4$ qubits (corresponding to Hilbert space dimension $D=16$) suffice to correct the damping to lowest order \cite{LeungAmpDampCode}. Because there are many physical modes (qubits), the noise channel naturally has increased entropy associated with (missing) information about the physical location of the errors. By way of comparison, the smallest single-mode bosonic code that corrects amplitude damping \cite{BinomialCodes} to lowest order uses only states with photon numbers $\le 4$ corresponding to a Hilbert space dimension of only $D=5$.
 When combined with necessary  strategies to determine the $n$ possible locations of various types of errors (e.g., the three possible Pauli errors on each qubit), reaching fault tolerance in qubit-based codes becomes extremely challenging. Algorithms such as Shor's factoring, which requires at least $\sim 4,000$ noiseless logical qubits to factor a 2048-bit integer, when considered practically, lead to resource estimations of three orders of magnitude more physical qubits \cite{gidney_how_2021}. One recent performance analysis of a repetition Cat Code QEC architecture \cite{PhysRevLett.131.040602} estimated that the computation of a 256-bit elliptic curve logarithm could be done in about 9 hours with approximately $125,000$ highly noise-biased cat qubits.

Bosonic quantum error correction has emerged as a promising strategy for detecting and correcting physical errors directly at the hardware level. By enabling local error correction, it reduces both the decoding complexity and resource overhead typically associated with fault-tolerant quantum computing~\cite{PhysRevLett.131.040602, raveendran2022finite}. This advantage arises from hardware-efficient syndrome extraction and correction protocols, made possible by the relatively simple error model of harmonic oscillators. Furthermore, the infinite-dimensional Hilbert space of bosonic systems provides a vast design space, allowing researchers to tailor encodings to the specific error mechanisms encountered in real-world hardware. Given that the choice of encoding fundamentally influences hardware control -- and, by extension, the ISA -- this section introduces an overview of the most popular bosonic encodings currently under investigation, with a particular emphasis on those demonstrated in experimental laboratories. Further details about compiling these codes can be found in Sec.~\ref{sec:qec-compilation}.

\subsubsection{Single-Mode Encodings}
The extension of QEC to bosonic modes was first carried out for amplitude damping (i.e., photon loss errors), using a multi-mode Fock basis scheme \cite{chuang_bosonic_1997} and then a superposition of coherent states (cat states) scheme \cite{cochrane_macroscopically_1999}. Following these, various codes were developed that leverage more complex phase-space patterns to protect against other oscillator decoherence mechanisms such as small displacements in Gottesman, Kitaev, and Preskill (GKP) codes \cite{GKP2001,Campagne-Ibarcq2020,HomeGKPQEC2022,Sivak_GKP_2022,lachance2024autonomous}, and against arbitrary polynomials of fixed degree in creation/annihilation operators with binomial codes \cite{BinomialCodes,LuyanSun2020,ni2022beating}.

\begin{figure}
    \centering
    \includegraphics[width=0.45\textwidth]{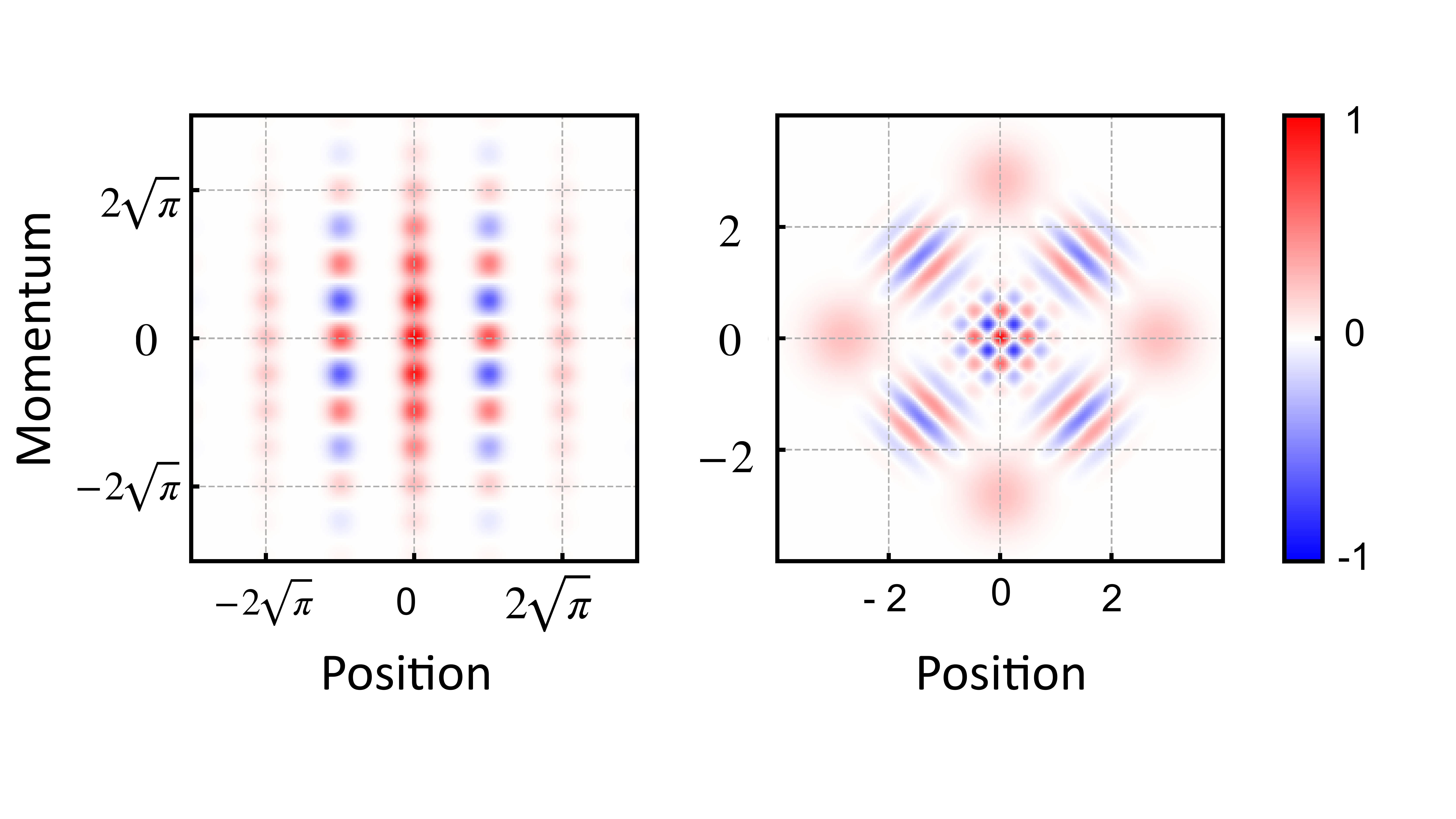}
    \caption{Wigner plots of two types of single-mode bosonic stabilizer codes, a translationally symmetric GKP code (left) and a rotationally symmetric code cat code (right). The horizontal axis represents the oscillator position (in standard units $\hat x=(a+a^\dagger)/\sqrt{2}$) and the vertical axis represents the oscillator momentum ($\hat p=(a-a^\dagger)/\sqrt{2}i$). The color bar represents the magnitude of the quasi-probability distribution in phase space. Details on Wigner function tomography can be found in App.~\ref{sec:characteristic-function}. Left panel: The logical $\ket{+Z_L}$ state of the finite-energy GKP code. In the limit of infinite lattice size (and therefore infinite energy), the states are invariant when shifted by $2\sqrt{\pi}$. This property holds for all (ideal) GKP code words, and is used for error correction against translation/displacement errors. A method to convert a bosonic damping channel to a displacement channel can be found in~\cite{GKP2001}. It has been shown that these codes are optimal in correcting for photon loss under ideal recovery \cite{albert_performance_2018}. Right panel: The logical $\ket{+Z_L}$ state of the four-legged cat code comprises a superposition of four coherent states.  These codes have a four-fold rotational symmetry. An $n$-legged cat state possesses an $n$-fold rotational symmetry. Such codes can efficiently correct for rotational errors (i.e., dephasing). For example, in Ref.~\cite{grimsmo_quantum_2020}, the authors propose the number-phase code, a rotationally symmetric code that is efficient in correcting for both oscillator dephasing and photon loss, under ideal recovery. 
    }
    \label{fig:single-mode-QEC}
\end{figure}
Cat codes and binomial codes are a subset of a code class with discrete rotation symmetry \cite{grimsmo_quantum_2020}, whereas GKP codes have discrete translation symmetry. Typically, all these bosonic codes perform better with increasing average photon number (even though the photon loss rate scales linearly with photon number) because a larger code space provides room for more redundancy.  This is analogous to qubit-based logical encodings where the error rate increases linearly with the number of physical qubits, but the logical error rate decreases (provided the physical error rate is below the threshold) because of the increasing code distance.  A study of the performance of cat, binomial, and GKP codes under the same photon loss error model \cite{albert_performance_2018} found that GKP codes significantly outperform other codes for most loss rates, and approaches (within a constant factor) the quantum capacity of the Gaussian thermal loss channel \cite{noh2018quantum}.

\subsubsection{Error Correction Techniques and Break-Even Performance:} 
Generally, QEC requires auxiliary registers for encoding, error detection, and recovery from errors, but these auxiliary elements also suffer from decoherence and excess measurement back-action. For an entr\'e into the literature on fault-tolerant control of  oscillators by qubits, see Refs.~\cite{xu2023faulttolerant,PathIndependentGatesPhysRevLett.125.110503,PathIndependentAlgebraicStructure,ReinholdErrorCorrectedGates,PhysRevApplied.23.024033}. One path forward is to use real-time measurement-based feedback  \cite{Ofek2016,hu_quantum_2019,Campagne-Ibarcq2020,ni2022beating}, or semi-autonomous \cite{Sivak_GKP_2022,schlegel2022quantum} or fully autonomous  \cite{KapitImprovedAutonomousQEC,lebreuilly2021autonomous,ChenWangAutonomous,li2023autonomous,li2023hardware,xu_autonomous_2023} feedback on the QEC circuit.

A more complete  unified approach is to consider the combined quantum channel of the encoding, error, and recovery steps and design recovery protocols that minimize hardware and measurement overheads. For example, autonomous QEC does not rely on syndrome error measurements but rather uses an engineered dissipation channel to stabilize some error-syndrome operators. Some protocols require only a single driven auxiliary non-linear element to create the appropriate dissipative bath  \cite{DissipativeCatproposal,touzard2018coherent,berdou2022second, Rglade2024}  or double-well Hamiltonian \cite{puri2017engineering,ShrutiBiasPreservingGates,grimm2020stabilization,PhysRevLett.128.110502,frattini2022squeezed,venkatraman2022quantum} to stabilize, for example, two-legged cat states which strongly and autonomously suppress certain errors. 
Refs.\ \cite{RoyerGKP1_2020,HomeGKPQEC2022} independently proposed an autonomous QEC protocol for GKP states, and \cite{HomeGKPQEC2022} have demonstrated a dissipative pumping technique for GKP code space stabilization in trapped ions, leading to a $\sim 3\times$ longer lifetime of the encoded state than without stabilization. (N.B.\ this is not however the error correction gain $G$.)  Ref.\ \cite{Sivak_GKP_2022} employed a semi-autonomous version of the protocol in \cite{RoyerGKP1_2020} that yielded an error correction gain (relative to the longest-lived  encoding [$0,1$ Fock] of the cavity) of $G\sim 2.3$.
As  decay errors in the auxiliary qubit used to detect error syndromes in the logical GKP qubit still limit the performance of GKP encodings, using a biased-noise auxiliary qubit like the Kerr-cat \cite{puri2017engineering, grimm2020stabilization} or fluxonium qubit \cite{fluxonium,somoroff2021millisecond} in which decay errors are exponentially suppressed compared to dephasing errors~\cite{GrimsmoPuri_GKP_2021} is helpful to further improve the error correction gain $G$. Recently, this gain was seen to be preserved in the first-error-corrected qudit encoding of $d=3,4$ levels using the GKP encoding~\cite{GKP_qudit_2024}.

\subsubsection{Multi-mode Encodings}
\label{sssec:multi-mode-encodings}

Encoding QEC codes across multiple oscillators offers additional advantages to tackle issues that arise specific to hardware implementation.
Recently discovered encodings of logical qubits in two oscillators~\cite{albert2019pair,royer2022encoding} have suggested ways around this problem. In Ref.~\cite{albert_performance_2018}, the authors encode a cat qubit in two oscillators where error syndromes can be detected with a circuit fault-tolerant to auxiliary qubit errors. Examples of 4D symplectic lattices for GKP encoding in two oscillators were studied in~\cite{royer2022encoding}. The authors show that the Tesseract lattice, where all stabilizers are represented by orthogonal translation vectors in 4D phase-space, can measure stabilizers using a `corner-circuit' which reduces the errors due to decay of the auxiliary qubit to essentially zero for an ideal GKP code. In practical (i.e., finite-energy) versions of this code, the auxiliary qubit contributes to logical error rates an amount which is inversely proportional to the finite-energy envelope size of the 4D GKP code.

Extensions of these encodings to more than two oscillators have been suggested for cat codes as well as GKP codes in Refs.~\cite{harrington2001achievable,albert2019pair,royer2022encoding, conrad2022gottesman, jain2023quantum, conrad2023good, lin2023closest} as a strategy for scaling up; however, a practical analysis with circuit-level noise is still pending  for such encodings. To use these multi-mode encodings realistically, one must limit the support of the stabilizers to nearest-neighbor modes, as in surface codes, or at least make the code rate constant, 
as in good LDPC codes (which are challenging because the stabilizers are not geographically local). 

Another prominent example of multi-mode encoding is the Schwinger boson representation of higher spin Hamiltonians for the SU($2$) group (and generalizations to the SU($N$) group \cite{schwinger1952on,mathur2010su(n)}) described in Sec.~\ref{sssec:many-spins}. 
Traditionally, such mappings have been important tools for formal analysis in low-energy \cite{gorshkov_two-orbital_2010,PhysRevLett.99.163002} and high-energy physics. As we discuss in Sec.~\ref{sec:application-ham-sim}, the inception of hybrid CV-DV quantum processors discussed in the present work can make good use of the Schwinger boson representations for efficient quantum simulation of spin Hamiltonians on bosonic hardware. Here, we focus on quantum error correction using the Schwinger boson representation of a spin-1/2 (i.e., a qubit), known in the optics community as the dual-rail encoding \cite{DualRailOriginalPhysRevA.52.3489,KLMdualrail,teoh2022dualrail,chou2023demonstrating,deGraaf2024midcircuit}. 

The dual-rail encoding has several interesting features, particularly in the circuit QED architecture where important gate, control, and measurement capabilities are more readily available than in traditional quantum optics.  First, as discussed in Sec.~\ref{sssec:many-spins}, arbitrary single-qubit rotations can be achieved using (microwave pulse activated) beam-splitter couplings between the two rails (described in Box~\ref{Box:beam-splitter}).  Second, for superconducting resonators the dominant intrinsic error is (slow) photon loss.  This leakage error out of the code space (span$\{|01\rangle,|10\rangle\}$) into the error state $|00\rangle$ can be efficiently converted to an erasure error (i.e., an error in a qubit whose location is heralded) by quantum non-demolition (QND) detection of the parity of the joint photon number \cite{Sun2014,Cat2Boxes,teoh2022dualrail} $\hat P = e^{i\pi(\hat n_1+\hat n_2)}$ (which changes from $-1$ in the code space to $+1$ in the error state).  This is important because erasure errors are much more readily corrected (by concatenation of the dual-rail qubits into a next-level code such as the surface code) than errors whose location is unknown \cite{PuriThompsonErasure,kang2023quantumerasure}, thus leading to substantially higher error-correction thresholds if erasure errors are dominant.

\subsubsection{Oscillator-to-Oscillator Encoding}
\label{sssec:oscillator-to-oscillator}
Rather than encoding discrete variable logical qubits in oscillators, an interesting but extremely challenging alternative would be to attempt to protect the entire bosonic Hilbert space from Gaussian errors by encoding an oscillator into  many oscillators. If this were possible, one could carry out error-corrected boson sampling for example.  It has been shown that Gaussian errors cannot be corrected using purely Gaussian quantum resources~\cite{Gaussian-no-go,vuillot_quantum_2019}. This no-go theorem stands in contrast to the result that in discrete variable error correction codes, Clifford operations and measurements are sufficient to correct both Clifford and non-Clifford errors.  To circumvent this no-go theorem, there has been a proposal to encode an oscillator into many oscillators by using GKP states as a non-Gaussian resource~\cite{noh2020encoding}, and using GKP codes as auxiliary qubits. In Ref.~\cite{wu2021continuous} this code was generalized to a GKP-Two-Mode-Squeezing code. The problem of optimal oscillator-to-oscillator encoding was addressed in~\cite{wu2022optimal,wu2021continuous}. However, Ref.~\cite{hanggli2021oscillator} shows that oscillator-oscillator code families do not have a threshold for the classical displacement channel, a noise model assumption used in Ref.~\cite{noh2020encoding}. This point is discussed further in the Sec.~\ref{sec:ConclusionsOutlook}.

\subsubsection{Code Concatenation}
\label{ssec:CodeConcatenation}
Among the leading candidates for discrete-variable codes in practice today are surface codes, which have been shown to possess good error-rate thresholds for fault-tolerant quantum computing~\cite{fowler2012surface}. A high threshold enables the multi-qubit code to achieve error correction gain with higher physical-level error rates. Recent demonstrations of beyond break-even error correction with GKP and binomial bosonic code qubits~\cite{Sivak_GKP_2022, ni2022beating} suggest that it would be fruitful to concatenate bosonic codes with discrete variable codes.  Using bosonic codes at the physical level could achieve physical error rates well below the threshold for the next-level discrete variable code allowing exponential reduction in the logical error rate with increasing size and distance of the discrete variable code, thereby yielding reduced resource overhead for fault-tolerant error correction. Ref.~\cite{vuillot_quantum_2019} proposed a decoding technique for GKP qubits concatenated with the surface code, while Refs.~\cite{noh_fault-tolerant_2020,lin2023closest,PRXQuantum.3.010315} studied how to increase the fault-tolerance threshold for GKP-surface encoding. In Refs.~\cite{terhal2020towards,GrimsmoPuri_GKP_2021}, the authors lay out perspectives and detailed approaches towards all GKP-surface codes and hybrid GKP-surface codes where the GKP states used at the base layer are stabilized by a biased-noise auxiliary qubit.

 This can further be extended to concatenation of cat codes with rectangular surface codes~\cite{PRXQuantum.2.030345}. Another advantage of CV-DV encoding was studied in Ref.~\cite{singh2022high} where, using biased noise Kerr-cat qubits in concatenation with the XZZX surface codes, the authors show a significant reduction in overhead for magic state distillation, a key ingredient for fault-tolerant quantum computing with discrete variable encoding. Concatenation of the biased-noise cat code into a repetition code was studied in \cite{PhysRevLett.131.040602,PRXQuantum.3.010329,PhysRevX.9.041053,putterman_hardware-efficient_2025}. Ref. \cite{raveendran2022finite} showed that concatenation of GKP code with generic quantum LDPC code can surpass the CSS Hamming bound by exploiting the analog information in the GKP code. Ref.~\cite{ruiz_ldpc-cat_2025} showed advantages of using cat codes with classical LDPC codes.

Despite the infinite Hilbert space of bosonic modes, the GKP code requires only two stabilizers to reduce this code subspace to dimension to $d=2$.  This is because the stabilizers (which are translations by the lattice constant in phase space) have a continuum of eigenvalues on the unit circle in the complex plane. Although CV codes seek to encode DV information into CV modes, the intrinsic analog character of CV error syndromes could be exploited in CV-DV code concatenation further to improve the overall performance of the hybrid system. Ref.~\cite{fukui_analog_2017} demonstrated a first example of this where analog information on the GKP qubits is used in concatenation with a Calderbank-Shor-Steane code using a maximum-likelihood method. They showed that the resulting CV-DV hybrid code can correct double-bit flip errors (as opposed to only single-bit flip errors in usual DV codes) in a three-qubit system. They also showed superior performance on quantum channel capacity with this concatenation scheme. Ref.~\cite{vuillot_quantum_2019} connected the analog character of the maximum likelihood  decoding problem for the GKP/toric code to a Feynman path integral for lattice gauge theory.

 In Ref.~\cite{xu2022qubit} CV-DV code concatenation was studied and compared with other types of concatenation using bosonic codes concatenated into smaller second-level codes such as the repetition code and the Steane code. The authors found that to encode quantum information in a way that can correct both logical $X$ and $Z$ errors, the above-mentioned CV-DV type encoding performs the best. Such code concatenation is yet to be compared with the practical multi-mode encoding discussed in the previous subsection. An analysis of this nature yields the optimal strategy for scaling up towards fault-tolerant error correction using oscillator codes.
 
\subsection{Logical ISA for Bosonic QEC}
\label{sec:qec-compilation}
In this section, we present compilation schemes for selected bosonic quantum error correction (QEC) codes  using the various hybrid instruction sets. Particular codes are most naturally implemented using particular ISAs. For example, translation symmetric codes, like the GKP codes shown in Fig.~\ref{fig:single-mode-QEC}(a) are most natural in the phase-space ISA while the rotation symmetric codes, like $4$-legged cat codes shown in Fig.~\ref{fig:single-mode-QEC}(b) are most natural in the Fock-space ISA. In addition to the $4$-legged cat codes we also discuss the binomial codes and dual-rail codes which fall under the umbrella of rotationally symmetric codes. We note that our classification of dual-rail codes as rotationally symmetric is not conventional; our reasoning for this classification lies in the fact that the logical codewords of dual-rail codes are based on Fock states which are rotationally symmetric states.

Importantly, throughout this section, we compile logical instructions using the hybrid CV-DV architecture where the qubits are used as ancillary systems and the oscillator as our encoding for data within the code space. Preliminary information required for the discussion in this section is given in Sec.~\ref{preliminary}. We present examples illustrating all components of bosonic QEC necessary for universal quantum computation with logical qubits: preparation (Sec.~\ref{sssec:qec-prep}), readout (Sec.~\ref{sssec:qec-readout}), stabilization (Sec.~\ref{sssec:qec-stabilization}), arbitrary single-qubit rotations (Sec.~\ref{sssec:qec-single-rot}) and entangling operations (Sec.~\ref{sssec:ent-gate}). The definitions of logical code words for each bosonic QEC code are given in Sec.~\ref{sssec:qec-prep}. 

We primarily review compilation schemes based on the phase-space ISA that 
utilize the methods of non-abelian QSP developed in Ref.~\cite{singh_towards_2025}. In addition, we give examples from Ref.~\cite{tsunoda2023error} of compilation schemes using the Fock-space ISA. For details on these schemes, we direct readers to these key references.

\subsubsection{Preliminary Information}\label{preliminary}

\subsubsubsection{Figure of merit.} In practice, the conditional displacements of the phase-space ISA generally involve displacing the oscillator by a large range of distances of order  $(-\sqrt{n}_\mathrm{max},\sqrt{n}_\mathrm{max})$, where $n_\mathrm{max}$ is the photon-number cutoff we impose on the oscillator. These gates thus require a large range of times during which errors on the oscillator-ancilla system occur at an approximately constant rate. Hence, the gate error probability is directly proportional to the duration of the gate. A good metric of efficiency for circuits involving such operations should take into account the total duration of the circuit  $T_\text{circuit}$ rather than just the gate count. This metric incorporates the fact that conditional displacement gates in the layers of Fig.~\ref{fig:numerical_constructions}b are not instantaneous, but require a time proportional to $|\gamma_i|$. Typical time scales \cite{EickbuschECD} for qubit rotations are $24\textrm{ ns}$ while the duration of conditional displacements 
$\mathrm{CD}(\alpha)$ is $\approx|\alpha|\text{ }\mu $s. The ratio of the two types of gate times ($T_R/T_{|\gamma_i|}$) is motivated from experimental parameters used in Ref.~\cite{EickbuschECD}. Similar arguments apply to other parameterized gates in the various ISAs.

\subsubsubsection{Non-Abelian QSP Primitives.} Many compilation techniques in this section use the entangling sequence described in the cat state preparation example in Sec.~\ref{sec:bivariable-qsp}. 
Here, in addition to cat codes, we consider other codes, such as GKP codes, whose code words are also represented by a sum of identical displaced Gaussian states $\ket{\mathcal{G}_i}$ with insignificant overlap (see Eqs.~(\ref{eq:GKP0}-\ref{eq:GKP}) further below)
\begin{align}
\ket{\psi}=\sum_i c_i\ket{\mathcal{G}_i}_\textrm{osc}. \label{Gauss_sum}    
\end{align}
We define the sequences $\mathcal{U}_v,\mathcal{E}_v$ that respectively disentangle and entangle the ancilla qubit with the oscillator state as
\begin{align}
    \mathcal{U}_v\left[\frac{\theta}{\alpha},\Delta\right]&=e^{i\frac{\theta}{\alpha}\hat v\otimes\vec{\sigma}\cdot {\hat n}_\phi}\times e^{i\frac{2\theta}{\alpha}\Delta^2{\hat v}_\perp\otimes\vec{\sigma}\cdot {\hat n}_\gamma},\\
     \mathcal{E}_v\left[\frac{\theta}{\alpha},\Delta\right]&=e^{i\frac{2\theta}{\alpha}\Delta^2{\hat v}_\perp\otimes\vec{\sigma}\cdot {\hat n}_\gamma}\times e^{i\frac{\theta}{\alpha}\hat v\otimes\vec{\sigma}\cdot {\hat n}_\phi},
\end{align}
 where 
 \begin{align}
 \hat v=\hat x \cos{\xi}+\hat p \sin{\xi},\quad
 {\hat v}_\perp=\hat p \cos{\xi}-\hat x \sin{\xi}
 \,
 \end{align}
 Here $\Delta^2=\langle {\hat v}^2\rangle -\langle \hat v\rangle^2$ is the variance of the individual Gaussian states in $\ket{\psi}$. In general, for a squeezed state, this depends on the magnitude and orientation of the squeezing. Note that the choice of $\xi$ is also determined by the orientation of the squeezing. The ancilla Bloch sphere vector ${\hat n}_\phi$ depends on the protocol. The vector ${\hat n}_\gamma$ is decided by the final (initial) state of the disentangled ancilla for $\mathcal{U}_v$ ($\mathcal{E}_v$). Formally, for $\mathcal{U}_v$ ($\mathcal{E}_v$) we need ${\hat n}_\gamma$ to satisfy $\vec{\sigma}\cdot {\hat n}_\gamma\ket{\zeta}=i\vec{\sigma}\cdot {\hat n}_\phi\ket{\zeta}$, given the ancilla final (initial) disentangled state is $\ket{\zeta}$.
\subsubsection{Logical State Preparation}\label{sssec:qec-prep} High-fidelity fast preparation of states encoded in the logical manifold, popularly known as logical states, is a key requirement for quantum error correction. Any error in this preparation leads to an additional burden on error correction. Here, we discuss the preparation of various bosonic code words using gates from the phase-space and Fock-space instruction sets. 

In using the phase-space instruction set, we more broadly employ the tricks used in the specific example of cat state preparation, discussed in Sec.~\ref{ssec:compilation-bosonic-qsp-qsvt}. The initial hybrid state for the preparation circuits is $\ket{0}_\textrm{vac}\otimes\ket{0}$. We use the definition of target state fidelity in this case as $\mathcal{F}=|\langle \phi|\psi,g\rangle|^2$ where $\ket{\psi}$ is the target oscillator state and $\phi$ is the hybrid oscillator-ancilla state at the end of the preparation circuit. Thus, the goal of the preparation circuits is to disentangle the oscillator from the ancilla qubit without measurements, such that the qubit is in a known state at the end of oscillator state preparation. Several of these preparation schemes are on par with the numerical circuits given in Sec.~\ref{sec:numerical-optimization} in terms of fidelity and circuit duration. In addition, analytical knowledge of these preparation schemes leads to additional applications in quantum error correction, as we discuss below. 

\subsubsubsection{GKP Code.}
\label{ssssec:GKPcomp}
The physically realistic normalizable version of the GKP code words defined in Eqs.~(\ref{eq:GKP0L}-\ref{eq:GKP1L}) (see the Wigner function plot for $\ket{0}_\textrm{GKP}$  in Fig.~\ref{fig:single-mode-QEC}) are defined as follows (where we have left the required normalization constants implicit):
\begin{align}
   \ket{0}_\textrm{GKP}&=\hat E_\Delta\sum_{m\in 2\mathbb Z} \ket{m\sqrt{\pi}}_x, \label{eq:GKP0}\\
   \ket{1}_\textrm{GKP}&=\hat E_\Delta\sum_{m\in 2\mathbb Z} \ket{(m+1)\sqrt{\pi}}_x ,\label{eq:GKP1}\\
   \hat E_\Delta&=e^{-\Delta^2\hat a^\dagger \hat a}\label{eq:GKP}
\end{align} 
Here $\ket{.}_x$ is the infinitely squeezed position eigenstates. Similarly, the $\pm$ logical basis is represented by the same superposition of momentum eigenstates (not shown here). The $\hat E_\Delta$ operator applies a Gaussian envelope on the infinite support of this state in phase space to restrict the energy of the system to a finite average photon number determined by $\Delta$. This operation takes the infinitely squeezed states $\ket{.}_x$ represented by Dirac-delta functions to finitely squeezed states represented by Gaussian functions with standard deviation equal to $\Delta$. Thus, in position basis, the GKP states are a superposition of multiple Gaussian wave functions with the same width displaced from the origin by even and odd integer multiples of $\sqrt{\pi}$.

To prepare these code words with the phase-space instruction set defined in Table.~\ref{tab:ISA_overview}, one can perform multiple rounds of the cat state preparation protocol with displacement $\alpha=\sqrt{\pi}$ which was discussed briefly in this work in Sec.~\ref{ssec:compilation-bosonic-qsp-qsvt}. However, for this preparation, the performance of the circuit is better when the protocol is applied to a squeezed vacuum. Ref.~\cite{singh_towards_2025} gives an efficient state-of-the-art construction for the preparation of squeezed states using phase-space ISA, so here we take the squeezed states as freely accessible. For logical $Z$ eigenstates, the circuit for $k^{th}$ round is given by,
\begin{align}
    \mathcal{C}_k=D_{c}(\sqrt{\pi},-\sqrt{\pi})\  \mathcal{U}_x\left[\frac{\sqrt{\pi}}{4k},\Delta\right ]\label{GKP_prep}.
\end{align}
Odd (even) $k$ yields a state that has high overlap with $\ket{0}_\textrm{GKP}$ ($\ket{1}_\textrm{GKP}$). $N$ applications of this protocol yields $N+1$ Gaussian wave functions in superposition. The Gaussian width $\Delta$ of each state in this superposition is given by the squeezing in the initial vacuum state. The circuit duration $T_\text{circuit}$ for the preparation scheme is decided by the target $\Delta$ and $\mathcal{U}_x$ is optimized for each $k$ in Eq.~(\ref{GKP_prep}). A squeezed vacuum can be achieved using the side-band ISA as well as the phase-space ISA in combination with various primitives described in Sec.~\ref{sec:compilation} such as Trotterization, BCH, etc. The squeezing required for the GKP code word is decided by the target envelope size $\Delta$ of the Gaussian functions in the superposition of the desired GKP state. For $\Delta=0.302$ with $N=3$ peaks, used in recent experiments~\cite{EickbuschECD,Sivak_GKP_2022}, this method yields an infidelity of $\mathcal{O}(10^{-2})$ using a circuit-depth of $T_\text{circuit}\approx 3\sqrt{\pi}(1+\Delta^2)\textrm{ }\mu\textrm{s}=5.84\textrm{ }\mu\textrm{s}$. However, it should be noted that since the ancilla is reset after every large conditional displacement by $\sqrt{\pi}$, post-selection can be performed from measurements of ancilla after every $1.95\textrm{ }\mu\textrm{s}$. This is the only fully deterministic scheme known to prepare GKP states without numerical optimizations. In addition, the authors show that this scheme can be modified to detect errors at various points in the circuit and, upon post-selection after each $C_k$, yield higher fidelity in the presence of faults. In Ref.~\cite{Hastrup_GKP_QSP}, the authors prepare GKP states using numerically optimized circuits that resemble the structure of this preparation scheme, that is, alternating conditional displacements and conditional momentum boosts. 

\subsubsubsection{Four-Legged Cat Code.} The four-legged cat code (see Fig.~\ref{fig:single-mode-QEC}), is a superposition of four coherent states positioned equidistant from the origin. The logical $Z$ code words are defined as cat states along the position axis and the momentum axis. Again with implicit normalization
\begin{align}
\ket{0}_\textrm{4C}=\ket{C_{0,\alpha,+}},\quad\ket{1}_\textrm{4C}=\ket{C_{0,i\alpha,+}}\label{4cat}
\end{align}
where $\ket{C_{\beta_1,\beta_2,\pm}}$ is an even $(+)$ or odd $(-)$ cat state of size $\beta_2$ displaced from the origin by $\beta_1$:
\begin{align}
\ket{C_{\beta_1,\beta_2,\pm}}=\ket{\beta_1+\beta_2}_\textrm{osc}\pm\ket{\beta_1-\beta_2}_\textrm{osc}.\label{eq:disp_cat}
\end{align}
The superposition states $\ket{\pm}_\textrm{4C}=\frac{\ket{0}_\textrm{4C}\pm\ket{1}_\textrm{4C}}{\sqrt{2}}$ are the X-basis four-legged cat codewords as shown in Fig.~\ref{fig:single-mode-QEC}.

Let us discuss the preparation of $\ket{+}_\textrm{4C}$ starting with the hybrid initial state of  $\ket{0}_\textrm{vac}\otimes\ket{0}$. Using the phase-space instruction set, we can create the four blobs in phase space as seen in Fig.~\ref{fig:single-mode-QEC} using a conditional displacement followed by a conditional momentum boost as follows:
 \begin{align}
  \ket{\psi_1}=e^{i\alpha(\hat x-\hat p)\otimes\sigma_z}  e^{i\alpha(\hat x+\hat p)\otimes\sigma_x} \ket{0}_\textrm{vac}\otimes\ket{0} \,.
 \end{align}
 Now, to disentangle the ancilla qubit from the oscillator we can apply the controlled rotation of the ancilla
 \begin{align}
 \ket{\psi_2}=e^{-i\frac{\pi}{4\alpha}(\hat x+\hat p)\sigma_x}\ket{\psi_1}\label{eq:4cat_prep} \,.
 \end{align}
 Following this step, $\ket{\psi_2}$ has a small residual entanglement between the two systems analogous to what is shown in Fig.~\ref{fig:QSPcat} for the cat state. For two-legged cat states, cancellation of this entanglement was accomplished using two methods: non-abelian QSP and a single-variable QSP sequence $\textrm{BB1}_{90}$ (see Fig.~\ref{fig:QSP_cat2} for a performance comparison of the two methods). In the present case, non-abelian QSP may not apply (see Ref.~\cite{singh_towards_2025} for details). We therefore use single-variable QSP with the $\textrm{BB1}_{90}$ scheme for disentangling the four-legged cat state from the ancilla. For $\alpha=5$, we obtain a fidelity to the target state of $\mathcal{F}=0.99$. The circuit depth for this sequence is $N=6$. Ref.~\cite{singh_towards_2025} shows that if the cat sizes of $\ket{0}_\textrm{4C}$ and $\ket{1}_\textrm{4C}$ in Eq.~(\ref{4cat}) are unequal, say $\alpha_1,\alpha_2$ such that $|\alpha_1|/|\alpha_2|$ is an even integer, a non-abelian QSP correction can be found.

\subsubsubsection{Binomial Code.} The simplest binomial code~\cite{BinomialCodes} can correct at most a single photon loss. It has logical code words defined as
\begin{align}
    \ket{0}_\textrm{bin}=\frac{\ket{0}_\textrm{osc}+\ket{4}_\textrm{osc}}{\sqrt{2}},\quad  \ket{1}_\textrm{bin}=\ket{2}_\textrm{osc}.
\end{align}
The preparation of these states can be achieved using the Law-Eberly protocol~\cite{law1996arbitrary} accessible via the sideband-ISA.  They are also efficiently realized in experiment \cite{EickbuschECD} using numerically optimized phase-space ISA gate sequences.

\subsubsubsection{Dual-rail Code.} The logical code words of dual-rail encoding~\cite{DualRailOriginalPhysRevA.52.3489,KLMdualrail} are represented by a single photon being in one of two distinct modes 
\begin{align}
    |0_\mathrm{L}\rangle &=|0\rangle_\textrm{osc}\otimes|1\rangle_\textrm{osc}\\
    |1_\mathrm{L}\rangle &=|1\rangle_\textrm{osc}\otimes|0\rangle_\textrm{osc}.
\end{align}
These can be distinct spatial or temporal modes of flying qubits, one standing mode in each of two separate resonators \cite{teoh2022dualrail,chou2023demonstrating,deGraaf2024midcircuit}, or two standing modes of a single resonator \cite{koottandavida2024erasure}.  The same type of encoding for pairs of two-level qubits has also recently been explored \cite{kubica2022erasure,levine2024demonstrating} but we focus here solely on the bosonic mode case. The preparation of these states can also be achieved using the Law-Eberly protocol accessible via the sideband-ISA or using non-abelian QSP via the phase-space ISA~\cite{EickbuschECD,singh_towards_2025}. 

\subsubsection{Logical Readout}\label{sssec:qec-readout}
A logical readout circuit can be realized by maximally entangling the logical code state in the oscillator with an auxiliary qubit  such that the measurement of the auxiliary qubit realizes the measurement of the logical code word.  
For example, consider the following maximally entangled state,
\begin{align}
    \ket{0}_L\ket{0}+\ket{1}_L\ket{1},
\end{align}
where $\ket{0}_L$ and $\ket{1}_L$ denote a logical bosonic qubit.  Measurement of the auxiliary $\{|0\rangle, |1\rangle\}$ qubit in the computational basis is then equivalent to a logical $Z$ measurement of the bosonic qubit. Thus, the task of readout can be accomplished by realizing an entangling sequence $\mathcal{E}$ between the bosonic qubit and the auxiliary qubit, followed by measurement of the qubit.  And assuming the measurement of the auxiliary qubit is perfect, the measurement failure probability is the probability of predicting the wrong logical state in the oscillator given a specific ancilla measurement outcome. 

The specific circuits that have been experimentally realized for entangling the auxiliary readout qubit with the logical bosonic qubit state for the dual-rail \cite{chou2023demonstrating,levine2024demonstrating}, binomial \cite{LuyanSun2020,ni2022beating}, four-legged cat \cite{Ofek2016}, and GKP codes \cite{Campagne-Ibarcq2020,Sivak_GKP_2022} are described in Sec.~\ref{sssec:qec-single-rot}. 

Here, as an example, we shall discuss the case of GKP logical readout in detail.  Importantly, as we shall see, the GKP codes are the only ones which can be read out in all three logical Pauli bases with equal ease.  The inverse of the disentangling gate in Eq.~(\ref{GKP_prep}) for $\mathcal{C}_1$ can be used as the entangling gate between the GKP qubit and the ancilla qubit.  

The GKP readout sequence is as follows. As explained in Sec.~\ref{sssec:qec-prep}, GKP $\ket{0}_L(\ket{1}_L)$ is a superposition of multiple Gaussian functions with peaks located at even (odd) multiples of $\sqrt{\pi}$. The logical GKP readout requires the ancilla qubit to rotate to orthogonal qubit states depending on whether the locations of the peaks in the wave function are at even or odd multiples of $\sqrt{\pi}$. Using the idea of conditional momentum boosts as rotations of the ancillary qubit depending on the position eigenvalue of the oscillator state, we can use $e^{\frac{i}{2}\sqrt{\pi}\hat x\sigma_x}$ for logical $Z$ readout. 

Just as in the case of cat states (Sec.~\ref{ssec:compilation-bosonic-qsp-qsvt}), we have rotation errors due to the uncertainty in the position of each Gaussian state in the superposition. These errors can be corrected using single-variable as well as non-abelian QSP. In this case, since the peaks have a standard deviation of $\Delta$, as fixed by the envelope parameter (see Eq.~\eqref{eq:GKP}), the pre-correction to reduce the rotation errors during logical $Z$ readout is achieved via $e^{i\sqrt{\pi}\Delta^2\hat x\sigma_y}$. Below, we give the non-abelian QSP sequences for logical Pauli readout of GKP codewords with envelope size $\Delta$, 
\begin{align}
    \mathcal{E}_p\left[\frac{\sqrt{\pi}}{2},\Delta\right] &=e^{-i\sqrt{\pi}\hat x\Delta^2\sigma_y}e^{i\frac{\sqrt{\pi}}{2}\hat p\sigma_x}\quad\textrm{X-basis},\label{eq:GKP_readoutX}\\
    \mathcal{E}_x\left[\frac{\sqrt{\pi}}{2},\Delta\right] &=e^{i\sqrt{\pi}\hat p\Delta^2\sigma_y}e^{i\frac{\sqrt{\pi}}{2}\hat x\sigma_x}\quad\textrm{Z-basis},\label{eq:GKP_readoutZ}\\
        \mathcal{E}_{\frac{x+p}{\sqrt{2}}}\left[\frac{\sqrt{\pi}}{\sqrt{2}},\Delta^2\right]&=e^{-i\sqrt{\pi}(\hat x-\hat p)\Delta^2\sigma_y}e^{i\frac{\sqrt{\pi}}{2}(\hat x+\hat p)\sigma_x}\quad\textrm{Y-basis}.\label{eq:GKP_readoutY}
 \end{align}
 
Notice that for a square GKP code defined in Eq.~(\ref{eq:GKP}) and shown in Fig.~\ref{fig:single-mode-QEC}, the displacement required to implement a logical $Y$ operation is longer than the corresponding displacements required for logical $X$ or $Z$ operations by a factor of $\sqrt{2}$. This fact accounts for the difference in the arguments of Y-basis readout in Eq.~(\ref{eq:GKP_readoutY}) and X/Z-basis readouts in Eq.~(\ref{eq:GKP_readoutX}-\ref{eq:GKP_readoutZ}).

Following the sequence in Eq.~(\ref{eq:GKP_readoutZ}), the $+1$ eigenstates of the GKP logical $Z$ operator rotates the ancilla by $2\pi$ while the $-1$ eigenstates rotate the ancilla by $\pi$ with high probability. Thus, for an arbitrary logical GKP state $\ket{\psi}_\textrm{GKP}=\alpha\ket{0}_\textrm{GKP}+\beta\ket{1}_\textrm{GKP}$, we can write
\begin{align}
    \mathcal{E}_x\left[\frac{\sqrt{\pi}}{2},\Delta\right]\ket{\psi}_\textrm{GKP}\ket{0}&=\alpha\ket{0}_\textrm{GKP}\ket{0}+\beta\ket{1}_\textrm{GKP}\ket{1}\nonumber\\&\quad\quad\quad+\mathcal{O}(\epsilon^3) \,,
\end{align}
where $\epsilon=\frac{\sqrt{\pi}}{2}\Delta^2$. At this point, measuring the ancilla realizes the measurement of the logical $Z$ operator. The $\textrm{Y-basis}$ readout is different due to the longer logical $Y$
 the operator of the GKP codes encoded in a square grid as shown in Fig.~\ref{fig:single-mode-QEC}. The failure probability for this readout is $\epsilon^3=10^{-4}$ for a GKP code with $\Delta=0.302$. The back action on the oscillator, however, scales as $\mathcal{O}(\epsilon^2)$. In practice, this error is negligible relative to those associated with faults in the auxiliary qubit during the entangling operation and subsequent measurement. In Ref.~\cite{Hastrup_GKP_QSP}, the authors numerically optimized the amplitude of this pre-correction using numerical methods and found it to be the same as  the analytical quantity given here, by the use of non-abelian QSP (see Sec.~\ref{ssec:compilation-bosonic-qsp-qsvt}). To date, this is the only known measurement circuit designed for finite-energy GKP logical readout and it works better than the simpler circuit designed for readout of the ideal (infinite-energy GKP) code words. For details, we direct the readers to the aforementioned references.

\subsubsection{Stabilization and Error Correction}\label{sssec:qec-stabilization}
Errors induced by interactions with the environment may increase or decrease the energy of the system (and/or introduce phase errors) leaving the oscillator in a state outside of the code space. Non-unitary channels involving measurements and qubit resets can be used to extract entropy from the system to bring the oscillator back to the code space. 
In this section, we  discuss how two of these {\em oscillator error correction} protocols can be achieved using hybrid instruction sets. Specifically, we highlight the methods of cat-state amplification and dissipative GKP stabilization.  These circuits extract entropy from the system via auxiliary qubit measurements (thereby converting the von Neumann entropy to Shannon entropy) or resets (which transfer the von Neumann entropy of the oscillator into the cold bath of the qubit).

\subsubsubsection{Cat State Amplification.} Energy relaxation extracts energy from an oscillator, shrinking its quantum phase space. Recall that two-legged even (+) and odd (-) cat states, with implicit re-normalization, are given by
\begin{align}
\ket{C_{0,\alpha,\pm}}=\ket{\alpha}_\textrm{osc}\pm\ket{-\alpha}_\textrm{osc}
\,,
\end{align}
using Eq.~(\ref{eq:disp_cat}).
Energy relaxation reduces the cat size $|\alpha|$ to $|\alpha^\prime|<|\alpha|$, bringing the two blobs closer to the origin. This is a deterministic error and one knows how much the cat size has shrunk at any given time since the coherent state amplitude obeys $\alpha(t)=\exp(-\frac{\kappa}{2}t)\alpha(0)$. 

The task of cat state amplification is to add energy to the system such that the cat state is boosted back to size $\alpha$. It is well-known in quantum optics that in the process of boosting the coherent state amplitude, even perfect quantum-limited amplifiers must either add noise (phase insensitive amplification entangles the signal mode with another idler mode) or cause squeezing (phase-sensitive amplification preserves phase-space volume so squeezes the quadrature not being amplified), thereby introducing errors. This fact can be recalled from the phase-space flow pictures for squeezing in  Fig.~\ref{fig:SqueezingPlot} and the desired phase-space flow picture in Fig.~\ref{fig:TanhFlowField-2Leg}. We could use quantum signal processing to create a polynomial approximation to the Hamiltonian of Eq.~\eqref{eq:Tanh2-LegHam} to approximate the phase-space flow in Fig.~\ref{fig:TanhFlowField-2Leg}.

\begin{figure}[htb]
    \centering
    \includegraphics[width=0.45\textwidth]{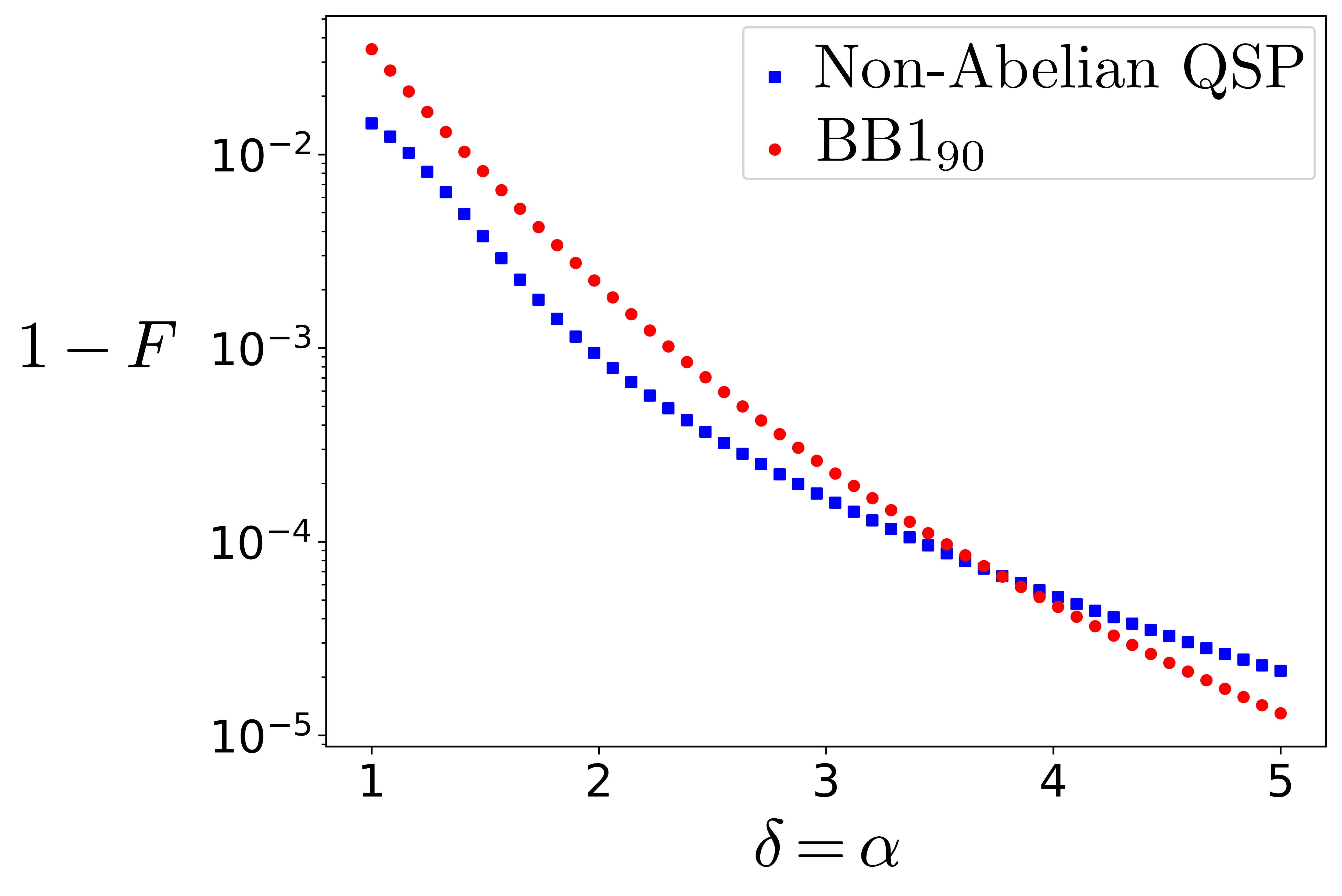}
    \caption{A direct comparison between the single-variable (red) and non-abelian QSP (blue) cat state amplification protocols. The plot shows the infidelity of the hybrid output state with the final cat state of size $\alpha+\delta=2\alpha$ using different cat state amplification protocols. We see that the two curves cross at approximately $\alpha+\delta\approx 7.5$, after which $\textrm{BB1}_{90}$ outperforms non-abelian QSP. It is useful to contrast the results of this plot with Fig.~\ref{fig:QSP_cat2} which compares the two protocols for cat state preparation from vacuum.}
    \label{fig:cat-state-ampl}
\end{figure}

Here, we describe a new method to achieve cat amplification using prior knowledge of the specific cat state to be amplified and the techniques we already developed to create the cat state. The cat state preparation scheme (Sec.~\ref{ssec:compilation-bosonic-qsp-qsvt}) uses non-abelian QSP to disentangle the qubit from the oscillator state. First, use the adjoint of that sequence to entangle the qubit with the individual blobs of a two-legged cat state of size $\alpha$:
\begin{align}
    \mathcal{E}_x\left[-\frac{\pi}{4\alpha},1\right]\ket{C_{0,\alpha,\pm}}\ket{+}=\ket{\alpha}_\textrm{osc}\ket{0}\pm\ket{-\alpha}_\textrm{osc}\ket{1}.
\end{align}
Then apply a conditional displacement to move the blobs farther apart by $\delta$. For real $\alpha$ this step is equivalent to
\begin{align}
    &e^{-i\delta\sqrt{2}\hat p\sigma_z}\ket{\alpha}_\textrm{osc}\ket{0}\pm\ket{-\alpha}_\textrm{osc}\ket{1}\nonumber\\
    &~~~~~~~~~~~~~~~~~~=\ket{\alpha+\delta}_\textrm{osc}\ket{0}\pm\ket{-\alpha-\delta}_\textrm{osc}\ket{1}.
\end{align}
After this operation, we can disentangle the oscillator from the ancilla, with a circuit whose parameters depend on the new size of the cat state $\alpha^\prime=\alpha+\delta$,
\begin{align}
    \mathcal{U}_x\left[\frac{\pi}{4\alpha^\prime},1\right]\ket{\alpha^\prime}_\textrm{osc}\ket{0}\pm\ket{-\alpha^\prime}_\textrm{osc}\ket{1}=\ket{C_{0,\alpha^\prime,\pm}}\ket{+}.
\end{align}
Similar to cat state preparation, the amplification scheme can alternatively be carried out using the BB1 scheme of single-variable QSP in place of the non-abelian QSP sequences $\mathcal{E},\mathcal{U}$. In Fig.~\ref{fig:cat-state-ampl} we show a comparison between the two schemes for the amplification of cat state with size $\alpha$ to size  $\alpha+\delta$ (here, $\delta=\alpha$). 

Amplification of four-legged cat states requires a scheme that moves all four blobs away from the origin. Such an operation could be done using two ancilla qubits (or a qudit with d=4) where each of the four ancilla states are matched with one blob of the four-legged cat state in the overall entangled state. The desired flow field for the 4-legged cat amplifier can be found in Fig.~\ref{fig:TanhFlowField-4Leg} in Sec.~\ref{subsec:flowsinphasespace}.  Again quantum signal processing could be invoked to achieve a polynomial approximation to the quantum version of the classical Hamiltonian of Eq.~\eqref{eq:4legflowfield} to achieve the desired flow field.

Here, we review steps to perform this amplification for the four-legged cat state using a single auxiliary qubit. For simplicity, we consider the case of a four-legged cat state rotated by an angle of $\frac{\pi}{4}$ in phase space. This allows us to see the X-basis codewords of four-legged cat code as a superposition of two two-legged cat states parallel to the position axis or momentum axis of the quantum phase space:
\begin{align}
\ket{+}_\textrm{4C}&=\ket{C_{\alpha,i\alpha,+}}+\ket{C_{-\alpha,i\alpha,+}}\\
    \ket{-}_\textrm{4C}&=\ket{C_{\alpha,i\alpha,+}}-\ket{C_{-\alpha,i\alpha,+}},
\end{align}
where $\ket{C_{\alpha,\beta,\pm}}$ is defined in Eq.~(\ref{eq:disp_cat}). In order to amplify $\ket{-}_\textrm{4C}$ we first prepare the entangled state
\begin{align}
\ket{C_{\alpha,i\alpha,+}}\ket{+}\ +\ \ket{C_{-\alpha,i\alpha,+}}\ket{-}
\end{align}
 using the single-variable QSP scheme described in Sec.~\ref{sssec:qec-prep} adapted to entangle two displaced cat states (see Ref.~\cite{singh_towards_2025}). We then amplify the state along the position quadrature to $\alpha^\prime=\alpha+\delta$ using $e^{-i2\delta\hat p\otimes\sigma_x}$:
 \begin{align}
\ket{C_{\alpha^\prime,i\alpha,+}}\ket{+}\ +\ \ket{C_{-\alpha^\prime,i\alpha,+}}\ket{-}
\,.
\end{align} 

 Next, we change the entanglement of the state to entangle cat states parallel to the position axis using non-abelian QSP\footnote{Disentangling the ancilla from  cat states parallel to position axis in phase space using non-abelian QSP entangles the ancilla with cat states parallel to the momentum axis.}:
 \begin{align}
\ket{C_{i\alpha,\alpha^\prime,+}}\ket{+}\ +\ \ket{C_{-i\alpha,\alpha^\prime,+}}\ket{-}
\,.
\end{align} 

 This allows us to amplify the four-legged cat state along the momentum quadrature using $e^{i2\delta\hat x\otimes\sigma_x}$:
 \begin{align}
\ket{C_{i\alpha^\prime,\alpha^\prime,+}}\ket{+}\ +\ \ket{C_{-i\alpha^\prime,\alpha^\prime,+}}\ket{-}
\,.
 \end{align}
Finally, we disentangle the qubit from the oscillator using the method described in Eq.~(\ref{eq:4cat_prep}) for the preparation of the new four-legged cat state sized $\alpha^\prime$:
\begin{align}
\ket{+}_\textrm{4C}=\ket{C_{\alpha^\prime,i\alpha^\prime,+}}+\ket{C_{-\alpha^\prime,i\alpha^\prime,+}}
\,.
\end{align}

\subsubsubsection{Dissipative GKP Stabilization.} 

\begin{figure}
    \centering
    \includegraphics[width=0.5\textwidth]{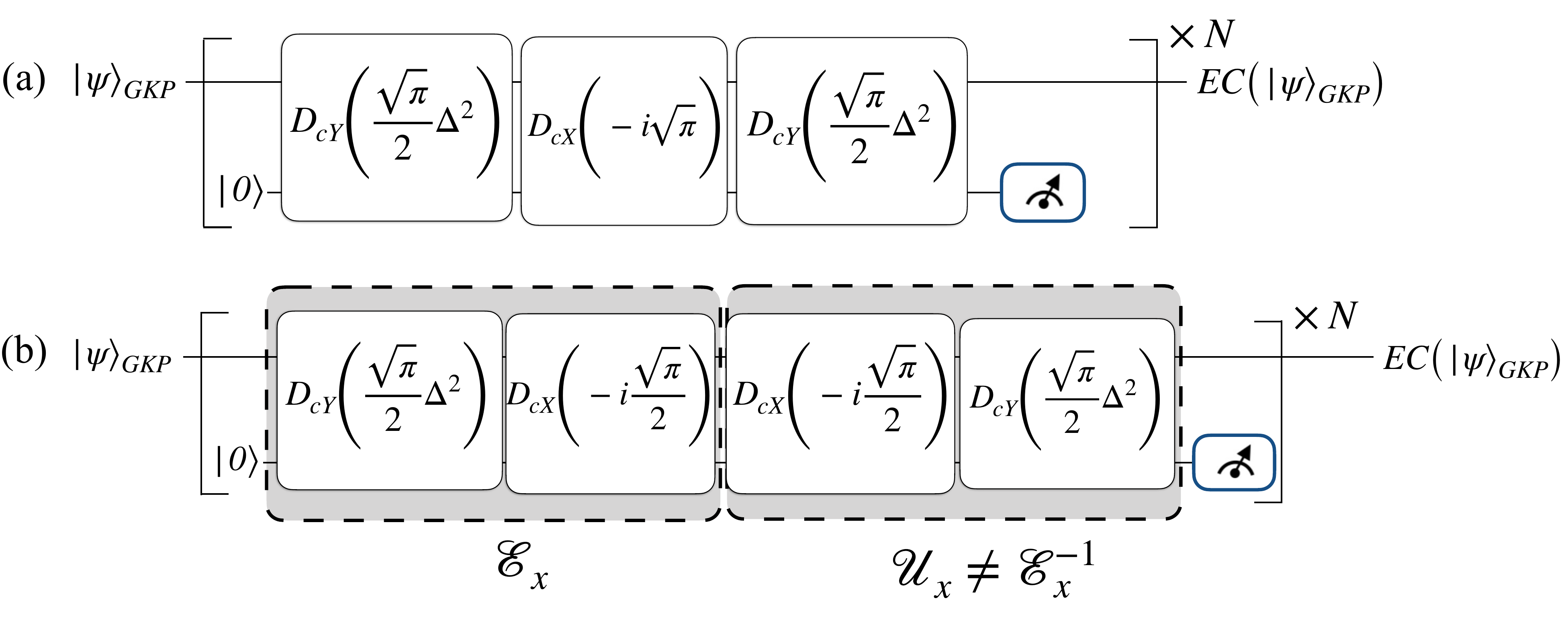}
    \caption{Dissipative GKP Stabilization. In this figure, we have used shorthand notations for symmetric $X-$ and $Y-$ conditional displacements, respectively, as $D_{cX}(\alpha)=e^{(\alpha\hat a^\dagger-\alpha^*\hat a)\otimes\sigma_x}$ and $D_{cY}(\alpha)=e^{(\alpha\hat a^\dagger-\alpha^*\hat a)\otimes\sigma_y}$. The input of the circuits shown here is an arbitrary logical GKP state $\ket{\psi}_{GKP}$ and the output $EC(\ket{\psi}_\textrm{GKP})$ is an error-corrected GKP qubit with the same logical information as the input. Here $EC(*)$ denotes that $*$ has been error-corrected. The circuit variable $\Delta$ is the finite-energy parameter that represents the envelope size of the GKP states. (a) Circuit representation of the small-big-small (sBs) scheme using the phase-space ISA, repeated for $N$ rounds. Here, higher $N$ yields a closer approximation to the continuous variable error correction process. 
    (b) Decomposition of the sBs scheme into an entangling sequence $\mathcal{E}_x$ and a disentangling sequence $\mathcal{U}_x$. Note that $\mathcal{U}_x\neq \mathcal{E}_x^{-1}$ enables the qubit to apply a corrective back action on the oscillator in the presence of errors.  
    This decomposition will be helpful in Sec.~\ref{sssec:qec-single-rot}. Figures (a) and (b) have been adapted from Refs.~\cite{RoyerGKP1_2020} and~\cite{singh_towards_2025}, respectively.}
    \label{fig:sBs}
\end{figure}

We review here recent theoretical and experimental progress in CV-DV control that has led to successful error correction gain for quantum memories based on the GKP code \cite{GKP2001}.
The stabilizers of the GKP states can be viewed as exponentiated modular quadratures -- for example, $e^{i\sqrt{\pi}(\hat x_m+i\Delta^2 \hat p_m)}$, where $\hat x_m=\hat x \quad\mathrm{mod}\sqrt{\pi}$, $\hat p_m=\hat p\quad\mathrm{mod}\sqrt{\pi}$. Here $\Delta$ is the Gaussian width of the target state as described in Eq.~\eqref{eq:GKP}. In Refs.~\cite{RoyerGKP1_2020,HomeGKPQEC2022} the authors show that this stabilizer can be extracted by engineering the dissipator $\hat d=\hat x_m+i\Delta^2 \hat p_m$. This is possible by using unitary evolution under 
$e^{i\sqrt{\pi}(\hat x\sigma_x+\Delta^2 \hat p\sigma_y)}$  
combined with qubit resets. Using the phase-space ISA, this dissipative quantum channel can be realized by a circuit known as the \emph{small-big-small} (sBs) scheme~\cite{RoyerGKP1_2020}, shown in Fig.~\ref{fig:sBs}a). This circuit uses three conditional displacements such that the first and last conditional displacements  have a small amplitude proportional to $\Delta^2$ while the middle gate has a relatively larger amplitude of $\sqrt{\pi}$, and thus the only gate that contributes to the circuit duration significantly is the middle gate with amplitude $\sqrt{\pi}\gg \Delta^2$. This circuit has been used in recent beyond-break-even QEC memory experiments for GKP qubits \cite{Sivak_GKP_2022} and qudits \cite{GKP_qudit_2024}. 

In Ref.~\cite{HomeGKPQEC2022} the authors demonstrated the construction of this channel with a trapped ions system using the sideband ISA. This dissipative channel method is an autonomous stabilization scheme where no additional correction needs to be applied based on an ancilla measurement outcome. The ancilla is reset at the end of the circuit. With high probability, each reset takes the oscillator closer to the GKP code space, removing (up to) one bit of entropy from the system. Fig.~\ref{fig:sBs} shows that this circuit representation is equivalent to a combination of an entangling sequence and disentangling sequence.

While this circuit corrects errors, it also  applies a deterministic logical Pauli operation on the GKP states\footnote{This is admissible since it can be tracked in software with a deterministic Pauli frame change after each stabilization step.}. The circuit applies an ancilla rotation by $2m\pi$ for each peak ($m$) in the GKP logical qubit. Hence the ancilla qubit learns no information about the GKP state of the system. However, if the oscillator is not in a GKP logical eigenstate the non-abelian QSP correction for the disentangling operation changes. Thus, the oscillator is not disentangled from ancilla. The qubit reset in this case applies a corrective back-action on the oscillator in addition to the deterministic Pauli logical operation.

\subsubsection{Logical Single-Qubit Rotations}

\label{sssec:qec-single-rot} 

\begin{figure}[htb]
    \centering
    \includegraphics[width=0.45\textwidth]{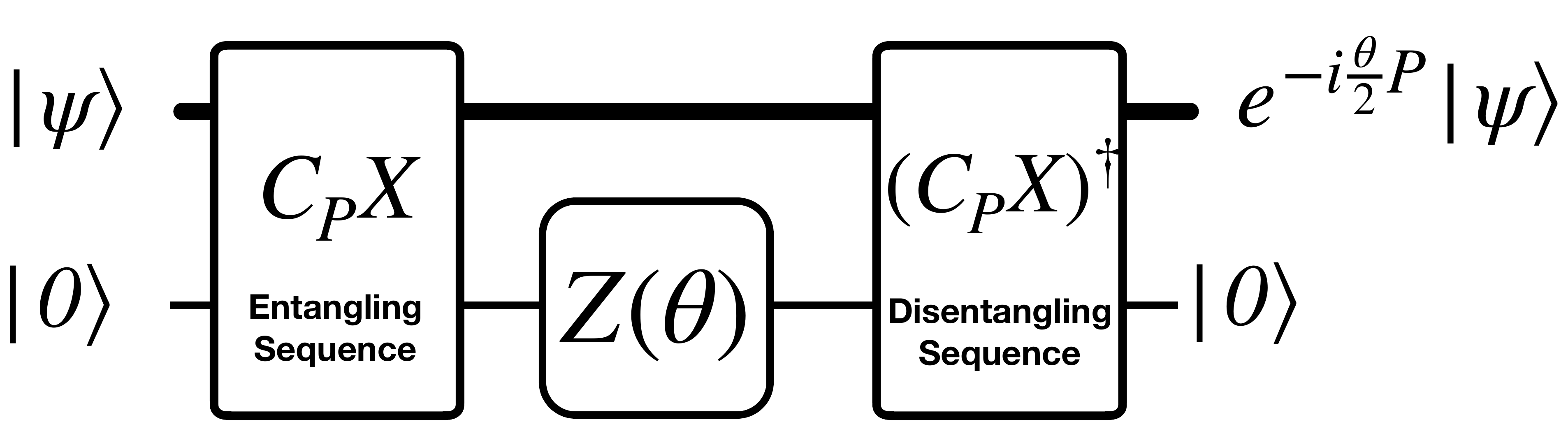}
    \caption{Phase Transfer circuit. Here $P\in\{X,Y,Z\}$. The oscillator-controlled Pauli gates $C_PX$ entangle the two wires in the state $\ket{P_+}_\textrm{osc}\ket{0}+\ket{P_-}_\textrm{osc}\ket{1}$, where thin line represent the ancilla and $\ket{P_+}_\textrm{osc}, \ket{P_-}_\textrm{osc}$ represent the $\pm 1$ eigenstates of $\hat P$ for the top bosonic qubit. $(C_PX)^\dagger$ is the adjoint of the operator $C_PX$. 
    } 
    \label{fig:exponentiation-gadget}
\end{figure}

This section reviews the construction of single-logical-qubit rotations on the logical Bloch sphere using gates from the various instruction sets in consideration. A majority of the examples discussed below teleport the fast two-level ancilla operations to bosonic logical qubits. Here, we use a special case of teleportation circuits, namely a phase transfer circuit (see Fig.~\ref{fig:exponentiation-gadget}) which can be advantageous in mitigating effects of ancilla errors~\cite{zhou2000logicalgate,tsunoda2023error,singh_towards_2025}. As noted above in Sec.~\ref{sssec:qec-readout}, the entangling circuits involved in the teleportation are also useful for mapping the logical state onto an ancilla qubit for the purpose of logical readout.

Let  $P\in\{X,Y,Z\}$ denote logical Pauli operations of the bosonic code under consideration, and $\ket{P_\pm}$ be the $\pm 1$ eigenstates of the logical $P$ operation.  The phase transfer circuit can be constructed straightforwardly with access to hybrid entangling operations that flip the ancilla conditioned on the state of the bosonic qubit. 
\begin{align}
    C_P X&=\ket{P_+}_\textrm{osc}\bra{P_+}_\textrm{osc}\otimes \hat I+\ket{P_-}_\textrm{osc}\bra{P_-}_\textrm{osc}\otimes X
    \label{eq:ancillacontrolledbosonicPauli}
\end{align}
as shown in Fig.~\ref{fig:exponentiation-gadget}. In this case, $\ket \psi=\alpha_1\ket 0_L+\alpha_2\ket 1_L$ in the top wire is an oscillator state encoded in a bosonic code, while the bottom wire is a two-level ancilla. The first $C_P X$ gate entangles the logical qubit in the oscillator and the ancilla qubit,
\begin{align}
    \alpha_1\ket{+P}_\textrm{osc}\ket{0}+\alpha_2\ket{-P}_\textrm{osc}\ket{1}
\end{align}
The rotation gate $Z(\theta)$ now adds a local phase to this entangled state.
\begin{align}
    \alpha_1\ket{+P}_\textrm{osc}\ket{0}+e^{i\theta}\alpha_2\ket{-P}_\textrm{osc}\ket{1}
\end{align}
Finally, the second $(C_PX)^\dagger$ gate disentangles the oscillator from the auxiliary qubit.
\begin{align}
    (\alpha_1\ket{+P}_\textrm{osc}+e^{i\theta}\alpha_2\ket{-P}_\textrm{osc})\ket{0}
\end{align}
Thus, the circuit performs a logical $\exp(-i\frac{\theta}{2} P) $ operation, realized from the oscillator-controlled operations $C_PX$ and a single-qubit ancilla rotation. One of the key advantages of this circuit is the fact that at the end of the circuit the ancilla qubit is in a deterministic state (here, $\ket{0}$), and thus certain ancilla errors are detectable using ancilla measurement. 

Below, we show examples of the construction of the entangling gates $C_PX$ and disentangling gates $(C_PX)^\dagger$ for various bosonic codes using gates available in different selected instruction sets. 

\subsubsubsection{GKP Code.} The logical Pauli operations on realistic GKP code words are equivalent to translations in phase space up to errors due to the finite energy cutoff ($\Delta\neq 0$). Assuming that finite-energy errors do not induce logical errors in the GKP codespace, phase-space translations can yield logical Pauli operations with classical controls on the oscillator. However, it is impossible to perform arbitrary Pauli rotation using classical controls since this requires creating a linear combination of identity and a Pauli translation operation. Hence we must teleport the gate into the system using an ancillary qubit and the phase transfer circuit. 

We have discussed the entangling operation between a two-level ancilla and GKP code word in the logical readout section. The sequences used for GKP logical $P$ readout in Eqs.~(\ref{eq:GKP_readoutX}-\ref{eq:GKP_readoutZ}) yield $C_XX,C_YX,C_ZX$. The disentangling operation can be chosen to be the inverse operations $(C_XX)^\dagger, (C_YX)^\dagger, (C_ZX)^\dagger$. An alternative to $(C_ZX)^\dagger$ for disentangling the oscillator and the ancilla qubit is to use the second half of the stabilization scheme, $\mathcal{U}_x$ (see Fig.~\ref{fig:sBs}b). The resulting phase transfer circuit not only teleports gates but also corrects errors in the GKP qubit at the same time.  This feature is a consequence of the fact that, in the absence of $Z(\theta)$ rotation, Fig~\ref{fig:exponentiation-gadget} is equal to the GKP stabilization circuit $\textrm{sBs}$\footnote{The sBs circuit rotates the ancilla state entangled with $\ket{P_+}_L$ ($\ket{P_-}_L$) by $4\pi$  ($2\pi$) Thus, the qubit is disentangled with an additional local phase of $\pi$ on the logical state. Hence, this operation applies a deterministic logical Pauli $P$ operation along with the logical single-qubit rotation. See~\cite{singh_towards_2025} for details.}.  The stabilization along each quadrature can be used to derive the phase transfer circuit to teleport universal single qubit rotations in the GKP Bloch sphere. Importantly, this gate is valid for not only GKP qubits but also can be, mutatis mutandis, used for GKP qudits encoding a $d$-level logical system.
The authors also illustrate a piece-wise circuit construction to reduce the effect of ancilla dephasing. With this, the protocol can yield high fidelity gate operation with biased noise ancilla same as claimed for the stabilization scheme in~\cite{RoyerGKP1_2020}.

\subsubsubsection{Four-Legged Cat Code.} Here, the entangling gate for logical $Z(\theta)$ rotations is given by the controlled rotation gate, 
\begin{align}
C_ZX= e^{i\frac{\pi}{4}a^\dagger a\otimes\sigma_x}.
\end{align}
 This enables single qubit logical $Z(\theta)$ rotations. In addition, one can use SNAP gates and displacements (see Sec.~\ref{sssec:SNAP}) from the Fock-space ISA to perform entangling operations which can correct for errors in the ancilla up to first-order. 
In~\cite{PathIndependentGatesPhysRevLett.125.110503,xu2023faulttolerant} the authors use this method to perform $Z(\theta)$ and $X(2\phi)$ rotations of the four-legged cat code. Under correctly chosen conditions, the logical rotations can be made fault tolerant by utilizing a three-level (qutrit) ancilla. This technique was first described in~\cite{ReinholdErrorCorrectedGates} for the control of binomial codes with a qubit ancilla. 

\subsubsubsection{Binomial Code.} The entangling operation between binomial codes and an ancilla qubit can be carried out using 
 \begin{align}
    C_ZX=e^{i\frac{\pi}{4}a^\dagger a\otimes\sigma_x}.   
 \end{align}
 This operation rotates the ancilla by an angle of $2\pi$ ($\pi$) to $\ket{0}$ ($\ket{1}$) if the oscillator state is $\ket{0}_\textrm{bin}$ ($\ket{1}_\textrm{bin}$). It should be noted that while the operation $C_ZX$ takes the logical qubit outside of the codespace ($\ket{0}+\ket{4}\rightarrow \ket{0}-\ket{4}$) but the disentangling operation (also, $C_ZX$) brings it back to the codespace. This circuit in Fig.~\ref{fig:exponentiation-gadget} can be easily modified to use the natively available controlled rotation gate in Box~\ref{Box:CRotation}. The circuit is an exact gate teleportation sequence and is on par with the numerically optimized GRAPE pulses used in~\cite{ReinholdErrorCorrectedGates,LuyanSun2020}. If either an ancilla error or a photon loss error in the binomial qubit occurs then at the end of this circuit the ancilla would be in state $\ket{1}$, heralding the error.  A better protocol from the perspective of fault-tolerance uses SNAP gates which is tolerant to single ancilla errors as described in~\cite{ReinholdErrorCorrectedGates,LuyanSun2020}. This work also uses a qutrit ancilla to achieve fault-tolerance by enabling ancilla error correction as in the four-legged cat example discussed above. Rotations around other axes on the logical Bloch sphere of the binomial code (required for universal control) are more complex and to date have required numerically optimized sequences of gates from one of the ISAs.  Finding efficient analytic gate sequences based on QSP remains an open problem.

\subsubsubsection{Dual-Rail Code.} The Dual-Rail qubit encodes quantum information equivalent to a single qubit in two physical systems. Consider the case where these two systems are oscillators in Fock states $\ket{n=0}$ and $\ket{n=1}$. The preparation of these states can be achieved simply using the sideband ISA. Note that, the beam-splitter gate acting on the dual-rail qubit provides universal control, and thus, we already have natively available logical Pauli rotations for this encoding. However, for the sake of completeness, we note that one could also teleport these single-qubit Pauli rotations. A hybrid controlled rotation gate from the Fock-space ISA is defined as (or equivalently a SNAP gate), 
\begin{align}
C_ZX=e^{i\frac{\pi}{2}a^\dagger a\otimes\sigma_x},
\end{align}
acting on one of the two oscillators. This gate can be used to entangle and disentangle the two-level ancilla from the dual-rail qubit, and teleport single-qubit $Z$ rotation using the circuit in Fig.~\ref{fig:exponentiation-gadget}. Again, the circuit in the figure can be easily modified to use the natively available gate in Box~\ref{Box:CRotation}. The same entangling gate can also be used for readout purposes.

\subsubsection{Entangling Bosonic Logical Qubits}

\label{sssec:ent-gate}

In this section we focus on entangling operations between two logical qubits (as opposed to the previously discussed entanglement between a logical qubit and an ancilla). These operations are necessary complements to single-qubit logical operations, preparation and measurement to achieve universal logical control. Primarily we focus on recent proposals~\cite{rojkov2023two,xu2023faulttolerant} (in addition to Refs.~\cite{tsunoda2023error,singh_towards_2025} cited in the introduction to this section) for two-qubit logical operations which use gates from the instruction sets discussed in this work. These operations require the multi-mode extensions of the ISA, which includes beam-splitters between oscillators or $\textrm{CPHASE}$ gates between ancilla qubits. We have already introduced several ISAs with beam-splitters in Table.~\ref{tab:ISA_overview}. The ISA where beam-splitters are replaced by $\textrm{CPHASE}$ is better suited for Rydberg atoms (see \ref{tab:ISA_neutral}).

\begin{figure}[htb]
    \centering
    \includegraphics[width=0.45\textwidth]{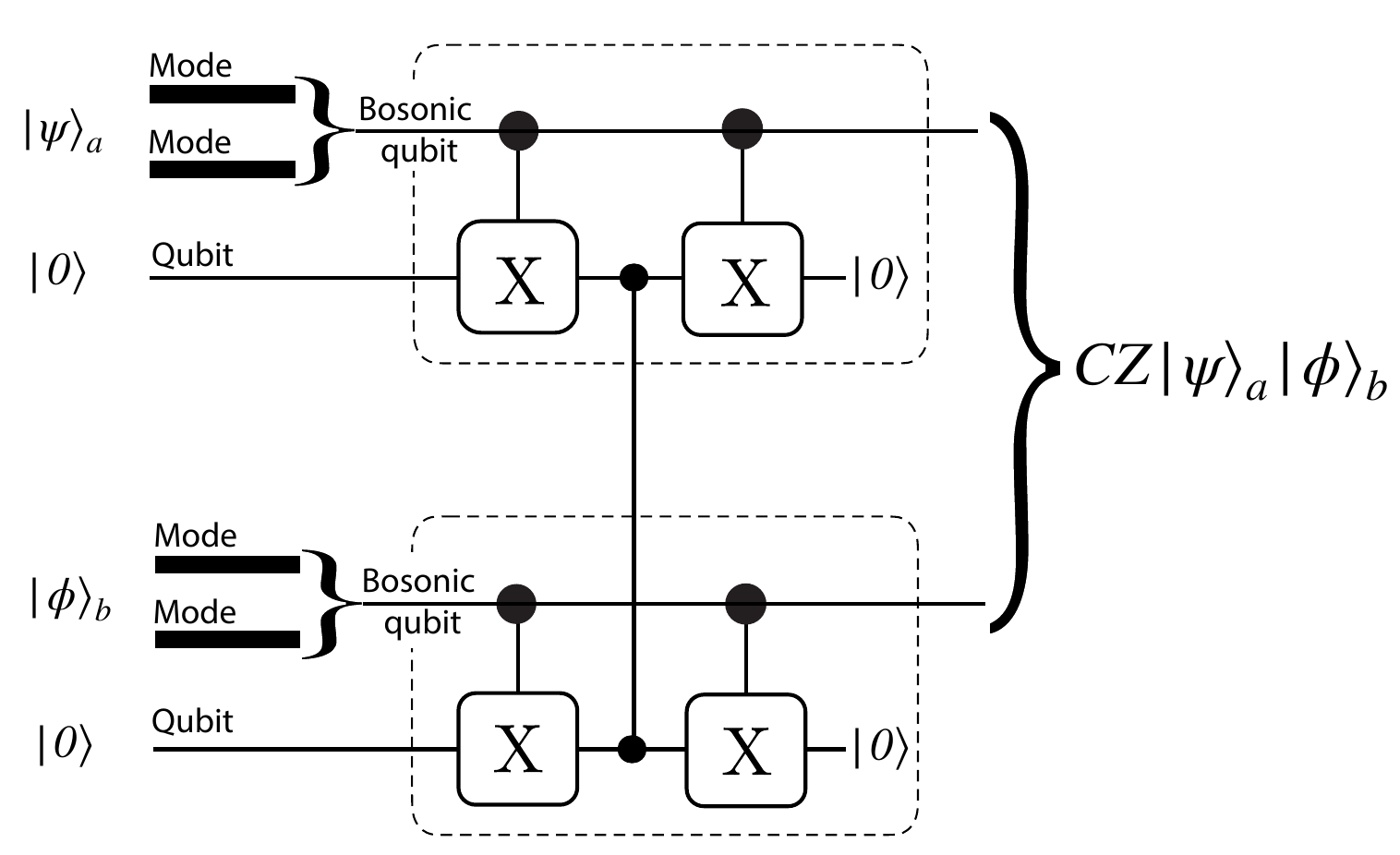}
    \caption{Entangling operation between two logical bosonic qubits using the oscillator-controlled $C_Z X$ gate on the ancilla and the $\textrm{CPHASE}$ gate to entangle the two ancillae. The top lines in both boxed figures represent bosonic logical qubits while the bottom lines represent the two-level ancillary qubits. The two thick lines indicate that more than one oscillator can be used to encode a single bosonic logical qubit, for example, a dual-rail qubit.
}
    \label{fig:entangling-logicals}
\end{figure}

First, we begin with the case where the ISA only includes $\textrm{CPHASE}$ gates. A simple circuit for bosonic logical qubit entangling task can be constructed using the entangling-disentangling gates $C_PX$ between two-level ancilla and bosonic code words discussed with respect to arbitrary single-qubit-logical rotations in Sec.~\ref{sssec:qec-single-rot}. One can entangle both logical code words $\ket{\psi}_a,\ket{\phi}_b$ that belong to logical qubits encoded in modes $a,b$ respectively, with one ancilla each,
\begin{align}
(\alpha_1\ket{0}_L\ket{0}+\beta_1\ket{1}_L\ket{1})\otimes(\alpha_2\ket{0}_L\ket{0}+\beta_2\ket{1}_L\ket{1}).
\end{align}
Then we entangle the two ancillary systems with a $\textrm{CPHASE}$ gate, 
\begin{align}
    &\alpha_1\ket{0}_L\ket{0}\otimes(\alpha_2\ket{0}_L\ket{0}+\beta_2\ket{1}_L\ket{1})\nonumber\\
    &+\beta_1\ket{1}_L\ket{1}\otimes(\alpha_2\ket{0}_L\ket{0}-\beta_2\ket{1}_L\ket{1}).
\end{align}
Finally, we disentangle the ancillary systems using the disentangling gates $C_ZX$.
\begin{align}
&\alpha_1\ket{0}_L\ket{0}\otimes(\alpha_2\ket{0}_L+\beta_2\ket{1}_L)\ket{0}\nonumber\\
    &+\beta_1\ket{1}_L\ket{0}\otimes(\alpha_2\ket{0}_L-\beta_2\ket{1}_L)\ket{0}\\
    =&\alpha_1\ket{0}_L\ket{0}\otimes\ket{\phi}_b\ket{0}+\beta_1\ket{1}_L\ket{0}\otimes(Z_L\ket{\phi}_b)\ket{0}\\
&\equiv \mathrm{CZ}_L^{a,b}\otimes I\otimes I(\ket{\psi}_a\ket{\phi}_b)\otimes\ket{0}\ket{0}
\end{align}
where $Z_L$ is the logical $Z$ operation on the bosonic codeword $\ket{\phi}_b$. This scheme is described in Fig.~\ref{fig:entangling-logicals}. 

This circuit can also detect ancilla errors since post measurement both auxiliary qubits are in a deterministic state $\ket{0}$. Thus, the gate could be made fault-tolerant if we assume an upper layer QEC code above bosonic modes~\cite{GrimsmoPuri_GKP_2021,teoh2022dualrail,kubica2022erasure}. While the above can be generalized easily using the discussion of hybrid entangling operations in the previous section (\ref{sssec:qec-single-rot}), it requires two extra auxiliary qubits to perform each entangling operation. 

Next, we discuss examples for the case where the ISA includes beam-splitters instead of $\textrm{CPHASE}$ gates. For this architecture, logical two-qubit entangling gates for bosonic code-words can be realized using a single ancilla qubit and high-fidelity beam-splitters~\cite{chapman2022high}. These examples discuss operations of type $C_PC_PX$, generalizations of $C_PX$ (see Eq.~(\ref{eq:ancillacontrolledbosonicPauli})), which directly entangle the two codewords $\ket{\psi}_a$ and $\ket{\phi}_b$ with a single auxiliary qubit in the initial state $\ket{0}$. The $C_PC_PX$ operation can then be used directly as shown in Fig.~\ref{fig:exponentiation-gadget}, replacing the top line with two thick lines representing the two codewords. With the available architecture, this method can not only yield logical entangling gates like $CZ_L$ or $CX_L$ but also two-qubit logical rotations like $ZZ(\theta)$ and $XX(\theta)$ gates. All operations discussed below can be realized using conditional beam-splitters, compilation of which can be realized natively, or using multi-mode Fock space or phase-space ISA (see Eq.~(\ref{eq:cbs})).

\subsubsubsection{GKP Code.} The logical gates for GKP qubits are obtained via the non-unitary gate $\hat{E}_\Delta A\hat{E}_\Delta^{-1}$ where $A$ is the gate for unrealistic infinite-energy GKP while $\hat{E}_\Delta$ is the envelope operator defined in Eq.~(\ref{eq:GKP}). The entangling gate $CZ_\textrm{L}$ such that $A=e^{i2\hat x\otimes \hat x}=e^{i2\hat x_1\hat x_2}$ takes the following form
\begin{align}
 \hat{E}_\Delta e^{i2\hat{x}\otimes\hat{x}}\hat{E}_\Delta^{-1} &
 \nonumber \\
& \hspace*{-6ex} =e^{i(2\cosh{\Delta^2}\hat{x}+i2\sinh{\Delta^2}\hat{p})\otimes(2\cosh{\Delta^2}\hat{x}+i2\sinh{\Delta^2}\hat{p})} \\    
    &   \hspace*{-6ex}
    \approx e^{i(2\hat{x}_1\hat{x}_2-2\Delta^4\hat{p}_1\hat{p}_2+i2\Delta^2(\hat{x}_1\hat{p}_2+2\hat{p}_1\hat{x}_2))}\quad \\
    & \text{(Finite-energy $CZ_{L}$ gate)}\nonumber
    \end{align}
These non-unitary gates can be approximated using an auxiliary qubit, where the approximations hold in the small $\Delta$ limit such that $\cosh{\Delta^2}\approx 1$ and $\sinh{\Delta^2}\approx \Delta^2$.  
\begin{align}
    CZ_{L}&\approx e^{i\Delta^2(\hat{x}_1\hat{p}_2+\hat{p}_1\hat{x}_2)\sigma_y}e^{i(2\hat{x}_1\hat{x}_2-2\Delta^4\hat{p}_1\hat{p}_2)\sigma_x}\times\nonumber\\&\quad e^{i\Delta^2(\hat{x}_1\hat{p}_2+\hat{p}_1\hat{x}_2)\sigma_y}\ket{\psi}_{\textrm{GKP}}\ket{0}
\end{align}
Ref.~\cite{rojkov2023two} proposes a circuit to perform these non-unitary gates using an auxiliary qubit through the dissipation model described in Ref.~\cite{RoyerGKP1_2020}. The circuit involves conditional beam-splitters. If the middle gate $e^{i(\hat x_1\hat x_2-\hat p_1\hat p_2)\otimes\sigma_x}$ is split into two halves just like the GKP stabilization circuit in Fig.~\ref{fig:sBs} then the first half of the circuit can be the entangling operation $C_PC_PX$  while the second half can be the disentangling operation. Thus, we can introduce a $Z(\theta)$ rotation on the qubit to achieve a $ZZ(\theta)_L$ rotation on the two qubits. The authors also show that this circuit can be achieved using non-abelian QSP, highlighting generalization of this QSP framework to two modes. In App.~\ref{sssec:ECBS}, we provide a compilation of the fast $e^{i(\hat x_1\hat x_2-\hat p_1\hat p_2)\otimes\sigma_x}$ gates in weak-dispersive regime, an artifact of phase-space ISA in superconducting circuits~\cite{EickbuschECD}.

\subsubsubsection{Four-Legged Cat Code.} The entangling operation $C_ZC_Z X$ for four-legged cats is given by
\begin{align}
e^{i\frac{\pi}{4}(a^\dagger a+b^\dagger b)\sigma_x}. 
\end{align}
Again, the circuit can be easily modified to use beam splitters and SQR gate from the Fock-space ISA discussed in Box~\ref{box:SQRgate}. A realization of this operation where a qutrit ancilla is used to make the circuit error-detectable against ancilla errors is possible. Alternatively, in~\cite{xu2023faulttolerant} the authors use SNAP gates and beam-splitter from the multi-mode Fock-space ISA to achieve an $XX(\theta)$ rotation for four-legged cat codes.

\subsubsubsection{Binomial Code.} Again, the trick from~\cite{tsunoda2023error} can again be used to construct $C_ZC_Z X=e^{i\frac{\pi}{4}(a^\dagger a+b^\dagger b)\sigma_x}$. It also shows methods to implement two-qubit $ZZ(\theta)$ gates using the same instruction set for all three bosonic codes. We can create arbitrary excitation-preserving two-qubit gates~\cite{Gao2019, tsunoda2023error} which can be achieved using the Fock-space ISA.
 
\subsubsubsection{Dual-Rail Code.} Given two dual-rail qubits with modes $\{a_1,a_2\}$ and $\{b_1,b_2\}$, the entangling operation $C_ZC_ZX$ is given by, 
\begin{align}
e^{i\frac{\pi}{2}({a_1}^\dagger {b_1}+{a_2}^\dagger {b_2})\sigma_x}.
\end{align}
which entangles the two dual-rail qubits (comprising of four modes in total) with an ancilla qubit. Note the change in the parameter of this gate from $\frac{\pi}{4}$ to $\frac{\pi}{2}$. This operation can be compiled with the multi-mode Fock-space ISA using a qutrit ancilla. The authors also give parameters to construct a controlled-SWAP operation using the same resources for dual-rail qubits, four-legged cat codes as well as binomial codes.

\section{Compilation Methods}
\label{sec:compilation}

Having established physical and logical ISAs introduced in Secs.~\ref{sec:instruction_set} and \ref{sec:qec-compilation}, a vital next question is how to compile programs for the three AMMs established in Sec.~\ref{ssec:amms-detail}. By definition, \emph{compilation} means \emph{translation} from one programming language to another. A related concept it \emph{transpilation}, which refers to the compilation of a high-level (compiler-facing) ISA into a low-level (transpiler-facing) ISA, the latter being aware of hardware-specific details such as qubit connectivity and the dominant error rates.  For simplicity, we broadly use the word compilation throughout this section for both of these tasks. 

The primary challenge is that we cannot directly rely on intuition from boolean logic to design unitaries acting on qumodes. A second challenge is the significant architectural differences across various hybrid CV-DV platforms in terms of hardware layout (Fig. \ref{fig:hybrid-processor-arch}) and native interactions -- see Table \ref{tab:free_interaction} for details. As a result, many compilation and synthesis strategies developed for trapped ions, for example, cannot be directly adapted to superconducting or neutral-atom platforms.

To jump start efforts toward tackling these challenges, we begin in Sec.~\ref{ssec:intro-overview-compilation-scheme} with a broad introduction to compilation and lay out a high-level compilation strategy for hybrid CV-DV processors. We then follow in Sec.~\ref{sec:algorithm-polynomial-description} with two complementary descriptions of an arbitrary single-qubit-oscillator unitary in terms of polynomials of different operators in the hybrid system. In Sec.~\ref{ssec:exact-analytical-qubit-gates}, we address the exact, analytic synthesis of multi-qubit and multi-oscillator entangling gates and present several novel compilation techniques tailored to the superconducting CV-DV architecture illustrated in Fig.~\ref{fig:hardware_layout}a. 

Sec.~\ref{ssec:approximate-1-qubit-oscillator-unitary} is devoted to the use of ideas from quantum simulation and quantum signal processing in DV systems for hybrid CV-DV compilation purposes, particularly toward the systematic control and synthesis of single qubit-oscillator unitaries. We note that the extension of product formulas (Sec.~\ref{sec:trotter-product}) and linear combination of unitaries techniques (Sec.~\ref{sec:lcu}) to bosonic platforms has previously discussed in Ref.~\cite{kang2023leveraging} and \cite{Chakraborty2024implementingany} respectively, and our treatment on these topics will therefore be brief and pedagogical. In contrast, we will provide a more detailed discussion on the generalization of quantum signal processing to hybrid CV-DV platforms (Sec.~\ref{ssec:compilation-bosonic-qsp-qsvt}), as this concerns new material. Finally, we discuss numerical compilation strategies in Sec.~\ref{sec:numerical-optimization} as a complement to the analytical compilation methods. For the interested reader, we provide further in-depth comparisons between numerical and analytical compilation techniques in App.~\ref{sec:circuit-compare}. 

For each compilation technique described in this section, we provide simple illustrative examples and whenever possible, present them in a ``human-readable'' fashion to provide physical insights. We hope that the techniques presented in this section (and the examples provided) spur further active research and development on hybrid CV-DV compilation.

\subsection{Introduction and Overview of Compilation Schemes}
\label{ssec:intro-overview-compilation-scheme}

Given the fact that the development of quantum programming languages is still in its early stage \cite{yuan2022twist}, especially for bosonic programming languages, a systematic discussion of bosonic compilation schemes in a rigorous fashion at the level of compilation schemes on classical computers \cite{aho2007compilers} is not yet available. At the core of compiling hybrid oscillator-qubit algorithms lie efficient methods to decompose a multiple-qubit-oscillator unitary operator into one of the hybrid ISAs defined in Sec.~\ref{sec:instruction_set}. While analytical constructions of arbitrary multi-qubit-oscillator unitaries are challenging \cite{becker2021energy,gschwendtner2021infinite}, it is the goal of this section to lay major frameworks to describe qubit-oscillator unitaries as well as ways to perform the compilation, with ample examples as demonstrations. For complex algorithms, it becomes increasingly difficult and less efficient to keep track of these compilation rules. Not surprisingly, modern compilers have embraced numerical optimization algorithms to find good  (yet usually not globally optimal) circuits, a method which we adopt as well  (\emph{mutatis mutandis}, since quantum instructions contain continuous parameters). 

\begin{figure}[htb]
    \centering
    \includegraphics[width=0.5\textwidth]{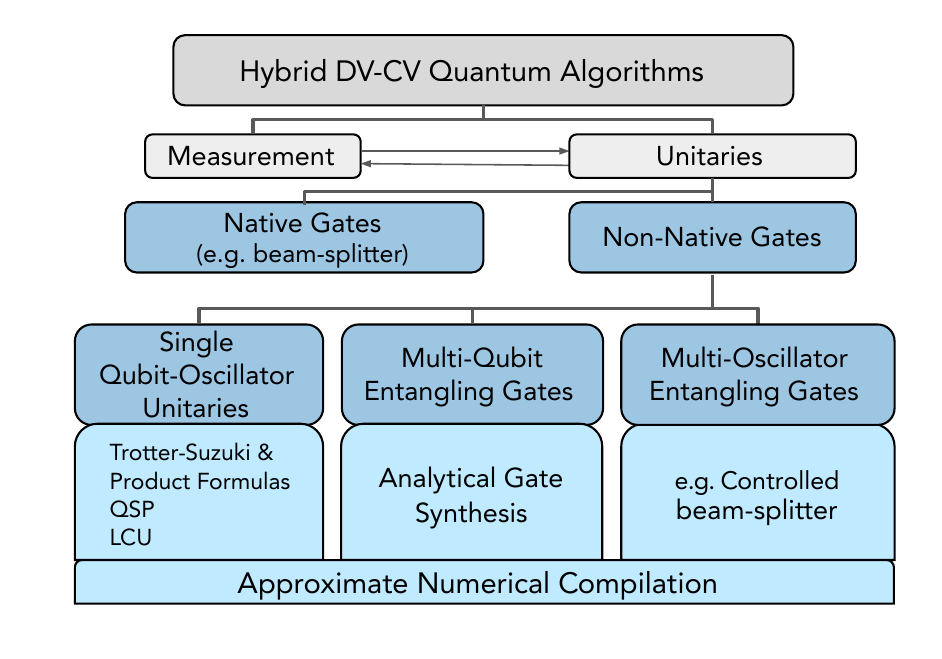}
    \caption{Compilation flowchart for hybrid CV-DV quantum operations on multiple qubits and oscillators. Starting from the top, a generic quantum operation consists of unitaries interleaved with measurements where feed-forward (unitaries conditioned on measurement results) is possible (denoted by the horizontal arrows in between). The multi-qubit-oscillator unitaries are composed of native (e.g., beam-splitter) and non-native gates (e.g., qubit-conditioned beam-splitter).  The latter can be further decomposed into native single-qubit oscillator unitaries and entangling gates between either multiple qubits or multiple oscillators.
    }
    \label{fig:compilation-scheme}
\end{figure}

The compilation scheme for hybrid CV-DV quantum computers is summarized in Fig.~\ref{fig:compilation-scheme}. A generic algorithm or operation is composed of multiple unitary operations interleaved with measurements and feed-forward to produce new unitaries conditioned on the measurement results. The hybrid architecture has some native multi-qubit-oscillator unitaries such as the beam-splitter that can serve as primitives to use in compilation. Non-native multi-qubit-oscillator unitaries may be decomposed into unitary operations on individual qubit-oscillator pairs combined with multi-qubit or multi-oscillator entangling gates, and we present examples to demonstrate compilation methods using these different categories.

We note that in actual systems, there will additionally be practical tradeoffs to consider due to the heterogeneous nature of lifetime and operation speed between qubits and oscillators. See Ref.~\cite{stein2023microarchitectures} for a recent work along this line.  Refs.~\cite{Sharma_2020,9773233,10.1145/3297858.3304075} discuss related ideas for noise-adaptive compilation schemes for qubit-based NISQ architectures. We also note that while transmon qubits in some superconducting cQED hardware platforms have direct near-neighbor couplings between them, the superconducting qubit/cavity architecture we have outlined in Fig.~\ref{fig:hardware_layout} intentionally does not for the purpose of avoiding cross-talk issues. Importantly, our novel compilation strategy (Sec.~\ref{ssec:exact-analytical-qubit-gates}) for multi-qubit entangling gates with cavity-mediated interactions enables high-level algorithms to treat qubits and oscillators on equal footing, using fast, high-fidelity SWAPs to achieve logical all-to-all connectivity despite near-neighbor physical constraints. The compilation strategies in the following sections are novel for superconducting cQED systems.

\subsection{Representations of Hybrid CV-DV Algorithms}
\label{sec:algorithm-polynomial-description}
To discuss compilation methods in detail, it is useful to develop a more formal description of generic unitaries on hybrid systems. In this section, we expand the earlier discussion on universality in Sec.~\ref{sec:instruction_set} and introduce two complementary pictures for understanding a generic multi-qubit-oscillator unitary, namely its Hermitian generator (i.e., the control Hamiltonians in an instruction set), and the unitary itself -- accessible via Hamiltonians in the algebra generated by the Lie brackets of the control set.

\subsubsection{Hermitian Generator Picture}
\label{sssec:hermitianpicture}
In the Hermitian generator picture, let us consider the Hermitian operators of a system with $n$ qubits and $m$ oscillators that generate its corresponding unitary evolution. We denote the Pauli operator on $n$ qubits as $P_k$ where $k = 0, 1, \cdots, 4^n - 1$. Each oscillator term can likewise be described by a finite-degree Hermitian polynomial $h(\hat{x},\hat{p})$. In particular, we have seen in Eq.~\eqref{eq:osc-qubit-control-H} 
that the most general single-qubit and single-oscillator generator is a sum of four different finite-degree polynomials in $\hat{x}, \hat{p}$. Similarly, a generic Hamiltonian on a multi-qubit and multi-oscillator system can be written as
\begin{align}
    H = \sum_{k=0}^{4^n - 1} P_k h_k(\hat{x}_1, \hat{p}_1, \hat{x}_2, \hat{p}_2, \cdots, \hat{x}_m, \hat{p}_m)
    \label{unitary-rep-hermitian-gen}
\end{align}
where the subscript $k$ runs over all hermitian basis $P_k$ of the $n$-qubit system. In the case of a single qubit $n=1$, $k=0,1,2,3$ and $P_k$ reduces to the identity ($P_0$) and the three Pauli matrices. $h_k(\cdot)$ is a generic polynomial of the $2m$ canonical variables on the $m$ oscillators, which can be of infinite degree (or infinitely smooth such that its arbitrary order derivative concerning $\hat{x}$ and $\hat{p}$ exists). In practice, it is necessary to truncate $h_k(\cdot)$ to a finite degree. More concretely, a degree-$d$ truncated representation can be written as
\begin{align}
     & h_k(\hat{x}_1, \hat{p}_1, \hat{x}_2, \hat{p}_2, \cdots, \hat{x}_m, \hat{p}_m) \nonumber \\
    =& \sum_{\substack{r_1,\cdots, r_m, s_1,\cdots, s_m =0 \\ \sum_j (r_j + s_j) \le d}}^d \left( c_{r_1,\cdots, r_m, s_1,\cdots, s_m } \prod_{j = 1}^m \hat{x}_j^{r_j} \hat{p}_j^{s_j} + \mathrm{h.c.} \right),
    \label{h-rep-polynomial}
\end{align}
for some constant coefficients $c_{r_1,\cdots, r_m, s_1,\cdots, s_m } \in \mathbb{C}$ and h.c. means hermitian conjugate. 

This form captures all possible terms of finite degree because $\hat{p}^{s} \hat{x}^{r} $ can always be rewritten as a linear combination of $\hat{x}^{r'} \hat{p}^{s'}$ without increasing the power, i.e., $\max\{|r'|, |s'| \} < \max\{|r|, |s|\}$ (see Eq.~\eqref{psxr-commutator} in App.~\ref{sec:hybrid-multi-universality}).

Under this representation, the goal of the compilation algorithm is to find a sequence of gates from the instruction set which results in a Hermitian matrix of the form Eq.~\eqref{unitary-rep-hermitian-gen} that generates the target unitary. As alluded to earlier in Sec.~\ref{subsec:flowsinphasespace}, one advantage of this representation is that the polynomial takes arguments as the canonical positions and momenta, so that the transformation on the oscillator may be viewed as flows in the $m$-oscillator phase space, in a similar spirit to the phase-space formulation of classical mechanics. It also admits a simple connection to the holomorphic (stellar) representation of CV computation \cite{chabaud2022holomorphic}, which is useful to quantify the non-Gaussian character of the transformation in terms of zeros of polynomials. We refer to Sec.~\ref{sssec:stellar-rep} for details.

\subsubsection{Unitary Picture}
\label{sssec:unitarypicture}

A complementary representation is to write the unitary matrix generated from Eq.~\eqref{unitary-rep-hermitian-gen} explicitly,
\begin{align}
    U =& A_0(\hat{w}_1, \hat{w}_2,\cdots,\hat{w}_m,\hat{v}_1,\hat{v}_2,\cdots,\hat{v}_m) \nonumber \\
       &+ i \sum_{k=1}^{4^n - 1} P_k A_k(\hat{w}_1, \hat{w}_2,\cdots,\hat{w}_m,\hat{v}_1,\hat{v}_2,\cdots,\hat{v}_m)
    \label{rep-unitary}
\end{align}
where $A_k$ ($k=0,\cdots,4^n-1$) are polynomials acting on the $m$-oscillators which satisfy certain constraints given by the unitarity of $U$. The arguments $w_j = e^{+i2k_j \hat{x}_j}$ and $\hat{v}_j = e^{-i2\lambda_j \hat{p}_j}$ are the unitaries that respectively boost the momentum of the $j$th oscillator by $k_j$ and displace the position by $\lambda_j$ (in the dimensionless Wigner units introduced in Sec.~\ref{sssec:xp-basis}).

The representation in Eq.~\eqref{rep-unitary} can be viewed as a generalized notation of quantum signal processing for single-qubit dynamics \cite{low2016methodology,low2017optimal}.  Here we adopt it to represent unitary dynamics in a coupled multi-qubit-oscillator system. Similar to the Hermitian picture representation, the polynomials $A_k$ are usually truncated to finite degrees. Moreover, an additional truncation is set by the scale of $k_j$ and $\lambda_j$. This truncation provides a coarse-grained description of the bosonic algorithm, which is necessary to avoid unphysical requirements during the compilation.

With this representation, the goal of the compilation is to find a sequence of hybrid qubit/bosonic gates from the instruction set such that the resulting unitary admits a given decomposition in Eq.~\eqref{rep-unitary}. One advantage of this representation is that results from single-qubit QSP may be generalized to bosonic systems. We will see in Sec.~\ref{ssec:compilation-bosonic-qsp-qsvt} that in the case of a single qubit-oscillator system, Eq.~\eqref{rep-unitary} essentially provides a Fourier series decomposition of phase space under plane waves with different wave vectors, similar in spirit to the plane wave basis used for modeling periodic systems \cite{singh2006planewaves}.

\subsection{Exact Analytical Compilation of Entangling Gates}
\label{ssec:exact-analytical-qubit-gates}

We begin by discussing useful techniques for exact analytical compilation of entangling gates. Throughout this section, we specialize to the superconducting hybrid oscillator-qubit hardware layout illustrated in Fig.~\ref{fig:hardware_layout}a, though note that many of the techniques discussed are extendable to other hybrid hardware CV-DV platforms. We emphasize that some of these techniques developed in this section, while novel in the context of the superconducting architecture of Fig.~\ref{fig:hardware_layout}a, draw direct inspiration from known methods in the trapped-ion community (the technique for compilation of multi-qubit entangling gates in Sec.~\ref{ssec:compilation-entangling}, reminiscient of M\o{}lmer-S\o{}renson gates ~\cite{sorensen2000entanglement, milburn1999simulating}, being a prime example). We will remark on these connections where appropriate.

\subsubsection{Useful Primitives}
\label{sssec:useful_primitives}
\subsubsubsection{Beam-Splitter SWAP Networks.}\label{ssssec:beamsplitter_swap_networks}
In a typical superconducting qubit-based quantum processor, lack of all-to-all connectivity inhibits the ability to natively perform entangling gates between geometrically nonlocal qubits. Instead, such gates are typically realized using SWAP networks, where information is routed to facilitate a geometrically local entangling operation.

A similar challenge persists in the superconducting hybrid oscillator-qubit hardware layout displayed in Fig.~\ref{fig:hardware_layout}a, where entanglement between sites is facilitated by beam-splitters between ``connected'' oscillators. Analogous to the qubit-only case, long-range entangling gates can be synthesized by leveraging a beam-splitter SWAP network, which we now describe. One advantageous property of bosonic SWAP networks is that the operations are agnostic to the contents of the bosonic modes. For example, each bosonic mode can store multiple bits of quantum information in a complex state, or each can encode a single logical qubit in one of a variety of bosonic error correcting codes. The latter possibility would allow for the  correction of errors occurring during the routing of the information across the hardware fabric. Recent experimental progress has dramatically enhanced the speed and fidelity of bosonic SWAP operations \cite{chapman2022high,lu2023highfidelity,chou2023demonstrating,deGraaf2024midcircuit}.

Defining a beam-splitter between oscillators at site $i$ and $j$ as
\begin{equation}
    \mathrm{BS}^{(i,j)}(\theta,\varphi) = e^{-i\frac{\theta}{2}[e^{i\varphi}a_i^\dagger a_j + e^{-i\varphi}a_i a_j^\dagger]},
\end{equation}
the key idea is to first note that $\textrm{SWAP} = \mathrm{BS}\left(\pi, \pi/2\right)$ can be used to realize the transformation
\begin{equation}
    \begin{split}
        \mathrm{BS}^{(i,j)\dagger}\left(\pi, \frac{\pi}{2}\right)a_i\mathrm{BS}^{(i,j)}\left(\pi, \frac{\pi}{2}\right) &= a_j\\
        \mathrm{BS}^{(i,j)\dagger}\left(\pi, \frac{\pi}{2}\right)a_j\mathrm{BS}^{(i,j)}\left(\pi, \frac{\pi}{2}\right) &= -a_i. \\
    \end{split}
\label{eq:ai_to_aj}
\end{equation}
We note that the above convention for SWAP differs from the choice of $\varphi=0$ used in the text surrounding Box \ref{Box:beam-splitter}. The net effect is a difference in global phase, and either can be used for the purpose of a beam-splitter SWAP network. An alternative option is to define SWAP by incorporating additional single-oscillator rotations to remove the phase entirely, as in Ref.~\cite{C2QA-LGTpaper}. Here, we opt to use the above definition and track the phase across compilation steps.

To see how this transformation can be leveraged for long-range entangling gates, for specificity let us imagine that we wish to synthesize a phase-space rotation of the $k$th oscillator controlled on the $j$th qubit,
\begin{equation}
\textrm{CR}^{(k,j)}(\theta) = e^{-i\frac{\theta}{2} Z_j a_k^\dagger a_k^{\phantom{\dagger}}}.
\label{eq:CRjk}
\end{equation}
By ``stacking'' beam-splitter transformations, this is achieved via the sequence
\begin{equation}
    \textrm{CR}^{(k,j)}(\theta) = U^\dagger \textrm{CR}^{(j,j)}(\theta) U \\
\end{equation}
where $U = \prod_{p=1}^{k-j}\mathrm{BS}^{(j,j+p)}(\pi, \pi/2)$ is a string of SWAPS along a path connecting oscillators $j$ and $k$.
The controlled rotation gate is just one example of the many possible hybrid oscillator-qubit gates that can be achieved using this technique.  Furthermore, it is not limited to hybrid oscillator-qubit gates and can also be used to compile non-local oscillator-oscillator entangling gates, such as beam-splitters, two-mode squeezers, and non-Gaussian gates.  As previously  noted, multiple SWAP strings can be parallelized and a scheduling protocol invoked to avoid collisions in the 2+1-dimensional space-time fabric.

\subsubsubsection{Synthesizing Hybrid Oscillator-Qubit Gates.}
The above SWAP network example illustrates the ability to synthesize hybrid oscillator-qubit gates even when the qubit and oscillator are not natively coupled. Similarly, we can condition natively available oscillator-only gates on a qubit or on multiple (possibly distant) qubits. As an example, let us consider the action of the controlled-parity gate
\begin{equation}
\textrm{CP}^{(j,k)} a_j \textrm{CP}^{(j,k)\dagger} = i Z_k a_j,
\label{eq:cparity_conj}
\end{equation}
where $\textrm{CP}^{(j,k)} = \textrm{CR}^{(j,k)}(\pi)$ is the controlled-parity gate between the qubit at site $k$ and oscillator at site $j$ (see Box \ref{Box:CRotation}). For $j=k$, this gate is natively realizable via the dispersive interaction, and for $j\neq k$, this gate can be synthesized by with the aid of beam-splitter SWAPs. Nonetheless, Eq.~(\ref{eq:cparity_conj}) provides a route for turning unconditional oscillator gates into conditional ones.  As described shortly below, this process can be iterated to add higher-weight Pauli strings.

A displacement of mode $j$ controlled by qubit $k$, realized via the sequence
\begin{equation}
    \begin{split}
        \mathrm{CD}^{(j,k)}(\alpha) &= \textrm{CP}^{(j,k)} D^{(j)}(i \alpha)\textrm{CP}^{(j,k)\dagger}\\
        &= e^{Z_k(\alpha a_j^\dagger - \alpha^* a_j)},
    \end{split}
    \label{eq:cond_disp_compilation}
\end{equation}
is a particularly useful equivalence for converting the Fock-space instruction set (where conditional displacements are not native primitives) into the phase-space instruction set.

Another useful gate is the conditional beam-splitter, with applications both toward quantum simulation (see, for example, Sec.~\ref{Z2section}), as well as for universal control and measurement of logical dual-rail qubits realized in superconducting cavities \cite{teoh2022dualrail, tsunoda2023error, chou2023demonstrating,deGraaf2024midcircuit}. It can be compiled as
\begin{equation}
    \begin{split}
        \textrm{CBS}^{(j,k,\ell)}(\theta,\varphi) &= \textrm{CP}^{(k,\ell)} \mathrm{BS}^{(j,k)}(\theta,\varphi-\pi/2)\textrm{CP}^{(k,\ell)\dagger}\\
        &= e^{-i Z_\ell \frac{\theta}{2}[e^{i\varphi}a_j^\dagger a_k^{\phantom{\dagger}} + e^{-i\varphi}a_j^{\phantom{\dagger}} a_k^\dagger]}.
    \end{split}
    \label{eq:cbs}
\end{equation}
Alternatively, this gate can be viewed as natively available in the phase-space ISA since it is readily compiled at the hardware level from a short pulse sequence taking advantage of the dispersive qubit-oscillator coupling (see Sec.~\ref{sssec:ECBS}). This latter strategy is particularly useful for optimizing gate speed in the weak dispersive regime in superconducting platforms.

Finally, we note the possibility of repeatedly using parity operations controlled on different qubits to realize oscillator operations conditioned on an arbitrary, high-weight Pauli operator. For example, it is possible to combine single qubit rotations, controlled parity operations, conditional displacements, and a mediating beam-splitter network to realize a displacement conditioned on a weight-$m$ Pauli operator via the sequence:
\begin{equation}
    \begin{split}
        \mathrm{CD}^{(k,\{\hat{n}_1,\hat{n}_2,\ldots\hat{n}_m\})}(\alpha)
        &= U_{\mathrm{seq}}^\dagger D(i^m \alpha)U_{\mathrm{seq}}\\
        &= e^{(\prod_{j=1}^m\hat{n}_j\cdot\vec{\sigma}_j)(\alpha a_k^\dagger - \alpha^* a_k)},
    \end{split}
    \label{eq:cond-disp-weight-m}
\end{equation}
where $U_{\textrm{seq}}=\prod_j^m e^{i\pi (\hat{n}_j\cdot\vec{\sigma}_j)a_k^\dagger a_k/2}$. The result is a very valuable resource that, for example, can be used to compile many-qubit entangling gates (see Sec.~\ref{ssec:compilation-entangling}, where we compile the Toffoli gate as an illustrative example).

\subsubsubsection{Composing Conditional Displacements.
}
\label{sssec:composingCDs}
The composition of conditional displacement gates offers a powerful resource for the control of hybrid oscillator-qubit systems. For example, sequences of conditional displacements are integral to the phase-space ISA, and underlie many examples and applications discussed throughout this work, including the readout, stabilization and logical gates for GKP error correcting codes in Sec.~\ref{sec:qec-compilation}, the compilation of Jaynes-Cummings and conditional squeezing gates in Sec.~\ref{sec:trotter-product}, the non-abelian QSP framework discussed in Sec.~\ref{ssec:compilation-bosonic-qsp-qsvt} and the realization of random quantum walks in Sec.~\ref{ssec:randomwalk}. However, inextricably tied to their utility, sequences of conditional displacements can combine to create extremely complex hybrid oscillator-qubit operations that are challenging to analyze. Here, our goal is to demystify some of this complexity by providing intuition for how such sequences act in phase space.

To begin, we recall the generalized conditional displacement operation of the form,
    \begin{equation} \mathrm{CD}_{\hat{b}\cdot\vec\sigma}(\alpha)=e^{\hat b\cdot\vec\sigma \otimes[\alpha a^\dagger -\alpha^* a]}
    \label{eq:gencdisp_cond}
\end{equation}
where $\hat b$ is a unit vector defining an axis on the qubit Bloch sphere. As discussed in Box~\ref{Box:c-displacement}, this general form can be synthesized by conjugating the standard controlled displacement operation $\mathrm{CD}(\alpha)$ by an unconditional rotation of the qubit. Next, we note that this conditional displacement can be cast into the form,
\begin{equation}  
    \mathrm{CD}_{\hat{b}\cdot\vec\sigma}(\alpha) = \sum_{\eta = \pm 1} \ket{\eta^{(\hat{b})}}\bra{\eta^{(\hat{b})}}\otimes D(\eta \alpha),
    \label{eq:cdisp_projform}
\end{equation}
where $\ket{\eta^{(\hat{b})}}\bra{\eta^{(\hat{b})}}=\frac{1}{2} \left[\hat I + \eta \hat b\cdot\vec \sigma \right]$ is a projector onto the eigenstate of $\hat{b}\cdot\vec{\sigma}$ with eigenvalue $\eta = \pm 1$.
We can compose two such conditional displacements with different displacement amplitudes and qubit axes,
\begin{equation}
    U =  \mathrm{CD}_{\hat{c}\cdot\vec{\sigma}}(\beta)\mathrm{CD}_{\hat{b}\cdot\vec{\sigma}}(\alpha).
\end{equation}
Using the form in Eq.~\eqref{eq:cdisp_projform}, this can be decomposed into a Feynman path summation of four distinct terms,
\begin{equation}
    \begin{split}
        U &= \sum_{\eta_1, \eta_{2} =\pm 1} W_{\eta_2,\eta_1} \otimes D(\eta_2 \beta)D(\eta_1 \alpha) \\
        &= \sum_{\eta_1, \eta_{2} =\pm 1} e^{2i c\eta_1\eta_2 \textrm{Im}(\beta \alpha^*)}W_{\eta_2,\eta_1} \otimes D(\eta_2 \beta + \eta_1 \alpha),
    \end{split}
    \label{eq:conddisp_sequence}
\end{equation}
where $c = 1$ in standard units and $c=\frac{1}{2}$ in Wigner units (i.e., $[\hat{p},\hat{x}]=-ic$). Here, the qubit transformation operator $W_{\eta_2,\eta_1}$ is defined as\footnote{Using vector identities, it can also be shown that $W_{\eta_2,\eta_1}=\frac{1}{4}[\hat{I}(1+\eta_1\eta_2\hat{c}\cdot\hat{b}) + \eta_2 \hat{c}\cdot\vec{\sigma} + \eta_1 \hat{b}\cdot\vec{\sigma} + i\eta_1\eta_2\hat{c}\times\hat{b}\cdot\vec{\sigma}]$.}
\begin{equation}
    \begin{split}
        W_{\eta_2,\eta_1} &= \ket{\eta_2^{(\hat{c})}}\braket{\eta_2^{(\hat{c})}}{\eta_1^{(\hat{b})}}\bra{\eta_1^{(\hat{b})}}.
    \end{split}
\end{equation}
This decomposition lends itself to an intuitive physical interpretation. In the case where $\hat{c}$ and $\hat{b}$ are orthogonal, $U$ enacts a superposition of four distinct oscillator displacements $D(\beta+\alpha)$, $D(\beta-\alpha)$, $D(-\beta+\alpha)$, and $D(-\beta-\alpha)$. Each is paired with a complicated qubit operation that sends eigenstates of $\hat{b}\cdot\vec{\sigma}$ to eigenstates of $\hat{c}\cdot\vec{\sigma}$ with weights given by the overlap between these eigenstates. In contrast, if $\hat{c}$ and $\hat{b}$ correspond to the same direction on the Bloch sphere, $U$ collapses into a single conditional displacement as only terms with $\eta_1 = \eta_2$ contribute.

More generally, a sequence of many conditional displacements yields a larger superposition of the form in Eq.~\eqref{eq:conddisp_sequence}, and each conditional displacement tends to double the number of terms. From a phase-space perspective, if the oscillator initially starts in the vacuum state, $n$ conditional displacements split the initial coherent state into $2^n$ Gaussian phase-space `blobs' (Sec. \ref{subsec:flowsinphasespace}).

These blobs can overlap, causing interference fringes in the final state (similar to the four-legged cat state shown in Fig.~\ref{fig:single-mode-QEC}). Generally, the qubit is entangled with the oscillator at the end of this sequence, though it is possible to disentangle the qubit through a carefully chosen series of displacements -- see the example of preparing a cat state using a $\textrm{BB1}_{90}$ composite pulse sequence in Fig.~\ref{fig:QSPcat}.

The simplicity of the above representation for composing conditional displacements allows for the efficient numerical simulation of sequences of many such gates. Rather than representing the controlled displacements numerically as matrices acting on a large Hilbert space, we can simply  keep track of how the phase space `blobs' split and move in phase space. Building upon the intuition presented above, each of the $2^n$ phase-space blobs is associated with a definite spin state. If the most recent displacement operation was conditioned on, say the Pauli $X$ operator (i.e., $\hat{b}=\hat{x}$), then every phase space blob corresponds to one of the two $X$ qubit eigenstates, $|\pm\rangle$.  However, if two blobs overlap, the phase of their superposition determines the orientation of the spin associated with that region of phase space. Ref.~\cite{dias2024classical} discusses an efficient method to analyze linear combinations of Gaussian states which is readily extensible to the present case of entangled qubit/oscillator states produced in the phase-space ISA.

\begin{figure*}[htb]
    \centering
    \includegraphics[width=0.98\linewidth]{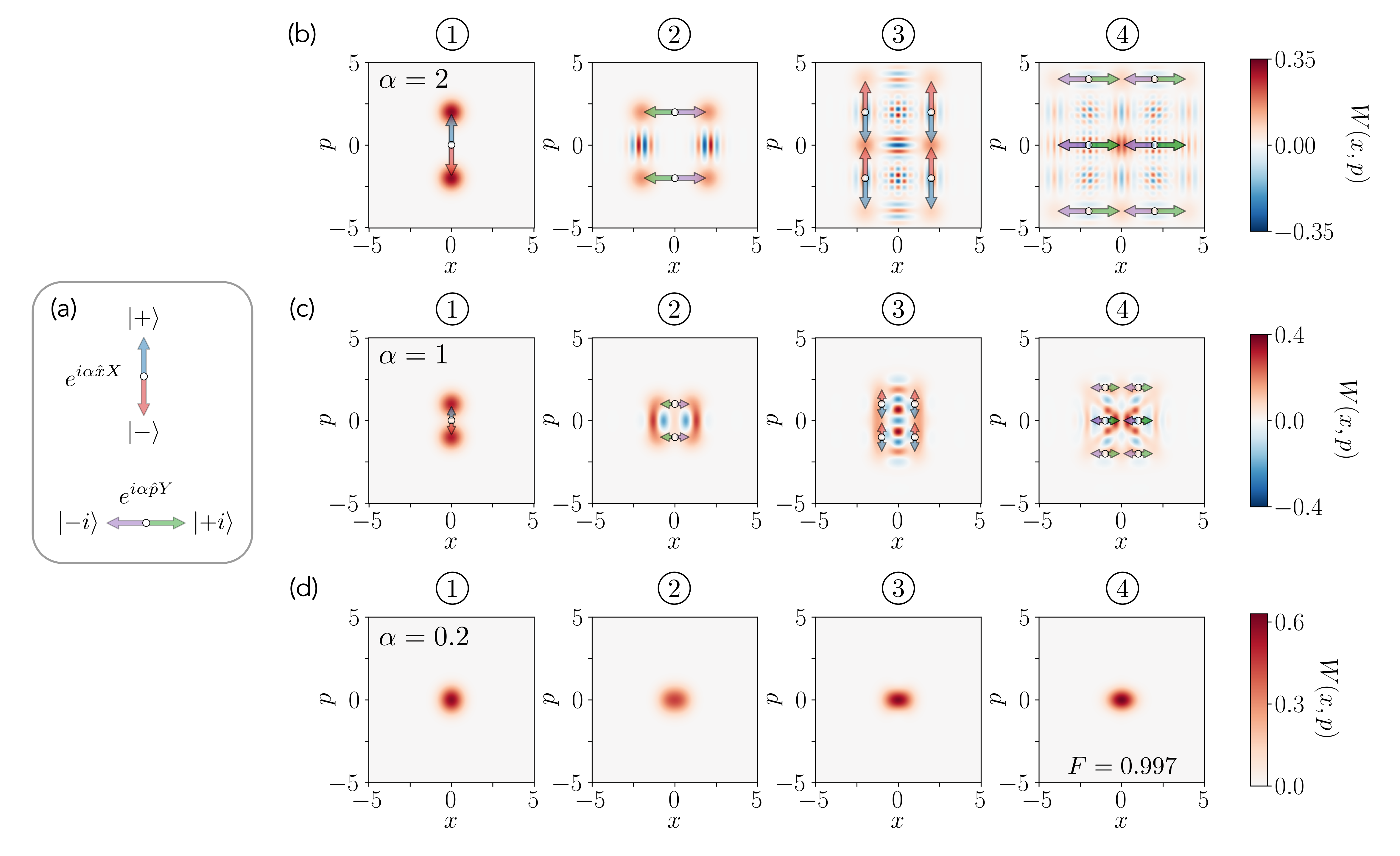}
    \caption{\textbf{Intuition for composed conditional displacements.}  (a) Graphical overlays are used to indicate $X$-controlled momentum boosts (top) and $Y$-controlled coordinate boosts (bottom) in the other panels. The white circle indicates the `starting point' for each blob, while the arrows indicate the displacement direction for each conditioned qubit state. All other panels: Wigner quasiprobability distributions for the oscillator after each successive displacement (from left-to-right) of the sequence in Eq.~\eqref{eq:figure_cd_sequence} applied to an oscillator and qubit in vacuum and $\ket{0}$, respectively. See also the `sum of paths' decomposition in Eq.~\eqref{eq:figure_cd_decomp}. Before computing each Wigner distribution, the qubit state is traced out. Interference fringes are therefore visible only between components of definite qubit state. For example, there are no fringes after the initial conditional displacement as the two blobs are entangled with the qubit but, after the second, each displaced two-legged cat is paired with a definite $X$-basis qubit state (i.e., there are two $|+\rangle$ state blobs that can interfere and two $|-\rangle$ state blobs that can interfere). (b) Large displacement $\alpha = 2$. In this case, distinct blobs are visible after each successive displacement, each corresponding to a unique composite displacement in the sum in Eq.~\eqref{eq:figure_cd_decomp}. While there are 16 total terms, only 9 contain unique displacements, leading to interference fringes within the blobs along the axes $x=0$ and $p=0$. (c) Intermediate displacement $\alpha = 1$. In this case, $\alpha$ is not large enough to resolve distinct blobs past the first displacement, leading to complicated interference patterns in the resulting Wigner function. (d) Small displacement $\alpha = 0.2$. As discussed in Sec.~\ref{sec:trotter-product}, the sequence in Eq.~\eqref{eq:figure_cd_sequence} approximates a conditional squeezing gate $\textrm{CS}(\zeta)$ gate with $\zeta = -4\alpha^2$ for small $\alpha$. We compute a fidelity of $F=0.997$ with exact application of $\textrm{CS}(\zeta)$. While the oscillator appears squeezed after the first displacement, we note that the purity $P_i = \textrm{Tr}(\rho_{\textrm{osc},i})$ of the oscillator state $\rho_{\textrm{osc},i}$ after the $i$th displacement reveals the oscillator and qubit to be entangled until the final step ($P_1=0.93$, $P_2=0.75$, $P_3=0.94$, $P_4=0.99$), highlighting the important role of each successive conditional displacement in not only squeezing the oscillator but leaving the hybrid oscillator-qubit system disentangled (in the case where the qubit is initialized to a $Z$ eigenstate).}
    \label{fig:composing_conditional_disp}
\end{figure*}

Thus far, our discussion has concerned the composition of conditional displacements in the most general case. To gain further intuition, and as a particular example, we consider a sequence of four conditional displacements,
\begin{equation}
    U = \mathrm{CD}_{\hat{y}\cdot\vec{\sigma}}(-\alpha)\mathrm{CD}_{\hat{x}\cdot\vec{\sigma}}(-i\alpha) \mathrm{CD}_{\hat{y}\cdot\vec{\sigma}}(\alpha) \mathrm{CD}_{\hat{x}\cdot\vec{\sigma}}(i\alpha)
    \label{eq:figure_cd_sequence}
\end{equation}
where $\alpha$ is a real-valued parameter. As will be discussed in Sec.~\ref{sec:trotter-product}, such a sequence approximates a conditional squeezing gate in the limit $\alpha \ll 1$ via a BCH formula, see Eq.~(\ref{eq:Trottersqueezing}). It is instructive, however, to examine the trajectory of the Wigner function after each conditional displacement for general $\alpha$. In Fig.~\ref{fig:composing_conditional_disp}, we show the phase-space trajectory for three cases: $\alpha=2$, $\alpha=1$, and $\alpha=0.3$. Each case produces qualitatively distinct behavior, highlighting the expressivity (and complexity) of sequences of conditional displacements.

To provide further intuition, it is helpful to decompose $U$ into a Feynman path summation of unconditional displacements as in Eq.~\eqref{eq:conddisp_sequence},
\begin{align}
\begin{split}
    U &= \sum_{\vec{\eta}} W_{\vec{\eta}}\otimes D(-\eta_4 \alpha)D(-i\eta_3 \alpha)D(\eta_2 \alpha)D(i\eta_1 \alpha).
\end{split}
\end{align}
where $\vec{\eta} = (\eta_1,\eta_2,\eta_3,\eta_4)$ is shorthand for a list of indices, each of which is summed over $\pm 1$. Furthermore, as before, we have adopted the shorthand,
\begin{equation}
W_{\vec{\eta}} = \ket{\eta_4^{(\hat{y})}}\braket{\eta_4^{(\hat{y})}}{\eta_3^{(\hat{x})}}\braket{\eta_3^{(\hat{x})}}{\eta_2^{(\hat{y})}}\braket{\eta_2^{(\hat{y})}}{\eta_1^{(\hat{x})}}\bra{\eta_1^{(\hat{x})}}
\end{equation}
for the weighted qubit transformation operator $W_{\vec{\eta}}$.

To facilitate computation of $W_{\vec{\eta}}$, it is helpful to  recognize that the qubit projectors can be conveniently expressed in the computational basis as
\begin{align}
    \ket{\eta_i^{(\hat{x})}}\bra{\eta_i^{(\hat{x})}}&=\frac{1}{2}\left(\begin{array}{cc} 1&\eta_i\\ \eta_i&1\end{array}\right)\\
    \ket{\eta_i^{(\hat{y})}}\bra{\eta_i^{(\hat{y})}}&=\frac{1}{2}\left(\begin{array}{cc} 1&-i\eta_i\\ +i\eta_i&1\end{array}\right).
\end{align}
In addition, similar to Eq.~\eqref{eq:conddisp_sequence}, the four unconditional displacements can be combined into a single unconditional displacement: using the convention $[\hat p,\hat x]=-ic$, we may repeatedly use the identities in Box~\ref{Box:UncondDispGate} to obtain
\begin{equation}
    \begin{split}
        U &= \sum_{\vec{\eta}} e^{i\phi_{\vec{\eta}}}W_{\vec{\eta}}\otimes e^{i2\alpha[(\eta_4-\eta_2)\hat p+(\eta_1-\eta_3)\hat x]},
    \end{split}
    \label{eq:figure_cd_decomp}
\end{equation}
where $\phi_{\vec{\eta}} = -2c\alpha^2(\eta_1\eta_2+\eta_2\eta_3+\eta_3\eta_4-\eta_4\eta_1)$.
Thus, we see that each of the 16 phase space displacements in the `sum over paths' is paired with an associated phase factor $e^{i\phi_{\vec{\eta}}}$ and qubit transformation operator $W_{\vec{\eta}}$.  We also see that some distinct terms contain identical displacements (e.g., $\vec{\eta}=(+1,+1,-1,+1)$ and $\vec{\eta}=(-1,+1,+1,+1)$) and therefore leads to interference effects arising from a superposition of qubit transformation operators associated with the same point in phase space -- see, for example, the center `blob' in the final step of Fig.~\ref{fig:composing_conditional_disp}(b). 

\subsubsection{Compilation of Entangling Gates between Physical Qubits}
\label{ssec:compilation-entangling}
A particularly important primitive for any quantum computation is the entangling gate. In the hybrid computational model where both oscillator modes and physical qubits are used for computation, it is imperative to not only have access (either natively or through compilation) to oscillator-oscillator entangling gates such as the two-mode beam-splitter (Box \ref{Box:beam-splitter}) and two-mode squeezing (Box \ref{Box:2-mode-squeezing}) gates, but also to entangling operations between physical qubits such as the CNOT and Toffoli gates. Ideally, such operations should be achievable independently of the state of the cavity modes to which they are coupled. 

Hybrid CV-DV compilation strategies for such entangling operations have a rich history in the trapped ions platform, where bosonic modes serve as mediators of effective qubit-qubit interactions. One of the earliest and most influential examples is the Mølmer-Sørensen (MS) scheme, which enables high-fidelity two-qubit gates even in the presence of thermally populated motional modes \cite{molmer1999quantum, molmer1999multiparticle,sorensen2000entanglement,milburn2000iontrap}. This approach has since been extensively generalized to include the use of local radial modes \cite{Serafini_2009}, the synthesis of multi-body spin interactions \cite{Katz_2023, katz2022nbody}, and enhancements through parametric driving \cite{burd_quantum_2021}. The reverse -- mediating oscillator interactions via Berry phases accumulated on spins -- has also been realized, leading to programmable, high-fidelity multi-mode operations \cite{katz2023programmable, gan2020hybrid, Slichter_coherent_2024}. While these strategies are mature and platform-optimized for trapped ions, they are largely unused or not directly applicable to other architectures, including superconducting, due to the fundamentally distinct native instructions and unique CV-DV connectivity of each physical platform.

To remedy this gap, in the following we present a flexible strategy for compiling arbitrary entangling gates between physical qubits coupled to independent oscillator modes spefically tailored to the superconducting architecture of Fig.~\ref{fig:hardware_layout}a. As each mode is coupled to a single physical qubit, we will heavily rely on the ability to natively realize or compile beam-splitters between any pair of modes (using, for example, the beam-splitter SWAP network described in Sec.~\ref{ssssec:beamsplitter_swap_networks}). For convenience, we work in the phase-space ISA, though we note that all assumed native operations can be compiled in any of the ISAs found in Table \ref{tab:ISA_overview} to achieve universal hybrid control. We begin by explaining the core principle using a single-qubit gate as an example, and follow this with extension to two- and multi-qubit gates and comment on optimization strategies. Finally, we conclude with a brief comparison to M{\o}lmer-S{\o}rensen gates \cite{molmer1999quantum, molmer1999multiparticle,milburn2000iontrap} (in trapped-ions) and resonator-induced phase gates \cite{Cross2015, Paik2016, Puri2016c} (in circuit QED), both of which provided inspiration for the scheme presented here.

\subsubsubsection{Single-Qubit Gate.}
To explain the general strategy, it is instructive to first consider the compilation of a single (physical) qubit gate. In particular, we demonstrate how one can synthesize $R_z(\theta) = e^{-i\theta Z/2}$ using only beam-splitters and displacements (both conditional and unconditional). We emphasize that this is not the most practical strategy for implementing a single-qubit rotation, an operation that can be much more efficiently realized natively without the need for (relatively time-consuming) oscillator-qubit entangling gates. However, the general idea lays the groundwork for the compilation of multi-qubit gates which cannot be natively realized.

First, recall that a series of displacements forming a closed loop in phase-space composed as 
\begin{equation}
    D(-\beta)D(-\alpha)D(\beta)D(\alpha) = e^{2iA}
    \label{eq:displacement_loop}
\end{equation} 
where $A = \Im{\alpha^* \beta}$ is the oriented area of the enclosed parallelogram (see Fig.~\ref{fig:parallelogram} and surrounding text). Consequently, this sequence of unconditional displacements imparts a phase independent of the state of the oscillator. Here, the idea is to condition these displacements on a qubit such that the oriented area, and therefore the imparted phase, depends on its state (but not that of the oscillator). The net effect is a single-qubit rotation.

\begin{figure}[htb]
    \includegraphics[width=0.8\linewidth]{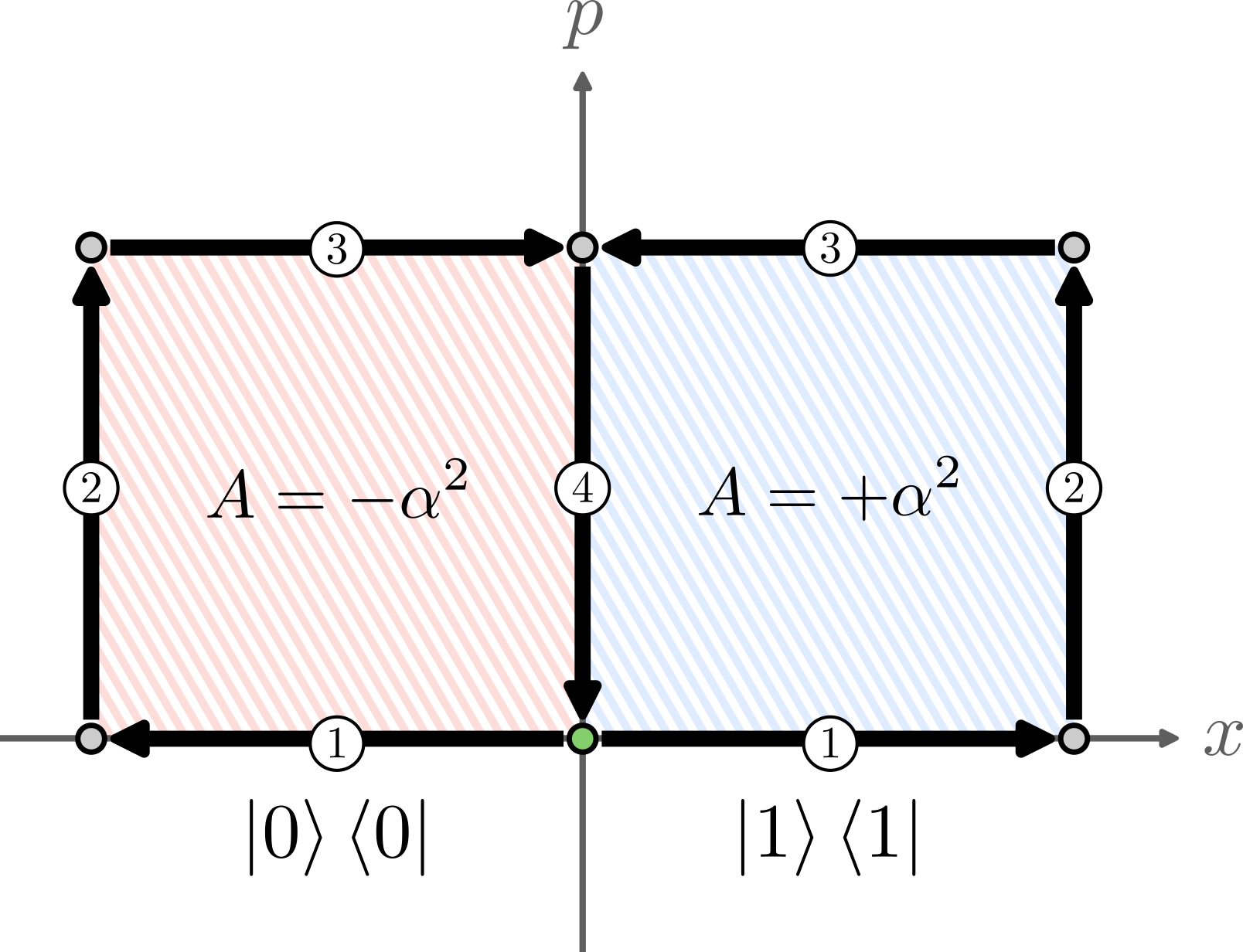}
    \caption{A conditional oscillator phase-space trajectory that realizes $R_z(\theta) = \textrm{exp}(-i\theta Z/2)$. The gate sequence is as follows: (1) Displacement by $\alpha$ or $-\alpha$, conditioned on the state of the qubit. (2) Unconditional displacement by $i\alpha$. (3) Displacement by $-\alpha$ or $\alpha$, conditioned on the state of the qubit. (4) Unconditional displacement by $-i\alpha$. In all, the sequence leaves the oscillator unchanged (mapping each point in phase-space back to itself), but imparts a phase that is dependent on the state of the qubit (but independent of the state of the resonator), equivalent to the application of $R_z(\theta)$ for the choice $\alpha = \sqrt{\theta}/2$.  }
    \label{fig:compilation_single_qubit_gate}
\end{figure}

One such example is the sequence
\begin{equation}
    \begin{split}
        U &= D(-i\alpha)D_c(\alpha, -\alpha)D(i\alpha)D_c(-\alpha, \alpha),
        \label{eq:calibration-displacement-sequences-1}
    \end{split}
\end{equation}
realizing the phase-space trajectory shown in Fig.~\ref{fig:compilation_single_qubit_gate}. Associated with each of the two possible ``pathways'' is an oriented area whose sign depends on the $Z$-basis state of the qubit, i.e., $A = -\alpha^2 Z$. This can equivalently be seen through brute-force simplification:
\begin{equation}
    \begin{split}
        U &= \ketbra{0}{0}e^{-2i\alpha^2} + \ketbra{1}{1}e^{2i\alpha^2}, \\
        &= e^{-2i\alpha^2 Z},
        \label{eq:calibration-displacement-sequences-2}
    \end{split}
\end{equation}
where we have used the definition of the conditional displacement (Box \ref{Box:c-displacement}) in combination with Eq.~\eqref{eq:displacement_loop}. Defining $\theta = 4\alpha^2$, we see that this amounts to a single-qubit rotation about the $Z$ axis, $U = R_z(\theta)$. This gate sequence is routinely used in experiments to calibrate the size of displacements and controlled-displacements in phase space (see for example, Fig. S4 in the supplementary material of \cite{Campagne-Ibarcq2020}).

\subsubsubsection{Two-Qubit Entangling Gates.} Building upon the core idea of the previous section, we now demonstrate the general recipe for compiling entangling gates between physical qubits. In particular, we show two examples of two-qubit gates: $R_{ZZ}(\theta) = e^{-i\frac{\theta}{2} Z_1\otimes Z_2}$ and $\textrm{CNOT} = e^{i\frac{\pi}{4}[I_1-Z_1] \otimes [I_2-X_2]}$. To that end, it is helpful to first define some notation. We envision a system comprising at least two modes (with annihilation operators $a_1$ and $a_2$) and two qubits (with Pauli-Z operators $Z_1$ and $Z_2$). Working within the phase-space ISA, we assume native access to controlled-displacements between the $i$th mode and $i$th qubit (denoted $D_c^{(i,i)}(\alpha,\beta)$), a beam-splitter between the two cavity modes (denoted $\mathrm{BS}^{(1,2)}(\theta,\varphi)$), and arbitrary single-qubit rotations $R_{\hat{n}\cdot\vec{\sigma}}(\theta)$. Furthermore, we also leverage the native dispersive interaction between the $i$th qubit and the $i$th oscillator whenever it is advantageous. As noted earlier, the remote beam-splitter can be achieved through a network of SWAPS.

\begin{figure*}[htb]
    \includegraphics[width=1\linewidth]{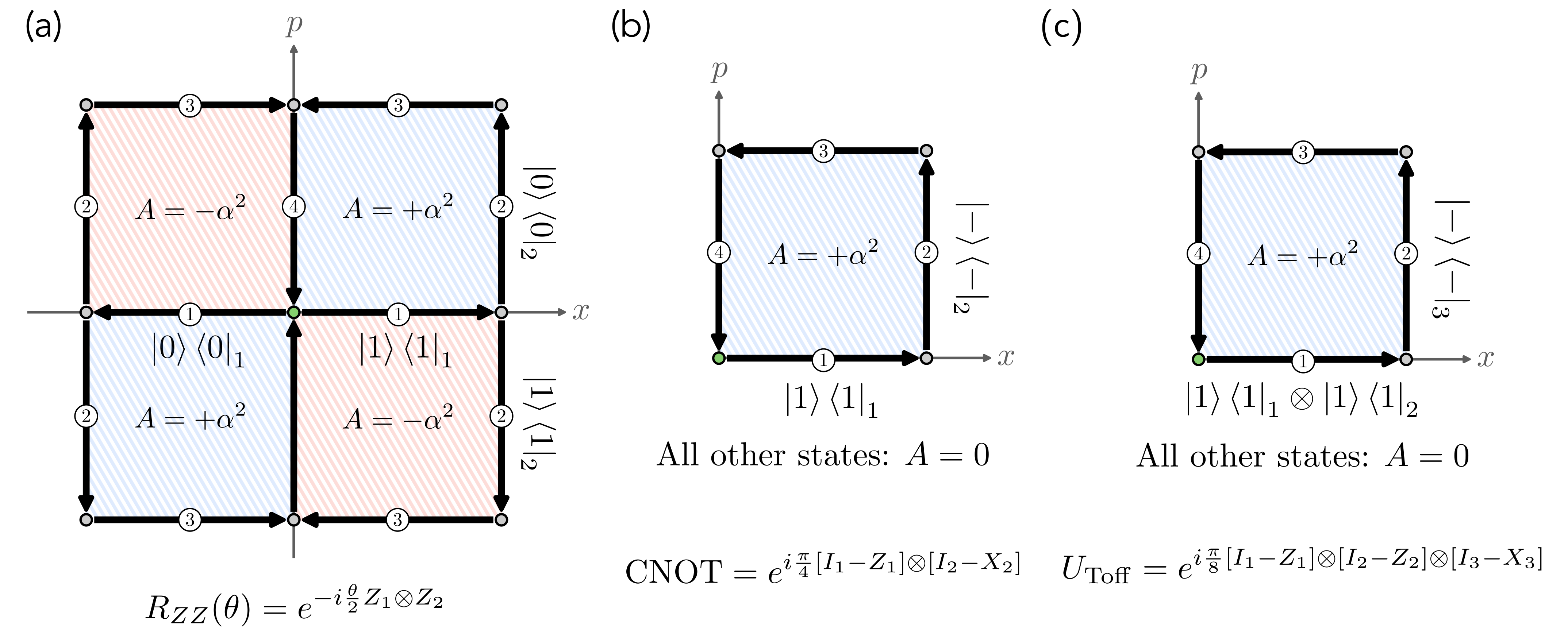}
    \caption{Phase-space trajectories that realize (a) $R_{ZZ}(\theta)$, (b) \textrm{CNOT}, and (c) \textrm{Toffoli}. In all cases, displacement conditions are indicated by projectors onto particular qubit states, with subscripts labeling the qubit index. We label only the conditions for steps 1 and 2, as steps 3 and 4 consist of applying the Hermitian conjugate operator of the former.}
    \label{fig:compilation_multi_qubit_gates}
\end{figure*}

En route to an entangling gate, it is helpful to first compile a conditional displacement between mode $1$ and qubit $2$ by leveraging mode SWAPs via beam splitters. Recalling the identity in Eq.~\eqref{eq:ai_to_aj}, this is achieved via the sequence
\begin{equation}
    \begin{split}
        D_c^{(1,2)}(\alpha,\beta) = \mathrm{BS}^{(1,2)\dagger}\left(\pi,-\frac{\pi}{2}\right) &D_c^{(2,2)}(\alpha,\beta) \\\times &\mathrm{BS}^{(1,2)}\left(\pi,-\frac{\pi}{2}\right).
    \end{split}
    \label{eq:beamsplitter_conjugation}
\end{equation}
To compile $R_{ZZ}(\theta)$, we can then follow the general strategy of the previous section, but with conditional displacements replacing their unconditional counterparts:
\begin{align}
        R_{ZZ}(\theta) &= D^{(1,2)}_c(-i\alpha, +i\alpha)D^{(1,1)}_c(\alpha, -\alpha)\nonumber\\
        &\times D^{(1,2)}_c(+i\alpha, -i\alpha)D^{(1,1)}_c(-\alpha, \alpha),
    \label{eq:RZZ}
\end{align}
with $\theta = 4\alpha^2$. As shown in Fig.~\ref{fig:compilation_multi_qubit_gates}a, the four possible eigenstates of $Z_1\otimes Z_2$ correspond to four distinct phase-space trajectories, enclosing an oriented area proportional to its associated eigenvalue, i.e., $A=-\alpha^2 Z_1\otimes Z_2$. 

Similarly, a CNOT gate can be compiled with the sequence
\begin{equation}
    \begin{split}
        \mathrm{CNOT} &= D^{(1,2)}_{cX}\left(0, -i\sqrt{\frac{\pi}{2}}\right)D^{(1,1)}_{c}\left(0, -\sqrt{\frac{\pi}{2}}\right)\\
                &\times D^{(1,2)}_{cX}\left(0, +i\sqrt{\frac{\pi}{2}}\right)D^{(1,1)}_c\left(0, \sqrt{\frac{\pi}{2}}\right),\label{DCX}
    \end{split}
\end{equation}
where we have employed the shorthand $D^{(1,2)}_{cX}(\alpha,\beta) = e^{-i\pi Y_2/4}D^{(1,2)}_{c}(\alpha,\beta)e^{+i\pi Y_2/4}$, converting a $Z-$conditional displacement into an $X-$conditional displacement. As illustrated in Fig.~\ref{fig:compilation_multi_qubit_gates}b, this sequence corresponds to the phase-space trajectory with oriented area $A = \pi [I_1-Z_1]\otimes[I_2 - X_2]/4$. In total, both the $R_{ZZ}(\theta)$ and CNOT gates require four beam-splitters and four conditional displacements.

\subsubsubsection{Many-Qubit Gates.}
We now demonstrate that this recipe can be extended beyond two-qubit gates and toward more general many-qubit entangling gates. To that end, we use the Toffoli gate as an illustrative example. The core strategy is similar to the previous examples but with additional synthesis of displacements conditioned on high-weight Pauli operators using the technique in Eq.~(\ref{eq:cond-disp-weight-m}).

To compile a many-qubit gate, we follow the same strategy as the two-qubit gates, but replace displacements conditioned on a single qubit by ones conditioned on an arbitrary number of qubits. In particular, in Eq.~(\ref{eq:cond-disp-weight-m}) we have already shown the ability to synthesize a displacement conditioned on an arbitrary, weight-$m$ Pauli operator. Consequently, this ability facilitates the ability to realize highly complex $n$-qubit entangling gates.

As a simple example, we now consider the Toffoli gate defined as
\begin{equation}
U_{\mathrm{Toff}} = e^{i\pi[I_1-Z_1]\otimes[I_2-Z_2]\otimes[I_3-X_3]/8}.
\end{equation}
There is freedom in choosing the particular sequence of conditional displacements. One solution is to rewrite $U_\textrm{Toff}$ as a product of displacements conditioned on (commuting) Pauli operators of weight 3 and realize each with Eq.~(\ref{eq:cond-disp-weight-m}). A more efficient solution, though, is to recognize that
\begin{equation}
    \begin{split}
        U_{1,2}(\alpha) &= e^{[I_1-Z_1][I_2-Z_2](\alpha a^\dagger - \alpha^* a)/4} \\
                &= D^{(2,2)}_c\left(0,\frac{\alpha}{2}\right)\mathrm{CP}^{(2,1)}D^{(2,2)}_c\left(0,-\frac{i\alpha}{2}\right)\mathrm{CP}^{(2,1)\dagger},
    \end{split}
    \label{eq:U12_doublecond_disp}
\end{equation}
where $\mathrm{CP}^{(j,k)}$ is a controlled-parity operation between mode $j$ and qubit $k$, natively realizable using the dispersive interaction (see Box~\ref{Box:CRotation}). Here, the general idea is that $U_{1,2}(\alpha)$ realizes a displacement of oscillator 2 conditioned on \emph{both} qubits 1 and 2 being in the state $\ket{1}$, as shown in the first step of Fig.~\ref{fig:compilation_multi_qubit_gates}c. This is realized through the combination of the controlled-parity technique described in Eq.~(\ref{eq:cond_disp_compilation}) and a final conditional displacement such that oscillator 2 is displaced only if both qubits are in the state $\ket{1}$.

Likewise, defining $U_{3}(\alpha) = D^{(2,3)}_{cX}(0,\alpha)$, the Toffoli gate can then be compiled as
\begin{equation}
    U_{\mathrm{Toff}} = U_{3}^\dagger\left(i\sqrt{\frac{\pi}{2}}\right)U_{1,2}^\dagger\left(\sqrt{\frac{\pi}{2}}\right)U_{3}\left(i\sqrt{\frac{\pi}{2}}\right)U_{1,2}\left(\sqrt{\frac{\pi}{2}}\right),
\end{equation}
 with the corresponding phase-space trajectory illustrated in Fig.~\ref{fig:compilation_multi_qubit_gates}c.

\subsubsubsection{Strategies for Optimization.}
\begin{figure*}[htb]
    \centering
    \includegraphics[width=0.75\linewidth]{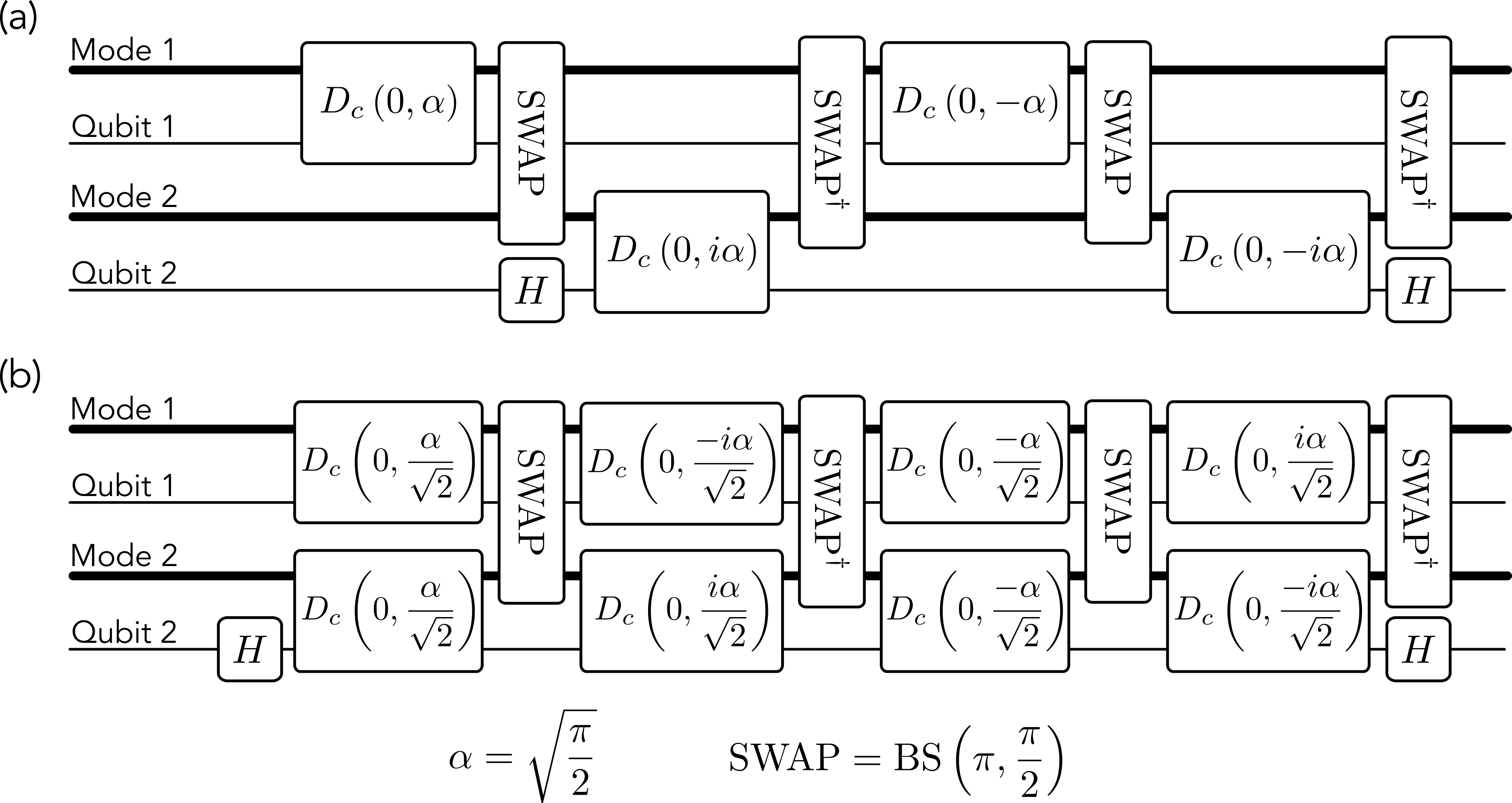}
    \caption{Two quantum circuits that realize a CNOT gate between qubits 1 and 2. Panel (a) corresponds to the phase-space trajectory in Fig.~\ref{fig:compilation_multi_qubit_gates}b, where the entirety of the requisite geometric phase is achieved through the phase-space area enclosed by a single mode. Panel (b) shows a more efficient strategy, where two modes are displaced in parallel enclosing an area such that each contributes half the requisite geometric phase. The benefit of such an approach is that the size (and consequently duration) of each displacement is reduced by a factor of $\sqrt{2}$.}
    \label{fig:CNOT_circuit}
\end{figure*}
 As a final remark on the compilation of multi-qubit gates, we note some additional strategies that can be implemented to reduce overall runtime and minimize decoherence. For one, in the above examples we have used a series of conditional displacements to accumulate a geometric phase on a single oscillator mode -- a choice largely made for ease of explanation. A more efficient strategy is to ``symmetrize'' the compilation such that all involved oscillator modes accumulate a geometric phase in parallel. In particular, involving $n$ modes allows us to impart a total phase of $A$ by enclosing a phase-space area of $A/n$ with each mode in parallel, leading to a $1/\sqrt{n}$ reduction in both the size and duration of each displacement relative to the single mode realization. Thus, the benefit is a reduction in the multi-qubit gate time by a factor of $\sqrt{n}$, while simultaneously reducing the total photon number per cavity throughout the gate. This last feature helps reduce the impact of non-idealities such as cavity self-Kerr, though the total number of photons across all modes remains the same, leaving the rate of photon loss unchanged in comparison. As a simple illustration of this strategy in practice, Fig.~\ref{fig:CNOT_circuit} compares two circuits realizing a CNOT gate: one using the single mode displacement sequence in Eq.~(\ref{eq:RZZ}), and the other using two modes in parallel.

In each of the previous examples, we have used a series of displacements of equal magnitude, thereby enclosing a square region of phase space. However, in certain contexts (and depending on the particulars of the experimental platform), it is advantageous to alter this approach and instead enclose a non-square region. To see this, let us take as an example the CNOT gate and note that there are multiple strategies for compilation, with Fig.~\ref{fig:compilation_multi_qubit_gates}b providing one such case where both position and momentum boosts are conditional. Alternatively, we could have performed coordinate boosts conditioned on both $\ket{1}\bra{1}_1$ and $\ket{-}\bra{-}_2$ (synthesized similarly to the first step of our $U_{\textrm{Toff}}$ compilation) with momentum boosts \emph{unconditional}. As large unconditional displacements can be carried out much faster than their conditional counterparts \cite{EickbuschECD}, the overall runtime can be decreased by enclosing a rectangular region of phase space using short conditional displacements alternated with long unconditional displacements. The trade-off of this strategy is a relative increase in photon number (and, correspondingly, photon loss) during the gate. The optimal balance of these competing effects will in general depend on the specifics of the experimental platform and the gate being compiled, and constitutes an interesting future direction.

\subsubsubsection{Advantages over MS and RIP Gates.}

In superconducting platforms, analogs to MS gates known as resonator-induced phase (RIP) gates have been proposed \cite{Cross2015, Paik2016, Puri2016c}. The core idea is to leverage an architecture similar to that of trapped ion case platforms in Fig.~\ref{fig:hardware_layout}b, where multiple qubits are statically coupled to a common microwave bus resonator. Implementation of the gate then follows in analogy to MS gates -- the resonator is driven in a qubit-conditioned closed phase-space loop, thereby enacting an entangling operation on the qubits. However, generalizing this gate beyond two-qubit gates to the multi-qubit case is a challenge, involving complex microwave pulse shapes \cite{Cross2015}.

A major weakness of MS and RIP gates is that both rely on a mediating motional or bus microwave mode that is \emph{statically} coupled to a particular subset of qubits, as in Fig.~\ref{fig:hardware_layout}b, leaving the limited ability to selectively entangle arbitrary combinations of qubits. In contrast, our proposed approach exploits the local oscillator-qubit connectivity of the proposed superconducting architecture in Fig.~\ref{fig:hardware_layout}a to realize a general multi-qubit entangling scheme with comparatively greater flexibility. While it relies on the same general principle as MS and RIP gates -- i.e., an oscillator that is driven through closed, qubit-state-dependent loops in phase space -- it is not limited to the use of motional or microwave modes with particular mode-qubit couplings and, instead, one can \emph{selectively} couple particular qubits to a chosen oscillator. In some sense, it may be viewed as a digital analog of MS/RIP gates with a programmable common mode, where one can choose the number and locations of qubits to couple to through judicious use of beam-splitters. Furthermore, as noted previously, the 2D connectivity graph of Fig.~\ref{fig:hardware_layout} can facilitate the implementation of many entangling gates in parallel given sufficient care is taken to avoid ``collisions'' between photons traveling via beam-splitter SWAPS.

Finally, we note that our scheme has potential advantages over conventional entangling gates in superconducting platforms that rely on direct coupling between superconducting qubits and are thus limited by cross-talk (see, for example, Ref.~\cite{sheldon2016procedure} for a discussion regarding the cross-resonance gate). In contrast, here we rely only on the dispersive coupling to individual oscillators and use beam-splitters to controllably route information, avoiding the problem of cross-talk while offering the flexibility of oscillator-mediated gates. An in-depth comparison between conventional entangling gates and the proposed oscillator-mediated scheme is beyond the scope of this work, however, and is left as a future direction.

\subsubsection{Compilation of Entangling Gates between Oscillators}
\label{ssec:compilation-entangling-oscillators}
While the superconducting architecture in Fig.~\ref{fig:hardware_layout}a includes beam-splitters as a native oscillator-oscillator entangling operation, systematic techniques to compile more complex entangling gates are desirable. In this section, we briefly review such a technique developed in Ref.~\cite{C2QA-LGTpaper} in the context of simulating lattice gauge theories. We refer to this work for details and merely summarize the basic concept here.

The core principle is to leverage an ancillary qubit to mediate interactions between two or more oscillators. Due to the mismatch in Hilbert space between the oscillators and the ancilla, this interaction is synthesized in a sequential, bit-wise fashion. To explain further, let us first assume access to the gate
\begin{equation}
    U_i(\theta) = e^{-i\theta Z_{\textrm{anc}}O_i},
\end{equation}
where the subscript `anc' denotes an ancillary qubit in a known initial eigenstate of $Z_{\textrm{anc}}$ and $O_i$ is any Hermitian operator acting on the $i$th mode. For example, $U_i$ can be a conditional displacement ($O_i\to e^{i\phi} a_i^\dagger + e^{-i\phi} a_i$), a controlled phase-space rotation ($O_i\to  a_i^\dagger a_i$), or even a SNAP gate (with $O_i\to\sum_n \varphi_n \ket{n}\bra{n})$. We assume $U_i(\theta)$ to be native or easily synthesized using the primitives discussed above.

Next, we can synthesize a variant of this gate that is conditioned on information contained in mode $j$. As this information is mediated by an ancillary qubit, it is passed one bit at a time. In that vein, let us define the projector $P_j$ onto a set of Fock states in the subspace $\mathcal{P}$,
\begin{equation}
    P_j = \sum_{n\in \mathcal{P}}\ket{n}\bra{n}_j.
\end{equation}
We can realize a qubit rotation conditioned on this projector with an SQR gate (Box~\ref{box:SQRgate}),
\begin{equation}
    \textrm{SQR}_{j}(\vec{\theta}_{P},\vec{0}) = e^{-i\frac{\pi}{2} X_{\textrm{anc}}P_j}\equiv V_{P_j},
\end{equation}
where $\vec{0} = \{0,0,0,\ldots\}$ and $\vec{\theta}_{P}=\{\theta_0,\theta_1,\theta_2,\ldots\}$ with
\begin{equation}
  \theta_n =
    \begin{cases}
      \pi & \text{if $n \in \mathcal{P}$}\\
      0 & \text{otherwise.}
    \end{cases}       
\end{equation}
Leveraging this gate, it can then be shown that
\begin{equation}
    \begin{split}
        \textrm{CU}_i^{P_j}(\theta) &= e^{-i\theta Z_{\textrm{anc}} O_i P_j} \\
        &= V_{P_j} U_i(\theta/2) V_{P_j}^{\dagger} U_i(-\theta/2).
    \end{split}
    \label{eq:proj_op}
\end{equation}
Therefore, we can synthesize the ``$P$-conditional'' unitary $\textrm{CU}^{P_j}_i$ using two SQR gates and two applications of $U_i$.

While $\textrm{CU}^{P_j}_i$ enacts an operation on the $i$th mode conditioned on a \emph{single} bit of information concerning the $j$th mode, it is possible to synthesize more complex entangling operations by iterating this gate for different choices of $P_j$. As a simple example, let us choose for the unconditioned unitary,
\begin{equation}
    \begin{split}
        U_i(\theta) &= \textrm{SNAP}_i(\vec{\varphi}(f,\theta)) \\ &= e^{-i\theta Z_{\textrm{anc}}f(\hat{n}_i)},
    \end{split}
\end{equation}
where $f(\hat{n}_i) = \sum_{k}(\varphi_k/\theta)\ket{k}\bra{k}_i$ is an arbitrary polynomial in $\hat{n}_i$ whose form depends on the chosen vector of angles $\vec{\varphi}=\{\varphi_0,\varphi_1,\varphi_2,\ldots\}$. Furthermore, let the projector for the $k$th iteration be that of the $k$th Fock state, $P_j^{(k)} = \ket{k}\bra{k}_j$. Then for an appropriately chosen cutoff $N_{\textrm{max}}$, one can realize
\begin{equation}
    \begin{split}
        \textrm{CU}_i^{h(\hat{n}_j)} &= e^{-i\theta Z_{\textrm{anc}}f(\hat{n}_i) h(\hat{n}_j)}\\
        &= \prod_{k=0}^{N_{\textrm{max}}-1}V_{\ket{k}\bra{k}_j} U_i(\theta_k/2) V_{\ket{k}\bra{k}_j}^{\dagger} U_i(-\theta_k/2).
    \end{split}
    \label{eq:CUh}
\end{equation}
Here, $h(\hat{n}_j) = \sum_k(\theta_k/\theta) \ket{n}\bra{n}_j$ is an arbitrary function of the number operator $\hat{n}_j$. The depth of this sequence scales as $O(N_{\textrm{max}})$. However, for certain functions $h(\hat{n}_i)$, a more efficient scheme is possible that requires $O(\log(N_{\textrm{max}}))$ iterations -- see Ref.~\cite{C2QA-LGTpaper} for details. Moreover, we emphasize that while the product $f(\hat{n}_i)h(\hat{n}_j)$ encompasses the synthesis of extremely complex, nonlinear entangling two-mode gates, it is a specialized form of the more general function $f(\hat{n}_i,\hat{n}_j)$. If the aim is to compile the latter more general form, this can also be done. In particular, it can be realized by iterating Eq.~\eqref{eq:CUh} with $f_{\ell}(\hat{n}_i)=\ket{\ell}\bra{\ell}_i$ for $\ell\in[0,N_{\textrm{max}})$, requiring a total depth $O(N_{\textrm{max}}^2)$ sequence.

Finally, we note that the examples above merely scratch the surface of the possible gates that can be analytically synthesized using the primitive in Eq.~\eqref{eq:proj_op}. For example, it is possible to iterate this scheme for more than two oscillators, enabling the systematic synthesis of multi-oscillator entangling gates mediated by an ancilla qubit -- a powerful resource for the OCMM (see Sec.~\ref{sssec:OCMM}). We note that unlike the oscillator-mediated multi-qubit gates of Sec.~\ref{ssec:compilation-entangling}, the qubit-mediated multi-oscillator gates proposed here require the ancillary qubit to be in a known eigenstate of $Z_{\textrm{anc}}$ if one wishes to leave the qubit and oscillators disentangled. However, in line with the HMM (see Sec.~\ref{sssec:HMM}), there may be instances in which this family of gates are useful for hybrid multi-oscillator-qubit operations where the qubit is not ancillary but instead used as a computational degree of freedom. Both of these possibilities are fruitful areas for further research.

\subsection{Systematic Compilation of Single-Qubit-Oscillator Unitaries}
\label{ssec:approximate-1-qubit-oscillator-unitary}
In the previous section, we presented exact analytic compilation schemes for some key compilation tasks.
In this section, we present systematic compilation techniques for single-qubit oscillator unitaries by leveraging methods in DV quantum simulation for compilation purposes. These techniques are analytic and formally exact, but only in the limit of an infinite number of infinitesimal compilation steps. Thus, while the gate sequences are ``human-readable,'' they are not necessarily efficient.  More concretely, Sec.~\ref{sec:trotter-product} discusses how to compile an algorithm into the bosonic ISA using Trotter-Suzuki discretization and product formulas \cite{borneman2012parallel,childs2013product,chen2022efficient,childs2021theory,kang2023leveraging}. We then present generalized version of Linear Combination of Unitary (LCU) techniques \cite{childs2012hamiltonian} on hybrid CV-DV systems in Sec.~\ref{sec:lcu} that can be applied on top of any unitaries compiled from product formulas or QSP to generate more sophisticated algorithms. We focus our discussion in Sec.~\ref{ssec:compilation-bosonic-qsp-qsvt} to hybrid CV-DV quantum signal processing (QSP) by generalizing the original QSP theorem developed for compiling single-qubit dynamics \cite{low2017optimal,GrandUnificationAlgos}.

\subsubsection{Trotter-Suzuki and Product Formulas}
\label{sec:trotter-product}
The first technique we discuss is the Trotter decomposition method and the product formulas, in general\cite{sefi2011how,borneman2012parallel,childs2013product,chen2022efficient,childs2021theory,concentration2021chen}. This section is a continued discussion of Ref. \cite{kang2023leveraging} and is largely pedagogical. The idea is to use Hamiltonian simulation and group commutator expansions to construct particular unitaries. In the case of hybrid oscillator-qubit operators, special attention is needed to take care of the unbounded operator norm for typical operators of the infinite-dimensional oscillator (e.g.,  the position, momentum, and photon number operators).

Specifically, imagine that we wish to add two operators with support on a qubit and an oscillator $A = \hat{b}_1\cdot \vec{\sigma} h_1(\hat{x},\hat{p})$ and $B = \hat{b}_2\cdot \vec{\sigma} h_2(\hat{x},\hat{p})$.  We can approximate the sum of the two operators using a first-order Trotter formula with $r$ time steps via
\begin{equation}
    e^{-i(A+B)} = (e^{-iA/r}e^{-iB/r})^r +O(\max(\|A\|,\|B\|)^2/r),
    \label{trotter-suzuki-1}
\end{equation}
or a second-order Trotter formula
\begin{align}
    e^{-i(A+B)} &= (e^{-iA/2r}e^{-iB/r}e^{-iA/2r})^r\nonumber\\
    &+ O(\max(\|A\|,\|B\|)^3/r^2),
    \label{trotter-suzuki-2SMG}
\end{align}
where the approximation error is given by the norm of the operator (while a more accurate commutator norm may be used to improve the bound~\cite{childs2021theory,PhysRevLett.127.020504}). 

Note that $\hat{x}$ and $\hat{p}$ have infinite operator norm, so $\|A\|, \|B\|$ in general are unbounded. In practice, however, the unitaries are applied to quantum states with bounded energy, thus producing states with bounded energy; therefore, we can loosely view the various relevant operator norms as effectively bounded.  Furthermore, while $\hat x$ and $\hat p$ are unbounded, we do not apply them directly.  Rather, we apply exponentials of $\hat x$ and $\hat p$ which, despite the unbounded nature of their generators, are themselves well-defined unitaries whose eigenvalues lie on the unit circle in the complex plane. Products of such unitaries obey a simple composition law whose analytic form is derived in Sec.~\ref{sssec:composingCDs} and this can be useful both for circuit synthesis and for estimation of errors.  

In addition to the Trotter-Suzuki method to \emph{add} Hamiltonian terms, BCH formulas are also useful to create commutators of two Hamiltonians \cite{sefi2011how,borneman2012parallel,childs2013product,chen2022efficient,childs2021theory,kang2023leveraging,concentration2021chen} 
\begin{equation}
    e^{it A}e^{it B} e^{-itA}e^{-itB} = e^{-t^2 [A,B]} + O(t^3).
    \label{trotter-suzuki-2}
\end{equation}
Using this, the \emph{product} between two operators can be approximated by exploiting the anti-commutation properties of Pauli operations.  This leads us to the following expansion
\begin{align}
    &e^{-i[h_1(\hat{x},\hat{p}) h_2(\hat{x},\hat{p}) + h_2(\hat{x},\hat{p}) h_1(\hat{x},\hat{p})] \otimes \sigma_z} \nonumber \\
    =& \left(e^{-i h_1 \otimes \sigma_x/\sqrt{r}}e^{-i h_2\otimes \sigma_y/\sqrt{r}} e^{ih_1\otimes \sigma_x/\sqrt{r}} e^{ih_2\otimes \sigma_y/\sqrt{r}}\right)^r \nonumber\\
    &+ O(\max(\|h_1\|,\|h_2\|)^2 \min(\|h_1\|,\|h_2\|)/\sqrt{r}).
    \label{trotter-suzuki-3}
\end{align}

Higher-order variants of both the formula for the sum and the formula for the product exist~\cite{suzuki1991general,childs2013product,chen2022efficient,childs2021theory,PhysRevLett.127.020504} and can be used to achieve sub-polynomial scaling of the error with $1/r$ in both cases.  These formulas are of particular interest in our case because gates such as conditional displacements and SNAP gates are controlled by an auxiliary qubit that can be manipulated through the use of single-qubit rotations, enabling us to transform from $\sigma_z$-controlled operations into  $\sigma_x$- or $\sigma_y$- controlled operations.  This ability allows us to leverage a single qubit to build arbitrary polynomial interactions in $\hat{x}$ and $\hat{p}$ (or $a^\dagger$ and $a$) on the qumode and thus achieve universality as discussed in Sec.~\ref{sec:instruction_set}. 

More formally, given an arbitrary degree-$d$ hermitian polynomial $h_d(\hat{x}, \hat{p})$ as in Eq.~\eqref{h-rep-polynomial}, we would like to find a sequence of $L_d$ controlled-displacements $\{ U_j \}$ ($j = 1, 2, \cdots, L_d$) such that
\begin{align}
    U_{L_d} U_{L_d-1} \cdots U_{2} U_{1} \approx e^{i \hat{b}_d \cdot \vec{\sigma} h_d(\hat{x}, \hat{p})},
    \label{trotter-compilation-1}
\end{align}
where $U_j = e^{i \hat{c}_j \cdot \vec{\sigma} (\alpha_j a^\dagger - \alpha_j^* a)}$ and $\hat{b}_d$ is an arbitrary unit vector. One may construct the sequence of $\{ U_j \}$ in reverse by multiplying a controlled displacement on the left of both sides of Eq.~\eqref{trotter-compilation-1}, such that the resulting unitary has a simpler generator.

\subsubsubsection{Example: Jaynes-Cummings Interaction.}
As a simple example, let us consider the approximate synthesis of the Jaynes-Cummings and anti-Jaynes-Cummings gates (see Table \ref{tab:gates-qubit-osc}),
\begin{align}
    \mathrm{JC}(\theta) &= e^{-i\theta (\sigma_- a^\dagger  + \sigma_+ a)},\\
    \mathrm{AJC}(\theta) &= e^{-i\theta (\sigma_+ a^\dagger  + \sigma_- a)},
\end{align}
where for simplicity we have restricted the parameter $\theta$ to be real.  (The reader is also directed to Sec.~\ref{sssec:Rabi} which discusses the Rabi model, which is the parent of these models.) The microscopic physical Rabi model Hamiltonian in circuit QED (see App.~\ref{app:PhysImp}) contains the sum of the JC and AJC Hamiltonians, and the corresponding unitaries are individually accessible within separate rotating wave approximations.  Hence we list these as natively available within the Sideband ISA.  

Our goal here is to approximately synthesize these sideband unitaries within the phase-space ISA and the Fock-space ISA.
 To that end, we first note that they can be cast into the form 
 \begin{align}
     \mathrm{JC}(\theta) &= e^{-i\theta(B-A)},\\
     \mathrm{AJC}(\theta) &= e^{-i\theta(A+B)},
 \end{align}
where
\begin{align}
        A &= \hat p\sigma_y,\label{eq:hatpsigy}\\
        B &= \hat x\sigma_x.\label{eq:hatxsigx}
\end{align}
Importantly, both $A$ and $B$ generate conditional displacements that, up to single-qubit rotations, can be directly realized either natively in the phase-space ISA or through a combination of controlled-parity gates and unconditional displacements in the Fock-space ISA. Thus, both $\mathrm{JC}(\theta)$ and $\mathrm{AJC}(\theta)$ can be approximately implemented within both of these ISAs via Eq.~(\ref{trotter-suzuki-1}) through an alternation of coordinate and momentum displacements, conditioned on orthogonal qubit bases. We give a comparison of first-order and second-order Suzuki-Trotter approximations to the unitary $\textrm{AJC}(\theta)$ using the phase-space ISA gates in App.~\ref{sec:circuit-compare}.

\subsubsubsection{Example: Conditional Squeezing Operation.}
\label{sssec:conditionalsqueeze}
Using Eq.~(\ref{trotter-suzuki-2}) with $t=\sqrt{\zeta}$ and $A,B$ as defined in Eqs.~(\ref{eq:hatpsigy}-\ref{eq:hatxsigx}) yields (for the case of $\zeta$ real and small)
\begin{align}
e^{i\sqrt{\zeta} A}e^{i\sqrt{\zeta} B} e^{-i\sqrt{\zeta}A}e^{-i\sqrt{\zeta}B} &\approx e^{-\zeta [A,B]}\nonumber\\
&\approx e^{i\zeta Z(\hat x\hat p + \hat p\hat x)},\label{eq:Trottersqueezing}
\end{align}
which matches the controlled squeezing unitary $\mathrm{CS}(\zeta)$ defined in Table \ref{tab:gates-qubit-osc}. For an illustration of how the Wigner function evolves after each successive conditional displacement, see Fig.~\ref{fig:composing_conditional_disp}(d). See also Sec.~\ref{sssec:composingCDs}, where we provide an analytical decomposition for sequences of conditional phase-space displacements of the form in Eq.~(\ref{eq:Trottersqueezing}), and provide intuition on the resulting composite operation when $\zeta$ is not infinitesimal. Separately, we note that in Ref.~\cite{Hastrup_Squeezed_State_QSP}, the authors numerically optimize circuits for unconditional squeezing using a sequence of alternating conditional displacements and momentum boosts. Ref.~\cite{EickbuschECD} experimentally demonstrates conditional squeezing using a numerically optimized sequence of conditional displacements and momentum boosts. Refs.~\cite{ayyash2024drivenmultiphotonqubitresonatorinteractions,ayyash2025multimodequbitconditionaloperationsgeneralized} present an analog scheme for producing conditional squeezing by driving a qubit/resonator system at twice the resonator frequency, while Ref.~\cite{hope2025preparationconditionallysqueezedstatesqubitoscillator} presents a scheme based on driving an ancilla qubit at the resonator frequency.

\subsubsubsection{Example: (Conditional) Cross-Kerr Interaction.}
While Trotter-Suzuki and BCH formulas are particularly useful for the compilation of unitaries acting on a single qubit-oscillator block, we emphasize that these methods are quite general and can also be used to synthesize non-trivial entangling gates. As a simple demonstration, we approximately synthesize a conditional cross-Kerr interaction between a pair of oscillators labeled $a_1$ and $a_2$. In particular, we envision the connectivity illustrated in Fig.~\ref{fig:hardware_layout}a, with the $i$th mode dispersively coupled to the $i$th qubit, and a tunable beam-splitter between the two modes of interest. 
Utilizing Eq.~(\ref{trotter-suzuki-2}), we make the choice $A = X_1 a_1^\dagger a_1$ and $B = Y_1 a_2^\dagger a_2$, with $X_1$ and $Y_1$ Pauli operators for qubit 1, $U_{A} = \textrm{exp}(-i \frac{\eta}{2} A)$ corresponding (up to single-qubit rotations) to a conditional cavity rotation gate $\mathrm{CR}(\eta)$, and similarly $U_{B} = \textrm{exp}(-i \frac{\eta}{2} B)$ is independently compiled by conjugating $U_A$ by appropriate single-qubit rotations and by a beam-splitter that SWAPs the two modes. In all, this yields the desired unitary up to an error scaling as $\eta^3$,
\begin{equation}
    U_A^\dagger U_B^\dagger U_A U_B = e^{-i\frac{\theta}{2} Z_1 a_1^\dagger a_1 a_2^\dagger a_2} + O(\eta^3),
    \label{eq:ccrosskerr}
\end{equation}
where $\theta =  \eta^2$.

By setting the auxiliary qubit to be in $|0\rangle$ (the +1 eigenstate of $Z$), this cross-Kerr interaction can be made useful in quantum simulations of the real-time dynamics (Hamiltonian evolution) of repulsively interacting bosons, for example.  The fact that the interaction is conditional on the auxiliary qubit state means that we can use the phase kick-back onto the qubit to measure the mean value of the interaction Hamiltonian term in a trial variational ground state produced by a VQE \cite{Peruzzo2014} or QAOA \cite{farhi2014quantum} algorithm. This conditional cross-Kerr interaction can also be useful in compiling the fSim gate given in Eq.~(\ref{eq:fSimFock}).

\subsubsection{LCU}
\label{sec:lcu}
The linear combination of unitaries (LCU) technique has been used in the context of DV systems for efficient Hamiltonian simulation \cite{childs2012hamiltonian}. In the standard LCU approach, a DV register composed of qubits is used to store the amplitudes defining the linear combination of unitaries. A final measurement on the auxiliary qubit register imposes a linear combination of multiple unitaries on a system DV register. 
Note that since an LCU is in general not a unitary operation, it can be realized only non-deterministically through the measurement of auxiliary variables.  In some cases it may be possible to use amplitude amplification techniques \cite{berry2015simulating,low2017hamiltonian,low2017optimal} to bring the measurement success probability close to unity.  

\begin{figure}[htbp]
\centering
    \includegraphics[width=1\linewidth]{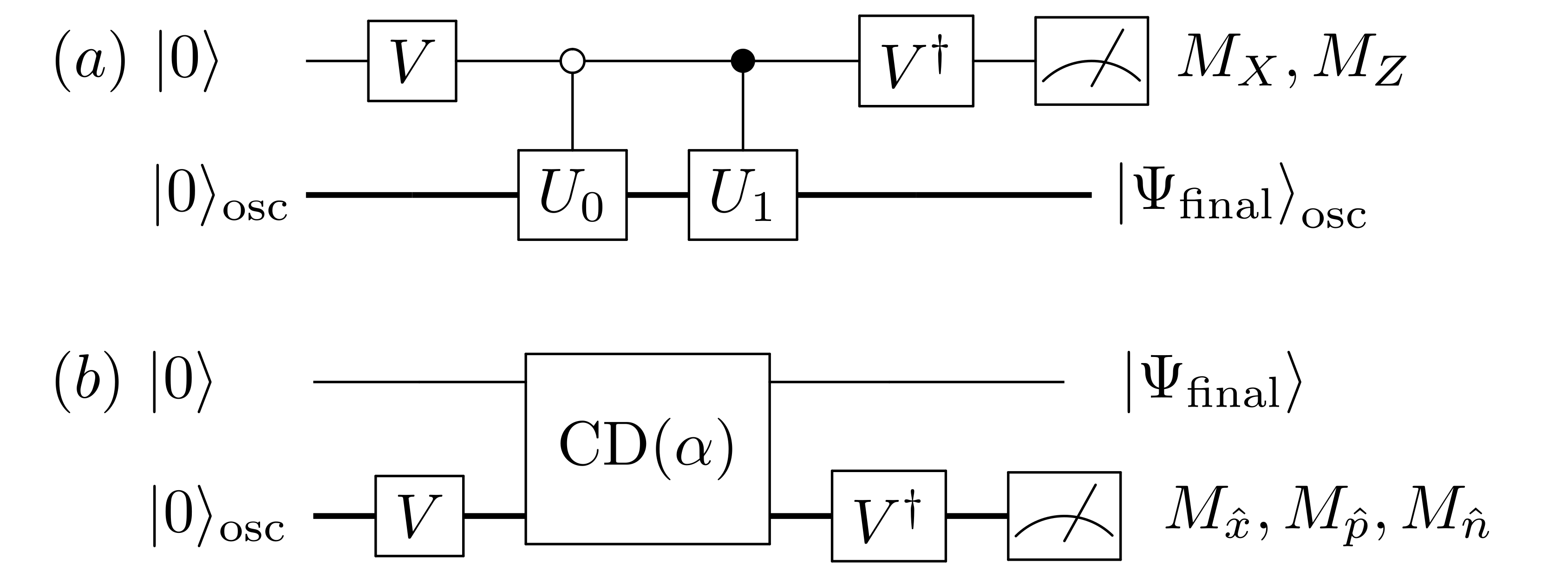}
            \caption{LCU in hybrid bosonic quantum processors using a) qubits or b) oscillators as auxiliary registers to store the linear combination amplitudes. The scheme can be easily generalized to multiple qubits and oscillators. In (a), the measurement on the qubit that post-selects the oscillator final state $\ket{\Psi_{\rm final}}_{\rm osc}$ can be made (for example) in either the $Z$ ($M_Z$) or $X$ ($M_X$) basis. Similarly in (b), the post-selection measurement on the oscillator can be made (for example) via homodyne measurement in one of the oscillator quadrature bases $M_{\hat{x}}, M_{\hat{p}}$ or the photon number basis $(M_{\hat{n}})$, each of which yields a different final state $\ket{\Psi_{\rm final}}$ on the qubit.
            }
            \label{fig:lcu-schematic}
\end{figure}

Here, we consider a generalization of LCU to hybrid bosonic quantum processors, where either the qubits or the oscillators can serve as auxiliary registers to store the linear combination coefficients in the LCU. Fig.~\ref{fig:lcu-schematic}a illustrates such a scheme for a single qubit-oscillator system, and the multi-qubit generalization for the auxiliary register immediately follows.

Using oscillators as auxiliary registers enables the implementation of a continuous integral of unitaries on the system register, instead of a discrete sum of unitaries as in the original formulation of LCU. In fact, we have already seen a prototypical example of this concept for the conditional displacement gate in Fig.~\ref{fig:CD-control-equiv}.  Ref.~\cite{bell2025codesigningeigensingularvaluetransformation} gives another powerful example of a continuous-variable LCU block-encoding of a Gaussian imaginary time evolution spectral filter and applies this to the task of eigenstate preparation in quantum spin models.

Fig.~\ref{fig:lcu-schematic}b shows that in the case of a single qubit and oscillator, a state preparation unitary $V$ can be applied, followed by an oscillator-controlled qubit rotation and the inverse preparation unitary $V^\dagger$. A linear combination of a continuous family of  qubit rotations can be applied providing that a measurement on the oscillators is performed for post-selection. In the case where the oscillator is projected to a position eigenstate $\ket{x''}$, the resulting qubit state is given by
\begin{align}
    \ket{\psi_{\rm final}} = \int_{-\infty}^{\infty} \mathrm{d}x ~c_{x''}(x) ~\hat{R}(x) \ket{0}
\end{align}
where 
\begin{align}
    c_{x''}(x) = \Psi(x) \bra{x''} V^\dagger \ket{x}
\end{align}
is an amplitude parameterized by a position $x$ when the oscillator is projected to $x''$, $\hat{R}(x)$ is a single qubit rotation parameterized by $x$, and $\Psi(x) = \bra{x} V \ket{0}_{\rm osc}$ is the wave function of the oscillator resulting from the preparation unitary $V$. Note that the oscillator-controlled qubit rotation in the middle of Fig.~\ref{fig:lcu-schematic}b applies the qubit rotation $\hat{R}(x)$ conditioned on the position $x$ of the oscillator.
\begin{align}
    \ket{x}_{\rm osc} \ket{0} \rightarrow \ket{x}_{\rm osc} \hat{R}(x) \ket{0}.
\end{align}

\begin{figure}[t]
            \centering
             \includegraphics[width=0.45\textwidth]{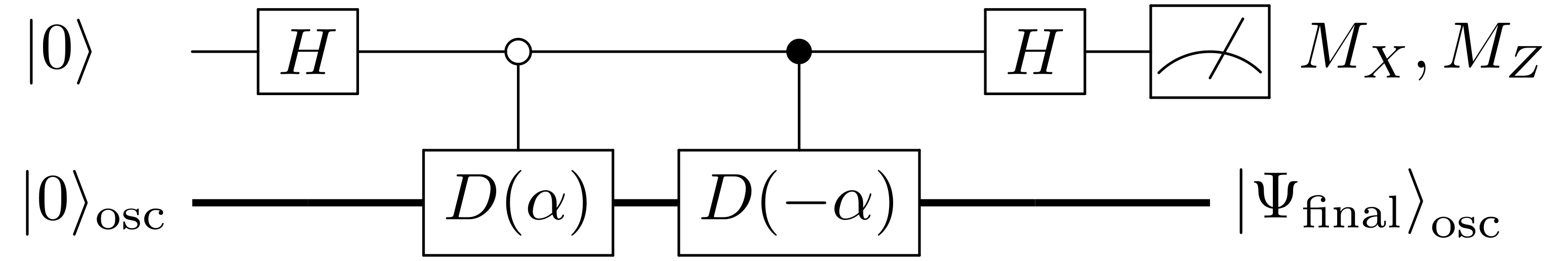}
             \includegraphics[width=0.5\textwidth]{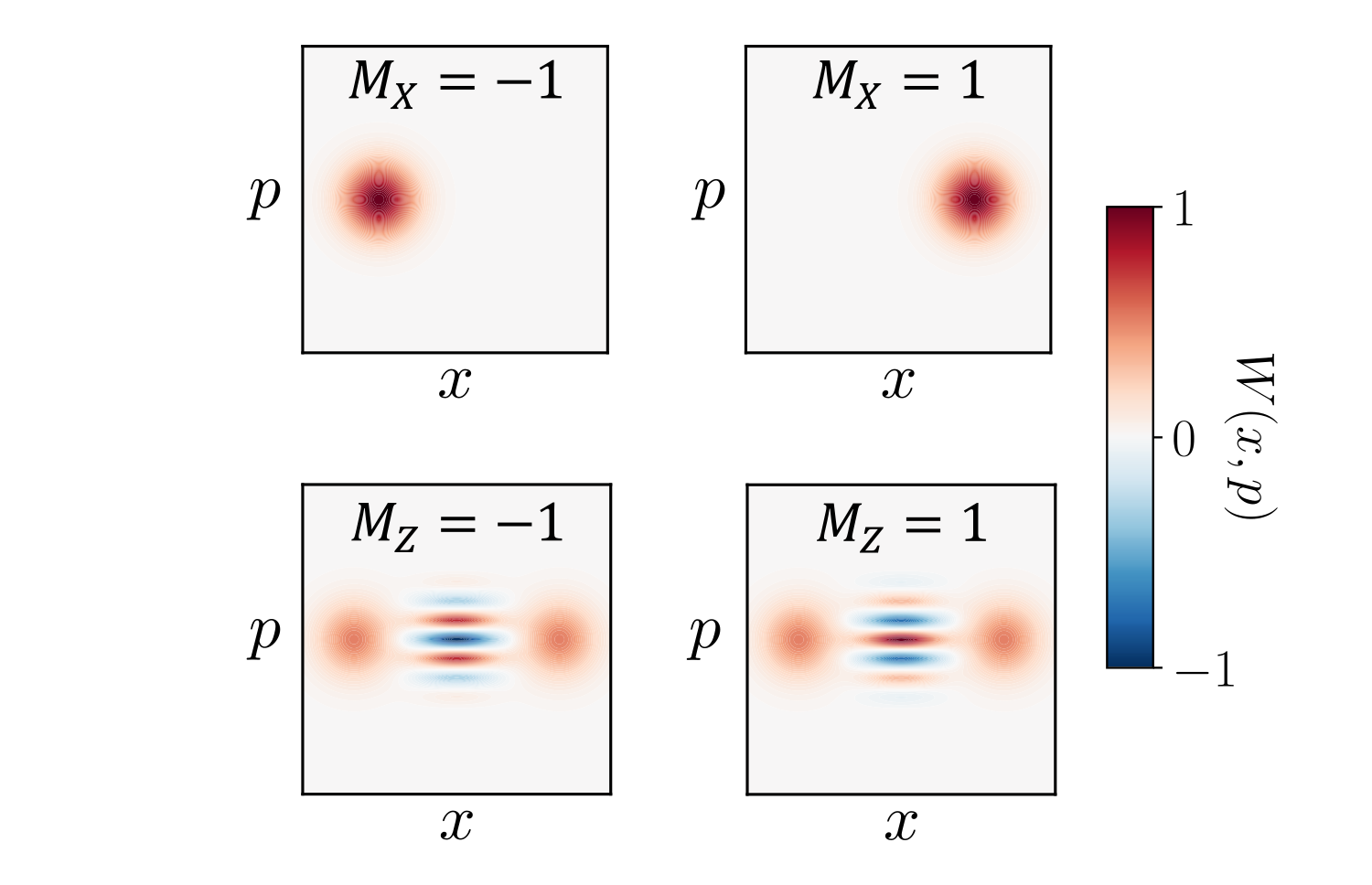}
            \caption{Example of the LCU circuits in panel (a) of Fig. \ref{fig:lcu-schematic} that produces simply a displaced coherent state $|\pm\alpha\rangle$ conditioned on the ancilla qubit measurement result in the $X$ basis (Wigner functions shown in the middle row), or produces a cat state if the qubit measurement is performed in the $Z$ basis (Wigner functions shown in the lower row).  Notice (from the color of the central interference fringe) that the parity of the cat is even (lower right panel) if the $Z$ measurement yields +1 and odd (lower left panel) if the $Z$ measurement yields -1.}
            \label{fig:lcu-cat-example}
\end{figure}

As a simple example let us consider the non-deterministic creation of an oscillator cat state by application of the controlled unitary given in Eq.~(\ref{entangled_cat}).  The circuit is illustrated in Fig.~\ref{fig:lcu-cat-example}.  Measurement of the qubit in the Z basis produces an ideal even or odd parity cat depending on the measurement result.  Measurement of the qubit in the $X$ basis produces a coherent state $|\pm\alpha\rangle$ instead of a cat state. These hybrid CV-DV circuits were implemented and the resulting theoretical Wigner functions shown in Fig. \ref{fig:lcu-cat-example} were obtained using Bosonic Qiskit \cite{Biskit,BiskitGitHub}. For experimental implementation of this protocol in circuit QED see Ref. \cite{Vlastakis2015}.

\subsubsection{Quantum Signal Processing for Hybrid Systems}
\label{ssec:compilation-bosonic-qsp-qsvt}

Quantum signal processing (QSP) and its cousin, the Quantum Singular Value Transformation (QSVT) \cite{low2017optimal,GrandUnificationAlgos,Gilyen2019}, are the second family of techniques that we consider for programming hybrid devices via unitary synthesis. The intellectual origins of the QSP concept are found in the robust spin rotation control techniques developed by the nuclear magnetic resonance (NMR) community. Today, QSP is the unifying principle behind many existing quantum algorithms for qubit-based platforms \cite{GrandUnificationAlgos} with many applications -- see, for example, Ref.~\cite{Stamatopoulos2024derivativepricing}). 

In this section, we describe how to extend these concepts from control problems where the spin rotation angles are \emph{classical} control parameters to the case where they are \emph{quantum} operators -- namely, the position and momentum operators of harmonic oscillators. Ultimately, this enables the extension of QSP unitary synthesis techniques to hybrid qubit-oscillator systems. We emphasize that this section assumes knowledge of qubit-based QSP; for a brief introduction, see App.~\ref{app:qsp-qsvt}.

Before formally defining hybrid CV-DV QSP, we will, as a first example, use a well-known composite pulse sequence from NMR, namely, $\mathrm{BB1}_{90,y}$~\cite{WimperisRobust,GrandUnificationAlgos} to illustrate how QSP can be applied to oscillator-qubit control. The BB1$_{90,y}$ (see also App.~\ref{app:qsp-qsvt}) composite pulse can be written as, \begin{align}
    \mathrm{BB1}_{90,y} 
    &= Z(\phi_0)\prod_{j=1}^d W_x(\theta)Z(\phi_j) \label{bb1a-main-text} \\
    &= e^{-i\frac{\pi}{2}\vec h(\theta)\cdot\vec \sigma},
    \label{bb1c-main-text}
\end{align}
where $ \vec h(\theta)=\big(h_0(\theta),h_x(\theta),h_y(\theta),h_z(\theta)\big)
$ is a vector whose components are four functions of the rotation angle $\theta$ and $\vec\sigma=(\sigma_0,\sigma_x,\sigma_y,\sigma_z)$.
The striking similarity between Eq.~(\ref{bb1c-main-text}) and the general qubit-oscillator hybrid Hamiltonian in Eq.~(\ref{eq:osc-qubit-control-H}) for controlling a hybrid qubit system inspires us to think about a method to promote $\theta$ from a noisy classical variable to a quantum operator  $
\hat{\theta}$ such as $\hat x$  acting on an oscillator.

Consider for example, the following controlled displacement operation (Box~\ref{Box:c-displacement}, Table \ref{tab:ISA_overview})
\begin{align}
    e^{-i\frac{k}{2} \hat{x}  \sigma_z} = 
    \begin{bmatrix}
        e^{-i\frac{k}{2}\hat{x}} & 0\\
        0 & e^{i\frac{k}{2}\hat{x}}
    \end{bmatrix} := W_z,
    \label{wz_ecd}
\end{align}
which imparts a qubit-dependent momentum boost on the oscillator.  Importantly, we can also interpret this as a rotation of the qubit about the $z$ axis by an angle linearly proportional to the position of the oscillator
\begin{align}
    W_z(\hat\theta)&=e^{-i\frac{\hat \theta}{2}\sigma_z}, \label{eq:WzCD}\\
    \hat\theta &\equiv k \hat x.\label{eq:hatthetaquantumangle}
\end{align}
Since the oscillator position is a quantum operator, it has quantum fluctuations analogous to the classical fluctuations of the control parameter $\theta$ we discussed in connection with composite pulses above.  Remarkably, as long as we deal only with a single operator $\hat x$ or $\hat p$ (or a fixed linear combination of them) then all of the rotation angles commute with each other, and the machinery of QSP developed for classical control variables can be applied without modification.   

The noise model we used in the classical BB1$_{90}$ composite pulse example was that we have perfect control over the axis of rotation on the Bloch sphere, but imperfect control over the rotation angle.  For the quantum case, we assume that we have perfect control over all qubit rotation axes and angles when they are created using standard classical control pulses directly applied to the qubit.  However, when we rotate the qubit by an angle 
$\hat\theta$ associated with the oscillator position, the quantum fluctuations of the oscillator play the role of noise in the rotation angle.  Within this model, we can convert the controlled momentum boost 
$W_z(\hat\theta)$ into the form compatible with the $\mathrm{BB1}_{90,y}$ protocol in Eq.~(\ref{bb1a-main-text})
\begin{align}
W_x(\hat\theta)&=R_{+\frac{\pi}{2}}\left (\frac{\pi}{2} \right ) \, W_z(\hat\theta)\, R_{+\frac{\pi}{2}} \left(-\frac{\pi}{2} \right) \nonumber\\
&=e^{-i\frac{k}{2} \hat{x}  \sigma_x} = 
    \begin{bmatrix}
       0 & e^{-i\frac{k}{2}\hat{x}} \\
     e^{i\frac{k}{2}\hat{x}} & 0
    \end{bmatrix} \label{eq:quantumWx}
\end{align}

As a simple illustration of the hybrid CV-DV QSP procedure, we use the $\mathrm{BB1}_{90,y}$ protocol within the phase-space ISA (Sec.~\ref{sssec:phasespaceISA}) to deterministically create a Schr\"odinger cat state in an oscillator.  Two-legged Schr\"odinger cat states are superpositions of coherent states (defined in Sec.~\ref{sssec:fock-basis})
$\ket{\pm\alpha}$ equidistant from the origin in phase space 
\begin{equation}  
    \ket{C}_\alpha=\mathcal{N} (\ket{\alpha}\pm\ket{-\alpha}),
\end{equation}
where $\mathcal{N} = \frac{1}{\sqrt{2+2e^{-2|\alpha|^2}}}$ and the $\pm$ sign determining whether the photon number parity is even (+) or odd (-).  This is useful for encoding logical information and for quantum sensing of displacements. For simplicity, we have chosen 
$\alpha$  to be real so that the cat is aligned along the position axis of phase space. 

\begin{figure}[htb]
    \centering
    \includegraphics[width=0.45\textwidth]{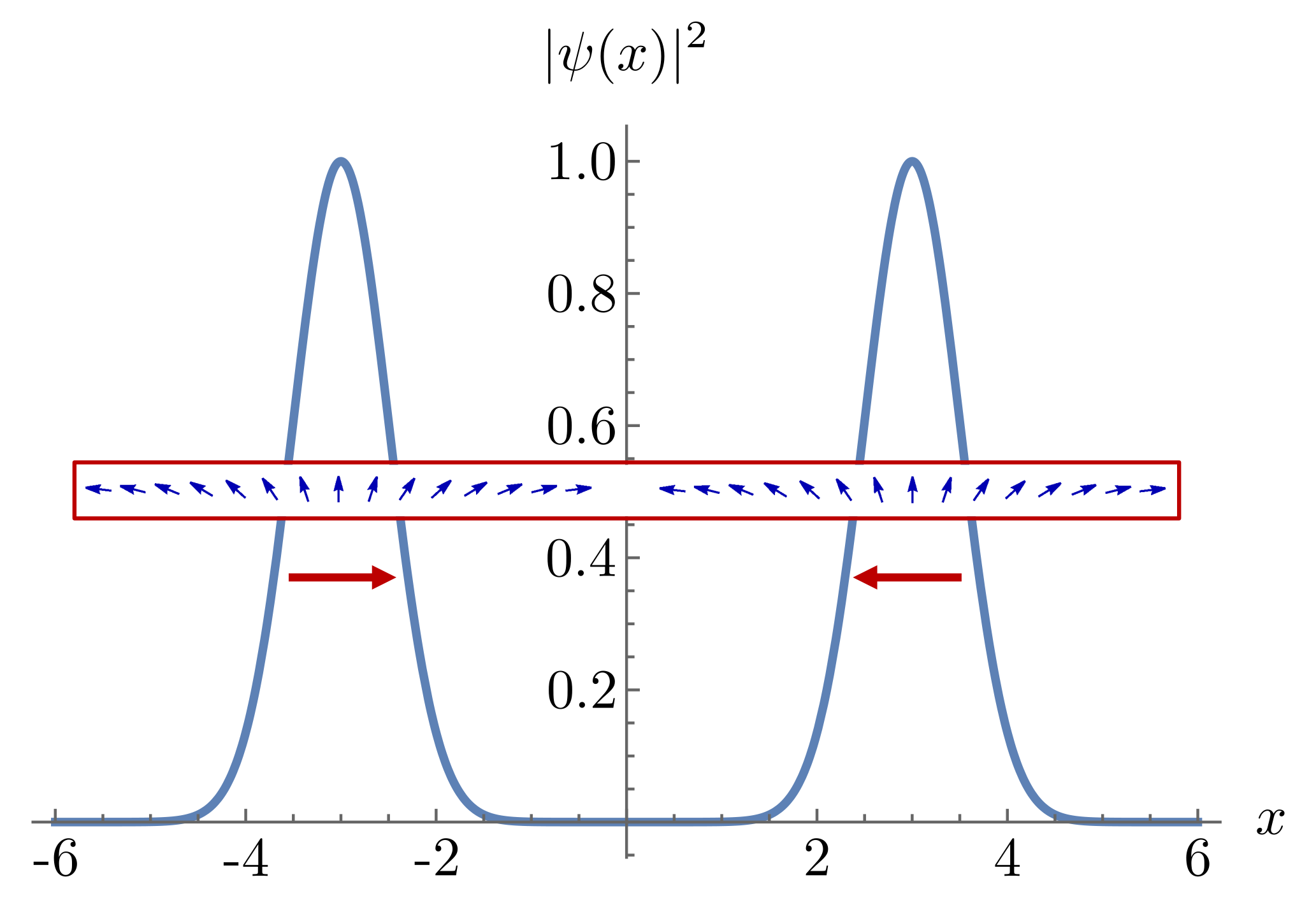}
    \caption{Probability density $|\psi(x)|^2$ for a cat of size $\alpha=3$ illustrating the quantum fluctuations of the position about the peak positions $x=\pm\alpha$. Large red arrows indicate the initial spin entanglement with position based on Eq.~(\ref{entangled_cat}). Blue arrows in the red box indicate the spin configuration after the application of the single qubit-controlled momentum boost (or equivalently an $x$-dependent qubit rotation) given in Eq.~(\ref{eq:simplerotation}).  In the position regime where the probability density is high, the spin orientation is approximately constant, indicating that the spin and the oscillator are largely (but not perfectly) disentangled.}
    \label{fig:QSPcat}
\end{figure}

Suppose we would like to prepare the two-legged cat state from the ground state of the qubit and oscillator without any measurement. Using conditional displacement ($e^{+i2\alpha \hat{p}\otimes\sigma_x}$, in Wigner units) from the phase-space instruction set, the initial hybrid state can be entangled with the qubit as follows
 \begin{equation}
    e^{-i2\alpha \hat{p}\otimes\sigma_x}\ket{0}_\mathrm{vac}\otimes\ket{0}=\frac{1}{\sqrt{2}}\left[ \rule{0pt}{2.4ex} |\alpha\rangle_\textrm{osc}|-\rangle + |-\alpha\rangle_\textrm{osc}|+\rangle\right]
    \label{entangled_cat}
\end{equation}
Here, $\ket{\pm}=\frac{1}{\sqrt{2}}(\ket{0}\pm\ket{1})$ are the qubit $\sigma_x$ eigenstates. For large enough  $\alpha$, we can safely assume that the Gaussian wave functions of $\ket{\pm\alpha}$ do not overlap as shown in Fig.~\ref{fig:QSPcat}. We can also see directly from this plot that the probability distribution for the position of the oscillator is strongly peaked near $=\pm\alpha$, but as expected, has quantum fluctuations around those nominal positions.

Now, to create an even-parity cat state, the qubit (see the red arrows in Fig.~\ref{fig:QSPcat}) needs to be disentangled from the oscillator state by a rotation about the $y$ axis by an angle $+\pi/2$ if the oscillator position, $\hat x$, is positive and $-\pi/2$ if the position is negative to create the target cat state
\begin{align}
    \frac{1}{\sqrt{2}}\left[ \rule{0pt}{2.4ex} |\alpha\rangle_\textrm{osc}+ |\! -\! \alpha\rangle_\textrm{osc}\right]\otimes |0\rangle.
\end{align}
This transformation is (essentially) unitary so long as the coherent states have negligible overlap.
As noted above, conditional momentum boosts can be viewed as a conditional rotation of the qubit based on the position of CV state. Thus, as a first approximation, we could use 
\begin{align}
e^{-i\frac{k}{2}\hat{x}\otimes\sigma_y}&=Z \left(+\frac{\pi}{2} \right) \, W_x(\hat\theta) \, Z \left(-\frac{\pi}{2} \right),\label{eq:simplerotation}
\\
\mathrm{with}\;\;\; k&=\frac{\pi}{2\alpha}\label{eq:rotrate}
\end{align}
to rotate the qubits correctly at positions $x = \pm\alpha$ corresponding to the peaks in the oscillator position probability distribution. The blue arrows inside the red box in Fig.~\ref{fig:QSPcat} show how the initial spin polarisation of the qubit (shown as red arrows) changes upon applying this oscillator position-dependent rotation to a (moderately) large cat state. 
 
 We see that the controlled momentum boost has approximately disentangled the qubit from the oscillator, thus moving the state closer to the target.  The disentanglement is not perfect however because for 
$|x|\neq\alpha$, there is some under-rotation or over-rotation due to the (quantum) uncertainty in the value of the position of the oscillator in the coherent states $\pm\ket{\alpha}$. The position uncertainty visible in Fig.~\ref{fig:QSPcat}  is set by the vacuum fluctuations and is independent of the displacement $\alpha$, while we see from Eq.~(\ref{eq:rotrate}) that the rate of spin rotation with position is inversely proportional to 
$\alpha$. Thus we expect that the rotation errors to decrease and thus the fidelity to the target state should improve with increasing $\alpha$.  Indeed, we find that the fidelity obeys
\begin{align}
   F_{\mathrm{cat}} \approx 1-\frac{\pi^2}{64\alpha^2}
\end{align}
for asymptotically large $\alpha$.

To increase this fidelity, we can directly apply the $\mathrm{BB1}_{90,y}$ protocol in Eq.~(\ref{bb1a-main-text}) to (largely) eliminate the over/under spin-rotation errors.  The only change that needs to be made is to replace the classically noisy rotation $W_x(\theta)$ in Eq.~(\ref{bb1a-main-text}) with the quantum noisy rotation $W_x(\hat\theta)$ in Eq.~(\ref{eq:quantumWx}).  The classical $\mathrm{BB1}_{90,y}$ protocol still works because the only quantum rotation angle is $\hat\theta$ (defined in Eq.~(\ref{eq:hatthetaquantumangle})) and so there are no non-commuting operators in the quantum version of the $\mathrm{BB1}_{90,y}$ sequence.

\begin{figure}[t]
    \centering
    \includegraphics[width=0.45\textwidth]{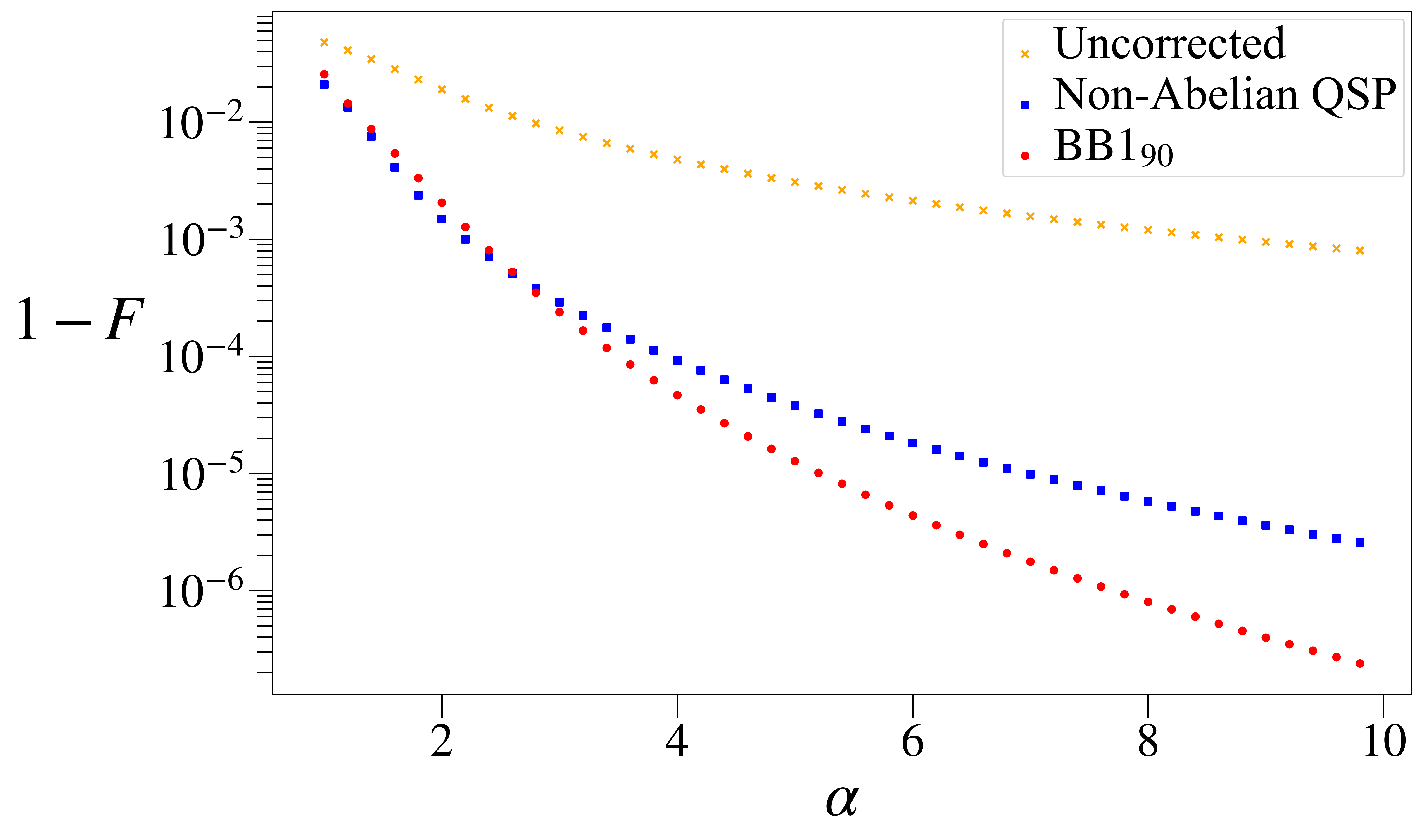}
    \caption{Infidelity of cat state preparation by unentangling the qubit from oscillator-state in Eq.~\eqref{entangled_cat} using the $\textrm{BB1}_{90}$ sequence of commuting variables and non-abelian QSP correction. For various cat sizes $\alpha$, we plot the infidelity of the output hybrid state with the desired state $\ket{C}_{\alpha}\otimes\ket{0}$. Here, `uncorrected' refers to the simple protocol using the operation described in Eq.~(\ref{eq:simplerotation}). The `non-abelian QSP' and `BB1$_90$' curves, respectively denote corrections over the simple protocol using operations described in Eq.~(\ref{eq:bb1y}) and Sec.~\ref{sec:bivariable-qsp}.}
    \label{fig:QSP_cat2}
\end{figure}

Fig.~\ref{fig:QSP_cat2} shows a comparison of the fidelities 
to the target state for three different state preparation protocols: the simple protocol in Eq.~(\ref{eq:rotrate}), the $\mathrm{BB1}_{90,y}$ composite pulse QSP protocol, and a bi-variable or `non-abelian QSP' protocol~\cite{singh_towards_2025} which is discussed further below.  We see that the $\mathrm{BB1}_{90,y}$ protocol is very efficient at correcting the spin rotation errors associated with quantum fluctuations in the oscillator position.

{\bf Formal Exposition of QSP for hybrid CV-DV systems.}  We turn now to a more formal exposition of the extension of QSP to the case where the rotation angles can be quantum operators acting on an oscillator, focusing on the unitary picture (Sec.~\ref{sssec:unitarypicture}) rather than the Hamiltonian picture of 
Eq.~(\ref{bb1c-main-text}) and Sec.~\ref{sssec:hermitianpicture}.  We explore here the single commuting variable case (e.g., all rotation angles proportional to $\hat x$) and in the next subsection, we consider how to extend QSP to the case of non-commuting rotation angles in which both variables $\hat x$ and $\hat p$ are involved.
One key intuition for extending qubit QSP to hybrid qubit-oscillator systems is to realize that the controlled displacement operation (conditional rotation) in the phase-space ISA in Table \ref{tab:ISA_overview} on the qubit-oscillator system can be viewed as a block-encoding of the displacement operator on the $2 \times 2$ matrix of the qubit subspace. 
In the spirit of QSP and QSVT, combining any block-encoding and single-qubit rotations that \emph{process} the block-encoded operator, a polynomial transformation on the block-encoded operator can be realized.

Viewing the controlled displacement $W_z$ in Eq.~(\ref{eq:WzCD}) as a block-encoding of the operator $w = e^{-i\frac{k}{2}\hat{x}}$, the following QSP sequence comprising $d$ steps (described in more detail in App.~\ref{app:qsp-qsvt}) produces an arbitrary polynomial transformation on $w$ 
\begin{align}
    e^{i\phi_0 \sigma_x} \prod_{j = 1}^d W_z e^{i\phi_j \sigma_x} =
    \begin{bmatrix}
        F(w) & iG(w) \\
        iG(w^{-1}) & F(w^{-1})
    \end{bmatrix}
    \label{qsp_wz}
\end{align}
where here we work in the z-rotation convention in which the roles of $x$ and $z$ rotations are interchanged relative to the $\mathrm{BB1}_{90,y}$ example above.  We again take the operator ordering to be left to right and
\begin{align}
    F(w) &= \sum_{n = -d}^d f_n w^n = \sum_{n = -d}^d f_n e^{-i\frac{nk}{2}\hat{x}} := f(\hat{x}), \nonumber \\
    G(w) &= \sum_{n = -d}^d g_n w^n = \sum_{n = -d}^d g_n e^{-i\frac{nk}{2}\hat{x}} := g(\hat{x}).
    \label{fwgw-poly}
\end{align} 
is a Laurent polynomial expansion of arbitrary functions $f(\hat{x})$ and $g(\hat{x})$ in canonical position space, and the coefficients $f_n, g_n \in \mathbb{R}$ can be computed by Fourier series of $f(x)$ and $g(x)$ under the discrete plane-wave basis $\{ e^{-i\frac{nk}{2} x} \}$ for $ -d < n < d$ and for a given fixed $k$. In addition, unitarity requires $F(w) F(w^{-1}) + G(w) G(w^{-1}) = I$, and $F(w), G(w) \in \mathbb{R}[w, w^{-1}]$ with degree $d$, where $\mathbb{R}[w, w^{-1}]$ represents the set of polynomials of $w, w^{-1}$ with real coefficients. 

Note that even though $k\hat x$ is a quantum operator rather than a classical rotation angle, this construction is the same as the $W_z$ convention of single-qubit QSP, and therefore the QSP theorem holds. The only difference is that the position variable of the oscillator is hidden inside a complex variable $w$. Such block-encoding of a unitary operator has also been discussed in Ref.~\cite{QSPGroundStateProjection_PRXQuantum.3.040305} to prepare the ground state and estimate the energy in standard DV quantum computation by transforming the eigenvalues of the encoded unitary matrix.

More concretely, the form of $F(w)$ and $G(w)$ as given in Eq.~\eqref{fwgw-poly} can be directly verified by explicitly evaluating the left-hand side of Eq.~\eqref{qsp_wz}. Conversely, given any Laurent polynomial $F(w)$ of degree-$d$ written in the form in Eq.~\eqref{fwgw-poly}, there always exists a set of phase angles $\{ \phi_0, \phi_1, ..., \phi_d \}$ such that the gate sequence as constructed in Eq.~\eqref{qsp_wz} block-encodes $F(w)$. See Ref.~\cite{chao2020finding} for a constructive proof on the existence of the QSP phases for an arbitrary Laurent polynomial $F(w)$. By comparing Eq.~\eqref{qsp_wz} and \eqref{rep-unitary}, we may readily identify the polynomials $A_0(w) = \frac{F(w) + F(w^{-1})}{2}, A_1(w) = \frac{G(w) + G(w^{-1})}{2}, A_2(w) = i \frac{G(w) - G(w^{-1})}{2}, A_3(w) = \frac{F(w)-F(w^{-1})}{2}$, corresponding to $I, \sigma_x, \sigma_y, \sigma_z$, respectively. 

This simple generalization of QSP to hybrid bosonic systems allows us to implement an arbitrary unitary operation on the oscillator if that unitary is the exponential of a purely imaginary (polynomial) function of $\hat{x}$. This can immediately be changed to a function only of momentum $\hat p$, 
by simply using a qubit-dependent position kick $W_z = e^{-i\frac{\lambda}{2} \hat{p} \sigma_z}$ in the construction of Eq.~\eqref{qsp_wz}. More generally, for an operator taking a fixed linear combination of $\hat{x}$ and $\hat{p}$, $\frac{k}{2} \hat{x} + \frac{\lambda}{2} \hat{p}$, one may use $W_z = e^{-i (\frac{k}{2} \hat{x} + \frac{\lambda}{2} \hat{p} )\sigma_z}$ in the construction. The ability of hybrid CV-DV QSP to prepare a wide range of oscillator states is useful for a variety of applications. For example, in Ref.~\cite{qspi2023}, hybrid QSP has been used to construct an interferometer for quantum sensing applications, where Heisenberg-like scaling is achieved for performing binary decisions on a displacement channel in the single-shot limit. It is also shown that such decision protocols can be chained together for efficient parameter estimation.

\subsubsubsection{Non-Abelian QSP.}
\label{sec:bivariable-qsp}
The QSP on hybrid CV-DV systems in the previous section is of course a limited transformation, as in general the position and momentum of the oscillator may be transformed independently by different polynomials

In this section, we provide a construction of a non-abelian QSP, discussed in~\cite{singh_towards_2025}, for the same task of cat state preparation and compare it to the single-variable case to highlight the advantages of processing both phase-space quadratures. Despite recent progress on multivariate QSP in the commuting case \cite{rossi2021mqsp,rossi2023modular}, a complete theory on polynomial transformation on the phase space for two non-commuting quadratures is yet to be developed \cite{nemeth2023variants,singh_towards_2025}.

Consider the controlled displacements along the momentum and position directions, 
\begin{align}
    e^{-i\frac{k}{2} \hat{x} \cdot \sigma_z} &= 
    \begin{bmatrix}
        e^{-i\frac{k}{2}\hat{x}} & \\
        & e^{i\frac{k}{2}\hat{x}}
    \end{bmatrix} = 
    \begin{bmatrix}
        w(\hat{x}) & \\
        & w^{-1}(\hat{x})
    \end{bmatrix}:= W_z^{(k)}(w),     \label{wz_k} \\
    e^{-i\frac{\lambda}{2} \hat{p} \cdot \sigma_z} &= 
    \begin{bmatrix}
        e^{-i\frac{\lambda}{2}\hat{p}} & \\
        & e^{i\frac{\lambda}{2}\hat{p}}
    \end{bmatrix}
    =
    \begin{bmatrix}
        v(\hat{p}) & \\
        & v^{-1}(\hat{p})
    \end{bmatrix}
    := W_z^{(\lambda)}(v),
    \label{wz_lambda}
\end{align}
which can be viewed respectively as a block-encoding of a momentum boost $W_z^{(k)}(w)$ or a position displacement $W_z^{(\lambda)}(v)$.

We may now extend this to obtain the following unitary constructed from the above unitary operator alternating with $x$-rotations on the qubit:
\begin{align}
    U_d 
    &= e^{i\phi_0 \sigma_x} \prod_{j = 1}^d W_z^{(k)} e^{i\phi_j^{(k)} \sigma_x} W_z^{(\lambda)} e^{i\phi_j^{(\lambda)} \sigma_x} \nonumber \\
    &=
    \begin{bmatrix}
        F_d(w, v) & iG_d(w, v) \\
        iG_d(v^{-1}, w^{-1}) & F_d(v^{-1}, w^{-1})
    \end{bmatrix}.
    \label{qsp_u_bivar}
\end{align}
parameterized by the set of phases $\{ \phi_{j}^{(k)}, \phi_{j}^{(\lambda)} \}$ $(j = 1,2,\cdots,d)$. It can be shown by direct multiplication of terms in Eq.~\eqref{qsp_u_bivar} that the above construction implements a bivariable Laurent polynomial transformation on non-commuting $w$ and $v$ with the following form:
\begin{align}
    F_d(w, v) := \sum_{r,s=-d}^d f_{rs} w^r v^s,~~
    G_d(w, v) := \sum_{r,s=-d}^d g_{rs} w^r v^s,
    \label{f-g_coeff}
\end{align}
where $f_{rs}$ and $g_{rs}$ are complex coefficients determined by the phase angles $\{ \phi_j^{(k)}, \phi_j^{(\lambda)} \}$. Note that because $w$ and $v$ do not commute, their order in the above expression matters. We define (somewhat arbitrarily) the way of writing out the bivariable Laurent polynomial to be \emph{canonical} when all $w$ are written to the left of $v$. Again, it can be seen that the unitary in Eq.~\eqref{qsp_wz} indeed complies with the form given in Eq.~\eqref{rep-unitary},  but in the non-abelian case, reduction to the Hamiltonian form in Eq.~(\ref{eq:osc-qubit-control-H}) may not be straightforward (since it constitutes the most hybrid general circuit compilation problem).  We can however use the rules for the composition of controlled displacements given in Sec.~\ref{sssec:composingCDs} to obtain the overall unitary as a sum of terms.  

To make this construction useful for compilation, it is desirable to show that for an arbitrary function $F_d$ given in Eq.~\eqref{f-g_coeff}, the QSP phases $\{ \phi_{j}^{(k)}, \phi_{j}^{(\lambda)} \}$ $(j = 1,2,\cdots,d)$ always exist. The bivariable nature makes the qubit QSP theorem in App.~\ref{app:qsp-qsvt} inapplicable. The non-commutative nature of $\hat{x}$ and $\hat{p}$ distinguishes the current context from the recent development of multi-variable QSP for commuting variables \cite{rossi2021mqsp}, and makes it challenging to prove a theorem governing efficient universality of non-abelian QSP. However, numerical evidence suggests that the non-abelian QSP construction can indeed efficiently approximate an arbitrary transformation of the phase space, with resolution set by the momentum kick $k$ and the position kick $\lambda$. Ref.~\cite{singh_towards_2025} shines a light on the analytical understanding of the non-abelian QSP construction towards control of bosonic modes.

To gain some intuition about the capabilities of bi-variable non-abelian QSP, let us revisit the two-legged cat state preparation example discussed above. There is no classical analog for this framework since the bi-variable non-abelian QSP  scheme makes full use of the non-commuting position and momentum variables in addition to the non-commuting Pauli matrices. In a recent work~\cite{singh_towards_2025}, a basic model of composite pulses in phase space has been introduced to correct for errors in Gaussian uncertainty of position variables using momentum variables, and vice-versa. The authors begin with the Schr\"odinger entangled cat state in Eq.~(\ref{entangled_cat}), but generalize it to allow for the possibility of squeezed states that appear in both squeezed cat  \cite{schlegel2022quantum,PhysRevLett.120.073603,xu_autonomous_2023,PhysRevA.107.032423}  and GKP error correction code states.
\begin{align}
    \psi(x)=\frac{1}{\sqrt{\pi\Delta^2}}\left[ e^{-\frac{(x-\alpha)^2}{\Delta^2}} |-\rangle  +  e^{-\frac{(x+\alpha)^2}{\Delta^2}} |+\rangle   \right],
\end{align}
where $\Delta^2$ is the position variance of the squeezed coherent state relative to the vacuum value of $\sigma_x^2=\frac{1}{4}$ found in Eq.~(\ref{eq:cohvarx}).

In the abelian QSP protocol, the next step would be to apply the controlled-momentum boost in Eq.~(\ref{eq:simplerotation}) which rotates the qubit by an angle linear in the oscillator position to (nearly) disentangle the spin from the cavity.  The quantum fluctuations in the position leave small residual residual over/under rotation errors illustrated in Fig.~\ref{fig:QSPcat}.  Near the peak positions $x=\pm\alpha$ the rotation error around the $y$ axis of the Bloch sphere has the form
\begin{align}
    \delta\hat\theta_\pm =k(\hat x\mp\alpha),
\end{align}
with the constant $k$ defined in Eq.~(\ref{eq:rotrate}).  In the  abelian QSP protocol, we substantially reduced these rotation errors by applying the three additional conditional momentum boosts of the $\mathrm{BB1}_{90,y}$ protocol.  

In the non-abelian scheme, Ref.~\cite{singh_towards_2025} showed that we have access to new spin rotation angles that are linear in the oscillator momentum. We can use this new freedom to preemptively correct this error to first order in the quantum fluctuations of the oscillator position. We see from Fig.~\ref{fig:QSPcat} that the residual rotation errors are the same for both peaks in the wave function. However, a little thought shows that if the correction is to be applied preemptively before the controlled-momentum boost in Eq.~(\ref{eq:simplerotation}), then we must apply the small corrective rotations in \emph{opposite} directions for positive and negative oscillator positions
\begin{align}
    \phi(x)=\frac{1}{\sqrt{\pi\Delta^2}}&\Bigg[ e^{+i\frac{\delta\hat\theta_+}{2}\sigma_y}e^{-\frac{(x-\alpha)^2}{\Delta^2}} |-\rangle \nonumber\\
    &+  e^{-i\frac{\delta\hat\theta_-}{2}\sigma_y}e^{-\frac{(x+\alpha)^2}{\Delta^2}} |+\rangle   \Bigg].
\end{align}
To achieve this (to the lowest order in $k$, we can take advantage of the fact that the derivative of a Gaussian has the right form (again in Wigner units)
\begin{align}
    \hat p e^{-\frac{(x \mp \alpha)^2}{\Delta^2}}=i\frac{(x\mp\alpha)}{\Delta^2}e^{-\frac{(x \mp \alpha)^2}{\Delta^2}},
\end{align}
and further, use $\sigma_y=-i\sigma_z\sigma_x$ to obtain the necessary sign difference for the two peaks
\begin{align}
    \sigma_y|\pm\rangle=\mp i\sigma_z|\pm\rangle.
\end{align}
Using these results we see that we can obtain the preemptively corrected state (accurate to first order in $k$) by
\begin{align}
    \phi(x)\approx e^{i\frac{\pi\Delta^2}{4\alpha}\hat p\sigma_z}\psi(x).
\end{align}
Applying the same controlled momentum boost in Eq.~(\ref{eq:simplerotation}) as before then yields the desired nearly disentangled state with the residual rotation errors eliminated to first order in $\frac{\Delta^2}{4\alpha}$.  While the $\mathrm{BB1}_{90,y}$ scheme also corrects second-order errors, this non-abelian QSP preparation scheme requires fewer operations than $\mathrm{BB1}_{90,y}$ and avoids the associated large momentum boosts making it more robust to ancilla and cavity decay errors and residual cavity anharmonicity. Essentially the same circuit is already in use experimentally for GKP state generation, stabilization, and measurement~\cite{RoyerGKP1_2020,Sivak_GKP_2022} and was independently discovered by numerical optimization in \cite{Hastrup_GKP_QSP}.  An interesting extension of the method would be to use non-abelian QSP techniques to modify this `Big-small' protocol to also cancel the rotation errors to second order while avoiding the large momentum boosts of $\mathrm{BB1}_{90,y}$.

Fig.~\ref{fig:QSP_cat2} shows a comparison of the fidelities for the abelian and non-abelian QSP protocols presented above. A more general discussion on the framework of composite pulses in phase space is described in \cite{singh_towards_2025}, where the authors also discuss robust protocols for other cases including cat states represented by overlapping Gaussian functions, GKP code states, and Fock states. 

Recall that the cat states realized by the LCU circuits in Sec.~\ref{sec:lcu} were exact but non-deterministic. The abelian and non-abelian QSP circuits here are deterministic but approximate.

Finally, we note that Ref.~\cite{rossi2023quantum} has proposed a different generalization of QSP to purely continuous-variable systems (rather than hybrid) based on the $\mathrm{SU}(1,1)$ group, which can be realized via beam-splitter and parametric amplifier interactions (single- or two-mode squeezing) between two bosonic modes and are therefore potentially useful for compiling multi-mode unitary operations. This reveals the richness of the synergy between QSP and continuous-variable quantum systems, and we anticipate more developments in the future.

\subsection{Approximate Numerical Compilation}
\label{sec:numerical-optimization}

\begin{figure}[htb]
            \includegraphics[width=0.45\textwidth]{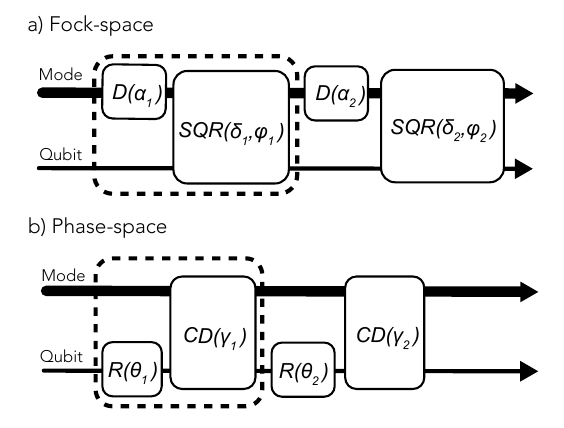}
            \caption{Examples of circuit constructions used in numerical optimization with the Fock-space (a) and phase-space (b) instruction sets. For each, a layer is shown in the dotted box, and contains enough operations for universality. The circuit depth (the number of layers) and circuit parameters are optimized to realize a target unitary or state transfer. The SNAP construction (not shown) is similar, comprised of alternating SNAP gates and oscillator displacements. Recall from Eq.~\eqref{eq:SNAPSQRSQR} that the SNAP gates discussed in the main text can be compiled from pairs of SQR gates.}
            \label{fig:numerical_constructions}
\end{figure}

 As a pedagogical example of numerical compilation for bosonic unitaries, we detail here the particular approach from Ref.~\cite{EickbuschECD} and further extend this framework to any universal instruction set developed in this work.
In many cases, a target bosonic operation (such as a unitary or state preparation) can be approximately synthesized from a universal gate set using numerical optimization. 
For example, Ref.~\cite{kan2024benchmarking} presents benchmarking techniques for variational quantum optimizers for bosonic state preparation. Separately, in \cite{sivak2021modelfree}, model-free reinforcement learning was applied to approximately synthesize target bosonic operations using SNAP and displacement gates. 
The model-free reinforcement learning method is particularly well-suited for in-situ optimization, where unknown experimental microwave transfer functions can be included. Other numerical SNAP optimization strategies were explored in Refs.~\cite{fosel2020efficient,Kundra_2022_Robust}. Similarly, in Ref.~\cite{EickbuschECD}, gradient-based optimization methods were developed that make use of automatic differentiation and gradient back-propagation to find gate parameters that approximately realize a target operation.

\begin{figure}[htb]
            \includegraphics[width=0.45\textwidth]{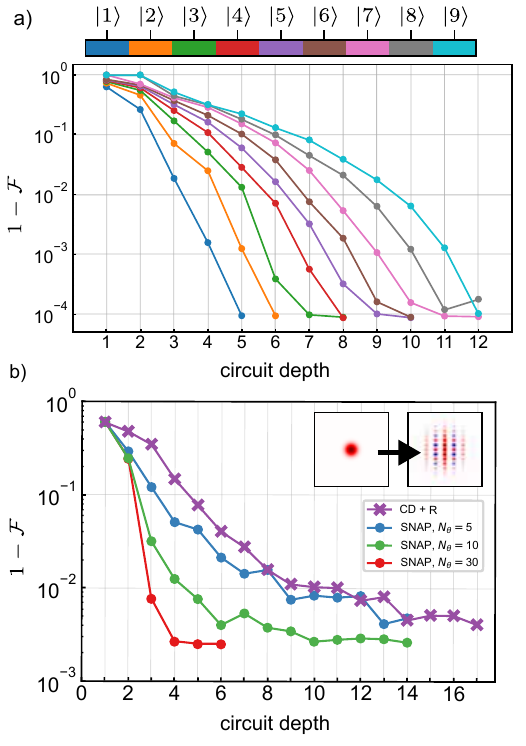}
            \caption{Examples of numerical circuit optimization for an oscillator coupled to a qubit. (a) Infidelity of Fock states optimized using the phase-space ISA as a function of circuit depth ($N$)~\footnote{We note through private communication that the fidelities presented here for points at circuit-depth of $N=2$ correct a minor simulation error that was later found for points at $N=2$ in Fig. 2(a) of Ref.~\cite{EickbuschECD}. To the best of our knowledge, at $N=2$ the results here represent the best known numerically optimized protocols.}. (b) Optimization of oscillator GKP $\ket{+Z}$ logical states with squeezing $\Delta = 0.32$ using either the phase-space ISA (crosses) or the Fock-space ISA via SNAP (circles). For the SNAP gates, the first $N_\theta$ phases are optimized, where $N_\theta$ is swept in the figure. As shown, including more phases in each SNAP gate leads to shorter-depth circuits.}
            \label{fig:numerical_example}
\end{figure}

To outline the optimization problem, we first define the fundamental unit of each bosonic circuit to be a `layer.'  Each layer should consist of at least one application of each gate in the generating set of a universal set. Examples of circuits used for the Fock-space and phase-space gate sets are shown as the dotted boxes in Fig.~\ref{fig:numerical_constructions}. The circuit depth, $N_\text{circuit}$, is the number of layers. For circuits acting on multiple oscillators with some well-defined connectivity, the layer should also include one application of entangling gates between each connected oscillator (for example, a beam-splitter between each pair of neighboring oscillators).  

To aid in computational efficiency, it is advantageous to optimize multiple randomly-initialized circuits in parallel; this set of circuits is called a batch. Let $B$ be the total number of independent circuits in a batch. We can denote $b_{jk}$ as the unitary associated with layer $k \in \left\{1,2,...,N_\text{layer}\right\}$ in circuit $j \in \left\{1,2,...,B\right\}$ of the batch. With this, the unitary generated by the $j$-th circuit is written as
\begin{equation}
    U(\vec{\theta}_j)=b_{jN}\ldots b_{j1}b_{j0},
\end{equation}
where $\vec{\theta}_j$ is the set of parameters specifying the $j^\text{th}$ circuit. This parameterized unitary can then be used to define a cost function to be minimized. By calculating gradients using back-propagation, the cost function can be chosen to handle different situations. In the case of a target unitary operation, the cost function can be defined using the trace fidelity. For the $j^{\text{th}}$ circuit, this is given by
\begin{equation}
    \label{trace fid}
    \mathcal{F}_j=\left| \frac{1}{\Tr{P}}\Tr{\mathcal{P}U_{\text{target}}^\dagger U_j }\right|^2,
\end{equation}
where $\mathcal{P}$ is a projector onto a subspace of interest \cite{SchulteHerbrggen2011, Ball2021}. For example, $\mathcal{P}$ could consist of the first $l$ Fock levels to implement a Fock truncation, $\mathcal{P}=\sum_{n=0}^{l-1} \ket{n}\bra{n}$.

If the goal is to instead prepare a target pure state $\ket{\psi_t}$, the fidelity of the $j^{\text{th}}$ circuit can be defined using the overlap, 
\begin{align}
\mathcal{F}_j =  \bra{\psi_t}\rho_j\ket{\psi_t},
\end{align}
where $\rho_j$ is the result of the optimized circuit applied to a specified initial state. For example, consider the setting of a single oscillator coupled to a qubit, with the goal being to prepare a target cavity state $\ket{\psi_t}$ of the oscillator starting from the composite oscillator-qubit state $\ket{0}_\mathrm{vac}\otimes\ket{0}$.  If we do not care about the final state of the qubit (only that it is disentangled from the oscillator), the qubit can be traced out to obtain $\rho_j = \text{Tr}_q\left(U_j [\ket{0}_\mathrm{vac}\bra{0}_\mathrm{vac} \otimes \ket{0}\bra{0}] U^\dag_j\right)$. The reduced density matrix $\rho_j$ can then be used to evaluate the fidelity $\mathcal{F}_j$ of the resulting oscillator state. 

From the fidelity $\mathcal{F}_j$ of the $j^{th}$ circuit, a total cost function can be defined using all $B$ circuits in the batch, allowing for parallel optimization. As an example, one option that has proven useful in practice is
\begin{equation}
\text{C} = \sum_{j = 1}^B \log ( 1 - \mathcal{F}_j).
\end{equation}
Since the cost function $C$ is a simple sum of independent logarithmic cost functions, gradient descent of $\text{C}$ realizes independent gradient descent of each circuit realization in parallel. The logarithm of $1-\mathcal{F}_j$ is used to aid with the problem of vanishing gradients as the solution infidelity approaches zero. 

Once the parameterized cost function is defined, it is optimized by calculating gradients using back-propagation. Gradient descent methods can be then be applied using techniques that are well established. For example, the optimizer called \textit{Adam} \cite{kingma2017adam}, commonly used to train neural networks, can be directly applied, making use of graphics processing units (GPUs). The optimization is stopped when any circuit fidelity $\mathcal{F}_j$ reaches the target fidelity, and the parameters from that circuit are selected. As an example of this, in Fig.~\ref{fig:numerical_example}, we show results of optimizing the preparation of (a) Fock and (b) GKP states using the phase-space or Fock-space ISAs.

It is not always guaranteed that random quantum circuits will converge to realize a target unitary operation. A complete formal complexity theoretic analysis of hybrid CV-DV circuit synthesis for tasks such as state transfer and general unitary synthesis does not yet exist. In many cases, optimization fails when the variance of the cost function gradient (calculated over the ensemble of randomly-initialized circuits in the batch) becomes exponentially small, and the circuit optimization becomes stuck due to so-called `barren plateaus'.

Ref.~\cite{zhang2023energydependent} has made progress in analyzing the barren plateau problem for numerical optimization of CV variational quantum circuits, demonstrating that the variance of the cost function gradient decays as $E^{-\nu M}$, and is thus exponential in the number of modes $M$ but polynomial
in the (per-mode) circuit energy $E$ (i.e., mean boson number). The exponent is $\nu=1$ for shallow circuits and $\nu=2$ for deep
circuits. The authors prove these results for state preparation of general Gaussian states and number states.  Separately, Ref.~\cite{PRXQuantum.4.030305barrenplateaus} has studied the barren plateau problem in the more general setting of quantum channels that include mid-circuit measurements and feedforward. One possible approach to the barren plateau problem is to replace the global fidelity cost function with local cost functions \cite{Khatri2019quantumassistedcompiling,cerezo_cost_2021}. Therefore, further study of the best local cost functions for hybrid CV-DV circuit synthesis is a promising direction for future study. 
In this section we have introduced strategies that we have developed to aid compilation, and in the next section, we introduce applications of these strategies in quantum algorithms.

\section{Algorithms and Applications}
\label{sec:apps}

Major quantum algorithms including the Deutsch-Josza \cite{pati2003deutsch}, Grover's search \cite{pati2000quantum}, and Shor's quantum factoring algorithm have been generalized from the discrete variable (DV) setting to the continuous variable (CV) setting \cite{lomonaco2002continuous}, and experimental implementations of these are now emerging \cite{wang2021implementing}.  In addition, novel CV-based quantum algorithms such as boson sampling protocols \cite{aaronson2011computational,lund2014boson,hamilton2017gaussian} have been devised and experimentally demonstrated in the optical domain \cite{tillmann2013experimental,thekkadath2022experimental,zhong2019experimental}. Apart from certain classically simulable regimes~\cite{GarciaPatron2019simulatingboson, Qi2020Regimes, oh2023tensor}, these algorithms open the door to scenarios where predicting experimental outcomes is difficult or intractable for present-day classical computers \cite{zhong2020quantum,madsen2022quantum}. Small-scale demonstrations of boson sampling \cite{HU2018293}, Gaussian boson sampling (i.e., including squeezing), and efficient photon number resolving detection \cite{Wang2020FCFs} have also been made in the microwave domain using circuit QED techniques.

The development of algorithms that uniquely take advantage of continuous-variable quantum systems has not yet flourished for several reasons. A common drawback of many CV algorithms is the lack of strict orthogonality in the computational basis (e.g., oscillator position states are not infinitely squeezed). As a result, only quasi-orthogonal states are available for use in computations, which complicates the analysis of the algorithm. Another issue is that there is no fault-tolerant threshold in CV systems unless a finite-dimensional logical qubit or qudit subspace is encoded into the infinite-dimensional CV system. This limits our ability to do fault-tolerant operations on oscillators (as full oscillators rather than qudits). Moreover, from an operational point of view, much less is known about how to control and implement quantum algorithms in CV systems. For these reasons, it is unclear how to implement such CV algorithms and evaluate their efficiency relative to their DV counterparts. 

The systematic establishment of CV-DV ISAs (Secs.~\ref{ssec:intro_isa} and \ref{sec:instruction_set}) allows for a more elaborate discussion of CV and hybrid CV-DV algorithms, addressing many of the challenges previously faced by algorithm developers. In this section, we illustrate potential applications of quantum CV-DV ISAs. We begin by presenting new algorithms for arbitrary CV-to-DV and DV-to-CV state transfer, alongside a review of an existing protocol (Sec.~\ref{sssec:qubit-osc-state-transfer}), and then introduce a new implementation of the quantum Fourier transform based on state transfer and oscillator dynamics (Sec.~\ref{sssec:qft}).
Sec.~\ref{sec:application-ham-sim} provides pedagogical examples of the  application of bosonic ISAs to the simulation of time-evolution dynamics of Hamiltonians containing bosonic degrees of freedom that are otherwise costly to simulate with standard methods using qubit-based fault-tolerant quantum computers.   Finally, we briefly review quantum random walks pointing out the role of the phase-space ISA in the implementation of such walks (Sec.~\ref{ssec:randomwalk}).

\subsection{Arbitrary Qubit-Oscillator State Transfer}
\label{sssec:qubit-osc-state-transfer}
We first present how to transfer a multi-qubit DV state to a CV state of a single oscillator, leaving the qubits and the oscillator disentangled at the end.  We present two different methods for such DV-to-CV state transfer, using single-variable hybrid QSP (Sec.~\ref{ssssection:single_var_QSP_state_transfer}), and non-Abelian QSP (Sec.~\ref{ssssection:bi_var_QSP_state_transfer}). In these protocols, each qubit basis state is directly mapped to a coherent state located at the position corresponding to the integer representation of the qubit basis state. As these protocols make use of both qubits and oscillators, they serve as useful subroutines for each the qubit-centric, oscillator-centric, and hybrid CV-DV AMMs. The presentation here is kept brief; for more details on these state transfer protocols, see their full presentation in  Ref.~\cite{qftwithoscillator}.

Formally, in the state transfer problem, we begin with an $n$-qubit state $|\psi \rangle_Q = \sum_{\mathbf{x}} c_{\mathbf{x}} |\mathbf{x}\rangle_Q $, where $\mathbf{x} = (x_1, x_2, ..., x_{n})$ is a Boolean-valued vector and $|\mathbf{x}\rangle_Q = |x_1\rangle |x_2\rangle ... |x_{n} \rangle$ is the corresponding qubit state. Using the binary representation of integers, let us denote the integer corresponding to $\mathbf{x}$ as $x: = \sum_{j=1}^n x_j\cdot 2^{n-j}$.
We wish to transfer $|\psi \rangle_Q$ to a corresponding oscillator state $|\psi \rangle_B = \sum_{\mathbf{x}} c_{\mathbf{x}} |  x, \Delta \rangle_B $, where the basis states $| x ,  \Delta \rangle_B $ now exist in continuous space and depend on $x$ and a spacing parameter $\Delta$. One can think of the state $| x ,  \Delta \rangle_B $ as being localized around the position $q=x \Delta$, such that adjacent basis states (i.e., $| x \pm 1 , \Delta \rangle_B $) are separated by $\Delta$; the exact form of these basis states depends on the specific state transfer protocol, as discussed below. Ultimately, we desire an $n$-qubit DV-to-CV \textit{state-transfer unitary} $U_{\text{st}}^{(n)}(\Delta)$, that obeys
\begin{align}
    U_{\text{st}}^{(n)}(\Delta) |\psi \rangle_Q |0, \Delta \rangle_B = |\mathbf{0}\rangle_Q | \psi \rangle_B,
\end{align}
where $\mathbf{0}$ is a vector of all $0$'s, and $|0, \Delta \rangle_B$ is the initial state of the oscillator (e.g., the vacuum state or a squeezed vacuum state).

\subsubsection{State Transfer Method 1: Single-variable QSP}
\label{ssssection:single_var_QSP_state_transfer}
\begin{figure}[tbp]
    \begin{center}
    \includegraphics[width=0.99\linewidth]{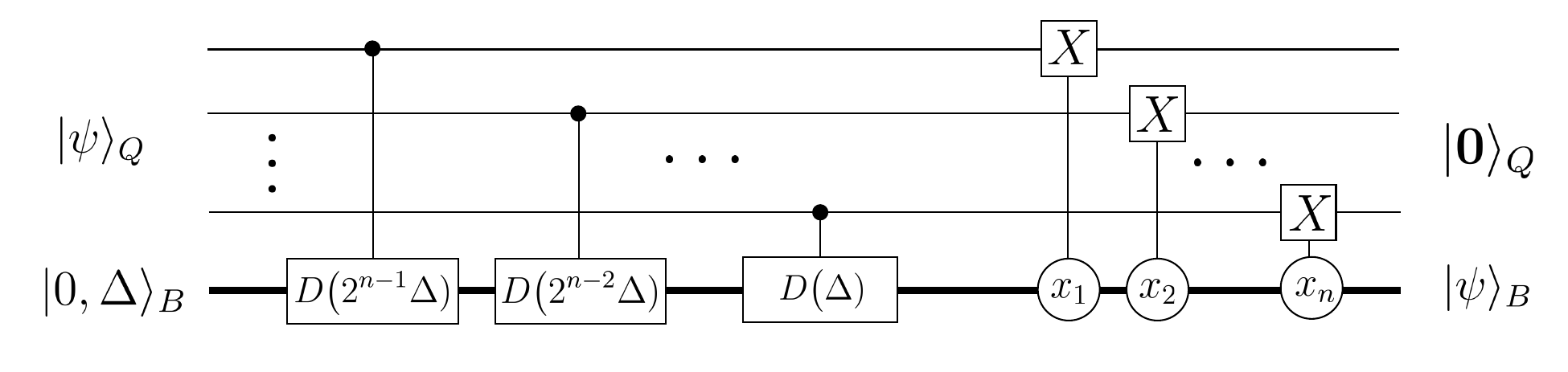}
    \end{center}
    \caption{A circuit that implements DV-to-CV state transfer with single-variable QSP.
    The initial $n$-qubit state is $|\psi\rangle_Q$, and the initial oscillator state is a squeezed vacuum as per Eq.~\eqref{eq:ST_basis_QSP}. The first stage applies a series of controlled displacements $D( 2^{n-j} \Delta)$. The second stage disentangles the state by flipping the $j^{\text{th}}$ qubit conditioned on the $j^{\text{th}}$ bit of the oscillator position. We illustrate the corresponding operations as $X$ gates conditioned on the bits $x_j$. These conditional bit-flips are implemented using a QSP approximation to a square wave filter in position space, which is depicted in Fig.~\ref{fig:SquareWaveIllustration} and implemented through the procedure described in Eq.~\eqref{eq:W_j_QSP}.
    }\label{fig:QSP_state_transfer}
\end{figure}

Our first DV-to-CV state transfer protocol employs quantum signal processing (QSP). This approach works by first applying a series of conditional displacements to the initial state, and then disentangling this intermediate state by applying a series of strategically-chosen QSP sequences. We depict the circuit that implements this protocol in Fig.~\ref{fig:QSP_state_transfer}. 

For this protocol, we would ideally want to take our oscillator basis states to be position eigenstates $|q\rangle_B$, with eigenvalues $q=x\Delta$ that are integer multiples of the spacing parameter. However, we cannot perform infinite squeezing to prepare a perfect position space eigenstate. Instead, we take our basis states to be Gaussians of width $\sigma$ centered around the value $x \Delta$: 
\begin{equation}\label{eq:ST_basis_QSP}
    |x, \Delta \rangle_B = \frac{1}{\sqrt{\sigma} (2\pi)^{1/4}} \int dq\ e^{-(q-x\Delta)^2/4\sigma^2} |q\rangle_B,
\end{equation}
which reduces to a position eigenstate as $\sigma \rightarrow 0$. In addition, we assume that the spacing and squeezing parameters are such that these basis states are nearly orthonormal:
\begin{equation}\label{eq:stateTransIP}
    _B\langle y, \Delta  |x, \Delta \rangle_B = e^{-\frac{\Delta^2}{\sigma^2} \frac{(x-y)^2}{8}},
\end{equation}
which approaches the Kronecker $\delta_{x,y}$ as $\sigma/\Delta \rightarrow 0$.

\textbf{Protocol:} In more detail, the state transfer protocol works by first preparing the qubits in the initial state $|\psi\rangle_Q$ and the oscillator in the state $|0, \Delta\rangle_B$. Then, in the first stage, we apply the following series of controlled displacements to the oscillator 
\begin{equation}
    \prod_{j=1}^{n} D_{c_j}(0, \Delta 2^{n-j}),
\end{equation}
where $D_{c_j}(0, \Delta 2^{n-j})$ is a conditional displacement as in Eq.~\eqref{eq:condDisp}, controlled by the $j^{\text{th}}$ qubit. Upon application to the initial state, this produces
\begin{equation}\label{eq:QSP1_intermediate_state}
   |\psi\rangle_Q |0, \Delta \rangle_B = \sum_{\mathbf{x}} c_{\mathbf{x}} |\mathbf{x}\rangle_Q |0, \Delta \rangle_B \mapsto\sum_{\mathbf{x}} c_{\mathbf{x}} |\mathbf{x}\rangle_Q |x , \Delta \rangle_B.
\end{equation}

If we can disentangle the qubits and oscillator as $|\mathbf{x}\rangle_Q |x , \Delta \rangle_B \mapsto  |0\rangle_Q |x , \Delta \rangle_B$, then up to the small errors created by non-orthogonality (see Eq.~\eqref{eq:stateTransIP}), we perform the ideal transfer from the qubits to the oscillator. 

\begin{figure}[tbp]
    \begin{center}
    \includegraphics[width=0.92\linewidth]{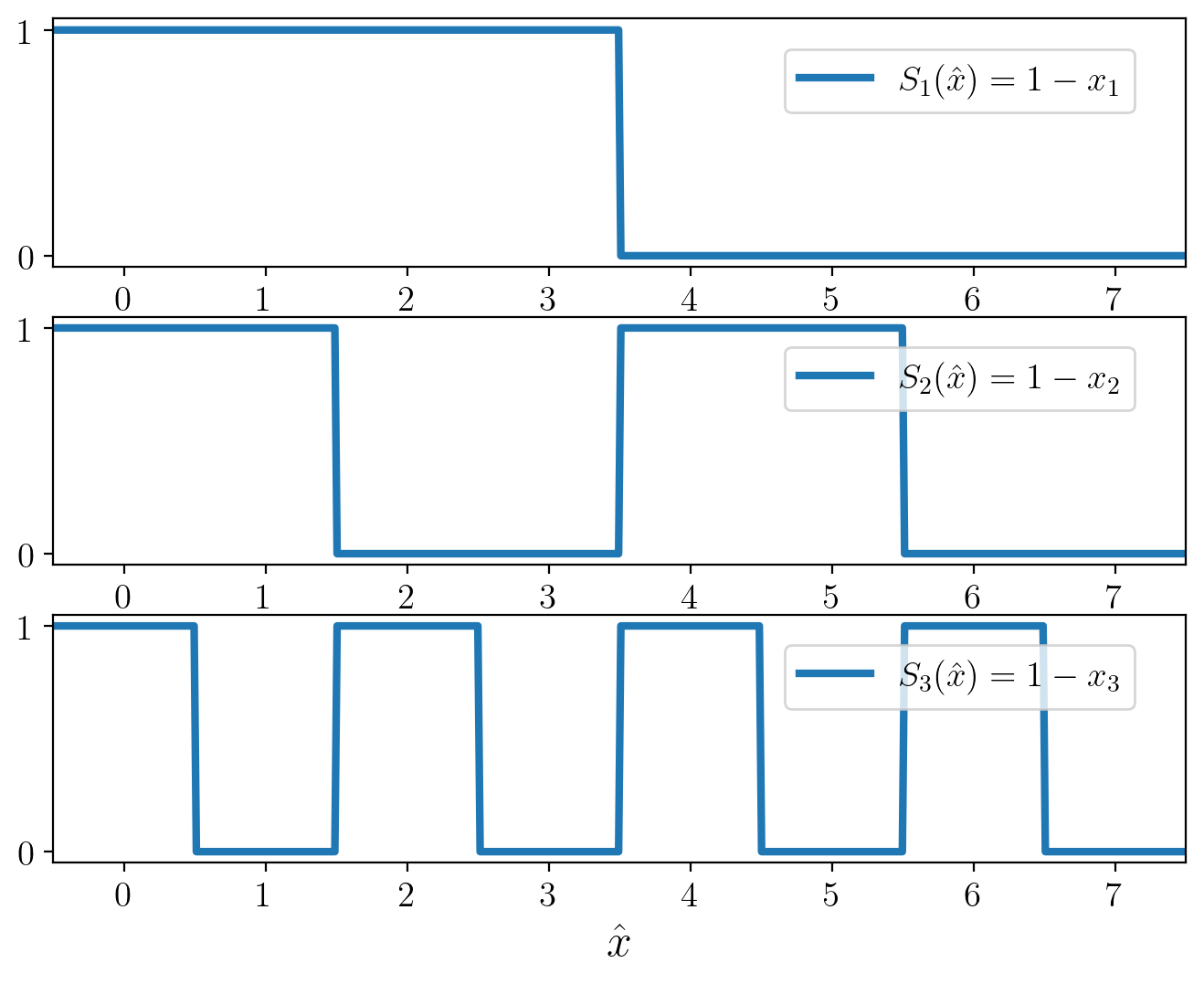}
    \end{center}
    \caption{An illustration of the square wave functions $S_j(x\Delta)$ from Eq.~\eqref{eq:square_wave_func} for $j=1,2,3$, with $n=3$ and $\Delta=1$. Observe how at integer values $x$, these square waves are equal to $1-x_j$, which enables one to read out the bits $\{ x_j \}$.}
    \label{fig:SquareWaveIllustration}
\end{figure}

We perform this disentangling procedure with single-variable QSP. The idea behind this is to devise a series of operations that flips the $j^{\text{th}}$ qubit conditioned on the $j^{\text{th}}$ bit $x_j$. 
This behavior requires that we determine the bit $x_j$ from the bosonic mode's binary representation $| x , \Delta \rangle_B$. Accordingly, observe that the bit $x_j$ of the position $q=x \Delta = \Delta \sum_{j=1}^n 2^{n-j} x_j $ can be read out by a ``square wave'' function: 
\begin{equation}\label{eq:square_wave_func}
    S_j(\hat{x}) := \Theta \Big(\cos\big[\tfrac{\pi}{2^{n-j}} \big(\tfrac{\hat{x}}{\Delta} - 2^{n-j-1}+\tfrac{1}{2}\big)\big] \Big) = 1-x_j,
\end{equation}
where $\Theta(\cdot)$ is the Heaviside step function, and $\hat{x}$ denotes the position operator. For visual intuition, we depict this function in Fig.~\ref{fig:SquareWaveIllustration}, which illustrates how such a square wave function can output the bits $x_j$. 

We may use this observation to our advantage by constructing the unitary
\begin{align}\label{eq:W_j_QSP}
    W_j(\hat{x}) &= 
    \begin{pmatrix}
        S_j(\hat{x}) & \sqrt{1-S_j(\hat{x})^2} \\
        -\sqrt{1-S_j(\hat{x})^2} & S_j(\hat{x})
    \end{pmatrix}\nonumber\\
    &= 
    \begin{cases}
    I & x_j = 0 \\
    iY & x_j = 1,
    \end{cases}
\end{align}
where $Y=\sigma_y$ here. This operation correctly flips the $j^{\text{th}}$ qubit conditioned on $x_j$: $W_j(\hat{x}) |x_j\rangle | x , \Delta \rangle_B = |0\rangle | x,  \Delta \rangle_B$.\footnote{Here we are taking $|x , \Delta \rangle_B$ to be an exact position eigenstate; we will remedy this assumption and evaluate performance on $|x,\Delta \rangle_B$ (which is not an exact position eigenstate) shortly.} Therefore, the sequence $\prod_{j=1}^{n} W_j(\hat{x})$ correctly disentangles all $n$ qubits.

Our strategy is therefore to approximate each $W_j(\hat{x})$ with a QSP sequence. 

To achieve this, we may apply QSP to the rotation $e^{iX \cdot \frac{\pi}{2^{n-j}}(\frac{\hat{x}}{\Delta} - 2^{n-j-1}+\frac{1}{2})}$, which block encodes $\hat{R}_j = \cos\big[\frac{\pi}{ 2^{n-j}}(\frac{\hat{x}}{\Delta} - 2^{n-j-1}+\frac{1}{2})\big] $ in its $|0\rangle \langle 0|$ component.
We appropriately select QSP phases to construct a real polynomial $P(\hat{x})$ that approximates the Heaviside step function. Upon conjugating the corresponding QSP unitary (see Eq.~\eqref{eq:QSP_sequence}) by a phase gate $S$, we obtain
\begin{equation}
\begin{aligned}
    & \begin{pmatrix}
        P(\hat{R}_j) & \sqrt{1-P(\hat{R}_j)^2} \\
        -\sqrt{1-P(\hat{R}_j)^2} & P(\hat{R}_j)
    \end{pmatrix} =: \tilde{W}_j(\hat{x}).
\end{aligned}
\end{equation}
This approximates $\tilde{W}_j(\hat{x}) \approx W_j(\hat{x})$ in Eq.~\eqref{eq:W_j_QSP} because $P(\hat{R}_j) \approx S_j(\hat{x})$, where the accuracy in this approximation is dictated by how well a polynomial can approximate the Heaviside step function. It is well established that a QSP polynomial can approximate the step function to within some error $\epsilon$, except within a region of width $w$ centered around the discontinuity, and that the corresponding polynomial has degree $\mathcal{O} \big( \frac{1}{w} \log(\frac{1}{\epsilon}) \big)$~\cite{low2017hamiltonian, martyn2023efficient}. 

Using this construction, the second stage of the protocol applies the series of QSP sequences $\prod_{j=1}^n \tilde{W}_j(\hat{x})$. This disentangles the intermediate state of Eq.~\eqref{eq:QSP1_intermediate_state} to (approximately) produce the desired final state $\sum_{\mathbf{x}} c_{\mathbf{x}}  |\mathbf{0}\rangle_Q |x , \Delta \rangle_B = |\mathbf{0}\rangle_Q |\psi\rangle_B $.

\textbf{Performance:} We illustrate the entire circuit that implements the state transfer protocol in Fig.~\ref{fig:QSP_state_transfer}, showcasing its decomposition into a stage of controlled displacements, followed by the series of QSP sequences. Let us now analyze its gate complexity and fidelity. 

The first stage requires $\mathcal{O}(n)$ displacement gates, of size $\Delta 2^{n-j}$. This translates to a total time complexity that scales as $T_{\rm CD} = \sum_{j=1}^{n} \Delta 2^{n-j}= \mathcal{O}(\Delta 2^n)$. This implies an overall time complexity that scales linearly with $2^n$ when we implement the controlled displacements with a fixed coupling between the qubits and oscillator.
However, this need not be exponential in general as $\Delta \in \mathcal{O}(2^{-n})$ is also a possibility if sufficient squeezing is available on our quantum device.

In the second stage (the QSP sequences), we take the polynomial implemented by the $j^{\text{th}}$ QSP sequence to be an approximation to the step function that suffers error at most $\epsilon$, outside of a region of width $w_j$ centered about the discontinuity. Each such QSP sequence requires $N_{\text{QSP},j}=\mathcal{O}\big(\frac{1}{w_j}\log(1/\epsilon)\big)$ gates on the qubit register~\cite{low2017optimal,GrandUnificationAlgos}. To ensure that the $j^{\text{th}}$ QSP sequence can discern the correct bit $x_j$ when acting on a state located at position $q=x\Delta$, we require that the width of the approximate step function needed in the $j^{\rm th}$ bit in our encoding of the binary number scales as $w_j \leq \mathcal{O}(\tfrac{1}{2^{n-j}})$. Therefore, adding together the gate complexities of all the QSP sequences, we obtain a total gate complexity of $\sum_j N_{\text{QSP},j} =\mathcal{O}\big( \sum_j \frac{1}{w_j}\log(1/\epsilon) \big) =\mathcal{O}(2^n \log(1/\epsilon))$.

Next, consider the time complexity of this algorithm. Each step of a QSP sequence at fixed $j$ requires application of the block encoding $\hat{R}_j$ and a qubit rotation, which takes time $\mathcal{O}(1)$. Therefore, as a QSP sequence consists $N_{\text{QSP},j}$ of these steps, its time complexity is $T_{\text{QSP},j} = \mathcal{O}(N_{\text{QSP},j})$. Accounting for all $n$ QSP sequences, we obtain a total time complexity of the QSP stage: $T_{\text{QSP}} = \sum_j T_{\text{QSP},j} = \mathcal{O}( 2^n \log(1/\epsilon) )$.
The time complexity of the overall algorithm is then
\begin{align}
    T_{\rm Tot}&=T_{\rm CD} + T_{\text{QSP}} = \mathcal{O}\Big(2^n\Big(\Delta + \log(1/\epsilon)\Big)\Big).
\end{align}
If $\Delta$ is at most a constant, this scales exponentially with the number of bits that we wish to encode in the oscillator. This is perhaps unsurprising as we are not using a multi-modal binary encoding, but instead are storing the information in the position of a single oscillator.  In principle, one could use binary encoding to more efficiently encode large numbers in multiple oscillator modes, but for simplicity, we focus on the case where all the information is encoded in a single mode.
\begin{figure}[tbp]
    \centering
    \includegraphics[width=\columnwidth]{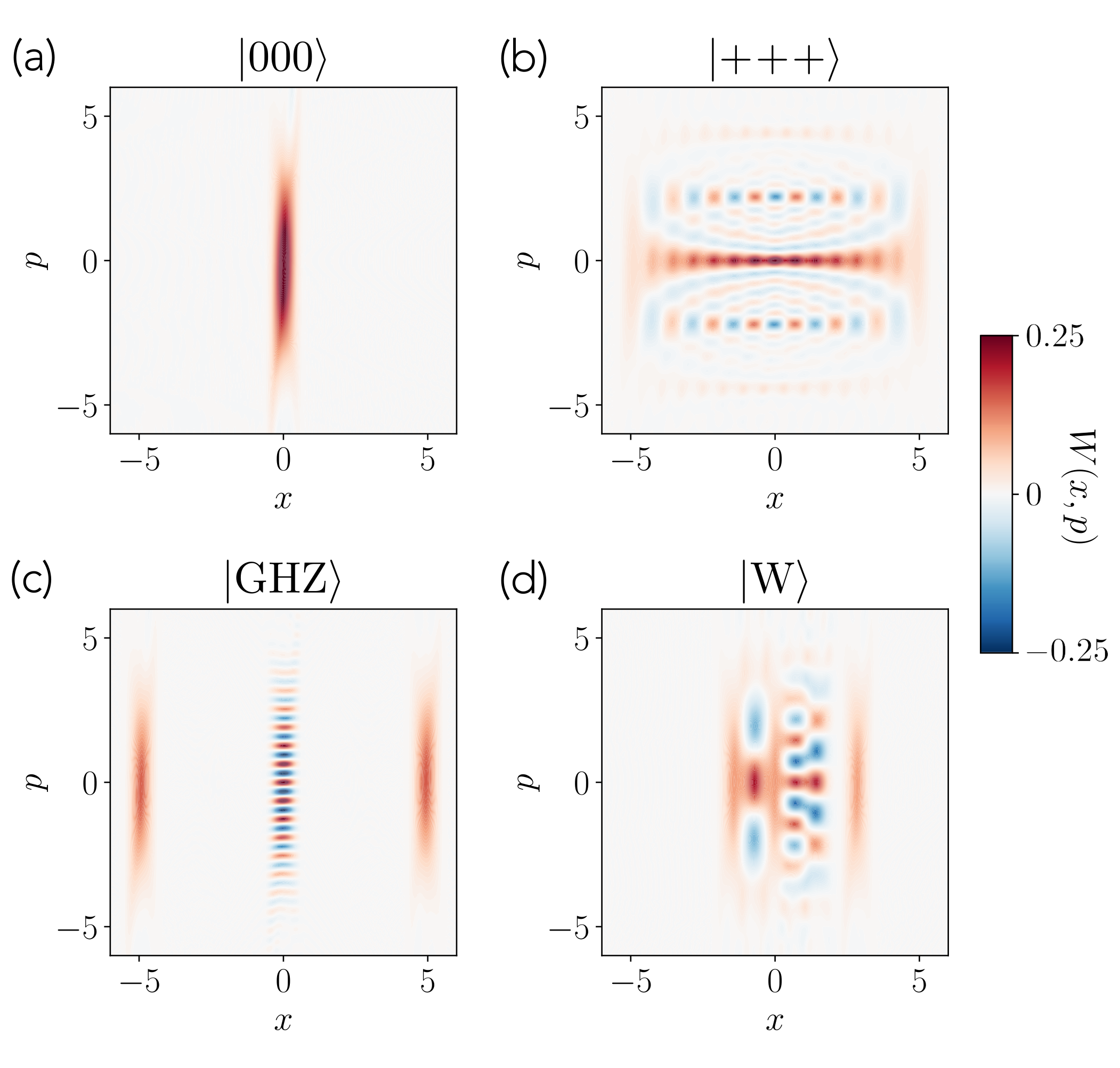}
    \caption{DV-to-CV state transfer for various three-qubit states, including $\ket{\textrm{GHZ}}=(\ket{000}+\ket{111})/\sqrt{2}$ and $\ket{\textrm{W}}=(\ket{001}+\ket{010}+\ket{100})/\sqrt{3}$. In all cases, we applied the protocol in Sec.~\ref{ssssection:single_var_QSP_state_transfer}, using a single-variable QSP sequence of depth 60 with $w=0.2$, $\Delta=1$, and $\sigma = \pi/16$. Fidelities were computed against the ideal state transfer, yielding (a) $F=0.970$ (b) $F=0.980$ (c) $F=0.972$, and (d) $F=0.982$. All simulations were carried out in Bosonic Qiskit \cite{BiskitGitHub}.
    }
    \label{fig_statetransfer_examples}
\end{figure}

Lastly, the values of $\Delta$ and $\epsilon$ also influence the fidelity of the state transfer. The protocol suffers from two possible types of errors: first, the bosonic basis states are not exact position eigenstates but instead Gaussians of finite width $\sigma$; and second, the QSP polynomial approximation is only accurate to within $\epsilon$, and fails within the region of width $w_j$ about the discontinuity of the step function. A careful analysis of these error modes, presented in Ref.~\cite{qftwithoscillator}, indicates that the total fidelity scales as 
\begin{align}
    & 1-\mathcal{O}(n\epsilon) - \mathcal{O}(e^{-\mathcal{O}(\Delta^2/\sigma^2)} ).
\end{align}
 Note that this depends on the ratio $\sigma/\Delta$, which represents the relative spacing between the basis states $|x,\Delta \rangle_B$. Therefore, one can improve fidelity by either squeezing the initial state to decrease $\sigma$, or selecting a larger separation $\Delta$. Numerical examples of the DV-to-CV state transfer protocol for various three-qubit states are given in Fig.~\ref{fig_statetransfer_examples}.

\subsubsection{State Transfer Method 2: Non-Abelian QSP}
\label{ssssection:bi_var_QSP_state_transfer}
\begin{figure}[tbp]
    \begin{center}
    \includegraphics[width=0.99\linewidth]{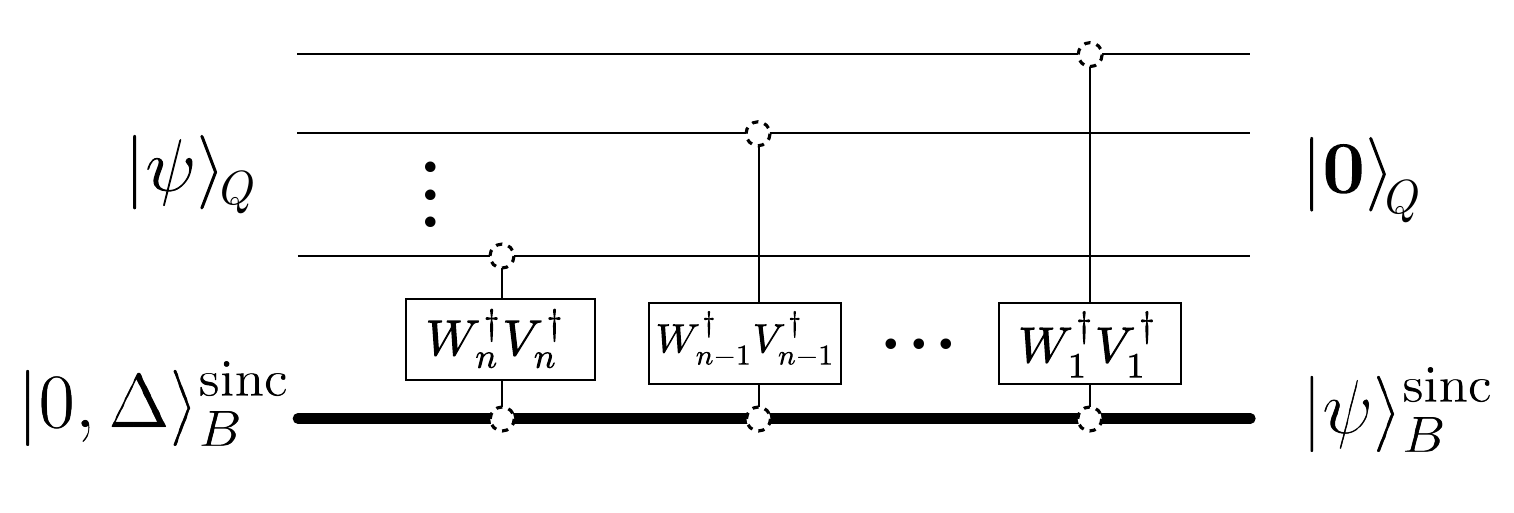}
    \end{center}
    \caption{The circuit that implements DV-to-CV state transfer, adapted from Ref.~\cite{hastrup2022universal}; crucially, the order of $W_n$ and $V_n$ here are flipped relative to Fig.\~1 of Ref.~\cite{hastrup2022universal}, which we believe to be a typo. The initial state of the qubits  is $|\psi\rangle_Q$, and the initial oscillator state is a sinc state $|0, \Delta \rangle_B = \frac{1}{\sqrt{\Delta}} \int dq \  \text{sinc}(\pi q /\Delta) |q\rangle_B$ with $\Delta = 2\lambda$. Then, one applies a series of operations $W_j^\dag V_j^\dag $ between the oscillator and the $j^{\text{th}}$ qubit (see Eq.~\eqref{eq:V_and_W}), which maps the initial state of the qubits to an equivalent oscillator state $|\psi\rangle_B$ encoded in a basis of displaced sinc states, as per Eq.~\eqref{eq:Hastrup_inverse_state_transfer}.}
    \label{fig:Hastrup_state_transfer}
\end{figure}

Recently, Refs.~\cite{hastrup2022universal,Hastrup_GKP_QSP} proposed an approach to quasi-deterministically transfer an oscillator state to an $n$-qubit state by repeatedly performing two controlled displacement operations between the oscillator and each qubit. This defines a CV-to-DV state transfer protocol, which in reverse furnishes a DV-to-CV state transfer protocol. For completeness, we depict the circuit that implements the DV-to-CV state transfer protocol in Fig.~\ref{fig:Hastrup_state_transfer}.

Below, we first review the protocol of Ref.~\cite{hastrup2022universal} and provide bounds on its performance. In addition, we show how this protocol can be viewed as a special case of non-Abelian QSP as discussed in Sec.~\ref{ssec:compilation-bosonic-qsp-qsvt}.

\textbf{Protocol:} Let us denote the CV-to-DV state transfer operation presented in Ref.~\cite{hastrup2022universal} by $U_{\text{st}}^{(n)}(2\lambda)^\dag$. This uses a spacing parameter $2\lambda$ (analogous to $\Delta$ above) and transfers an oscillator state to a state on $n$ qubits. Explicitly, $U_{\text{st}}^{(n)}(2\lambda)^\dag$ is the unitary operation defined by
\begin{equation}\label{eq:Hastrup_state_tr_operator}
    U_{\text{st}}^{(n)}(2\lambda)^\dag = \prod_{j = n}^{1} W_j V_j = W_n V_n \cdots W_1 V_1 
\end{equation}
where 
\begin{equation}\label{eq:V_and_W}
    V_j = e^{i \frac{\pi}{2^{j+1} \lambda} \hat{x}\hat{\sigma}_y^{(j)}}, \quad 
    W_j = 
    \begin{cases} 
        e^{i \lambda 2^{j-1}\hat{p}\hat{\sigma}_x^{(j)}} & j<n, \\
        e^{-i \lambda 2^{j-1}\hat{p}\hat{\sigma}_x^{(j)}} & j=n,
    \end{cases}
\end{equation}
are momentum boosts and displacements of the oscillator conditioned on the $j^{\text{th}}$ qubit. Note that we have defined this operation in terms of the Hermitian conjugate $U_{\text{st}}^{(n)}(2\lambda)^\dag$, such that its inverse $U_{\text{st}}^{(n)}(2\lambda)$ provides a DV-to-CV state transfer protocol, as per our conventions.

The input state to this protocol is $|\mathbf{0}\rangle_Q |\psi\rangle_B $, where $|\psi\rangle_B = \int dq \ \psi(q) dq |q\rangle_B$ is a state on the oscillator to be transferred to the $n$ qubits. Intuitively, the arrangement of momentum boosts and displacements in Eq.~\eqref{eq:Hastrup_state_tr_operator} are carefully chosen to conspire together to map the CV wave function $\psi(q)$, evaluated at a discrete set of positions $q_{\mathbf{s}}$, onto the amplitudes of a DV quantum state, thus performing state transfer. As shown in Ref.~\cite{hastrup2022universal}, application of $U_{\text{st}}^{(n)}(2\lambda)^\dag$ maps the initial state to
\begin{align}\label{eq:HastrupTransferedState}
&U_{\text{st}}^{(n)}(2\lambda)^\dag \ket{\bf 0}_Q\ket{\psi}_B
    \approx \nonumber\\
    &\quad \sum_{\mathbf{s} \in \{ -1, +1\}^n } \sqrt{2\lambda} \psi(q_{\mathbf{s}}) |\phi_{\mathbf{s}} \rangle_Q  \otimes \frac{1}{\sqrt{2\lambda}} \int dq \ \text{sinc}(\tfrac{\pi q}{2 \lambda}) |q\rangle_B,
\end{align}
where the basis states here are 
\begin{equation}
\begin{aligned}
    &|\phi_{\textbf{s}} \rangle = (-1)^{\gamma_{\textbf{s}}} \cdot \bigotimes_{j=1}^n \big[ Z^{(1-s_j)/2}|+\rangle \big] , \\
    &\gamma_{\mathbf{s}} = \sum_{j=1}^{n-2} \frac{1}{2}(s_j + s_{j+1}) + \frac{1}{2}(s_{n-1} - s_n) . 
\end{aligned}
\end{equation}
For instance, if $\mathbf{s} = (1,-1,1,-1)$, then $\gamma_{\mathbf{s}} = 1$, and $|\phi_{\mathbf{s}} \rangle = (-1)\cdot |+\rangle |-\rangle |+\rangle |-\rangle$. Finally, the value $q_{\textbf{s}}$ in Eq.~\eqref{eq:HastrupTransferedState} is 
\begin{equation}\label{eq:q_s}
    q_{\textbf{s}} = \sum_{j=1}^{n-1} s_j w_j - s_n w_n,
\end{equation} 
where $w_j = \lambda 2^{j-1}$. This quantity takes $2^n$ discrete values in the range $[-\lambda (2^n-1), \lambda (2^n-1)]$, with each possible value equally spaced by $2\lambda$.

Two approximations are used in Ref.~\cite{hastrup2022universal} to obtain Eq.~(\ref{eq:HastrupTransferedState}). It is assumed that the support of $\psi(q)$ is limited to $|q| \leq \lambda (2^n-1)$, and that $\psi(q)$ is slowly varying as $|\tfrac{d\psi}{dq}| \ll 1/\lambda$. This approximately decouples the oscillator from the qubits, thus transferring the initial CV state $ \psi(q) $ to a corresponding state of the qubits $\sum_{\mathbf{s}} \sqrt{2\lambda} \psi(q_{\mathbf{s}}) |\phi_{\mathbf{s}} \rangle_Q$.

Moreover, this protocol can also be run in reverse to furnish a DV-to-CV state transfer protocol. In this direction, one first prepares the qubits in the state $\sum_{\mathbf{s}} c_{\mathbf{s}} |\phi_{\mathbf{s}} \rangle$, and the oscillator in the state $|0, 2\lambda\rangle_B = \frac{1}{\sqrt{2\lambda}} \int dq \ \text{sinc}(\tfrac{\pi q}{2 \lambda}) |q\rangle$.\footnote{As mentioned in Ref.~\cite{hastrup2022universal}, this exact state is unphysical because it has infinite energy; however, it can be well approximated by a squeezed vacuum.} Then, executing $U_{\text{st}}^{(n)}(2\lambda) = \prod_{j = n}^{1} W_j^\dag V_j^\dag $ (approximately) outputs the state 
\begin{equation}\label{eq:Hastrup_inverse_state_transfer}
\begin{aligned}
    &|\mathbf{0}\rangle_Q \otimes \sum_{\mathbf{s}}  c_{\mathbf{s}} \frac{1}{\sqrt{2\lambda}} \int dq \ \text{sinc} \big( \tfrac{\pi (q-q_{\mathbf{s}})}{2\lambda} \big) |q\rangle_B = | \mathbf{0}\rangle_Q
    |\psi\rangle_B  .
\end{aligned}
\end{equation}
This has effectively transferred the state of the qubits to an oscillator state, now encoded in the basis of displaced ``sinc states''$|\frac{q_{\mathbf{s}}}{2\lambda}, 2\lambda \rangle_B := \frac{1}{\sqrt{2\lambda}} \int dq \ \text{sinc} \big( \tfrac{\pi (q-q_{\mathbf{s}})}{2\lambda} \big) |q\rangle_B$. A sinc state represents a state in continuous space that is peaked around $q=q_s$, with peaks of adjacent sinc states separated by $2\lambda$. Moreover, the sinc states are orthonormal: $_B  \langle \frac{q_{\mathbf{s}}}{2\lambda}, 2\lambda  | \frac{q_{\mathbf{s}'}}{2\lambda}, 2\lambda \rangle_B = \delta_{\mathbf{s} \mathbf{s}'} $. 

\textbf{Performance:} We illustrate the circuit of this DV-to-CV state transfer protocol in Fig.~\ref{fig:Hastrup_state_transfer}. This showcases the series of $V_j^\dag W_j^\dag $ operations acting between the oscillator and the $j^{\text{th}}$ qubit. Let us now study the gate complexity and performance of this state transfer protocol.

Consider first the gate complexity of this protocol. According to Eq.~(\ref{eq:Hastrup_state_tr_operator}), this protocol requires $\mathcal{O}(n)$ gates. However, collectively these gates require a total displacement $ \sum_{j=1}^n \mathcal{O}(2\lambda 2^j) = \mathcal{O}(\lambda 2^n)$. As in the previous state transfer algorithm, this implies an overall time complexity of $\mathcal{O}(\lambda 2^n)$ when we implement the displacements with a fixed coupling between the qubits and oscillator. But, with a sufficiently tunable and strong coupling between the qubits and the oscillator, one can adjust the coupling strength to reduce this exponential overhead.

Next, let us consider the fidelity of this protocol. Ref.~\cite{hastrup2022universal} presents numerical results on the fidelity achieved in the CV-to-DV direction. For example, in transferring the Fock state $|3\rangle$ onto $n$ qubits, the protocol achieves infidelity $\approx 0.2$ for $n=4$ and $\approx 9\cdot 10^{-4}$ for $n=10$, indicating how the performance improves with increasing $n$. Achieving this performance, however, requires that $\lambda$ be carefully tuned for each $n$ to maximize the fidelity. While no explicit bounds on fidelity are provided in Ref.~\cite{hastrup2022universal} to guide this tuning, in Ref.~\cite{qftwithoscillator} we derive bounds on the fidelity of both the DV-to-CV and CV-to-DV directions.

First consider the fidelity of the CV-to-DV direction, in which case the output state is approximately that of Eq.~\eqref{eq:HastrupTransferedState}. The approximations used in reaching this expression require that $\psi(q)$ have support limited to $|q|\leq \lambda (2^n-1)$, and be slowly varying such that $|\frac{d\psi}{dq}| \ll 1/\lambda$. A careful analysis of this protocol, presented in Ref.~\cite{qftwithoscillator}, indicates that the infidelity between these two states, and equivalently the infidelity of the CV-to-DV direction, is
\begin{equation}\label{eq:non-abelian-fidelity}
\begin{aligned}
     \mathcal{O} \Bigg( \int_{|q|\geq \lambda (2^n-1)} dq |\psi(q)|^2 + \lambda \int_{-\lambda(2^n-1)}^{\lambda(2^n-1)} dq \big| \tfrac{d}{dq}|\psi(q)|^2 \big| \Bigg).
\end{aligned}
\end{equation}
Notably, the two contributions to the infidelity stem from the support of $\psi(q)$ outside of $|q|\leq \lambda (2^n-1)$ and the derivative of $\psi(q)$, arising precisely from the two approximations used in Ref.~\cite{hastrup2022universal} to obtain Eq.~\eqref{eq:HastrupTransferedState}.

Moreover, the action of the DV-to-CV state transfer is defined by Eq.~\eqref{eq:Hastrup_inverse_state_transfer}. By a similar analysis, also presented in Ref.~\cite{qftwithoscillator}, we find that the infidelity of the DV-to-CV state transfer protocol is
\begin{equation}
    \mathcal{O}\Bigg( \int_{|q|\geq \lambda(2^n-1)} dq |\psi(q)|^2 \Bigg),
\end{equation}
where now $\psi(q) = \frac{1}{\sqrt{2\lambda}} \sum_{\mathbf{s}} c_{\mathbf{s}} \text{sinc} \big( \tfrac{\pi (q-q_{\mathbf{s}})}{2\lambda} \big)$ is the wave function of the state of the qubits transferred to the basis of sinc states. Evidently, in this direction, the fidelity is primarily impeded by the support of $\psi(q)$ outside $|q| < \lambda (2^n-1)$. The additional term of Eq.~\eqref{eq:non-abelian-fidelity} that depends on $\tfrac{d}{dq}\psi(q)$ is absent because the approximation that produces this contribution is naturally satisfied in the DV-to-CV direction; see Ref.~\cite{qftwithoscillator} for details.

\textbf{Recontextualization of DV-to-CV State Transfer as Non-Abelian QSP:}
Let us take a closer look at this DV-to-CV state transfer protocol, and illustrate how it can be re-expressed in the language of non-Abelian QSP. If we rewrite an individual term $V_j^\dag W_j^\dag$ using the conventions of non-Abelian QSP outlined in Sec.~\ref{ssec:compilation-bosonic-qsp-qsvt}, the following identity can be verified
\begin{align}
     & \hspace*{-3ex} e^{-i\frac{\pi}{4} \sigma_y^{(j)}} (V_j^\dag W_j^\dag
       ) e^{i\frac{\pi}{4} \sigma_y^{(j)}}    \nn \\
    & =  e^{-i \frac{\pi}{\lambda 2^{j+1} } \hat{x}\hat{\sigma}_y^{(j)}} e^{\mp i \lambda 2^{j-1}\hat{p}\hat{\sigma}_z^{(j)}} 
        \\
    & = e^{i\frac{\pi}{4} \sigma_x^{(j)}} e^{-i \frac{\pi}{\lambda 2^{j+1}} \hat{x}\hat{\sigma}_z^{(j)}}  e^{-i\frac{\pi}{4} \sigma_x^{(j)}} e^{\mp i \lambda 2^{j-1}\hat{p}\hat{\sigma}_z^{(j)}},
    \label{qft_readin-qsp}
\end{align}
where the $-$ sign is taken for $j<n$, and the $+$ sign for $j = n$.

By comparing this expression to the non-Abelian QSP sequence of Eq.~\eqref{qsp_u_bivar}, it is readily identified that the product $V_j^\dag W_j^\dag$, upon conjugation by $e^{-i\frac{\pi}{4} \sigma_y^{(j)}}$, corresponds to a degree-1 non-Abelian QSP sequence with momentum and positions boosts 
\begin{equation}
    W_z^{(k)}(w) = e^{-i \frac{\pi}{\lambda 2^{j+1}} \hat{x} \cdot \sigma_z}, \qquad  
    W_z^{(\lambda)}(v) = e^{\mp i \lambda 2^{j-1} \hat{p} \cdot \sigma_z},
\end{equation}
for the $j^{\text{th}}$ qubit, and with QSP phases
\begin{align}
    \phi_0 = \frac{\pi}{4}, ~~~\phi_1^{(k)} = - \frac{\pi}{4}, ~~~\phi_1^{(\lambda)} = 0
\end{align}
for all $j$ qubits. In this incarnation, the DV-to-CV state transfer protocol of this section may be interpreted as a product of $n$ degree-$1$ non-Abelian QSP sequences, where each sequence acts between the oscillator and the $j^{\text{th}}$ qubit. Accordingly, this protocol could admit a generalization by using a higher degree non-Abelian QSP sequence for each qubit during state transfer.

\subsection{Quantum Fourier Transform}
\label{sssec:qft}
The quantum Fourier transform (QFT) is an important quantum subroutine ubiquitous in many quantum algorithms, such as Shor's algorithm~\cite{shor}, phase estimation~\cite{Nielsen_Chuang} and quantum gradient estimation~\cite{gilyen2019optimizing}. It is defined on an $n$-qubit state $|\psi \rangle_Q = \sum_{\mathbf{x}} c_{\mathbf{x}} |\mathbf{x}\rangle_Q $ as the unitary transformation 
\begin{equation}
    U_{\text{QFT}} |\psi \rangle_Q = \sum_{\mathbf{x}} \Bigg[ \sum_{\mathbf{y}} \frac{1}{\sqrt{2^n}}  c_{\mathbf{y}} e^{2\pi i xy/2^n} \Bigg] |\mathbf{x}\rangle_Q ,
\end{equation} 
which effectively implements a discrete Fourier transform on the coefficients of the initial state. 

Here we provide an implementation of the QFT on the qubit register using bosonic operations, incorporating the state transfer protocols presented in Sec.~\ref{sssec:qubit-osc-state-transfer}. In particular, we assume access to a DV-to-CV state transfer unitary $U_{\text{st}}^{(n)}(\Delta)$ and its inverse, as well as traditional bosonic gates like free evolution and displacement. Because these state transfer protocols incorporate qubits and oscillators, this implementation of the QFT is a useful subroutine for each of the three AMMs: qubit-centric, oscillator-centric, and hybrid CV-DV. For more specific details on the implementation of the QFT protocol, see Ref.~\cite{qftwithoscillator}.

To preface our construction, recall that in phase space the free evolution of an oscillator naturally swaps position and momentum, thus enacting a continuous Fourier transform on the underlying wave function. Using this intuition, we show that by mapping an initial DV state to an oscillator, letting it undergo free evolution, and finally transferring the state back to qubits, the QFT of the DV state can be extracted.

\subsubsection{Continuous-Discrete Fourier Transform Correspondence} 
A vital component of our construction of the QFT is a correspondence between the continuous Fourier transform and the discrete Fourier transform. Consider a discrete signal $c_x$ for $x \in [0, ..., N-1]$, that is made periodic such that $c_x = c_{x \ \text{mod} \ N}$. Let this discrete signal be encoded in a continuous function $f(q)$ as
\begin{equation}\label{eq:wave_func_post_transfer}
    f(q) = \sum_{x \in \mathbb{Z}} c_x g(q-x\Delta),
\end{equation}
where $g(q)$ is a basis function localized about $q=0$ and $\Delta$ is the spacing between basis states. The continuous Fourier transform of this function evaluates to
\begin{equation}
    \tilde{f}(p) = \sum_{x \in \mathbb{Z}} c_x \tilde{g}(p) e^{ipx\Delta} = \tilde{g}(p) \sum_{k\in \mathbb{Z}} e^{ipNk \Delta}  \sum_{y=0}^{N-1} c_y e^{ipy \Delta},
\end{equation}
where $\tilde{g}(p)$ is the Fourier transform of $g(q)$, and we have split the index $x$ into $x = N k+y$ for $k \in \mathbb{Z}$ and $y \in [0,1,...,N-1]$.
Noting that the quantity $\sum_{k\in \mathbb{Z}} e^{ipNk \Delta}$ equates to a Dirac comb $\sum_{l\in \mathbb{Z}} \delta(\tfrac{pN\Delta}{2\pi}-l)$, this expression simplifies to 
\begin{equation}\label{eq:FT_correspondence}
\begin{aligned}
    \sum_{l \in \mathbb{Z}} \frac{2 \pi}{\Delta} \tilde{g}(\tfrac{2\pi }{ \Delta} \tfrac{l}{N}) \delta(p - \tfrac{2\pi }{ \Delta} \tfrac{l}{N}) \cdot \tilde{c}_l
\end{aligned}
\end{equation}
where $\tilde{c}_l = \sum_{y=0}^{N-1}  c_y e^{i 2\pi y l /N}$ is the discrete Fourier transform of $c_y$. Therefore, by taking the continuous Fourier transform of a discrete signal that is encoded periodically in some basis function, we obtain a sum over the discrete Fourier transform of the signal, with coefficients proportional to the Fourier transform of the basis function.

This correspondence can be used to perform the quantum Fourier transform by letting $c_x$ be the coefficients of an initial state on qubits. Then, $f(q)$ represents the wave function of an oscillator after transferring the state of the qubits to the oscillator and encoding in the basis $g(q)$. By enacting a continuous Fourier transform on the oscillator (i.e. free evolution), the new wave function given by Eq.~\eqref{eq:FT_correspondence} plucks out the states $|q = \tfrac{2\pi l}{N \Delta} \rangle$ with coefficients proportional to the discrete Fourier transform of $c_x$. Because the discrete Fourier transform of these coefficients is equivalent to the coefficients of QFT, we find that appropriately transferring this state back to qubits produces the QFT of the initial state. This is of course dependent on choosing a suitable basis function $g(q)$ whose Fourier transform behaves favorably.

\subsubsection{QFT Method 1: Single-variable QSP}

\begin{figure*}[htbp]
    \begin{center}
    \includegraphics[width=0.95\linewidth]{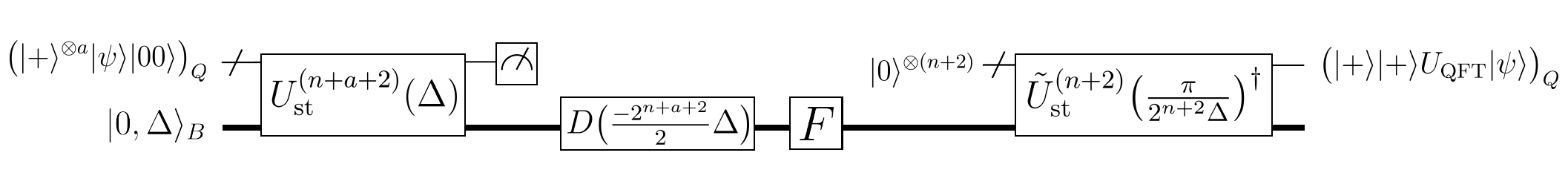}
    \end{center}
    \caption{The circuit used to implement the quantum Fourier transform of an $n$-qubit state $|\psi\rangle$ using the single-variable QSP state transfer unitary of Sec.~\ref{ssssection:single_var_QSP_state_transfer}, denoted $U_{\text{st}}$. Here, $D \big( \tfrac{-2^{n+a+2}}{2} \Delta \big)$ is a displacement operation, and $F$ is the Fourier transform operation $F = U(\pi/2)$ from Box~\ref{Box:phase-space-rotation} (i.e., oscillator free-evolution). The ancilla qubits $|+\rangle^{\otimes a}$ and $|00\rangle$ are appended to the initial state to increase the fidelity with the exact QFT, as discussed below. The result of this circuit is the state $|+\rangle |+\rangle U_{\text{QFT}}|\psi\rangle$ in the qubit register, from which the QFT may be obtained.} 
    \label{fig:QFT_QSP1}
\end{figure*}

We first present a construction of the QFT using the single-variable QSP state transfer protocol of Sec.~\ref{ssssection:single_var_QSP_state_transfer} (i.e., the circuit presented in Fig.~\ref{fig:QSP_state_transfer}). For illustrative purposes, we present the corresponding QFT circuit in Fig.~\ref{fig:QFT_QSP1}.

\textbf{Protocol:} In our construction, we first append the initial state with an appropriate number of ancilla qubits in the $|+\rangle$ state so as to make the initial state's coefficients periodic over a range $O(2^a)$, so that we can leverage the above continuous-discrete Fourier transform correspondence to generate the QFT. Then, we transfer this state to an equivalent oscillator state with basis spacing $\Delta$, and apply a displacement operation to make the state symmetric about $q=0$ in position space. Subsequently, the Fourier gate is applied to the oscillator to enact a continuous Fourier transform, and then the state is transferred back to qubits using a reciprocal spacing $\frac{\pi}{2^n \Delta}$. Crucially, this reciprocal spacing is chosen precisely to produce the correct phases of the QFT (i.e., $2\pi xy/2^n$). For more details, we present a pedagogical walk-through of each stage of this protocol in App.~\ref{app:QFT}.

\textbf{Performance:} The bulk of the gate count in this QFT protocol comes from the use of the single-variable QSP state transfer of Sec.~\ref{ssssection:single_var_QSP_state_transfer}. Copying over the gate count of that protocol, we have that the total gate count of the QFT here is $\mathcal{O}(2^{n+a}\log(1/\epsilon))$. Similarly, the time complexity is $\mathcal{O}(2^{n+a}(\Delta+\log(1/\epsilon)))$.

Next, the steps of this protocol can each suffer errors that decrease the fidelity with the target QFT. By aggregating together the infidelities suffered at each step, which are explicated in App.~\ref{app:QFT}, we find that the total fidelity of this protocol is 
\begin{equation}
    F = 1 - \mathcal{O}((n+a)\epsilon) - \mathcal{O}(\sigma/\Delta) - \mathcal{O}(1/2^a),
\end{equation}
where $\epsilon$ is the error in the QSP polynomial approximation to a square wave function. Understandably, the fidelity is maximized in the limit of small QSP error, large relative spacing between the oscillator basis states, and a large number of ancilla qubits.

\subsubsection{QFT Method 2: Non-Abelian QSP}
\begin{figure*}[htbp]
    \begin{center}
    \includegraphics[width=0.90\linewidth]{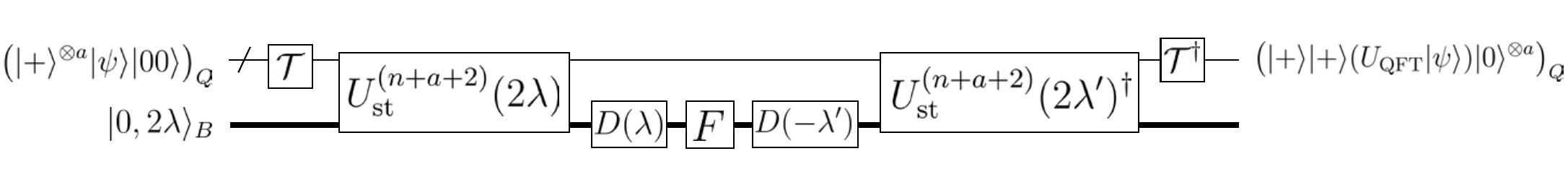}
    \end{center}
    \caption{The circuit used to implement the quantum Fourier transform of an $n$-qubit state $|\psi\rangle$ using the non-Abelian QSP state transfer unitary of Sec.~\ref{ssssection:bi_var_QSP_state_transfer}, denoted $U_{\text{st}}$. Here, $\mathcal{T}$ is a basis transformation from the computational basis to the basis $|\phi_s\rangle$, which is easily realized with simple qubit gates as discussed in the text. The operation $D ( \cdot )$ is a displacement operation, and $F$ is the Fourier gate $F = U(\pi/2)$ (oscillator free-evolution). Ancilla qubits $|+\rangle^{\otimes a}$ are prepended to the initial state to increase the fidelity with the exact QFT. The circuit outputs the state $|+\rangle |+\rangle U_{\text{QFT}}|\psi\rangle |0\rangle^{\otimes a} $ in the qubits register, from which the QFT may be extracted.} 
    \label{fig:QFT_QSP2}
\end{figure*}

Analogous to the presentation above, we can also implement the QFT using the non-Abelian QSP state transfer protocol of Sec.~\ref{ssssection:bi_var_QSP_state_transfer}. We depict the circuit that implements this algorithm in Fig.~\ref{fig:QFT_QSP2}.

\textbf{Protocol:} Our approach is to first append the initial state with ancilla qubits in the $|+\rangle$ state, such that we make the state's coefficients periodic and can use the continuous-discrete Fourier transform correspondence to generate the QFT. Then, we perform a change of basis from the computational basis to the basis $\{ | \phi_{\mathbf{s}} \rangle \}$ used in the non-Abelian QSP state transfer protocol; this is achieved using an operation denoted $\mathcal{T}$ in Fig.~\ref{fig:QFT_QSP2}, and can be shown to be implemented with simple local operations (see App.~\ref{app:QFT}). Subsequently, we transfer the state of the qubits to the oscillator, symmetrize the state about $q=0$ in position space, apply the Fourier gate, and finally transfer the oscillator state back to the qubits with an appropriately chosen reciprocal spacing. Upon transforming back to the computational basis, we obtain the QFT of the initial state. We provide a step-by-step analysis of this protocol in App.~\ref{app:QFT}.

\textbf{Performance:} Consider first the gate complexity of this QFT protocol, the bulk of which is due to the non-Abelian state transfer protocol. According to the performance bounds in Sec.~\ref{ssssection:bi_var_QSP_state_transfer}, the gate count scales as $\mathcal{O}(n+a)$. However, these gates require a total displacement $\mathcal{O}(\lambda 2^{n+a})$, implying a time complexity of $\mathcal{O}(\lambda 2^{n+a})$ when displacements are implemented with a fixed coupling between the qubits and the oscillator, but this can be reduced with a sufficiently tunable and strong coupling.

Next, consider the fidelity of this protocol. Again, each step of this protocol can incur errors that decease the fidelity with the target QFT, and we analyze these error modes in App.~\ref{app:QFT}. Upon aggregating these error contributions, we find that the overall fidelity of this QFT protocol is $F = 1- \mathcal{O}(1/2^a)$.

\subsection{Quantum Simulation}
\label{sec:application-ham-sim}

In this section, we provide a brief overview of techniques for quantum simulation using hybrid CV-DV hardware. For an in-depth study of the simulation of fermions, bosons, and gauge fields in CV-DV hardware, we refer the reader to Ref.~\cite{C2QA-LGTpaper}.

Quantum simulation aims to solve classically intractable physical problems with controllable quantum devices and has been identified as a likely arena for the first demonstration of \textit{practical} advantage over existing classical methods~\cite{quantum_initiative_US,quantum_flagship_Europe, Bauer_SimulationHEP_2022, Cao2019, RevModPhys.86.153}. Exact classical simulation of quantum systems is often intractable due to the exponential growth of Hilbert space with system size. As a result, classical methods often resort to approximations that involve either truncation or finite sampling methods to remove the exponential computational overhead at the price of introducing approximation errors that may or may not be acceptable for a given application. Such approximations include the use of mean-field theory \cite{meyer1990multi,marques2004time} and semiclassical approximations of quantum phase space~\cite{POLKOVNIKOV20101790} that do not capture particle-particle correlations.  Other representative approximations that truncate the Hilbert space such as many-body perturbation theory, coupled cluster theory \cite{bartlett1981many,bartlett2007coupled}, and tensor network-based methods \cite{banuls2023tensor} face difficulties with strongly correlated systems, as they are capable of only limited treatment of many-body correlations. Separately, sampling-based methods, such as quantum Monte Carlo, suffer from sign or phase problems for frustrated spin systems and interacting fermionic matter \cite{foulkes2001quantum,MonteCarlo_QCD_DANZER2009240,MonteCarlo_QCD_Bongiovanni_2016}.

As originally noted by Richard Feynman \cite{feynman1985quantum}, quantum hardware, unlike classical hardware, is naturally suited to the simulation of quantum physics \cite{RevModPhys.86.153}.\footnote{It is important to distinguish between two cases: (a) Simulation of the time evolution of a quantum system with a specified Hamiltonian and a specified starting state; and (b) Finding the ground state (or other eigenstate) of a specified Hamiltonian.  For most physical systems, the former is in the complexity class \BQP~and the latter is in \QMA~which is a quantum analog of \NP~and is believed to be exponentially more challenging than \BQP~\cite{Nielsen_Chuang}.}   To that end, there are two popular approaches: analog quantum simulation and digital quantum simulation. The former corresponds to the \emph{native} realization of a Hamiltonian in a controlled quantum system, while the latter corresponds to the \emph{synthesis} of a Hamiltonian using gate-based instructions on a universal quantum computer (or if not universal, at least a programmable quantum simulator).

While both approaches have proven promising in recent years \cite{altman2021quantum, daley2022practical}, most experiments have thus far focused on ``single-species'' quantum systems. For example, demonstrations of digital qubit-based quantum simulation have typically concerned either spin \cite{salathe2015digital, kim2023evidence} or fermionic \cite{barends2015digital} systems (though we note the recent proposal for hybrid fermion-qudit processors in Ref.~\cite{zache2023fermion}). Similarly, analog quantum simulators have successfully implemented interacting spin \cite{simon2011quantum, monroe2021programmable}, all-bosonic \cite{Greiner2002, katz2023programmable}, and all-fermionic \cite{mazurenko2017cold} systems.  However, only a few experiments have demonstrated the capability to simulate bosons and fermions within a single system~\cite{Zhang2018, Mil2020}. However, multi-species quantum Hamiltonians are among the most important and are crucial to the development of fundamental science across a range of energy scales, from the interactions between force-carrying bosons and fermionic elementary particles underlying the standard model to the nuclear vibrational-electron interactions underpinning quantum chemical phenomena such as photosynthesis \cite{domcke2012role}. Consequently, this motivates the need for flexible approaches to quantum simulation that enable hardware-efficient simulation of different species within a single platform.

We focus on the problem of digital quantum simulation here as it is most pertinent to the scope of this work, but also make a comparison to analog quantum simulation where applicable. Turning first to conventional qubit-based hardware, we note that it is possible to encode fermionic~\cite{JordanWigner1928}, bosonic~\cite{shaw2020quantum,jordan2012quantum}, and spin-$S$ \cite{KevinAKLT_PRXQuantum.4.020315} degrees of freedom using many qubits, in principle enabling multi-species digital quantum simulation. In practice, however, bosonic degrees of freedom and the corresponding bosonic operations are costly to implement in all-qubit hardware, as discussed in Sec.~\ref{sssec:complexity}. For example, the implementation of a single displacement gate on a qubit-encoded-oscillator containing 10 bosons requires on the order of $100$ two-qubit gates using the Fock-binary encoding~\cite{shaw2020quantum} (see Sec.~\ref{app:compilation-cv-to-qubits}). This inefficiency is particularly problematic in cases where it becomes important to represent high mean boson numbers with large density fluctuations, regardless of the choice of encoding. Ref.~\cite{bureik_suppression_2025} provides an example from cold-atom physics of such a situation where the density fluctuations are large and non-Gaussian.

Hybrid oscillator-qubit platforms can simulate bosonic degrees of freedom natively while retaining the capabilities of qubit-based hardware. Such hybrid platforms show immense promise for the study of both single- and multi-species models containing spin, fermionic, and bosonic degrees of freedom. Furthermore, the universal ISAs presented in this work enable the implementation of arbitrary Hamiltonian terms needed to simulate strongly correlated matter and facilitate measurement of both local and nonlocal observables, the latter playing a crucial role in the identification of exotic phases of matter \cite{Kitaev2006, levin2006detecting}.

In the remainder of this section, we present a high-level survey of the opportunities provided by digital hybrid oscillator-qubit hardware for quantum simulation. For a more in-depth investigation into compilation techniques for the simulation of fermion-boson models and lattice gauge theories, we refer the reader to Ref.~\cite{C2QA-LGTpaper}. 

We begin by first describing high-level techniques that enable the study of dynamics (Sec.~\ref{sssec:dynamics}), ground and excited states (Sec.~\ref{sssec:state-prep}), and observables (Sec.~\ref{observablessection}). Following this, we provide several explicit examples of physically relevant models with bosonic modes, spins, or some combination of the two. These include the Rabi model (Sec.~\ref{sssec:Rabi}), the Bose-Hubbard model (Sec.~\ref{BoseHubbard}), large spin models (Sec.~\ref{sssec:many-spins}), a $\mathbb{Z}_2$ lattice gauge theory (Sec.~\ref{Z2section}), and fermion-boson problems (Sec.~\ref{sssec:fermion-boson-problems}). In each case, we demonstrate how one can synthesize the Hamiltonian terms of interest for each case using the techniques outlined in Sec.~\ref{sec:compilation}. Many of the specific compilation ingredients discussed throughout are summarized in Table ~\ref{tab:simulation}, which lists a selection of common bosonic Hamiltonian terms and strategies for realizing the unitaries they generate with native instructions. We then conclude with comments on the potential for simulation of fermion-boson problems.
\def\arraystretch{2}
\begin{table*}[tb]
    \centering
    \begin{tabular}{|c|c|c|c|}
    \hline 
        \textbf{Term} $H_i$ & \textbf{Description} & {\bf Synthesis of $e^{-i H_i t}$} & {\bf Example Models}\\
        \hline \hline
        $a_i^{\dagger} a_i$ & \makecell{Mode energy, \\ Chemical potential} & $R(\theta)$ & \makecell{Jaynes-Cummings~\cite{JaynesCummings}, \\Bose-Hubbard~\cite{Bose_Hubbard}}\\
        \hline
        $\hat{n}_i \left( \hat{n}_i - 1 \right)$ & \makecell{On-site interactions} & $\textrm{SNAP}(\vec{\varphi})$ & Bose-Hubbard~\cite{Bose_Hubbard} \\
        \hline
        $e^{i\phi}\hat{a}_i^{\dagger}\hat{a}_j + \mathrm{h.c.}$ & Hopping &  $\mathrm{BS}(\theta,\varphi)$ & \makecell{Bose-Hubbard~\cite{Bose_Hubbard}, \\ Su–Schrieffer–Heeger \cite{Su1979}}\\
        \hline
        $Z_i (a + a^{\dagger})$ & Spin-dependent displacement & $\mathrm{CD}(\alpha)$ & \makecell{Rabi~\cite{Rabi1937}, Dicke~\cite{Dicke_model}, \\ Hubbard-Holstein~\cite{HOLSTEIN}} \\
        \hline
        $Z_i Z_j (a+a^\dagger)$ & Spin-spin-dependent displacement & \makecell{$\mathrm{CD}(\alpha)$, $\textrm{SWAP}$, $\textrm{CP}$ \\ (Sec.~\ref{sssec:useful_primitives})}& Photosynthetic complex \cite{arsenault2021vibronic} \\
        \hline
        $Z_{i} (e^{i\varphi} a^\dagger_i a_j + \mathrm{h.c.})$ & Spin-dependent hopping & $\textrm{CBS}(\theta,\varphi)$  & Bosonic $\mathbb{Z}_2$ LGT~\cite{Schweizer2019} \\
        \hline
         $a_i^\dagger a_j a_k$, $a^\dagger_ia^\dagger_ja_k a_l$, etc. & Higher-order terms & \makecell{$R(\theta)$, $D(\alpha)$, $\textrm{TMS}(r,\varphi)$, \\ $\mathrm{BS}(\theta,\varphi)$ + Approx.} & \makecell{Bosonic SYK \cite{murugan2017more,fu2016numerical}} \\
        \hline
        $S^z_i S^z_j$, ($\vec{S}_i\cdot \vec{S}_j)^k$,  etc. & (Large)-spin interactions & $R(\theta)$, $\mathrm{BS}(\theta,\varphi)$ + Approx. & \makecell{AKLT \cite{Affleck1987}} \\
        \hline
        $f_i^\dagger f_j$, $f_i^\dagger f_j^\dagger f_k f_l$, etc. & Fermionic matter & \makecell{Multi-qubit gates\\ (Sec.~\ref{ssec:compilation-entangling} )} & \makecell{Electronic structure \cite{szabo2012modern}, \\ Hubbard-Holstein~\cite{HOLSTEIN} \\ Fermi-Hubbard \cite{arovas2022hubbard}} \\
        \hline
    \end{tabular}
    \caption{Common native and synthesized Hamiltonian terms in condensed matter physics and quantum chemistry. `Higher-order' terms refer to those composed of $>2$ creation or annihilation operators multiplied together (i.e., polynomials of order 3 or higher in terms of $\hat x$ and $\hat p$). $f_i^\dagger, f_i$ are fermionic creation and annihilation operators, realized via fermion-to-qubit mappings such as Jordan-Wigner or Bravyi-Kitaev \cite{tranter2018comparison}. We note that there will, in general, be multiple avenues to synthesize $e^{-i H_i t}$, and the aim here is to give a rough, non-exhaustive guide to the native gates and techniques needed. Here, `Approx.' refers to the use of approximate compilation techniques, such as BCH and Trotter-Suzuki formulas (see Sec.~\ref{ssec:approximate-1-qubit-oscillator-unitary}). See Table~\ref{tab:gates-qubit-osc} for definitions of common hybrid gates listed in the Hamiltonian synthesis column. See Ref.~\cite{C2QA-LGTpaper} for a thorough manual on how to simulate fermions, bosons, and gauge fields in CV-DV hardware. 
    }
    \label{tab:simulation}
\end{table*}

\subsubsection{Dynamics}
\label{sssec:dynamics}
Time dynamics involves the study of a system evolving under the time-evolution operator $U(t) = e^{-i H t}$, where $H$ is some model Hamiltonian of interest. For digital quantum simulation, the goal is therefore to synthesize and enact $U(t)$ through decomposition into a discrete gate set. Many strategies exist to achieve this, including Trotter-Suzuki and product formulas \cite{Lloyd1073,suzuki1991general,childs2021theory}, linear combinations of unitaries \cite{childs2012hamiltonian,berry2015simulating} and multi-product formulas \cite{low2019well}, quantum walks \cite{Childs_2009,berry2012black,Berry_2015_walks}, qubitization \cite{Low2019}, and various randomized approaches \cite{campbell2019random,faehrmann2022randomizing,concentration2021chen}.
Hybrid oscillator-qubit hardware may facilitate many of these methods. For further discussion of Trotterization in the context of compilation, see Sec.~\ref{sec:trotter-product}.

In the Trotter-Suzuki approach, the unitary time evolution operator $U(t)$ corresponding to the Hamiltonian $H=\sum_j H_j$ is written as
\begin{equation}
    \begin{split}
        U(t) &= e^{-i \sum_{j=1}^M H_j t} \\
        &= \left(\prod_{j=1}^M e^{-i H_j t/r}\right)^r + O\left(\sum^M_{j,k}\|[H_j,H_k] \| t^2/r\right)
    \end{split}
    \label{eq:time_evolution_trotter}
\end{equation}
where $r$ is a positive integer introduced to decrease the error in the approximation by increasing the number of timesteps, and $M$ is the total number of Hamiltonian terms. This decomposition reduces the problem to the implementation of $e^{-i H_j t/r}$ for each $j$, via either a single native gate or through compilation using a particular instruction set (see Sec. \ref{sec:compilation}). For example, a chemical potential term $H_j = e^{-i \mu \hat{n}_k (t/r)}$ can be implemented via a phase-space rotation $R(\mu t /r)$ (see Box \ref{Box:phase-space-rotation}), while higher-order interactions such $H_j=Z_ia_i^\dagger a_i a_k^\dagger a_k$ can be synthesized using BCH methods (see the examples portion of Sec.~\ref{ssec:approximate-1-qubit-oscillator-unitary}). Higher-order variants of Eq.~(\ref{eq:time_evolution_trotter}) exist, such as the second-order (or symmetric) Trotter formula that achieves an improved error scaling at the cost of higher gate count \cite{childs2021theory}.

\subsubsection{State Preparation}
\label{sssec:state-prep}
Initial state preparation is a key component for both static and dynamic simulations.  For static problems, state preparation is used to construct a prior quantum distribution over eigenstates that can be refined through variational techniques~\cite{Peruzzo2014} or quantum phase estimation~\cite{kitaev1995quantum}.  In general, the question of whether or not quantum computers can  efficiently prepare high-fidelity approximations to ground states of physically relevant systems remains an open question; however, if a generic quantum algorithm existed for preparing such states then a consequence would be that $\BQP = \QMA$.  This is not believed to be true, as $\BQP$ and $\QMA$ are the quantum analogs of $\P$ and $\NP$ for classical computation~\cite{Nielsen_Chuang}.

For calculating dynamics, quantum states need to be provided that correspond to the initial quantum state $\ket{\psi(0)}$ that we wish to evolve into $\ket{\psi(t)} = U(t) \ket{\psi(0)}$.  It is likely, however, that calculating observables of thermal states relevant to material simulations is easier than for ground states~\cite{Bravyi2022, PRXQuantum.2.020321, PhysRevB.107.L140410}.

Digital methods capable of determining low energy states using only shallow-depth quantum circuits are of particular interest for simulation of statics in the NISQ-era, with the variational quantum eigensolver (VQE)~\cite{Peruzzo2014} and the quantum approximate optimization algorithm (QAOA)~\cite{farhi2014quantum} being two examples that are particularly well suited to near-term digital oscillator-qubit hardware~\cite{C2QA-LGTpaper,zhang2023energydependent,zhang2023sequential}.  Annealing, or adiabatic state preparation approaches can also be used to solve such problems or as intuition for an ansatz for QAOA or VQE~\cite{farhi2014quantum,morita2008mathematical}.  { However, the cost of these methods scales with the energy gap between the ground state and the first excited state.  This means that while adiabatic algorithms can be used to construct arbitrarily accurate  approximations to the quantum ground state for gapped systems, the fact that the excitation gap  typically shrinks exponentially with the system size causes such approaches to be inefficient in general.}

In the fault-tolerant regime, we do not usually assume that fundamental limitations exist on gate depth and several qubits/qumodes.  This allows us to use alternative methods that have exact bounds on success probability and fidelity (or provable speedups) that can be implemented on hybrid oscillator-qubit hardware, with the instruction sets presented in this work enabling the implementation of controlled-unitaries implementing $e^{-i\theta H}$ or various energy filter functions. These include quantum phase estimation (QPE) \cite{kitaev1995quantum} and quantum signal processing (QSP)~\cite{GrandUnificationAlgos} discussed in Sec.~\ref{sec:bivariable-qsp} and applied to Hamiltonian simulation in hybrid qubit-oscillator systems in \cite{C2QA-LGTpaper}.

With some modifications, the above preparation methods can be used to prepare either ground or excited states. For example, in the context of variational methods, instead of minimizing $\langle\hat{H}\rangle$ to target the ground state, we can instead
minimize $\langle (E-\hat{H})^2\rangle$ with $E$ the estimated energy of a target excited state. In the context of fault-tolerant algorithms, QPE for example can be used to prepare excited states as long as the initial trial state has non-negligible overlap with the target excited state of interest\cite{lee2023evaluating}. In the case of QSP or qubitization, a filter function centering at the excited state energy can be implemented to facilitate excited state preparation \cite{GrandUnificationAlgos}.

\subsubsection{Observables}\label{observablessection}
Extracting useful information from a quantum state requires the ability to make measurements and estimate observables. These range from local observables, such as the energy density of a particular state or a time-ordered correlation function of a spin, to nonlocal observables crucial to the identification of exotic phases of matter and topological properties ~\cite{pollmann2012symmetry,Kitaev2006, levin2006detecting}.   For qubit systems, measurements only in the computational basis are inadequate to determine the complex phase factors in superposition states that determine the expectation values of observables that have off-diagonal components. Thus, if measurements in the computational basis are the only ones possible, it is essential to be able to rotate the state before measurement to map off-diagonal observables into the computational basis, thereby effectively creating the ability to measure in other bases. Similarly, in CV systems, one needs to be able to make measurements in various bases (e.g., the Fock (boson number) basis and in the position and momentum quadrature bases).

Using the gate sets and techniques presented in this work, a wide range of observables can be measured in hybrid oscillator-qubit hardware. We note that, particularly in the NISQ-era, error mitigation strategies \cite{endo2018practical, cai2023quantum} such as zero-noise extrapolation (ZNE) and probabilistic error cancellation (PEC) play an important role in the determination of observables from noisy CV \cite{su2020error,Taylor_2024} and DV \cite{kim2023evidence} hardware. Separately, classical shadow methods \cite{huang2020predicting} -- a class of techniques to estimate observables from a small number of measurements -- have garnered much recent interest in the literature and have recently been extended to CV systems \cite{gandhari2022continuous, becker2022classical}. We leave a more complete description of these techniques to the above-cited works, and here simply note their importance for the estimation of observables. 

In the following paragraphs, we provide a few illustrative (but non-exhaustive) examples of how one can measure Hamiltonian terms of interest in hybrid oscillator-qubit hardware.

\subsubsubsection{Number Operator.}
The occupation of a mode can be measured in multiple ways. First, one can probe whether or not a mode is in a particular Fock state $n$ by, for example, using an \textrm{SQR} gate (Box~\ref{box:SQRgate}) and encoding the answer in the state of a qubit. However, asking this question for all $n$ (up to some cutoff $n_{\textrm{max}}$) is costly, requiring a number of measurements that is linear in $n_{\textrm{max}}$. Alternatively, it has been demonstrated experimentally \cite{Wang2020FCFs} that one can make a binary search using successive Hadamard tests with the parity operator, where the $k$th round reveals the $k$th bit of $n$ (i.e., $\lfloor n/2^{k}\rfloor \textrm{ mod } 2)$ via phase-kickback on the qubit \cite{Wang2020FCFs}. This latter approach scales logarithmically in $n_{\textrm{max}}$. For details, see App.~\ref{app:measurement-tomography}. See also App.~\ref{SNAPECDCompilation} where this approach is recompiled into the phase-space ISA.

Either technique by extension enables the measurement of operators that are functions of $n$, such as the chemical potential $\langle n \rangle$, onsite interaction $\langle n^2 \rangle$, or other operator strings that include $\hat{n}$ such as the Hamiltonian terms discussed in Sec.~\ref{sssec:many-spins}.

\subsubsubsection{Field Operators.}
\label{sssec:fieldops}
We can directly synthesize measurements of the resonator field operators $a=\hat x+i\hat p$ (here in Wigner units) via measurements of $\hat x$ and $\hat p$ \cite{strandberg2023digital}.  To achieve this, recall from Eqs.~(\ref{wz_ecd}-\ref{eq:hatthetaquantumangle}) that a conditional momentum boost $\mathrm{CD}(ik)$ (Box~\ref{Box:c-displacement}) 
can be viewed either as a momentum boost $\pm k$ conditioned on $\sigma_z$ of the qubit or as a qubit rotation about the $z$ axis by an angle $\theta = -4k\hat x$ that is linearly dependent on the position of the oscillator. Taking the latter viewpoint, by starting the ancilla in the $|+\rangle$ state and then measuring $\sigma_y$ after application of the conditional momentum boost, one obtains a one-bit estimate of $\sin (4k\hat x)$.  For sufficiently small $k$ such that the sine can be linearized, this is essentially a homodyne measurement of $\langle\hat x\rangle$. This is similar to what has been achieved for flying microwave photons using traditional microwave homodyne techniques (and huge amounts of careful averaging to eliminate amplifier/detector noise) \cite{Eichler_PhysRevLett.106.220503}. In Ref.~\cite{C2QA-LGTpaper}, it was shown that this technique can be leveraged to measure non-local observables such as string order correlators.

\subsubsubsection{Hopping Terms.}
Many bosonic models include terms of the form $T_{ij} \propto  a^\dagger_ia_j + a_i a^\dagger_j$ describing hopping of a single particle from site $i$ to site $j$. Notably, such a term exactly generates a beam-splitter gate (Box \ref{Box:beam-splitter}). To measure this term, we note that conjugation by an appropriately chosen beam-splitter  gate maps $T_{ij}$ onto the photon number difference between sites $i$ and $j$ via what is effectively interferometry,
\begin{eqnarray}
\mathrm{BS}^\dagger\left(\frac{\pi}{2},-\frac{\pi}{2}\right )T_{ij}\mathrm{BS}\left(\frac{\pi}{2},-\frac{\pi}{2}\right ) &=& a^\dagger_i a^{\phantom{\dagger}}_i - a^\dagger_j a^{\phantom{\dagger}}_j.
\end{eqnarray}
Consequently, measurement of $T_{ij}$ is achieved by first applying the beam-splitter $\mathrm{BS}^\dagger\left(\pi/2,-\pi/2\right )$ followed by measurement of the number operator at sites $i$ and $j$, in close analogy to the combination of a Hadamard gate and $Z$-basis measurement facilitating an $X$-basis measurement for qubits.

\subsubsubsection{Bosonic Interaction Terms.}
\label{sssec:bosonicinteractionterms}
Another useful capability is the measurement of interaction terms involving multiple modes. For example, consider an interaction term of the form $V_{ij} \propto a_i^\dagger a_i a_j^\dagger a_j$. To measure $V_{ij}$, we can proceed in multiple ways. First, we can use the previously described strategy for measuring the number operator to independently determine $n_i$ and $n_j$ and, over repeated shots, correlate the results to determine $\langle n_i n_j\rangle$ (which, importantly, differs from $\langle n_i\rangle \langle n_j\rangle$ when $n_i$ and $n_j$ are correlated). While sufficient as a terminal measurement (i.e., at the very end of a circuit), this strategy extracts more information than needed, collapsing each mode onto a Fock state in the process. In situations where we wish to measure $V_{ij}$ without disentangling modes $i$ and $j$ (for example, if we expect $V_{ij}$ to be a conserved quantity and want to check this mid-experiment), the conditional cross-Kerr gate synthesized in Eq.~(\ref{eq:ccrosskerr}) provides an alternative approach strategy, as it rotates an ancilla by an amount proportional to $V_{ij}$. In principle, one could leverage this to carry out phase estimation to do an efficient binary search for the value of the integer $n_in_j$. However, in practice, this requires repeated fast and accurate synthesis of the cross-Kerr unitary in Eq.~(\ref{eq:ccrosskerr}) for a variety of different rotation angles.

Next, we provide several explicit examples of physically interesting models that can be implemented efficiently using hybrid oscillator-qubit platforms.

\subsubsection{One Qubit, One Mode: The Rabi Model}
\label{sssec:Rabi}
One of the simplest yet most profound models in quantum physics is the Rabi model \cite{Rabi1937, Braak2011}. It describes the interactions between a harmonic oscillator and a single spin-1/2 system. Its applications are wide-ranging, from the fundamental description of quantum optics and molecular physics to real-world technologies such as nuclear magnetic imaging. Along with the Jaynes-Cummings model ~\cite{JaynesCummings} (to which, as discussed below, the Rabi model reduces in a certain parameter regime), it is widely used to describe light-matter interactions under the dipole approximation and consequently serves as the foundation for the fields of cavity and circuit quantum electrodynamics -- the latter of which forms the very basis for the superconducting hardware considered in this paper \cite{Blais2004, BlaiscQEDReviewRMP2020}.

The Rabi model is described by the Hamiltonian
\begin{equation}
    H_{\textrm{Rabi}}=\hbar\omega_m a^{\dagger} a+\frac{\hbar \omega_q}{2} \sigma_z + g\left(a^\dagger +a\right) \sigma_x,
\end{equation}
where $\omega_m$ and $\omega_q$ are the frequencies of the oscillator and the two-level system, respectively, while $g$ is the coupling strength between the two, here taken (via a gauge choice) to be real. The first and second terms of $H_{\textrm{Rabi}}$ correspond to the free evolution of the oscillator and qubit, while the final term describes interactions between the two. This is, to a good approximation, the natural physical Hamiltonian natively available (and always `on') in circuit QED settings.  However, we may wish to simulate a Rabi Hamiltonian with different (programmable and possibly time-varying) parameters than are natively available in circuit QED setups, trapped ions, or neutral atoms.  Doing so within the instruction sets presented in this work is straightforward: the first two terms can be implemented via oscillator and single-qubit rotations (in software via simple rotating frame changes), respectively, while the last is exactly the generator of a conditional displacement, $\mathrm{CD}(\alpha)$, up to single qubit rotations. 

As discussed in App.~\ref{app:PhysImp}, in the parameter regime where $g \ll \{\omega_m, \omega_q\}$, one can make the replacement $\sigma_x \to \sigma_+ + \sigma_-$ and discard the rapidly-rotating, non-energy-conserving terms, $a^\dagger \sigma_+$ and $a \sigma_-$, under what is known as the rotating-wave approximation. This results in the comparatively simpler Jaynes-Cummings model,
\begin{equation}
    H_{\textrm{JC}}=\hbar \omega_m a^{\dagger} a+\hbar \omega_b \sigma_z + g\left(a\sigma_+ + a^{\dagger} \sigma_-\right),
\end{equation}
which reduces light-matter interactions to the number-conserving exchange of mode and qubit excitations, e.g., absorption ($a\sigma^+$) and emission ($a^\dagger \sigma^-$) of a single photon. However, as one increases the coupling such that $g \ll \{\omega_m, \omega_q\}$ no longer holds, the system enters the so-called ultra-strong coupling regime ($g \gtrsim 0.1 \{\omega_m, \omega_q\}$) and, for even larger $g$, the deep strong coupling regime ($g \gtrsim  \{\omega_m, \omega_q\}$). It is predicted that these parameter regimes display exotic behavior such as ground state virtual excitations, and as such have been the topic of both theoretical and experimental investigations over the past decade \cite{Kockum2019, forn2019ultrastrong}. 

\subsubsection{Many Modes: The Bose-Hubbard Model and the Fractional Quantum Hall Effect}
\label{BoseHubbard}

Several models in condensed matter, optical, and nuclear physics are purely bosonic and therefore well-suited for simulation using hardware with native bosonic modes. For example, boson sampling, in which  photons pass through a linear-optical network before measurement in the number basis, is a candidate for quantum advantage~\cite{Aaronson2013,PhysRevLett.131.150601,thekkadath2022experimental,zhong2019experimental,zhong2020quantum}. Furthermore, the Bose-Hubbard model has become a testbed for phenomena such as the well-known superfluid-to-Mott-insulator transition, whose first observation spurred forward the field of quantum simulation with cold atoms in optical lattices~\cite{Greiner2002}. Moreover, it is conjectured that the Bose-Hubbard model hosts topological phases with novel dynamics akin to the ones encountered in the fractional quantum Hall effect~\cite{girvin1999LesHouches_FQHE} (FQHE) for electrons and which could be observed in circuit QED systems \cite{PhysRevA.87.062336,PhysRevB.90.060503,Owens_FQHE_2018,owens_chiral_2022,10.21468/SciPostPhys.13.5.107}. 

While these models can often be implemented in analog quantum simulators, such approaches are limited in terms of Hamiltonian terms that can be realized and observables that can be measured. For example, the complex phases in the hopping term that incorporate effects of an external magnetic field (required, for example, in the description of the FQHE) are challenging to implement in cold atoms due to their neutrality. Synthetic magnetic fields have been employed in past work ~\cite{Aidelsburger2016}, but an FQHE-like state has only so far been observed in a few small-scale experiments in cold atom ~\cite{Leonard_FQHE_2022} and cavity/circuit QED platforms~\cite{Clark2020, Owens_FQHE_2018,wang2024realization}. Furthermore, entanglement-based measures, important for verifying the topological properties of candidate FQHE states~\cite{levin2006detecting,Kitaev2006}, are challenging to obtain; analog cold atom experiments, for example, have typically relied on measurement of occupations which supplies no information regarding coherences~\cite{Leonard_FQHE_2022} (though we note a recent proposal for measuring arbitrary properties in analog simulators~\cite{tran2022measuring}).

Digital simulation with oscillator-qubit hardware presents advantages on all of these fronts. Not only does the hardware natively support bosonic modes and operations, but arbitrary Hamiltonian terms, such as magnetic field terms crucial to the observation of the FQHE can be synthesized by adjusting the phases of the beam-splitter couplings on each link of the lattice, using the instruction set architecture presented in this work. The programmability of such simulators allows one to easily tune Hamiltonian parameters and synthesize additional terms. Finally, as previously discussed, digital hardware offers the ability to measure a wide range of observables that otherwise present a challenge in analog simulations.

\begin{figure}[htb]
    \centering
    \includegraphics[width=0.98\linewidth]{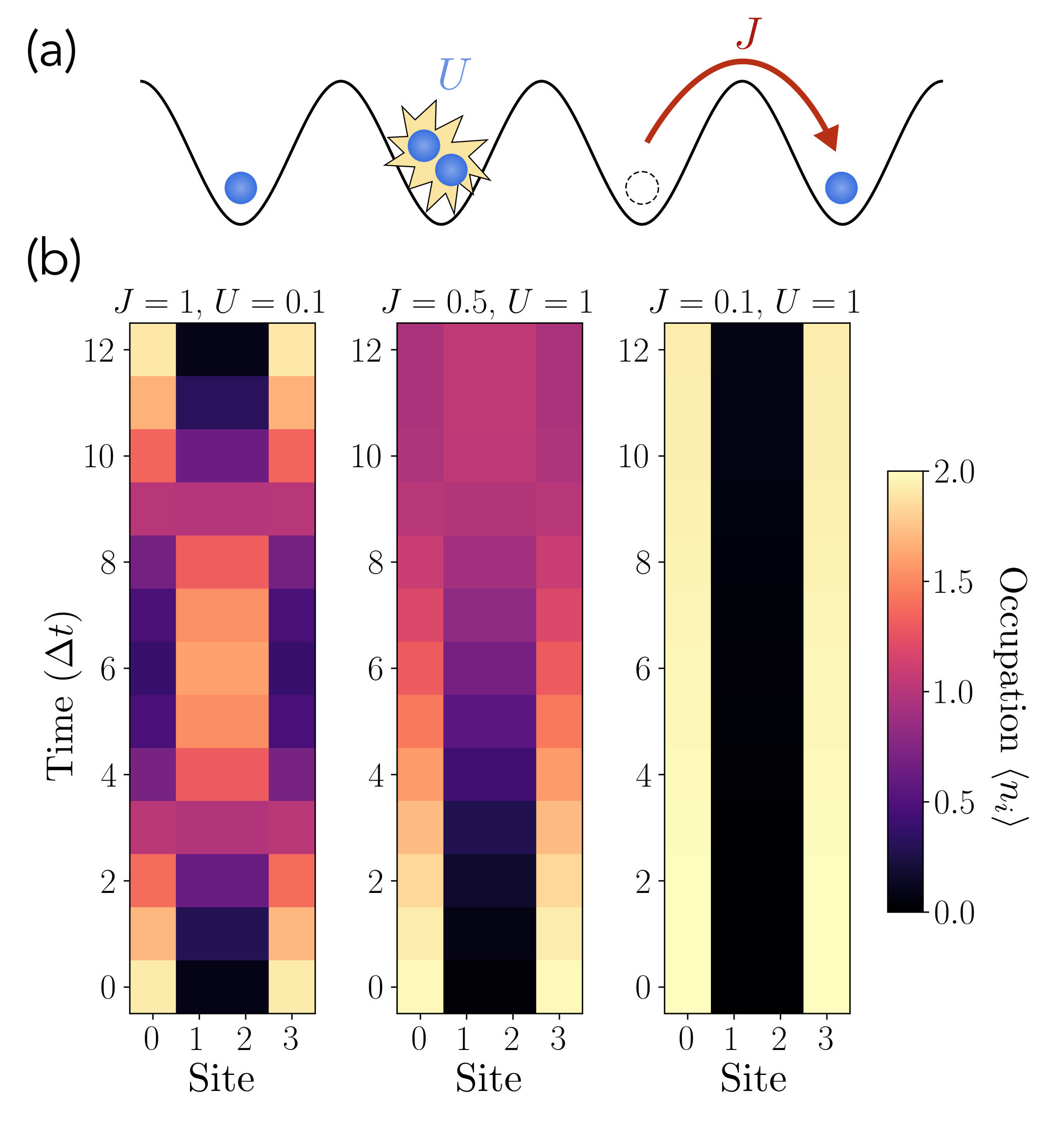}
    \caption{Simulation of a four-site Bose-Hubbard model using Bosonic Qiskit. (a) A schematic of the Bose-Hubbard model, illustrating nearest-neighbor hopping with strength $J$ and on-site repulsion of strength $U>0$. This competition between these Hamiltonian terms leads to a superfluid-to-Mott-insulator phase transition \cite{Greiner2002}. (b) Occupancy as a function of time for various choices of $J$ and $U$. In all cases, a time step of $\Delta t = 0.2$ was used and the system was initialized to have two bosons at each edge site, with intermediate sites vacant.}
    \label{fig:BoseHubbard}
\end{figure}

With the above motivations in mind, we now turn explicitly to the Bose-Hubbard model as an illustrative example. Its Hamiltonian (for zero magnetic field) is given by
\begin{equation}
    \begin{split}
        H_{\textrm{BH}}&=-\mu \sum_i \hat{n}_i + \frac{U}{2} \sum_i \hat{n}_i \left( \hat{n}_i - 1 \right) + H_{\textrm{hop}} \\ H_{\textrm{hop}} &= -J\sum_i \left( a_i^{\dagger}a_{i+1} + \mathrm{h.c.} \right ),
    \end{split}
\end{equation}
where $\mu$ is the chemical potential, $J$ the hopping strength, and $U$ the interaction strength (with $U>1$ repulsive and $U<1$ attractive). Time evolution $e^{-i H_i \Delta t}$ under each term $H_i$ for a duration $\Delta t$ can be directly realized in the Fock-space ISA~\cite{Biskit, C2QA-LGTpaper}: $H_{\textrm{hop}}$ generates a beam-splitter $\mathrm{BS}(-2J\Delta t ,0)$, the chemical potential term is realized using a phase-space rotation gate $R(-\mu \Delta t)$, and the onsite interactions can be implemented using a SNAP gate (Box \ref{Box:SNAP}) by choosing $\varphi_n = Un(n-1)\Delta t/2 $, such that 
\begin{equation}
    \begin{split}
        \textrm{SNAP}(\varphi_0, \varphi_1, \varphi_2, \ldots) &= e^{-i\sum_n U n (n-1) \Delta t \ket{n}\bra{n}/2}\\
        &= e^{-i\sum_n U \hat{n} (\hat{n}-1) \Delta t/2}.
    \end{split}
\end{equation}
As a simple illustration, Fig.~\ref{fig:BoseHubbard} displays the Trotterized evolution of the mode occupation implemented in Bosonic Qiskit \cite{Biskit}. 

While here we have simplified the case where $U$, $\mu$, and $J$ are identical across all sites, we note that additional interesting possibilities emerge by varying these parameters across the lattice. For example, it has been shown in cold atomic platforms that the 1D Bose-Hubbard model with the added disorder in the chemical potential $\mu_i$ leads to many-body localization~\cite{Lukin2019}. Separately, alternation of the hopping strength $J_i$ leads to an (interacting) extension of the Su–Schrieffer–Heeger model, known to display topological features such as edge states~\cite{Su1979}.

Moreover, magnetic fields and strong interactions in the Bose-Hubbard model give rise to fractionally charged quasi-particles, long-ranged entanglement, and anyonic exchange statistics~\cite{girvin1999LesHouches_FQHE,palmer2006high}. In this direction, the selection of the beam-splitter phase $\phi$ allows one to implement hopping in a magnetic field, thus facilitating potential investigation of the FQHE. In all, this multitude of possibilities constitutes a rich and promising playground for quantum simulation of many-body bosonic models using hybrid oscillator-qubit hardware.

\subsubsection{Large Spin Models: The Transverse-field Ising Model} \label{sssec:many-spins}

Several large-spin models, i.e., those with spin-$S$ degrees of freedom with $S>1/2$, are of interest in condensed matter physics~\cite{Affleck1987,KevinAKLT_PRXQuantum.4.020315} and chemistry~\cite{Tabrizi2016}. While individual qubits are naturally suited to represent spin-1/2 degrees of freedom, encoding a generic spin-$S$ system in qubit-based hardware requires $\log(2S+1)$ qubits. Though not prohibitively expensive in terms of qubit resources, this approach comes at an additional cost in the two-qubit gate count required to implement both single- and multi-spin-$S$ operations, prohibiting straightforward simulation, particularly on near-term devices. 

In contrast, in hybrid oscillator-qubit hardware, one can make use of the larger Hilbert space of the bosonic modes to represent $d$-level systems in a hardware-efficient manner~\cite{C2QA-LGTpaper, QuditsfromOscillatorsPhysRevA.104.032605}. Here we discuss the Schwinger-boson representation, an encoding that \emph{directly} implements a spin-$S$ and its defining commutation relations in terms of simple combinations of bosonic operators -- a possibility first noted by Julian Schwinger \cite{schwinger2001angular}. Previously, the Schwinger-boson representation has been proposed for the realization of large spin $U(1)$ lattice gauge theories using qubit-oscillator hardware \cite{Wiese2013, C2QA-LGTpaper}. Separately, it is the basis for the oscillator-based dual-rail representation of a qubit in, for example, circuit QED experiments~\cite{teoh2022dualrail}.

The core idea of the Schwinger-boson representation is the following: if one restricts to the subspace of total photon number $N$ (consisting of $N+1$ total states) for two oscillators with annihilation operators $a$ and $b$, then the combinations $a^\dagger b$ and $a b^\dagger$ act \emph{exactly} as spin-$S$ raising and lowering operators $S^+$ and $S^-$ within this subspace, with the total spin being given by $S = N/2$. For example, if we take $N=1$, this subsystem consists of the Fock states $\ket{1,0}$ and $\ket{0,1}$, with $S^+ = a^\dagger b$ and $S^- = a b^\dagger$ acting exactly as raising and lowering operators for an effective spin-1/2 system. Notably, this case is precisely the idea behind the dual-rail qubit in a pair of oscillators (see Sec.~\ref{sssec:multi-mode-encodings}). 

More generally, one can represent spin-$S$ systems of arbitrary $S$ by restricting to the subspace of $N = 2S$ photons distributed between a pair of oscillators (representing a spin-$S$ in a single oscillator is also possible \cite{roy2024synthetic,champion2024multifrequency}). Formally, the mapping is as follows:
\begin{equation}
    \begin{split}
        S^x &= (a^\dagger b +ab^\dagger)/2\\
        S^y &= (a^\dagger b-a b^\dagger)/2i\\
        S^z &= (a^\dagger a - b^\dagger b)/2,\\
    \end{split}
\end{equation}
which exactly obeys the commutation relations
\begin{equation}
    [S^x,S^y]=iS^z \,\,\,\mathrm{et\ cyc.}
\end{equation}
We also see that the beam-splitter (Box \ref{Box:beam-splitter}) can be written as
\begin{equation}
    \mathrm{BS}(\theta, \phi) = e^{-i\theta [\cos(\phi)S_x - \sin(\phi)S_y]}. \label{eq:BS_SxSy}
\end{equation}
Beam-splitters alone therefore give us access to two generators of rotations and allow for arbitrary SU(2) rotations of the two-mode qudit. In particular, we note that rotations based on the third SU(2) generator, $S^z$, can be implemented using a pair of beam splitter gates 
\begin{align}
  e^{-i\phi S^z} =  \mathrm{BS}(\pi, 0)\mathrm{BS}(\pi, \phi).
\end{align}
This could also be implemented `in software' with phase-space rotations (see Box \ref{Box:phase-space-rotation}). 

For the case of a spin-$1/2$ ($N=1$), the two modes form a two-level qubit and SU(2) rotations consequently provide universal single-spin operations. 
However, for $S>1/2$ ($N>1$), the most general unitary single-spin operation lives in SU($d$) with $d=N+1>2$, and therefore beam splitters alone do not provide universal control. Instead, we need additional hybrid oscillator-qubit operators to synthesize generators nonlinear in the spin operators. To that end, one can recognize that any polynomial of $S^z$ is diagonal in the Fock basis. Consequently, it is possible to realize nonlinear unitaries of the form $U=\textrm{exp}\,(-i\theta f(S^z))$ via a single SNAP gate on mode $a$ which was proposed in~\cite{Biskit, C2QA-LGTpaper}. Here, $f(S^z)$ is an arbitrary polynomial in $S^z$, and we have additionally used the fact that $S^z = (a^\dagger a - b^\dagger b)/2 = a^\dagger a + N/2$. Together with beam-splitter enabled $SU(2)$ rotations, this nonlinear gate provides universal control for $SU(d)$.  This is analogous to the universal control of high-spin atoms in atomic physics utilizing the nonlinear Zeeman effect \cite{PhysRevA.68.062320,PhysRevLett.99.163002,PhysRevLett.114.240401}.

Measurement of the spin-$S$ system in the $S^z$ basis has a natural implementation in the Fock-space ISA -- one must simply resolve the number operator $a^\dagger a$ (or $b^\dagger b$), which can then be used to determine $S^z$ (see App. \ref{app:measurement-tomography} for details on photon number measurement). By extension, this ability enables the measurement of the spin along any direction, $\hat{n}\cdot \vec{S}$, by first rotating to the $\hat{z}$-basis with an appropriate beam-splitter (much like a Hadamard gate prepending a $Z$-basis measurement enables an $X$-basis measurement for a qubit). With this primitive, one can more generally carry out quantum state tomography on a spin-$S$ system by measuring $\hat{n}\cdot \vec{S}$ along a minimum of $4S + 1$ distinct directions of $\hat{n}$ \cite{newton1968measurability}. To reduce statistical uncertainty in the state construction, however, it is highly beneficial to increase the number of distinct measurement axes beyond the bare minimum \cite{perlin2021spin}. Notably, the tomographic protocols described here require only beam-splitters and photon-number measurements, independent of the value of $S$.

As a simple example of how the Schwinger-boson representation can be leveraged for quantum simulation, let us consider the 1D transverse-field Ising model of spin-$S$. The model Hamiltonian is
\begin{align}
H_{\textrm{TFIM}} = -\sum_i S_i^z S_{i+1}^z - \lambda\sum_i S^x_i,
\end{align}
where the first and second terms correspond to spin-spin interactions and coupling to a transverse field, respectively. To implement this model, we envision that the spin at site $i$ is composed of two oscillators with corresponding annihilation operators $a_i$ and $b_i$. Unitary evolution under the transverse field terms can be implemented via beam-splitter gates of the form $\mathrm{BS}(\theta, 0)$ using Eq.~(\ref{eq:BS_SxSy}). The spin-spin term, on the other hand, requires a non-Gaussian gate due to being quadratic in spin operators. To see how to proceed, we can first recast in terms of bosonic operators,
\begin{equation}
        S_i^z S_{i+1}^z = \frac{1}{4}(a_i^\dagger a_{i} - b_i^\dagger b_{i})(a_{i+1}^\dagger a_{i+1} - b_{i+1}^\dagger b_{i+1}),
\end{equation}
and note that the four above terms commute, such that
\begin{equation}
    \begin{split}
        e^{-i\theta S_i^z S_{i+1}^z} &= e^{-i\frac{\theta}{4}\hat{n}_{a_i} \hat{n}_{a_{i+1}}}
        e^{-i\frac{\theta}{4}\hat{n}_{b_i} \hat{n}_{b_{i+1}}}\\
        &\times e^{i\frac{\theta}{4}\hat{n}_{a_i} \hat{n}_{b_{i+1}}}
        e^{-i\frac{\theta}{4}\hat{n}_{b_i} \hat{n}_{a_{i+1}}}.
    \end{split}
\end{equation}
Taking advantage of the fact that $S=(\hat n_{a_j}+\hat n_{b_j})/2$ is conserved, we can simplify this to
\begin{align}
e^{-i\theta S_i^z S_{i+1}^z} &=e^{-i\theta (\hat n_{a_i}-S)(\hat n_{a_{i+1}}-S)}.
\end{align}
Therefore, the synthesis of $\textrm{exp}(-i\theta S_i^z S^z_{i+1})$ reduces to that of generators that are products of number operators of distinct oscillators. Although challenging to implement exactly, we can note that each unitary is equivalent to the conditional cross-Kerr gate approximately synthesized using BCH techniques in Eq.~(\ref{eq:ccrosskerr}) if we use an auxiliary qubit in the state $\ket{0}$. Thus, the digital simulation of the transverse-field Ising model is straightforward using the gate sets and compilation techniques presented in this work.

The strategy of using the Schwinger boson representation to simulate spin systems is of much wider interest than this simple example, with possible applications of classical and quantum spin models spanning the fields of magnetism \cite{girvin2019modern}, materials \cite{evans2014atomistic}, artificial intelligence \cite{hopfield1982neural}, and even neural processing \cite{fisher2015quantum}. Furthermore, because each spin is entirely encoded in a photon-number-preserving subspace, the possibility of detecting errors emerges; similarly to the dual-rail qubit, one can detect photon loss errors by making measurements of the joint parity in the corresponding pair of oscillators \cite{teoh2022dualrail, tsunoda2023error}. Thus, studying spin models using the Schwinger boson representation in hybrid oscillator-qubit hardware not only provides an avenue for direct study of a wide range of interesting models but does so in a particularly promising NISQ-friendly manner by offering a degree of noise resilience through post-selection. For an in-depth example, see Ref.~\cite{C2QA-LGTpaper} on implementing a $U(1)$ lattice gauge theory in hybrid qubit-oscillator hardware.

\subsubsection{Many Qubits, Many Modes: $\mathbb{Z}_2$ Lattice Gauge Theory}\label{Z2section}

A variety of interesting Hamiltonians include both qubits and bosonic modes and are thus a natural fit for the hybrid oscillator-qubit hardware. Examples include multispin-boson~\cite{winter2014quantum} models, bosonic lattice gauge theories~\cite{Schweizer2019}, the Jaynes-Cummings-Hubbard model describing many-body states of (dressed) light \cite{schmidt2009strong, smith2021exact}, and the exciton-vibrational Hamiltonians governing energy transfer and relaxation pathways in molecules and materials \cite{kundu2022intramolecular,wasielewski2020exploiting}. Such Hamiltonians are challenging to implement both on qubit-based hardware, due to the inefficiency of implementing bosonic gates (see Sec.~\ref{app:compilation-cv-to-qubits} and Ref.~\cite{C2QA-LGTpaper}), and analog quantum simulators, due to the multi-species nature of the models and the lack of universality of the operations. Hybrid oscillator-qubit platforms provide a promising hardware-efficient approach for the quantum simulation of these Hamiltonians due to the ease with which multi-oscillator, multi-qubit, and hybrid oscillator-qubit gates can be implemented.

\begin{figure}[htb]
\centering
\includegraphics[width=\columnwidth]{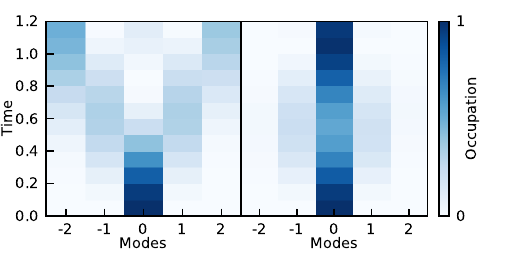}
\caption{\label{fig_trotterisation_LGT}
Implementation of the $\mathbb{Z}_2$ lattice gauge theory in Bosonic Qiskit~\cite{Biskit}. Both panels display the average occupation at each site as a function of time. The initial state is chosen such that there is one boson at the central site, and all others vacant. For Gauss's law, we fix $G_j=+1$. In the left panel, we show results for $g/J = 0.1$ (hopping-dominated) and observe a light-cone spread of the boson. In the right panel, we show results for $g/J=5$ (field-dominated) and observe that the boson is (mostly) trapped at the central site. In both cases, a step size of 0.1$J$ was used.}
\end{figure}

Here, we take as an illustrative example a 1D bosonic $\mathbb{Z}_2$ lattice gauge theory~\cite{Wiese2013}. The implementation of this model in the proposed hybrid oscillator-qubit architecture, along with its extension to 2D, was studied in Ref.~\cite{C2QA-LGTpaper}. Here, we provide a high-level review of the general strategy and refer the reader to the aforementioned work for an in-depth discussion.

The $\mathbb{Z}_2$ lattice gauge model describes bosonic matter that hops along a 1D chain of sites alternated by qubits that represent a $\mathbb{Z}_2$ gauge field living on the links of the lattice. The Hamiltonian of this model is given by
\begin{equation}
    \begin{split}
        H_{\textrm{LGT}}&= H_{\textrm{hop}} + H_{\textrm{field}}\\
        H_{\textrm{hop}} &= -J \sum_i a^\dagger_i Z_i a_{i+1} + a_i Z_i a_{i+1}^\dagger\\
        H_{\textrm{field}} &=  -g \sum_i X_i,
    \end{split}
\end{equation}
where $a_i$ ($a_i^\dagger$) removes (adds) a particle at site $i$, and $X_i$ and $Z_i$ are Pauli operators for the qubit at link $i$ (which connects site $i$ and site $i+1$). To better understand this model (and the competition between $H_{\textrm{hop}}$ and $H_{\textrm{field}}$ it describes), it is helpful to first imagine all qubits are initialized in an $X$ eigenstate. $H_{\textrm{field}}$ describes the energy of the field and is minimized by aligning all qubits in the $+X$ direction. In contrast, $H_{\textrm{hop}}$ describes bosons hopping between sites, causing a phase-flip (sending $+X$ to $-X$ and vice-versa) at each intermediary link, altering the field energy. Consequently, the model displays drastically different behavior in opposite parameter regimes, with $|J/g| \gg 1$ leading to a ``deconfined'' dynamics where matter freely hops (or more precisely, the confinement length greatly exceeds the finite sample size), and $|J/g| \ll 1$ leading to ``confined'' dynamics where the large field inhibits hopping~\cite{Schweizer2019}. This is a statement about the dynamics - the ground-state phase is always confining for $g/J>0$~\cite{Kebric2021}.

Much like in electrodynamics, we can define an analog of Gauss's law that relates the number of bosons (or \emph{charges}) at site $i$ with the product of ``left'' and ``right'' field values $X_i X_{i+1}$ (akin to a field integral in electrodynamics). In particular, we can identify this gauge symmetry with the generators
\begin{equation}
    \hat{G}_j = X_j e^{i\pi a^\dagger_j a^{\phantom{\dagger}}_j} X_{j+1}.
\end{equation}

To see that the generators $\hat{G}_j$ correspond to a conserved quantity, note that when a boson hops from site $j$ to site $j+1$, the factors of $\hat{G}_j$ change as follows: (1) $a^\dagger_j a_j$ is decremented by one, flipping the sign of the photon number parity $e^{i\pi a^\dagger_j a_j}$, and (2) the presence of $Z_{j}$ in the hopping amplitude flips the sign of $X_{j}$, leaving $\hat{G}_j$ unchanged overall. In other words, this means that $[H_{\textrm{hop}},\hat{G}_j]=0$. Noting that $\hat{G}_j$ trivially commutes with $H_{\textrm{field}}$, $\hat{G}_j$ is therefore a conserved quantity. Such a feature is advantageous for simulations on noisy hardware, as we can think of the generators $\hat{G}_j$ as stabilizers in an error-detecting code in the sense that we can perform error mitigation by post-selection of results that obey the Gauss law constraints.

Both $H_{\textrm{field}}$ and $H_{\textrm{hop}}$ can be implemented directly with gates previously discussed: $H_{\textrm{field}}$ is simply a single-qubit rotation, while $H_{\textrm{hop}}$ corresponds to a controlled beam-splitter. As a demonstration, we implement Trotterized time dynamics of $H_{\textrm{LGT}}$ in Bosonic Qiskit for both hopping-dominated and field-dominated regimes in a five-site system. We initialize a boson at the central site, with all others vacant. 

Fig.~\ref{fig_trotterisation_LGT} displays the resulting time-evolved occupancy at each site. The competing order between $J$ and $g$ is evident, as the hopping-dominated regime gives rise to a light-cone spread of a boson initialized at the central site, whereas the field-dominated regime prohibits hopping due to the prohibitive cost to flip link qubits. For a numerical simulation of dynamics and ground state preparation of this model and others in the presence of noise, see Ref.~\cite{C2QA-LGTpaper}.

\subsubsection{Outlook: Fermion-Boson Problems}
\label{sssec:fermion-boson-problems}

Classically intractable fermion-boson problems include non-equilibrium dynamics in QCD relevant to heavy-ion collisions~\cite{Ratti2018}, calculation of vibrational spectra of large anharmonic molecules coupled to an environment for identification of unknown samples~\cite{Cao2019}, and the study of phonon dynamics in strongly correlated materials such as high-temperature superconductors~\cite{FLORESLIVAS20201}. Hybrid oscillator-qubit hardware is well-suited to such problems, as the native oscillators naturally encode bosonic excitations, while fermionic degrees of freedom can either be mapped onto physical or logical qubits or to high-$Q$ oscillators \cite{dutta2024chemistry,dutta2024electronic}, depending on the specifics of the problem at hand. Furthermore, the instruction set architectures we have presented offer universal control, such that arbitrary fermionic and boson gates can be synthesized. 

A wide range of encodings can facilitate the mapping of fermions onto qubits. Well-known options include Jordan-Wigner~\cite{JordanWigner1928}, Bravyi-Kitaev~\cite{Bravyi2002}, Verstraete-Cirac~\cite{Verstraete2005}, and Derby-Klassen~\cite{Derby_2021} encodings, all of which enable the encoding of fermionic degrees of freedom using either logical qubits derived from CV-to-DV, DV-to-DV, or concatenated codes (see Sec. \ref{sec:bosonic-QEC}). For near-term applications, one could also use the physical qubits themselves. Notably, fermion-to-qubit mappings require non-local qubit-qubit gates which, in this case, can be implemented at the physical layer using the oscillator-assisted multi-qubit entangling gate techniques described in Sec.~\ref{ssec:compilation-entangling}. However, in the case of superconducting hardware, it is important to note that transmon qubit lifetimes are typically on the order of $\sim 100\mu$s, whereas 3D microwave cavity modes have far longer lifetimes on the order of $\sim1$ ms~\cite{Reagor_memory_2016} or more \cite{Rosenblum_TensofMilliseconds_PRXQuantum.4.030336,PhysRevApplied.13.034032}. Therefore, even without proper error correction, it may be advantageous to consider encoding the fermions into oscillators using, for example, the lowest-lying Fock states of cavity modes or dual-rail/Schwinger-boson qubits. 

As a simple demonstration, we now illustrate the compilation of the so-called fermionic simulation gate \cite{PhysRevLett.120.110501,Algaba_2024},
\begin{equation}
    \textrm{fSim}(\theta,\phi) = \left(\begin{matrix}1 & 0 & 0 & 0 \\
    0 & \cos(\theta) & -i\sin(\theta) & 0 \\ 0 & -i\sin(\theta) & \cos(\theta) & 0 \\ 0 & 0 & 0 & e^{-i\phi}\end{matrix}\right),
\end{equation}
a particular application of which is the fermionic-SWAP or $\textrm{fSWAP}$ gate, equivalent to a SWAP gate up to an additional minus sign applied to the $\ket{11}$ state to account for antisymmetric fermionic exchange statistics (see Table \ref{tab:gates-qubit}). Specializing to the case where we encode qubits in the $\ket{0}$ and $\ket{1}$ Fock states of a pair of oscillators indexed by $j$ and $k$, the fSim gate can be compiled as
\begin{equation}
    \begin{split}
    \label{eq:fSimFock}
        \textrm{fSim}(\theta,\phi) &=\textrm{exp}(-i\phi \sigma_z a_j^\dagger a_j a_k^\dagger a_k)\times\mathrm{BS}\left(2\theta,0 \right),
    \end{split}
\end{equation}
where the first term is a cross-Kerr gate, synthesized using BCH methods in Eq.~(\ref{eq:ccrosskerr}) and controlled on an ancillary qubit prepared in the state $\ket{0}$. Alternatively, one could replace this approximate cross-Kerr gate with a $\textrm{CPHASE}(\phi)$ gate realized using the methods of Ref.~\cite{tsunoda2023error}. 

The fermionic SWAP operator can be implemented via
\begin{equation}
    \begin{split}
        \textrm{fSWAP} &= \left[R\left(-\frac{\pi}{2}\right)\otimes R\left(-\frac{\pi}{2}\right)\right]\times\textrm{fSim}\left(\frac{\pi}{2},0\right)\\
        &= \begin{pmatrix}1 & 0 & 0 & 0 \\[-10pt]
    0 & 0 & 1 & 0 \\[-10pt] 0 & 1 & 0 & 0 \\[-10pt] 0 & 0 & 0 & -1\end{pmatrix},
    \end{split}
\end{equation}
where $R(\theta)$ corresponds to a phase-space rotation of the oscillator, and we have furthermore displayed only the portion of the unitary acting within the encoded two-qubit subspace. 

In summary, hybrid CV-DV hardware offers interesting opportunities for quantum simulation of coupled systems of spins, fermions, and bosons.
Here we have briefly discussed qubits and general spin $S$ qudits encoded in a pair of cavities using the Schwinger boson representation and presented a possible compilation of fermionic SWAP gates. 
For an in-depth manual for using CV-DV hardware for simulating fermions and bosons, including gauge theories in 2D, see Ref.~\cite{C2QA-LGTpaper}.

\subsection{Quantum Random Walks}
\label{ssec:randomwalk}

Quantum random walks (QRWs)~\cite{aharonov1993quantum} provide a generalization of classical random walks, and serve many purposes in quantum computation, quantum simulation and algorithm design.  They capture the physics of a particle moving in space over a graph or lattice, acting as a quantum analog of diffusion.  QRWs also provide a paradigm for quantum algorithms, offering speedups through reduced hit times of random walks, which can be polynomially (or sometimes exponentially) faster than their classical counterparts.  

Traditional quantum random walks employ qubits, but using an oscillator in place of qubits can offer unique advantages and access to richer dynamics.  Specifically, random walks often involve electromagnetic fields or motional modes which are naturally bosonic.  Walk dynamics with oscillators may also capture continuous-valued displacements, rather than discrete changes, making such hybrid quantum walks closer to analog quantum simulators than just qubit-based walks.  And a bosonic walker may represent many walker positions (in superposition), compared with a qubit walker.

Here, we briefly review for pedagogical purposes how QRWs 
with a bosonic random walker
can go beyond qubit systems and be realized using a hybrid CV-DV architecture.
In this setting, QRWs involve repeated
conditional displacements (CD$(\epsilon)$, see Box~\ref{Box:c-displacement}) of the oscillator, each followed by a multi-dimensional coin toss which decides the basis in which the auxiliary qubit is measured. In other words, the multi-dimensional coin toss decides the auxiliary qubit rotations about the vectors on the $XY$ plane of the Bloch sphere ($R_\phi(\theta)$, see Table~\ref{tab:gates-qubit}) before measuring in the $Z$ basis. Thus, the phase-space ISA can be exploited to straightforwardly implement QRWs in hybrid CV-DV quantum computing architectures. 

Ref.~\cite{aharonov1993quantum} showed that QRWs could be used to study non-trivial effects in quantum mechanics that arise due to quantum interference. For example, let us conditionally displace an oscillator by a very small amount $|\epsilon|\ll 1$. Using the definitions from Eq.~\eqref{eq:disp_cat} (but now including the normalization factors for the cat states), we can write,
\begin{align}
    &\mathrm{CD}(\epsilon)\ket{0}_\mathrm{osc}\ket{+}_\mathrm{qubit}\nonumber\\
    &=\left (  \rule{0pt}{2.4ex} \ket{\epsilon}\ket{0}\ +\ \ket{-\epsilon}\ket{1} \right)/\sqrt{2}\\
&=\frac{\mathcal{N}_{\text{even}}}{2}\ket{\mathcal{C}_{0,\epsilon,+}}_\mathrm{osc}\ket{+}_\mathrm{qubit}\  +\ \frac{\mathcal{N}_{\text{odd}}}{2}\ket{\mathcal{C}_{0,\epsilon,-}}_\mathrm{osc}\ket{-}\\
    &\approx \frac{\mathcal{N}_{\text{even}}}{2}\ket{0}_\mathrm{osc}\ket{+}_\mathrm{qubit}\ +\ \frac{\mathcal{N}_{\text{odd}}}{2}\ket{1}_\mathrm{osc}\ket{-}_\mathrm{qubit}~\label{eq:QRW}.
\end{align}
The final equation uses an approximation in which we have equated a small even (odd) cat state with Fock state $\ket{0}$ ($\ket{1}$). It is important to note that the definitions used here for cat states $\ket{\mathcal{C}_{0,\epsilon,\pm}}$ include their normalization prefactors, giving rise to the unequal amplitudes for each term in the last two equations. More specifically, the probability of each Fock state depends on the normalization constants of even ($\mathcal{N}_\text{even}$) and odd ($\mathcal{N}_\text{even}$) cat states of size $\epsilon$, 
\begin{align}
    \mathcal{N}_\text{odd}&=\sqrt{2(1-e^{-2\epsilon^2})},\\     \mathcal{N}_\text{even}&=\sqrt{2(1+e^{-2\epsilon^2})}.
\end{align}
The fidelity of even and odd cat states with the Fock states $\ket{0}$ and $\ket{1}$ improves with decreasing value of $\epsilon$ as long as $\epsilon\neq 0$, as shown by Eq.~\eqref{eq:fid0} and~\eqref{eq:fid1}.
\begin{align}
    F_0&=|\langle 0|\mathcal{C}_{0,\epsilon,+}\rangle |^2=4\frac{e^{-\epsilon^2}}{\mathcal{N}^2_\text{even}},\quad &\alpha\in\mathbb{R}~\label{eq:fid0},\\
    F_1&=|\langle 1|\mathcal{C}_{0,\epsilon,-}\rangle |^2=4\epsilon^{2}\frac{e^{-\epsilon^2}}{\mathcal{N}_\text{odd}^2},\quad &\alpha\in\mathbb{R}~\label{eq:fid1}.
\end{align}

\begin{figure}
    \centering
    \includegraphics[width=0.45\textwidth]{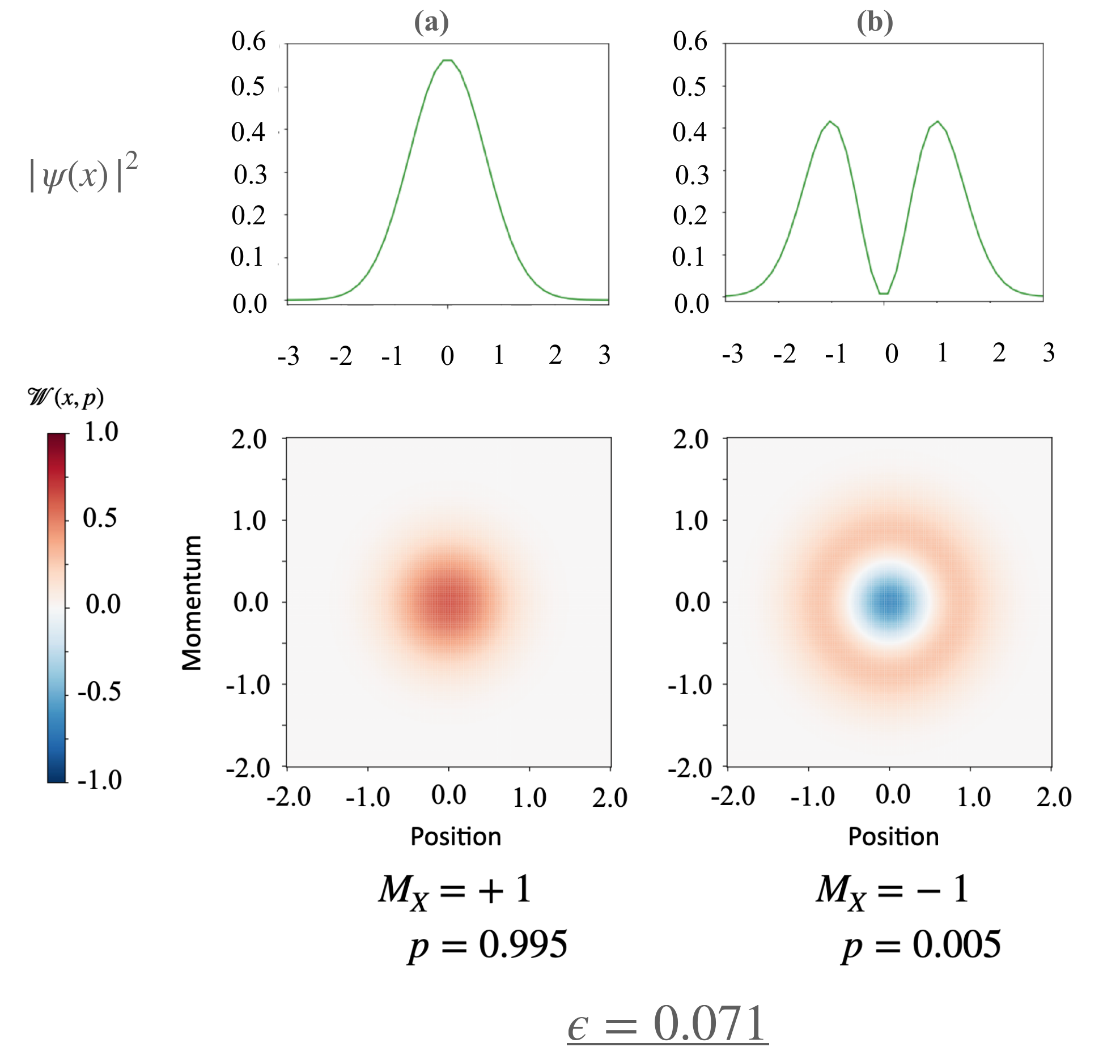}
    \caption{ Marginal probabilities ($|\psi(x)|^2$) and Wigner distribution ($\mathcal{W}(x,p)$) of the oscillator states entangled with orthogonal qubit states $\ket{\pm}$ after small conditional displacements $\mathrm{CD}(\epsilon=0.071)$ as shown in Eq.~\eqref{eq:QRW}. (Left) Oscillator state entangled with $\ket{+}$ qubit state. This state closely resembles the vacuum state $\ket{0}$  and occurs for qubit measurement $M_X=+1$ with a high probability $p=0.995$. (Right) Oscillator state entangled with $\ket{-}$ qubit state. This state is well-approximated by Fock state $\ket{1}$, and occurs for qubit measurement $M_X=-1$ with a low probability $p=0.005$. The large change in the position of the peaks, shown in the marginal probability distribution, between the two states is achieved merely by a small displacement of $\epsilon-0.071$. This phenomenon is an artifact of quantum interference and can be observed using quantum random walks as described in Ref.~\cite{aharonov1993quantum}. }
    \label{fig:QRW}
\end{figure}

Note that, for $\epsilon\rightarrow 0$, the small odd cat has a high overlap with the Fock $\ket{1}$ state (see Eq.~\eqref{eq:fid1}) in contrast to the small even cat which has a large overlap with $\ket{0}$ (see Eq.~\eqref{eq:fid0}) largely resembling the initial vacuum state. Thus, Eq.~\eqref{eq:QRW} shows that there can be a non-trivial unit increase in the photon number of the oscillator upon qubit measurement after an extremely small conditional displacement. We show the marginal distribution in position $|\psi(x)|^2$ and the phase-space quasi-probability distribution $\mathcal{W}(x,p)$ in Fig.~\ref{fig:QRW} for pictorial intuition. A qubit measurement in the $X$-basis collapses the oscillator onto one of the two states shown in this figure.

Let us discuss the consequences of this on a position basis. The even cat (left) resembles a vacuum and has its peak centered at $x=0$ while the odd cat (right) has two peaks centered at $x=\pm 1$. If the qubit measurement collapses the oscillator onto the odd cat state, the position measurement would yield $|x|=1$ with maximum probability, hence yielding a net displacement of $|\delta x_-|\sim 1$. Here the subscript `$-$' denotes displacement upon position measurement of the oscillator state conditioned on the qubit measurement yielding the $\ket{-}$ state. 

Thus, it is evident that upon measuring the position of the oscillator there is a non-zero probability that this measurement yields an effective displacement of $\delta x_{-}\sim\mathcal{O}(1)$ in position, much larger than the original displacement of $\epsilon$.  This is a unique and powerful quantum mechanical effect that is only possible due to a local phase of $\pi$ incurring a minus sign between the two slightly displaced coherent states $\ket{\pm\epsilon}$. The effect of this minus sign is to subtract the wave functions corresponding to the two displaced coherent states. While highly non-trivial, the collapse of the oscillator to this `$-$' superposition is also an extremely rare event because the ratio,
\begin{align}
    \mathcal{N}_\text{odd}/\mathcal{N}_\text{even}&=\sqrt{\frac{1-e^{-2\epsilon^2}}{1+e^{-2\epsilon^2}}} = \epsilon +O(\epsilon^5).
\end{align}
vanishes as $\epsilon\to 0$.

The probability of observing a large state change (displacement in phase space) induced by the measurement back-action can be increased by changing the qubit measurement axis, a task addressed by a quantum random walk. To see this, let us rewrite the state in Eq.~\eqref{eq:QRW} in terms of an arbitrary auxiliary qubit basis $\vec{\sigma}_\theta=\cos{\theta}\sigma_z+\sin{\theta}\sigma_x$, the $\pm 1$ eigenstates of which are denoted $\ket{\pm \theta}$.  Alternatively, we can rotate the qubit by $R_Y\big({\frac{\theta}{2}}\big)$ and then measure it in $Z$-basis. 
In this case, the composite oscillator-qubit state just before measurement is equal to (with implicit normalization),
\begin{align}
    \ket{\psi}&=(\ket{\epsilon}+\tan{(\theta/2)}\ket{-\epsilon})\ket{+\theta}\nonumber\\
    & ~~~~ +(\tan{(\theta/2)}\ket{\epsilon}-\ket{-\epsilon})\ket{-\theta}.\label{eq:final_state2}
\end{align}
Ref.~\cite{aharonov1993quantum} shows that under the assumption $\epsilon\rightarrow 0$, we can use $e^{\pm i\epsilon\hat p}\approx I\pm i\epsilon\hat p$ and approximate the final state in Eq.~\eqref{eq:final_state2} effectively as the state after an asymmetric conditional displacement $D_{\theta}(\delta x_+,\delta x_-)$ of the vacuum:

\begin{align}
    \ket{\psi}&\approx D_{\theta}(\delta x_+,\delta x_-)\ket{0}_\text{osc}\ket{0}_\mathrm{qub}\\
    &= \sqrt{\mathcal{P_{+}}}\ket{\delta x_+}\ket{+\theta}+\sqrt{\mathcal{P_{-}}}\ket{\delta x_-}\ket{-\theta},\\
    \text{where}\quad & \delta x_{\pm}=\frac{1\mp \tan{(\theta/2)}}{1\pm \tan{(\theta/2)}}\epsilon,\\
    D_{\theta}(\delta x_+,\delta x_-)&=e^{-i\delta x_+\hat p}\ket{+\theta}\bra{+\theta}+e^{-i\delta x_-\hat p}\ket{-\theta}\bra{-\theta},
\end{align}
and $\mathcal{P}_{\pm}$ denotes the probabilities of collapsing the oscillator onto coherent states $\ket{\delta x_{\pm}}$. This approximation holds as long as $|\delta x_{\pm}|\ll\Delta_x$, where $\Delta_x$ is the position uncertainty of the initial state
(here assumed to be $\Delta_x=\frac{1}{2}$, in correspondence with an initial vacuum state).

Despite this restriction, it is possible to observe large displacements, either $|\delta x_{+}|\gg \epsilon$ or $|\delta x_{-}|\gg \epsilon$, with significant probability for appropriate choice of $\theta$. 
For example, choosing $\tan{(\theta/2)}=1+\gamma$ such that $\epsilon/\Delta_x\ll |\gamma| \ll 1$, yields $\delta x_{-}=-2\epsilon/\gamma$ which satisfies the requirement $\epsilon \ll |\delta x_{-}|\ll \Delta_x$. That is, the effective displacement of the oscillator $|\delta x_-|$, upon collapse to $\ket{-\theta}$ qubit state, is much larger than $\epsilon$, while being much smaller than the initial position uncertainty $\Delta_x$. The probability of this event is $P_{-}=\gamma^2/4$. Note that for a pre-measurement rotation of $R_Y(\pi/2)$ such that $\gamma=0$, the probability of collapsing to $\ket{\delta x_{-}}$ under these assumptions vanishes. In contrast, by choosing a non-zero $\gamma$, and thus rotating the qubit by $R_Y(2\tan^{-1}{(1+\gamma)})$, the probability of observing a non-trivial displacement of the oscillator is enhanced. Ref.~\cite{aharonov1993quantum} shows that for $\epsilon=0.01$ and $\gamma=-0.1$, this effect is apparent after repeating the protocol for 10 rounds with post-selection upon $-1$ measurement outcomes, and the qubit reinitialized to $\ket{+}$ after each measurement. The authors also show another example of an interesting quantum mechanical observation using photon numbers and the side-band ISA: the proposed random walk indicates that the photon number of the oscillator can be changed by a large amount even if the interaction only allows for the addition or removal of a single photon.

Another interesting scenario is to carry out the above procedure but skip the `measurement + post-selection' step. In this case, the iterative procedure corresponds to a random walk for the composite system with $N$ pairs of conditional displacements and random qubit rotations in the equatorial plane. Such a sequence can be described by a unitary channel that produces a superposition of $2^N$ different state trajectories. To this end, we note a close similarity between QRWs and the phase-space trajectory illustrated in Fig.~\ref{fig:composing_conditional_disp}(b). Ref.~\cite{flurin2017observing} experimentally demonstrated such quantum walks via unitary channels composed entirely out of the phase-space ISA and used them to explore a novel topological effect, which they argued represents a potential technique for the study of topological quantum materials. 

We note that by swapping the roles of qubits and oscillators in QRWs, the coin toss can instead be performed via an oscillator measurement under the Fock or position/momentum basis (or any other basis). This measurement back-action will induce a ``walk'' of the qubit state according to an infinite-dimensional configuration space parameterized by the oscillator. A coherent version of this can also be constructed by concatenating iteratively a random coin toss on the oscillator and a qubit rotation. This can produce a channel on the qubit which is a superposition of a set of trajectories with a continuous distribution. In the case of a multi-qubit system coupled to an oscillator such as in Refs.~\cite{Liu2016e,bell2025codesigningeigensingularvaluetransformation}, this QRW construction is more interesting. The flexibility of measuring the oscillator under various bases can enact nontrivial dynamics on the multi-qubit systems.

In summary, quantum random walks are an application of the phase-space ISA (see Sec.~\ref{sssec:phasespaceISA}) that paves the way for observations of novel quantum mechanical phenomena using hybrid CV-DV architectures.

\section{Advantages over Qubit-only Architectures}~\label{sec:advantages}

In this final section, we illustrate the advantages of hybrid CV-DV hardware compared to DV-only hardware by comparing the gate count of simulating bosonic systems with CV-DV hardware to DV-only hardware. We first discuss in Sec.~\ref{sec:mode-to-qubit} different schemes to map oscillators into qubits, before compiling a displacement gate into qubit gates in a qubit-efficient encoding in Sec.~\ref{app:compilation-cv-to-qubits}. We close by discussing the gate complexity in the context of other compilation aspects in Sec.~\ref{sssec:complexity}.

\subsection{Mapping between Oscillators and Qubits}
\label{sec:mode-to-qubit}

\renewcommand{\arraystretch}{1.2}

\begin{table*}
\centering
\begin{tabular}{|c|p{4cm}|p{4cm}|c|c|}
\hline
\textbf{Encoding} & \textbf{Advantages} & \textbf{Disadvantages} & \textbf{Qubit Count} & \textbf{Representation} \\
\hline
\textbf{Unary} & Simplifies hardware implementation due to one-hot representation which could allow for repetition code detection of bit flip errors. 
& Highly inefficient for high boson occupation due to linear scaling of qubits. Resulting poor scalability for large Fock states. 
& $O(N_{\textrm{max}})$ & Fock space \\
\hline
\textbf{Gray}~\cite{gray1953pulse} 
& Bit-flip errors represent small boson occupation differences. 
& Limited benefit for dense matrices or operators with high Hamming distances. 
& $O(\log(N_{\textrm{max}}))$ & Fock space \\
\hline
\textbf{Fock Binary}~\cite{shaw2020quantum} 
& Very compact representation. Efficient for diagonal operators. 
& Susceptible to bit-flip errors since adjacent Fock states are not adjacent in binary representation. 
& $O(\log(N_{\textrm{max}}))$ & Fock space \\
\hline
\textbf{JLP}~\cite{jordan2012quantum, jordan2014quantum} 
& Enables representation of bosonic modes in phase space. 
& High qubit and gate count overhead. 
& $O(\log(\varphi_{\rm max} / \delta_\varphi))$ & Phase space \\
\hline
\end{tabular}
\caption{Overview of different ways to map an oscillator state to qubits. In the last row, JLP stands for Jordan-Lee-Preskill, where $\varphi_{\rm max}$ is the maximum field amplitude (boundary of truncated phase space), and $\delta_\varphi$ is the spacing between nearest grid points in the discrete representation of the phase space. $N_\mathrm{max}$ is the maximum Fock state number encoded by the encoding.}
\label{tab:encodings}
\end{table*}

An oscillator has an in-principle infinite-dimensional Hilbert space, whereas a qubit is a two-level system. We therefore need to use a mapping of the Hilbert space of an oscillator and the Hilbert space of a collection of qubits, which requires a truncation of the oscillator Hilbert space.

We first give two simple mappings as examples in Sec.~\ref{sec:hardware-efficient-mapping} and Sec.~\ref{sec:redundant-mapping}. Other approaches such as the field basis representation used by Jordan, Lee, and Preskill \cite{jordan2012quantum,jordan2014quantum} and the Gray code \cite{gray1953pulse} can also be useful and may be more efficient for implementation of certain unitary operations. See Ref.~\cite{sawaya2020resource} for a detailed comparison of simulating $d$-level system on qubits using different encodings and Ref.~\cite{C2QA-LGTpaper} for a comparison of the quantum simulation of qubit-boson models on DV vs. CV-DV hardware. Table \ref{tab:encodings} summarizes all mappings discussed in this section.

\subsubsection{Hardware-Efficient Mapping (Fock Binary Encoding)}
\label{sec:hardware-efficient-mapping}

Notice the important (and not accidental) fact that  in the multi-qubit states defined in Eqs.~(\ref{eq:psi0}-\ref{eq:psi3}), the ordinal number of the basis vector (numbering the list 0-3 from the top) matches the spin state interpreted as a binary number, e.g. $|11\rangle$ corresponds to the 3rd basis vector (starting from $0$). More generally for $L$ qubits, we have for the $j$th standard basis state
\begin{equation}
    |\psi_j\rangle=|b_{L-1} b_{L-2}\ldots b_1 b_0\rangle,\,\,\,\,   j\in [0,2^L-1] \label{eq:psijbinaryrep}
\end{equation}
where $b_k=0,1$ labels the standard basis state of the $k$th qubit and
\begin{equation}
    j=\sum_{k=0}^{L-1}2^kb_k,
    \label{eq:binary_rep}
\end{equation}
is the integer corresponding to the binary number labeling the qubit state.

As we have seen in Sec.~\ref{sec:hilbert-space}, the Fock basis states of an oscillator are also labeled by an integer (beginning at zero). Thus we can very conveniently choose to view the qubit state $|\psi_j\rangle$ as representing the state of an oscillator containing $j$ excitations (bosons), the Fock state $\ket{j}$. A register of $L$ qubits can thus be used to represent the lowest $2^L$ states of an oscillator (representing boson numbers 0 to $N_\mathrm{max}=2^L-1$). This provides a mapping from an oscillator to (logarithmically) many qubits, where every Fock state of the oscillator corresponds to a multi-qubit state. This mapping allows us to simulate bosonic modes in classical simulator codes such as Qiskit that are designed for qubits only, a strategy that we have implemented in Bosonic Qiskit \cite{Biskit,BiskitGitHub,BiskitBlogPost}.  This representation is very convenient because the Fock basis matrix representations of the boson operators $a$ and $a^\dagger$ given in Eqs.~(\ref{eq:amatrix}-\ref{eq:adaggermatrix}) are \emph{unchanged} when we translate them into the multi-qubit basis.

Conversely, a single resonator with boson number cutoff $N_\mathrm{max}=2^L-1$ in a hybrid architecture can replace $L$ qubits leading to considerable hardware efficiency\footnote{Note however that $L$ cannot be made too large because the maximum energy stored in the oscillator (and thus its decay rate) grows exponentially with $L$.  See Ref.~\cite{ClimbingMountScalable} for a general discussion of why the Hilbert space of quantum computers must be built up from tensor products of multiple subsystems.}.  We thus have two ways to realize a Hilbert space of dimension $2^L$--using $L$ physical qubits or $2^L$ states of a single oscillator.  Either way, to have universal control, we must be able to perform arbitrary unitary transformations which are $2^L\times 2^L$ matrices.  There are `natural' operations in each of the two physical realizations, and those may be quite unnatural in the other realization.  In particular, as hinted at in the introduction, it is important to understand that implementing natural bosonic operations on qubit-only quantum hardware can be very difficult.  The qubit cost is only logarithmic in $N_\mathrm{max}$ and so is not the problem -- it is instead the matrix elements of the raising and lowering operators given in Eq.~(\ref{eq_sqrt1}) and (\ref{eq_sqrt2}) containing square root factors.  While these are easy to implement \cite{Biskit,BiskitGitHub,BiskitBlogPost} in a classical simulator designed for qubits only, they are very difficult to realize on quantum qubit hardware because of the quantum arithmetic needed to produce the square root factors \cite{sawaya2019quantum,shaw2020quantum} -- see also Sec.~\ref{app:compilation-cv-to-qubits} for details.  It is here that we see the main advantage of replacing qubit-only hardware with hybrid oscillator/qubit hardware where the oscillator raising and lowering operators are natively available.

Separately, it is instructive to compare the impact of amplitude damping on a single oscillator to the impact of amplitude damping on the $L$ qubits representing the oscillator up to a photon number cutoff of $N_\mathrm{max}=2^L-1$. 
Recall that the physical oscillator with damping parameter $\kappa$, the probability to be in an initially prepared Fock state $\ket{n}$ decays at a rate $n\kappa$, and thus the mean photon number obeys Eq.~(\ref{eq:SHOdampingrate}).   
\begin{align}
    \frac{d}{dt}\bar n = -\kappa\bar n.
    \label{eq:cav_amp_damp}
\end{align}
Let us assume that all $L$ qubits used to represent the oscillator have the same decay rate $\gamma$ out of their excited states.  The qubit representation of the oscillator seems advantageous since the error rate in the qubit register representing the oscillator will scale only logarithmically, whereas in an actual oscillator the dominant error (photon loss) scales linearly with the number of photons.
For example, the rate of decay out of Fock state $n=2^k$ is only  $\gamma$, since only one qubit is in the excited state. In the worst case ($n=N_\mathrm{max}$) where all $L$ qubits are in the excited state, the decay rate is still only $L\gamma$, suggesting possible advantages for the qubit encoding. 

On the other hand, one might suspect that the loss rate of the mean boson number of the qubit encoding is significantly worse than the one of an oscillator, as $2^j$ bosons are lost at once if the $j$th qubit decays. 
However, a little thought shows that the mean occupation number in the qubit representation still obeys Eq.~(\ref{eq:cav_amp_damp}), except with the cavity damping rate $\kappa$ replaced by the qubit decay rate $\gamma$. This is
 because the mean boson number is a linear function of the qubit excitations -- see Eq.~(\ref{eq:binary_rep}). 
 Typically, the lifetime of state-of-the-art qubits ($\gamma^{-1}\sim100-500$ $\mu$s) is substantially shorter than that of state-of-the-art cavities ($\kappa^{-1}\sim 10^4-10^6$ $\mu$s).  This demonstrates that while the rate of leaving a given state for the qubits is bounded above by $\gamma L$ (which can in some cases be much smaller than $\kappa \langle \hat n\rangle$), the rate of change of the mean photon number represented by the qubits is always worse than $\kappa \langle \hat n\rangle$ (assuming $\gamma>\kappa$).  For example, in Ref.~\cite{Rosenblum_TensofMilliseconds_PRXQuantum.4.030336} a transmon controller with a lifetime of $T_1=110$ $\mu$s was coupled to a cavity with a lifetime of $T_1=35$ ms and used to produce a $10^3$-photon cat state.  The phase coherence time of this cat state due to photon loss (which produces cat parity jumps) was $\sim T_1/\langle\hat n\rangle\approx 35$ $\mu$s. A superposition of $0$ and $10^3$ photons would decohere with precisely the same lifetime. It would take a register of approximately $L=10$ qubits to simulate this bosonic state and the corresponding lifetime would be only $1/(\gamma L)\sim T_1/10\sim 11$ $\mu$s.  
 We emphasize that beating these constraints on the qubit representation of oscillators would require the use of error-corrected logical qubits that are capable of performing deep and complex circuits with multi-qubit gates.
 
 These considerations show that the way decay affects oscillators and qubit representations of oscillators is significantly different. In particular, if the goal were a simulation of an open (i.e., damped) bosonic system, the use of a physical oscillator would be much more natural as the qubit noise processes are a bad representation of bosonic noise. We expect that these considerations concerning noise change substantially  for other encodings of oscillators into qubits (e.g., Jordan-Lee-Preskill \cite{jordan2012quantum,jordan2014quantum}), but the essential conclusion remains the same, namely that qubit noise processes produce a bad representation of the physical oscillator noise processes. 
 
 Finally, we stress that to thoroughly address the question of whether a single noisy oscillator holds an advantage over logarithmically-many noisy qubits, it is crucial to not only consider the timescale at which errors occur ($\tau_{\textrm{error}}$), but also how this timescale relates to gate duration for the $2^L$ system ($\tau_{\textrm{gates}}$), as it is comparison of the ratio $\tau_{\textrm{error}}/\tau_{\textrm{gates}}$ which determines which implementation holds a practical advantage. This type of analysis was recently carried out in Ref.~\cite{Jankovic2024} to compare noisy $d$-level qudits to logarithmically-many qubits undergoing dephasing errors. There, the authors found that for certain regimes of $\tau_{\textrm{error}}$, $\tau_{\textrm{gates}}$, and $d$, single noisy qudits indeed hold an advantage, despite the increased complexity of their dephasing error channel over their many-qubit counterparts. We expect this advantage to be even more pronounced for oscillators due to (i) the simplicity of their dominant error channel and, (ii) their substantial advantage in gate complexity (and consequently $\tau_{\textrm{gates}}$) for applications where bosonic gates are desired -- see Sec.~\ref{app:compilation-cv-to-qubits}. Nonetheless, it would be an interesting direction to rigorously extend the analysis of Ref.~\cite{Jankovic2024} for noisy oscillators vs.\ many-qubits for the case of amplitude damping. We leave this as a direction for future work.

\subsubsection{Mapping with Redundancy (Unary Encoding)}
\label{sec:redundant-mapping}

A different approach to mapping from oscillators to qubits is to forego the notion of hardware efficiency, and instead introduce redundancy. This may seem inefficient at first glance, but such redundancy is crucial to preserve quantum information in noisy quantum systems. Redundancy can be introduced to both the qubit states and the oscillator states, see Sec.~\ref{sec:bosonic-QEC} for a thorough discussion of oscillator redundancy for bosonic quantum error correction.  In this section, we instead discuss a mapping of oscillators to qubits that uses redundancy in the qubit states. 

One simple mapping with redundancy on the qubits uses the unary representation in a register of $d=N_\mathrm{max}+1$ qubits
\begin{eqnarray}
|0\rangle_\mathrm{b} &=& |0\ldots0001\rangle_\mathrm{q}\\
|1\rangle_\mathrm{b} &=& |0\ldots0010\rangle_\mathrm{q}\\
|2\rangle_\mathrm{b} &=& |0\ldots0100\rangle_\mathrm{q}\\
|3\rangle_\mathrm{b} &=& |0\ldots1000\rangle_\mathrm{q},
\end{eqnarray}
where the subscript $b$ labels the bosonic mode Fock states and the subscript $q$ denotes the qubit register.  
This means that $d$ qubits are needed to represent the lowest $d$ levels of an oscillator.
We number the qubits from right to left starting with zero. In this encoding, only a single qubit is in the 1 state and the location label of that qubit represents the number of photons in the cavity.

Note that this is an exponentially redundant mapping in the sense that the qubit Hilbert space dimensionality is $2^d$ but the relevant oscillator levels only have dimension $d$. However, this mapping has some fault-tolerance built into the qubit states. For example, for Fock state $\ket{3}_b$ if the qubit state is changed from $\ket{0\cdots 1000}_q$ to $\ket{0\cdots 1010}_q$ due to some noise by flipping the second qubit from 0 to 1, the resulting state $\ket{0\cdots 1010}_q$ is outside of the computational state space and thus can be detected.  Conversely, a spin flip leading to $\ket{0\cdots 0000}_q$ cannot be detected. Simple repetition encoding of this representation would allow for bit-flip error correction and of course, more sophisticated and powerful encodings could be used.

\subsection{Compilation of Bosonic Operators into and from Standard Qubit Gate Sets}
\label{app:compilation-cv-to-qubits}

In this section, we present a quantitative analysis on the overhead of simulating bosonic hardware on qubit hardware by estimating the CNOT cost for simulating a single displacement operator. For a comparison of different mappings, we refer the reader to Ref.~\cite{sawaya2020resource}. Specifically, we introduce a novel compilation technique suited for arbitrary bosonic operations which relies on calculating the square root matrix elements using Newton iterations. We explicitly calculate the gate count of this method for a displacement and compare to brute-force compilation (which is not scalable beyond a handful of qubits).

The destruction operator $a$ is a natural operator for bosonic systems.  Let us consider its matrix elements in the qubit representation.  For example, the simple Fock basis expression
\begin{equation}
    a|8\rangle=\sqrt{8}|7\rangle
\end{equation}
tells us that applying the destruction operator to the state with 8 bosons yields the state with 7 bosons.  This is equally simple on the standard multi-qubit basis as well
\begin{equation}
    a|\psi_8\rangle=\sqrt{8}|\psi_7\rangle.
\end{equation}

This should be all one needs to tell a qubit compiler [by giving it the matrix representation of $a$ in Eq.~(\ref{eq:amatrix})].
However it is important to note $a$ becomes a non-trivial multi-qubit operation in the binary basis that labels the states of the individual qubits, for example:
\begin{equation}
    a|1000\rangle=\sqrt{8}|0111\rangle
    \label{eq:binarycode8}
\end{equation}
Thus $a$ is an unnatural weight-4 operator (in this particular example) for qubits because it flips all four qubits.
The full matrix representation of the destruction operator in terms of direct products of single-qubit operators is complex because computing which bits have to be flipped requires knowing all the bits.  For example, the particular term in $a$ shown in Eq.~(\ref{eq:binarycode8}) is given in terms of qubit operators by a complicated (and difficult to physically implement) weight-8 expression
\begin{equation}
    X_3X_2X_1X_0[\hat I_3-Z_3][\hat I_2+Z_2][\hat I_1+Z_1][\hat I_0+Z_1]/16,
\end{equation}
and there are 15 other similar terms needed to fully represent the operator.  From this, we see immediately the huge potential efficiency of having boson degrees of freedom available natively in the hardware, if we are interested in simulating a physical system with bosonic modes. 

Conversely, if we would like to use resonators to replace multiple qubits in a quantum simulator or computer, we need to reckon with the fact that simple qubit gates typically correspond to non-trivial unitaries acting on the oscillator.  As we discuss further below, oscillators coupled to transmon qubits are `controllable' in the formal language of quantum control theory \cite{Heeres2015} and recent experiments have demonstrated an impressive level of control of such oscillators \cite{Heeres2015,Ofek2016,Rosenblum2018,hu_quantum_2019,Campagne-Ibarcq2020,LuyanSun2020}.

A natural alternative to the Fock-Binary encoding is the Gray Code\footnote{https://en.wikipedia.org/wiki/Gray$\_$code} which is a permutation of the binary encoding with the property that only one bit flips at a time as the number the code represents increases or decreases by 1.  This would seem to eliminate the multi-qubit flips we saw in Eq.~(\ref{eq:binarycode8}).  However which single bit flips is a complicated function of all the bits so we still have a high-weight operator and there is no simplification to be gained by using the Gray code.
Hence, we restrict our attention to the binary code for simplicity.

The implementation of Trotter steps in Hamiltonian simulation on a qubit-based quantum computer requires mapping of the bosonic Hamiltonian to the qubit spin-subspace, mapping of the bosonic state to the qubits, and then mapping the resulting state back to a bosonic state (during classical post-processing) for interpretation of the results.

Within the context of Trotter-Suzuki-based simulations, one of the most efficient ways of simulating bosonic systems in qubits is via the decomposition of the Hamiltonian into a sum of one-sparse Hamiltonians. One-sparse Hamiltonians are arguably the broadest family of quantum Hamiltonians that can be directly compiled into quantum circuits with nominally zero error.  These Hamiltonians correspond to matrices that have at most one non-zero matrix element in every row and column.  Most of the common matrices that we use in quantum computing are one-sparse, including the Pauli operators, single qubit $Z$-rotations (S, P, T gates), controlled-NOT, controlled-Z, SWAP, and Toffoli -- but not the Hadamard gate.  

For bosonic systems, however, many of the gates are not one-sparse when compiled into qubit-only hardware. 
In the following, we calculate the number of qubit gates required to implement a single bosonic displacement operation in a scalable fashion in an all-qubit device.

Assuming that a boson number cutoff of $K=2^{n}-1$ is used, our strategy is to encode such a bosonic state using $n$ qubits while using one additional qubit to ease the calculation of the square root factor from bosonic dynamics. As we see in the following, the additional qubit allows us to encode coefficients of adjacent Fock levels into it for easy computation of the square root operation as multi-qubit controlled single-qubit rotations. 

The intuition of why the additional qubit helps to compute the square root is as follows. For the displacement operator, the Hamiltonian of interest is 
\begin{align}
    a+a^\dagger &= \left[ \begin{array}{ccccc}
    0&1&0&0&\\
    1&0&\sqrt{2}&0&\\
    0&\sqrt{2}&0&\sqrt{3}&\cdots\\
    0&0&\sqrt{3}&0&\\
    &&\vdots&&\ddots
    \end{array} \right]
    \label{eq:a-plus-adagger}
\end{align}
which we would like to encode on qubits. Observing the self-repeating (though overlapping) $2\times 2$ block structure 
\begin{align}
\begin{bmatrix} 0 & \sqrt{k} \\ \sqrt{k} & 0 \end{bmatrix}
\end{align}
and that, for example
\begin{align}
    X\oplus \sqrt{3}X=\left[\begin{array}{cccc}
    0&1&0&0\\
    1&0&0&0\\
    0&0&0&\sqrt{3}\\
    0&0&\sqrt{3}&0   
    \end{array}\right],
\end{align}
we can write $a^\dagger + a$ as follows:
\begin{align}
    a+a^\dagger & = 
    [0]\oplus \bigoplus_{j=1}^{\infty} \begin{bmatrix} 0 & \sqrt{2j}\\ \sqrt{2j} & 0 \end{bmatrix} \nonumber\\ \quad & 
    ~~~~~~~~~~~ + \bigoplus_{j=0}^{\infty} \begin{bmatrix} 0 & \sqrt{2j+1}\\ \sqrt{2j+1} & 0 \end{bmatrix} \\
    &= H_{\rm even} + H_{\rm odd}.
\end{align}
Here, the ``even'' and ``odd'' subscripts denote the parity of the value inside the square-root. We can therefore simulate the quantum dynamics using gates analogous to $x$-rotations $R_0(-2it\sqrt{2j})$ or $R_0(-2it\sqrt{2j+1})$ (where the single-qubit rotation $R_\varphi(\theta)$ is a rotation through an angle $\theta$ about an axis $\phi$, formally defined in Table \ref{tab:gates-qubit} of App.~\ref{app:PhysImp}).

In the following, we describe in detail the steps necessary to use the above idea to simulate the displacement operator on qubits, the overview of which is given in Algorithm \ref{alg_Fock_Binary}.  We denote bosonic Fock states by $|k\rangle_\mathrm{B}$ where the non-negative integer $k$ is the boson number.  The corresponding representation of this Fock state in the $n$ qubit register is denoted $\ket{{\rm binary}(k)}$ without the subscript 
$\mathrm{B}$.

\begin{algorithm}
\caption{Fock-Binary Simulation of Bosonic Displacement Gate with Qubits}\label{alg_Fock_Binary}
\SetKwInOut{Input}{Input}\SetKwInOut{Output}{Output}
\Input{ A bosonic state defined in the Fock basis by $\ket{\psi}_B = \sum_{k=0}^K a_k \ket{k}_B$ and an $n$-qubit register augmented by a single additional qubit (hereafter called the main qubit), prepare the initial state on the register of $n$ qubits as $\left( \sum_{k=0}^K a_k \ket{\mathrm{binary}(k)} \right) \otimes \ket{0}$, a displacement $\alpha\in \mathbb{R}$ and an error tolerance $\epsilon$ }
\Output{A quantum state $V(\alpha) \ket{\psi_b}$ such that $\|V(\alpha) - D(\alpha)\|\le \epsilon$}
\BlankLine
Subdivide Hamiltonian into $H= H_{\rm even} + H_{\rm odd}$\;
Provide a Trotter decomposition of the form $V(\alpha)=\prod_{j=1}^{N_{\exp}/2} e^{-i \alpha H_{\rm odd} \tau_{2j-1}}e^{-i \alpha H_{\rm even} \tau_{2j}}$ such that $\|V(\alpha) - D(\alpha)\| \le \epsilon/2$\;
\For{$e^{-i H_p t_p} \in {\rm terms\ of\ } V(\alpha)$:{}}{
Simulate $e^{-i H_p t_p}$ by transferring the quantum amplitudes to the main qubit (parameters given for $H_{\rm odd}$): $\sum_{j=0}^{(K-1)/2} \ket{2j+1} \left( a_{2j} \ket{0} + a_{2j+1} \ket{1}\right)$ for odd cutoff $K$ \;
Compute square-roots using reversible operations within error $\epsilon/(2N_{\exp}\alpha)$\; 
Act with a series of register-controlled ancilla rotations to implement Hamiltonian\;
Transfer amplitudes back to the qubit register.
}
\end{algorithm}

\subsubsubsection{Mapping the Bosonic State to Two-Dimensional Subspaces and Back.}
Let us write the truncated bosonic state as a sum over even and odd Fock states:
\begin{equation}
	\ket{\psi}_\mathrm{B} = \sum_{j=0}^{(K-1)/2} \left( a_{2j} \ket{2j}_\mathrm{B} + a_{2j+1} \ket{2j+1}_\mathrm{B} \right).
\end{equation}
First, we assume that it is possible to encode this truncated bosonic state onto a register of $n$ qubits using a binary representation of the Fock states 
\begin{align}
    \ket{\psi} = \sum_{j=0}^{(K-1)/2} \left( 
    \rule{0pt}{2.4ex}
    a_{2j} \ket{{\rm binary}(2j)} + a_{2j+1} \ket{{\rm binary}(2j+1)} \right).
    \label{eq:qubit-rep-binary}
\end{align}
For example, an arbitrary state with a maximum of three bosons
\begin{align}
    \ket{\psi}_\mathrm{B} &= \alpha \ket{0}_\mathrm{B}+\beta\ket{1}_\mathrm{B}+\gamma\ket{2}_\mathrm{B}+\delta\ket{3}_\mathrm{B}
    \end{align}
would require a register of $n=2$ qubits to represent it:
    \begin{align}
    \ket{\psi} &= \alpha \ket{00}+\beta\ket{01}+\gamma\ket{10}+\delta\ket{11}.\label{eq_ini_st_FB}
\end{align}
The preparation of this qubit state is in general not a trivial task; however, we do not consider the complexity of this task here.

Second, by coupling one additional (`main') qubit in the last position,  we store the coefficients on the main qubit:
\begin{align}
	\ket{\psi'} &= \sum_{j=0}^{(K-1)/2} \ket{\mathrm{binary}(2j+1)}\otimes \ket{\psi^\prime_{2j+1}} \\
 \ket{\psi^\prime_{2j+1}}&\equiv a_{2j} \ket{0} + a_{2j+1} \ket{1},\label{eq_ini_qb_rep}
\end{align}
such that we can then evolve the system under the displacement Hamiltonian using a single unitary acting in parallel on all $2\times 2$ subspaces, each labeled by the projector $\ket{2j+1}\bra{2j+1}$
\begin{equation}
	\sum_{j=0}^{(K-1)/2} \ket{2j+1}\bra{2j+1}\otimes R_0(-i2t \sqrt{2j+1}),\label{eq:sqrtrotation}
\end{equation}
where the first part of the tensor product acts on the $n$ qubit register and the $R_0$ gate (Table \ref{tab:gates-qubit}) rotates the main qubit around the $x$ axis.

To prepare $\ket{\psi'}$ in Eq.~\eqref{eq_ini_qb_rep} from $\ket{\psi}\otimes \ket{0}$ for $\ket{\psi}$ given in Eq.~\eqref{eq:qubit-rep-binary}, we use the knowledge that in binary, an odd number ends in 1: we use a CNOT gate whose target is the main qubit and whose control is the least significant bit (LSB) in the  $n$-qubit register. This flips the main qubit if the state in the first $n$-qubit (system register) represents an odd state. We obtain
\begin{align}
    a\ket{2j}\ket{0} + b\ket{2j+1} \ket{0} &\mapsto a\ket{2j}\ket{0} + b\ket{2j+1} \ket{1}.
    \label{eq:cnot-odd-to-main-qubit}
\end{align}
Applying an incrementer operation on the system register, controlled on the main qubit being in $\ket{0}$, we obtain
\begin{equation}
    a\ket{2j}\ket{0} + b\ket{2j+1} \ket{1} \mapsto \ket{2j+1}(a\ket{0} + b\ket{1}).
    \label{eq:main-qubit-controlled-incrementer}
\end{equation}
For the example initial state in Eq.~\eqref{eq_ini_st_FB}, we now have:
\begin{equation}
    \ket{01}\left( \alpha \ket{0} + \beta \ket{1}\right)+\ket{11}\left( \gamma \ket{0} + \delta \ket{1}\right)
\end{equation}

In the following, we calculate the two-qubit entangling gates needed to operate Eqs.~\eqref{eq:cnot-odd-to-main-qubit} and \eqref{eq:main-qubit-controlled-incrementer} for encoding the coefficients to the main qubit. We need one CNOT gate and a single-qubit controlled $n$-qubit incrementer.

An incrementer, in which we do not reset the additional carry qubits required for the computation of the incrementation, takes $(n-2)$ Toffoli gates and $n$ CNOT gates (with an $n-1$ qubit overhead)~\cite{Li_incrementer_2014}. There are multiple ways to implement Toffoli gates, but the standard decomposition uses $6$ CNOT gates~\cite{Nielsen_Chuang}. Assuming that decomposition, an incrementer takes $7n-12$ CNOT gates.

To implement a controlled incrementer, we can use an extra register of qubits: we carry out a controlled SWAP between the system register and the extra register, then implement the controlled adder on the extra register, and carry out another controlled SWAP to return the incremented value to the system register. A controlled SWAP gate between the two registers requires $n$ Toffoli and $2n$ CNOT gates.  This is because a SWAP gate between qubits $i$ and $j$ can be implemented using ${\rm CNOT}_{ij} {\rm CNOT}_{ji} {\rm CNOT}_{ij}$ and this circuit can be transformed into a controlled SWAP by only controlling the ${\rm CNOT}_{ji}$ gate.  This implies that two CNOTs and one Toffoli are needed to do a controlled SWAP, which corresponds to $8$ CNOT gates. Therefore, with the given construction, a controlled incrementer requires $23n-12$ CNOT gates.

Because we need to map the amplitudes to the main qubit to make the computation and back to the system register to interpret the results, this requires two controlled-NOT gates, one controlled incrementer and one controlled decrementer, and the aforementioned controlled rotation for the implementation of the Hamiltonian. The complexity of a decrementer is the same as that of an adder (up to single qubit gates).

Therefore, for this construction, the total number of CNOT gates required to transfer two neighboring Fock state amplitudes from the bosonic state (represented in binary in an $n$ qubit register) onto the main qubit, and return that information to the binary qubit register representing the Fock state is
\begin{align}
    N_{\textrm{CNOT/transfer}}&= 2\times(23n-12 + 1) \nonumber \\
    &= 46n - 22.
    \label{Ncnottr}
 \end{align}

The system register requires $n$ qubits, where $n$ is the number of qubits required to represent the Fock state up to Fock level $K$ in binary. Although all amplitudes can be stored in a single qubit, each incrementer requires $n-1$ extra qubits, and each controlled SWAP operation requires $n$ extra qubits. We also count the main qubit. Therefore a total of $n + n-1 + n + 1 = 3n$ qubits are needed.

\subsubsubsection{Computation of the Square-Roots.}
One of the main costs regarding the entangling gate counts for computation of the dynamics relates to the implementation of the $R_0(-i2t\sqrt{2j+1})$ gate, which maps 
\begin{align}
&\ket{\mathrm{binary}(2j+1)} \ket{\psi_{2j+1}^\prime}\nonumber\\
\mapsto &\ket{\mathrm{binary}(2j+1)} R_0(-i2t\sqrt{2j+1})\ket{\psi_{2j+1}^\prime}.
\end{align}
Similarly, we can address the states within the even subspaces by computing instead the square root of $2j$ in analogy to the above mapping. 
Since the square root needs to be computed for each different value of  $2j$ and $2j+1$ in the superposition, we cannot construct a fixed circuit for it. It would be possible to calculate the square roots classically in advance and store them in a register; however, given the number of values we need in cases with large cutoffs, the cost may become prohibitive. Several approaches are commonly taken to implement functions such as this. The first approach directly converts classical networks for these operations into quantum gates through the use of Bennett pebbling~\cite{PhysRevA.52.3457}. This approach is time-consuming, but more importantly, often requires hundreds of qubits to maintain the result.

The cost of performing the square root transformation can be upper bounded using the cost of a reciprocal square root operation.  The reciprocal square root can be readily computed using Newton iterations, which converge quadratically, meaning that the number of iterations needed scales only doubly logarithmically with the desired error tolerance.  The square root can then be computed via
\begin{equation}
    \sqrt{x} = \frac{x}{\sqrt{x}}.
\end{equation}
Thus we can use this tactic to take advantage of the rapid convergence of the Newton iterations to the reciprocal square root by following the reversible computation of the reciprocal square root with a single multiplication.

We represent the Fock state value (an integer) using $n$ bits. We denote $m$ to be the number of iterations needed to compute the reciprocal square root, for example, results with $10^{-8}$ error have been obtained using $m=3$ Newton iterations~\cite{haner2018optimizing}. In the following, the iterates are no longer integers, therefore, we denote $p$ the point precision and choose to set it to $n$ such that there are as many qubits representing the number before the floating point as there 
are qubits representing the number after. As in fixed-point arithmetic, we now write the formulae with the new number of bits $n' = 2n$.

The cost of computing a square root in terms of the number of Toffoli gates (without un-computing the iterates) is~\cite{haner2018optimizing}
\begin{equation}
    \begin{split}
    n'^2\left(\frac{15}{2} m + 3\right) + 15n'pm + n'\left(\frac{23}{2} m + 5\right)& \\ - 15p^2m + 15pm - 2m &.
    \end{split}
\end{equation}
Re-written in terms of $n$, we now have
\begin{equation}
    45 n^2m + 12 n^2 + 38 nm +10n-2m.
\end{equation}

The cost of a multiplication of two $n$-bit numbers with point precision $p$ is~\cite{haner2018optimizing}
\begin{equation}\label{eq:cnotMult}
    N_{\rm CNOT/\times}=\frac{3}{2} n'^2 + 3n'p + \frac{3}{2} n' - 3p^2 + 3p
\end{equation}
which in terms of $n$ is
\begin{equation}
    9 n^2 + 6 n.
\end{equation}

If we use the standard $6$ CNOT per Toffoli gate construction~\cite{Nielsen_Chuang} then the number of CNOT gates needed to implement the square root calculation
\begin{equation}
    \ket{x}\ket{0} \mapsto \ket{x} \ket{\sqrt{x}} \ket{\psi_{\rm anc}}
\end{equation}
to be
\begin{align}
    &N_\mathrm{CNOT / SQRT}\nonumber\\
    & = 6(45 n^2m + 12 n^2 + 38 nm +10n-2m) + 6(9 n^2 + 6 n)\\
    &= \left(270m+126\right)n^2 + \left(228m+96\right)n-12m.
    \label{Ncnotsqrt}
\end{align}

If $m$ iterations are needed for the computation of the reciprocal square root, $m$ qubit registers will be needed to store the values of the reciprocal square root in addition to any ancillae needed in the implementation of any binary adders that appear in the construction. Each Newton iteration requires 3 ancilla registers (which are cleaned up after each round) as well as a register to hold the initial guess, totaling $(m+4)n'$ qubits. Therefore, the total number of ancillae needed to compute the square root to $n'$ bits of precision is 
\begin{align}
    n_{\textrm{qubits}} &= n' + (m+4)n'\\
    &= 2n(m+5),
\end{align}
as an additional register is required to hold the initial value to carry out the multiplication giving the square root.

\subsubsubsection{Controlled Rotations (CR).} Having calculated the square-roots, controlled rotations must be employed to apply the rotation.  That is to say if $(\sqrt{x})_j$ represents the value of the $j^{\rm th}$ bit in the fixed point representation of the square root in increasing order of significance then
\begin{equation}
    e^{-i \sqrt{x} \sigma_x t} = H e^{-i \sqrt{x} \sigma_z t} H =H \prod_{j=0}^{n-1} e^{-i (\sqrt{x})_j 2^j \sigma_z t} H .
\end{equation}
Thus we can implement $e^{-i H_{\rm even} t}$ or $e^{-i H_{\rm odd} t}$ by applying this circuit to the ancilla qubit in Eq.~\eqref{eq_ini_qb_rep}.  A controlled single-qubit rotation can be implemented using two controlled NOT gates and single-qubit rotations.  Thus the cost of the simulation of this term would also involve another $2n$ single-qubit rotations and CNOT gates
\begin{align}
    N_{\mathrm{CNOT/CR}}= 2n.
    \label{Ncnotcr}
 \end{align}  This cost is, however, negligible in most settings compared to the $O(n^2)$ Toffoli gates needed to compute the square root.

\begin{figure}[t]
    \centering
    \includegraphics[width=\columnwidth]{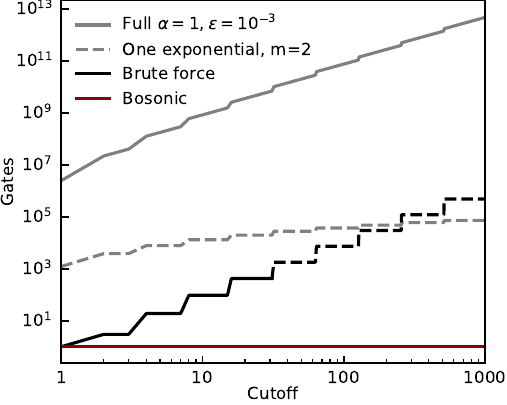}
    \caption{CNOT gate count of implementing a displacement operation with qubit-only quantum computers in the Fock-binary encoding. The gray dashed line represents the CNOT count per exponential in Eq.~\eqref{eq:total-cnot} which appears in our estimate for the displacement. It consists of transferring the amplitude from the $n$-qubit system register to a single main qubit and back,  Eq.~\eqref{Ncnottr}, and the cost of calculating square-roots using Newton iterations, in this case for $m=2$ iterations. The gray line is our full gate estimate for a displacement for target error $\epsilon=0.1\%$ and $\alpha=1$. The black line is the currently best-known brute-force decomposition algorithm, see Ref.~\cite{krol2024}. It constructs the $2^n\times 2^n$ unitary corresponding to the displacement and decomposes it into two and one gates. From about five qubits, the classical resources to actually perform the compilation is prohibitively expensive, but in this regime the algorithm still provides an upper bound to how many gates are actually required, indicated by a dashed line. The red line is at cost 1, representing the single displacement gate needed in bosonic hardware. This unit circuit depth metric does not take into account the fact that the time cost of the displacement is linear in the size of the displacement. We also neglect here the subtle effects of noise in the complexity considerations (see discussion in Sec.~\ref{sec:hardware-efficient-mapping}).
    }
    \label{fig_comp2}
\end{figure}

\subsubsubsection{Cost of simulating a single exponential.}
Combining Eq.~\eqref{Ncnottr}, \eqref{Ncnotsqrt}, and \eqref{Ncnotcr}, the number of two-qubit operations sufficient to implement a single step $e^{-iH_p t_p}$ in Algorithm~\ref{alg_Fock_Binary} is
\begin{align}
    N_\mathrm{CNOT} =&N_\mathrm{CNOT/H_\mathrm{even}} + N_\mathrm{CNOT/H_\mathrm{odd}} \nonumber\\
    =& 2N_\mathrm{CNOT/transfer} + 4N_\mathrm{CNOT/SQRT} +  2N_\mathrm{CNOT/CR} \nonumber\\
    =& (270 m + 126)n^2+(144+228m)n-12m-22,
    \label{eq:total-cnot}
\end{align}
 where $n$ is the size of the qubit register required to represent the largest initial Fock state in binary, and $m$ is the number of Newton iterations required to calculate the inverse square root.

 The number of qubits required for a system with the largest Fock state $N$, requiring $n$ bits for binary representation, and $m$ Newton iterations would be
\begin{equation}
    N_\mathrm{qubits} = 2n(m+5) + 3n.
\end{equation}

If we wish to represent all occupations up to a Fock state cutoff $K+1=64$, for example,  a 7-qubit register is required for the binary representation: $n=7$. In the case where two Newton iterations $m=2$ are sufficient for our precision goals -- the level of approximation error that this corresponds to due to Newton iterations is on the order of $10^{-4}$~\cite{haner2018optimizing}-- this evaluates to
\begin{align}
    N_\mathrm{CNOT} = 36788.
\end{align}
and
\begin{equation}
    N_\mathrm{qubits} = 119.
\end{equation}

In order to calculate the full cost of a single displacement, we need to find the number of exponentials needed in the Trotter-like decomposition in Algorithm~\ref{alg_Fock_Binary}. The number of exponentials is determined by the Trotter error, which is difficult to accurately estimate as the error depends sensitively on the input state and the signs of the commutators that arise in a Trotter-Suzuki decomposition.  As a first (most likely grossly overestimated) estimate, the number of exponentials $N_{\exp}$ needed in a Trotter-Suzuki formula of order $2k$ to ensure that the error is $\epsilon/2$-small is~\cite{wiebe2010higher} 
\begin{equation}\label{eq:Trotterbd}
    N_{\exp} \le \left\lceil 3K\alpha k \left(\frac{25}{3} \right)^{k}\left( \frac{2\alpha K}{\epsilon} \right)^{1/2k} \right \rceil. 
\end{equation}
From Box 4.1 of Nielsen and Chuang~\cite{Nielsen_Chuang}, we then see that if we denote the idealized case where the square roots are taken to be exact as $V_0(\alpha)$, and take the error in the evaluation of the square roots to be $\epsilon/(2N_\mathrm{exp}\alpha)$, we have that the total error for approximating a displacement $D(\alpha)$ by a Trotter-approximated and square-root-approximated unitary $V(\alpha)$ is bounded above by $\epsilon$ from our above error choices and the following argument:
\begin{align}
    \|V(\alpha) - D(\alpha)\|&\le \|V(\alpha) -V_0(\alpha)\| + \|V_0(\alpha) - D(\alpha)\|\nonumber
    \\
&\le \frac{\epsilon}{2} + N_{\exp} \alpha\left(\frac{\epsilon}{2\alpha N_{\exp}} \right)\le \epsilon,
\end{align}
where the first $\epsilon/2$ is the Trotter error and the second term is the square-root approximation error. For the latter, we have multiplied the per-exponential error $\epsilon/(2N_\mathrm{exp}\alpha)$ by the number of Trotter exponentials $ N_{\exp}$ and the magnitude of the displacement $\alpha$.

\subsubsubsection{Cost of Simulating the Displacement Operator.} Putting everything together, we now calculate the CNOT cost of a single displacement. It is given by the product of our estimate of the number of exponentials in Eq.~\eqref{eq:Trotterbd} with the number of CNOT gates per exponential in Eq.~\eqref{eq:total-cnot}. The latter implicitly depends on the number of exponentials due to the fact that the number of Newton iterations per square root needs to be smaller than $\epsilon/(2N_\mathrm{exp}\alpha)$. We therefore calculate the total CNOT cost as follows. First, we calculate the number of exponentials for a given cutoff, displacement amount $\alpha$, target error $\epsilon$ and Trotter order $k$ from Eq.~\eqref{eq:Trotterbd}. We then determine the target square-root error by $\epsilon/(2N_\mathrm{exp}\alpha)$. Finally we determine the number of Newton iterations $m$ from the numbers shown in Ref.~\cite{haner2018optimizing}: for a square-root error of below $10^{-4}$, $10^{-6}$, $10^{-12}$, $m=2,3,4$ is required. Finally, we total number of CNOT gates is the product of our estimate of the number of exponentials in Eq.~\eqref{eq:Trotterbd} with the number of CNOT gates per exponential in Eq.~\eqref{eq:total-cnot}. 

We show the resulting estimate in Fig.~\ref{fig_comp2} as a function of the cutoff $K$. We find that more than a million CNOT gates are needed in our estimate even for small cutoffs, and an approximately linear increase in the number of gates with cutoff. We also find a significant gap between our gate estimate and the worst-case brute-force decomposition CNOT upper bounds in Ref.~\cite{krol2024}. This implies that our estimate is far from optimal. Most importantly, our estimate for the number of Trotter steps needed most likely is a gross overestimation and a proper use of commutator bounds would improve on this~\cite{childs2021theory}.

Finally, our gate count estimate does not imply that this represents a fundamental gap between qubit-only and qubit-oscillator gate sets. For example, Ref.~\cite{sawaya2020resource} showed more efficient methods which however were not scalable and did not result in significantly better results than brute-force decomposition. Instead, our estimate shows that there are some examples where a straightforward application of qubit compilation techniques leads to a cost that is far greater than that needed by the qubit-oscillator system.  Regardless, the large gap between the two suggests that applications likely exist where the costs of performing a computation in this hybrid instruction set are lower than that of a traditional qubit-based computer.

\subsection{Efficiency of Bosonic Instruction Set Relative to Qubit-only Simulations} \label{sssec:complexity}

As we have seen in Sec.~\ref{app:compilation-cv-to-qubits}, the number of CNOT gates required to carry out a single displacement on a single oscillator with maximum Fock state $63$ (requiring six qubits), in the case where two Newton-Raphson iterations $m=2$ (which corresponds to errors on the order of $10^{-4}$ ~\cite{haner2018optimizing}) are sufficient for our precision goals, is $N_\mathrm{CNOT} = 36788$,
using the described procedure.
In contrast, the number of qubit-controlled operations needed in the bosonic instruction set is 1, because this operation corresponds to a native gate in the instruction set.  In addition to the circuit complexity issues discussed here, the reader is reminded of the important complications associated with qubit damping errors in the representation of oscillators with qubits previously discussed in Sec.~\ref{sec:hardware-efficient-mapping}.

The above analysis demonstrates a significant overhead to simulate a hybrid instruction set using qubit-only hardware. However, it is by no means optimal. Alternative mappings exist that may privilege efficiency in terms of gate count rather than qubit number for example (see Table~\ref{tab:encodings}).
In contrast to the Fock encoding used above, some encodings (e.g., Jordan-Lee-Preskill \cite{jordan2012quantum}) privilege representation of oscillator position and momentum and thus might be better for the particular case of simple displacements, but worse for representing interactions best described in the Fock basis. Many encodings have been compared in a comprehensive review \cite{sawaya2020resource} with optimized qubit gate counts, and a method for decomposing arbitrary continuous variable quantum operations via linear combinations of commutation relations is described in Ref.~\cite{sefi2011how}. Ref.~\cite{Kane2025blockencodingbosons} discusses block encoding bosonic operators into qubits using the Jordan-Lee-Preskill representation.  Ref.~\cite{hanada2025exponentialimprovementquantumsimulations} discusses optimizing the efficiency of qubit simulations of bosonic lattice gauge theories.

To better reveal the value of hybrid CV-DV instruction sets, a systematic study of protocols to simulate bosonic instruction sets using qubits in terms of entangling gates, qubit count, and number of samples required, and conversely using bosonic instruction sets to simulate qubits, needs to be performed. Furthermore, theory-experiment collaboration to include realistic gate infidelities in these analyses is another important aspect to facilitate device engineering. Lastly, cross-platform simulation at the system- and algorithm-level (beyond the gate level) targeting specific applications such as Hamiltonian simulation will deliver a more practical understanding of the advantages and limitations of such hybrid qubit/bosonic hardware.

\section{Conclusions and Outlook}
\label{sec:ConclusionsOutlook}

The primary focus of this work is pedagogical introduction of explicit and experimentally relevant instruction set architectures (ISAs) and abstract machine models (AMMs) for hybrid CV-DV quantum processors composed of qubits and quantized bosonic modes. We start from the physical behavior of bosonic states and systematically develop bosonic gate models, and hybrid architectures including error correction strategies, compilation theory and techniques, as well as high-level proxy algorithms and applications. We have presented a few examples illustrating how such hybrid quantum processors can be utilized to solve practical problems, including DV-CV state transfer, the quantum Fourier transform, bosonic quantum error correction, and quantum simulation of matter with bosonic degrees of freedom.  We show for the latter that hybrid digital architectures may provide substantial computational advantages over existing approaches that use only qubits because of the ability to avoid costly arithmetic operations in favor of equivalent native bosonic operations.  Below, we close with some perspective on challenges and opportunities and thoughts on directions for the future.

\subsection{Challenges and Opportunities for AMMs and ISAs}
The challenges of, and opportunities for, defining meaningful AMMs and designing effective ISAs for hybrid quantum CV-DV arise concerning five aspects: (i) the choice of how quantum information is represented using bosonic states, (ii) the universal bosonic gate set employed, (iii) the variety of hardware and instruction set architectures available for hybrid oscillator-qubit systems, (iv) effective compilation methods, and (v) use cases for hybrid quantum processors. In this work, we have presented and analyzed AMMs and ISAs for hybrid CV-DV quantum hardware by addressing these five aspects. We touch upon each of these below.

Unlike a qubit, which admits a natural representation analogous to a binary bit, bosonic systems offer many different potential representations for encoding state information. Since the bosonic system is a simple harmonic oscillator, the position or momentum of the state can be natural state variables, with continuous values. Alternatively, since the energy of the oscillator is quantized, the Fock basis given by the discrete number of energy quanta may also be a natural choice. Understanding and taking advantage of these state representation options can be complex, particularly when it is desirable to be able to map quantum information between oscillators and qubits, as would be natural within a hybrid CV-DV AMM.

Sec.~\ref{sec:basics} discussed how quantum logic gate sets acting on bosonic systems are much richer than those for qubits. Available ``analog'' bosonic gates include not just operations on single oscillators such as displacement and squeezing, but also mixing operations such as beam-splitters, between multiple oscillators. And despite their analog nature, the simplest bosonic gates generated by quadratic oscillator Hamiltonians can be understood as being directly analogous to Clifford gates, in that they form a discrete subspace of unitary transforms governed by the Gottesman-Knill theorem and are easy to simulate classically on Gaussian input states. But many bosonic gates, and especially gates that couple bosonic systems to a qubit, are non-Clifford. 

Noise models and quantum error correction constructs also depend on the expressibility of a chosen gate set. It is thus challenging to choose appropriate gate sets to make meaningful ISAs for hybrid quantum CV-DV systems. The ISAs that we present are quite small and can each be viewed as a kind of Reduced Instruction Set Computer (RISC) ISA \cite{patterson-riscv} having a small number of instructions defined with a set of continuous parameters accompanying them. Such RISC instruction set architectures greatly ease the problem of translation down the stack from applications at the top to the calibrated microwave pulses that need to be applied to the hardware at the bottom.

Efficient compilation of desired unitary transformations into the instructions for a hybrid CV-DV architecture is also challenging, and two routes are generally available. First, numerical optimization of the parameters for sequences of instructions can be performed to effect the circuit compilation. This is computationally feasible for defining `assembly language' style instructions at small scales for portions of the hardware but is completely infeasible for compiling entire algorithms, and results in circuits that are not `human-readable' and provide little intuition for generalization of hybrid CV-DV hardware utility. A second route is via analytical circuit synthesis. This is possible for specific families of hybrid CV-DV circuits, such as quantum signal processing, as we showed. While the resulting circuits are not necessarily as efficient as numerically synthesized circuits, they are human-readable and provide significant insights into the power of hybrid hardware. They also may prove useful as `seed' circuits and motifs for further numerical optimization.

The fifth challenge for these AMMs and ISAs is their use case: for what end-applications might hybrid quantum CV-DV be particularly well suited? As we discussed, systems composed of spins and oscillators are natural candidates for quantum simulation with hybrid CV-DV hardware, due to the expected efficiency of mapping possible, compared with using solely qubit-based hardware, or solely oscillator-based hardware. The natural physical dynamics of oscillators also present intriguing possibilities to be harnessed, as the Hamiltonian for a simple harmonic oscillator is known to evolve position eigenstates into momentum eigenstates, embodying a Fourier transform. We studied both quantum simulation and the quantum Fourier transforms using hybrid hardware in this paper. More applications will arise, and we expect their discovery to be accelerated by the formulation of accessible abstract machine models and instruction set architectures for hybrid quantum CV-DV hardware -- precisely our goal here.

The physical-level interactions that enable the proposed ISAs and AMMs can be realized on quantum hardware containing both qubits and bosonic degrees of freedom, and we have discussed here superconducting, trapped ion, and neutral atom platforms. Dissipation and quantum noise on such hardware can be corrected or suppressed with bosonic quantum error-correcting codes (potentially further concatenated with qubit codes). These physical interactions allow us to define several instruction sets summarized in Table \ref{tab:ISA_overview}, including among others, the Fock-space, phase-space, and sideband ISAs. We have also shown explicitly how instructions from these ISAs can be composed to achieve universal hybrid CV-DV quantum computation and further demonstrate how to characterize the quantum device with a continuous variable analog of gate set tomography. 

Systematic compilation techniques compose these universal gates into useful primitives and algorithms that solve practical problems, as we discussed in Sec.~\ref{sec:compilation}.  These compilation techniques include strategies for compiling hybrid single-qubit oscillator unitaries, which are obtained by generalizing the well-known Trotter-Suzuki and product formulas, quantum signal processing, and linear combination of unitaries techniques from DV systems to hybrid CV-DV systems. We presented exact analytical gate synthesis methods for multi-qubit entangling gates and multi-oscillator entangling gates in superconducting systems that are inspired by the  M{\o}lmer-S{\o}rensen gate for ion traps.

At the system level, we defined three key quantum AMMs: 1) qubit-centric, 2) oscillator-centric, and 3) hybrid oscillator-qubit, which all have universal computation capabilities. These AMMs can represent abstractions of the machine for a high level of the stack that is user-facing, defining the mathematical abstract Hilbert space where the computation happens.  Alternatively, they can be lower-level abstractions that are transpiler-visible,  presenting to the transpiler what physical degrees of freedom are available as computation components and what their connectivity and allowed gates are. The formal establishment of these AMMs should prove particularly useful for resource estimation on hybrid CV-DV quantum processors. 

We further examined applications of quantum computing, such as quantum simulation, illustrating how quantum electrodynamics and related lattice gauge Hamiltonians can be compiled into hybrid qubit/bosonic operations.  We find evidence of advantage for near-term applications of hybrid systems over standard methods for purely qubit-based simulations. In addition, we presented extensions of quantum signal processing to the non-abelian case, for which the qubit rotation angles are functions of the non-commuting position and momentum variables of bosonic modes.  With this tool in hand, we examined applications of quantum signal processing within these architectures and discussed explicit approaches to implement the quantum Fourier transform. 

In conclusion, the ISAs and AMMs proposed here significantly expand our understanding of how distinct quantum resources can be leveraged in a unified computational framework and demonstrate the potential of DV-CV hybrid quantum processors for the upcoming post-NISQ era of early-stage fault tolerance.

\subsection{Open Questions and Directions for Future Work}
Despite these new understandings and achievements, there are still many open questions ranging from the fundamental side to more practical aspects of hybrid DV-CV quantum systems that remain to be addressed in the future.

First and foremost, as we have seen in Sec.~\ref{sec:apps}, hybrid quantum hardware opens many possibilities for interesting applications, but it is worth approaching the potential with careful consideration. After all, classical digital computation has advanced just fine without needing assistance from embedded classical harmonic oscillators. But quantum computing hardware is much less mature, and early applications might benefit from hardware capabilities which are closer to application models. Identifying more use cases where CV-DV hardware can really deliver an advantage over qubit-only or classical computers will be important. Given NISQ needs, examples of such applications likely include quantum simulation of coupled nuclei-electron processes \cite{weinberg2012proton,hammes2015proton} or light-matter interactions \cite{bloch2022strongly,ribeiro2018polariton} relevant to chemical reactions and materials science applications. As another example, the hybrid CV-DV quantum signal processing framework established in the current work serves as a potentially ideal tool for processing quantum signals.  Classical signal processing techniques from electrical engineering for filter design and classical  CV-DV conversion \cite{oppenheim1999discrete} can provide much intuition and guidance for further developments in the quantum case.

And while the major quantum algorithms become united in the framework of quantum signal processing and quantum singular value transforms, the newly emerging hybrid CV-DV quantum signal processing techniques extend possibilities to the non-abelian case, for which a complete theory remains to be discovered. From a NISQ viewpoint, further development is required for variational CV-DV algorithms beyond Sec.~\ref{sec:numerical-optimization} that can be run on such hybrid processors \cite{zhang2023energydependent}. Such a set of tools would dramatically increase the scope of the signal-processing transformations that can be implemented on input parameters and allow more powerful analytic compilation methods for general multi-variable transformations based on unitary transformations of bosonic modes.  While we describe some QSP methods in this work, the presentation is mathematically abstract, and higher-level CV-DV quantum programming languages such as Bosonic Qiskit~\cite{Biskit} are needed to simplify the programming process for a broader community of users.

We have presented various CV-DV ISAs based on discrete sets of control Hamiltonians but continuously parameterized gates. An open question in this direction is to address the complexity of generating arbitrary unitaries in CV systems.  While the Trotter-Suzuki-based approaches considered here and in~\cite{kang_seeking_2024} can provide sub-polynomial scaling with the error tolerance for the synthesized gates, existing approaches fall short of the scaling provided by circuit synthesis methods such as the Solovay-Kitaev theorem~\cite{dawson2006solovay,kliuchnikov2013asymptotically} which achieves poly-logarithmic scaling in the error tolerance when compiling over a discrete gate set.   The development of better CV gate synthesis methods may not only open up new possibilities in compiling unitaries but also open up the door to the development of bosonic error correcting codes where the physical and logical operations are natural bosonic operations such as discrete displacement or squeezing operations.

The non-Abelian QSP techniques used in this work involve composite pulses in phase space designed to cancel first-order errors. Generalizations of these composite pulses similar to Ref.~\cite{low2016methodology,low2017optimal,low2017hamiltonian} are required to develop optimal sequences for arbitrary CV Hamiltonian construction. Formally, this task is equivalent to designing a constructive algorithm that can find optimal non-abelian QSP sequences for arbitrary polynomials $f(\hat x,\hat p)$~\cite{qspi2023}. In Refs.~\cite{haah2019product} the authors find a constructive algorithm to generate arbitrary polynomials for the case of single-variable QSP. The generalization of this scheme to QSP using two commuting variables was shown in Ref.~\cite{Gilyen2019}. Such a construction for non-abelian QSP extends to the case of multi-variable QSP with non-commuting variables, and thus, remains an open question. The multiple non-commuting variables for the case of the phase-space ISA will be the numerous quantum phase space vectors $\hat v=\alpha \hat x+\beta \hat p$. The examples from Ref.~\cite{singh_towards_2025} discussed in Secs.~\ref{ssec:compilation-bosonic-qsp-qsvt} and~\ref{sec:qec-compilation}  belong to a subclass of this framework, QSP with just two non-commuting variables. A hierarchy of various QSP frameworks developed with an eye towards the hybrid phase-space instruction set can be found in Ref.~\cite{singh_towards_2025}.

A related aspect of algorithm design is to identify new useful universal instruction sets for such hybrid quantum processors. In particular, we have seen in Sec.~\ref{sssec:stellar-rep} that the stellar representation provides an intuitive picture of how CV states and operations can be characterized in terms of the roots of complex analytic functions and their associated time dynamics. Designing an ISA that works directly within the stellar representation picture would be very interesting and useful.

From an architectural perspective, we need more complete ISAs that include a memory hierarchy and sophisticated quantum algorithmic and arithmetic logic units (ALUs). In terms of memory, a mixed system of qubits and bosonic modes provides opportunities by balancing the fast gates and high connectivity of qubits with the long lifetimes and novel gate sets of bosonic modes.  The needed high connectivity can be obtained through quantum communication using bosonic modes, for example by the cavity SWAP gates discussed in Sec.~\ref{ssec:exact-analytical-qubit-gates} and/or through use of the LAQCC model described in Sec.~\ref{ssec:amms-detail} and Sec.~\ref{sssec:control-flow} to efficiently produce qubit GHZ states that can be used for quantum teleportation. 

At the simplest level, recent architectures \cite{stein2023microarchitectures} have proposed using bosonic modes as small register memories, embedding them in architectures specifically designed to match the connectivity requirements of algorithms. Recent work with multi-mode bosonic systems has shown their utility as memories for factoring~\cite{2021PhRvL.127n0503G} and unique error correction architectures~\cite{9251988}. This is particularly important if proposed memory hierarchies are to be realized~\cite{10.1109/ISCA.2006.32} as in classical systems. Building upon these memory hierarchies, more sophisticated quantum ALUs that take advantage of the hybrid CV-DV gates (Table \ref{tab:gates-osc} and \ref{tab:gates-qubit-osc}) and the continuous-variable nature of oscillator dynamics need to be developed. Sec.~\ref{sec:apps} demonstrated key quantum primitives including the Fourier transform and Hamiltonian simulation, but creating more advanced primitives and proxy applications, such as modular exponentiation and matrix arithmetic, remains to be done. Instruction- and processor-level parallelism often used in modern classical computers \cite{hennessy2012computer} could also be deployed based on a more sophisticated ISA and memory hierarchy.

Qubit gate fidelities can be efficiently measured using randomized benchmarking and related techniques.   To date, however, there are no corresponding systematic methods for CV operations, due to the absence of unitary $2$-designs for CV systems.  While we provide  in Sec.~\ref{sssec:benchmarking} an approach to gate set tomography using Wigner function tomography (see App. \ref{sec:characteristic-function}), this approach cannot be generalized to arbitrarily high dimensions without making strong assumptions about the quantum channels being learned. Such channel tomography estimates will be necessary for us to assess the degree to which these systems will need quantum error correction. 

Regarding the topic of noise and reliability, we have discussed examples of bosonic error correction codes that encode discrete logical qubits into oscillators.  Creation and quantum control of such bosonic logical qubits requires non-Gaussian operations that in turn require the presence of a transmon or other auxiliary qubit or anharmonic element which typically has poorer coherence properties than the oscillator they are controlling.  This raises fundamental fault-tolerant control and measurement challenges which are only now beginning to be addressed \cite{pietikäinen2024strategies,PhysRevX.10.011001,teoh2022dualrail,ReinholdErrorCorrectedGates,PathIndependentGatesPhysRevLett.125.110503,PathIndependentAlgebraicStructure} and many opportunities remain for further progress.  An interesting future direction would be to build fault tolerance into the cost function when designing  numerically optimized quantum control sequences.

To bring out the role of bosonic error correction in terms of fault tolerance, we have discussed this direction in Sec.~\ref{sec:bosonic-QEC}. We discuss how oscillators can be useful in improving the cost of magic state distillation, a resource-intensive component of fault-tolerant quantum computing (FTQC). So far, all versions of quantum error correction have deemed magic state distillation to be a necessary component to implement non-Clifford gates, an essential requirement for quantum advantage. We pointed out in Sec.~\ref{ssec:qec-overview} that oscillator-based codes can quadratically reduce the overhead cost associated with this FTQC component by changing the noise profile of the underlying qubit~\cite{singh2022high}. In addition, we highlighted that there are efficient ways to realize an error-corrected qudit in an oscillator. Qudits can be helpful in various ways including reduction of the computation cost and one may expect that low error qudits will be high in demand in the future. The ability to engineer high-fidelity qudits via error correction in the oscillator also takes improvement in magic state distillation several steps further. It has been shown that qudits can directly improve the resource overhead of this preparation by several orders of magnitude by directly enabling more efficient distillation schemes~\cite{krishna2019towards,campbell2012magic}.

We also briefly discussed the more challenging question of error correction of oscillators (\textit{qua} oscillators, and not as logical qubits). So far there has been only one attempt at such a construction, given by Ref.~\cite{noh2020encoding}, which corrects errors under a classical displacement channel. Soon after, in Ref.~\cite{hanggli2021oscillator} it was shown that code families designed under this noise model assumption do not have a threshold. Assuming that the true noise model of an oscillator is dominantly photon loss, it remains an open question if there is an oscillator code family that can protect oscillators against this more realistic noise.  Formally if we introduce a cutoff on the boson number the problem reduces to qudit error correction for large qudit dimension $d$, but it seems extremely challenging to efficiently preserve the oscillator states and operator algebra for large $d$. If such a construction exists, it would be a revolutionary step in reducing the complexity of many quantum simulations and CV algorithms potentially yielding a quantum advantage.  However, as analogous ideas for classical computing have failed to deliver advantage thus far~\cite{Yamamoto2017,Wang2021Solving,Bonnin2022} a complete investigation of all resources needed to achieve a computational advantage in near-term quantum devices will likely be needed to more fully understand the benefits and drawbacks of this approach.

Finally, while this effort seeks to build bridges translating between the languages and worlds of physics and computer science, the core ideas of compilation techniques need to evolve further beyond those presented here.  We have demonstrated how techniques such as the Trotter-Suzuki formula and a linear combination of unitaries can be used to achieve effective compilation of single qubit-oscillator operations and multi-qubit-oscillator entangling gates (Sec.~\ref{sec:compilation}). However, more work is required to develop additional compilation strategies. Recently, ZX-calculus -- a useful graphical tool for reasoning about DV circuits~\cite{Coecke_2011,Wetering2020} -- was formulated for CV quantum circuits~\cite{Nagayoshi2024}. An extension to hybrid CV-DV circuits constitutes an interesting future direction. Separately, as already noted, a complete theory of non-abelian hybrid QSP does not yet exist, but its formulation would be very helpful for analytical compilation. More generally, a complete formal complexity-theoretic analysis of hybrid CV-DV circuit synthesis for tasks such as state transfer and general unitary synthesis does not yet exist. In App.~\ref{app:cross-compilation} we present cross-compilation techniques among different pairs of ISAs, but much work remains to be done to improve the efficiency and practicality of cross-compilation. 

At a fine-grain level, compilation techniques capable of identifying long idle times and inserting SWAPS to memory are crucial. This may be performed in conjunction with the use of flag qubits~\cite{2018Quant...2...53C} or single-shot error mitigation~\cite{2022arXiv221203937V}. Furthermore, hybrid architectures will require transpilation techniques that map programmatic elements with long idle times to qubits near memories. Compilers should also be adapted to take advantage of the new sets of native, low-cost gates afforded by bosonic hardware. More practically, it would be very useful to create a software library or toolkit built on top of these analytical and numerical compilation techniques that can perform application-tailored resource estimation in terms of appropriate metrics such as the gate count in hybrid ISAs.

Looking further into the future, optimization of quantum compilation using quantum computers \cite{Khatri2019quantumassistedcompiling} is an exciting possibility.  Another potentially important direction that would apply to algorithm development for both traditional and hybrid hardware would be the extension of the computation model for quantum computers beyond algorithms based on purely unitary operations and final state measurements to more general quantum channel operations. Such more general quantum channels for algorithms could be created by including engineered dissipation and/or mid-circuit measurements and feed-forward. 

The rate of current experimental and theoretical progress in the field is astounding, yet many problems, both fundamental and engineering, remain to be solved before we fully escape the NISQ era and move into the era of useful, practical, and fault-tolerant quantum computation and simulation.  We hope that the hybrid architectures and compilation methods described in this work will help accelerate progress across multiple platforms  towards this important goal.

\begin{acknowledgments}
    This research was funded by the U.S. Department of Energy, Office of Science, National Quantum Information Science Research Centers, Co-design Center for Quantum Advantage (C$^2$QA) under contract number DE-SC0012704 (Basic Energy Sciences, PNNL FWP 76274). C$^2$QA led this research. Shraddha Singh and Steven Girvin acknowledge support by the Army Research Office (ARO) under Grant Number W911NF-23-1-0051. Jasmine Sinanan-Singh was supported in part by the Army Research Office under the CVQC project W911NF-17-1-0481. Jasmine Sinanan-Singh and Yuan Liu were supported in part by NTT Research. Yuan Liu acknowledges startup funding support from North Carolina State University. Alexander Schuckert acknowledges funding from the U.S. Department of Energy, Office of Science, National Quantum Information Science Research Centers, and Quantum Systems Accelerator. 
John M. Martyn acknowledges support from the National Science Foundation Graduate Research Fellowship under Grant No. 2141064.
Micheline B. Soley was supported by the Yale Quantum Institute Postdoctoral Fellowship and the Office of the Vice Chancellor for Research and Graduate Education at the University of Wisconsin-Madison with funding from the Wisconsin Alumni Research Foundation.
Isaac L. Chuang was supported in part by the NSF Frontier Center for Ultracold Atoms (grant number PHY- 2317134).
The views and conclusions contained in this document are those of the authors and should not be interpreted as representing the official policies, 
either expressed or implied, of the Army Research Office (ARO), or the U.S. Government. 
The U.S. Government is authorized to reproduce and distribute reprints for Government purposes 
notwithstanding any copyright notation herein. 

The authors acknowledge helpful discussions with Jacob Curtis, Michael DeMarco, Shruti Puri, Yongshan Ding, Robert Schoelkopf, Cindy Regal, Baptiste Royer, Timothy Stavenger, and Ang Li, and thank Luke Bell for helpful discussions and verification of some of the key equations presented in the compilation section.

External interest disclosure: SMG receives consulting fees and is an equity holder in Quantum Circuits, Inc.
\end{acknowledgments}

\bigskip\bigskip\bigskip
\centerline{\bf APPENDICES}

\appendix

\section{Physical Implementations}
\label{app:PhysImp}

Here we present a physical-level discussion of quantum control of hybrid systems consisting of bosonic modes and qubits, which serves as background information for the hybrid gates presented in Sec.~\ref{sec:non-Gaussian_and_Hybrid}. Table \ref{tab:gates-qubit} summarizes all qubit-only gates available in superconducting circuits and ion trap systems and Table \ref{tab:free_interaction} presents the microscopic Hamiltonian terms freely available in these systems.

 \begin{table*}[htb]
\centering
\begin{tabular}{|c|c|c|c|c|}

                                              \hline &                                    & \textbf{Gate Name}                            & \textbf{Parameters} & \textbf{Definition} 

                                              \\ \hline
\multirow{11}{*}{\rotatebox[origin=c]{90}{\textbf{Qubit-only}}}                  & \multirow{3}{*}{\rotatebox[origin=c]{90}{ Single-}} 
& Qubit $x, y$ rotation\textsuperscript{1} & $\theta \in [0,4\pi),\varphi \in [0,2\pi)$ & $R_\varphi\left(\theta\right) = \exp\left(-i \frac{\theta}{2}\sigma_\varphi\right)$        \\
& &  &  & $\sigma_\varphi = \sigma_x\cos\varphi + \sigma_y \sin \varphi$     \\ \cline{3-5} 
                                             &                                    & Qubit $z$ rotation\textsuperscript{1}                     & $\theta \in [0,4\pi)$      & $R_z(\theta) = \exp\left(-i\frac{\theta}{2}\sigma_z\right)$       \\ \cline{2-5} 
                                             & \multirow{8}{*}{\rotatebox[origin=c]{90}{ Multi-}}             & $ZZ$ interaction                       & $\theta \in [0,2\pi)$     & $ZZ(\theta) = \exp\left( -i \frac{\theta}{2} \sigma_z \otimes \sigma_z \right)$      \\ \cline{3-5} 
                                             &                                    &   CNOT    (aka C$X$)                               &            &    $\frac{I+\sigma_z}{2} \otimes I + \frac{I-\sigma_z}{2} \otimes \sigma_x$        \\ \cline{3-5} 
                         &                    &   CPHASE                              &  $\varphi \in [0,2\pi)$           &   $\frac{I+\sigma_z}{2} \otimes I + \frac{I-\sigma_z}{2} \otimes e^{i\varphi\frac{I-\sigma_z}{2}}$\\ 
                         &                           &   C$Z$                                &   $\varphi = \pi$         &    
                         $\frac{I+\sigma_z}{2} \otimes I + \frac{I-\sigma_z}{2} \otimes \sigma_z$
                            \\ \cline{3-5}                                        
                                             &                                    &   $i$SWAP                              &  $\theta \in [0,2\pi) $         &   $\exp(i\frac{\theta}{2}(\sigma_x \sigma_x + \sigma_y \sigma_y))$          \\ 
                                            &                                    &                                 &   $\theta=\pi/2$         &   $\exp(i\frac{\pi}{4}(\sigma_x \sigma_x + \sigma_y \sigma_y))$          \\ \cline{3-5} 
                                             &                                    &   fSim                             &     $\theta \in [0,2\pi),\varphi\in [0,2\pi)$       &   $i$SWAP($-\theta$) $\cdot$ CPHASE($\varphi$)         \\  
                                             &                                    &   fSWAP                              &     &   $i \left[R_z(\frac{\pi}{2})\otimes R_z(\frac{\pi}{2})\right] \textrm{fSim}(\frac{\pi}{2},0)$     \\
                                             \hline 
\end{tabular}
\caption{\textbf{Common qubit-only gates.} The single-qubit gates are typically natively available. If not natively available, the multi-qubit gates can be readily synthesized through their coupling to the bosonic modes as discussed in Sec.~\ref{ssec:compilation-entangling}. \textsuperscript{1}In general, qubit rotations about any axis of the Bloch-sphere are given by $R_{\hat{n}}\left(\theta\right) = \exp\left(-i\frac{\theta}{2} \hat{n} \cdot \vec \sigma\right)$ where $\vec{\sigma} = \left(\sigma_x,\sigma_y, \sigma_z\right)$ and $\hat{n}$ is a unit-vector, and this gate can be implemented directly on some physical qubits with x-y-z control. However, for many qubits (such as fixed-frequency transmons), only x-y control is native, and the z rotations are implemented \textit{in software} \cite{EfficientZgates_PhysRevA.96.022330}, hence our choice to separate qubit rotations into either x-y or z rotations. fSim (Sec.~\ref{sssec:fermion-boson-problems}) is the fermion simulation gate \cite{PhysRevLett.120.110501}, while fSWAP (Sec.~\ref{sssec:fermion-boson-problems}) is the fermionic SWAP gate, which swaps the state of two qubits and adds a minus sign to the state $\ket{11}$ to account for antisymmetric fermionic statistics.
}
\label{tab:gates-qubit}
\end{table*}

\begin{table*}[htb]
    \centering    
    \begin{tabular}{|c|c|c|}
    \hline
   \textbf{Interaction} & \textbf{Hamiltonian} & \textbf{Platform}\\
    \hline
    Dispersive Coupling & $a^\dagger a\sigma_z$ & Superconducting Circuits\\
    \hline
    Sideband Interaction & $a\sigma_+ +a^\dagger\sigma_-$ & Trapped Ions, Superconducting Circuits, Neutral Atoms\\
    \hline
    Linear Drive & $\alpha a^\dagger - \alpha^* a$ & Superconducting Circuits, Trapped Ions\\
    \hline
    Parametric Drive & $\zeta^* a^2 - \zeta a^{\dag 2}$ & Superconducting Circuits, Trapped Ions\\
    \hline
    Controlled Squeezing & $\sigma_x ((a^\dagger)^2 + a^2)$ & Trapped Ions, Neutral Atoms\\
    \hline
    Free Oscillator Evolution & $a^\dagger a$ & Trapped Ions, Superconducting Circuits\\
    \hline
    Controlled Beam-splitter & $\sigma_x (a^\dagger b + b^\dagger a)$ & Trapped Ions\\
    \hline
    \end{tabular}
    \caption{Natively available interactions on various quantum computing platforms.  Not listed are the Gaussian oscillator-only gates available in the superconducting platform:  displacement, single-mode squeezing, two-mode squeezing, two-mode SUM gates, and two-mode beam-splitter operations are all natively available through classical microwave drives applied to directly to the oscillators and/or to parametric elements (i.e., 3- and 4-wave mixing elements) coupled to the oscillators.}
    \label{tab:free_interaction}
\end{table*}

\subsection{Superconducting and Cavity QED}
\label{app:cQED}

\subsubsection{Fundamentals}
Let us consider a single oscillator coupled to a single transmon qubit via the Hamiltonian \cite{Blais2004,Koch2007,Blais2020,BlaiscQEDReviewRMP2020}
\begin{eqnarray}
    H&=&H_0+H_1+V_0,\\
    H_0&=&\bar\omega_\mathrm{R}a^\dagger a + \bar\omega_\mathrm{Q}b^\dagger b,\\
    H_1&=&-\frac{\bar K}{2}b^\dagger b^\dagger bb,\\
    V_0&=&g (a^\dagger + a)(b^\dagger + b).
\end{eqnarray}
Here, $H_0$ is the Hamiltonian of the resonator with `bare' (i.e., uncoupled) frequency $\bar\omega_\mathrm{R}$ and the transmon with bare frequency $\bar \omega_\mathrm{Q}$ and $H_1$ describes the (bare, negative) anharmonicity $-\bar K$ of the transmon. The microscopic Hamiltonian of the transmon involves a cosine of the phase difference across the Josephson junction.  $H_1$ is an approximation to the transmon Hamiltonian that results from expanding the cosine to fourth order and making a rotating wave approximation (RWA) \cite{Koch2007}.  Because of the RWA, $H_1$ is diagonal in the Fock (excitation number) basis and the eigenstates of $H_0+H_1$ are unchanged by $H_1$.  We still have harmonic oscillator eigenstates, but for the transmon, the energy level spacing is not uniform due to the presence of the anharmonicity from $H_1$.  The spectrum of the exact Hamiltonian is of course bounded from below.  The approximate Hamiltonian containing $H_1$ is not bounded from below, implying that it must eventually break down at large excitation numbers where the truncation of the cosine potential of the Josephson junction ceases to be a valid approximation.  We ignore this complication here.

$V_0$ is the coupling between the charge polarization (dipole moment) of the transmon and the cavity electric field.  The coupling constant $g$ is known in the cavity- and circuit-QED literature as the vacuum Rabi coupling \cite{Blais2004,Wallraff_cQED_2004,Blais2020,BlaiscQEDReviewRMP2020}.

If we assume that the transmon is always within the manifold spanned by its ground state $|g\rangle$ and its first excited state $|e\rangle$, we can make the two-level qubit approximation, mapping the transmon operators onto spin operators.  The mapping of the lowest two transmon states onto a qubit spin is accomplished by 
\begin{eqnarray}
b^\dagger  b &\rightarrow& |e\rangle\langle e|=\frac{\hat I+\sigma_z}{2},\label{eq:bdagbisZ}\\
b&\rightarrow& \sigma^-;\,\,\, b^\dagger\rightarrow \sigma^+\label{eq:bissigminus},\\
H_1&\rightarrow& 0,\\
H&\rightarrow&\omega_\mathrm{R} a^\dagger a +\frac{\omega_\mathrm{Q}}{2}\sigma_z+g\left(a+a^\dagger\right)\sigma_x,
\label{eq:RabiHam}
\end{eqnarray}
where we have neglected an irrelevant constant in the energy in the last expression. Eq.~(\ref{eq:RabiHam}) describes the so-called Rabi Hamiltonian.  

The Rabi Hamiltonian can be further simplified if we assume that the cavity--qubit detuning $\Delta= \bar\omega_\mathrm{R}-\bar\omega_\mathrm{Q}$ is small enough relative to the cavity frequency $\bar \omega_\mathrm{R}$ that the rotating wave approximation can be made to further simplify $V_0$ to
\begin{equation}
    V_0=g(a^\dagger b+ab^\dagger),
\end{equation}
which yields the celebrated Jaynes-Cummings Hamiltonian \cite{Blais2004}
\begin{equation}
    H=\bar\omega_\mathrm{R} a^\dagger a +\frac{\bar\omega_\mathrm{Q}}{2}\sigma_z+g\left(a\sigma^+ + a^\dagger\sigma^-\right).\label{eq:JCphysical}
\end{equation}
This `bare' Jaynes-Cummings Hamiltonian has its parameters fixed at the time of hardware fabrication and is not programmable, though in some circuit QED architectures, the transmon frequency $\bar\omega_\mathrm{Q}$ can be tuned by application of a magnetic field that threads flux through a SQUID loop.
Typical values for the anharmonicity of the transmon are $\bar K/(2\pi)\sim 50-200$ MHz and the vacuum Rabi coupling can be up to $g/(2\pi) \sim 150$ MHz.  Typical values for the operating frequencies of transmons and cavities are in the range $\bar\omega_\mathrm{Q}/(2\pi), \bar\omega_\mathrm{R}/(2\pi)\sim 5-10$ GHz.

An alternate approach to the Hamiltonian of a qubit coupled to a cavity that is especially useful when higher levels of the transmon come into play, or when the cavity and transmon are detuned from each other (`dispersive regime') is described below.  In the dispersive regime, we find that it is possible to apply continuous microwave drives to the qubit and the cavity which yields an effective Jaynes-Cummings model with adjustable parameters.

$H_0+V_0$ is the harmonic part of the total Hamiltonian and is of course exactly soluble.  We use the strategy of expressing the anharmonic term, $H_1$ based on exact normal modes of the harmonic part of the Hamiltonian.  For simplicity, we make the rotating wave approximation for $V_0$ so that the harmonic Hamiltonian can be written
\begin{eqnarray}
    H_0+V_0&=&\begin{array}{c}\left(\begin{array}{cc}a^\dagger&b^\dagger\end{array}\right)\\ \phantom{a}\end{array}
    \left(\begin{array}{cc}M_{11}&M_{12}\\M_{21}&M_{22}\end{array}\right)\left(\begin{array}{c}a\\b\end{array}\right),\nonumber\\
    &\quad&   
\end{eqnarray}
where
\begin{equation}
M   =\left(\begin{array}{cc}\bar\omega_\mathrm{R}&g\\g&\bar\omega_\mathrm{Q}\end{array}\right)
=\frac{\bar\omega_\mathrm{R}+\bar\omega_\mathrm{Q}}{2}\hat I+ \left(\begin{array}{cc}\frac{\Delta}{2}&g\\g&-\frac{\Delta}{2}\end{array}\right).
\end{equation}
From this, we see that the two  normal modes of the coupled oscillators have frequencies
\begin{equation}
    \omega_\pm=\frac{\bar\omega_\mathrm{R}+\bar\omega_\mathrm{Q}}{2}\pm\sqrt{(\Delta/2)^2+g^2},\label{eq:normmodefreqs}
\end{equation}
and the normal mode operators are
\begin{eqnarray}
\left(\begin{array}{c}A\\B\end{array}\right)=\left(\begin{array}{cc}\cos\frac{\theta}{2}&+\sin\frac{\theta}{2}\\-\sin\frac{\theta}{2}&\cos\frac{\theta}{2}\end{array}\right) \left(\begin{array}{c}a\\b\end{array}\right),
\end{eqnarray}
where $\tan\theta=\frac{2g}{\Delta}$, and
\begin{eqnarray}
\sin\frac{\theta}{2} &=&\sqrt{\frac{1-(\Delta/2)/\sqrt{g^2+(\Delta/2)^2}}{2}}    , \\
\cos\frac{\theta}{2} &=&   \sqrt{\frac{1+(\Delta/2)/\sqrt{g^2+(\Delta/2)^2}}{2}}.
\end{eqnarray}

To express the anharmonic term $H_1$ in the normal-mode basis, it is convenient to have the inverse transform
\begin{eqnarray}
\left(\begin{array}{c}a\\b\end{array}\right)=\left(\begin{array}{cc}\cos\frac{\theta}{2}&-\sin\frac{\theta}{2}\\+\sin\frac{\theta}{2}&\cos\frac{\theta}{2}\end{array}\right) \left(\begin{array}{c}A\\B\end{array}\right).
\end{eqnarray}
Making the rotating wave approximation we have
\begin{eqnarray}
H_1&=&-\frac{\bar K}{2}b^\dagger b^\dagger bb,\\
&\approx&-\big{[}\chi_{BB}B^\dagger B^\dagger BB+\chi_{AB} B^\dagger BA^\dagger A
\nonumber\\&\quad& ~~~~~~ + \chi_{AA}A^\dagger A^\dagger AA\big{]},
\label{eq:dispHRWA}
\end{eqnarray}
where the renormalized transmon self-Kerr (anharmonicity), the transmon-cavity cross-Kerr (dispersive interaction), and the self-Kerr induced in the cavity are respectively
\begin{eqnarray}
\chi_{BB}&=&\frac{K}{2}\cos^4(\theta/2),\\
\chi_{AB}&=&2K\cos^2(\theta/2)\sin^2(\theta/2),\\
\chi_{AA}&=&\frac{K}{2}\sin^4(\theta/2).
\end{eqnarray}

In the so-called dispersive regime, $g\ll |\Delta|$, $\theta\ll 1$ and hence $\chi_{BB}\gg \chi_{AB}\gg \chi_{AA}.$ In the so-called `strong-dispersive regime,' the cross-Kerr coupling is much larger than $\kappa$, the energy dissipation rate of the cavity and $\gamma$, the energy relaxation rate of the qubit $\chi_{AB}\gg \kappa,\gamma$ \cite{Gambetta_2006_PhysRevA.74.042318,Schuster2007a}.

If we now make the two-level qubit approximation for the transmon using the analog of Eqs.~(\ref{eq:bdagbisZ}-\ref{eq:bissigminus}), we obtain the Hamiltonian (up to an irrelevant constant)
\begin{eqnarray}
H&=&H_\mathrm{D}+V,\\
H_\mathrm{D}&=&\omega_\mathrm{R}A^\dagger A + \frac{\omega_\mathrm{Q}}{2}\sigma_z+\frac{\chi}{2}\sigma_z A^\dagger A \label{eq:dispersiveHD},\\
V&=&-\frac{K_{A}}{2} A^\dagger A^\dagger AA\label{eq:cavityKerr},
\end{eqnarray}
where $\chi\equiv \chi_\mathrm{AB}$ is the dispersive shift and $K_A\equiv 2\chi_\mathrm{AA}$ represents the weak negative anharmonicity inherited by the cavity normal mode due to the coupling to the transmons. Here, the renormalized cavity frequency is given by Eq.~(\ref{eq:normmodefreqs}), with $\omega_\mathrm{R}=\omega_+$ if the detuning $\Delta>0$, and $\omega_\mathrm{R}=\omega_-$ if $\Delta<0$.  The renormalized qubit frequency is the remaining eigenvalue, $\omega_\mathrm{Q}=\omega_-$ if the detuning $\Delta>0$, and $\omega_\mathrm{Q}=\omega_+$ if $\Delta<0$. 

Because $K_A\ll\chi$, we make the simplifying assumption that we can neglect the induced cavity self-Kerr term $V$, and work only with the dispersive Hamiltonian $H_\mathrm{D}$.
The dispersive Hamiltonian $V_0$ can be interpreted two different ways.  First, it tells us that the resonant frequency of the cavity takes on different values, $\omega_\mathrm{R}\pm \frac{\chi}{2}$, depending on the state of the qubit.  The frequency of the cavity is thus a proxy that can be used to make a high-fidelity QND dispersive measurement of the state of the qubit \cite{Blais2004,Wallraff_cQED_2004}.

In the so-called `strong-dispersive regime,' \cite{Gambetta_2006_PhysRevA.74.042318,Schuster2007a} where $\chi$ greatly exceeds the damping rate of the cavity and the transmon, the two resonance peaks of the cavity are spectrally well-resolved.  We see shortly that this allows us to make operations on the cavity that are conditioned on the qubit state, thus creating entanglement between the cavity and qubit.  

The second interpretation of the dispersive Hamiltonian is that the qubit frequency shifts by $\chi$ for each photon that is added to the cavity.  We will see shortly that this allows us to make rotations of the qubit conditioned on the precise number of photons in the cavity, giving us a second way to entangle the qubit and cavity.  In particular, it gives us the ability to apply selective number-dependent arbitrary phase (SNAP, Sec.~\ref{sssec:SNAP}) gates \cite{Heeres2015} to each Fock state of the cavity. 

Together, these two resources, qubit-conditioned cavity displacements and cavity-conditioned qubit rotations, grant us full quantum control of the combined system \cite{Krastanov2015,Heeres2015,Heeres2017}.  While some gates require only simple and intuitive microwave control pulses, the execution of more general unitaries at the fastest possible rates requires numerical determination of the control pulses, using for example the GRAPE algorithm from optimal control theory (OCT), which are difficult to interpret.  For example, Heeres Ref.~\cite{Heeres2017} carried out a highly non-Gaussian operation that executed a state transfer from the  $n=0$ Fock state of the cavity to the $n=6$ Fock state, using OCT pulses applied simultaneously to the transmon and to the cavity.  Efficiency gains from bypassing qubit and gate abstractions to directly compile quantum algorithm steps at the control pulse level have been discussed recently in Ref.~\cite{Breaking_Abstractions_Chong_2020}.

\subsubsection{Conditional Cavity Gates}

\subsubsubsection{Conditional Displacements.}
 Qubit-controlled cavity displacement operations have been previously discussed in \cite{qcMAP,Campagne-Ibarcq2020, EickbuschECD, diringer2023conditional}. In this subsection, we aim to sketch at a high level the implementation of the conditional displacement gate $\mathrm{CD}(\alpha)$, an integral gate to the phase-space ISA defined in Box \ref{Box:c-displacement}.

In Ref.~\cite{Campagne-Ibarcq2020, EickbuschECD, diringer2023conditional}, a clever echo sequence is used to make rapid conditional displacements with a strong drive and simultaneously eliminate unwanted Stark shift rotations of the ancilla.  Adding a drive to the cavity centered on frequency $\omega_\mathrm{R}$ and having an envelope $\epsilon(t)$,  the Hamiltonian in Eq.~(\ref{eq:dispersiveHD}) in the frame doubly rotating at the qubit and cavity frequencies becomes
\begin{equation}
H=\frac{\chi}{2}\sigma_z a^\dagger a+i\epsilon(t) a^\dagger -i \epsilon^*(t)  a,
\end{equation}
where for notational simplicity and to match the usage in the main text, we shall henceforth substitute $A\rightarrow a$.
For the case where the complex drive phase is time-independent, we can remove the drive using a time-dependent unitary transformation 
\begin{equation}
    U(t)=e^{-\beta(t)a^\dagger+\beta^*(t)a}
\end{equation}
that shifts the oscillator to a displaced frame
\begin{equation}
    a\rightarrow a+\beta(t),
\end{equation}
where
\begin{equation}
    \beta(t)=\int_0^t d\tau\, \epsilon(\tau).
\end{equation}
The Hamiltonian in the new frame becomes
\begin{eqnarray}
    H&=&\frac{\chi}{2}\sigma_z [a^\dagger+\beta^*(t)][a+\beta(t)]\\
    &=&\frac{\chi}{2}\sigma_z\left( \rule{0pt}{2.4ex} \left [a^\dagger a \right] + \left [\beta^*(t)a+\beta(t)a^\dagger \right]+\left [|\beta(t)|^2 \right ]\right ).\nonumber\\&\quad&
\end{eqnarray}
The middle term is the generator of the desired conditional displacement.
The first and third terms are spurious but can be eliminated by a `double echo' sequence in which we first drive $\beta$ to a large positive value, then flip the ancilla state and drive $\beta$ to a large negative value.  The details are described in the Supplementary Material of Ref.~\cite{EickbuschECD}, where the authors call the sequence an \emph{echoed-conditional-displacement (ECD) gate}.  For the appropriate choice of $\beta(t)$, this gate is equivalent (up to a bit-flip on the ancilla) to the conditional displacement $\mathrm{CD}(\alpha)$ defined in Box \ref{Box:c-displacement} and used throughout this work.

\subsubsubsection{Conditional Oscillator-Oscillator Entangling Gates in the Weak Dispersive Regime}\label{sssec:ECBS}.
In Sec.~\ref{sssec:useful_primitives}, we demonstrate a useful trick for compiling conditional oscillator-oscillator entangling gates using controlled-parity gates (see the controlled beam-splitter gate in Eq.~(\ref{eq:cbs}) as an example). Importantly, controlled parity gates are realized by allowing the dispersive interaction $e^{-i\frac{\chi}{2} t a^\dagger a\sigma_z}$ to act for time $ t=\pi/\chi$. However, in the weak-dispersive regime (where dispersive coupling $\chi$ is extremely small), this strategy is undesirable because it results in a long gate duration. Here, we show an alternative method to compile a conditional oscillator-oscillator entangling gate, namely, a conditional two-mode squeezing, in the weak dispersive regime. To achieve this, we leverage uncontrolled two-mode squeezing operations, ${\rm TMS}(r,\phi)$, as defined in Table~\ref{tab:gates-osc}.

Using the definition of ${\rm TMS}(r,\phi)$, we first note that 
\begin{align}
     {\rm TMS}(\alpha,\pi)~ a~ {\rm TMS}^\dagger(\alpha,\pi)
     = a\cosh{\alpha}+b^\dagger \sinh{\alpha} \,.
\end{align}
This in turn implies
\begin{align}
&{\rm TMS}(\alpha,\pi) e^{-i\chi ta^\dagger a\sigma_z} {\rm TMS}^\dagger(\alpha,\pi) \nonumber \\
&=\mathrm{exp} \left [  \rule{0pt}{2.4ex} -i\chi t \left (  \rule{0pt}{2.4ex} \cosh^2{(\alpha)}a^\dagger a+\sinh^2{(\alpha)}bb^\dagger \right. \right. \nonumber\\
    & ~~~~~~~~~~~~~~~ \left.\left. +\frac{1}{2}\sinh{(2\alpha)}(a^\dagger b^\dagger +ab) \right) \sigma_z  \rule{0pt}{2.4ex} \right]\label{eq:TMS-rot}
\end{align}
The next steps of the method are inspired by the construction of echoed-conditional displacements (ECD) described above. Notice that the first two terms in Eq.~\eqref{eq:TMS-rot} do not change signs with $\alpha$, whereas the last two will. Hence, running the pulse shown below yields an echoed two-mode squeezing, in close analogy with the echoed displacement gate previously discussed,
\begin{align}
    &{\rm TMS}(\alpha,\pi)e^{-i\chi ta^\dagger a\sigma_Z} {\rm TMS}^\dagger(\alpha,\pi)\times\sigma_x\nonumber\\&\times {\rm TMS}(-\alpha,\pi)e^{-i\chi ta^\dagger a\sigma_Z} {\rm TMS}^\dagger(-\alpha,\pi)\nonumber\\
    &~~~~~ = e^{-i\chi t\sinh{2\alpha}(a^\dagger b^\dagger +ab)\sigma_z}
    \label{eq:Echoed-TMS}
\end{align}
The speed of the conditional two-mode squeezing gate, in this case, is decided by $\chi t\sinh{2\alpha}$ instead of $\chi t$ as in the case where this gate is synthesized with controlled parity gates. As $\sinh{2\alpha}$ is an unbounded function, we can in principle increase it to extremely large values by varying $\alpha$. Thus, in the weak dispersive regime, we can achieve fast conditional oscillator-oscillator entangling gates by leveraging unconditional two-mode squeezing with large $\alpha$ as a resource.  Hamiltonian amplification techniques \cite{arenz2020amplification} have been used in ion traps as well, see Ref.~\cite{burd2019quantum,burd_quantum_2021}.

\subsubsection{Cavity-Controlled Qubit Gates}\label{app:cavitycontrolledqubitgates}
\subsubsubsection{SQR Gates.}
A microwave electric field applied to the qubit couples to the electric dipole moment of the qubit and causes transitions between the ground and excited state. If the drive is applied at the qubit frequency it would coherently rotate the qubit around an axis that is determined by the phase of the microwave drive.  Notice from the dispersive Hamiltonian $H_\mathrm{D}$ for the coupled qubit-cavity system in Eq.~(\ref{eq:dispersiveHD}) that the qubit transition frequency suffers a quantized light shift depending on the photon number in the cavity
\begin{equation}
    \omega_Q+\chi \hat n.
\end{equation}
Hence we are able to execute a qubit gate conditioned on the photon number in the cavity. If we apply the qubit drive at frequency $\omega_Q(m)= \omega_Q+m\chi$, it is on resonance and therefore effective at rotating the qubit on the Bloch sphere, if and only if, there are exactly $m$ photons in the cavity (assuming $\chi$ is sufficiently large relative to the drive amplitude).  We now derive this in detail.

Starting from the driven-dispersive Hamiltonian in the co-rotating frame of the oscillator and qubit,
\begin{align}
H = \frac{\chi}{2} a^\dag a \sigma_z + \Omega^*(t)\sigma_- + \Omega(t)\sigma_+,
\end{align}
we perform a unitary frame transformation to eliminate the dynamics of the dispersive term by using the unitary $U_\chi=\exp\left(-i\frac{\chi}{2} t a^\dag a \sigma_z\right)$. The transformed Hamiltonian $\tilde{H} = U_\chi^\dag H U_\chi + i \left(\partial_t U_\chi^\dag \right) U_\chi$ is
\begin{align}
\tilde{H} = \sum_{n=0}^{\infty} \ket{n}\bra{n} \left(\Omega^*(t) \sigma_- e^{-i\chi n t } + \Omega(t) \sigma_+ e^{+i\chi n t } \right).
\end{align}
For selective qubit rotation gates up to Fock state $N$, we drive the qubit resonant with each number-shifted transition.  This drive takes the form $\Omega(t) = \sum_{k=0}^{N} \Omega_k(t) e^{-i k \chi t}$ where $\Omega_k(t)$ are slowly varying (complex-valued) envelopes. With this drive, the Hamiltonian becomes 
\begin{align}
    \tilde{H} &= \sum_{n=0}^{\infty} \ket{n}\bra{n}\nonumber\\
    &\times \sum_{k=0}^{N} \left(\Omega^*_k(t) \sigma_- e^{-i\chi \left(n-k\right) t } + \Omega_k(t) \sigma_+ e^{+i\chi \left(n-k\right) t } \right).
\end{align}
Under the condition of slow pulses (total time $T_\text{gate} > 2\pi/\chi$) $\Omega_k(t)$ will have a small bandwidth compared to $\chi$, and the first order rotating wave approximation (RWA) can be invoked. The implications of this approximation, along with mitigation techniques for faster pulses, are discussed in \cite{sivak2021modelfree}. The RWA results in 
\begin{align}
    \tilde{H}_{\text{RWA}} = \sum_{n=0}^{N} \ket{n}\bra{n} \left(\Omega^*_n(t) \sigma_-  + \Omega_n(t) \sigma_+  \right).
\end{align}
To realize an SQR gate, we pick $\Omega_n(t) = s(t) \frac{\theta_n}{2} e^{-i\varphi_n}$ where $s(t)$ is a real-valued envelope function with $\int_0^{T_\text{gate}}s(\tau) d\tau = 1$ so that
\begin{align}
   \text{SQR}\left(\vec{\theta},\vec{\varphi}\right)= \mathcal{T}\exp{\left(-i\int_0^{T_\text{gate}} \tilde{H}_\text{RWA} dt\right)}.
\end{align}
By transforming the operator back to the initial frame, the total unitary after time $T_\text{gate}$ is
\begin{align}
&U = U_\chi\, \text{SQR}\left(\vec{\theta},\vec{\varphi}\right) \\
&= \text{CR}\left(\chi T_\text{gate}\right) \text{SQR}\left(\vec{\theta},\vec{\varphi}\right).
\end{align} 
As indicated, the full gate in the lab frame includes an additional conditional cavity rotation, which accounts for the dispersive interaction during the gate. This CR can be canceled by either applying a counter $\text{CR}$ gate after the evolution, by constructing a suitable qubit echo sequence similar to the ECD construction \cite{EickbuschECD} or, in cases where the initial and final qubit states are known eigenstates of $\sigma_z$ (such as when implementing the SNAP gate), can be handled simply by tracking the cavity frame change in software.

\subsubsubsection{SNAP Gates.}
\label{sssec:SNAPgatesApp}
As described in Box \ref{Box:SNAP}, two SQR gates can be combined to produce the photon-number selective phase (SNAP) gate.
One interesting choice of phases is
\begin{equation}
    \varphi_m=m\varphi 
\end{equation}
which simply rotates the cavity state in phase space by angle $-\varphi$.  Since the Berry phase is equal and opposite if the qubit starts and ends in the excited state, one can achieve the qubit-dependent cavity rotation gate defined in Box \ref{Box:CRotation}
\begin{equation}
    \mathrm{CR}(2\varphi)=e^{-i\varphi\hat n\sigma_z},
\end{equation}
which can be used either to enhance or compensate for the naturally occurring time evolution under the dispersive Hamiltonian in Eq.~(\ref{eq:dispersiveHD}).

For quantum simulation and control purposes, another interesting choice of phases is
\begin{equation}
    \theta_m=\lambda m^2,
\end{equation}
which yields
\begin{equation}
    U_\mathrm{Kerr}(\lambda)=e^{i\lambda\hat n^2\sigma_z},
\end{equation}
which can be used to dynamically cancel the undesired residual Kerr interaction in the cavity found in Eq.~(\ref{eq:cavityKerr}).  This gate can also be used to enhance the cavity self-Kerr to provide a programmable interaction for simulations of the boson Hubbard model as discussed in Sec.~\ref{BoseHubbard}.

\subsection{Trapped Ion Quantum Processors}
\label{app:trapped-ions}

In ion traps \cite{bruzewicz2019trapped}, one can produce the phase-space instructions set using the interaction between ions and lasers or microwaves.   As discussed in Fig.~\ref{fig:hardware_layout}, the mechanical oscillation modes of the ions form the bosonic modes. Unlike in the superconducting case where the microwave resonators typically have GHz frequencies, the mechanical modes in ion traps typically have frequencies in the MHz range.  Beyond the various gates we discuss below, multi-chromatic laser drives that simultaneously couple many qubits to many mechanical modes have been developed \cite{PhysRevA.101.032330}.

\subsubsection{Qubit Rotations}

The qubits in ion-trap systems can be encoded in optical transitions of the ions, but can also be encoded in hyperfine levels resulting from the coupling between electron and nuclear spins in the ground state manifold of the ions. The hyperfine energy level spacing is usually on the order of MHz and the transitions can be driven using microwaves to achieve single-qubit rotations much as in NMR experiments \cite{gershenfeld1997bulk}. Given their long wavelength, microwaves can be difficult to focus to address individual ions, but by engineering near-field microwave gradients, individual addressability can be achieved \cite{warring2013individual}. An alternative way to drive the qubit hyperfine transition is to use two detuned laser beams which have a frequency difference that matches the hyperfine splitting to drive the qubit transitions via a stimulated Raman process (STIRAP). Because the stimulated Raman transition is 2nd-order (i.e., requires one-photon absorption and one emission), it generally requires relatively strong laser intensities.

\subsubsection{Sideband Gates}
\label{subsec:sidebandgates}
A sideband gate $\mathrm{JC}(\theta, \varphi)$ (see Table~\ref{tab:gates-qubit-osc} ) that flips one of the qubits and emits or absorbs one mechanical quantum (phonon) is (for the case of a red sideband)
\begin{align}
   \text{JC}(\theta,\varphi) = \exp\left[-i\theta\left( e^{i\varphi} \sigma_- a^\dag  + e^{-i\varphi}\sigma_+ a\right)\right],
    \label{sideband-gate}
\end{align}
which is universal when combined with an arbitrary single-qubit rotation $R_\varphi(\theta)$ in the XY plane (see Table \ref{tab:gates-qubit}) \cite{QuditsfromOscillatorsPhysRevA.104.032605,qudit2013mischuck,sutherland2021universal}. This gives the ISA listed in Table~\ref{tab:ISA_trapped-ion}.

Similarly, a blue sideband gate $\text{AJC}(\theta,\varphi)$ (Table \ref{tab:gates-qubit-osc}) could be used, and when combined with a red sideband gate produces the controlled displacement gate 
$\mathrm{CD}(\alpha)$ of the phase-space  ISA, as discussed below.
We note that by use of a pulse shaping technique, a uniform sideband gate has been realized wherein the Rabi frequency between nearest neighbor Fock levels does not scale as the Fock number but is a constant \cite{um2016phonon,Kihwan_Kim_an_experimental_2015}. 

In the ion-trap setting, the finite momentum kick from the absorption  or emission of an optical photon allows for the possibility of higher-order sideband transitions in which a spin flip is coupled to the creation or destruction of two or more mechanical oscillation quanta (phonons).  While such higher sideband Hamiltonians are weaker than the lowest sideband coupling by powers of the Lamb-Dicke parameter, they present interesting possibilities for additional control schemes and instruction sets.
However, we leave these for future work and not discuss them further here.

We are not aware of a SNAP gate realization on trapped ion platforms because of the lack of dispersive coupling between the qubits and the mechanical modes of the system.  Such a coupling can however be synthesized from sideband transitions \cite{sutherland2021universal}.  Beam-splitters can also be synthesized outside this sideband gate set without using any lasers by parametrically modulating the electric fields used to create the trapping potential seen by the ions \cite{gorman2014twomode,sutherland2021universal}. Alternatively, the second sidebands include beam-splitter interactions, which are however weaker by an additional power of the Lamb-Dicke factor. Moreover, arbitrary all-to-all connected beam-splitter Hamiltonians can be implemented using multi-tone driving~\cite{katz2023programmable}.

\begin{table}[htb]
    \centering
    \begin{tabular}{c|c}
    \hline \hline
        \multicolumn{2}{c}{Trapped Ion ISA} \\
        \hline
        Red sideband gate (Table \ref{tab:gates-qubit-osc}) & $\text{JC}(\theta,\varphi)$ \\
        Qubit rotation (Table \ref{tab:gates-qubit}) & $R_\varphi(\theta)$\\

        \hline \hline
         \end{tabular}
    \caption{Minimal instruction set for a trapped ion platform. This is the sideband ISA defined in Table \ref{tab:ISA_overview} with the important difference that any qubit can have JC coupling with any motional mode in which it participates. Application of the sideband gates to different qubits allows them to be entangled through the exchange of quanta of one of the bosonic modes. A beam splitter gate (see Box~\ref{Box:beam-splitter}) to couple different mechanical modes, $a$ and $b$, can be synthesized from this gate set or can be created using RF control of the trap parameters as noted in the caption of Fig.~\ref{fig:hardware_layout}.  Blue sideband gates (AJC, Table \ref{tab:gates-qubit-osc}) are equally easy to perform and are a convenient natively available gate that can be added to this minimal set.
    }
    \label{tab:ISA_trapped-ion}
\end{table}

\subsubsection{Conditional Displacements}

Recall that the red side-band produces coupling Hamiltonian in the form of $\sigma^+ a + \sigma^- a^\dagger$, whereas a blue side-band produces coupling in the form of $\sigma^- a + \sigma^+ a^\dagger$. If a bichromatic laser drive \cite{haljan2005spin} is applied so that both sideband drives are switched on simultaneously and with equal strengths, we can realize the following Hamiltonian as a sum of the two:
\begin{align}
     (\sigma^+ a + \sigma^- a^\dagger) + (\sigma^- a + \sigma^+ a^\dagger) ~~~~~~~~~~~~~~~~~~~~~~ \nonumber \\
    ~~~ = \sigma^+ (a + a^\dag) + \sigma^- (a + a^\dagger) ~~ \propto ~~ \sigma_x  \hat{x} \,.
\end{align}
This is nothing but a conditional displacement Hamiltonian, conditioned on the eigenstates of the qubit Pauli-X operator. We may further recast this into the familiar controlled displacement, $\mathrm{CD}(\alpha)$ (Table \ref{tab:gates-qubit-osc}), by conjugating the qubit using a Hadamard gate since $\sigma_z = H \sigma_x H$.

Unconditional displacements can be achieved by using one of the qubits as an ancilla that is not involved in the circuit. Initializing this spare qubit in $\ket{+}$ and driving a conditional displacement realizes an unconditional displacement.

\subsection{Neutral Atom Quantum Processors}
\label{sct:neutralatoms}
The motion of single atoms in optical tweezers~\cite{Schlosser2001, sortais2007,Kaufman2012} is also approximately that of a quantum harmonic oscillator. In contrast to ions, each atom is coupled to a maximum of three modes, where in practice one of those modes is coupled to the atomic qubit, as illustrated in Fig.~\ref{fig:hardware_layout}. This means that there is no mode-crowding effect restricting the scaling to larger systems, but this comes at the cost of reduced connectivity among the modes. Moreover, due to the much weaker dipole force confining the atoms compared to the electrostatic force confining ions, the motional mode frequency $\omega_r$ is smaller, usually in the tens of kHz. Qubits may be encoded in hyperfine levels (in alkali atoms such as rubidium) or optical clock qubits (in strontium). By qubit-state-selective of the atoms into Rydberg states, the Rydberg blockade effect can be used to create two-qubit entangling gates \cite{Evered2023}. Recently, combined qubit-boson operations have been demonstrated in a tweezer array~\cite{scholl2023}, showing coherence times of the modes over $100$ms and non-classical motion of the atoms due to Fock-state preparation has been shown~\cite{brown2022}.  There have been recent theory proposals for realization of bosonic quantum error correction codes in mechanical vibration states of neutral atoms \cite{10.1063/5.0197119,bohnmann2024bosonicquantumerrorcorrection}.

Contrary to ions, the tweezer potential has a sizeable nonlinearity. The maximum boson number occupation is given by the trap depth $V_0$, which in current implementations is on the order of 1-10 MHz such that the maximum accessible Fock state is $100$-$1000$. Another effect of the finite trap depth is a finite Kerr nonlinearity of the potential, which is $0.1\%$-$1\%$ of the mode frequency (coincidentally similar to superconducting transmon qubits). The nonlinearity is in principle tunable by the power of the trapping laser. However, despite this sizeable nonlinearity, qubit-boson gates are not substantially altered since the qubit-boson coupling is so large.

\begin{table}[htb]
    \centering
    \begin{tabular}{c|c}
    \hline \hline
        \multicolumn{2}{c}{Neutral atom ISA} \\
        \hline
        Red sideband gate (Table \ref{tab:gates-qubit-osc}) & $\text{JC}(\theta,\varphi)$ \\
        Qubit rotation (Table \ref{tab:gates-qubit}) & $R_\varphi(\theta)$\\
        Qubit-qubit gate (Table \ref{tab:gates-qubit}) &  CZ \\
        \hline \hline
         \end{tabular}
    \caption{Minimal instruction set for neutral atom platforms. Sideband gates couple a single qubit to a single mode.  Beam splitter gates (see Box~\ref{Box:beam-splitter}) to a couple of different mechanical modes, $a$ and $b$, can be synthesized via the qubit-qubit Rydberg CZ gate.  
    }
    \label{tab:ISA_neutral}
\end{table}

\subsubsection{Single Qubit Gates}

Single-qubit gates are implemented just as in qubit-only applications of neutral atoms by either Raman transitions~\cite{Graham2022} or microwaves combined with AC Stark shifts~\cite{Wang2016}. 

\subsubsection{Multi Qubit Gates}

Multi-qubit gates are implemented by using a Rydberg level of the atoms. High fidelity two-qubit CZ and even n-qubit multi-controlled Z gates have been demonstrated \cite{Evered2023} by using the blockade mechanism, which forbids two atoms (within the blockade radius of each other) to be transferred to the Rydberg state simultaneously \cite{Jaksch2000}. When two atoms are simultaneously in the Rydberg state, they produce forces that modify the bosonic state, as discussed in the next subsection. However, during the Rydberg blockade entangling gates, the atoms do not simultaneously occupy Rydberg states, thus preventing any impact on the motional states.

\subsubsection{Single Mode Gates}   
Phase space rotations are naturally implemented by the free evolution under the tweezer potential. 
   Moreover, the nonlinearity enables universal single-mode control by modulating the tweezer potential~\cite{Grochowski2023}.
   
\subsubsection{Multi Mode Gates}

\textbf{Using two-qubit gates.} Inter-mode gates such as beam-splitters can be compiled using sideband gates and CZ gates. 
We employ the BCH formula
\begin{equation}
	e^{i\hat A\theta}e^{i\hat B\theta}e^{-i\hat A\theta}e^{-i\hat B\theta}=e^{-[\hat A,\hat B]\theta^2+O(\theta^3)}.
\end{equation}
First, we create a doubly-controlled displacement by choosing $\hat A=\hat Z_1\hat Z_2$ and $\hat B=\hat X_1(\hat a+\hat a^\dagger)$. This way we get $-[\hat A, \hat B]\theta^2=2i\hat Y_1\hat Z_2 (\hat a+\hat a^\dagger)\theta^2\equiv i\hat C\tilde \theta$ with $\hat C=\hat Y_1\hat Z_2 (\hat a+\hat a^\dagger)$ and $\tilde \theta=\sqrt{2}\theta$. Then, we can repeat the BCH trick, conjugating $\hat C$ with $\hat D=\hat X_2(\hat a+\hat a^\dagger)$. From this concatenation, we obtain
\begin{equation}
	e^{-[\hat C,D]\tilde \theta^2}= e^{i\hat Y_1\hat Y_2(\hat a +\hat a^\dagger)(\hat b +\hat b^\dagger)4\tilde \theta^4}.
\end{equation}
Hence, we have realized a doubly qubit-controlled beam-splitter/two-mode-squeezing gate. The two-mode-squeezing or beam-splitter terms can be singled out by applying the same sequence with different phases and employing the Trotter formula.

\textbf{Using dipolar forces in Rydberg state.} Two neutral atoms excited to Rydberg states exert strong dipolar forces on each other. By ``dressing'' the ground state with the Rydberg state and modulating the tweezer potentials, beam-splitters as well as two-mode squeezing can be realized~\cite{Buchmann2017}.

\textbf{Using tunneling.} Another approach is to use tunneling between tweezers, generalizing the approach discussed in the context of qubits encoded in the lowest two levels of each tweezer~\cite{Eckert2002}.

\subsubsection{Sideband Gates}

Qubit-boson coupling is implemented using sideband gates as in ions, which in the Lamb-Dicke regime $\eta \sqrt{n}\ll 1$ is given by
\begin{equation}
	\hat H = \frac{1}{2} \Omega \eta \left(\hat a^\dagger \hat \sigma^- + \hat a \hat \sigma^+ \right)
\end{equation} 
with $\Omega$ the Rabi frequency, and the Lamb-Dicke factor $\eta$ is given in terms of the momentum transfer $\delta k$ of the Raman beams and the atomic mass $m$ by $\eta=\delta k/\sqrt{2m\omega_r}$. Due to the essentially unlimited Rabi frequency in this context, the main restriction to the coupling $\Omega \eta$ of these gates is given by the boson mode frequency - to resolve the sideband transitions it needs to fulfill $\Omega \eta < \omega_r$.

\subsubsection{Alternative Architectures}
\label{app:neutral_alternatives}
\textbf{Motional states in optical lattices.} Alternatively, to the tweezer platform, the motional modes in optical lattices can be similarly used. Lattice shaking can then be used to excite oscillators, which may also be done in a qubit-dependent way. Superlattices can be used to engineer tunneling interactions for beam-splitters.

\textbf{Bosonic atoms.} A completely different approach is to use bosonic atoms in optical lattices or tweezers, similar to the recent proposal for fermion-qubit digital computation~\cite{GonzlezCuadra2023}. Recently, boson sampling has been realized this way~\cite{young2023atomic}. This approach is most naturally suited for problems conserving the total number of bosons. However, displacements may be engineered by coupling the tweezers to a Bose-Einstein-Condensate.

\section{Bloch-Messiah Decomposition}\label{app:bloch-messiah}

In the last section we learned that a unitary Gaussian transformation on the infinite Hilbert space of $N$ bosonic modes corresponds to a transformation of the $2N$ mode positions and coordinates (or equivalently the raising and lowering mode operators), which must be symplectic to preserve the commutation relations.
The Bloch-Messiah decomposition \cite{Braunstein2005squeezing,Weedbrook2012,PhysRevA.94.062109} is essentially a singular-value decomposition for these $2N\times 2N$ symplectic matrices.  Any $2N\times 2N$ real symplectic matrix $\mathbf{M}$ can be decomposed into the form
\begin{align}
    \mathbf{M}&=\mathbf{O_1}\mathbf{Z}\mathbf{O_2},
\end{align}
where $\mathbf{O_1}$ and $\mathbf{O_2}$ are $2N\times 2N$ orthogonal symplectic matrices, and $\mathbf{Z}$ is a $2N\times 2N$ diagonal squeezing matrix containing the singular values in the form
\begin{align}
    \mathbf{Z}&=\mathrm{Diag}\left(e^{r_1},e^{-r_1},e^{r_2},e^{-r_2},\ldots,e^{r_N},e^{-r_N}\right).
\end{align}
Physically this decomposition can be realized by having a system of beam splitters applied before ($\mathbf{O_2}$) and after ($\mathbf{O_1}$) a middle step $\mathbf{Z}$ involving only single-mode squeezing (by varying amounts on each mode). We make use of this decomposition in the following derivations.

There are two reasons that this decomposition is important for experiments. The first is that it allows synthesis of two-mode squeezing using single-mode squeezing, which is often substantially easier to realize. The second is that very high-fidelity beam-splitter gates between neighboring qubits now exist and can be used to synthesize the $\mathbf{O}_1$ and $\mathbf{O}_2$ operations which in general involve beam splitters between distant resonators.  This in turn requires SWAP networks (again based on beam splitters, see Sec.~\ref{ssssec:beamsplitter_swap_networks}) to be able to achieve geographically non-local beam splitter gates.

We turn now to some specific examples.

\subsection{SUM Gate}
The two-mode SUM gate given in Eq.~(\ref{eq:SUMGate}) is a unitary operation that can be represented by the corresponding $4\times 4$ symplectic matrix (in the mode operator basis)
\begin{align}
    \mathbf{SUM}(\lambda) &=  \left(\begin{array}{cccc} 1&0&-\lambda/2&+\lambda/2\\
    0&1&+\lambda/2&-\lambda/2\\
    +\lambda/2&+\lambda/2&1&0\\
    +\lambda/2&+\lambda/2&0&1
    \end{array} \right),\label{eq:SUMsymplectic}
\end{align}
acting on the vector of mode operators
\begin{align}
    \left(\begin{array}{c}a^{\phantom{\dag}}\\a^\dag\\b^{\phantom{\dag}}\\b^\dag  \end{array} \right).
\end{align}
The transformation obeys Eq.~(\ref{eq:symplectic})

\begin{align}
    \mathbf{SUM}^\mathrm{T}(\lambda)&\left(\begin{array}{cccc}
 0& +1& 0& 0\\
 -1& 0& 0& 0\\
 0& 0& 0& +1\\
 0& 0& -1& 0
\end{array}\right)\mathbf{SUM}(\lambda)  \nonumber\\
=&
\left(\begin{array}{cccc}
 0& +1& 0& 0\\
 -1& 0& 0& 0\\
 0& 0& 0& +1\\
 0& 0& -1& 0
\end{array}\right)
\end{align}

and is therefore symplectic, proving that it preserves the commutation relations among the mode operators as required.

The Bloch-Messiah decomposition given in Eq.~(\ref{eq:SUMBlochMessiah}) can be written in this form using the symplectic matrix representation of the two-mode beam-splitter, whose unitary is given in Eq.~(\ref{eq:BSgateunitary}) as:
\begin{align}
    &\mathbf{BS}(\theta,\varphi)=\nonumber\\
    &\left(\begin{array}{cccc}
    \cos\frac{\theta}{2}&0& -ie^{i\varphi}\sin\frac{\theta}{2}&0\\
     0&\cos\frac{\theta}{2}&0& ie^{-i\varphi}\sin\frac{\theta}{2}\\
     -ie^{-i\varphi}\sin\frac{\theta}{2}&0&\cos\frac{\theta}{2}&0\\
     0&ie^{+i\varphi}\sin\frac{\theta}{2}&0&\cos\frac{\theta}{2}
    \end{array}\right).\label{eq:BSsymplec}
\end{align}
Similarly, the symplectic matrix representation of the pair of single-mode squeezers (with the upper arm of the interferometer being antisqueezed by $s(-r)$ and the lower arm squeezed by $s(r)$) is
\begin{align}
    \mathbf{s}(r)\boldsymbol{\otimes} \mathbf{s}(-r)&=
    \left (\begin {array} {cccc}   
    \cosh r&+\sinh r& 0& 0\\
    +\sinh r& \cosh r &0 &0\\
    0&0&\cosh r&-\sinh r\\
    0&0&-\sinh r&\cosh r
    \end {array} \right).
\end{align}
Plugging these into Eq.~(\ref{eq:SUMBlochMessiah}) yields the following expression

\begin{widetext}
\begin{align}
   \mathbf{SUM}(\lambda)=\left(\begin{array}{cccc}
   -\cosh(r)\sin(2\theta)&0&-\cosh(r)\cos(2\theta)&\sinh(r)\\
   0& -\cosh(r)\sin(2\theta)&\sinh(r)& -\cosh(r)\cos(2\theta)\\
   \cosh(r)\cos(2\theta)&\sinh(r)&-\cosh(r)\sin(2\theta)&0\\
   \sinh(r)&\cosh(r)\cos(2\theta)&0&-\cosh(r)\sin(2\theta)
   \end{array}\right).
\end{align}
\end{widetext}
Choosing the parameter values for $\theta$ and $r$ as in Eqs.~(\ref{eq:SUMBM2}-\ref{eq:SUMBM4}) yields the desired transformation given in Eq.~(\ref{eq:SUMsymplectic}).

\subsection{Two-Mode Squeezing}
The two-mode squeezing gate in Box~\ref{Box:2-mode-squeezing} has the $4\times 4$ symplectic representation given in Eq.~(\ref{eq:TMSsymplectic})
\begin{align}
   &\mathbf{TMS}(r,\varphi)=\nonumber\\
&\quad\left(\begin{array}{cccc}
 \cosh (r) & 0 & 0 & e^{i \varphi } \sinh (r) \\
 0 & \cosh (r) & e^{-i \varphi } \sinh (r) & 0 \\
 0 & e^{i \varphi } \sinh (r) & \cosh (r) & 0 \\
 e^{-i \varphi } \sinh (r) & 0 & 0 & \cosh (r) \\
\end{array}
\right).\label{eq:TMS4x4}
\end{align}

It can be synthesized in a similar manner using the same, rather than opposite, single-mode squeezing operations in each arm of the interferometer, as illustrated in Fig.~\ref{fig:TMS} and Eq.~(\ref{eq:BMdecompTMS}).  These squeezing operations are represented by the symplectic matrix
\begin{align}
    \mathbf{s}(r)\mathbf{\otimes} \mathbf{s}(r)&=\nonumber\\
    &\quad\left (\begin {array} {cccc}   
    \cosh r&-\sinh r& 0& 0\\
    -\sinh r& \cosh r &0 &0\\
    0&0&\cosh r&-\sinh r\\
    0&0&-\sinh r&\cosh r
    \end {array} \right).
\end{align}
Plugging this expression and Eq.~(\ref{eq:BSsymplec}) into Eq.~(\ref{eq:BMdecompTMS}) correctly yields

\begin{align}
  &\mathbf{TMS}(r,\pi/2) \nonumber \\
  &= \left(
\begin{array}{cccc}
 \cosh (r) & 0 & 0 & i \sinh (r) \\
 0 & \cosh (r) & -i \sinh (r) & 0 \\
 0 & i \sinh (r) & \cosh (r) & 0 \\
 -i \sinh (r) & 0 & 0 & \cosh (r) \\
\end{array}
\right)
\end{align}
in agreement with Eq.~(\ref{eq:TMS4x4}).

 The corresponding symplectic transformation $\mathrm{TMS}(r,0)$ of the quadrature coordinates is
 \begin{align}
     (x^\prime_a+x^\prime_b)&= e^{+r}(x_a+x_b)\\
     (p^\prime_a+p^\prime_b)&=e^{-r}(p_a+p_b)\\
          (x^\prime_a-x^\prime_b)&= e^{-r}(x_a-x_b)\\
     (p^\prime_a-p^\prime_b)&=e^{+r}(p_a-p_b),
 \end{align}
 showing that (for large $r$ and the phase choice $\varphi=0$), the position quadratures of output modes become strongly correlated while the momentum quadratures become strongly anti-correlated.

\section{Oscillator Noise Models}
\label{app:Noise}
We discuss a few common noise models for bosonic systems in different experimental platforms in this section. 
\subsection{Kraus Operator Representation of an Error Channel}

A quantum channel is the most general operation that can be applied to a density matrix that creates another physically allowed density matrix.  Because physically allowed density matrices must be positive semi-definite and have unit trace, this mapping on the space of density matrices must be completely positive and trace preserving (CPTP). The Kraus representation of a quantum channel has the form
\begin{equation}
  \hat\rho(t+\tau)=\sum_{\ell=0}^{D_K-1}\,\hat K^{\phantom\dagger}_\ell\hat\rho(t)\hat K^\dagger_\ell,  
\end{equation}
where the different Kraus  operators $\hat K_\ell$ can be unitary or non-unitary and describe the action of the time evolution of the system under its own Hamiltonian plus the couplings to the dissipative bath.  The number of such operators $D_K$ needed to describe the quantum channel is called the Kraus dimension (or rank) of the map.  For a single qubit $D_K\le 4$, while for a bosonic mode, the number of operators is formally infinite.

The form of the Kraus map guarantees that it is completely positive. To guarantee that it is trace-preserving, the Kraus operators must obey the constraint
\begin{equation}
    \sum_{\ell=0}^{D_K-1}\hat K^\dagger_\ell\hat K^{\phantom\dagger}_\ell=\hat I,
    \label{eq:Kraustoidentity}
\end{equation}
where $\hat I$ is the identity operator.

\subsection{Amplitude Damping}\label{app:ampdamp}

In Sec.~\ref{sec:dissipation} we briefly introduced the amplitude damping channel for bosonic modes by using the master equation formalism. Here, we briefly discuss its Kraus representation. Following Ref.~\cite{BinomialCodes} we can choose to label the Kraus operators in by the total number of quanta (photons or phonons) $\ell$ lost during time $\tau$
\begin{equation}
    \hat K_\ell =\sqrt{\frac{(1-e^{-\kappa \tau})^\ell}{\ell !}} e^{-\frac{\kappa}{2}\tau \hat n}a^\ell;\,\,\ell=0,1,2,3,\ldots
\end{equation}
where $\hat n=a^\dagger a$ is the photon number operator and $\kappa$ is the energy relaxation (photon loss) rate from Eq.~(\ref{eq:SHOdampingrate}). 
We see that the $a^\ell$ term in $\hat K_\ell$ destroys $\ell$ photons, but that even if by chance no photons are lost, the $\ell=0$ Kraus operator $\hat K_0$ enacts a `no-jump backaction,' causing the probability amplitude for Fock states with large photon numbers to be reduced.  This is essentially a Bayesian probability update -- given the additional information that no photons decayed into the environment, we must revise downward our prior estimate of the initial photon number.
 
  For small values of $\kappa\tau$, the probability of losing $\ell$ photons in a short time interval scales as $(\kappa\tau)^\ell$ and becomes small for large $\ell$.  Hence we can safely truncate the Kraus dimension to some value $\ell_\mathrm{max}$ if needed.  However, the truncated series of Kraus operators will not be trace-preserving.  We can fix this by modifying the `no-jump' operator $\hat K_0$ to be
  \begin{equation}
       K_0^\prime=\sqrt{\hat I - \sum_{\ell=1}^{\ell_\mathrm{max}}\hat K^\dagger_\ell\hat K^{\phantom\dagger}_\ell}.
  \end{equation}
  Because the operator inside the square root is diagonal (and each diagonal element is non-negative) in the Fock basis, the square root of the matrix can be found by taking the element-wise square root.  
  
\subsection{Dephasing}

The action of the bosonic dephasing channel can be represented as a sum over continuously many Kraus operators, each representing a phase-space rotation by some random angle $\theta$ via
\begin{equation}
    \mathcal{E}(\rho)=\int_{-\pi}^{\pi} d\theta\ p(\theta) e^{i\theta a^\dagger a}\rho e^{-i\theta a^\dagger a} \,.
\end{equation}
The channel randomizes the phase according to the probability distribution $p(\theta)$. When $\rho$ is expressed in the photon number basis
\begin{equation}
   \mathcal{E}(\rho)=\sum_{m,n}\bra{m}\rho\ket{n}\int_{-\pi}^{\pi} d\theta\  p(\theta) e^{i\theta(n-m)}\ket{n}\bra{m} \,, 
\end{equation}
we see that the channel preserves the diagonal elements but reduces the magnitude of the off-diagonal elements, indicating decoherence. 

The exact solution for the quantum capacity of the bosonic dephasing channel was recently given in~\cite{lami2023exact}. For microwave oscillators realized as three-dimensional superconducting cavities, intrinsic dephasing rates (i.e., rates not associated with state changes in the auxiliary qubit coupled to the cavity) are known to be substantially smaller than the energy relaxation rate $\kappa$ (unlike superconducting qubits where the rates are often similar) and have been measured to be below $\kappa_\phi \lesssim 1/(\SI{10}{ms})$ \cite{rosenblum_fault-tolerant_2018,Reagor_memory_2016}, and more recently below $\kappa_\phi \lesssim 1/(\SI{0.5}{s})$ \cite{Rosenblum_TensofMilliseconds_PRXQuantum.4.030336}. 

In ion traps, the mechanical mode frequencies are sensitive to electric field noise which can lead to significant dephasing. Similarly, the tweezer beams trapping neutral atoms may have amplitude or phase fluctuations, which leads to mode dephasing. 

\subsection{Heating}
Microwave resonators are generally in good thermodynamic equilibrium with the cryogenic refrigeration system, meaning that they are cold enough that the resonator is rarely thermally excited out of the vacuum state.  This is not the case for ion traps as we discuss next. Further below, we also discuss heating in neutral atom tweezer arrays.

Instead of operating at a typical frequency of GHz as in cQED hardware, the AC electric field that traps ions in ion traps produces mechanical resonance frequencies in the MHz regime, and they are not generally operated at cryogenic temperatures. As a result, the dominant quantum noise in the trapped ion platform is the heating of ions' motional vibrational modes \cite{brownnutt2015ion,turchette2000heating,bruzewicz2019trapped}, which can be described by the time-derivative of the frequency-dependent average Fock number, i.e., the population of a given mode, $\dot{\bar{n}} $. The microscopic origin of the heating mechanism remains an open question and has been historically under debate, such that it is often dubbed `anomalous heating.'

It has been demonstrated that the heating rate depends on the trap frequency and geometry as well as temperature. Because the ions mostly interact with the AC trapping electric field, it is clear that the fluctuations in the electric field can significantly affect the heating rate. From the fluctuation-dissipation theorem and the fact that the ions couple linearly to the fluctuating electric field, the heating rate $\dot{\bar{n}} $ can be quantified by \cite{turchette2000heating}
\begin{align}
    \dot{\bar{n}}(\omega, d, T) = \frac{q^2}{2m_I \hbar \omega} S_E(\omega, d, T),
    \label{eq:heating-rate-def}
\end{align}
where $S_E$ and $\omega$ ($\sim$ MHz) are the noise-spectra density and frequency of the trapping electric field, and $m_I$ is the ion's mass. Additionally, $S_E$ depends on the distance $d$ between the ion and the trap surface ($\sim$10-100 $\mu m$) as well as the temperature of the operation $T$, which is found experimentally to obey the following empirical expression \cite{brownnutt2015ion}
\begin{align}
    S_E (\omega, d, T) \propto \omega^{-\alpha} d^{-\beta} f(T),
\end{align}
where most experiments suggest $\alpha \sim 1$ and $\beta \sim 4$, and $f(T)$ is a function of temperature that typically exhibits a thermal activation behavior. The distance dependence $d^4$ excludes the possibility of Johnson noise which should have $d^2$ dependence.
The reason that $S_E$ has such dependence on $\omega, d, T$ is still under debate, 

but there have been many theoretical studies ranging from phenomenological models \cite{noel2019electric,dubessy2009electric,low2011finite,Kumph2016electric,teller2021heating} to \emph{ab initio} electronic structure and molecular dynamics simulation \cite{safavi2011microscopic,foulon2022omega,Ray2019van,kim2017electric} to reveal more microscopic details of the heating mechanism. One possibility is that the molecular absorbates on the trap surface serve as a thin layer of lossy dielectric medium that induces dissipation, where the temperature dependence of the motion of the absorbed molecules can contribute to the temperature scaling. Typical values for the heating rate $\dot{\bar{n}}$ in Eq.~\eqref{eq:heating-rate-def} ranges from $\sim 10^{1}$ phonons/sec at 4 K to $10^{3}$ phonons/sec at room temperature, for trapping frequency $~ 1$ MHz and a trapping distance of $~ 50~ \mu m$ \cite{bruzewicz2015measurement}. Also note that because the electric field fluctuations have a long wavelength, they mostly couple to the center of mass mode. Modes other than the center of mass mode have generally much lower heating rates~\cite{brownnutt2015ion}.

In neutral atom arrays, heating is mainly caused by atom movement~\cite{Bluvstein2022}, which can be reduced by making the atom movement slower. Moreover, in the future, optimal control movement schemes may be developed to optimize the transport of oscillator states~\cite{Pagano2024}. Finally, atom movement may be avoided altogether by using a fixed architecture and using SWAP gates to entangle distant tweezers.

\section{Universality of ISAs and Cross Compilation}
\label{app:cross-compilation}

Here we demonstrate explicit cross-compilation techniques between various ISAs, including the phase-space and Fock-space ISAs (App.~\ref{sec:cross-compilation-SQR-phase-space}), the sideband and phase-space ISAs (App.~\ref{sec:cross-compilation-sideband-phase-space}), as well as the sideband and Fock-space ISAs (App.~\ref{sec:cross-compilation-sideband-SQR}). These cross-compilation schemes also serve as a universality proof for all three ISAs on single qubit-oscillator systems, since at least one of them is universal in previous works \cite{QuditsfromOscillatorsPhysRevA.104.032605,Krastanov2015}.

Building upon the single qubit-oscillator universality, we provide a recursive proof of universality for multi-qubit-oscillator systems (App.~\ref{sec:hybrid-multi-universality}) by adding the analytical compilation of multi-qubit entangling gates as described in Sec.~\ref{ssec:exact-analytical-qubit-gates}.
The cross compilation schemes and proofs presented in this section are far from optimal, but we hope the materials presented here spurs more efficient ways for cross-compilation in future works.

\subsection{Compilation between the Fock-Space ISA and the Phase-Space ISA}
\label{sec:cross-compilation-SQR-phase-space}

In this section, we explore cross-compilation between the two most important ISAs in the current paper, the phase-space and the Fock-space ISAs. Sec.~\ref{sssec:fockISA-from-phaseISA-intuition} provides physical intuition on compiling the SNAP gate from CD gates. App.~\ref{SNAPECDCompilation} leverages Hamiltonian simulation techniques to compile the SNAP gate in the Fock-space ISA using the phase-space ISA. Finally, App.~\ref{sssec:phaseISA-from-fockISA} presents the reverse direction of compiling a CD gate using the Fock space ISA.

\subsubsection{Fock-Space ISA from Phase-Space ISA: Intuition}
\label{sssec:fockISA-from-phaseISA-intuition}

SNAP gates (described in Box.~\ref{Box:SNAP}) are cavity-controlled qubit rotations which apply a separate Berry phase to each Fock state in the cavity.  These gates plus unconditional cavity displacements constitute a universal Fock-space gate set which is dual to the phase-space ISA in the sense that in the phase-space ISA, it is the cavity operations that are controlled by the qubit and the qubit rotations are unconditional. As described in Sec.~\ref{sssec:SNAP}), cavity-controlled qubit operations are most easily engineered in physical architectures with large dispersive coupling between the ancilla qubit and the cavity.  Conversely the advantage of the phase-space ISA is that its gates can be rapidly executed (using large displacements) even when the dispersive coupling is relatively small.  An interesting computational task is to attempt to compile each of these ISAs in terms of the other and analyze the complexity of this task.

Physically we know how to implement SNAP gates using the strong-dispersive Hamiltonian (see App.~\ref{app:cavitycontrolledqubitgates})
\begin{equation}
    H=\chi a^\dagger a\sigma_z.\label{eq:chidispersive}
\end{equation}
The phase-space ISA is based on this very Hamiltonian but operates experimentally with a small value of the dispersive coupling $\chi$.  However, we could use the phase-space ISA with Trotter-Suzuki to synthesize time evolution under this Hamiltonian with a larger effective value of $\chi$. Then we can use controlled displacements with this to literally do a quantum simulation of the SNAP gate dynamics.  Although slow, this strategy is, principle, sufficient to compile a SNAP gate as we outline below.

\subsubsection{SNAP Gate Compilation by Quantum Hamiltonian Simulation}
\label{SNAPECDCompilation}

For this compilation, we need to use the phase-space ISA to simulate the dispersive Hamiltonian given in Eq.~(\ref{eq:chidispersive}) with the physical dispersive coupling $\chi$ replaced by a larger effective value, 
$\chi_\mathrm{eff}$.

The time evolution unitary corresponds to a conditional cavity rotation gate
\begin{equation}
    \mathrm{CR}(\theta)=e^{-i \frac{\theta}{2} \hat n\sigma_z}
\end{equation}

which we can synthesize from CD gates via Trotter-Suzuki. Importantly, this conditional cavity rotation gate can also be interpreted as a rotation of the qubit around the $z$ axis by an angle that is proportional to the number of photons in the cavity:  $\theta_n=\theta\hat n$.

The SNAP gate (Box.~\ref{Box:SNAP}) is realized within the Fock state ISA via a pair of SQR gates (Box.~\ref{box:SQRgate}).  In the phase-space ISA we make use of a Rabi drive to rotate the qubit gate
\begin{equation}
    R_\varphi(\delta)=e^{-i\frac{\delta}{2}[\cos\varphi\sigma_x + \sin\varphi\sigma_y]}.
\end{equation}
To make the qubit rotation be conditioned on the photon number being $m$, let us choose $\theta,\delta\ll 1$ and repeatedly execute the gate sequence
\begin{align}
U_m(\theta,\delta)&=e^{+i\frac{\theta}{2}m\sigma_z}\mathrm{CR}(\theta) R_\varphi(\delta)\nonumber \\
&=e^{-i\frac{\theta}{2} [\hat n-m]\sigma_z}R_\varphi(\delta).
\label{eq:U_m}
\end{align}
If, for example, we wish to rotate the qubit through angle $\pi$ iff the photon number is $m$, then we repeat this $\pi/\delta$ times. If the photon number is not equal to $m$, then the rapid rotation of the qubit about the $z$ axis simulates the strong dispersive coupling that makes the Rabi drive off-resonant and hence ineffective at rotating the qubit.  Thus we have produced the SQR gate that selectively rotates the qubit conditioned on the photon number.  A second SQR gate with a different Rabi phase $\varphi$ then imparts the desired phase on cavity Fock state $m$ yielding the SNAP gate. 

If we simply rotate the qubit through $2\pi$ conditioned on the photon number being $m$, we have a SNAP gate that produces a Grover oracle that adds a minus sign to Fock state $|m\rangle$
\begin{align}
    G=\hat I -2|m\rangle\langle m|.
\end{align}
If we choose special values of $\theta,\delta$ that, rather than being very small, are rational multiples of $\pi$, we can achieve other novel gates that, for example, execute a `superparity' operation.
Executing the unitary $N$ times with $\delta=\frac{2\pi}{N}$ for selected values of $\theta$ applies a minus sign to every Fock state satisfying the condition $\hat n =0\mod \ell$ shown in Table~\ref{table:superparity}
\begin{align}
    G_\ell=\hat I -2\sum_m |m\ell\rangle\langle m\ell|.
\end{align}
All the gates shown in Table~\ref{table:superparity} disentangle the ancilla nearly perfectly from the cavity and leave the ancilla in its starting state. However, the phases for those Fock states that are supposed to be 0 can be erroneous.  The worst-case phase error is shown the last column.  For this simple algorithm, the error scales towards zero only as $1/N$.  There is doubtless a more efficient Hamiltonian simulation scheme that can do better (though perhaps ruining the precise modular in $n$ feature of this simple version).  Note that the errors in the QSP realization of the dispersive Hamiltonian evolution have not been included here.

\begin{table}[tbhp]

\centering
\caption{QSP for `superparity' operations.  Applying the unitary $U_0(\theta,\delta)$ found in Eq.~(\ref{eq:U_m}) $N$ times with $m=0$ and $\delta=\frac{2\pi}{N}$, applies a minus sign to every Fock state satisfying the condition shown in the first column.  The phases applied to the other states are $0$ to varying degrees of approximation. The worst-case error is shown in the last column and scales to zero like $1/N$. Note that the errors in the QSP realization of the dispersive Hamiltonian evolution have not been included here.  
\label{table:superparity}
} 
\begin{tabular}{|c|c|c|c|}
 \hline
 \makecell{Condition\\ $n=0 \mod \ell$} & $\theta$ &$N$ &\makecell{Max phase error\\ (radians)}   \\
   \hline
   \hline
$\mod 2$  & $\pi/2$ & $4$ & $0$ \\
  \hline
$\mod 4$ &  $\pi/4$ & $32$  & $0.152$ \\
 \hline
$\mod 4$& $\pi/4$  & $64$ & $0.077$ \\
   \hline
$\mod 4$ & $\pi/4$ & $128$ & $0.0385$ \\
  \hline
$\mod 8$ & $\pi/8$ & $64$ & $0.184$ \\
   \hline
$\mod 8$ &  $\pi/8$ & $128$ & $0.093$ \\
 \hline
 $\mod 8$  &  $\pi/8$ & $256$ & $0.047$ \\
 \hline
  $\mod 16$ &  $\pi/16$  & $128$ & $0.192$ \\
 \hline
 $\mod 16$ &   $\pi/16$ & $256$  & $0.097$\\
 \hline
\end{tabular}
\end{table}

\subsubsection{Phase-Space ISA from Fock-Space ISA}
\label{sssec:phaseISA-from-fockISA}

We now turn to the problem of compiling the phase-space ISA in terms of the Fock-space ISA. In particular, this requires the compilation of single-qubit rotations $R_\varphi (\theta)$ and conditional displacements $\mathrm{CD}(\alpha)$ in terms of selective qubit rotations $\textrm{SQR}(\vec{\theta},\vec{\varphi})$ and unconditional displacements $D(\alpha)$. Realizing the first of these is particularly straightforward: one just needs to choose $\theta_n = \theta$ and $\varphi_n = \varphi$, such that the same qubit rotation is applied for every Fock state, reducing the $\textrm{SQR}$ gate to a single qubit rotation $R_\varphi (\theta)$.

To synthesize a controlled displacement, we next recognize that an $\textrm{SQR}$ gate with $\theta_n = n \pi$ conjugated by single qubit rotations $R_y(\pi/2)$ realizes a controlled parity operation $\textrm{CP}$ (Box \ref{Box:CRotation}). We note that one can more efficiently realize this gate using the native dispersive interaction, though here we fix ourselves to gates within the phase-space ISA to prove cross-compilation capability. With the controlled parity gate as a primitive, we can now use the hybrid oscillator-qubit gate synthesis technique outlined in Sec.~\ref{sssec:useful_primitives} (see the text surrounding Eq.~(\ref{eq:cond_disp_compilation})) to realize a conditional displacement,
\begin{equation}
\mathrm{CD}(\alpha) = \textrm{CP}\  D(i\alpha) \ \textrm{CP}^\dagger,
\end{equation}
thus completing the compilation of the phase-space ISA in terms of the Fock-space ISA.

\subsection{Compilation between Sideband ISA and Phase-Space ISA}
\label{sec:cross-compilation-sideband-phase-space}

In this section, we show that cross compilation between the sideband gates and the controlled displacement gates is possible via Trotterization.

\subsubsection{Sideband Gates from Controlled Displacement}
We first show that sideband gates can be compiled from controlled displacements. Redefining $\alpha = -i\frac{\theta}{2} e^{i\phi}$, the red sideband gate in Eq.~\eqref{sideband-gate} can be rewritten as
\begin{align}
  \text{JC}(\theta,\varphi)
    &= e^{\alpha \sigma^+ a - \alpha^* \sigma^- a^\dagger} \\
    &= \left(e^{\delta \sigma^+ a - \delta^* \sigma^- a^\dagger}\right)^L
\end{align}
where we have Trotterized into a product of $L$ smaller sideband gates with $\delta = \alpha / L$. Each small sideband gate admits the following decomposition
\begin{align}
    e^{\delta \sigma^+ a - \delta^* \sigma^- a^\dagger} 
    &= \exp\left[ \frac{\sigma_x}{2} (\delta a - \delta^* a^\dagger) + \frac{i\sigma_y}{2} (\delta a + \delta^* a^\dagger) \right] \nonumber \\
    &= e^{\frac{\sigma_x}{2} (\delta a - \delta^* a^\dagger)} e^{\frac{i\sigma_y}{2} (\delta a + \delta^* a^\dagger)} + O(|\delta|^2),
    \label{sideband2ecd-product}
\end{align}
where the final line corresponds to two controlled displacement gates. Therefore, we can compile a single sideband gate with parameter $\alpha$ into $2L$ controlled-displacement gates with an error of $L \cdot O(|\delta|^2) = O(|\alpha|^2/L)$. Note that higher-order product formulas can be used in Eq.~\eqref{sideband2ecd-product} to improve the compilation accuracy.

\subsubsection{Controlled Displacement Gates from Sideband Gates}
Next, we show the reverse direction of compiling controlled displacement gates into sideband gates by again relying on Trotterization and product formulas. Without loss of generality, any controlled displacement gate can always be written as $e^{i\sigma_x (\alpha^* a^\dagger + \alpha a)}$ upon conjugation by single-qubit rotations. In addition, we can express this conditional displacement as a product of $L$ small conditional displacements via Trotterization,
\begin{equation}
    e^{i\sigma_x (\alpha^* a^\dagger + \alpha a)} = \left(e^{i\sigma_x (\delta^* a^\dagger + \delta a)}\right)^L,
\end{equation}
where $\delta = \alpha/ L$. Using the identity $\sigma_x = \sigma^+ + \sigma^-$ in combination with the first-order product formula, each small conditional displacement by $\delta$ can be rewritten as
\begin{align}
    e^{i\sigma_x (\delta a^\dagger + \delta a)} = e^{i\delta \sigma^+ a + i \delta^* \sigma^- a^\dagger } e^{i\delta^* \sigma^+ a^\dagger + i \delta \sigma^- a }  + O(|\delta|^2).
    \label{ecd2sideband-product}
\end{align}
We can then note that the first term in Eq.~\eqref{ecd2sideband-product} is a red sideband gate as given in Eq.~\eqref{sideband-gate}, while the second term is a blue sideband gate that (if not natively available) can be synthesized via a red sideband gate and single-qubit rotations,
\begin{align}
    e^{i\delta^* \sigma^+ a^\dagger + i \delta \sigma^- a } = e^{i\frac{\pi}{2} \sigma_x} \left[ e^{i\delta^* \sigma^- a^\dagger + i \delta \sigma^+ a } \right] e^{-i\frac{\pi}{2}\sigma_x}.
    \label{blue-from-red-sideband}
\end{align}
Therefore, we can express ``small'' controlled displacements by $\delta$ as two sideband gates interleaved by $\pi$ rotations on the qubit,
\begin{align}
    e^{i\sigma_x (\delta a^\dagger + \delta a)} = V_s(i\delta) e^{i\frac{\pi}{2} \sigma_x} V_s(i\delta) e^{-i\frac{\pi}{2} \sigma_x} + O(|\delta|^2).
    \label{eq:CDfromSideband}
\end{align}
Consequently, a ``large'' controlled displacement by $\alpha$ can be compiled as a product of $2L$ red and blue sideband gates interleaved with single-qubit rotations, with an overall error of $O(|\alpha|^2/L)$. 
We note that the red sideband gates can be alternatively replaced by blue sideband gates, again at the expense of additional single-qubit rotations.  We also note the analogy here with controlled displacement gates realized in ion traps by simultaneous red and blue sideband drives. Here we have demonstrated cross-compilation using first-order product formulas, but generalization to higher-order formulas is possible for more favorable error scaling.

\subsection{Compilation between the Sideband ISA and the Fock-Space ISA}
\label{sec:cross-compilation-sideband-SQR}

In this section, we demonstrate the ability to cross-compile between the sideband ISA and the Fock-space ISA. We note that having already shown how to cross-compile between (i) the Fock-space and phase-space ISAs and (ii) the sideband and phase-space ISAs, it is possible to cross-compile between sideband and Fock-space ISAs by employing the phase-space ISA as an intermediary. 
While this is a valid approach that proves cross-compilation is possible, in this section we discuss strategies for \emph{direct} cross-compilation between the sideband and Fock-space ISAs to provide further insight. 

\subsubsection{Sideband Gates from Fock-Space ISA}
To compile sideband operations in terms of the Fock-space ISA, it is necessary to synthesize both single-qubit rotations and sideband operations. The first of these is straightforward: by setting all rotation angles $\theta_n = \theta$ and $\phi_n = \phi$ for $\forall~n$ to be identical, the SQR gate $\mathrm{SQR}(\vec{\theta},\vec{\phi})$ (Box \ref{box:SQRgate}) naturally reduces to a single-qubit rotation,
\begin{align}
    e^{-i\sum_{n=0}^{n_{\rm max}} \frac{\theta}{2} \vec{b} \cdot \vec{\sigma} \ket{n}\bra{n}} &= e^{-i \frac{\theta}{2} \vec{b} \cdot \vec{\sigma} \sum_n \ket{n}\bra{n}} \nonumber\\
    &= e^{-i \frac{\theta}{2} \vec{b} \cdot \vec{\sigma} \otimes I} = R(\theta,\phi)
\end{align}
where $\vec{b}$ is determined by the phase $\phi$.

The remaining task is to compile the sideband gate $\text{JC}(\theta,\varphi)$ in Eq.~\eqref{sideband-gate} in terms of SQR gate and unconditional displacement. We can modify the technique developed in Ref.~\cite{Krastanov2015} to achieve this.  We begin by defining a particular SQR gate
\begin{align}
   U_n(\varphi) &=e^{-i\frac{\theta}{2}\sigma_\varphi \hat Q_n},\\
   \hat Q_n&=\sum_{n^\prime=0}^n|n\rangle\langle n|.
\end{align}
Now note the following useful commutators
\begin{align}
   \hat J_n &= [(a^\dagger - a),\hat Q_n] = \sqrt{n+1}\big(|n+1\rangle\langle n|+|n\rangle\langle n+1|\big)\\
    \hat L_n&=[i(a^\dagger + a),\hat Q_n]=i\sqrt{n+1}\big(|n+1\rangle\langle n|-|n\rangle\langle n+1|\big).
\end{align}
Using unconditional displacements and these commutation relations we obtain from BCH for small $\alpha \in \mathbb{R}$
\begin{align}
   \Omega_n= D(\alpha)U_n(0)D(-\alpha)U^\dagger_n(0)&\approx e^{-i\frac{\alpha\theta}{2}\sigma_x\hat J_n},\\
   \Xi_n =  D(i\alpha)U_n \left(\frac{\pi}{2} \right) D(-i\alpha)U^\dagger_n \left(\frac{\pi}{2} \right)&\approx e^{-i\frac{\alpha\theta}{2}\sigma_y\hat L_n}.
\end{align}
Using Trotter to add these two exponents for all $n$ yields the blue side band Hamiltonian evolution operator
\begin{align}
  \prod_{n=0}^{n_\mathrm{max}}\Omega_n\Xi_n  
  &\approx e^{-i\alpha\theta\sum_{n=0}^{n_\mathrm{max}}\sqrt{n+1}\left(\sigma^+|n+1\rangle\langle n|+\sigma^-|n\rangle\langle n+1|\right)}.
  \label{eq:bsb-from-sqr-d}
\end{align}
  To realize the red sideband we would use
\begin{align}
 \prod_{n=0}^{n_\mathrm{max}}\Omega_n\Xi^\dagger_n 
  &\approx e^{-i\alpha\theta\sum_{n=0}^{n_\mathrm{max}}\sqrt{n+1}\left(\sigma^-|n+1\rangle\langle n|+\sigma^+|n\rangle\langle n+1|\right)}.
  \label{eq:rsb-from-sqr-d}
\end{align}
Each gate in Eqs.~\eqref{eq:bsb-from-sqr-d} and \eqref{eq:rsb-from-sqr-d} costs $4n_{\rm max}$ unconditional displacement gates and $4 n_{\rm max}$ SQR gates to synthesize. Having synthesized unconditional qubit rotations and side band unitaries we have formally compiled the sideband gate set using only SQR gates and uncontrolled cavity displacements from the Fock space gate set, without going through an intermediate compilation into the phase-space gate set.    While this compilation is formally correct and convergent (for a finite cutoff, $n_\mathrm{max}$ on the photon number), the accuracy could be improved by using higher-order product formulas. Finding a highly efficient cross-compilation with shallow circuit depth remains an open problem.

\subsubsection{Fock-Space ISA from Sideband Gates}

To compile the SQR gate sets from the sideband gate sets, we may use the intermediate constructions in Ref.~\cite{QuditsfromOscillatorsPhysRevA.104.032605}. For the SQR gate, note that Ref.~\cite{QuditsfromOscillatorsPhysRevA.104.032605} constructed arbitrary clean SU(2) carrier rotations in the subspace $\{ \ket{0,n}, \ket{1,n}\}$ for any Fock level $n$ expanded by the upper and lower states of the qubit. Therefore, multiplying many such gates together directly leads to a SQR gate.

For the unconditional displacement gate, note that it is also possible to realize arbitrary SU(2) rotations between any pair of oscillator Fock levels \cite{QuditsfromOscillatorsPhysRevA.104.032605}. Therefore, using many such gates can immediately realize an unconditional displacement gate on the oscillators.

\subsection{Universality of Multi-Qubit-Oscillator Systems}
\label{sec:hybrid-multi-universality} 

In this section, we show that the phase-space ISA is universal on $n$-qubit-$m$-oscillator systems for $\forall n, m \in \mathbb{N}^+$ by explicitly constructing all possible generators of it as given in Eqs.~\eqref{unitary-rep-hermitian-gen} and \eqref{h-rep-polynomial}. The idea is to start from the 1-qubit-1-oscillator case and then repeatedly couple one qubit at a time to first realize an $n$-qubit-1-oscillator universality. Next, by repeatedly coupling one oscillator at a time to the $n$-qubit-1-oscillator system, universality can be shown successively on an $n$-qubit-$j$-oscillator system for $j=2,3,\cdots,m$ until finally the $n$-qubit-$m$-oscillator universality is achieved. We outline the details in the following. We note a recent work has demonstrated universal multi-mode control by coupling a single qubit to the modes using a conditional NOT displacement gate \cite{diringer2023conditional} which is in a similar spirit as the CD gate defined in Box \ref{Box:c-displacement}. 

\subsubsection{From $1$-Qubit-$1$-Oscillator to $n$-Qubit-$1$-Oscillator Universality} 
Consider an arbitrary generator on an $n$-qubit-$1$-oscillator system of the form $G_Q h(\hat{x},\hat{p})$ where $h(\hat{x},\hat{p}) = \sum_{r,s} c_{r,s} \hat{x}^r \hat{p}^s + \mathrm{h.c.}$ acts on the oscillator and $G_Q$ acts on the qubits. One way to construct the qubit generators is via tensor product of individual Pauli matrices $\bigotimes_{j=1}^n \sigma_{j,\alpha_j}$ for $\alpha_j = 0, x, y, z$ where $\sigma_{j,0}$ is the identity matrix for the $j$th qubit. However, a better set of generators are the generalized Gell-Mann matrices  \cite{bertlmann2008bloch,QuditsfromOscillatorsPhysRevA.104.032605} which possess cleaner commutation relationships. In the following proof, the notation $G_Q$ is reserved for these matrices.
Since $h(\hat{x},\hat{p})$ can be written as a sum of different terms, we only need to focus on generating one such term, say $(\bigotimes_{j=1}^n \sigma_{j,\alpha_j}) \hat{x}^r \hat{p}^s$ for $\forall j, \alpha_j, r, s$.

From the 1-qubit-1-oscillator universality, we can first compile the generator $\sigma_{1,\alpha_1} \hat{x}^r \hat{p}^s$ using qubit rotations on qubit 1 and conditional displacements. The qubit-qubit entangling gate in Sec.~\ref{ssec:exact-analytical-qubit-gates} can be used to first entangle a second qubit with qubit 1 to obtain the generator $(\sigma_{1,\alpha_1} \otimes \sigma_{2,\alpha_2}) \hat{x}^r \hat{p}^s$ (proper single-qubit rotation on qubit 2 is needed to precisely realize the desired Pauli $\sigma_{2,\alpha_2}$). This process can be continued to entangle more qubits one by one until we achieve the desired $(\bigotimes_{j=1}^n \sigma_{j,\alpha}) \hat{x}^r \hat{p}^s$. A linear combination of these then gives us the desired generator $G_Q \hat{x}^r \hat{p}^s$.

\subsubsection{From $n$-qubit-$1$-Oscillator to $n$-Qubit-$m$-Oscillator Universality}

Now we consider entangling more oscillators with the above $n$-qubit-$1$-oscillator system to create the desired generators. Let us first note the following identity when commuting products of high-powers of $\hat{x}$ and $\hat{p}$
\begin{align}
    \hat{p}^s \hat{x}^r = \hat{x}^r \hat{p}^s - i s \sum_{u=0}^{r-1} \hat{x}^{r-1-u} \hat{p}^{s-1} \hat{x}^u.
    \label{psxr-commutator}
\end{align}
Recursively applying Eq.~\eqref{psxr-commutator} to terms on the RHS to push any powers of $\hat{p}$ to the rightmost position, it can be seen that $\hat{p}^s \hat{x}^r$ can be written as a linear combination of $\hat{x}^{r'} \hat{p}^{s'}$ where the powers $r'$ and $s'$ is no larger than $r$ and $s$.

Now consider adding one more oscillator to achieve an arbitrary generator on the $n$-qubit-$2$-oscillator system $G_Q h(\hat{x}_1, \hat{p}_1, \hat{x}_2, \hat{p}_2)$ where $h(\hat{x}_1, \hat{p}_1, \hat{x}_2, \hat{p}_2)$ can be written in the form of Eq.~\eqref{h-rep-polynomial}. Again, we only need to focus on one term in the series sum of \eqref{h-rep-polynomial}, say $G_Q (\hat{x}_1^{r_1} \hat{p}_1^{s_1} \hat{x}_2^{r_2} \hat{p}_2^{s_2} + \mathrm{h.c.})$. Our goal is to generate this from the $n$-qubit-$1$-oscillator generator with additional gates in the phase-space ISA.

Due to the closure of the $n$-qubit generator set, it is always possible to find another two elements $G_Q'$ and $G_Q''$ such that $[G_Q', G_Q''] = i G_Q $. Now consider the following four Hermitian generators for small $t$
\begin{align}
    h_1 &= t G_Q' (\hat{x}_1^{r_1} \hat{p}_1^{s_1} + \mathrm{h.c.}) ,\\
    h_2 &= t G_Q'' (\hat{x}_2^{r_2} \hat{p}_2^{s_2} + \mathrm{h.c.}) ,\\
    h_3 &= i t G_Q' (\hat{x}_1^{r_1} \hat{p}_1^{s_1} - \mathrm{h.c.}) ,\\
    h_4 &= i t G_Q'' (\hat{x}_2^{r_2} \hat{p}_2^{s_2} - \mathrm{h.c.}) ,
\end{align}
each of which can be realized following from the $n$-qubit-$1$-oscillator universality. More specifically, $h_1$ and $h_3$ can be realized from universality on $n$-qubits with oscillator 1, whereas $h_2$ and $h_4$ follow from universality on $n$-qubits with oscillator 2. Using product formulas we have
\begin{align}
      (e^{ih_1} e^{ih_2} e^{-ih_1} e^{-ih_2}) (e^{ih_4} e^{ih_3} e^{-ih_4} e^{-ih_3}) 
     ~~~~~~~~~~~~~~~
     \nonumber \\ 
    ~~~ \approx \left(e^{-i t^2 G_Q (\hat{x}_1^{r_1} \hat{p}_1^{s_1} \hat{x}_2^{r_2} \hat{p}_2^{s_2} + \hat{x}_1^{r_1} \hat{p}_1^{s_1} \hat{p}_2^{s_2} \hat{x}_2^{r_2} + \mathrm{h.c.})} \right) 
    ~~~~~~~
    \nonumber \\
    ~~~~~~~~~~~~~~~ \times
    \left(e^{-i t^2 G_Q (\hat{x}_1^{r_1} \hat{p}_1^{s_1} \hat{x}_2^{r_2} \hat{p}_2^{s_2} - \hat{x}_1^{r_1} \hat{p}_1^{s_1} \hat{p}_2^{s_2} \hat{x}_2^{r_2} + \mathrm{h.c.})} \right) \nonumber \\
    \approx ~ e^{-i 2 t^2 G_Q (\hat{x}_1^{r_1} \hat{p}_1^{s_1} \hat{x}_2^{r_2} \hat{p}_2^{s_2} + \mathrm{h.c.})} ~~~~~~~~~~~~~~~~~~~~~~~~~
    \label{adding-osc-1to2}
\end{align}
where the generator on the RHS is the desired generator on the $n$-qubit-2-oscillator system.

More generally, to construct a desired generator on an $n$-qubit-$m'$-oscillator system $(m'>2)$ given by the form $G_Q (\prod_{j=1}^{m'} \hat{x}_j^{r_j} \hat{p}_j^{s_j} + \mathrm{h.c.})$, we use the universality of an $n$-qubit-$(m'-1)$-oscillator to obtain 
\begin{align}
    h_1 &= t G_Q' \left (\prod_{j=1}^{m'-1} \hat{x}_j^{r_j} \hat{p}_j^{s_j} + \mathrm{h.c.} \right) \label{h1-m} \\
    h_3 &= i t G_Q' \left(\prod_{j=1}^{m'-1} \hat{x}_j^{r_j} \hat{p}_j^{s_j} - \mathrm{h.c.} \right)
    \label{h3-m}
\end{align}
and use the $n$-qubit-1-oscillator universality for oscillator $m'$ to obtain
\begin{align}
    h_2 &= t G_Q'' (\hat{x}_{m'}^{r_{m'}} \hat{p}_{m'}^{s_{m'}} + \mathrm{h.c.}) \label{h2-m} \\
    h_4 &= t G_Q'' (\hat{x}_{m'}^{r_{m'}} \hat{p}_{m'}^{s_{m'}} - \mathrm{h.c.}). \label{h4-m}
\end{align}
Combining Eqs.~\eqref{h1-m} - \eqref{h4-m}, and using the same product formula as in Eq.~\eqref{adding-osc-1to2}, we obtain
\begin{align}
    (e^{ih_1} e^{ih_2} e^{-ih_1} e^{-ih_2}) (e^{ih_4} e^{ih_3} e^{-ih_4} e^{-ih_3}) 
    ~~~~~~~~~~
    \nonumber \\ 
    ~~~\approx  e^{-2i t^2 G_Q (\prod_{j=1}^{m'} \hat{x}_{m'}^{r_{m'}} \hat{p}_{m'}^{s_{m'}} + \mathrm{h.c.})}.
    \label{adding-osc-m'}
\end{align}
This accomplishes the induction step of adding one oscillator. When $m'=m$, we therefore have succeeded in constructing an arbitrary $n$-qubit-$m$-oscillator generator $G_Q (\prod_{j=1}^{m'} \hat{x}_{m'}^{r_{m'}} \hat{p}_{m'}^{s_{m'}} + \mathrm{h.c.})$ as given in Eq.~\eqref{unitary-rep-hermitian-gen} for any term in Eq.~\eqref{h-rep-polynomial}.

\subsection{Oscillator-only Universality: Controllability and the Cubic Instruction Set}\label{app:cubic_IS} 
This set comprises the Gaussian instruction set augmented by one operator polynomial that is cubic in $\hat x$ and $\hat p$ (or $a^\dagger$ and $a$). Let us see how the presence of a single control term higher-order than quadratic is sufficient to allow universal oscillator control \cite{Braunstein2005,hillmann2020universal,eriksson2023universal}.

Formal discussions of universal control for systems containing oscillators require dealing with the subtleties associated with the (countably) infinite dimension of the Hilbert space and the fact that essential  operators like position, momentum, and photon number are unbounded.  For simplicity, we will not go into these questions here, but rather simply define quantum (operator) controllability \cite{D'Alessandro1220755,Dong_2010} to mean that for a hybrid system consisting of an oscillator and a qubit, we have access to (i.e., can effectively generate) arbitrary unitary time evolution of the form
\begin{align}
    U(t)=e^{-ih_0(\hat x,\hat p)t}
    \label{eq:osc-control-H}
\end{align}
where the (effective) Hamiltonian $h_0$ is an arbitrary (Hermitian) polynomial function of position and momentum with a large but bounded degree. 

Such a general Hamiltonian is not always natively available, so our task is to understand how to generate the corresponding unitary evolution from a given finite (but continuously parameterized) set of primitive instructions. Let us assume that we can write the physically available  Hamiltonian $h_\mathrm{c}$  in the analog (time-dependent) control form
\begin{align}
    h_\mathrm{c}(\hat x,\hat p) &=\sum_{k=1}^M \mu_k(t)V_k,
\end{align}
where $G=\{V_k; k=1,\ldots M\}$ is a discrete set of (low-order polynomial) oscillator control operators to which we have access and the $\mu_k(t)$ are temporal modulation parameters available to us.  

For the qubit case, a set of $n$ qubits has a finite Hilbert space dimension $D=2^n$, and the fundamental test of full controllability is whether the Lie algebra generated by the control set $G$ is full rank; i.e., is the Lie Algebra $\mathfrak{s u}(D)$ \cite{D'Alessandro1220755,Dong_2010}.  If so, then there exist modulation functions $\{\mu_k(t);k=1,\ldots,M\}$ such that time evolution under $h_c$ generates the full Lie group SU$(D)$.  The connection between the Lie algebra based on operator commutators and the Lie group of unitary transformations can be seen from the presence of a commutator in the Heisenberg equation of motion for an operator.  More specifically, it is useful to consider the Baker-Campbell-Hausdorff (BCH) identity
\begin{align}
    e^{i\theta_jV_j}e^{i\theta_kV_k}e^{-i\theta_jV_j}&=
    \exp\{i\theta_k\{V_k+(i\theta_j)[V_j,V_k]\nonumber\\
    &+\frac{1}{2!}(i\theta_j)^2[V_j,[V_j,V_k]]+\dots\}\},
\end{align}
from which it follows (for small angles $\theta_j,\theta_k$) that
\begin{align}
    e^{i\theta_j V_j}e^{i\theta_kV_k}e^{-i\theta_jV_j}e^{-i\theta_kV_k}&\approx
    \exp\{(\theta_k\theta_j)[V_k,V_j]\}.
\end{align}
Such so-called `product formulas' \cite{sefi2011how,borneman2012parallel,childs2013product,chen2022efficient,childs2021theory,kang2023leveraging,concentration2021chen}  allow us to multiply group generators (in the form of commutators) to generate new effective Hamiltonian terms.  Equivalently, if the individual unitaries on the LHS above are taken from a discrete (but parameterized) instruction set, this equation shows us how to chain together elements of the instruction set to create new and more complex unitaries.  Similarly various Trotter-Suzuki formulas \cite{Suzuki_Generalized_Trotter} such as
\begin{align}
      e^{i\frac{\theta_j}{2}V_j}e^{i\theta_kV_k}e^{+i\frac{\theta_j}{2}V_j}\approx e^{i(\theta_jV_j+\theta_kV_k)}
\end{align}
allow us to sum Hamiltonian terms corresponding to different elements of the instruction set.

With these methods, expanded in Sec.~\ref{sec:compilation}, one can create (approximations to) any unitary in the group SU$(D)$, where $D=2^n$ is the Hilbert space of $n$ qubits, provided the set $G$ of control terms (or equivalently, the instruction set) is sufficiently rich to give controllability. That is, being able to create $D^2-1$ linearly independent traceless Hamiltonian terms (traceless since adding a constant to the Hamiltonian does not grant additional controllability). 

The case of infinite Hilbert space dimension is more subtle.  Nominally one only needs the algebra generated from the Lie brackets of the control Hamiltonians to be infinite, but this does not preclude the possibility of missing important terms.  Even if this algebra is infinite, there could be one or more symmetry operators that commute with all the generators.  For example, if the generators happen to commute with photon number parity you will have control over only half of the infinite number of states. Since we are defining controllability here as being able to generate effective Hamiltonian terms that are arbitrary polynomials of some bounded degree, this subtlety can be explicitly checked and avoided.  For an in-depth discussion of these matters, we refer to 
Ref.~\cite{bruschi2023deciding} which explicitly considers situations where the operator algebra closes.

As mentioned above, any non-quadratic Hamiltonian term is enough to achieve controllability of an oscillator \cite{Braunstein2005,hillmann2020universal,eriksson2023universal}.  Perhaps the simplest such term is $\hat x^3$ which, together with Gaussian operations, defines the Cubic instruction set.  
To understand why a single term in the Hamiltonian higher than quadratic is sufficient for universality, we can examine the commutation relations for different degrees of polynomial (using standard dimensionless units for which $[\hat x,\hat p]=+i$)
    \begin{align}
        [\hat{x},\hat{x}^m\hat{p}^n]&= in\hat{x}^m\hat{p}^{n-1},\\
        [\hat{x}^2,\hat{x}^m\hat{p}^n]&= \hat{x}[\hat{x},\hat{x}^m\hat{p}^n]+[\hat{x},\hat{x}^m\hat{p}^n]\hat x\nonumber\\
        &=in(\hat{x}^{m+1}\hat{p}^{n-1}+\hat{x}^m\hat{p}^{n-1}\hat{x}),\\
        [\hat{x}^3,\hat{x}^m\hat{p}^n]&= \hat{x}[\hat{x}^2,\hat{x}^m\hat{p}^n]+[\hat{x^2},\hat{x}^m\hat{p}^n]\hat x,\\
        &=in(\hat{x}^{m+2}\hat{p}^{n-1}+\hat{x}^{m+1}\hat{p}^{n-1}\hat{x}\nonumber\\
        &\quad+\hat{x}^{m+1}\hat{p}^{n-1}\hat{x}+\hat{x}^{m}\hat{p}^{n-1}\hat{x}^2).
    \end{align}
    As is evident, the degree of a polynomial in $\hat x,\hat p$ is preserved by its commutator with a quadratic term $\hat x^2$. In contrast, the commutator of a degree $m+n$ polynomial with a cubic term $\hat x^3$ yields a polynomial of degree $m+n+1$. Similarly, we can prove the same properties for $\hat p, \hat p^2, \hat p^3$ and $\hat x^m \hat p^n$. Thus, the algebra generated by the Lie brackets of these terms is infinite in the presence of a cubic or higher-order polynomial. As a specific example, consider
    \begin{align}
       [\hat x^3,\hat p^2]=i3(\hat p\hat x^2+\hat x^2\hat p),\\
       [\hat x^3,[\hat x^3,\hat p^2]]=-18\hat x^4.
    \end{align}
Using this inductive proof we have shown that, the algebra generated by the control Hamiltonian set which includes quadratic Hamiltonians with only one cubic Hamiltonian can generate arbitrary order polynomials enabling universal control. 

The above discussion shows us how to generate the oscillator-only control unitary in Eq.~(\ref{eq:osc-control-H}). From an experimental physics perspective, we require only the harmonic Hamiltonian $\hat p^2+\hat x^2$, linear drives on the oscillator that couple to $\hat x$ (and/or $\hat p$), and one additional higher-order term such as $\hat x^3$ to generate an infinite algebra which can synthesize arbitrary unitary operations.  The presence of the $\hat x^3$ (or higher) term makes the harmonic oscillator anharmonic so that each transition between adjacent levels has a different frequency making the transitions individually addressable through resonant drives of selected frequencies, amplitudes, and phases.  This physical picture of the origin of controllability gives nice intuition to accompany the mathematics of the infinite algebra generated by the Lie bracket of the control set.  Experimental oscillator-only control using cubic interactions 
was demonstrated for microwave circuits in~\cite{hillmann2020universal,eriksson2023universal}. 

Similar arguments to those above show that multi-mode universal control requires only the addition of Gaussian beam-splitter gates between physically adjacent oscillators (since the contents of any two oscillators can be made physically adjacent through SWAP gates built from the beam-splitter interaction).  The beam-splitter allows one to perform coherent rotations between mode operators (see Box~\ref{Box:beam-splitter}), and hence create arbitrary polynomials of multiple oscillator variables.  We have thus (non-rigorously) established the universality of the Cubic instruction set for oscillator-only control.

\section{Measurements and Tomography}
\label{app:measurement-tomography}

We have established instruction sets from which arbitrary universal operations on the hybrid bosonic quantum processor can be achieved, and which enable useful computation and information processing tasks to be accomplished. To extract information from the bosonic quantum processor after the computation, measurements are needed. In this section, we shall delve more into aspects of measurement and discuss how measurements may be performed on these hybrid systems. 

Measurements can be carried out on either qubits or oscillators, although direct measurements on oscillators are difficult because photon number detectors might be required. Surprisingly, by leveraging the native couplings between the qubit and oscillator, indirect measurement of the oscillator can be achieved by simple qubit measurement. We start by presenting a few basic measurement schemes based on qubit measurement in Apps.\ \ref{sec:hybridqubitcavitymeasurements}, \ref{subsec:dispersivereadout}, \ref{subsec:cavityphotonnumbermeasurements},  \ref{ssec:superparity} to extract oscillator photon numbers and parities. We then discuss a qubit-free homodyne measurement scheme in App.~\ref{sec:homodyne-detection} that utilizes a simple oscillator-oscillator gate. In cases where more information in the phase space of oscillators needs to be readout, Apps.\ \ref{sec:characteristic-function} and  \ref{sec:husimi-q} describe three different ways of performing tomography of the phase space. 

Finally, just as in DV system where random Clifford circuits combined with measurement are used to benchmark the performance of the circuit, Gaussian operations combined with the measurement techniques in this section can be used to benchmark CV systems. We discuss random benchmarking of CV systems in App.~\ref{sec:benchmarking-oscillators}. We note that these measurement protocols can be reformulated in the framework of classical shadow tomography for continuous-variable systems \cite{gandhari2022continuous,iosue2024continuous} and optimizations in the case of Gaussian states can also be considered~\cite{mele2024learning}.  App.~\ref{sec:benchmarking-oscillators} and Ref.~\cite{valahu2024benchmarking} considered benchmarking of displacement operations using a random sequence of oscillator displacements, followed by an inverting displacement that nominally returns the system to its starting state.

\subsection{Hybrid Qubit-Cavity Measurements}
\label{sec:hybridqubitcavitymeasurements}

The dispersive Hamiltonian for the coupled qubit-cavity system in Eq.~(\ref{eq:dispersiveHD}) commutes with both $\sigma_z$ and $\hat n=A^\dagger A$ and can therefore be used for quantum non-demolition (QND) read out of both the state of the qubit and the state of the cavity.  Indeed, one of the major advantages of storing quantum information in the large Hilbert space of an oscillator is that state tomography is much easier and more accurate than it would be (at least with present technology) in a qubit system with the corresponding Hilbert space dimension.

\subsection{Dispersive Readout of the Qubit}
\label{subsec:dispersivereadout}
The dispersive Hamiltonian in Eq.~(\ref{eq:dispersiveHD}) tells us that the cavity resonance frequency is shifted to
\begin{equation}
    \omega_R\pm \frac{\chi}{2},
\end{equation}
depending on the state of the qubit.  By measuring the phase shift of a microwave pulse reflected from the cavity one can infer the resonance frequency of the cavity and hence determine the state of the qubit (i.e.,\ measure $\sigma_z$) with high ($\sim 99\%-99.9\%$) fidelity \cite{Blais2004,Blais2020,Wallraff2005,Walter_PhysRevApplied.7.054020,kurilovich2025highfrequencyreadoutfreetransmon}.

\subsection{Cavity Photon Number Measurement}
\label{subsec:cavityphotonnumbermeasurements}
\begin{figure*}[tbh]
    \centering
    \includegraphics[width=0.75\textwidth]{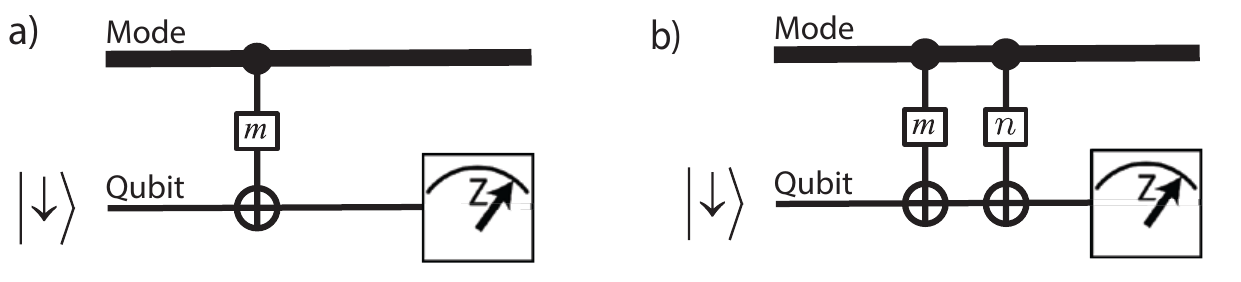}
    \caption{(a) Quantum circuit using the cavity conditioned qubit rotation gate $U_m(\pi/2,0)$ and a $Z$ measurement of the ancilla qubit to measure the operator $\hat P_m$ that answers the question `Does the cavity contain $m$ photons?' (b) Circuit using $U_n(\pi/2,0)U_m(\pi/2,0)$ and a $Z$ measurement of the ancilla qubit to measure the operator $\hat P_m+\hat P_n$ that answers the question `Does the cavity contain \emph{either} $m$ or $n$ photons?' }
    \label{fig:CavityPhotonNumMeas}
\end{figure*}

We can ask the question, `Is the cavity photon number m?' by measuring the value of the projector onto Fock state $|m\rangle$
\begin{equation}
    \hat P_m=|m\rangle\langle m|.
\end{equation}
This can be achieved in a two-step process \cite{Wang2020FCFs} illustrated in Fig.~\ref{fig:CavityPhotonNumMeas}a in which we first apply the SQR gate (Box.~\ref{box:SQRgate})
\begin{equation}
    U_m(\pi,0)=e^{i\frac{\pi}{2} \sigma_x \hat P_m}=(\hat I-\hat P_m)+i\sigma_x \hat P_m,\label{U_mgatespecial}
\end{equation}
which flips the qubit only if there are $m$ photons in the cavity.  Then we read out the qubit state using the resulting dispersive shift in the (possibly separate readout) cavity frequency.

Suppose that the initial cavity state is
\begin{equation}
    |\psi\rangle =\sum_{j=0}^{N_\mathrm{max}} \psi_j|j\rangle.
\end{equation}
If the measurement result on the ancilla qubit is $Z=+1$ the state collapses to $|\uparrow\rangle|m\rangle$.  If the measurement result on the ancilla qubit is $Z=-1$, the state collapses to
\begin{equation}
     |\psi\rangle =|\downarrow\rangle\frac{1}{\sqrt{1-|\psi_m|^2}}\sum_{j\ne m} \psi_j|j\rangle.
\end{equation}
These gates have been used in Sec.~\ref{sec:qec-compilation} to present schemes that readout or entangle dual-rail qubits using an ancilla qubit. Interestingly, we can apply multiple versions of the cavity-conditioned bit-flip gate to ask other questions about the cavity state.  As illustrated in Fig.~\ref{fig:CavityPhotonNumMeas}b, the compound operator
\begin{equation}
   U_n(\pi,0) U_m(\pi,0)=e^{i\frac{\pi}{2} \sigma_x [\hat P_m+\hat P_n]}
   \label{U_nmgatespecial},
\end{equation}
flips the qubit if the photon number is \emph{either} $n$ or $m$ (since it can't be both).  These two unitaries can be executed in parallel by applying two simultaneous drive tones on the ancilla qubit. Generalizing to a product of many unitaries, the most general projector we can measure is
\begin{equation}
    \hat P_{\vec c}=\sum_{m=0}^{N_\mathrm{max}} c_m\hat P_m,
    \label{eq:arbbinaryvectormeas}
\end{equation}
where $\vec c=(c_0,c_1,\ldots,c_{N_\mathrm{max}})$ is a binary-valued vector, with $c_j\in \{0,1\}; j=0,N_\mathrm{max}$.  

This is a powerful gate. One could use it, for example, to flip the ancilla bit if the photon number is less than (or greater than) a selected value. One could then use the qubit-conditioned cavity displacement gate (Box.~\ref{Box:c-displacement}) to displace the cavity conditioned on the photon number being less than (or greater than) the selected value.  As we describe further below, it can also be used in a highly efficient binary search for the photon number.

\subsubsection{Cavity Photon Number Parity Measurement}
A particularly interesting and useful example of the general projectors available in the class $\hat P_{\vec c}$ is the photon number parity operator
\begin{equation}
    \hat \Pi = e^{i\pi\hat n}=\sum_{m}(-1)^m\hat P_m,
    \label{eq:number-parity-operator}
\end{equation}
which can be mapped onto the qubit for measurement using the unitary
\begin{align}
    U_{\vec c}=e^{i\frac{\pi}{2} \sigma_x \hat P_{\vec c}}
    \label{eq:Uvecc}
\end{align}
with $\vec c=(0,1,0,1,0,1,0\ldots)$ so that the qubit is flipped only if the photon number parity is odd.
The photon parity operator plays an essential role in cavity state tomography (to be described below) and in quantum error correction using bosonic codes \cite{Ofek2016,BinomialCodes}.  The actual pulse sequence used in practice to measure the parity \cite{Sun2014} is different from that outlined above but produces the same result (see Sec.~\ref{sec:qec-compilation}).  The circuit used takes advantage of the conditional cavity rotation afforded directly under time evolution with the dispersive Hamiltonian in Eq.~(\ref{eq:dispersiveHD}).  Time evolution under the dispersive coupling
\begin{equation}
    \frac{\chi}{2}\sigma_z A^\dagger A
\end{equation}
for a time $t=-\theta/\chi$ ($\chi$ is typically negative) is described by the `conditional rotation' unitary (Box \ref{Box:CRotation})
\begin{equation}
    \mathrm{CR}(\theta)=e^{-i\frac{\theta}{2}\sigma_z\hat n}. \label{eq:cRot}
\end{equation}
Applying the phase estimation protocol to this unitary will give us a powerful way to measure the full photon number distribution along with photon number parity and `super-parity' measurements \cite{Wang2020FCFs} described further below in App.~\ref{ssec:superparity}.

\begin{figure}[tbh]
    \centering
    \includegraphics[width=0.3\textwidth]{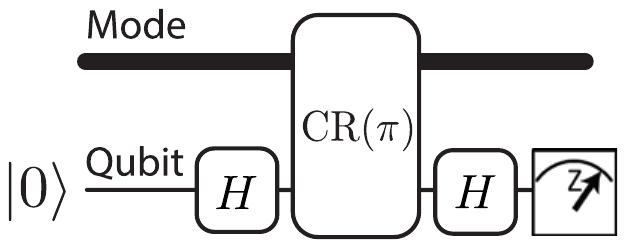}
    \caption{Quantum circuit using phase kickback from the controlled oscillator rotation, 
    ${\rm CR}(\pi)$, to entangle the ancilla qubit state with the photon number parity of the mode.  Measurement of the qubit state then yields a QND measurement of the parity of the photon number.   \label{fig:CavityParityMeas}}
\end{figure}

To see the connection to the parity operator, consider the conditional rotation gate for $\theta = \pi$ (Box \ref{Box:CRotation}) 
\begin{equation}
    \mathrm{CR}\left(\pi\right)=e^{-i\frac{\pi}{2}\hat n}\,\mathrm{CP},\label{eq:CPi1}
\end{equation}
where
\begin{equation}
   \mathrm{CP} =e^{i\pi |e\rangle\langle e|\hat n}=|g\rangle\langle g|\hat I +|e\rangle\langle e|\hat\Pi\label{eq:CPi2}
\end{equation}
is the qubit-controlled cavity parity operation which is active only if the qubit is in its excited state $|e\rangle$. The additional rotation of the cavity state in Eq.~(\ref{eq:CPi1}) corresponds for our purposes to an irrelevant frame choice.  The circuit shown in Fig.~\ref{fig:CavityParityMeas}, familiar from the phase estimation problem, takes advantage of the phase kickback on the ancilla from the controlled parity unitary to entangle the qubit state with the cavity parity, thus enabling the cavity parity to be measured by reading out the qubit state. Effectively, the controlled parity gate rotates the qubit state by angle $\hat n\pi$ around the $z$ axis taking it from $|+\rangle$ to $|-\rangle$ if the photon number is odd and returning it to $|+\rangle$ if the photon number is even. The measurement result $z_0=\pm 1$ after the final Hadamard gate determines the parity bit of the photon number via
\begin{equation}
    b_0=\frac{1-z_0}{2}.
\end{equation}

\subsection{Cavity Photon Number `Super-Parity' Measurement}
\label{ssec:superparity}
We can use phase estimation to generalize the above result to obtain the `super-parities' $b_j$ that yield the full binary number representation of the photon number
\begin{equation}
    n=\sum_{j=0}^{j_\mathrm{max}}2^j b_j.
\end{equation}
If the measurement result for $b_0$ is known, then 
we can determine the next binary digit by measuring $b_1=\frac{[(n-b_0)\, \mathrm{mod}\, 4]}{2}$.  This utilizes the same circuit in Fig.~\ref{fig:CavityParityMeas} with the ${\rm CR(\pi)}$ gate replaced by the following unitary 
\begin{equation}
    \mathrm{CR}\left(\frac{\pi}{2}\right)R_z\left(-b_0\frac{\pi}{2}\right),
\end{equation}
where the $R_z$ is the single-qubit rotation in Table \ref{tab:gates-qubit}
(again executable `in software' via a frame change)
\begin{equation}
    R_z(\theta)=e^{-i\frac{\theta}{2}\sigma_z}.
\end{equation}
It can be shown that 
\begin{align}
    b_1 = \frac{1- z_1}{2}
\end{align} 
where $z_1$ is the measurement outcome of the qubit.

Similarly, once the measurement results for $b_0$ and $b_1$ are known, $b_2$ is obtained using the unitary
\begin{equation}
    {\rm CR}\left(\frac{\pi}{4}\right)R_z\left(-[b_0+2b_1]\frac{\pi}{4}\right),
\end{equation}
and $b_3$ is obtained using the unitary
\begin{equation}
    {\rm CR}\left(\frac{\pi}{8}\right)R_z\left(-[b_0+2b_1+4b_2]\frac{\pi}{8}\right).
\end{equation}
In general, $b_k$ can be obtained using the unitary 
\begin{align}
    {\rm CR}\left(\frac{\pi}{2^k}\right)R_z\left(-\left[\sum_{j=0}^{k-1} 2^{j} b_j\right]\frac{\pi}{2^k}\right).
\end{align}

\begin{figure*}[tbh]
    \centering
    \includegraphics[width=0.85\textwidth]{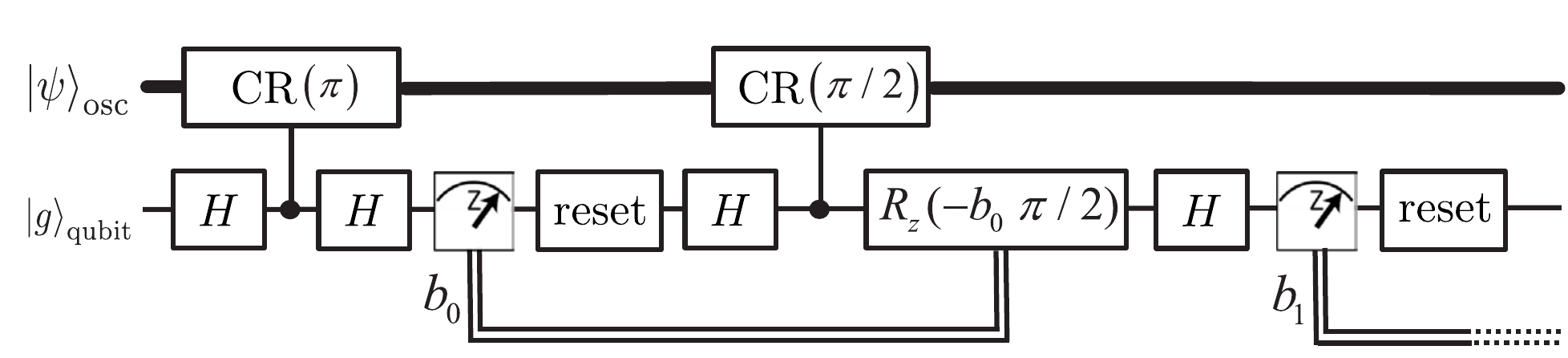}
    \caption{Quantum circuit with classical feed-forward of the ancilla measurement results $b_0,b_1,\ldots$ to carry out the phase estimation algorithm to measure the binary representation $(\ldots,b_1,b_0)$ of the photon number.}
    \label{fig:FeedforwardPhasEstimation}
\end{figure*}

Importantly, since all these unitaries commute with each other we can make a QND measurement of all the binary digits in a single shot (single state preparation) using the quantum circuit with classical feed-forward shown in Fig.~\ref{fig:FeedforwardPhasEstimation}.  This is very powerful because it effectively performs a binary search to obtain the value of $n$ in only $\sim\log N_\mathrm{max}$ steps.  It is exponentially more efficient than using the circuit in Fig.~\ref{fig:CavityPhotonNumMeas}a to sequentially ask, `Is $n$ equal to zero?, Is $n$ equal to 1? Is $n$ equal to two? ...' and so on.

As an alternative to the superparity measurements which determine the least-significant bit (LSB) first, we can perform a more traditional binary search for the photon number value by starting from the most-significant bit (MSB). If we consider a range of photon numbers $[0,N_\mathrm{max}]$, with $N_\mathrm{max}+1$ being an integer power of 2,
then we could use the projector in Eq.~(\ref{eq:arbbinaryvectormeas}) to ask, `Is the photon number in the upper half of the range? (i.e.\ in the interval $[(N_\mathrm{max}+1)/2,N_\mathrm{max}]$).  Given the answer, we then recursively ask,`Is the photon number in the upper half of the interval containing the photon number?' This is just like classical binary search.

\begin{figure}[tbh]
    \centering
    \includegraphics[width=0.45\textwidth]{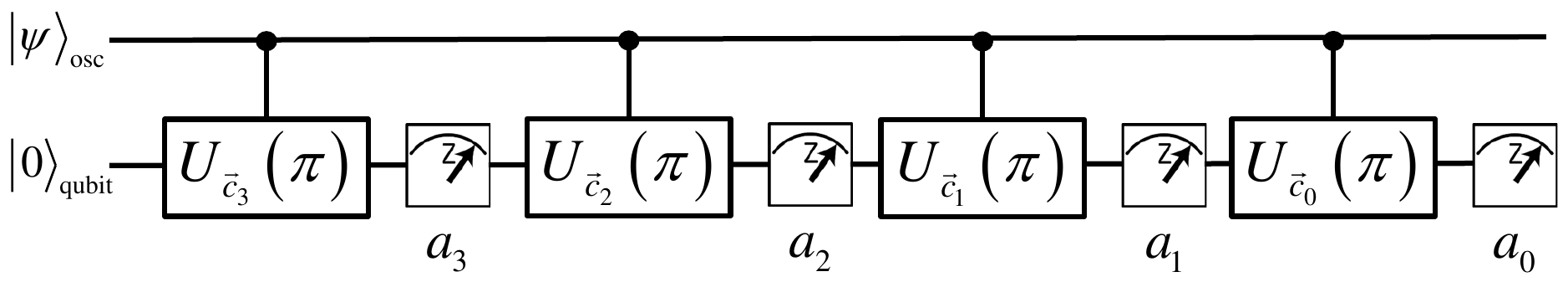}
    \caption{Quantum circuit for binary search to measure photon number $\hat n$.  Unlike the phase estimation protocol, this does not require with classical feed-forward of the ancilla measurement results $a_0,a_1,a_2,\ldots$. The binary representation of the photon number can be determined directly from the sequence of measurement results.  Note that the auxiliary qubit is assumed to be initialized in $|g\rangle=|0\rangle$, but is not reset after each measurement. Ref.~\cite{Wang2020FCFs} uses essentially this protocol but \emph{does} reset the qubit after each measurement.}
    \label{fig:BinarySearch_for_n}
\end{figure}

Proceeding in the spirit of Eq.~(\ref{eq:Uvecc}), let us define a generalized CNOT gate controlled by the state of the cavity using a family of binary vectors
\begin{eqnarray}
\vec c_2&=&(00001111)\\
\vec c_1&=&(00110011)\\
\vec c_0&=&(01010101).
\end{eqnarray}
In this simple example, we take the maximum photon number to be $N_\mathrm{max}=7$ and the vector $\vec{c}_j$ to be the $j$th  basis vector of the Walsh-Hadamard transform.
Now consider the use of this gate in the circuit depicted in Fig.~\ref{fig:BinarySearch_for_n}.  Note that in analyzing this sequence of measurements, we will assume that the auxiliary qubit is initialized in state $|0\rangle$ at the beginning of the protocol but is \emph{not} reset after each measurement.

From the vectors $\{\vec c_2,\vec c_1, \vec c_0\}$ we see that the measurement result $a_2$ tells us whether the photon number is in the upper half $[4, 7]$ ($a_2=1, \ket{e}_{\rm qubit}$) or the lower half $[0, 3]$ ($a_2 = 0, \ket{g}_{\rm qubit}$) of the considered range of $[0, 7]$. 

If the measurement result in $a_1$ is different than $a_2$, the ancilla spin has flipped and therefore the photon number is in the upper half of the relevant range.  Similarly, if the measurement result $a_0$ is different than $a_1$ the spin has again flipped and the photon number parity must be odd.  Thus we obtain the binary digits representing the photon number
\begin{eqnarray}
b_2&=&a_2,\\
b_1&=&a_2\oplus a_1\\
b_0&=&a_1\oplus a_0,
\end{eqnarray}
where $\oplus$ corresponds to addition modulo 2.

The fact that active feed-forward is not required is also a significant advantage.  This is (essentially) the technique used in the boson sampling experiment of Ref.~\cite{Wang2020FCFs}, except those authors used numerically generated optimal control pulses on the qubit to map the binary bits in the photon number onto an ancilla.  The other difference is in Ref.~\cite{Wang2020FCFs}, the authors reset the ancilla to its ground state after each measurement to minimize errors from the energy relaxation of the ancilla.

\subsection{Homodyne Detection}
\label{sec:homodyne-detection}
\begin{figure}[tbh]
    \centering
    \includegraphics[width=0.50\textwidth]{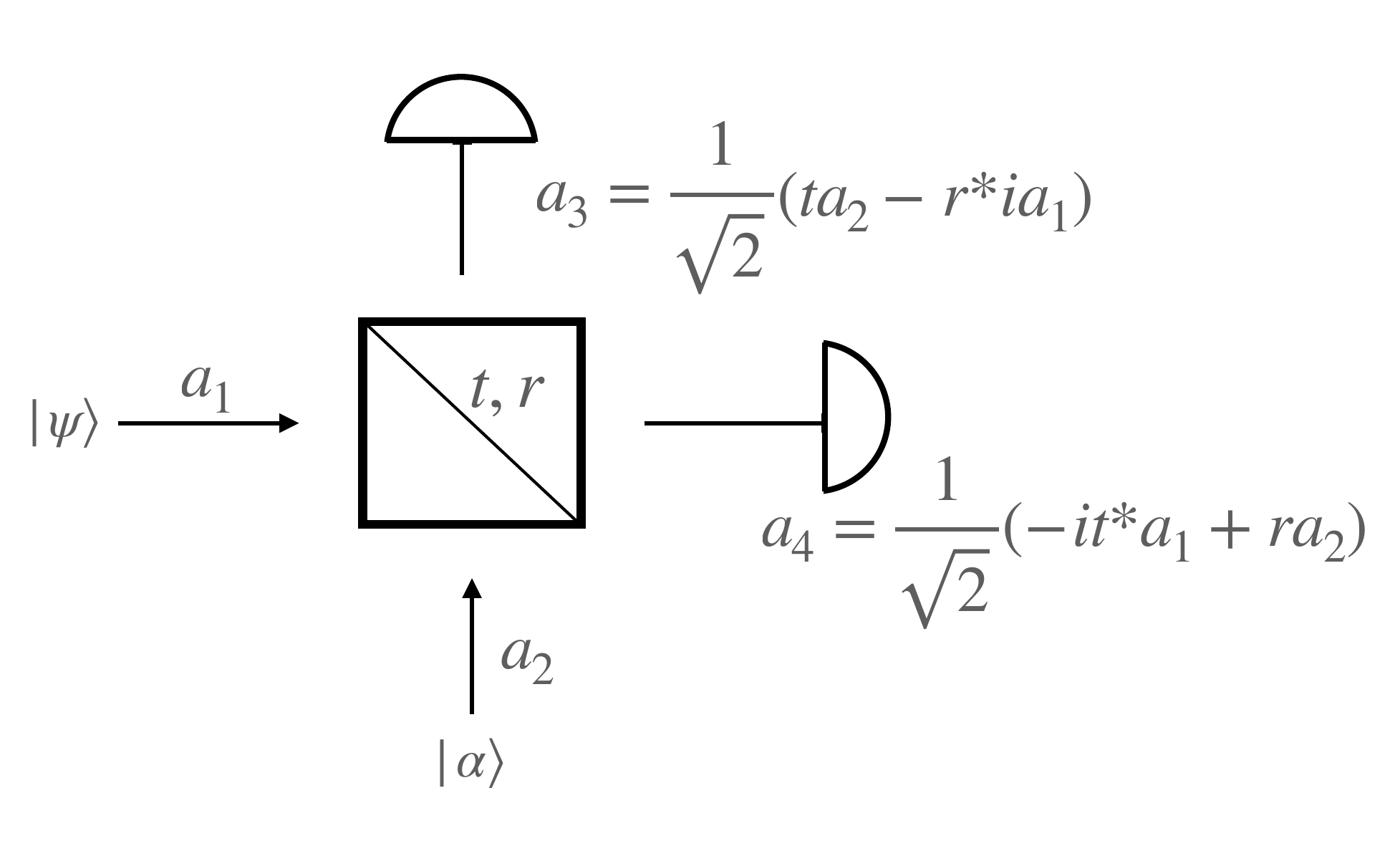}
    \caption{Homodyne detection involves a beam splitter (BS) with transmission coefficient $t$ and reflection coefficient $r$, and two detectors at the output ports of the BS. For balanced homodyne detection, $|t|^2=|r|^2=\frac{1}{2}$. }
    \label{fig:BS}
\end{figure}

Homodyne detection can be seen as a projective measurement of the phase-space quadratures. The setup for homodyne detection includes a beam splitter with transmission coefficient ($t$) and reflection coefficient ($r$), and two photodetectors (see Fig.~\ref{fig:BS}). The quantity of interest here is
\begin{equation}
    \langle n_3\rangle -\langle n_4\rangle,
\end{equation}
the difference of mean photon numbers detected by the two detectors for a given input state.
The state to be measured ($\ket{\psi}$) is sent through one port while the coherent state ($\ket{\alpha}$) is sent through other input port. In the Heisenberg picture, the photon number operators at the output ports are, 
\begin{eqnarray}
    n_3&=&\frac{1}{2} \left[|t|^2 n_1+|r|^2 n_2-i(t^*ra_1^\dagger a_2-r^*ta_2^\dagger a_1)\right ]
    \nonumber\\&&\\
    n_4&=&\frac{1}{2}\left [|t|^2 n_2+|r|^2 n_1-i(t^*ra_2^\dagger a_1-r^*ta_1^\dagger a_2) \right]
    \,,\nonumber\\    
\end{eqnarray}
and the mean photon number is given by
\begin{eqnarray}
    \langle n\rangle &=&\bra{\psi,\alpha}n\ket{\psi,\alpha},\\
    \langle n_3\rangle &=&\frac{1}{2}\left [
    \rule{0pt}{2.4ex}
    |t|^2\bra{\psi}n_1\ket{\psi}+|r|^2 |\alpha|^2
    \right. ~~~~~~~
    \nonumber\\
    && ~~~~~~ \left. -i \bra{\psi}(tr^*\alpha) a_1^\dagger-(tr^*\alpha)^* a_1\ket{\psi} 
    \rule{0pt}{2.4ex} \right]
    \nonumber.\\&&
\end{eqnarray}
Assuming a balanced homodyne detection with $t=r$ and $\alpha=|\alpha|e^{i\phi}$
\begin{equation}
     \langle n_3\rangle-\langle n_4\rangle=|\alpha|\bra{\psi} (a^\dagger e^{i\phi}-ae^{-i\phi})\ket{\psi}.
\end{equation}
Thus, the mean photon number measures the mean value of a phase-space quadrature along an axis dependent on the phase of the coherent state. This is a multi-purpose method. Homodyne detection on identically prepared states reconstructs the quadrature probability distribution. Repeating these measurements for all different phases is equivalent to computing the marginal integrals of Wigner function, an important visualization tool for quantum states, which then can be used to obtain the Wigner function (and the density operator) via a Radon transform \cite{PhysRevA.40.2847,PhysRevA.62.043814,PhysRevA.67.043815,PhysRevLett.96.213601}.

Homodyne detection of traveling photons at optical frequencies can be relatively efficient, but at microwave frequencies, even nearly quantum-limited linear amplifiers add enough noise to complicate state reconstruction, but with sufficient care, this can be done relatively well \cite{Eichler_PhysRevLett.106.220503}. 
For microwave photons trapped in stationary modes in high $Q$ resonators, it is possible to avoid these difficulties by synthesizing homodyne and heterodyne measurements by mapping the cavity state onto an auxiliary qubit and measuring the qubit \cite{strandberg2023digital}.

\subsection{Wigner and Characteristic Function Tomography}
\label{sec:characteristic-function}
A key step in validating quantum processes is to estimate the Wigner function for a quantum state.  This allows us to assess whether there are errors in our quantum circuits. In this section and the following, we review processes for estimating the characteristic function, the Wigner function, and the Husimi $Q$-function.

The phase-space density functions $W$ (Wigner), $Q$, and $P$ are different Fourier transforms of characteristic functions. The characteristic function for a bosonic system is the expectation value of the displacement operator at each point in phase space for a given density matrix $\rho$,
\begin{equation}
    C(\alpha)=\mathrm{Tr} \left [ \rule{0pt}{2.4ex} \rho D(\alpha) \right],
\end{equation}
where $\mathcal{D}(\alpha)$

is the (unconditional) displacement operator defined in Box~\ref{Box:UncondDispGate}. It can be seen that for a pure state, this quantity is an auto-correlation function of the quantum state in phase space. 

How does one measure the overlap of a state with a displaced version of itself?  This can be done by replacing the unconditional displacement operator with a conditional displacement operator controlled by the $Z$ value of an ancilla qubit. Measurement of the phase kick-back on the ancilla yields a one-bit estimator expectation value of the displacement operator. Initializing the ancilla in $|+\rangle$ and measuring $\langle X\rangle$ ($\langle Y\rangle$) gives a one-bit estimate of the real (imaginary) part of the characteristic function.  Repeated shots of these two types of measurements can be averaged to obtain the complex expectation value $C(\alpha)$ \cite{HomeCharacteristicFunction,EickbuschECD}.

$C(\alpha)$ is often referred to as the symmetric characteristic function to distinguish it from its normal ($C_n$) and anti-normal ($C_{an}$) order counterparts,
\begin{eqnarray}
    C_n(\alpha)=\mathrm{Tr}[\rho e^{\alpha a^\dagger}e^{-\alpha^* a}], \label{eq:characteristic-function-cn} \\
    C_{an}(\alpha)=\mathrm{Tr}[\rho e^{-\alpha^* a}e^{\alpha a^\dagger}]. \label{eq:characteristic-function-can}
\end{eqnarray}

The Fourier transform of the symmetric characteristic function $C(\alpha)$ is called the Wigner function and is defined in Eqs.~(\ref{eq:wigner1mainbody}-\ref{eq:wigner2mainbody}) in the main body of the text and is restated here for convenience:
\begin{equation}
    W(\beta)=\frac{1}{\pi^2}\int d^2\alpha \ C(\alpha)e^{\beta\alpha^*-\alpha \beta^*}.
\end{equation}
The density matrix $\rho$ of a system contains the full information needed to predict the (statistical) outcomes of any measurement on that system.  The Wigner function is a quasi-probability distribution in phase (position/momentum) space that contains the equivalent information to the density matrix but is easier to measure.  Moreover, practical methods for Wigner function estimation have been developed in the literature~\cite{he2023efficient}.  

In the same dimensionless complex coordinates that we use to describe the amplitude of a coherent state, the relationship of the Wigner function to the density matrix is given by the following remarkable identity due to Lutterbach and Davidovich, \cite{Lutterbach_DavidovichPhysRevLett.78.2547}
\begin{equation}
    W(\beta)=\frac{2}{\pi}\mathrm{Tr}\left [ 
    \rho{\mathcal D}^\dagger(-\beta)\ \hat \Pi \ 
    {\mathcal D}(-\beta)\right ]
    \,.
    \label{eq:wigner-app-g}
\end{equation}
where $\hat{\Pi}$ is the photon number parity operator as defined in Eq.~\eqref{eq:number-parity-operator}.

This formula tells us that the value of the Wigner function at the point $\beta$ in the complex plane is determined by displacing that point to the origin and then measuring the expectation value of the photon number parity.  This remarkable result tells us that state tomography on the oscillator is relatively easy to obtain, even though the effective dimension of the Hilbert space $N_\mathrm{max}+1$ can be quite large (though see the following paragraph for an important caveat regarding sampling complexity). In addition, the basic procedure does not change as $N_\mathrm{max}$ increases. It is worth noting that Eq.~\eqref{eq:wigner-app-g} shows that determining only the photon number parity, instead of the full photon number distribution, suffices for Wigner function tomography. However, Refs.~\cite{shen2016optimized,PRXQuantum.6.010303} demonstrated that the redundant information available in the full photon number distribution can significantly improve the robustness of state tomography via the Wigner and characteristic functions.

Despite the simplicity of determining $W(\beta)$ for a single point in phase space, the cost of performing full Wigner function characterization using the above relation faces challenges if we wish to learn the entire Wigner function of even a single oscillator. The reason for this is that an infinite number of values of $\beta$ are needed to ensure that the Wigner function can be accurately estimated at any $\beta$.  This means that we will ultimately be forced to estimate the value of the Wigner function at a given point by using a scheme such as a Lagrange interpolation, wherein  the value of the Wigner function at an arbitrary point is estimated from values learned at points on a grid. Assuming the Wigner function is differentiable $n+1$ times, from interpolation theory (in essence Taylor's remainder theorem), the error in estimating $W(\beta)$ at a particular point $\beta$ from surrounding gridpoints with a polynomial function $P_n(\beta)$ with constant grid spacing $h$ is

\begin{equation}
    |W(\beta) - P_n(\beta)| \le \frac{\max_{y,\alpha} |D^{n+1}_{\alpha} W(y)| h^n}{(n+1)!},
\end{equation}
where $D_\alpha$ is the directional derivative operator in the direction $\alpha$.
This suggests that if we wish to estimate the Wigner function at an arbitrary point within error $\epsilon$ for an unknown Wigner function, that is differentiable at least $n+1$ times then it suffices to choose
\begin{equation}
    h \in O\left(\frac{n  \epsilon^{1/n} }{\max_{y,\alpha} | D_{\alpha}^{n+1} W(y)|^{1/n}}  \right).
\end{equation}
If we assume that the Wigner function is (within sufficiently small error) supported on a compact region $[-\beta_{\max},\beta_{\max}]^{\otimes 2}$, then since the Wigner function for a single oscillator is a two-dimensional function, the total number of grid points needed within this space scales as
\begin{equation}
    N_{1}=\left(\frac{\beta_{\max}}{h} \right)^2 \subseteq O\left(\left(\frac{\beta_{\max}\max_{y,\alpha} | D_{\alpha}^{n+1} W(y)|^{1/n}}{n  \epsilon^{1/n} } \right)^2\right)
\end{equation}

 Repeating the same analysis for a system with $p\ge 1$ oscillators (but taking $y,\beta,\alpha$ now to be vectors in a $2p-$dimensional space), we conclude that the number of points needed to extrapolate the Wigner distribution is

\begin{align}
    N_{p}&=\left(\frac{\beta_{\max}}{h} \right)^{2p} \nonumber\\
    &\subseteq O\left(\left(\frac{p\beta_{\max}\max_{y,\alpha} | D_{\alpha}^{n+1} W(y)|^{1/n}}{n  \epsilon^{1/n} } \right)^{2p}\right).
\end{align}
Choosing for simplicity $n=2p$, we see that the scaling is
\begin{align}
    N_p &= O\left(\frac{\max_{y,\alpha} | D_{\alpha}^{2p+1} W(y)|}{\epsilon}\left(\frac{\beta_{\max}}{2}\right)^{2p} \right).
\end{align}
We see from this that unless the derivatives of the Wigner function become very small as $p$ increases, the number of points that need to be accurately estimated to approximately learn the Wigner function on $p$ oscillators scales exponentially with $p$. This can be problematic even for physical states.  For example, if we consider a coherent state in one dimension of the form 
\begin{equation}
\ket{\psi}^{\otimes j} =\left(\int d q \ e^{-q^2/2\sigma^2} \ket{q} / \sqrt{2\pi}\sigma\right)^{\otimes j}
\end{equation}
then for small $\sigma$ we have that the norm of the derivatives of the state grows exponentially with the number of copies $j$.  This illustrates the often-observed fact with conventional tomography that the cost grows exponentially with the number of systems.

This exponential increase of the number of measurement settings with the number of qumodes in the system exactly matches what  we would expect from qubit tomography of a density matrix (with no promises about the rank of the density operator)  that  $4^p$ distinct sets of measurements of Pauli operators would be needed to reconstruct a $p$-qubit state~\cite{flammia2012quantum}.  This suggests that the situation is analogous, if more complex, to the qubit case as we have to approximately learn the density matrix because we do not have a tomographically complete set of measurements for our continuous scheme.  Thus while individual components of the Wigner function can be easily estimated, learning the overall Wigner function for a generic $p$-mode oscillator system will likely be forever out of reach owing to the previous argument about the scaling with the derivatives of the tensor products between coherent states.  However, for cases where prior knowledge is provided about the form of the Wigner function, more efficient reconstruction schemes can be found and the optimization of such reconstruction schemes remains an important open question.

        \subsection{The Husimi Q-Function}
\label{sec:husimi-q}

The Fourier transform of the anti-normal order characteristic function $C_{an}(\alpha)$ in Eq.~\eqref{eq:characteristic-function-can} is called the Husimi-$Q$ function, and is defined as
\begin{equation}
    Q(\beta)=\frac{1}{\pi^2}\int d^2\alpha \ C_{an}(\alpha)e^{\beta\alpha^*-\alpha \beta^*} \,.
\end{equation}
The Husimi-$Q$ function can be more convenient to use in certain cases since, unlike $W$, this quantity is always positive (but therefore is less-suited to study the non-classicality of quantum states). In terms of the density matrix, it is given by
\begin{equation}
    Q(\beta)=\frac{1}{\pi}\mathrm{Tr}\left[ \rule{0pt}{2.4ex}  \ket{0} \bra{0}{\mathcal D}^\dagger(-\beta)\  \rho\  {\mathcal D}(-\beta)\right ].
\end{equation}
Thus, it is the average of the projector onto the vacuum state of the state displaced by $\beta$.

Similar to the definition of the $Q$-function, the $P$-function is the Fourier transform of $C_n(\alpha)$ in Eq.~\eqref{eq:characteristic-function-cn}. It is unnormalizable and difficult to use, and hence will not be discussed in this document.

\subsection{Calibrating Controlled Displacements.}
\label{sec:benchmarking-oscillators}

As discussed in Sec.~\ref{sssec:benchmarking}, benchmarking hybrid quantum processors is not as easy as qubit systems. However, some simple calibrations on hybrid CV/DV gates can still be performed.
One example is that the gate sequence used in Eq.~\eqref{eq:calibration-displacement-sequences-1} in Sec.~\ref{sec:compilation} can be used to calibrate the controlled displacement gate.

\begin{figure}[t]
    \centering
    \includegraphics[width=0.45\textwidth]{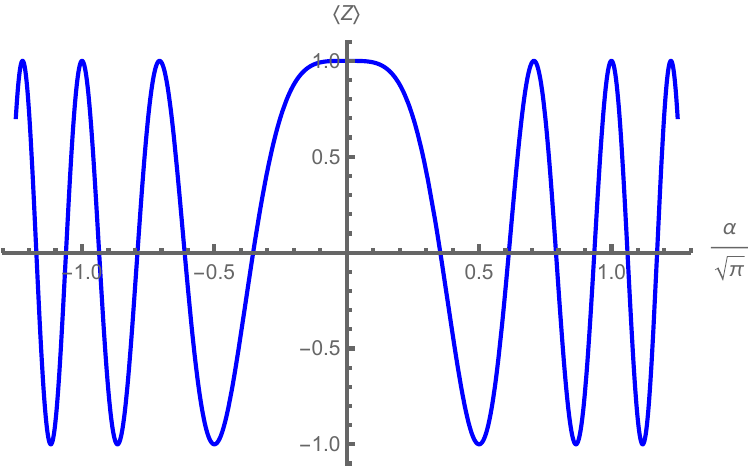}
    \caption{Average measurement results for the qubit measurement outcome using the calibration given in Eq.~\eqref{eq:Ucalibration}, demonstrating `Berry fringe' oscillations due to the phase kickback of the non-commuting displacements in phase space. The known form of these oscillations from Eq.~(\ref{eq:BerryFringes}) can be used to accurately calibrate the size of displacements in phase space.}
        \label{fig:BerryFringes}
\end{figure}

It works as follows. Consider conjugating Eq.~\eqref{eq:calibration-displacement-sequences-1} using Hadamard gates to construct the following $U_{\rm calibration}$
\begin{align}
    U_{\rm calibration} = H \left[ \rule{0pt}{2.4ex} D(-i\alpha)D_c(\alpha, -\alpha)D(i\alpha)D_c(-\alpha, \alpha) \right] H.
    \label{eq:Ucalibration}
\end{align}
As we have seen from Eq.~\eqref{eq:calibration-displacement-sequences-2} and Fig. \ref{fig:compilation_single_qubit_gate}, the $|0\rangle$ state of the qubit causes the displacements to drive the oscillator around the counterclockwise closed loop in phase space, which means that the $|0\rangle$ state acquires phase $\varphi_{|0\rangle}=-2|\alpha|^2$. Similarly, the $|1\rangle$ state of the qubit causes the displacements to drive the oscillator around the clockwise closed loop in phase space, which means that the $|1\rangle$ state acquires phase $\varphi_{|1\rangle}=+2|\alpha|^2$. The net effect is that the qubit undergoes a rotation around the $z$ axis through an angle $\varphi=\varphi_{|0\rangle}-\varphi_{|1\rangle}=-4|\alpha|^2$.  The conjugation by the Hadamard gates converts this to a rotation around the $x$ axis leading to the `Berry fringe' oscillations 
\begin{equation}
    \langle Z\rangle = \cos\left(4\alpha^2\right),
    \label{eq:BerryFringes}
\end{equation}
as is shown in Fig.~\ref{fig:BerryFringes}. Therefore, by accumulating enough shots on the qubit measurement, we can get an estimation of $\langle Z \rangle$ to infer the actual displacement $\alpha$ performed by the displacement gates.  Note that if we make the unconditional displacements in Eq.~\eqref{eq:Ucalibration} be constant $\pm i\beta$ independent of $\alpha$ then we obtain a simple sinusoidal oscillation
\begin{equation}
    \langle Z\rangle = \cos\left(4\alpha\beta\right).
    \label{eq:BerryFringes2}
\end{equation}

Ref.~\cite{valahu2024benchmarking} takes a different approach, considering a set of random displacements applied to an initial vacuum state.  This is followed by an inverting displacement that nominally returns the system to the vacuum state.  Measurement of the probability of being in the vacuum state (or more generally, measurement of the photon number distribution) provides information on noise and errors in the displacement operations.

\section{Comparison between Analytical and Numerical Compilation}\label{sec:circuit-compare}

In this Appendix, we compare the cost of using numerically optimized circuits described in Sec.~\ref{sec:numerical-optimization} against analytical compilations using approximate techniques discussed in Sec.~\ref{ssec:approximate-1-qubit-oscillator-unitary}. For this comparison, we employ two figures of merit,
\begin{itemize}
    \item circuit-depth or number of layers $N_\text{circuit}$used in the numerical optimization scheme in Sec.~\ref{sec:numerical-optimization}, and
    \item  circuit-duration or the total time taken by the circuit $T_\text{circuit}$ used in Sec.~\ref{sec:qec-compilation}.
    Note that, duration of the conditional displacements are lower bounded by $T_{|\gamma_i|<0.024}=48\text{ }$ns. This duration includes the necessary components for an echoed conditional displacement, an unconditional displacement $|\alpha_0|$ ($24$ ns) and a mid-circuit qubit rotation ($24$ ns). For details see App.~\ref{app:PhysImp} and Ref.~\cite{EickbuschECD}.
\end{itemize}
We use the preparation of Fock state $\ket{1}$ and squeezed vacuum as examples to study the efficiency of various approximate methods against numerical methods using phase-space ISA. Such comparisons can be used as a figure of merit to benchmark the analytical schemes.
\begin{figure}[b]
    \centering
    \includegraphics[width=0.5\textwidth]{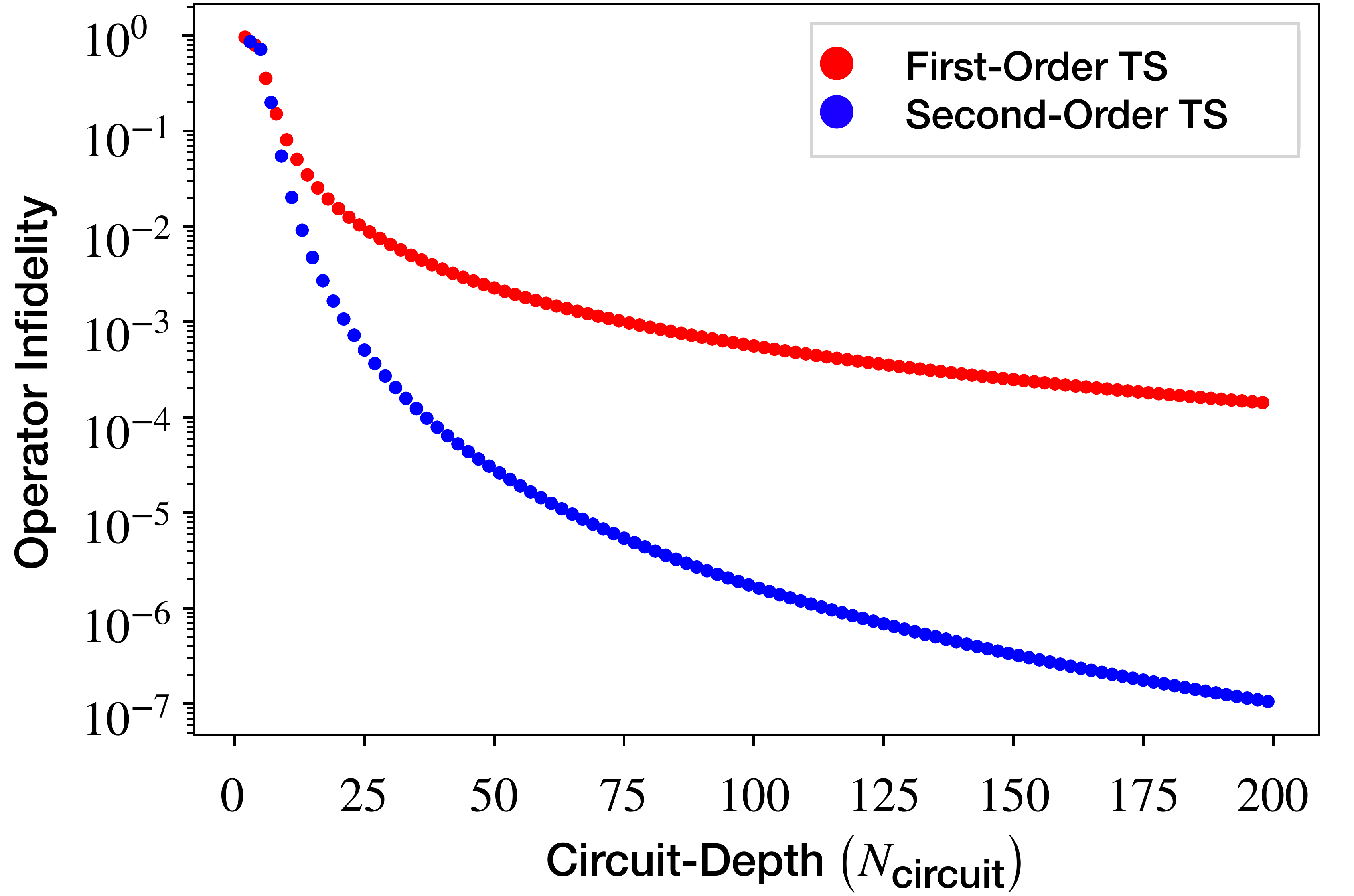}
    \caption{Fidelity of an approximate compilation of the Anti-Jaynes Cummings Interaction using the phase-space ISA. The $y$-axis gives the operator infidelity, defined using Hilbert-Schmidt product: $1-|\frac{1}{2d}\mathrm{Tr}(PU^\dagger V)|^2$, between the exact operator $U=\textrm{AJC}(\theta)$ and the approximate phase-space ISA sequence $V$. This operator fidelity is computed on the oscillator-qubit subspace with projector $P=\sum_{\ell=0}^{d-1} \ket{\ell}\bra{\ell}\otimes \sum_{q=0}^1 \ket{q}\bra{q}$. This is the projector on the joint subspace of a truncated oscillator with $d=15$ and a qubit. The $x$-axis corresponds to the circuit depth $N_\text{circuit}$ of the different Trotter-Suzuki (TS) sequences used. See App.~\ref{sec:circuit-compare} for the definition of $N_\text{circuit}$. (Red) First-order TS approximation $V=(e^{-i\frac{\theta}{r} \textrm{A}}e^{-i\frac{\theta}{r} \textrm B})^r$. Circuit depth in this case is $N_\text{circuit}=2r$. (Blue) Second-order TS (see Eq.~(\ref{trotter-suzuki-2SMG})) approximation $V=(e^{-i\frac{\theta}{2r} A}e^{-i\frac{\theta}{r} B}e^{-i\frac{\theta}{2r} A})^r$. Circuit depth in this case is $N_\text{circuit}=2r+1$. The AJC, and JC Hamiltonians can be used to prepare arbitrary superpositions of Fock states and hence can be employed for universal oscillator state preparation via the Law-Eberly protocol~\cite{law1996arbitrary}. The value of $\theta=\frac{\pi}{{\sqrt{2}}}$ chosen for this comparison is suitable for the preparation of Fock  state $\ket{1}$. The Hilbert space of the oscillator used to compute the operators $U,V$ is $N_\text{dim}=50\gg d$ dimensional and we have checked that the results are unaffected upon a further increase in $N_\text{dim}$.}
    \label{fig:AJC_sim}
\end{figure}

\begin{figure}
    \centering
    \includegraphics[width=0.45\textwidth]{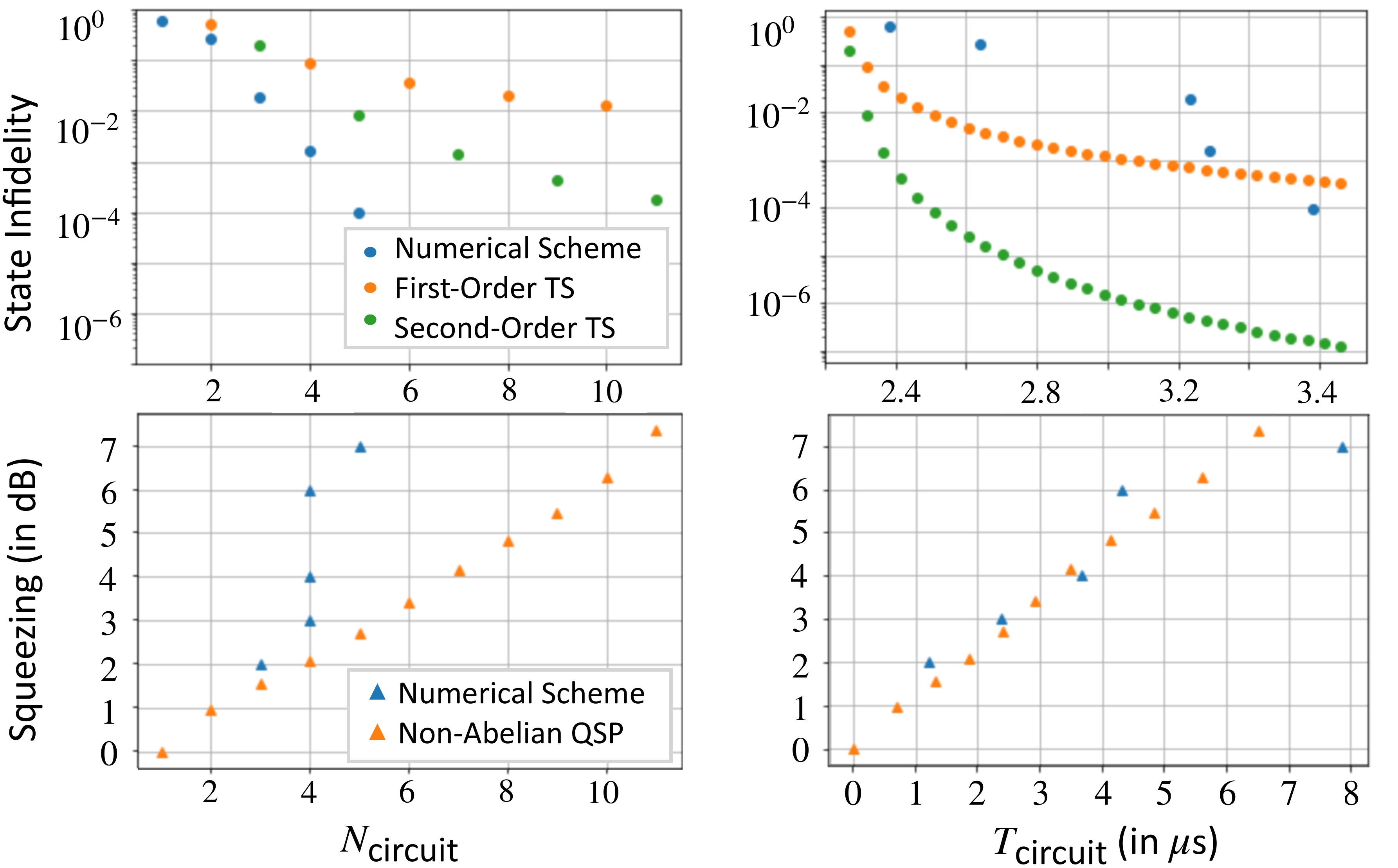}
    \caption{Comparison between numerically optimized and analytically derived circuits using techniques like Trotter-Suzuki (TS) and non-abelian QSP. (Top) Infidelity of Fock state $\ket{1}$ preparation from the Law-Eberly protocol~\cite{law1996arbitrary} using the JC Hamiltonian synthesized via TS methods, as discussed in Sec.~\ref{sec:trotter-product}. On the $x$-axis, we vary two metrics to improve these quantities. The left plot shows variation in the circuit depth or $N_\text{circuit}$, and the right plot shows varying circuit duration or $T_\textrm{circuit}$. (Bottom) Squeezing of vacuum achieved with fidelity $\mathcal{F}>0.99$ using non-abelian QSP as discussed in Ref.~\cite{singh_towards_2025}. As before, the $x$-axis shows varying $N_\text{circuit}$ in the left plot and varying $T_\text{circuit}$ in the 
 right plot.}
    \label{fig:circuit-compare}
\end{figure}

\subsubsubsection{Trotterization:}
We can prepare Fock states with access to the JC Hamiltonian. For the phase-space ISA, the
JC Hamiltonian can be engineered as discussed in Sec.~\ref{sec:trotter-product} via Trotterization. We show the operator fidelity to achieve this using first-order and second-order Trotter-Suzuki formulas (TS) in Fig.~\ref{fig:AJC_sim}. Once available, the JC Hamiltonian can be used to prepare arbitrary states from the vacuum state $\ket{0}_\textrm{vac}$ using the Law-Eberly protocol~\cite{law1996arbitrary}. The simplest of these tasks is to prepare a Fock $\ket{1}$ state. The first row of Fig.~\ref{fig:circuit-compare} shows a comparison between the Law-Eberly protocol using the JC Hamiltonian realized in this manner and numerical optimization. From the plots in the top row in Fig.~\ref{fig:circuit-compare}, we can see that the numerical scheme outperforms the Trotterization-based scheme in terms of circuit depth. On the other hand, the top right plot shows that the numerical scheme requires longer circuit duration for the same infidelity. We emphasize that the numerical circuits were optimized on circuit depth and not duration, which is why they do not converge to the Trotterized results. However, it would be an interesting future direction to develop a numerical scheme optimized on circuit duration, as this more accurately reflects the true cost of the circuit.

\subsubsubsection{QSP:} 
Ref.~\cite{singh_towards_2025} describes an analytical scheme to achieve the squeezed vacuum state using non-abelian QSP within the phase-space ISA. This scheme was optimized for the circuit duration ($T_\textrm{circuit}$) and performs better than the state-of-the-art partially numerical scheme in Ref.~\cite{Hastrup_Squeezed_State_QSP}. While we direct the readers to Ref.~\cite{singh_towards_2025} for the details of this preparation, here we compare it against the squeezed vacuum achieved numerically in Ref.~\cite{EickbuschECD} using the method employed to construct Fig.~\ref{fig:numerical_example}. In the bottom row of Fig.~\ref{fig:circuit-compare} we plot the circuit depth (left) and duration (right) required to achieve increased squeezing of the vacuum with fidelity $\mathcal{F}>0.99$. We can see that the squeezing achieved with this analytical scheme is on par with the numerical scheme in terms of circuit duration. The circuit depth variation is optimal for the numerical scheme because this is not the metric on which the QSP scheme was optimized. Notably, the protocol in Ref.~\cite{Hastrup_Squeezed_State_QSP} used circuit depth as a metric and produced states with worse fidelity compared to the QSP scheme. For example, the QSP-based Ref.~\cite{singh_towards_2025} achieves a squeezing of $8.5$ dB and an anti-squeezing of $-8.40$ dB with $T_\textrm{circuit}\approx 7.6$ with a fidelity of $0.9978$. With comparable $T_\textrm{circuit}$, partial numerical optimization in Ref.~\cite{Hastrup_Squeezed_State_QSP} based on circuit-depth yields a squeezing of $8.5$ dB and an anti-squeezing of $-9.9$ dB with a fidelity of $0.99$.

These comparisons indicate the need to optimize the numerical circuits not only using gate counts but also the lengths of conditional displacements. A simple way to resolve this issue would be to optimize the circuits using $T_\textrm{circuit}$ as a metric, thus taking into account the time cost of performing large conditional displacements $T_{|\gamma_i|}$ as well that of a large number of qubit rotations $T_R$. Such a metric would improve upon the numerical methods and would facilitate direct comparison to analytical schemes like QSP-based squeezing. Changing the cost function of the optimizer from circuit-depth to circuit duration may be more expensive in computational time. However, it would be interesting to use the various analytical schemes presented in this manuscript as starting templates for numerical optimization of circuit duration. This exercise would help us understand if these analytic methods are close to or at a local minimum.

\section{Brief Review of QSP, QSVT Formalism for Qubits}
\label{app:qsp-qsvt}
Quantum signal processing (QSP) combines non-commutative qubit operations to realize some desired polynomial transformation of an unknown signal encoded by a $2\times 2$ unitary matrix \cite{low2017optimal,GrandUnificationAlgos}.  The quantum singular value transformation (QSVT)\cite{Gilyen2019} generalizes QSP by enabling a polynomial transformation of the singular values of larger matrices.  For QSVT, this matrix need not be $2\times 2$ and the matrix need not even be square.  This is a powerful capability, since such a transformation ordinarily requires the ability to perform a singular value decomposition, which has a $O(2^{3n})$ cost for an $n$-qubit operator.  In contrast, QSVT realizes transformations of singular values {\em without} needing to perform the singular value decomposition itself.  The challenge for QSVT is that the matrix to be transformed must be ``block-encoded'' into a unitary operator.  For example, to realize quantum simulation of Hamiltonian $H$ necessitates block encoding of a positive semi-definite Hamiltonian inside a unitary matrix. The theory of QSP and QSVT comes from composite pulses in NMR, which we review below in App. \ref{app:composite-pulse}. A formal overview of qubit QSP is presented in App. \ref{app:qubit-qsp}.

\subsection{Intuition from Composite Pulses}
\label{app:composite-pulse}

A composite pulse is a sequence of qubit rotations which accomplishes the same thing as an ideal single qubit rotation, but behaves differently away from the ideal control point.  For example, composite pulses can be highly insensitive to pulse amplitude errors (e.g. to improve robustness to hardware imperfections).

Thus, composite pulses are frequently used in NMR \cite{WimperisRobust}, trapped-ion \cite{ChuangRobust,ChuangRobustErratum} and neutral atom \cite{gong2023robust} systems for optical or magnetic spin control to correct for classical errors in these single-qubit rotations associated with fluctuations in the frequency, phase and amplitude of the control pulses (e.g., due to laser noise in optical spin control).  

The mathematical framework for composite pulses is as follows.  Let us consider a single-qubit rotation of the form (see Table \ref{tab:gates-qubit})
\begin{equation}
R_{\varphi}(\theta)=e^{-i\frac{\theta}{2}[\cos(\varphi) X+\sin(\varphi) Y]}
\label{rot}
\end{equation} 
defined by the angle $\theta$ (determined by the drive pulse amplitude and duration) and the angle $\varphi$ (determined by the drive pulse phase relative to a master clock).
Suppose that we wish to implement the rotation $\theta=\pi/2$ and $\varphi=0$, but the control system is imperfect due to control pulse amplitude fluctuations that cause $\theta=(\pi/2)(1+\epsilon)$, where $\epsilon$ quantifies the over/under rotation error.  We may assume for simplicity that the phases $\varphi$ of the pulses are perfectly controlled, for this pedagogical example.

A composite pulse scheme known as BB1$_{90,x}$~\cite{WimperisRobust} corrects for the small errors $\epsilon$ in the classical parameter $\theta$ by appending to the nominal rotation $R_0(\theta)$ (rotation by angle $\theta$ about the $x$ axis) three subsequent extra rotations
\begin{equation}
    \mathrm{BB1}_{90,x}=R_{\varphi_1}(2\theta)R_{\varphi_2}(4\theta)R_{\varphi_1}(2\theta)R_{0}(\theta),
    \label{eq:bb1x}
\end{equation}
where $\varphi_1 = \cos^{-1}(-1/8)$ and $\varphi_2 = 3 \varphi_1$.  For the cat state preparation example we consider below, it is useful to have the analogous composite pulse for performing rotations about the $y$ axis
\begin{equation}
    \mathrm{BB1}_{90,y}=Z\left(+\frac{\pi}{2} \right)\mathrm{BB1}_{90,x}Z \left(-\frac{\pi}{2} \right)
    \,,
    \label{eq:bb1y}
\end{equation}
where $Z(\theta)=e^{-i\frac{\theta}{2} Z}$ is a rotation about the $z$ axis by angle $\phi$.

It is important to note that the error $\epsilon$ is presumed to be sufficiently slowly varying in time that it can be treated as a constant during any given short sequence of pulses.  Hence, each of the four pulses in this sequence suffers the same angle scaling factor $1+\epsilon$.  It is clear from inspection that if $\epsilon=0$, the four successively applied rotation angles are $\theta=\pi/2,2\theta=\pi$, $4\theta=2\pi$ and $2\theta=\pi$ and thus the three additional rotations combine to form the identity operation and the composite sequence of four pulses exactly yields the target unitary $R_0(\pi/2)$. Remarkably, by correctly choosing the azimuthal angles $\varphi_1,\varphi_2$, the errors in the three additional pulses combine to nearly perfectly cancel (for small $\epsilon)$ the error in the first pulse, giving an excellent approximation to the target unitary despite the presence of the error.

\begin{figure}[tb]
    \centering
    \includegraphics[width=0.45\textwidth]{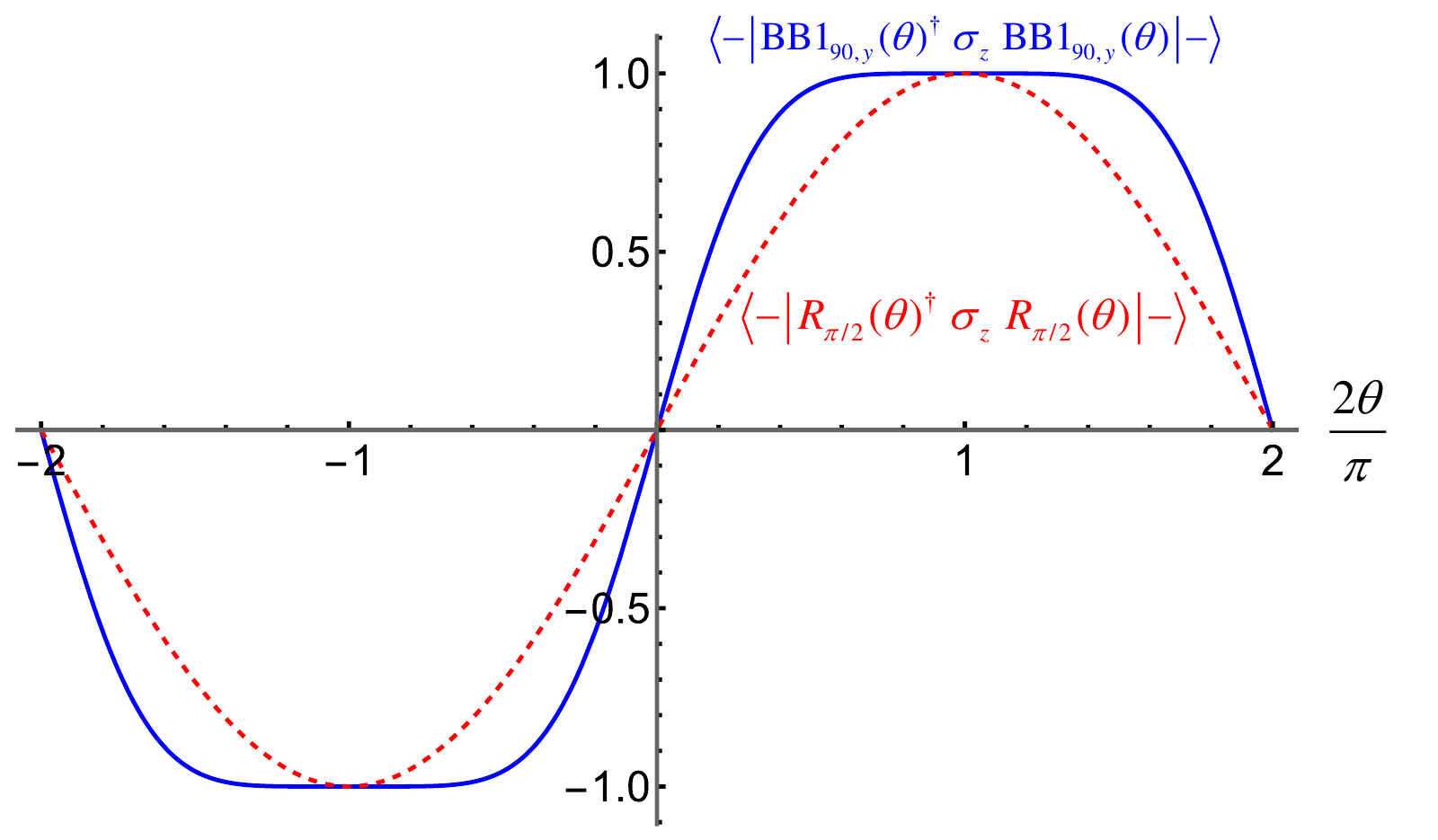}
    \caption{The BB1$_{90,y}$($\theta$) composite pulse sequence is robust against large over- and under-rotation errors in the angular parameter $\theta$.  An initial state $|-\rangle$ is accurately rotated to the state $|0\rangle$ as evidenced by this plot of the expectation value of $\sigma_z$ versus rotation angle $\theta$. The solid (blue) line is for the BB1$_{90,y}$($\theta$) composite pulse sequence and the dashed (red) line is for the ordinary single pulse rotating the spin by angle $\theta$ about the $y$ axis. Even relatively far from the nominal value of $\theta=\pi/2$, the qubit polarization in the $z$ direction on the Bloch sphere is close to unity for the BB1$_{90,y}$($\theta$) pulse sequence. 
    }
    \label{fig:BB1(90)FidelityFigure}
\end{figure}

This strong robustness against fluctuations in the rotation angle $\theta$ away from the nominal value $\theta=\pi/2$ is clearly illustrated in Fig.~\ref{fig:BB1(90)FidelityFigure}.  For small $\epsilon$, the polarization (and hence the gate fidelity decay away from unity as $\langle\sigma_z\rangle\approx 1-3.7\epsilon^6$, indicating that the rotation angle error is $\mathcal{O}(\epsilon^3)$.

For later more formal purposes, it is useful to note here that inspection of Eq.~(\ref{eq:bb1y}) shows that it can be rewritten in a `canonical form' as a sequence of $d=9$ identical rotations $W_x\equiv R_0(\theta)$ by angle $\theta$ around the $x$ axis, alternating with varying rotations $Z(\phi_j)$ around the $z$ axis
\begin{align}
    \mathrm{BB1}_{90,y}&=Z(\phi_0)\prod_{j=1}^d W_x(\theta)Z(\phi_j),
    \label{bb1a}
\end{align}
where we order the product of operators so that $j=1$ is the leftmost and $j=d$ is the rightmost, and the set of $d+1$ angles for the $z$ rotations is given by 
\begin{align}
     \vec v_\phi & \equiv  (\phi_0,\phi_1,\ldots,\phi_d)\nonumber\\
    & \!\!\!\!\! = \left(\varphi_1+\frac{\pi}{2},0,-\varphi_1+\varphi_2,0,0,0,-\varphi_2+\varphi_1,0,-\varphi_1,-\frac{\pi}{2} \right)
     \label{bb1b}
\end{align}
 where $\varphi_1,\varphi_2$ were defined below Eq.~(\ref{eq:bb1x}).
Furthermore, since the transformation is a product of qubit rotations it can be written as a single rotation having the generic form
\begin{align}
    \mathrm{BB1}_{90,y} = e^{-i\frac{\pi}{2}\vec h(\theta)\cdot\vec \sigma},
    \label{bb1c}
\end{align}
where 
\begin{align}
    \vec h(\theta)=\big(h_0(\theta),h_x(\theta),h_y(\theta),h_z(\theta)\big)
\end{align}
is a vector whose components are four functions of the rotation angle $\theta$ and $\vec\sigma=(\sigma_0,\sigma_x,\sigma_y,\sigma_z)$.  

As discussed in the introduction, in NMR composite pulse design, a common goal is to make these functions as constant as possible near the target value of $\theta$ so that the rotation matches the target and is robust against noise in the parameter $\theta$.  One could also imagine the opposite situation in a quantum sensing application where one is interested in detecting small changes in $\theta$ by making $\vec h(\theta)$ have a strong derivative.  This and other types of `filter' response functions are discussed in \cite{GrandUnificationAlgos,vandersypen2005nmr}.

\subsection{Qubit QSP}
\label{app:qubit-qsp}
Quantum signal processing (QSP) provides a way to generalize the intuition behind the ${\rm BB1}_{90,y}$ to transform arbitrary input rotations in SU(2).
The simplest case of QSP is the case where the matrix is an SU(2) matrix such as an ordinary single qubit rotation.  Specifically, assume that $x\in [-1,1]$ and let $U\in \mathbb{C}^{2\times 2}$ be
\begin{equation}
    U:= \begin{bmatrix} x & i\sqrt{1-x^2} \\ i\sqrt{1-x^2} &x \end{bmatrix} = e^{iX\arccos(x)}
\end{equation}
The aim of QSP is to find a transformation such that
\begin{equation}
    \Lambda: U \mapsto \begin{bmatrix} P(x) & i \sqrt{1-x^2} Q(x) \\ i\sqrt{1-x^2} Q^*(x) & P^*(x) \end{bmatrix}
\end{equation}
for polynomial functions $P(x)$ and $Q(x)$.  

Although not all polynomial transformations can be directly constructed using this approach,  the family of attainable transformations is broad.  Specifically, assume that we have a sequence of rotation angles $\{\phi_0,\ldots, \phi_k\}$ such that
\begin{equation} \label{eq:QSP_sequence}
    e^{i\phi_0 Z} \prod_{j=1}^k U(x) e^{i \phi_j Z} = \begin{bmatrix} P(x) & i \sqrt{1-x^2} Q(x) \\ i\sqrt{1-x^2} Q^*(x) & P^*(x) \end{bmatrix}.
\end{equation}
Such a set of angles can be found if
\begin{enumerate}
    \item ${\rm deg}(P(x)) \le k$ and ${\rm deg}(Q(x)) \le k-1$
    \item ${\rm Parity}(P(x)) = k\mod 2$ and ${\rm Parity}(Q(x)) = k-1\mod 2$
    \item $|P(x)|^2 +(1-x^2)|Q(x)|^2 = 1$
\end{enumerate}
There exist efficient classical algorithms to find these QSP angles \cite{ying2022stable,chao2020finding,WhaleyRobust,haah2019product,wang2022energy}. A generalization of QSP from a single-qubit to multi-qubit settings where the dynamics takes place in direct sums of two-level subspaces can be achieved by the qubitization technique. See Ref.~\cite{GrandUnificationAlgos} for a tutorial on more details. In the present work, we extend the QSP formulation to bosonic systems and exploit it for compilation on hybrid oscillator-qubit hardware.

In the main body of the paper, we discuss an alternative convention for QSP wherein for any $y$
\begin{equation}
U\rightarrow U_z(y) := \begin{bmatrix}
    e^{iy} & 0 \\ 0 &e^{-iy}
\end{bmatrix}
\end{equation}
We can use the previous results to see that the Hadamard transform of the original matrix can be written as
\begin{align} 
        U_{\bm \phi} &= H e^{i\phi_0 Z} \prod_{j=1}^k U(x) e^{i \phi_j Z}H\nonumber  \\ &= e^{i\phi_0 X} \prod_{j=1}^k U_z(\cos^{-1}(x)) e^{i \phi_j X} 
\label{eq:QSP_sequence2}
\end{align}
We then have using the Hadamard transform, for $z=\cos^{-1}(x)$, that
\begin{widetext}
\begin{equation}
U_{\bm \phi} = \begin{bmatrix} i{\rm Re}(Q(z))\sqrt{1-z^2} + {\rm Re}(P(z))&{\rm Im}(Q(z))\sqrt{1-z^2}+ i{\rm Im}(P(z))\\
-{\rm Im}(Q(z))\sqrt{1-z^2}+ i{\rm Im}(P(z))& -i{\rm Re}(Q(z))\sqrt{1-z^2} + {\rm Re}(P(z))
\end{bmatrix}.
\end{equation}
\end{widetext}
We can see from the constraints on $Q$ and $P$ that the norm of each column and row is 
\begin{equation}
|P(z)|^2 +(1-z^2)|Q(z)|^2 = 1
\end{equation}
and thus the resulting matrix is manifestly unitary.  In this form for the polynomials $Q(z)$ and $P(z)$, it is clear that in this formalism we can block-encode a polynomial of degree at most $k$ subject to the above constraints on $P$ and $Q$.

\section{Details on Quantum Fourier Transform Implementation}
\label{app:QFT}
In this section, we provide further details on the realization of the quantum Fourier transform on a CV-DV device, as presented in Sec.~\ref{sssec:qft}.

\subsection{Details on QFT Method 1: Single-Variable QSP}
\begin{figure*}[htbp]
    \begin{center}
    \includegraphics[width=0.95\linewidth]{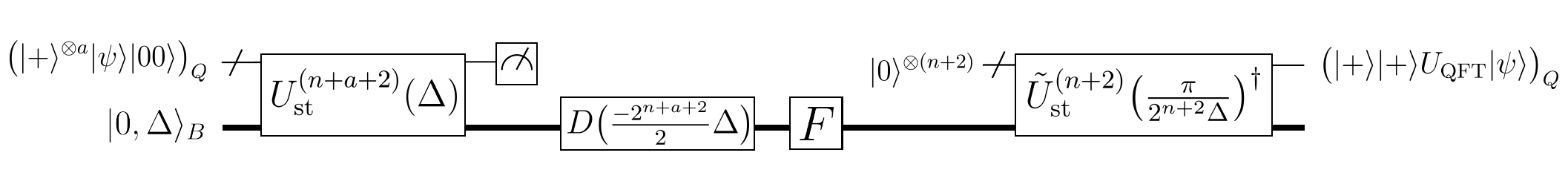}
    \end{center}
    \caption{The circuit used to implement the quantum Fourier transform of an $n$-qubit state $|\psi\rangle$ using the single-variable QSP state transfer unitary of Sec.~\ref{ssssection:single_var_QSP_state_transfer}, denoted $U_{\text{st}}$. Here, $D \big( \tfrac{-2^{n+a+2}}{2} \Delta \big)$ is a displacement operation, and $F$ is the Fourier transform operation $F = U(\pi/2)$ (i.e., oscillator free-evolution).} 
    \label{fig:QFT_QSP1_app}
\end{figure*}

For intuitive purposes, we reproduce the circuit that realizes the QFT here in Fig.~\ref{fig:QFT_QSP1_app}. Overall, the QFT is achieved in our construction by first transferring the qubit state to an equivalent oscillator state with basis spacing $\Delta$, and then applying a displacement operation to make the state symmetric about $q = 0$ in position space. Subsequently, the Fourier gate is applied to the oscillator to enact a continuous Fourier transform, and then the state is transferred back to qubits by using a \textit{modified} state transfer unitary $\tilde{U}_{\text{st}}^{(n)} (\frac{\pi}{ 2^n \Delta})$ with reciprocal spacing $\frac{\pi}{2^n \Delta}$. Importantly, this operation is modified such that the controlled displacements that comprise it (i.e., the first stage in Fig.~\ref{fig:QSP_state_transfer}) are each amplified by a factor of 2, which implements the correct phases corresponding to the QFT. Intuitively, it ensures that the product of the displacements used in each state transfer step equates to $(x \Delta) \cdot (y\cdot 2\cdot \frac{\pi}{2^n \Delta}) = 2\pi xy/2^n$, which is the correct phase of the QFT. Throughout this construction, ancilla qubits are appended to the initial state to make the underlying state periodic and increase fidelity with the exact QFT. 

We verify the performance of this QFT construction by considering its action on the input qubit state $|\psi \rangle_Q = \sum_{\mathbf{x}} c_{\mathbf{x}} |{\mathbf{x}}\rangle_Q $, where $| {\mathbf{x}}\rangle = |x_1 \rangle  |x_2 \rangle ... | x_{n} \rangle$. Let us go through this verification step-by-step. 

\textbf{Step 1:} Our first step is to prepend $|\psi\rangle_Q$ by $a$ ancilla qubits in the $|+\rangle_Q$ state, generating the state:
\begin{equation}\label{eq:QSPTransfer}
\begin{aligned}
    \big(|+\rangle^{\otimes a} |\psi\rangle \big)_Q  & =\frac{1}{\sqrt{2^a}}\sum_{k=0}^{2^a-1}\sum_{x=0}^{2^n-1} c_{\mathbf{x}} |2^n\cdot k + x \rangle_Q 
    \\
    & = \sum_{r=0}^{2^{n+a}-1} C_r | r \rangle_Q \,.
\end{aligned}
\end{equation}
The coefficients $C_r$ repeat over $2^a$ periods of size $2^n$: $C_r = C_{r+2^n \text{ (mod } 2^{n+a} )}$. This makes the initial state coefficients periodic from $r=0$ to $r=2^{n+a}-1$ and allows us to use the above continuous-discrete Fourier transform correspondence to implement the QFT. 

Additionally, we note that in the circuit of Fig.~\ref{fig:QFT_QSP1}, the initial state is also appended by a state $|00\rangle$. As we will see later, this is incorporated to improve the fidelity with the exact QFT. To simplify the current presentation, we exclude these qubits for now and reintroduce them at a later step when they are necessary.

\textbf{Step 2:} The next step is to use the single-variable QSP state transfer protocol of Sec.~\ref{ssssection:single_var_QSP_state_transfer} to transfer this state to the oscillator, encoded in a Gaussian basis of width $\sigma$. Applying the state transfer unitary $U^{(n+a)}_{\rm st}(\Delta)$ produces the state $|\mathbf{0}\rangle_Q \sum_{r} C_{r} |r, \Delta \rangle_B$ and suffers infidelity $\mathcal{O}((n+a)\epsilon) + e^{-\mathcal{O}(\Delta^2/\sigma^2)} $, where $\epsilon$ is the error in the QSP polynomial approximation to the square wave function.

\textbf{Step 3:} At this stage, the initial state has been mapped to an oscillator state, encoded in a basis of states peaked around $q=r \Delta$, from $q=0 $ to $q=(2^{n+a}-1)\cdot \Delta$. Next, we symmetrize these basis states about $q=0$ by applying a displacement $D(\tfrac{-2^{n+a} }{2} \Delta )$, producing the state
\begin{equation}\label{eq:state_step3_qft1}
\begin{aligned}
    &\sum_{r=0}^{2^{n+a}-1} C_{r} |r-\tfrac{2^{a+n}}{2}, \Delta \rangle_B = \sum_{k=-2^a/2}^{2^a/2-1} \sum_{x=0}^{2^n-1} c_{\mathbf{x}} |x + 2^n k, \Delta \rangle_B,
\end{aligned}
\end{equation}
where we have reintroduced the indices $x$ and $k$, and now symmetrized the sum over $k$.

\textbf{Step 4:} The next step is to apply the Fourier gate (i.e. free evolution, provided through the Bosonic ISA in Box~\ref{Box:phase-space-rotation}) to this state, which enacts the continuous Fourier transform on the wave function. As the basis states in Eq.~\eqref{eq:state_step3_qft1} are Gaussians, this produces the state
\begin{equation}\label{eq:step4_state_0}
\begin{aligned}
    &\sqrt{\sigma \sqrt{\tfrac{2}{\pi}}} \int dq  \sum_{\mathbf{x}} c_{\mathbf{x}}  e^{iq \Delta x} \\
    & \qquad \qquad \qquad \qquad \times \Big[ \frac{1}{\sqrt{2^a}} \sum_{k=-2^{a-1}}^{2^{a-1}-1} e^{iq 2^n k \Delta } \Big] e^{-\sigma^2 q^2} |q\rangle_B.
\end{aligned}
\end{equation} 

The factor $\sum_{k=-2^a/2}^{2^a/2-1} e^{iq\Delta 2^n k}$ in brackets is known as the Dirichlet kernel. In the limit of large $a$, it behaves as the sum of delta functions $\sum_{l \in \mathbb{Z}} \delta(\frac{q\Delta 2^n}{2\pi} - l)$, i.e., a Dirac comb. At finite $a$, its behavior under an integral deviates from that of the Dirac comb by an error $\mathcal{O}(1/2^a)$. Taking the large $a$ limit in Eq.~\eqref{eq:step4_state_0}, this factor selects out values $q_l :=l\frac{2\pi}{\Delta 2^n}$ from the integral, such that we can re-express the state as 
\begin{equation}\label{eq:step4_state_approx}
\begin{aligned}
    & \sqrt{ \frac{2\pi \sigma }{\Delta 2^n} \sqrt{\frac{2}{\pi}}} \sum_l   \sum_{\mathbf{x}} c_{\mathbf{x}}  e^{i 2\pi l x /2^n} e^{-\sigma^2 q_l^2} |\tilde{q}=q_l \rangle_B,
\end{aligned}
\end{equation}
where $|\tilde{q}=q_l \rangle_B$ is a normalized state concentrated around $q=q_l$ (see Ref.~\cite{qftwithoscillator} for exact expression), and the infidelity in reaching Eq.~\eqref{eq:step4_state_approx} is $\mathcal{O}(1/2^a)$. Also observe that taking this limit has produced a phase $e^{i2\pi lx/2^n}$ resembling of the QFT, which arises from the continuous-discrete Fourier transform correspondence.

\textbf{Step 5:}
Lastly, we introduce $n$ qubits $|\mathbf{0}\rangle_Q^{\otimes n}$ and transfer the oscillator state back to the qubits. For this, we use the inverse state transfer operation with an appropriately chosen reciprocal spacing $\tilde{U}_{\text{st}}^{(n)}(\frac{\pi}{2^{n} \Delta})^\dag$. Note that as we mentioned earlier, this unitary is modified such that the controlled displacements that comprise it are each amplified by a factor of $2$. 

The application of this operation is rather involved, and its analysis is fully presented in Ref.~\cite{qftwithoscillator}. The final result is that the output is approximately a product state $|\alpha\rangle_Q |\beta\rangle_B$, with an infidelity $\mathcal{O}(\sigma/ \Delta)$. Here the state of the qubits is
\begin{equation}
\begin{aligned}
    & |\alpha\rangle_Q = \sum_{\mathbf{x}}  c_{\mathbf{x}} e^{-i\pi x/2} U_{\text{QFT}}| \mathbf{x} \rangle_Q ,
\end{aligned}
\end{equation}
which is nearly the QFT, and the oscillator state is approximately a Gaussian state but will not concern us here.

The state of the qubits is the QFT of the initial state $|\psi\rangle_Q$ up to a phase factor $e^{-i \pi x/2 }$. This phase factor arises because the action of free evolution on a state symmetrized about position $q=0$ (as we have here) implements a symmetric QFT, with coefficients given by a symmetrized sum: $\sum_{y=-2^n/2}^{2^n/2-1} c_y e^{2\pi i x y / 2^n}$. This can be mapped to the usual QFT by shifting $x \mapsto x+2^n/2$, as the expense of incurring a phase factor $e^{-i \pi x/2 }$.

Fortunately, this phase factor is inconsequential and can be removed by appending $|00\rangle$ to the initial state as $ (|\psi\rangle |00\rangle)_Q$. This guarantees that $x \ \text{mod} \  4 = 0$ for any nonzero $c_{\mathbf{x}}$, such that each phase factor is always $e^{-i\pi x/2} = 1$. Then, performing the inverse state transfer step onto $n+2$ qubits yields the state \begin{equation}
    U_{\text{QFT}}(|\psi\rangle|00\rangle) = |+\rangle |+\rangle U_{\text{QFT}}(|\psi\rangle ),
\end{equation}
from which $U_{\text{QFT}}(|\psi\rangle )$ may be extracted. This addresses the additional ancilla qubits $| 00\rangle$ depicted in Fig.~\ref{fig:QFT_QSP1}.

\textbf{Fidelity Analysis:} By aggregating together the infidelities suffered at each step above, we find that the total fidelity of this entire protocol is
\begin{equation}
    F = 1 - \mathcal{O}((n + a)\epsilon) - \mathcal{O}(\sigma/\Delta) - \mathcal{O}(1/2^a)
\end{equation}
where $\epsilon$ is the error in the QSP polynomial approximation to a square wave function.

\subsection{Details on QFT Method 2: Non-Abelian QSP}

\begin{figure*}[htbp]
    \begin{center}
    \includegraphics[width=0.90\linewidth]{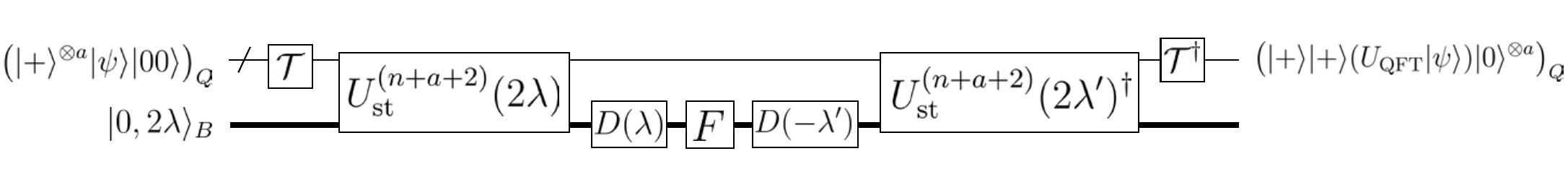}
    \end{center}
    \caption{The circuit used to implement the quantum Fourier transform of an $n$-qubit state $|\psi\rangle$ using the non-Abelian QSP state transfer unitary of Sec.~\ref{ssssection:bi_var_QSP_state_transfer}, denoted $U_{\text{st}}$. Here, $\mathcal{T}$ is a basis transformation from the computational basis to the basis $|\phi_s\rangle$, which is easily realized with simple qubit gates as discussed in the text. The operation $D ( \cdot )$ is a displacement operation, and $F$ is the Fourier gate (i.e., oscillator free-evolution). } 
    \label{fig:QFT_QSP2_app}
\end{figure*}

We again reproduce the figure that realizes the QFT in Fig.~\ref{fig:QFT_QSP2_app}. Our approach is to first perform a change of basis from the computational basis to the basis $\{ | \phi_{\mathbf{s}} \rangle \}$ used in the non-Abelian QSP state transfer protocol; this is achieved using an operation denoted $\mathcal{T}$ in Fig.~\ref{fig:QFT_QSP2}, which we show can be shown to be implemented with simple local operations. Subsequently, we transfer the state of the qubits to the oscillator, symmetrize the state about $q=0$ in position space, apply the Fourier gate, and finally transfer the oscillator state back to the qubits with an appropriately chosen reciprocal spacing. Upon transforming back to the computational basis, we obtain the QFT of the initial state. We again incorporate ancilla qubits throughout this protocol to improve the fidelity with the exact QFT. 

Let's again study this algorithm on an initial $n$-qubit state $|\psi\rangle_Q = \sum_{\mathbf{x}} c_{\mathbf{x}} |\mathbf{x} \rangle_Q$.

\textbf{Step 1:} As in the previous QFT construction, we first prepend $|\psi\rangle$ by $a$ ancilla qubits in the $|+\rangle$ state to make the coefficients periodic: $\big( |+\rangle^{\otimes a} |\psi\rangle\big)_Q = \sum_{r=0}^{2^{n+a}-1} C_r | r \rangle_Q$, where as before, the coefficients repeat over $2^a$ periods of size $2^n$: $C_r = C_{r+2^n \text{ (mod } 2^{n+a} )}$.

\textbf{Step 2:} The next step is to map the previous state from the computational basis to the basis $\{ |\phi_{\mathbf{s}} \rangle \} $ used in the non-Abelian state transfer protocol. 

To specify an appropriate correspondence between basis states, recall that in this context the string $\mathbf{s} \in \{+1, -1 \}^{n+a}$ is associated with a position $q_{\mathbf{s}}$ that takes discrete values in the range $[-\lambda(2^{n+a}-1), \lambda(2^{n+a}-1) ]$ with spacing $2\lambda$ (see Eq.~(\ref{eq:q_s})). It is therefore suitable to index each such possible value by an integer $s$, defined as
\begin{equation}
\begin{aligned}
    s :&= \tfrac{1}{2\lambda} \big( q_\mathbf{s} + \lambda(2^{n+a}-1)\big) \in \{0, 1, ..., 2^{n+a}-1\}.
\end{aligned}
\end{equation}

Thus, we assume access to a unitary change of basis operation $\mathcal{T}$ that acts as $\mathcal{T}|s\rangle = | \phi_{\mathbf{s}} \rangle$, such that its application to the state from Step 1 generates the state $\sum_{s=0}^{2^{n+a}-1} C_s | \phi_{{s}} \rangle$, where we are now indexing by the associated integer $s$ to simplify analysis. As shown in Ref.~\cite{qftwithoscillator}, $\mathcal{T}$ can be easily implemented as a product of local operations:
\begin{equation}
    \mathcal{T} = (HX)^{\otimes (n+a-1)} \cdot H_{n+a} \cdot Z_1 Z_{n+a} \cdot \mathcal{P}_{\leftrightarrow}, 
\end{equation}
where $\mathcal{P}_{\leftrightarrow}$ is a permutation operation that reverses the order of qubits (e.g., $\mathcal{P}_{\leftrightarrow} |0011\rangle = |1100\rangle$)

\textbf{Step 3:} The next step uses the non-Abelian state transfer protocol to transfer the state to an oscillator as per Eq.~(\ref{eq:Hastrup_inverse_state_transfer}). To achieve this, we prepare the oscillator in the sinc state $|0, 2\lambda\rangle_B = \tfrac{1}{\sqrt{2\lambda}} \int dq \ \text{sinc}\big( \tfrac{\pi q}{2\lambda } \big) |q\rangle_B$, and apply $U_{\text{st}}^{(n+a)}(2\lambda)$ to the joint system, producing 
\begin{equation}
\begin{aligned}
    |\mathbf{0} \rangle_Q \otimes \sum_{s=0}^{2^{n+a}-1} C_{s} \frac{1}{\sqrt{2\lambda}} \int dq \ \text{sinc} \big( \tfrac{\pi (q-q_{\mathbf{s}})}{2\lambda} \big) |q\rangle_B . 
\end{aligned}
\end{equation}
The infidelity suffered here is quantified by Eq.~(\ref{eq:non-abelian-fidelity}).

\textbf{Step 4:} Next, we symmetrize this state about $q=0$ by applying a displacement operation $D(\lambda)$. Using the identities $s = 2^n\cdot k +x $ and $q_s = 2\lambda (s-\tfrac{1}{2}2^{n+a}-\tfrac{1}{2}) = 2\lambda (2^n \cdot [k-\tfrac{1}{2}2^{a}]+x -\tfrac{1}{2})$, this yields the wave function 
\begin{equation}
\begin{aligned}
    & \sum_{x=0}^{2^n-1} \sum_{k=-2^a/2}^{2^a/2-1} \frac{1}{\sqrt{2^a}} c_{\mathbf{x}} \frac{1}{\sqrt{2\lambda}} \text{sinc}\big( \tfrac{\pi(q - 2\lambda (2^n\cdot k + x))}{2\lambda } \big).
\end{aligned}
\end{equation}

\textbf{Step 5:} Subsequently, we apply the Fourier gate to the oscillator, which enacts a continuous Fourier transform on the wave function. Noting the Fourier transform relation $\frac{1}{\sqrt{2\pi}} \int dq' e^{iqq'} \frac{1}{\sqrt{2\lambda }} \text{sinc}\big( \tfrac{\pi (q'-b)}{2\lambda } \big) = \sqrt{\frac{\lambda}{\pi}} 1_{|q| \leq \pi/2\lambda} e^{iq b}$, where $1_{|q| \leq \pi/2\lambda}$ is an indicator function (i.e. a box function in this context), this transforms the above wave function to
\begin{equation}
\begin{aligned}
    &\sqrt{\tfrac{\lambda }{\pi} } 1_{|q| \leq \pi/2\lambda} \sum_{x} c_j e^{iq 2\lambda x } \Big[ \tfrac{1}{\sqrt{2^a}} \sum_{k} e^{iq 2\lambda 2^n k} \Big] =: \tilde{\psi}(q).
\end{aligned}
\end{equation}

\textbf{Step 6:} After free evolution, we perform the inverse of Steps 4 and 5, but now with reciprocal spacing $2\lambda' = \tfrac{2\pi}{2^{n+a}} \tfrac{1}{2\lambda} $ chosen to produce the QFT. We first enact the inverse of step 5, which consists of applying the displacement $D(-\lambda')$ to map the wave function to $\tilde{\psi}(q+\lambda')$.

\textbf{Step 7:} Next, we transfer the oscillator state back to $n+a$ qubits by using $U_{\text{st}}^{(n+a)} (2\lambda')^\dag$. As per Eq.~(\ref{eq:HastrupTransferedState}), this maps the above state to
\begin{equation}\label{eq:psi_6}
\begin{aligned}
    &\sum_{s=0}^{2^{n+a}-1} \int dq \ \tilde{\psi}(q+q_s' + \lambda) \text{sinc}(\tfrac{\pi q}{2 \lambda }) |\phi_{\mathbf{s}}\rangle_Q |q\rangle_B  = \\
    &\sum_{s=0}^{2^{n+a}-1} \sum_{x=0}^{2^n-1} \sqrt{\frac{\lambda }{\pi} } \int_{-\pi/2\lambda}^{\pi/2\lambda} dq \ c_{\mathbf{x}} e^{i(q+q_s' + \lambda)2\lambda x} \\ 
    & \qquad \quad \times \Big[ \frac{1}{\sqrt{2^a}} \sum_k e^{i(q+q_s' + \lambda) 2\lambda 2^n k} \Big] \text{sinc}(\tfrac{\pi q}{2 \lambda' }) |\phi_{\mathbf{s}}\rangle_Q  |q\rangle_B , 
\end{aligned}
\end{equation}
where $q_s' = 2\lambda's-\lambda'(2^{n+a}-1)$. Note that we have not used any approximations to simplify this expression, as was done in Sec.~\ref{ssssection:bi_var_QSP_state_transfer}.

As in the previous QFT protocol, the quantity in brackets is the Dirichlet kernel, which behaves as $\sum_{l \in \mathbb{Z}} \delta(\frac{(q+q'_s+\lambda')\Delta 2^n}{2\pi} - l)$, with an overall error $\mathcal{O}(1/2^a)$ at finite $a$. One can take this limit to drastically simplify this state; the full technical analysis of this step is presented in Ref.~\cite{qftwithoscillator}. The final result is that the state of the system can be simplified to the product state 
\begin{equation}\label{eq:final_state}
\begin{aligned}
    &\Bigg[\sum_{l=0}^{2^n-1} \Big( \frac{1}{\sqrt{2^n}} \sum_{x=0}^{2^n-1} c_j e^{i 2\pi l x / 2^n} e^{-i\pi x/2} \Big) |\phi_{s=2^a l}\rangle_Q \Bigg] |\tilde{q}=0\rangle_B .
\end{aligned}
\end{equation}
where $|\tilde{q}=0\rangle_B$ is a state concentrated around $q=0$, and the infidelity suffered in reaching this expression is $\mathcal{O}(1/2^a)$. The oscillator and qubits are now decoupled, and the state of the qubits is nearly the quantum Fourier transform evaluated in the basis $|\phi_{s=2^a l} \rangle$. Again, the presence of the phase $e^{i 2\pi l x / 2^n}$
arises from the continuous-discrete correspondence of the Fourier transform discussed earlier.

\textbf{Step 8:} To extract the quantum Fourier transform from this state, the last step is to transform from the $\{ | \phi_s \rangle \}$ basis back to the computational basis using $\mathcal{T}^\dag$. This acts as $\mathcal{T}^\dag |\phi_{s=2^a l} \rangle = |2^a l\rangle = |l\rangle |0\rangle^{\otimes a} $, such that its application to the previous state produces
\begin{equation}
\begin{aligned}
    \left[\sum_{x=0}^{2^n-1} e^{-i\pi x /2 } c_{\mathbf{x}} U_{\text{QFT}} \big( |x \rangle \big) |0\rangle^{\otimes a} \right]_Q |\tilde{q}=0\rangle_B
\end{aligned}
\end{equation}
The remaining state on the first $n$ qubits is nearly our target QFT, up to a phase factor $e^{-i\pi x/2}$.

As before, this phase factor arises because this protocol really implements a symmetric QFT, which maps to the usual QFT at the expense of a factor $e^{-i\pi x/2}$. Again, this phase factor is inconsequential and can be removed by appending $|00\rangle$ to the initial state, such that $e^{-i\pi x/2}=1$ for all $x$ with nonzero coefficients $c_{\mathbf{x}}$. This then outputs the target QFT as $U_{\text{QFT}}(|\psi\rangle|00\rangle) = |+\rangle |+\rangle U_{\text{QFT}}(|\psi\rangle )$.

\textbf{Fidelity Analysis:} Consider the fidelity of this protocol. The first source of infidelity occurs in Step 3 due to the usage of the non-Abelian QSP state transfer protocol. The infidelity of this operation is quantified in Eq.~(\ref{eq:non-abelian-fidelity}); a careful analysis of this scenario, presented in Ref.~\cite{qftwithoscillator}, indicates that the resulting infidelity scales as $\mathcal{O}(1/2^a)$. Moreover, the other source of infidelity arises in Step 6 when taking the large $a$ limit, which also contributes an infidelity $\mathcal{O}(1/2^a)$. Lastly, we note that no infidelity is suffered in the inverse state transfer step, as we do not use any approximation to simplify this step. Therefore, summing these contributions to the infidelity, we find the overall fidelity of this QFT protocol is $F = 1- \mathcal{O}(1/2^a)$.

\pagestyle{refstyle}

\bibliography{BibliographycQEDBosons_small_cleaned}
\end{document}